\documentclass[12pt,letterpaper]{report}  

\usepackage{titlesec}
\titleformat{\chapter}{\normalfont\large}{Chapter \thechapter:}{1em}{}
\usepackage{graphicx}
\usepackage{cite}
\usepackage{lscape}
\usepackage{indentfirst}
\usepackage{latexsym}
\usepackage{multirow}
\usepackage{tabls}
\usepackage{wrapfig}
\usepackage{longtable}
\usepackage{supertabular}
\usepackage[onehalfspacing]{setspace}
\usepackage[table]{xcolor}
\usepackage[colorlinks=true,bookmarks=false]{hyperref}
\usepackage{amsmath}
\usepackage{cleveref}
\usepackage{amssymb}
\usepackage{graphicx}
\usepackage{placeins}
\usepackage{caption}
\usepackage{subcaption}
\usepackage{bm}
\usepackage{printlen}
\usepackage{amsthm}
\usepackage{commath}
\usepackage{setspace}
\usepackage{mathtools}
\usepackage{cancel}

\usepackage{sectsty}

\chapternumberfont{\LARGE} 
\chaptertitlefont{\LARGE}

\usepackage[square,numbers]{natbib}
\newcommand{\partder}[2]{\frac{\partial #1}{\partial #2}}
\newcommand{\der}[2]{\frac{d #1}{d #2}}

\usepackage{cleveref}
\crefname{equation}{}{}
\crefname{figure}{figure}{figures}

\usepackage[left=1in, right=1in, top=1in]{geometry}
\usepackage{enumerate}

\begin{document}

\pagestyle{plain}
\cleardoublepage
\singlespacing

\hbox{\ }

\renewcommand{\baselinestretch}{1}
\small \normalsize

\begin{center}
\large{{ABSTRACT}} 

\vspace{3em} 

\end{center}
\hspace{-.15in}
\begin{tabular}{ll}
Title of dissertation:    & {\large  ADJOINT METHODS FOR}\\
&                      {\large  STELLARATOR SHAPE OPTIMIZATION} \\
& {\large AND SENSITIVITY ANALYSIS}
 \\
 \\
&  {\large Elizabeth Joy Paul}\\
&  {\large Doctor of Philosophy, 2020} \\
\ \\ 

Dissertation directed by: & {\large  Professor William Dorland} \\
& {\large     Department of Physics } \\
\end{tabular}

\vspace{3em}

\renewcommand{\baselinestretch}{2}
\large \normalsize
\singlespacing

Stellarators are a class of device for the magnetic confinement of plasmas without toroidal symmetry. As the confining magnetic field is produced by clever shaping of external electro-magnetic coils rather than through internal plasma currents, stellarators enjoy enhanced stability properties over their two-dimensional counterpart, the tokamak. However, the design of a stellarator with acceptable confinement properties requires numerical optimization of the magnetic field in the non-convex, high-dimensional spaces describing their geometry. Another major challenge facing the stellarator program is the sensitive dependence of confinement properties on electro-magnetic coil shapes, necessitating the construction of the coils under tight tolerances. In this Thesis, we address these challenges with the application of adjoint methods and shape sensitivity analysis. 

Adjoint methods enable the efficient computation of the gradient of a function that depends on the solution to a system of equations, such as linear or nonlinear PDEs. Rather than perform a finite-difference step with respect to each parameter, one additional adjoint PDE is solved to compute the derivative with respect to any parameter. This enables gradient-based optimization in high-dimensional spaces and efficient sensitivity analysis. We present the first applications of adjoint methods for stellarator shape optimization. 

The first example we discuss is the optimization of coil shapes based on the generalization of a continuous current potential model. We optimize the geometry of the coil-winding surface using an adjoint-based method, producing coil shapes that can be more easily constructed. Understanding the sensitivity of coil metrics to perturbations of the winding surface allows us to gain intuition about features of configurations that enable simpler coils. We next consider solutions of the drift-kinetic equation, a kinetic model for collisional transport in curved magnetic fields. An adjoint drift-kinetic equation is derived based on the self-adjointness property of the Fokker-Planck collision operator. This adjoint method allows us to understand the sensitivity of neoclassical quantities, such as the radial collisional transport and self-driven plasma current, to perturbations of the magnetic field strength. Finally, we consider functions that depend on solutions of the magneto-hydrodynamic (MHD) equilibrium equations. We generalize the well-known self-adjointness property of the MHD force operator to include perturbations of the rotational transform and the currents outside the confinement region. This self-adjointness property is applied to develop an adjoint method for computing the derivatives of such functions with respect to perturbations of coil shapes or the plasma boundary. We present a method of solution for the adjoint equations based on a variational principle used in MHD stability analysis. 

\thispagestyle{empty}
\hbox{\ }
\vspace{1in}
\renewcommand{\baselinestretch}{1}
\small\normalsize
\begin{center}

\large{{ADJOINT METHODS FOR STELLARATOR SHAPE OPTIMIZATION AND SENSITIVITY ANALYSIS}}\\
\ \\
\ \\
\large{by} \\
\ \\
\large{Elizabeth Joy Paul}
\ \\
\ \\
\ \\
\ \\
\normalsize
Dissertation submitted to the Faculty of the Graduate School of the \\
University of Maryland, College Park in partial fulfillment \\
of the requirements for the degree of \\
Doctor of Philosophy \\
2020
\end{center}

\vspace{7.5em}

\noindent Advisory Committee: \\
Professor William Dorland, Chair/Advisor \\
Dr. Matthew Landreman, Co-Advisor \\
Professor Thomas M. Antonsen, Jr. \\
Professor Adil Hassam \\
Professor Ricardo Nochetto
 

\thispagestyle{empty}
\hbox{\ }

\vfill
\renewcommand{\baselinestretch}{1}
\small\normalsize

\vspace{-.65in}

\begin{center}
\large{\copyright \hbox{ }Copyright by\\
Elizabeth Joy Paul 
\\
2020}
\end{center}

\vfill
 
\pagestyle{plain} \pagenumbering{roman} \setcounter{page}{2}

\renewcommand{\baselinestretch}{2}
\small\normalsize
\hbox{\ }
 
\vspace{-.65in}

\begin{center}
\large{Preface} 
\end{center} 

\singlespacing

In an effort to promote open science, all data and the associated post-processing scripts used to produce the figures in this Thesis have been preserved in a \\
\href{https://zenodo.org/record/3745635\#.Xo5x1ZNKjxg}{Zenodo archive} with citeable DOI
10.5281/zenodo.3745635.


\renewcommand{\baselinestretch}{2}
\small\normalsize
\hbox{\ }
 
\vspace{-.65in}

\begin{center}
\large{Acknowledgments} 
\end{center} 

\vspace{1ex}

\singlespacing

I owe many thanks to the individuals who have made my graduate career fruitful and enjoyable. Most importantly, I would like to thank my advisors, Bill Dorland and Matt Landreman, who guided me toward interesting and important physics problems and made the completion of this Thesis possible. Bill, your positive outlook on life and constant curiosity are an inspiration to me. I walk away from every interaction with you with a smile on my face and a new interesting idea in my head. Matt, thank you for your generosity and meticulous attention to detail. From deriving the drift-kinetic equation on the board to providing detailed comments on every manuscript, I could never thank you enough for your investment in my graduate career. As an incoming graduate student I took a bit of a leap of faith when I decided to come to Maryland, and I could not have asked for a better pair of (award-winning!) advisors. Thank you for believing in me and supporting my career at every step of the way.

Many thanks goes to the other members of the dissertation committee. To Tom Antonsen, for giving me the opportunity to teach plasma physics and contributing to our games of ``dungeons and plasmas" with your top-secret notes. I feel honored to be able to work with a great mind such as yours. I hope we can continue to collaborate and spread the good news about ALPO. To Adil Hassam, for never ceasing to ask thought-provoking questions during group meeting. Your math methods course laid the perfect foundation for plasma physics research. To Ricardo Nochetto, for introducing our group to the methods of shape optimization. I appreciate the time you took in making the mathematical literature accessible to us physicists. Our interactions have contributed to much of the work in this Thesis. Thank you all for agreeing to serve on my committee. 

I would also like to give a special acknowledgement to Ian Abel, who introduced our group to adjoint methods which formed the basis for this Thesis work. 








This work was supported by the ARCS Foundation and the US Department of Energy FES grants DE-FG02-93ER-54197 and DE-FC02-08ER-54964. The computations presented in this Thesis have used resources at the National Energy Research Scientific Computing Center (NERSC).
 

\hbox{\ }
\vspace{-.65in}

\begin{center}
\large{Publication List}
\end{center} 

\begin{enumerate}
\item L. M. Imbert-Gerard, E. J. Paul, and A. Wright, ``An introduction to symmetries and stellarators,'' \textit{in preparation} (2019). \href{https://arxiv.org/abs/1908.05360}{(link to preprint)}
\item E. J. Paul, T. Antonsen, Jr., M. Landreman, and W. A. Cooper, ``Adjoint approach to calculating shape gradients for 3D magnetic confinement equilibria,'' \textit{Journal of Plasma Physics} 86, 905860103 (2020). \href{https://arxiv.org/abs/1910.14144}{(link to preprint)}
\item E. J. Paul, I. G. Abel, M. Landreman, and W. Dorland, ``An adjoint method for neoclassical stellarator optimization,'' \textit{Journal of Plasma Physics} 85, 795850501 (2019). \href{https://arxiv.org/pdf/1812.06154.pdf}{(link to preprint)}
\item T. Antonsen, Jr., E. J. Paul, and M. Landreman, ``Adjoint approach to calculating shape gradients for 3D magnetic confinement equilibria," \textit{Journal of Plasma Physics} 85, 905850207 (2019). \href{https://arxiv.org/pdf/1812.06154.pdf}{(link to preprint)}
\item M. Landreman and E. J. Paul, ``Computing local sensitivity and tolerances for stellarator physics properties using shape gradients,"  \textit{Nuclear Fusion} 58, 076023 (2018). \href{https://arxiv.org/pdf/1803.03187.pdf}{(link to preprint)}
\item E. J. Paul, M. Landreman, A. Bader, and W. Dorland, ``An adjoint method for gradient-based optimization of stellarator coil shapes," \textit{Nuclear Fusion} 58, 076015 (2018). \href{https://arxiv.org/pdf/1801.04317.pdf}{(link to preprint)}
\item E. J. Paul, M. Landreman, F. M. Poli, D. A. Spong, H. M. Smith, and W. Dorland, ``Rotation and neoclassical ripple transport in ITER," \textit{Nuclear Fusion} 57, 116044 (2017). \href{https://arxiv.org/pdf/1703.06129.pdf}{(link to preprint)}
\end{enumerate}
\phantomsection
\renewcommand{\contentsname}{Table of Contents}
\renewcommand{\baselinestretch}{1}
\small\normalsize
\tableofcontents
\newpage

\setlength{\abovedisplayskip}{6pt} 
\setlength{\belowdisplayskip}{6pt}
\setcounter{page}{1}
\pagestyle{plain}\pagenumbering{arabic}
\renewcommand{\thechapter}{1}

\chapter{Introduction}

This Chapter aims to motivate and place in context the work of this Thesis. We begin with an introduction to the stellarator concept of toroidal confinement in Section \ref{sec:stellarator_concept}, including the necessity of optimization of the magnetic field. We then discuss important properties of a stellarator device in Section \ref{sec:stellarator_properties}. To put stellarator optimization in perspective, we briefly discuss the relevant history in Section \ref{sec:stellarator_history}. We then, in Section \ref{sec:stellarator_optimization}, provide a detailed introduction to stellarator optimization, including typical assumptions, numerical methods, and associated challenges. We conclude with an overview of this Thesis in Section \ref{sec:overview}.

Throughout this Chapter, we use terminology related to magnetic field geometry and toroidal coordinate systems, which are introduced in Appendix \ref{app:toroidal_coordinates}.

\section{The stellarator concept}
\label{sec:stellarator_concept}

The fusion community must face several significant scientific challenges to demonstrate a viable magnetic fusion reactor. A large fraction of the present research in magnetic fusion is dedicated to the tokamak, a concept that relies on a large plasma current for confinement. Driving such a current requires a significant amount of recirculated power and necessitates either pulsed operation or non-inductive current drive, both of which are disadvantageous for a fusion reactor. This large current makes them susceptible to current-driven instabilities that can limit plasma performance. These instabilities, such as tearing and kink instabilities, can result in catastrophic terminations of the discharge (Chapter 7.9 in \cite{Wesson2011}). Runaway electrons formed due to disruptions can be accelerated by the inductive electric field, possibly causing damage to plasma-facing components and applying large electro-magnetic forces to the vacuum vessel. The effect of runaway electrons will be much more harmful in large reactor-scale tokamaks due to the exponential dependence of the density of relativistic electrons on the plasma current \cite{Hender2007}. Thus in a reactor, disruptions must be mitigated by active feedback and operation within a safe margin of stability limits. However, such control will be difficult when alpha particles provide a significant fraction of the heating power \cite{Hawryluk2019}.

\begin{figure}
    \centering
    \begin{subfigure}[b]{0.49\textwidth}
    \includegraphics[trim=2cm 8cm 1cm 7cm,clip,width=0.8\textwidth]{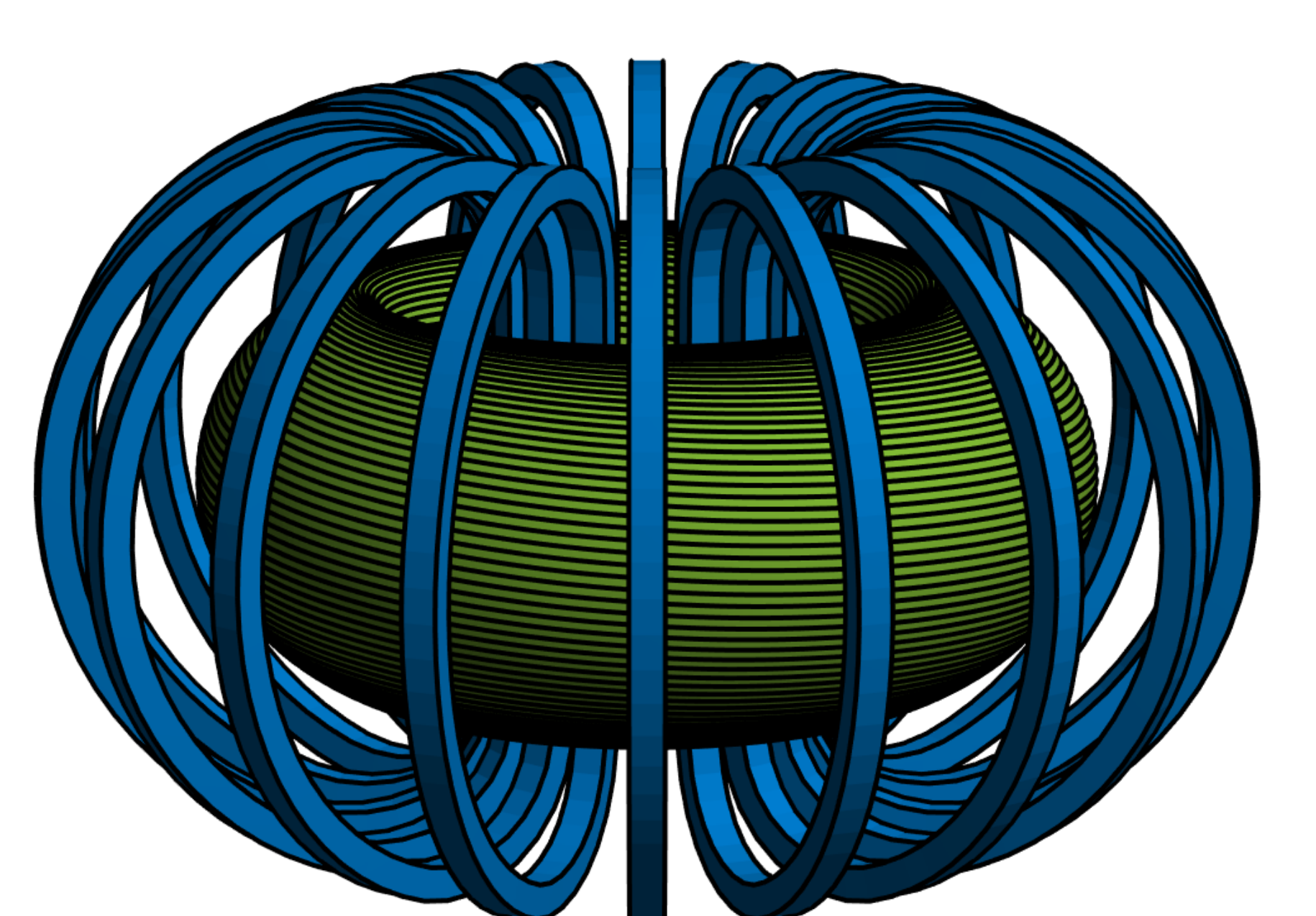}
    \caption{}
    \end{subfigure}
    \begin{subfigure}[b]{0.49\textwidth}
    \centering
    \includegraphics[width=1.0\textwidth,trim=8cm 6cm 8cm 5cm,clip]{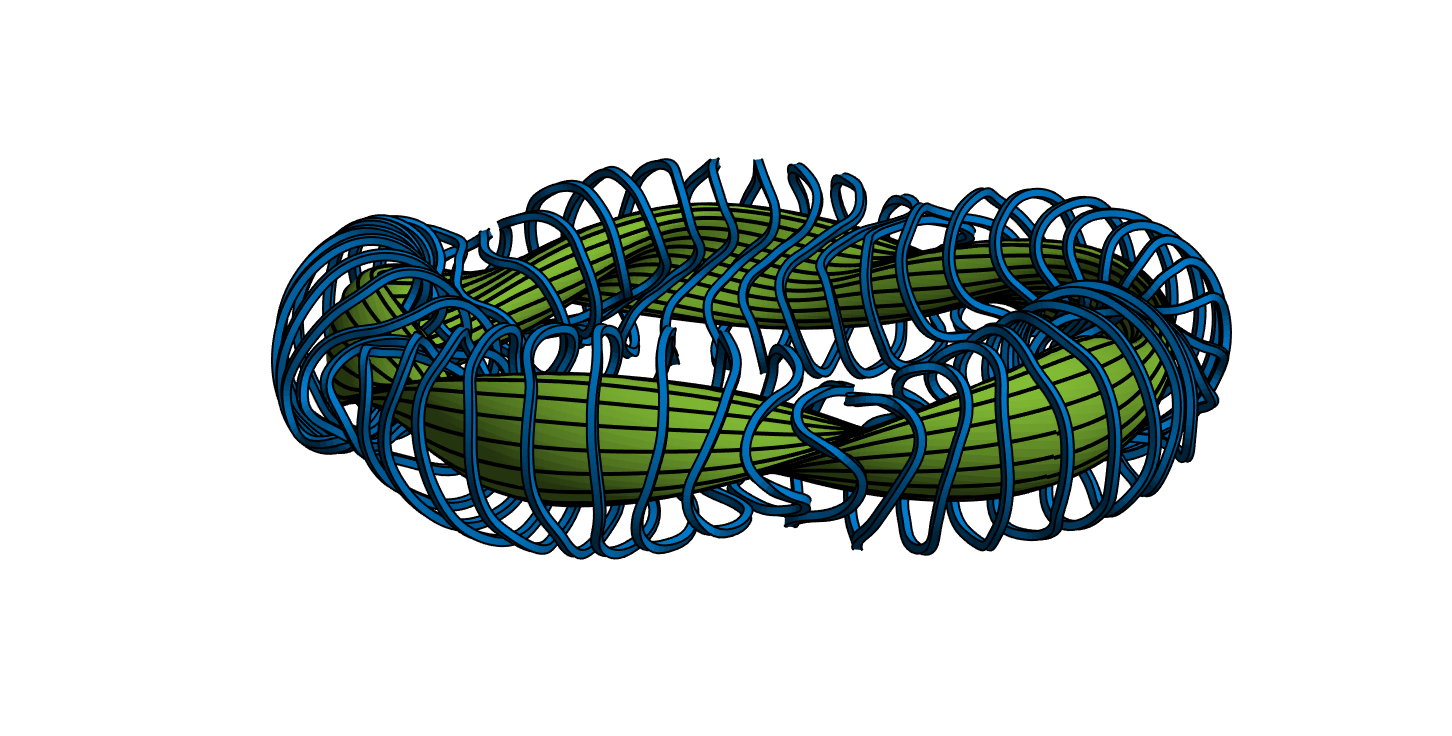}
    \caption{}
    \end{subfigure}
    \caption{A schematic image of a tokamak (a) and stellarator (b). The electro-magnetic coils are shown in blue, and the plasma domain is shown in green. Magnetic field lines lying on the outermost magnetic surface are shown in black.}
    \label{fig:tokamak_stellarator}
\end{figure}

Remarkably, Lyman Spitzer predicted these possible difficulties of tokamak confinement in 1952 \cite{Spitzer1952}, before the first toroidal confinement experiment,
\begin{quote}
    ``... a large induced current is open to the two practical objectives that it cannot be sustained in a steady equilibrium and that the rapid generation of such a current is likely to lead to plasma oscillations."
\end{quote}
These observations led to the development of the stellarator concept. In contrast to the tokamak, a stellarator generates a poloidal magnetic field through clever shaping by external currents rather than internal plasma currents. A small amount of current in the plasma is self-driven due to pressure gradients, though this is typically not large enough to result in significant MHD modes. There is some experimental evidence that stellarator configurations may be able to operate above the linear MHD stability pressure threshold \cite{Weller2006} rather than being terminated by a disruption. The Large Helical Device (LHD) has operated up to a volume-averaged $\beta$ of $5\%$ without any disruptive MHD phenomena, though the heat transport increases due to low-$n$ mode activity \cite{Sakakibara2008}. Here $\beta = p/(B^2/(2\mu_0))$ is the ratio of the plasma pressure, $p$, to the magnetic pressure, and $n$ is the toroidal mode number. Similarly, high-beta discharges in the Wendelstein 7-Advanced Stellarator (W7-AS) have shown saturation of low-$n$ and interchange modes at a low level that merely slowly degrades confinement \cite{Weller2006}. Stellarators can also operate at higher density than tokamaks due to the absence of the Greenwald limit \cite{Gates2012}. While in tokamaks, the limits on the density and pressure due to the Greenwald and MHD stability limits set hard boundaries on the operating points, in a stellarator much softer limits exist. Performance at high beta is often instead limited by equilibrium properties, such as magnetic field stochasticity near the edge. For example, if the Shafranov shift becomes comparable to the minor radius of the plasma, this can lead to loss of magnetic surfaces \cite{Spong2010}. The ability to operate at high beta is critical for an economical fusion reactor: in the temperature range of 10-20 keV, the fusion power density scales as $P \sim \beta^2 B^4$ \cite{Smith2010}. See Figure \ref{fig:tokamak_stellarator} for schematics of a tokamak and stellarator configuration.

\begin{figure}
    \centering
    \includegraphics[width=0.8\textwidth]{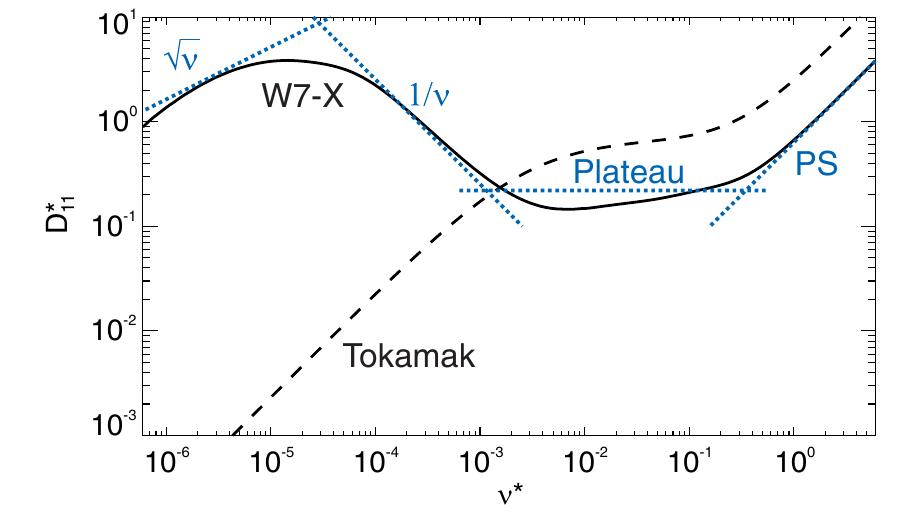}
    \caption{The neoclassical diffusion coefficient, $D_{11}^*$, as a function of the normalized collisionality, $\nu_* = \nu R/(\iota v)$, where $\nu$ is the collision frequency, $\iota$ is the rotational transform, $v$ is the speed, and $R$ is the major radius. An axisymmetric field exhibits a low-collisionality regime in which $D_{11}^* \sim \nu$, while a stellarator exhibits $D_{11}^* \sim 1/\nu$. Thus the neoclassical transport in a general three-dimensional field can be especially deleterious at low collisionality. Figure reproduced from \cite{Helander2012} with permission.}
    \label{fig:collisionality}
\end{figure}

Despite these clear advantages, much care must be taken to design a stellarator with acceptable confinement properties. Due to its continuous toroidal symmetry, the tokamak enjoys confinement of collisionless single-particle trajectories and the existence of closed, nested magnetic surfaces. However, in the general three-dimensional field of a stellarator, these properties are not always present. The trajectories of energetic ions, such as the alpha particles produced in a fusion reaction, may therefore be lost, resulting in damage to material surfaces. Stellarators can experience enhanced neoclassical transport, the collisional transport of thermal particles due to the magnetic field geometry, leading to increased transport of heat and particles, especially at low collisionality (Figure \ref{fig:collisionality}). The presence of large magnetic islands or chaotic regions in a three-dimensional field can also severely limit performance by locally flattening the temperature profile. 

However, none of these challenges appear to be showstoppers for stellarator confinement. The success of modern stellarators can be attributed to the ability to design the magnetic field with numerical optimization. While tokamak optimization is also possible \cite{Highcock2018}, it is much more difficult as confinement properties become very sensitive to the current density and pressure profiles. These profiles can be determined with multi-scale modeling on turbulent and transport time scales, which is very computationally intensive. On the other hand, the physical properties of stellarators are relatively insensitive to these profiles, as they primarily rely on the externally produced magnetic field for confinement \cite{Boozer2019}. Given the ability to numerically optimize the magnetic field of a stellarator, in Section  \ref{sec:stellarator_properties}, we discuss the properties one should consider in a design. 

\section{What makes a good stellarator?}
\label{sec:stellarator_properties}

We now outline the desired physical properties of a stellarator and standard proxy functions applied during their design. We will reserve any discussion of coils, the external currents that produce the magnetic field, until Section \ref{sec:coil_optimization}.

\noindent \textbf{\textit{Equilibrium properties}}

\begin{figure}
    \centering
    \begin{subfigure}[b]{0.49\textwidth}
    \includegraphics[trim=8cm 3cm 12cm 3cm ,clip,width=0.9\textwidth]{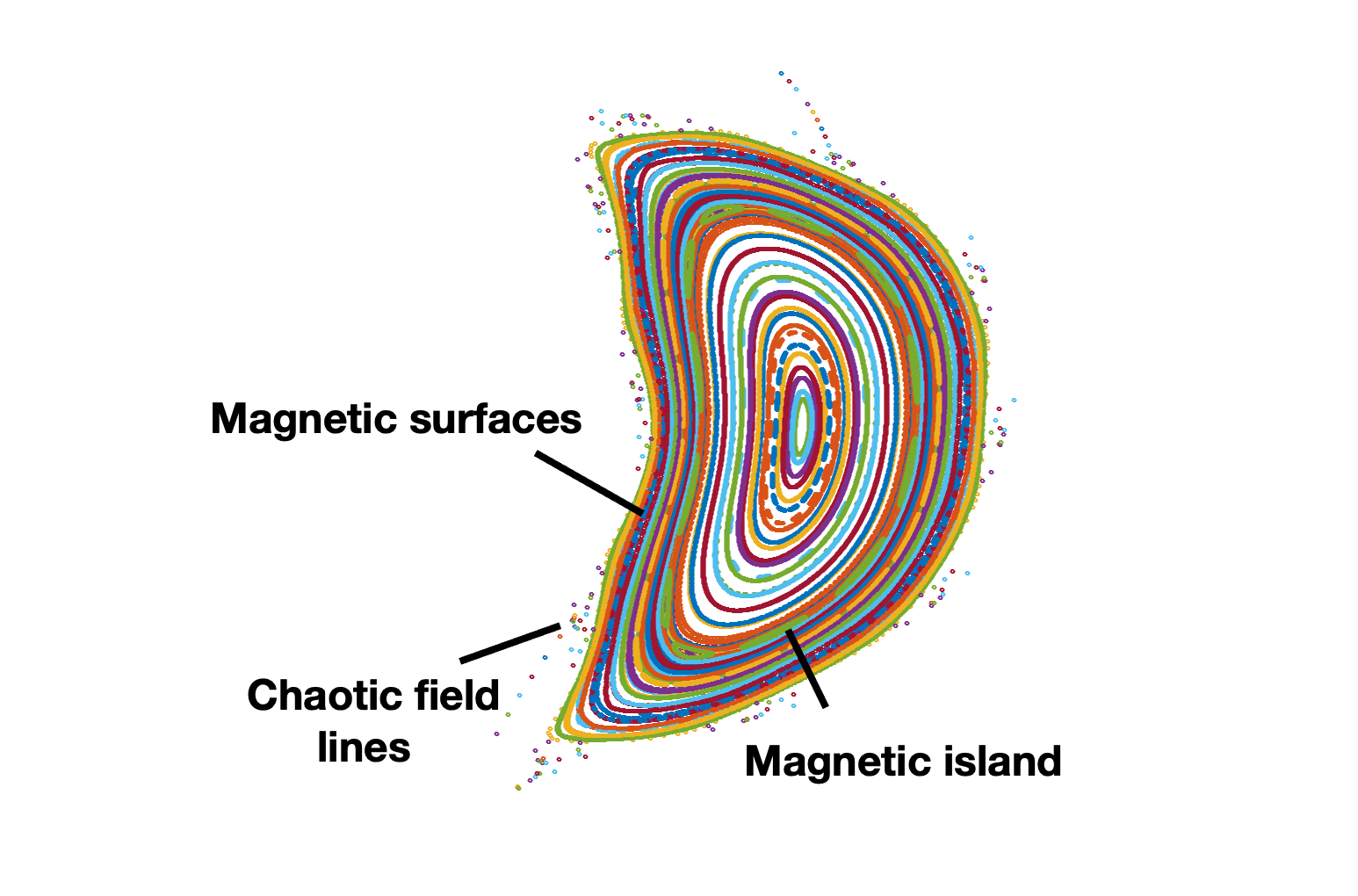}
    \end{subfigure}
    \begin{subfigure}[b]{0.49\textwidth}
    \includegraphics[trim=9cm 3cm 13cm 7cm ,clip,width=0.7\textwidth]{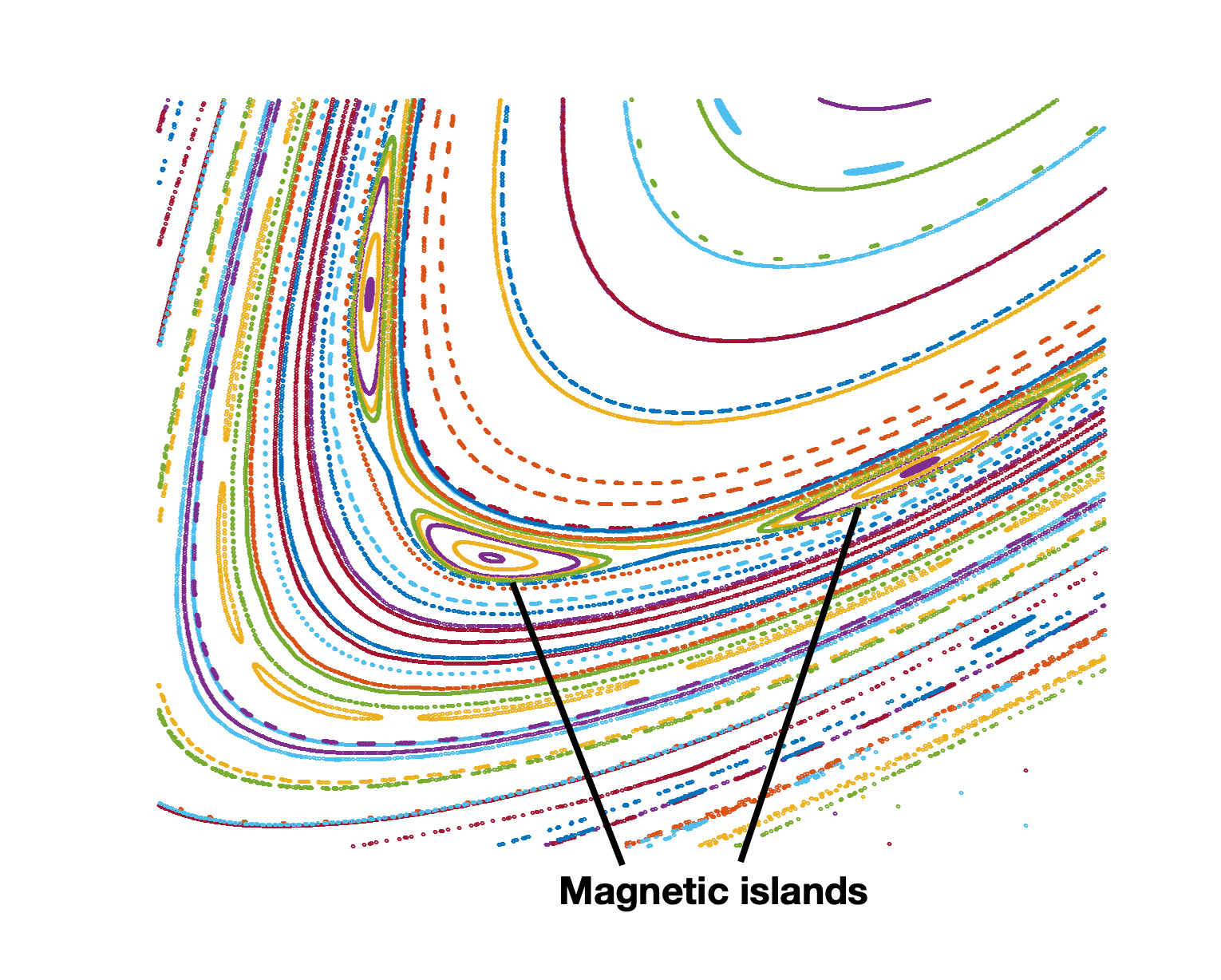}
    \end{subfigure}
    \caption{A Poincare surface computed from the NCSX coil shapes \cite{Williamson2005}. To produce this Figure, magnetic field lines are integrated toroidally around the device. Each time they hit a plane at constant toroidal angle, a point is plotted with color indicating the field line. A general 3D field contains regions of chaotic field lines and magnetic island chains along with a volume of nested toroidal magnetic surfaces. Figure adapted from \cite{Imbert2019}.}
    \label{fig:Poincare}
\end{figure}

The operating space of stellarators is often restricted due to MHD equilibrium properties rather than stability limits. For example, when $\beta \sim \epsilon \iota^2/2$ where $\epsilon$ is the inverse aspect ratio and $\iota$ is the rotational transform, the Shafranov shift becomes comparable to the minor radius, which may result in flux-surface break-up \cite{Spong2010,Helander2014}. There is a tendency of the edge magnetic field to become stochastic at large beta \cite{Sakakibara2008}, so a design should try to maximize the volume of continuously nested flux surfaces \cite{Hudson2002}. One should also minimize the island width at low-order rational surfaces, which can be estimated using analytic expressions \cite{Lee1990,Cary1991}, assuming the magnetic field is close to having perfect magnetic surfaces. Such islands can also be minimized by controlling the rotational transform, either by maintaining low magnetic shear and eliminating low-order rational surfaces altogether or by taking advantage of large magnetic shear, as the magnetic island width scales as $1/\sqrt{\iota'(\psi)}$ \cite{Boozer2015}. See Figure \ref{fig:Poincare} for a visualization of magnetic surfaces, magnetic islands, and chaotic field lines in the NCSX stellarator.

\noindent \textbf{\textit{Pressure-driven currents}} 

There are several sources of self-driven plasma current \cite{Helander2014}: the parallel bootstrap current arises due to collisions between trapped and passing particles in the presence of density and temperature gradients, and the parallel Pfirsch-Sch\"{u}ter and perpendicular diamagnetic currents occur due to equilibrium pressure gradients. The bootstrap current can cause shifts in the rotational transform toward low-order rational values, which must especially be avoided in low-shear devices. Control of the edge rotational transform is also vital for designs with an island divertor \cite{Geiger2010}. In the presence of reduced bootstrap current, the magnetic field structure becomes less sensitive to changes in beta. For these reasons, the Wendelstein 7-X (W7-X) configuration was designed for minimal bootstrap current \cite{Grieger1992}. Often optimization is performed with a low-collisionality semi-analytic bootstrap current model \cite{Shaing1989}. Bootstrap current optimization will be described further in Chapter \ref{ch:adjoint_neoclassical}. 
The Pfirsch-Schl\"{u}ter current does not provide any \textit{net} current and therefore does not shift the rotational transform. However, it can give rise to a Shafranov shift and thus affect the equilibrium beta limit \cite{Wagner1998}. The Pfirsch-Schl\"{u}ter current can be reduced by minimizing the magnitude of the geodesic curvature. The net diamagnetic current will only be non-zero in the presence of another source of net current; thus, the reduction of the bootstrap current will automatically reduce the diamagnetic current. 

While the presence of self-driven current can give rise to unfavorable shifts in the rotational transform, there are situations in which significant bootstrap current may be desirable. If the bootstrap current provides a source of rotational transform in addition to the external coils, the coil complexity may be reduced and a more compact device may be possible. Plasma current can also provide island healing \cite{Hegna1998}, reducing the width of islands in comparison with those in the vacuum configuration. For these reasons, the National Compact Stellarator Experiment (NCSX) was designed to be quasi-axisymmetric with a significant fraction of rotational transform provided by the plasma current \cite{Hirshman1999}. 

\noindent \textbf{\textit{Energetic-particle confinement}} 

A successful stellarator reactor must confine energetic alpha particles for at least their slowing-down time such that their energy can be deposited with the thermal population. Prompt losses of fast particles should especially be avoided because they can lead to damage to material surfaces. Collisional diffusion and deflection are minimal at energies near the birth energy of $3.5$ MeV for a D-T reaction (Chapter 3 in \cite{Helander2005}), so collisionless guiding center orbits are an informative metric of energetic particle confinement. If the collision frequency is small enough that energetic ions can complete their bounce or transit orbits, then the parallel adiabatic invariant, 
\begin{align}
J = \oint dl \,v_{||}, 
\label{eq:adiabatic_invariant}
\end{align}
is a conserved quantity, where $v_{||}$ is the velocity parallel to the magnetic field and $l$ measures length along a field line. For trapped particles, the integral is taken along a closed trajectory between bounce points. For passing particles, it is taken along a field line until it comes infinitesimally close to its starting point. If $J$ is constant on a magnetic surface, then the collisionless trajectories will experience no net radial drift, a property known as omnigeneity \cite{Cary1997}. Thus several properties involving $J$, such as its variation within a flux surface, have been considered during the design process \cite{Spong1998,Drevlak2014}. There is evidence that targeting quasi-symmetry (defined shortly) near the half-radius may also improve energetic particle confinement \cite{Henneberg2019b}. 
\vspace{1cm}

\noindent \textbf{\textit{Quasi-symmetry}} 

Quasi-symmetric magnetic fields are a subset of omnigeneous magnetic fields. A quasi-symmetric magnetic field possesses a symmetry direction of the magnetic field strength when expressed in Boozer coordinates (Appendix \ref{sec:boozer_coordinates}),
\begin{align}
    B(\psi,\vartheta_B,\varphi_B) = B(\psi,M\vartheta_B - N \varphi_B),
\end{align}
for fixed integers $M$ and $N$. If $M = 0$, the contours of the magnetic field strength close poloidally, known as quasi-poloidal symmetry. If $N = 0$, the contours of the magnetic field strength close toroidally, known as quasi-axisymmetry. If both $M$ and $N$ are non-zero, known as quasi-helical symmetry, the contours of the field strength close both toroidally and poloidally. 

This symmetry implies guiding center confinement \cite{Boozer1983} and neoclassical properties that are comparable to those of an equivalent tokamak \cite{Helander2014}, including the ability to rotate in the direction of quasi-symmetry \cite{Helander2008}. A quasi-symmetric field is omnigeneous, though the converse is not necessarily true. Quasi-symmetry is typically targeted by minimizing the symmetry-breaking Fourier harmonics of the magnetic field strength. 

\noindent \textbf{\textit{Neoclassical transport}}

Stellarators experience enhanced neoclassical transport at low collisionality in comparison with tokamaks (Figure \ref{fig:collisionality}). Neoclassical transport is typically the dominant transport channel in classical (unoptimized) stellarators. It is common to employ the effective ripple ($\epsilon_{\text{eff}}$) proxy, which quantifies the geometric dependence of the radial fluxes in the low-collisionality $1/\nu$ regime \cite{Nemov1999}. A discussion of $\epsilon_{\text{eff}}$ and neoclassical diffusion in the $1/\nu$ regime is given in Chapter \ref{ch:adjoint_MHD} and Appendix \ref{app:1_over_nu}. Neoclassical optimization will be discussed in more depth in Chapter \ref{ch:adjoint_neoclassical}. A review of neoclassical optimization strategies is given in \cite{Mynick2006}.

\noindent \textbf{\textit{Stability}}

Although stellarators may be able to operate above linear MHD stability limits, it is desirable to design a stellarator with an increased beta limit to reduce enhanced transport caused by MHD modes. It is common to employ the magnetic well \cite{Greene1997} (discussed in Chapter \ref{ch:adjoint_MHD}) or Mercier criterion \cite{Mercier1974} as proxies for the stability of low-$n$ interchange modes. One can also try to increase magnetic shear, the radial derivative of the rotational transform $\iota'(\psi)$, to improve large $n$ ballooning stability and Mercier stability \cite{Hegna1998}. It appears that stellarators can also be designed with reduced microturbulence, though turbulence optimization has yet to be demonstrated experimentally. Some proxies have been proposed, such as reducing the overlap between bad curvature and trapping regions \cite{Xanthopoulos2014} or increasing nonlinear energy transfer between unstable and damped modes \cite{Hegna2018}.

\section{A brief history of the stellarator}
\label{sec:stellarator_history}

\begin{figure}
    \centering
    \includegraphics[width=0.8\textwidth]{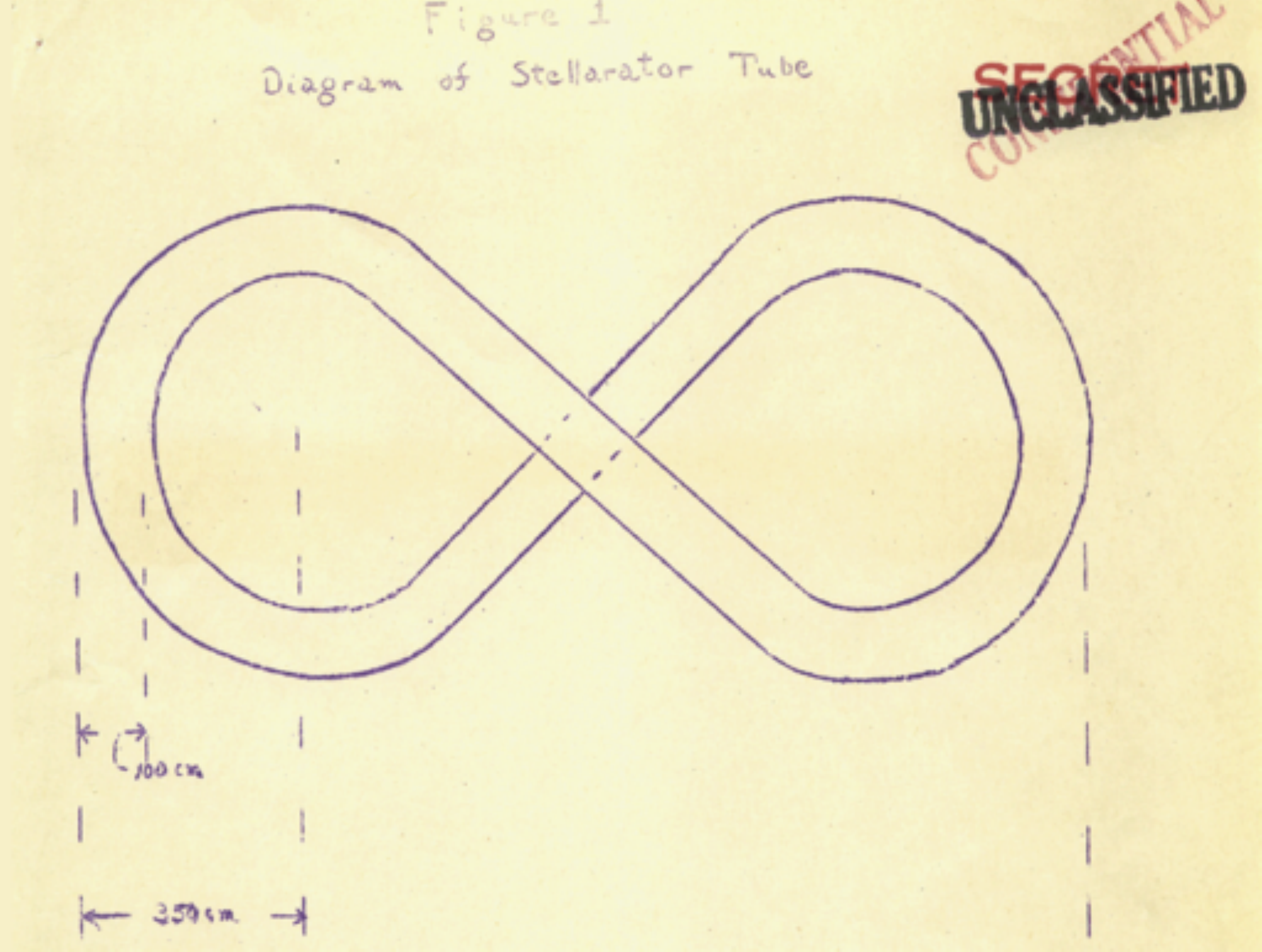}
    \caption{A diagram of the figure-eight stellarator design from Lyman Spitzer's 1951 Project Matterhorn report. Figure reproduced from \cite{Spitzer1951}.}
    \label{fig:spitzer_stellarator}
\end{figure}

Lyman Spitzer's first stellarator concept used a simple figure-eight design (Figure \ref{fig:spitzer_stellarator}), which produced rotational transform by ``twisting the torus out of the plane" \cite{Spitzer1958}. Spitzer and his team experimentally demonstrated that external shaping could produce rotational transform in a vacuum field with the Model A, B, and C series stellarators at Princeton \cite{Stix1998}. Results from the Model B1 demonstrated confinement of energetic electrons for several milliseconds, much longer than would be possible with a purely toroidal field. However, the observed diffusion of thermal particles was much larger than that predicted from Bohm scaling \cite{Coor1958}. The Model C, using a racetrack configuration with helically wound coils, was able to demonstrate the existence of nested magnetic surfaces \cite{Sinclair1970}. Nonetheless, the Model C experienced poor confinement with Bohm-like diffusion \cite{Yoshikawa1985}. These early stellarator experiments operated until the late 1960s when promising results from the Soviet T-3 tokamak became available, and it was decided that Princeton's Model C would be converted to a tokamak \cite{PPPLTimeline}. 

Meanwhile, the Wendelstein line of stellarators was active at IPP Garching, initially adopting Princeton's racetrack design. Experiments on WII-A provided insight into the benefits of low magnetic shear and accurate construction of the coil system for avoiding magnetic islands \cite{Berkl1968}. The performance continued, however, to be limited by neoclassical transport at low collisionality and low equilibrium pressure limits due to the Shafranov shift \cite{Hirsch2008}. 

\begin{figure}
    \centering
    \includegraphics[width=0.8\textwidth]{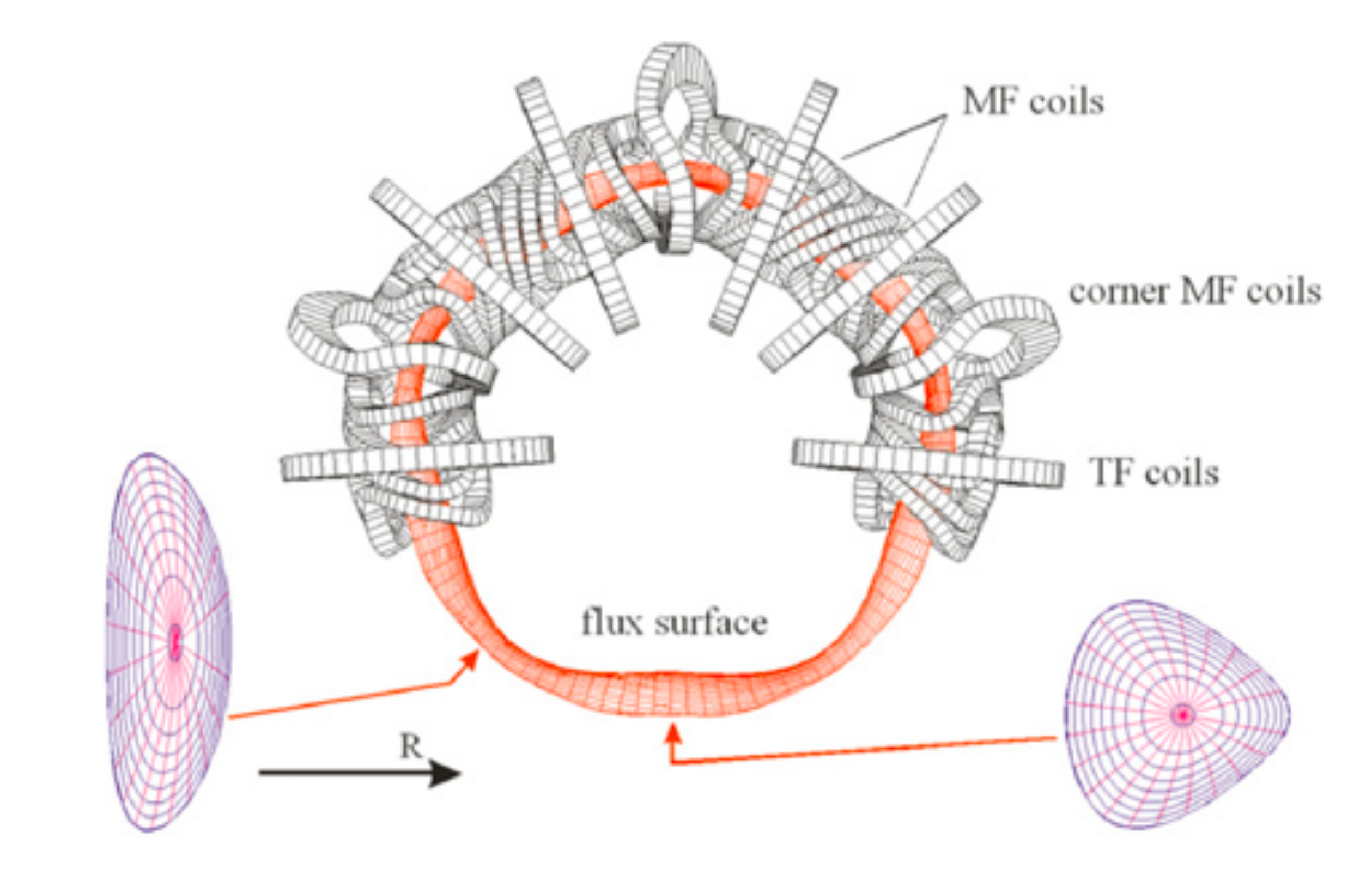}
    \caption{The modular field (MF) coils, toroidal field (TF) coils, and flux surfaces of the W7-AS stellarator. Figure reproduced from \cite{Hirsch2008} with permission.}
    \label{fig:w7_AS}
\end{figure}

A significant breakthrough in the stellarator program came with the design of W7-AS, which aimed to improve confinement with equilibrium optimization. To demonstrate the stellarator optimization concept, W7-AS was partially optimized for minimal geodesic curvature. Such an objective was predicted to minimize radial magnetic drifts and pressure-driven parallel currents.
For the first time, the magnetic field shaping was supplied by non-planar, modular coils (Figure \ref{fig:w7_AS}) that provided the freedom to tailor the magnetic field more carefully than helical coils. The experiment operated from 1988 to 2002, demonstrating the improved equilibrium and stability properties and reduction of neoclassical transport enabled through equilibrium optimization \cite{Hirsch2008,Hofmann1996}. 

The success of W7-AS paved the way for the W7-X experiment \cite{Wagner2005}, which was fully optimized for nested magnetic surfaces, fast-particle confinement, reduced parallel currents, minimal neoclassical transport at low collisionality, and MHD stability up to an average $\beta$ of $5\%$ \cite{Beidler1990}. The early optimization efforts of the Wendelstein team benefited greatly from the discovery that guiding center confinement could be achieved with a quasi-symmetric \cite{Boozer1983} magnetic field. N\"{u}hrenberg and Zille of the Wendelstein team then demonstrated that quasi-symmetric equilibria could be obtained from numerical optimization of MHD equilibria \cite{Nuhrenberg1988}. The W7-X configuration was designed based on one of their quasi-helical configurations, modified to achieve the objectives outlined above. The resulting configuration was quasi-isodynamic, a quasi-omnigenous magnetic field with poloidally closed contours of the magnetic field strength \cite{Nuhrenberg1994,Helander2009}. Experiments from the initial campaigns of W7-X have demonstrated the success of the stellarator equilibrium optimization concept, confirming the desired magnetic topology to within a tolerance of $10^{-5}$ \cite{Pedersen2016}. High-beta operation will not be demonstrated until an actively-cooled divertor is installed for the next operating campaign. However, there is initial evidence that recent high-performance shots could not have been achieved without neoclassical optimization \cite{Wolf2019}.

\begin{figure}
    \centering
    \includegraphics[width=0.8\textwidth]{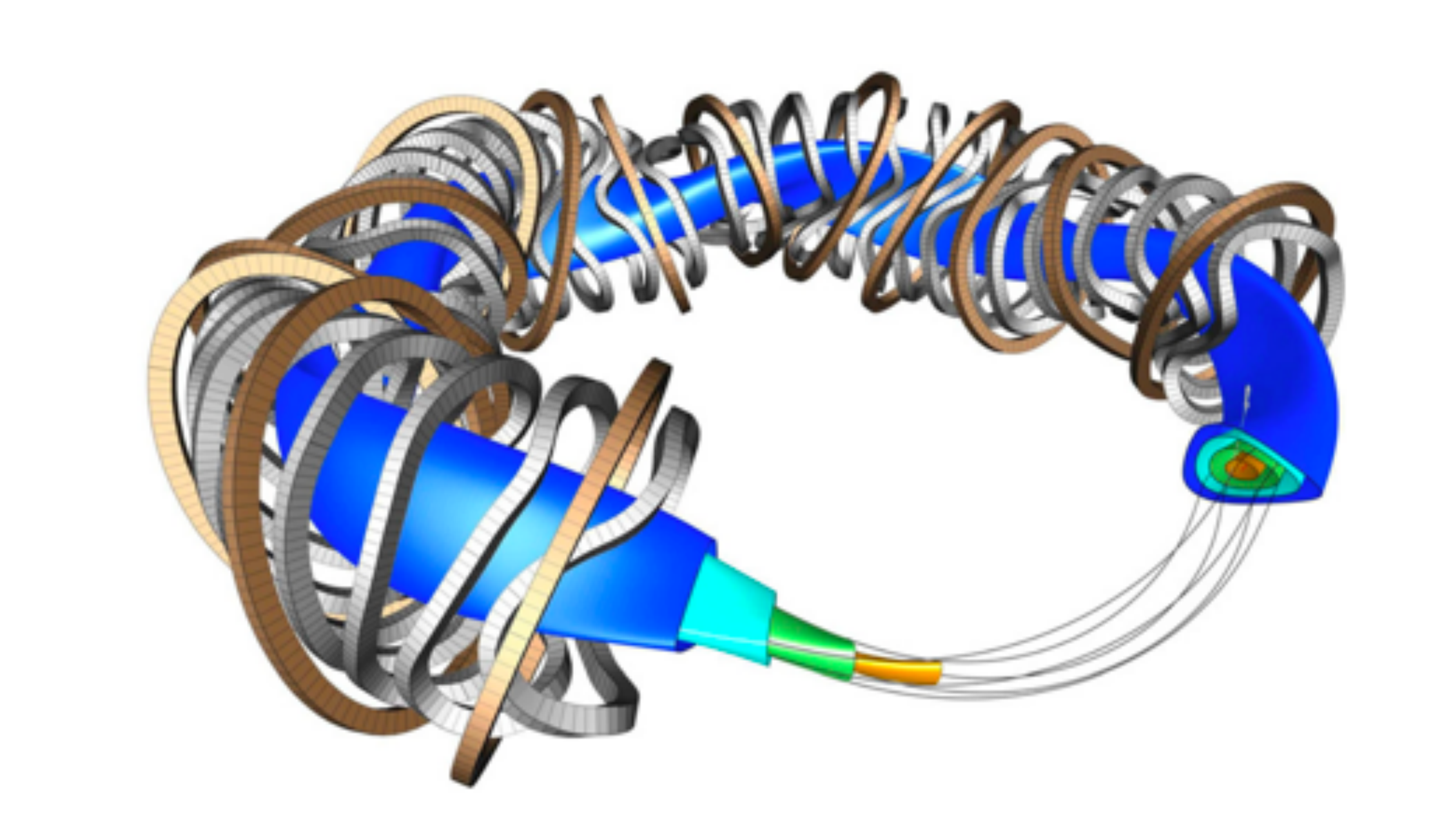}
    \caption{Modular field coils (silver), toroidal field coils (bronze), and magnetic surfaces of the W7-X stellarator. Figure reproduced from \cite{Pedersen2017} with permission.}
    \label{fig:w7x}
\end{figure}

W7-X was not, however, the first experimental demonstration of a fully optimized stellarator. The Helically Symmetric eXperiment (HSX) was designed to have quasi-helical symmetry, Mercier stability, and low magnetic shear \cite{Anderson1995} using the equilibrium optimization tools developed by the Wendelstein team \cite{Anderson2019}. HSX has demonstrated a reduction of electron thermal diffusivity \cite{Canik2007} due to the decrease in neoclassical transport and a reduction of flow damping in the symmetry direction \cite{Gerhardt2005}. The inward-shifted configuration of LHD was partially optimized for reduced neoclassical transport and energetic particle confinement \cite{Murakami2002}, though its ideal MHD stability is worsened in comparison with the standard configuration. Experiments have demonstrated higher electron temperatures and improved energetic ion confinement in the inward-shifted configuration as compared with the standard configuration \cite{Murakami2004}.

There continues to be an effort toward advanced stellarator designs. Construction has commenced for the Chinese First Quasi-symmetric Stellarator (CFQS) \cite{Shimizu2018}, which will be the first quasi-axisymmetric device in operation. The quasi-axisymmetric NCSX \cite{Zarnstorff2001} was designed and partially constructed at the Princeton Plasma Physics Laboratory (PPPL), but its funding was terminated before its
completion. As the field of stellarator optimization has developed, several other stellarator equilibria have been optimized to be quasi-symmetric \cite{Nelson2003,Ku2008,Garabedian2009,Ku2011,Drevlak2013,Henneberg2019,Bader2019} and quasi-omnigeneous \cite{Isaev2003,Mikhailov2012}.

\section{Stellarator optimization}
\label{sec:stellarator_optimization}

Historically, stellarator optimization has largely used a two-staged approach: in the first step, the magnetic field in the confinement region is optimized to obtain the desirable physics properties. The magnetic field must satisfy the MHD equilibrium equations; thus this task amounts to optimization in the space of free parameters that describe the MHD equilibrium. Often a fixed-boundary MHD calculation is performed, in which an outer flux surface is prescribed, as opposed to a free-boundary calculation, in which the currents in the vacuum region are prescribed. As a second step, the currents in the vacuum region are optimized to be consistent with the boundary obtained in the first step. As numerical MHD equilibrium calculations form the foundation of stellarator optimization, these will be described in Section \ref{sec:equilibrium_calculation}. The two stages of the optimization process are described in Sections \ref{sec:equilibrium_optimization} and \ref{sec:coil_optimization}. We will conclude with a discussion of the present challenges associated with the design of stellarators and how this Thesis will address them in Section \ref{sec:challenges_outlook}. 

\subsection{MHD equilibrium calculations}
\label{sec:equilibrium_calculation}


The MHD equilibrium equations,
\begin{subequations}
\begin{align}
   \textbf{J} \times \textbf{B}  &= \nabla p \label{eq:MHD_force_balance} \\
   \nabla \times \textbf{B} &= \mu_0 \textbf{J} \label{eq:MHD_ampere} \\
      \nabla \cdot \textbf{B} &= 0,
\end{align}
\label{eq:MHD_equilibrium}
\end{subequations}
describe the steady-state behavior of the magnetic field in strongly magnetized plasmas.
Many assumptions are made in arriving at \eqref{eq:MHD_equilibrium}, such as small plasma resistivity, low frequency in comparison with the cyclotron and collision frequencies, and small electron inertia. In practice, these equations describe the long-wavelength, low-frequency behavior of magnetic fusion plasma very well \cite{Freidberg2014}. 

Finding solutions to \eqref{eq:MHD_equilibrium} is non-trivial in a general three-dimensional field, as well-posedness requires a set of constraints to be satisfied on every closed field line unless the pressure profile is locally flattened (\cite{Grad1967}, Section 10.3 in \cite{Imbert2019}). An alternative is to rely on the assumption that there exists a set of continuously nested toroidal magnetic surfaces, $\Gamma(\psi)$, labeled by the toroidal flux label, $\psi$. Although magnetic surfaces are not guaranteed to exist in general three-dimensional geometry, any stellarator configuration of physical interest will possess a large region of continuously nested surfaces, and making this assumption will allow for tractable MHD equilibrium calculations.

Under the assumption of continuously nested toroidal magnetic surfaces, \eqref{eq:MHD_equilibrium} can be shown to be stationary points of an energy functional \cite{Kruskal1958}, 
\begin{align}
W[\textbf{B}] = \int_{V_P} d^3 x \, \left(\frac{B^2}{2\mu_0} - p\right),
\label{eq:energy_functional}
\end{align}
where $V_P$ is the volume of the confinement region bounded by a magnetic surface $S_P$. Variations of $W$ are computed at prescribed and fixed pressure ($p(\psi)$), rotational transform ($\iota(\psi)$), and the toroidal flux label on $S_P$ ($\psi_0$) (\cite{Helander2014}, Section 11.1 in \cite{Imbert2019}). 
Solutions to \eqref{eq:MHD_equilibrium} under these assumptions can be computed efficiently and robustly using gradient-descent methods to obtain local minima of $W[\textbf{B}]$. This approach is implemented in the VMEC \cite{Hirshman1983} and NSTAB \cite{Garabedian2002} codes. 

Sometimes another function of flux is prescribed instead of the rotational transform, such as the net toroidal current inside a constant $\psi$ surface,
\begin{align}
I_T(\psi) = \int_{\mathcal{S}_T(\psi)} d^2 x \, \textbf{J} \cdot \hat{\textbf{n}},
\end{align}
where $\mathcal{S}_T(\psi)$ is a surface at constant toroidal angle bounded by $\Gamma(\psi)$ (Figure \ref{fig:toroidal_flux}) and $\hat{\textbf{n}}$ is the unit normal. This choice of flux function is more common in the context of optimization, as $I_T(\psi)$ can be chosen to vanish for a vacuum field or to be consistent with a bootrstrap current model at finite pressure \cite{Spong2001,Shimizu2018}. 

We can consider \eqref{eq:MHD_equilibrium} to be an equation determining the magnetic field $\textbf{B}$, as the current density is computed from Ampere's law \eqref{eq:MHD_ampere} and the pressure is given as a function of flux, $p(\psi)$. The MHD equilibrium equations are solved with a Dirichlet boundary condition,
\begin{align}
    \textbf{B} \cdot \hat{\textbf{n}} \rvert_{S_P} = 0.
    \label{eq:MHD_bc}
\end{align}
In the fixed-boundary approach, $S_P$ is given and fixed during the equilibrium calculation. The relevant equations for a fixed-boundary calculation are summarized in Table \ref{table:fixed_boundary}. 
\begin{table}
\centering
    \begin{tabular}{|c|c|c|}
    \hline 
    PDE & BC & Given \\
         \hline \hline
         $\left(\nabla \times \textbf{B}\right) \times \textbf{B} = \mu_0 \nabla p(\psi)$ & $\textbf{B} \cdot \hat{\textbf{n}} \rvert_{S_P} = 0$ & $p(\psi)$, $\psi_0$, \& $S_P$ \\
         $\nabla \cdot \textbf{B} = 0$ & & $\iota(\psi)$ or $I_T(\psi)$ \\
         \hline 
    \end{tabular}
    \caption{Summary of fixed-boundary equilibrium PDE.}
    \label{table:fixed_boundary}
\end{table}

In the free-boundary approach, the current density, $\textbf{J}_C$, in the vacuum region, $\mathbb{R}^3 \backslash V_P$, is prescribed instead of $S_P$. The magnetic field due to this current is computed from the Biot-Savart law,
\begin{align}
    \textbf{B}_C(\textbf{x}) = \frac{\mu_0}{4\pi}\int_{\mathbb{R}^3 \backslash V_P} d^3 x' \, \frac{\textbf{J}_C(\textbf{x}')\times(\textbf{x}-\textbf{x}')}{|\textbf{x}-\textbf{x}'|^3}.
    \label{eq:Biot_Savart}
\end{align}

For a given $S_P$, the plasma current, $\textbf{J}_P$, is computed from \eqref{eq:MHD_equilibrium}. The magnetic field due to the plasma current can similarly be computed from the Biot-Savart law or more efficiently with the application of the virtual casing principle \cite{Lazerson2012}. The total magnetic field must be tangent to the boundary, 
\begin{align}
    \left(\textbf{B}_P + \textbf{B}_C\right) \cdot \hat{\textbf{n}}\rvert_{S_P} = 0.
    \label{eq:Bnormal_free}
\end{align}
Furthermore, the total pressure must be continuous across $S_P$,
\begin{align}
   \left[\left[ B^2/(2\mu_0) + p \right]\right]_{S_P} = 0,
   \label{eq:jump_condition_free}
\end{align}
to ensure force balance.

\begin{figure}
    \centering
    \includegraphics[width=0.7\textwidth]{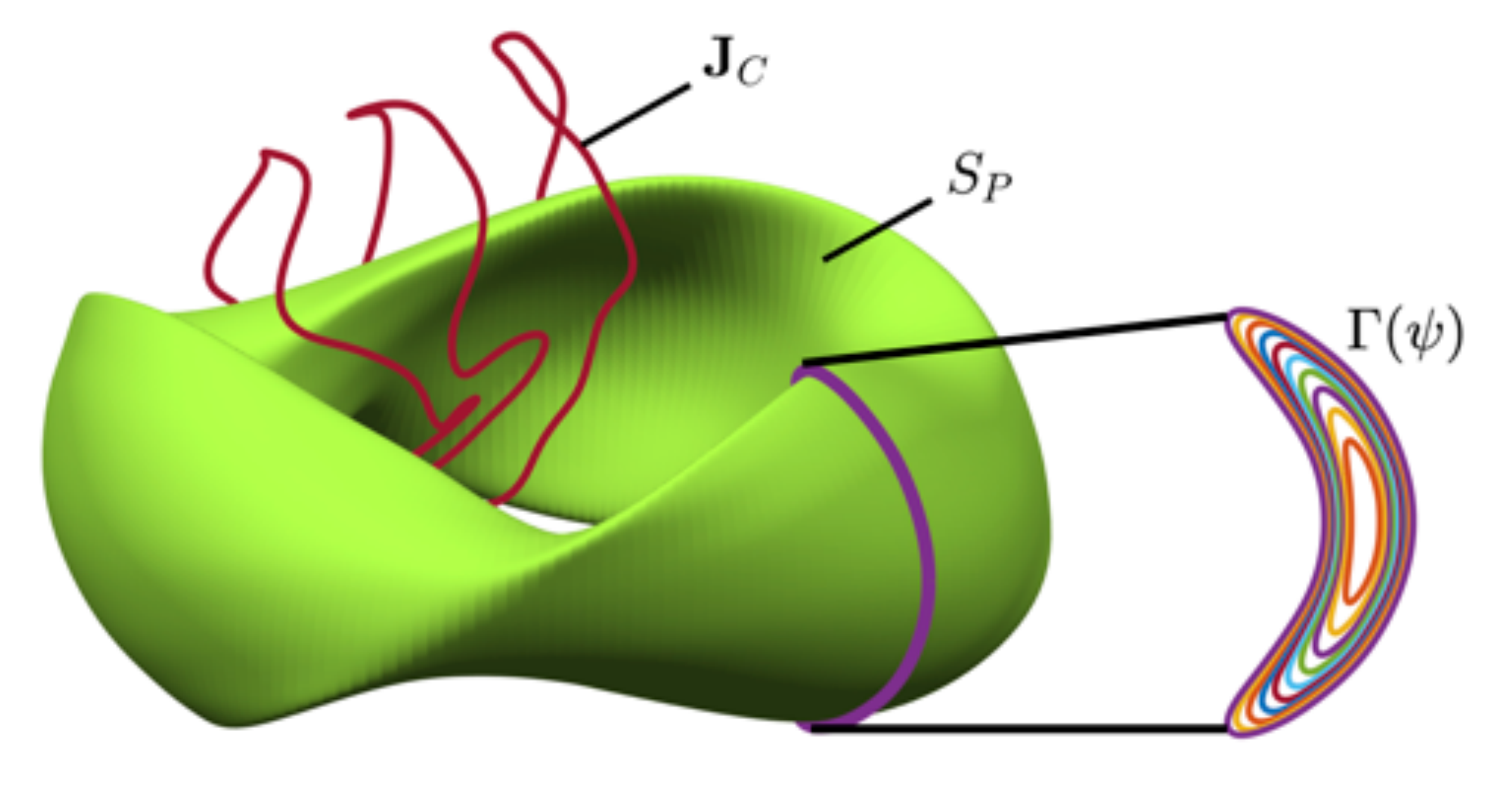}
    \caption{An equilibrium is computed with a fixed plasma boundary, $S_P$, or prescribed external currents, $\textbf{J}_C$. We assume the existence of a set of closed, nested toroidal surfaces, $\Gamma(\psi)$.}
    \label{fig:equilibrium}
\end{figure}

In the free-boundary approach, $S_P$ is varied until \eqref{eq:Bnormal_free} and \eqref{eq:jump_condition_free} are satisfied. These conditions \eqref{eq:Bnormal_free}-\eqref{eq:jump_condition_free} can also be obtained from a variational principle similar to \eqref{eq:energy_functional} including the vacuum region \cite{Bauer2012}. The free-boundary equilibrium problem is summarized in Table \ref{table:free_boundary}. Figure \ref{fig:equilibrium} shows the geometry of equilibrium calculations.

\begin{table}
\centering
    \begin{tabular}{|c|c|c|}
    \hline 
    PDE & BC & Given \\
         \hline \hline
         $\left(\nabla \times \textbf{B}\right) \times \textbf{B} = \mu_0 \nabla p(\psi)$ & $\textbf{B} \cdot \hat{\textbf{n}} \rvert_{S_P} = 0$ & $p(\psi)$, $\psi_0$, \& $\textbf{J}_C$ \\
         $\nabla \cdot \textbf{B} = 0$ & $S_P$ s.t.   $\left\{\begin{array}{c}(\textbf{B}_P + \textbf{B}_C) \cdot \hat{\textbf{n}} \rvert_{S_P} = 0  \\ \left[\left[ B^2/(2\mu_0) + p \right]\right]_{S_P} = 0  \end{array}\right.$  & $\iota(\psi)$ or $I_T(\psi)$ \\
         \hline 
    \end{tabular}
    \caption{Summary of free-boundary equilibrium PDEs. The magnetic field due to the plasma current, $\textbf{B}_P$, is computed from the Biot-Savart law \eqref{eq:Biot_Savart} or the virtual casing principle. The magnetic field due to the coil current, $\textbf{B}_C$, is computed from the Biot-Savart law.}
    \label{table:free_boundary}
\end{table}

Due to its efficiency and robustness, equilibrium optimization has primarily relied on this variational approach. There are several alternative approaches to obtaining numerical solutions to \eqref{eq:MHD_equilibrium} in a three-dimensional field. For example, sometimes the pressure is assumed to be piece-wise constant \cite{Hudson2011}, or the magnetic field is taken to resistively relax to an equilibrium \cite{Hirshman2011,Harafuji1989}. For a review of other 3D equilibrium models, see Chapter 11 in \cite{Imbert2019}.  

\subsection{Equilibrium optimization}
\label{sec:equilibrium_optimization}

The goal of stellarator optimization is ultimately to obtain the currents in the vacuum region needed to produce a stellarator configuration with desired physical properties. In this sense, it is logical to optimize the coils directly based on a free-boundary equilibrium. However, fixed-boundary optimization has been predominantly used for several practical reasons. Free-boundary equilibrium calculations tend to be more expensive, as they require iterations between an equilibrium solve and vacuum field calculations. This iterative scheme will not always converge in practice, hence the historical use of the more robust fixed-boundary method. It has also been suggested that fixed-boundary optimization may yield better equilibrium properties, as the model assumes the existence of at least one magnetic surface. With this approach, considerations of the physics properties of a configuration are largely decoupled from engineering considerations of the coils. As a second step, the electro-magnetic coils are designed, as described in Section \ref{sec:coil_optimization}.

The fixed-boundary optimization problem is,
\begin{align}
   \min_{S_P} f(S_P,\textbf{B}(S_P)),
   \label{eq:equilibrium_optimization}
\end{align}
where $\textbf{B}$ is seen as a function of $S_P$ through the fixed-boundary equations (Table \eqref{table:fixed_boundary}). Here, the objective function, $f$, quantifies physics or engineering properties of an equilibrium, such as those outlined in Section \ref{sec:stellarator_properties}. It is common to consider several objectives during an optimization, taking the objective function to be a sum of squares,
\begin{align}
    f(S_P,\textbf{B}(S_P)) = \sum_i \frac{\left(f_i(S_P,\textbf{B}(S_P))-f_i^{\text{target}}\right)^2}{\sigma_i^2}.
\end{align}
Here $f_i^{\text{target}}$ is the target value for objective $i$ and the $\sigma_i$ parameters quantify the relative weighting of the objectives.

Sometimes additional equality or inequality constraints are imposed,
\begin{subequations}
\begin{align}
    g(S_P, \textbf{B}(S_P)) &= 0 \\
    h(S_P, \textbf{B}(S_P)) &\le 0.
\end{align}
\end{subequations}
For example, the rotational transform might be constrained to be equal to a target value, or a maximum plasma volume may be imposed. Depending on the choice of optimization method, a local or global minimum will be sought. We will delay discussion of specific optimization algorithms until Section \ref{sec:challenges_outlook}. The fixed-boundary optimization method is implemented in the STELLOPT \citep{Spong1998,Reiman1999} and ROSE codes \cite{Drevlak2018}.

\subsection{Coil optimization}
\label{sec:coil_optimization}

Once a target plasma boundary, $S_P$, and equilibrium magnetic field, $\textbf{B}_0$, are identified from equilibrium optimization, electro-magnetic coils that are consistent with this equilibrium must be identified. 
The total magnetic field, $\textbf{B}$, can be decomposed into that which results from the target equilibrium plasma current, $\textbf{B}_0^P$, and that which results from the coil currents, $\textbf{B}_C$, computed from the Biot-Savart law. If the two are consistent, then the following relation will be satisfied,
\begin{align}
  0 = \textbf{B}_0^P(\textbf{x}) \cdot \hat{\textbf{n}}(\textbf{x}) + \frac{\mu_0}{4\pi}\int_{\mathbb{R}^3 \backslash V_P} d^3 x' \, \frac{\textbf{J}_C(\textbf{x}')\times(\textbf{x}-\textbf{x}') \cdot \hat{\textbf{n}}(\textbf{x})}{|\textbf{x}-\textbf{x}'|^3}, 
  \label{eq:coil_problem}
\end{align}
for all $\textbf{x} \in S_P$. In other words, the coils must be consistent with the last magnetic surface of the target equilibrium.

We note that the above is in the form of an integral equation of the first kind,
\begin{align}
       g(t) = \int_a^b \, ds K(t,s) f(s),
    \label{eq:integral_first_kind}
\end{align}
where $g(t)$ is given in some domain $t \in [c,d]$, $K(t,s)$ is a known kernel function, and $f(s)$ must be inferred. It is well-known that such problems are ill-posed \cite{Kress1989}, in the sense that small changes in the prescribed data, $g(t)$, result in large changes in the solution, $f(s)$, and a unique solution may not exist. 

Thus finding a solution for $\textbf{J}_C$ in \eqref{eq:coil_problem} is not well-posed. In some ways, this is advantageous, as there may be \textit{many} possible coil arrangements that provide the desired plasma configuration, and the one with the most favorable engineering properties can be chosen. However, one must be careful when obtaining numerical solutions to this problem so that noise in the prescribed data is not amplified. A classical technique for such problems is Tikhonov regularization \cite{Tikhonov1963}, in which \eqref{eq:integral_first_kind} is replaced by the optimization problem, 
\begin{align}
    \min_{f(t)} \left( \int_{c}^d \, dt \, \left(\int_a^b ds \,  K(t,s)f(s) - g(t) \right)^2 + \lambda \int_a^b ds \, \left(f(s)\right)^2 \right).
    \label{eq:tikhonov}
\end{align}
When $\lambda = 0$, the above is equivalent to \eqref{eq:integral_first_kind}. In order for the problem to be well-posed, additional information about the nature of the solution is provided. In \eqref{eq:tikhonov}, the assumption is made that the norm of the solution will be small. The regularization parameter, $\lambda$, describes the trade-off between obtaining a solution of \eqref{eq:integral_first_kind} and satisfying the expected or desired behavior of the solution. The regularized problem now has a unique solution and depends continuously on $g(t)$ for all $\lambda > 0$. 

In the context of coil optimization, we can choose the regularization term to coincide with the desired properties of our coils, such as small curvature or length. In this way, we seek coils that can be constructed more feasibly. We schematically write the modified coils problem as,
\begin{align}
    \min_{\textbf{J}_C} \left( \int_{S_P} d^2 x \, \left(\left(\textbf{B}_0^P + \textbf{B}_C\right)\cdot \hat{\textbf{n}} \right)^2 + \lambda \int_{\mathbb{R}^3\backslash V_P} d^3 x \, F(\textbf{J}_C)^2 \right),
    \label{eq:regularized_coil_problem}
\end{align}
where $\textbf{B}_C$ is the magnetic field due to $\textbf{J}_C$ computed from the Biot-Savart law \eqref{eq:Biot_Savart} and $F(\textbf{J}_C)$ is some function of the coil currents that characterizes desired engineering properties.

\subsubsection{Coil properties}
\label{sec:coil_properties}

Given the freedom inherent in designing stellarator coils, we now outline some desired properties for a set of stellarator coils. 
\begin{itemize}
    \item \textit{Physics objectives} - Our primary interest is to find a coil set consistent with our target fixed-boundary equilibrium. This objective is typically quantified by the error in obtaining the last magnetic surface, as in \eqref{eq:coil_problem}. In practice, some physics metrics depend very sensitively on coil perturbations, so other critical physics properties of the equilibrium can be included in the coil optimization, such as the magnetic ripple on axis (a measure of quasi-symmetry) or the rotational transform \cite{Drevlak1999}.
    \item \textit{Manufacturability} - Coil shapes have a minimum allowable radius of curvature due to their finite build, and overly-complex coils may be difficult to manufacture without excessive cost \cite{Strykowsky2009}. There are many metrics suggested for quantifying complexity, such as length \cite{Zhu2018}, torsion \cite{Hudson2018}, and curvature \cite{Brown2015}.
    \item \textit{Stresses} - Complex support structures must be built to maintain coil locations and shapes under their large electro-magnetic, thermal, and gravitational stresses.
    As coils tend to become more circular and planar under electro-magnetic stresses \cite{Kisslinger1999}, it is advantageous to minimize curvature and non-planarity when possible.  
    \item \textit{Access to the plasma chamber} - There should be sufficient distance between coils to allow for diagnostic ports and ease of machine assembly and maintenance. Coils with relatively straight sections on the outboard side may particularly provide improved access \cite{Brown2015}.
    \item \textit{Coil-plasma separation} - In a reactor, coils should be designed sufficiently far from the plasma boundary to allow space for neutron shielding, a blanket, the first wall, coil casing, and the vacuum vessel. Increased coil-plasma distance can also reduce the magnetic field ripple due to the finite number of coils. The minimum coil-plasma distance effectively sets the required size of a reactor, as $\approx 1.3$ m is needed for the breeding module \cite{Najmabadi2008}. Achieving a sufficient coil-plasma distance is difficult in practice: coils that are very far from the plasma may become overly-complex, as shaping components of the magnetic field decay rapidly with distance \cite{Landreman2016}. 
\end{itemize}

Several approaches to achieve such objectives are described in Section \ref{sec:current_potential} and Section \ref{sec:filamentary_methods}.

\subsubsection{Current potential methods}
\label{sec:current_potential}

\begin{figure}
    \centering
    \begin{subfigure}[b]{0.49\textwidth}
    \includegraphics[trim=1cm 6cm 2cm 6cm,clip,width=1.0\textwidth]{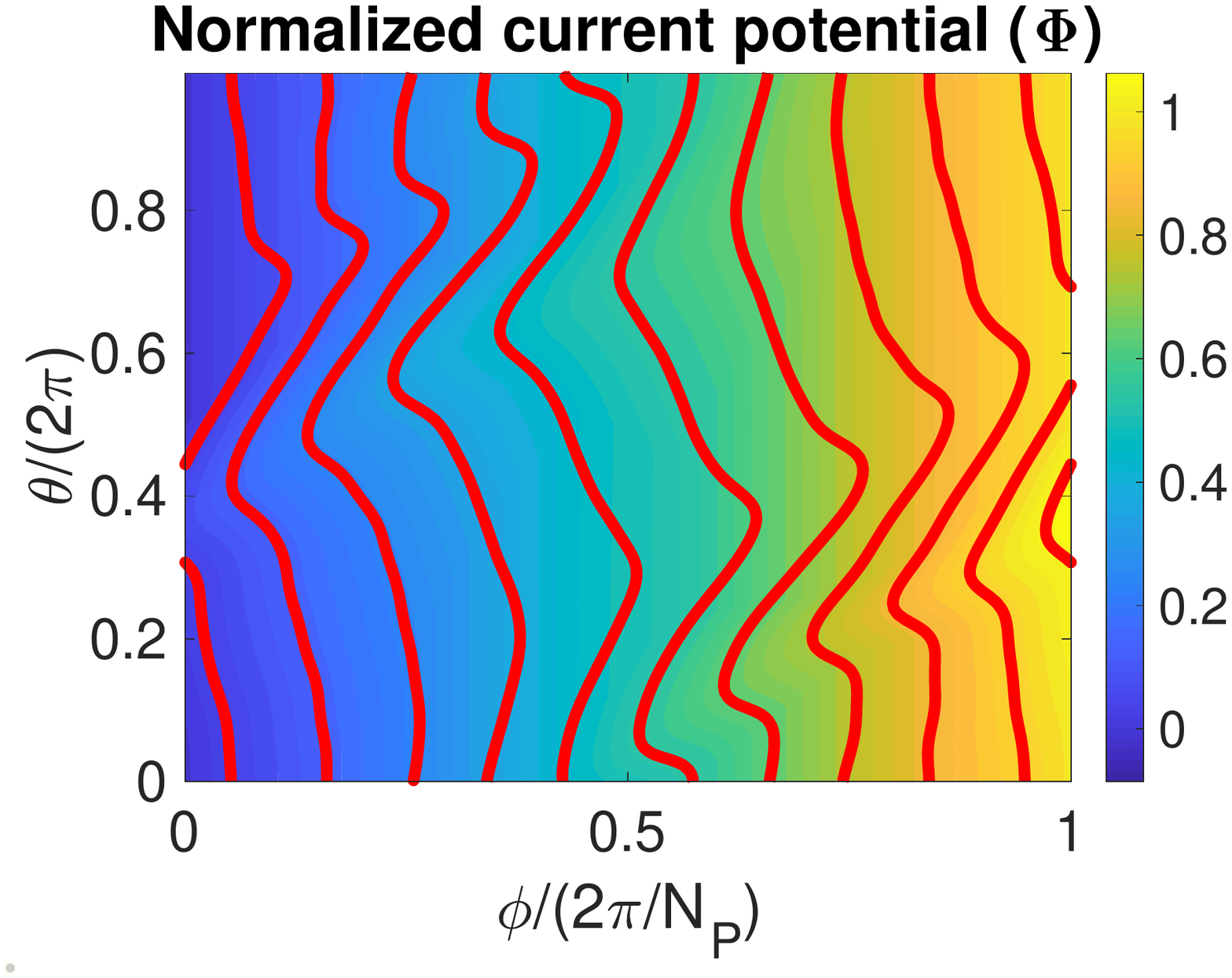}
    \caption{}
    \end{subfigure}
    \begin{subfigure}[b]{0.49\textwidth}
    \includegraphics[trim=6cm 10cm 6cm 10cm,clip,width=1.0\textwidth]{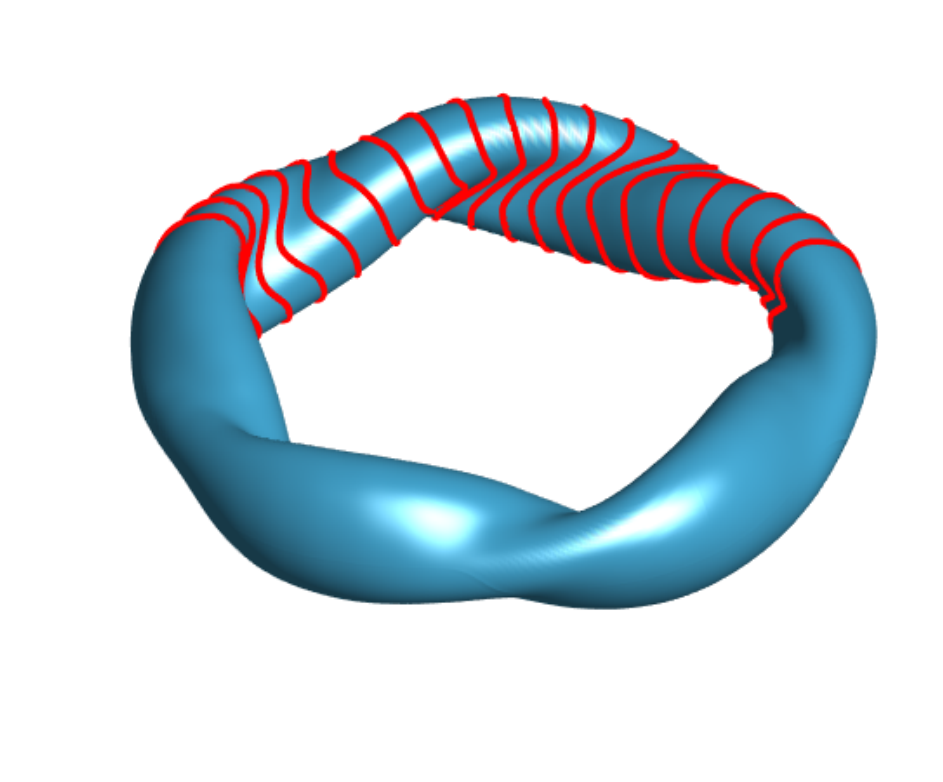}
    \caption{}
    \end{subfigure}
    \begin{subfigure}[b]{0.49\textwidth}
    \includegraphics[trim=6cm 10cm 6cm 10cm,clip,width=1.0\textwidth]{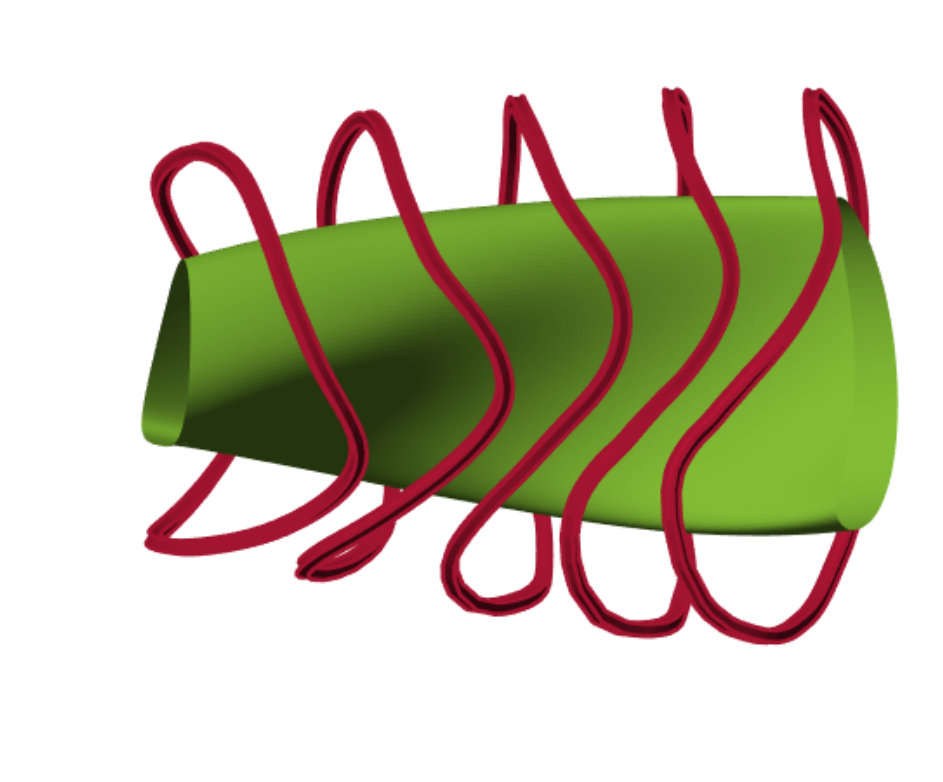}
    \caption{}
    \end{subfigure}
    \caption{An example of a REGCOIL calculation for the W7-X standard configuration equilibrium. The winding surface is taken to be a surface uniformly offset from $S_P$ by 0.5 m. (a) The current potential and the uniformly-spaced contours taken for the coil set. (b) The coil set computed from the contours on the winding surface. (c) The 5 unique coils in one half period and the plasma surface.}
    \label{fig:regcoil}
\end{figure}

The first stellarator coil design code, NESCOIL \cite{Merkel1987}, assumes that all currents in the vacuum region lie on a closed toroidal surface called the winding surface, $S_C$. This method was used to design the modular coils of W7-AS \cite{Hirsch2008}, W7-X \cite{Beidler1990}, and HSX \cite{Almagri1998} and was later generalized to include regularization in the REGCOIL \cite{Landreman2017} code. 
In the limit of a large number of coils, we can describe a set of discrete coils by a continuous current density on $S_C$, 
\begin{align}
    \textbf{J} = \delta(b(\textbf{x})) \textbf{J}_C(\theta,\phi).
\end{align}
Here $b(\textbf{x})$ is the signed-distance function \cite{Osher2004},
\begin{align}
    b(\textbf{x}) = \left\{ \begin{array}{c}
    -d(\textbf{x},S_C) \hspace{0.73cm} \textbf{x} \in V_C \\
    0 \hspace{2.1cm} \textbf{x} \in S_C \\
    d(\textbf{x},S_C) \hspace{1.0cm} \textbf{x} \not \in V_C
    \end{array}\right. .
\end{align}
The volume enclosed by $S_C$ is $V_C$ and $d(\textbf{x},S_C)$ is the shortest distance from $\textbf{x}$ to any point on $S_C$. The signed distance function is also discussed in Section \ref{sec:shape_optimization}. The surface current $\textbf{J}_C$ is a function of the two angles, $\theta$ and $\phi$, parameterizing the position on $S_C$. As a consequence of Ampere's law (Appendix \ref{app:current_potential}), the continuous surface current can be written as,
\begin{align}
    \textbf{J}_C = \hat{\textbf{n}} \times \nabla \Phi.
\end{align}
We can note that current will flow along the contours of $\Phi$, as $\textbf{J}_C \cdot \nabla \Phi = 0$. In this way, once $\Phi$ is computed, the coil shapes can be chosen to be a set of the contours of $\Phi$. As we will see in Section \ref{ch:adjoint_winding_surface}, it is possible to construct an objective function that is a convex function of $\Phi$, possessing a unique global minimum that can be obtained through linear least-squares. Thus current potential methods are particularly robust and efficient, though based on some severe assumptions. Coil complexity can be approximated from the properties of the current potential. In REGCOIL, this is done with the norm of the current density,
\begin{align}
    \chi^2_J = \int_{S_C} d^2 x \, |\textbf{J}_C|^2,
\end{align}
as large values of $\chi^2_J$ indicate small coil-coil spacing. An example REGCOIL calculation is shown in Figure \ref{fig:regcoil}.

\subsubsection{Filamentary methods}
\label{sec:filamentary_methods}

Other coil design codes instead assume that all currents in the vacuum region are confined to filamentary lines, $\{C_k\}$, taken to be the center of each winding pack. This assumption is again an idealization, as stellarator coils have a finite build consisting of several layers, each with several turns of the conducting material. However, the filamentary method is more realistic than current potential methods, as it accounts for the ripple due to the finite nature of coils. The lines and the current through each are optimized to minimize some objective function that includes the normal field error on $S_P$ in addition to engineering objectives, which serve as a form of regularization. For example, the FOCUS code \cite{Zhu2018} uses the coil length as a form of regularization, and the COILOPT code \cite{Strickler2002} includes the coil-plasma separation, coil-coil separation, and the coil curvature. These optimization problems are generally nonlinear and non-convex so that the resulting local minimum will depend on the initial guess. For this reason, a current potential solution can be used to initialize the optimization with filamentary methods.

\subsection{Challenges and outlook}
\label{sec:challenges_outlook}

\begin{figure}
    \centering
    \begin{subfigure}[b]{0.49\textwidth}
    \centering
    \includegraphics[trim=0cm 0cm 1cm 0cm,clip,width=1.0\textwidth]{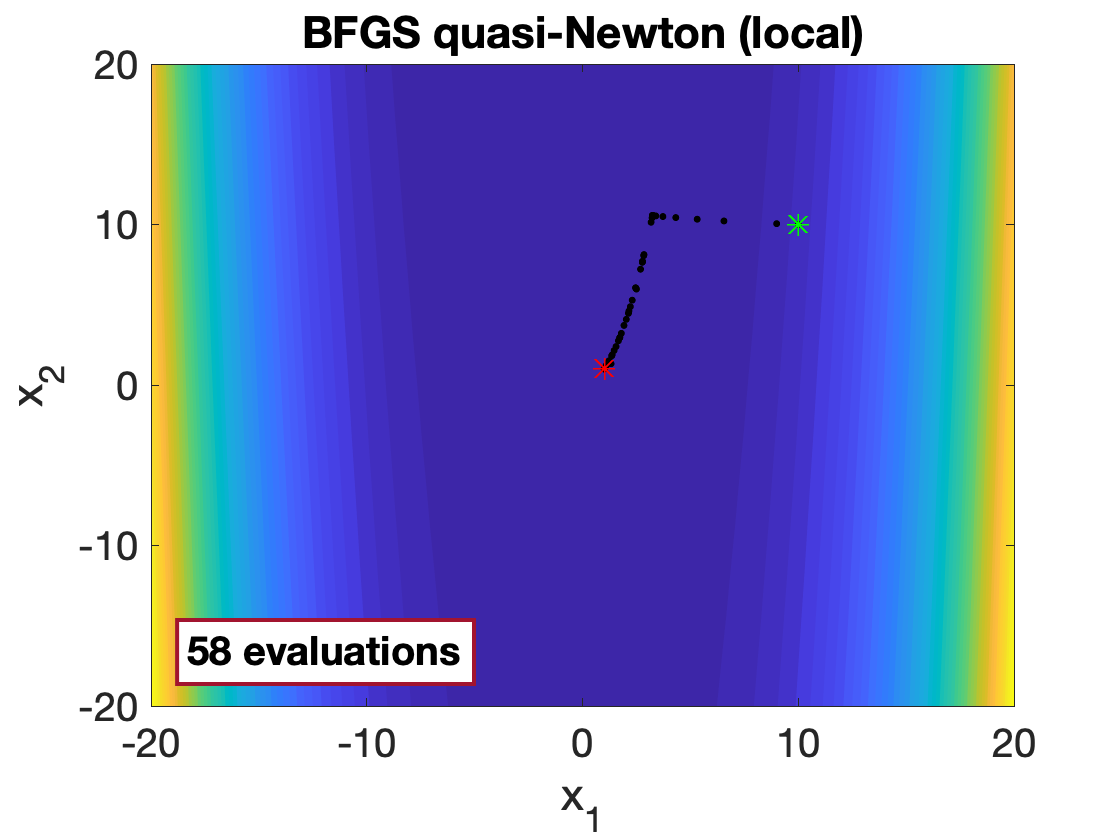}
    \end{subfigure}
    \begin{subfigure}[b]{0.49\textwidth}
    \centering
    \includegraphics[trim=0cm 0cm 1cm 0cm,clip,width=1.0\textwidth]{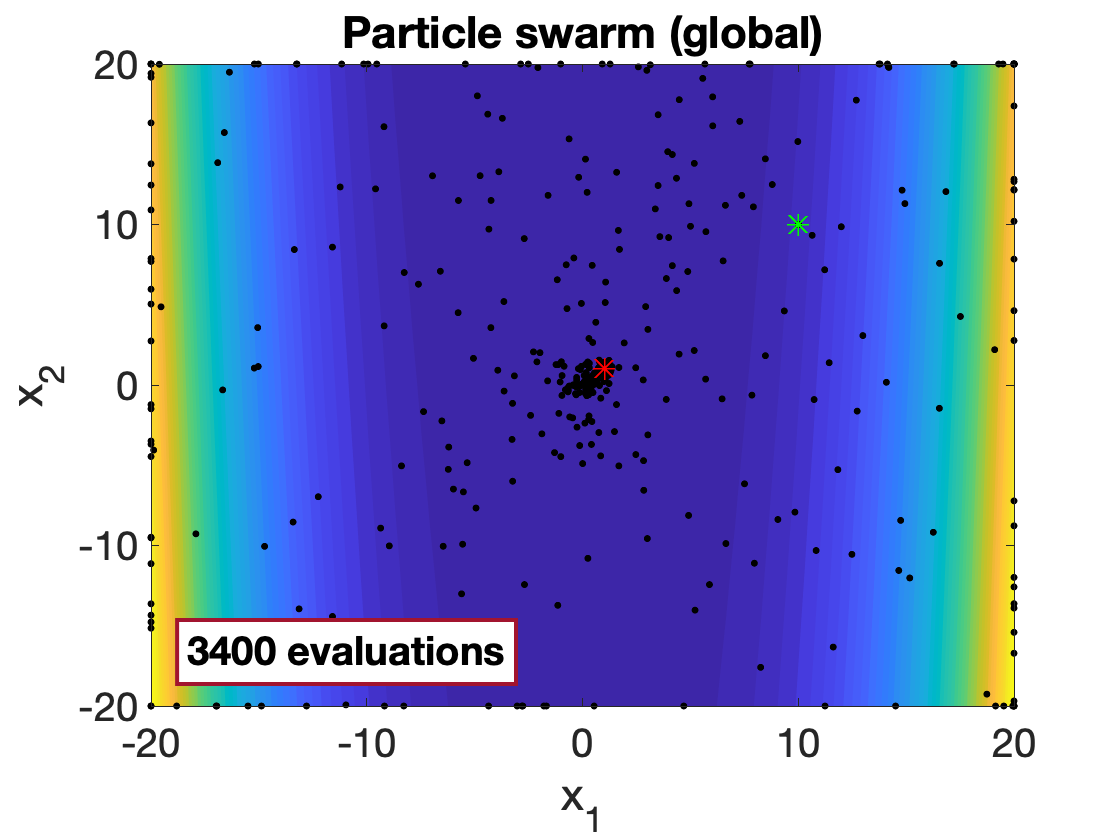}
    \end{subfigure}
    \caption{The optimization path of the gradient-based BFGS quasi-Newton local optimization method and gradient-free particle swarm global optimization method on the 2D Rosenbrock function \eqref{eq:rosenbrock}. The BFGS optimization is initialized at $(x_1,x_2) = (10,10)$ and converges to the optimum at $(1,1)$ in 58 function evaluations, using an analytic gradient to obtain the descent direction. The particle swarm optimization is initialized with a swarm of 20 particles at $(10,10)$ and converges to the optimum at $(1,1)$ in 3400 evaluations. The gradient-based method converges more directly toward a minimum, while the gradient-free method converges in a scattered way requiring excessive function evaluations. For (a), the optimization was terminated when the maximum of the absolute value of the gradient elements was less than $10^{-8}$, and for (b), the optimizations was terminated when the relative change in the objective function over the previous 20 iterations was less than $10^{-8}$.}
    \label{fig:Rosenbrock_1}
\end{figure}

\begin{figure}
    \centering
    \includegraphics[width=0.6\textwidth]{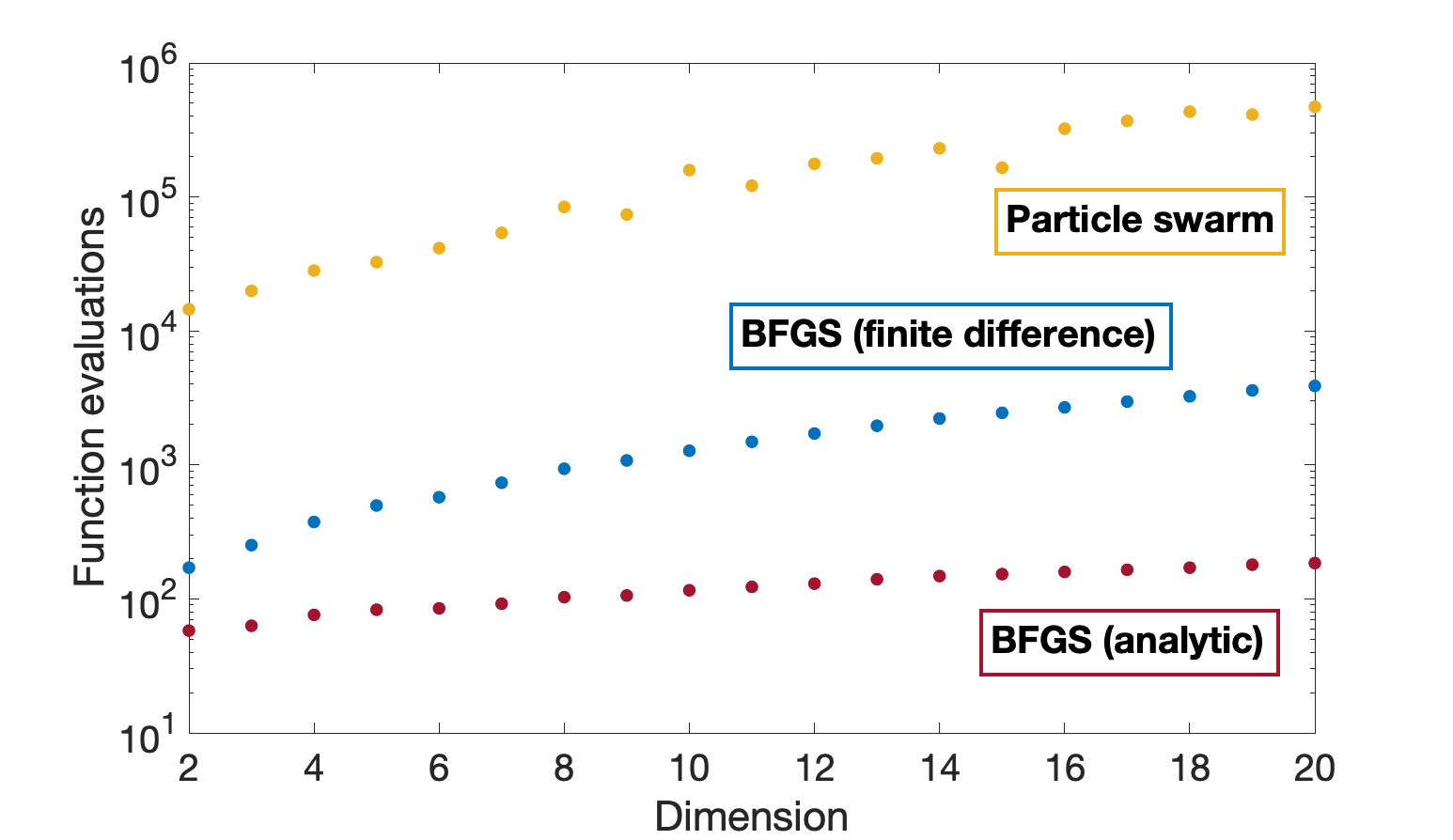}
    \caption{The number of function evaluations required for convergence to the minimum of the $N$-dimensional Rosenbrock function \eqref{eq:rosenbrock} as a function of the dimension. Results are shown for the gradient-based BFGS algorithm with finite-difference and analytic gradients and the gradient-free particle swarm method. We note that the gradient-free and finite-difference gradient-based methods scale poorly with the dimension. Knowledge of analytic gradients reduces the associated cost by several orders of magnitude in comparison. The cost reduction provided by analytic derivatives increases with increasing dimension. For the BFGS algorithm the optimization was terminated when the maximum of the absolute value of the gradient elements was less than $10^{-8}$, and for the particle swarm algorithm the optimizations was terminated when the relative change in the objective function over the previous 20 iterations was less than $10^{-8}$.}
    \label{fig:Rosenbrock_2}
\end{figure}

Although there have arguably been significant successes in optimized stellarator design, there is still room for improvement in the algorithms and numerical methods. Specifically, we aim to address several major challenges that arise in the optimization of stellarator configurations. 
\begin{enumerate}
    \item \textit{Coil complexity} - In the standard two-step approach, coil design is decoupled from equilibrium optimization. While this may allow for improved physics properties, the resulting equilibrium may require overly-complex coils that cannot be manufactured economically or are not consistent with engineering constraints. As was stated in the 2018 report of the National Stellarator Coordinating Committee \cite{Gates2018}, \begin{quote}
    ``The highest priority for technology is to better integrate the engineering design with the physics design at the earliest possible stage." 
    \end{quote}
    For this reason, it is favorable to include coil complexity metrics in equilibrium optimization. As an example, one approach is to compute the properties of the current potential (Section \ref{sec:current_potential}) on a winding surface that is uniformly offset from the plasma surface \cite{Drevlak2018} during fixed-boundary optimization. It has also been proposed that properties of the optimal filamentary coils for a given plasma boundary be included in equilibrium optimization \cite{Hudson2018}. Alternatively, the coils can be directly optimized with a free-boundary method. This approach was implemented in the late stages of the NCSX design  \cite{Hudson2002,Strickler2003} and in the QPS (Quasi-Poloidally Symmetric Stellarator) design \cite{Strickler2004}, resulting in simultaneous attainment of engineering feasibility and desired plasma properties. Another tactic to reduce coil complexity is replacing non-planar modular coils by permanent magnets \cite{Helander2019,Zhu2019b}.
    \item \textit{Non-convexity} - The optimization problems that arise in stellarator design are often non-convex (except for the current potential methods described in Section \ref{sec:current_potential}). While convex optimization problems can be solved in polynomial time (Chapter 1 in \cite{Boyd2004}), obtaining the global optimum of a non-convex optimization problem is generally $NP$-hard. As global optima are difficult to locate, it is common to apply algorithms that instead converge to local optima. Such methods are sensitive to the initial conditions and tend to get ``stuck" in small local minima or saddle points. For this reason, it is very valuable to have initial configurations that are close to the desired configuration. One approach is to begin with an analytic construction of an equilibrium close to quasi-symmetry or omnigeneity by employing an expansion about the magnetic axis \cite{Landreman2019,Landreman2018b,Plunk2019}. 
    
    \indent Gradient information is invaluable for obtaining the local minimum of an objective function. While there are some algorithms for derivative-free local optimization, they typically are only effective for small problems (Chapter 9 in \cite{Nocedal2006}). Gradient information is also useful for global optimization; for example, with a multi-start approach, many local optimization problems are solved to approximately obtain the global minimum. As considerations of the gradient will be central to this Thesis, we will discuss this topic further in Chapter \ref{ch:mathematical_fundamentals}. 
    
   In Figure \ref{fig:Rosenbrock_1} we show a benchmark of several optimization problems on the Rosenbrock function, 
    \begin{align}
       f(\{x_i\}_{i=1}^N) = \sum_{i=1}^{N-1} 100 (x_{i+1}-x_i^2)^2 + (x_i-1)^2,
    \label{eq:rosenbrock}
    \end{align}
    with $N = 2$, a non-convex function with a long, thin valley that is often used to benchmark optimization algorithms. 
    We can note that the gradient-based BFGS method converges rather directly toward the optimum. In contrast, the gradient-free particle swarm method takes a scattered trajectory and requires many additional functional evaluations. 
    \item \textit{High-dimensionality} - Often, the optimization problems that arise in stellarator design require navigation through the high-dimensional spaces that describe the outer boundary of the plasma or coil shapes. While such shapes are infinite-dimensional in reality, often they are parameterized with Fourier series, and only a finite number of modes are retained during the optimization. The number of parameters used in practice to describe such shapes is typically $\mathcal{O}(10^2)$ \cite{Zarnstorff2001}. We show a benchmark of the $N$-dimensional Rosenbrock function \eqref{eq:rosenbrock} in Figure \ref{fig:Rosenbrock_2}, noting that the number of function evaluations required to obtain the optimum scales poorly with $N$ for the gradient-free methods and finite difference based gradient-free methods. As computing the gradient with a finite-difference method requires $\mathcal{O}(N)$ function evaluations, the associated cost is reduced significantly if analytic derivatives are available. Stellarator equilibrium optimization has historically proceeded with gradient-free methods, such as genetic algorithms \cite{Miner2001} and the Brent algorithm \cite{Drevlak2018}, or gradient-based methods with finite-difference gradient calculations \cite{Spong1998}. Recently, gradient-based optimization of coils shapes has begun to take advantage of analytic gradient and Hessian calculations \cite{Zhu2018,Zhu2018b}. However, for many functions of interest, it is not so simple to compute the analytic derivative, as the objective function may depend on the solution to a system of equations. For such objectives, analytic derivatives can be computed with an adjoint method. This topic will be discussed in detail in Chapter \ref{ch:mathematical_fundamentals} and throughout the Thesis. 
    \item \textit{Tight engineering tolerances} - Once an optimal design is identified, engineering and metrology coil tolerances must be determined from the allowable deviations of physics parameters. In the NCSX design, it was determined that coil tolerances of $\approx1.5$ mm were required to achieve good flux surfaces in 90\% of the plasma volume \cite{Brooks2003}. These tight modular coil tolerances were identified as the largest contributor to the cost growth of the project, ultimately leading to the termination of its funding \cite{Strykowsky2009}. The first recommendation that came out of an analysis of the NCSX project was,
    \begin{quote}
    ``Be critical and surgical in requiring either small tolerances or low magnetic permeability requirements\dots The impact is not only in increased cost but schedule stretch-out which has a large management overhead cost." 
    \end{quote}
    One approach to address this challenge is to optimize the expected value of an objective function over a distribution of possible deviations, known as stochastic optimization. This technique has been shown to increase the tolerances of an optimized coil set \cite{Lobsien2018,Lobsien2020}. There has also been a recent development of tools for the efficient evaluation of tolerance information to avoid costly parameter scans or Monte Carlo sampling methods \cite{Brooks2003,Hammond2016}. The eigenvectors of the Hessian matrix illuminate the most sensitive perturbation directions at a local minimum \cite{Zhu2018,Zhu2019}, and in this Thesis, we will discuss the shape gradient approach \cite{Landreman2018}.
\end{enumerate}

\section{Overview of this Thesis}
\label{sec:overview}

This Thesis aims to address each of the challenges outlined in the previous Section. The focus will be on adjoint methods, which allow for efficient analytic gradient calculations. With such gradient information available, we can navigate through high-dimensional, non-convex spaces that arise in stellarator design with gradient-based methods, addressing objectives 2 and 3. Derivatives obtained from the adjoint method can also be used to analyze local sensitivity to perturbations using the shape gradient, addressing objective 4. Specific applications of the adjoint method described in this Thesis will enable efficient free-boundary coil optimization or coupled coil-plasma optimization, addressing objective 1. 

We begin in Chapter \ref{ch:mathematical_fundamentals} with an introduction to some mathematical fundamentals that lay the groundwork for this Thesis, including an overview of
shape optimization and adjoint methods. Chapter \ref{ch:adjoint_winding_surface} describes an adjoint method for the optimization of the coil winding surface for minimal coil complexity. Chapter \ref{ch:adjoint_neoclassical} describes an adjoint method for the optimization of several neoclassical figures of merit local to a magnetic surface, including radial fluxes and the bootstrap current. Chapter \ref{ch:adjoint_MHD} describes an adjoint method for the optimization of functions which depend on MHD equilibrium solutions, such as those that arise in fixed and free-boundary optimization. The adjoint method discussed in Chapter \ref{ch:adjoint_MHD} requires the solution of linearized MHD equilibrium equations, which are discussed in Chapter \ref{ch:linearized_mhd}. In Chapter \ref{ch:conclusions}, we summarize and discuss ongoing and future research related to this Thesis.

\renewcommand{\thechapter}{2}

\chapter{Mathematical fundamentals}
\label{ch:mathematical_fundamentals}

\section{Shape optimization}
\label{sec:shape_optimization}

The design of a stellarator requires optimizing in the space of shapes: equilibrium design involves optimization of the shape of the plasma boundary, $S_P$, and coil design involves optimization of the shapes of filamentary coils or toroidal winding surfaces. The mathematical field of shape optimization has developed to study such problems, contributing to the design of aerodynamic car bodies \cite{Othmer2014} and airplane wings with increased lift \cite{Mohammadi2004}. In this Section, we briefly outline several concepts from this field. We refer to several fundamental textbooks \cite{Pironneau1982,Haslinger2003,Choi2006,Delfour2011} and a Ph.D. thesis with a gentler introduction \cite{Dekeyser2014}.

\subsection{Definitions and identities}

Consider some functional, $f$, which depends on the shape of some domain, $\Gamma$. In order to compute the derivative of $f$, we must first identify a deformation field, $\delta \textbf{x}$, which describes the change of the shape. If the shape begins in a state $\Gamma$, the shape deformed in the direction $\delta \textbf{x}$ by magnitude $\epsilon$ is $\Gamma_{\epsilon} = \{\textbf{x}_0 + \epsilon \delta \textbf{x}(\textbf{x}_0) : \textbf{x}_0 \in \Gamma \}$. In this way, we can define the shape derivative of $f$ as,
\begin{align}
    \delta f(\Gamma;\delta \textbf{x}) \equiv \lim_{\epsilon \rightarrow 0} \frac{f(\Gamma_{\epsilon})-f(\Gamma)}{\epsilon}. 
\end{align}
This is a functional derivative in the direction $\delta \textbf{x}$ (a Gateaux functional derivative).

We can prove some useful properties of the shape derivative for specific choices of functional,
\begin{subequations}
\begin{align}
    J_1(\Gamma) &= \int_{\Gamma} d^3 x \, j_1(\Gamma) \label{eq:volume_functional} \\
    J_2(\Gamma) &= \int_{\partial \Gamma} d^2 x \, j_2(\Gamma) , \label{eq:surface_functional} 
\end{align}
\end{subequations}
volume and surface integrals. 

\begin{figure}
\begin{subfigure}[b]{0.49\textwidth}
\centering
\includegraphics[trim=35.5cm 10cm 31cm 12cm,clip,width=1.0\textwidth]{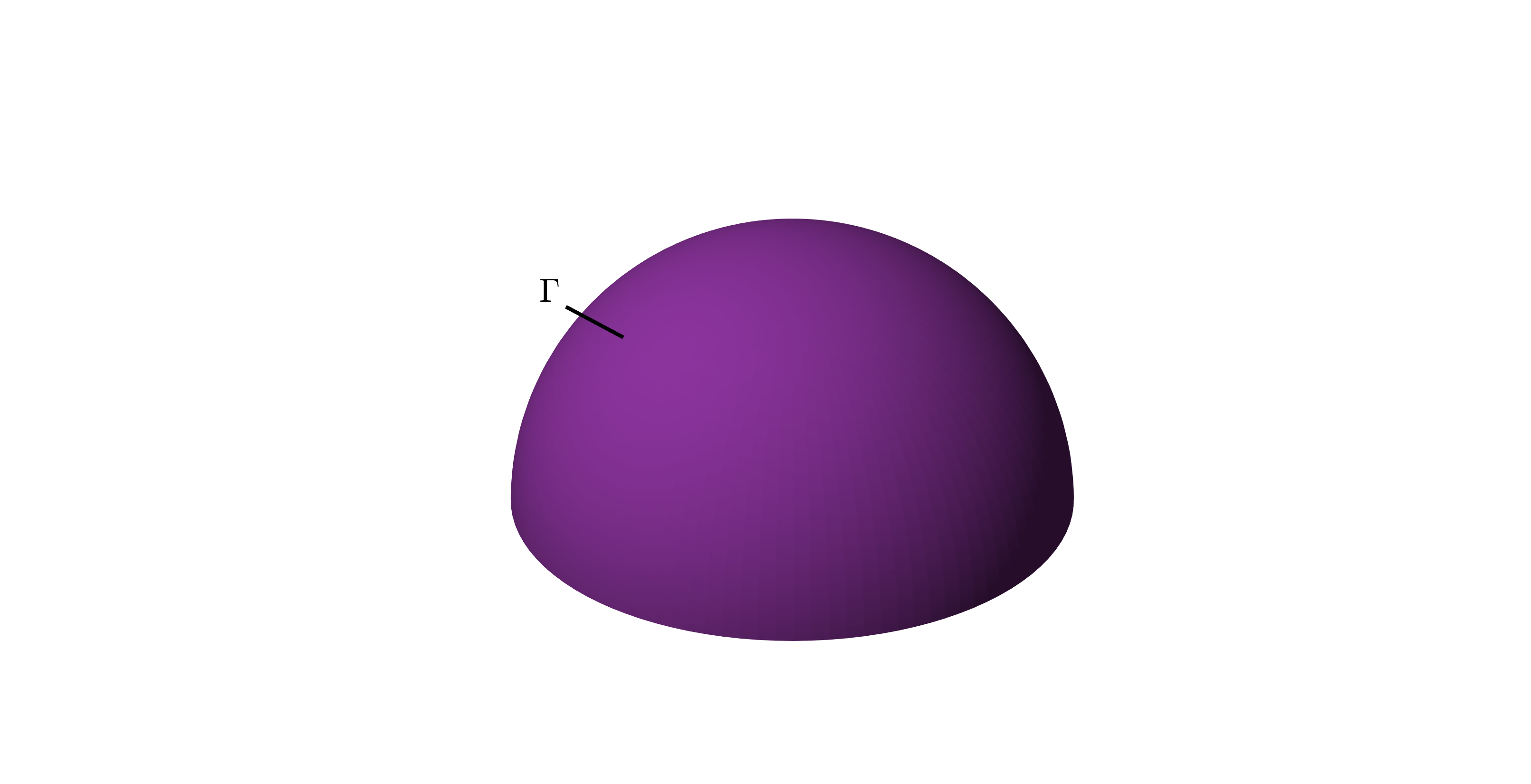}
\caption{}
\end{subfigure}
\begin{subfigure}[b]{0.49\textwidth}
\includegraphics[trim=37cm 10cm 36.4cm 12cm,clip,width=1.0\textwidth]{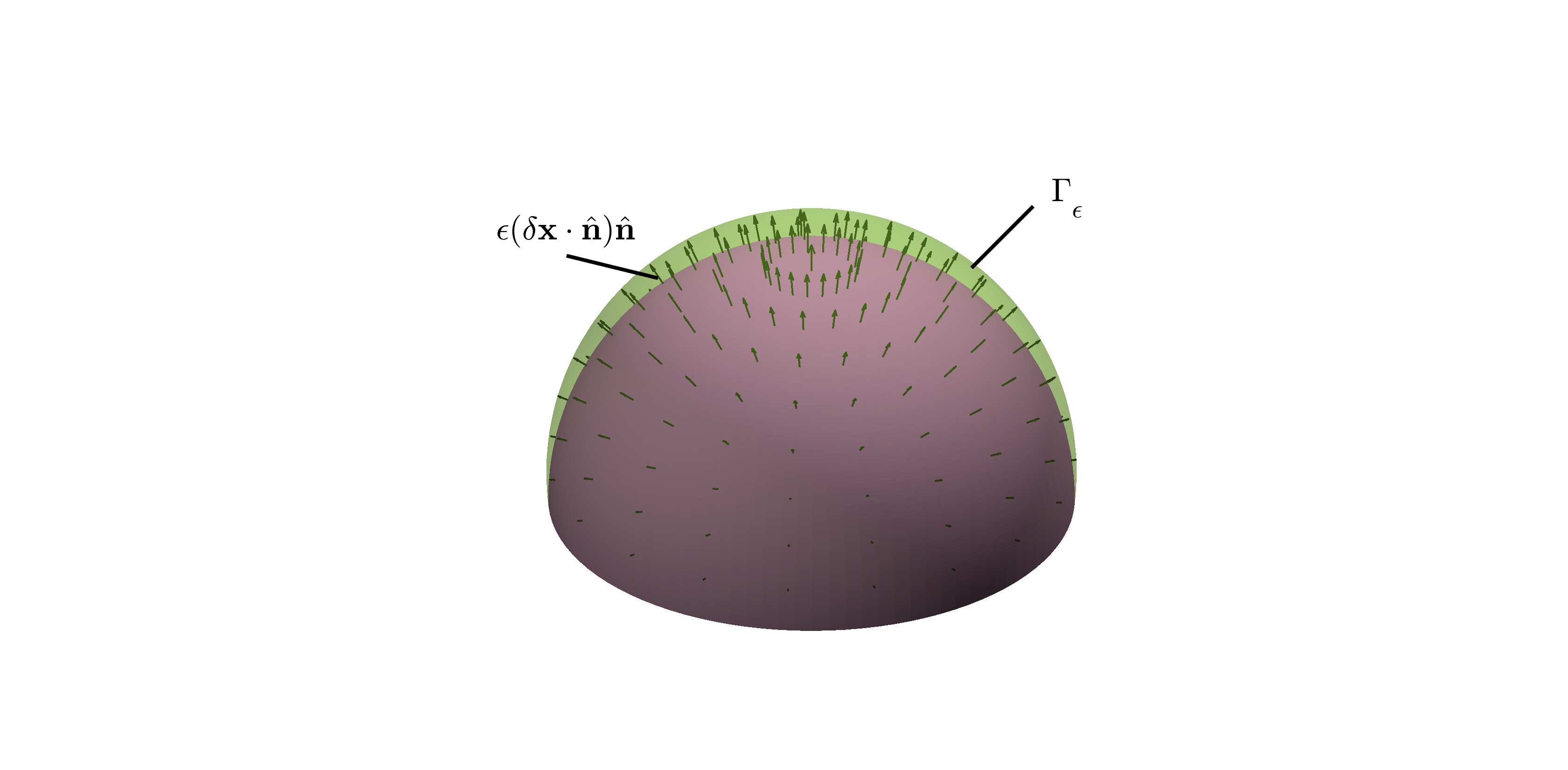}
\caption{}
\end{subfigure}
\caption{(a) An unperturbed volume, $\Gamma$. (b) The normal perturbation field of magnitude $\epsilon \delta \textbf{x} \cdot \hat{\textbf{n}}$ (black) and the perturbed volume, $\Gamma_{\epsilon}$ (green). We can see that the linear change in volume associated with the perturbation field is $\delta V = \int_{\partial \Gamma} d^2 x \, \delta \textbf{x} \cdot \hat{\textbf{n}}$.}
\label{fig:Reynolds}
\end{figure}

For volume-integrated functionals, the shape derivative can be evaluated by noting the Jacobian of the transformation $\textbf{x} \in \Gamma \rightarrow \textbf{x} \in \Gamma_{\epsilon}$ is given by $\textbf{I} + \epsilon \nabla \delta \textbf{x}$, where $\textbf{I}$ is the identity tensor. This allows us to relate the volume integral over $\Gamma_{\epsilon}$ to a volume integral over $\Gamma$,
\begin{align}
  \delta J_1(\Gamma;\delta \textbf{x}) &= \lim_{\epsilon\rightarrow 0} \frac{1}{\epsilon}\left(\int_{\Gamma_{\epsilon}} d^3 x \, j_1(\Gamma_{\epsilon}) - \int_{\Gamma} d^3 x \, j_1(\Gamma) \right)  \nonumber \\
  &= \lim_{\epsilon\rightarrow 0} \frac{1}{\epsilon}\int_{\Gamma} d^3 x \, \left[\det\left(\textbf{I} + \epsilon \nabla \delta \textbf{x} \right) j_1(\Gamma_{\epsilon}) \rvert_{\textbf{x} + \epsilon \delta \textbf{x}} - j_1(\Gamma)\right]. 
\end{align}
Noting that $j_1(\Gamma_{\epsilon}) \rvert_{\textbf{x} + \epsilon \delta \textbf{x}} = j_1(\Gamma) \rvert_{\textbf{x}} + \epsilon \delta j_1(\Gamma;\delta \textbf{x}) + \epsilon \delta \textbf{x} \cdot \nabla j_1(\Gamma) + \mathcal{O}(\epsilon^2)$ we have,
\begin{align}
\delta J_1(\Gamma;\delta \textbf{x}) &= \int_{\Gamma} d^3 x \, \left(\delta j_1(\Gamma;\delta \textbf{x}) + \delta \textbf{x} \cdot \nabla j_1(\Gamma) + \der{}{\epsilon} \left(\det(\textbf{I} + \epsilon \nabla \delta \textbf{x})\right)\bigg \rvert_{\epsilon = 0}j_1(\Gamma) \right). 
\end{align}
The derivative of the determinant of a matrix can be computed from Jacobi's formula, $d/dt \left(\det(A(t))\right) = \det(A(t)) \text{tr}( A(t)^{-1} A'(t))$,
\begin{align}
    \delta J_1(\Gamma;\delta \textbf{x}) &= \int_{\Gamma} d^3 x \, \left[\delta j_1(\Gamma;\delta \textbf{x}) + \delta \textbf{x} \cdot \nabla j_1(\Gamma) +  \left(\nabla \cdot \delta \textbf{x}\right) j_1(\Gamma) \right]. 
\end{align}
From the divergence theorem, we arrive at the following form for the shape derivative of volume-integrated functionals,
\begin{align}
    \delta J_1(\Gamma;\delta \textbf{x}) = \int_{\Gamma} d^3 x \, \delta j_1(\Gamma;\delta \textbf{x}) +
    \int_{\partial \Gamma} d^2 x \,  \delta \textbf{x} \cdot \hat{\textbf{n}} j_1(\Gamma).
    \label{eq:transport_theorem_volume}
\end{align}
The first term accounts for the Eulerian change to $j_1$ while the second term accounts for the motion of the boundary. In fluid mechanics, this relation is sometimes referred to as the Reynolds transport theorem (Chapter 2 in \cite{Leal2007}), which describes the time derivative of integrated quantities associated with a moving fluid. A physical picture of this result is given in Figure \ref{fig:Reynolds}.

We can now use \eqref{eq:transport_theorem_volume} to obtain the shape derivative of the surface-integrated functional \eqref{eq:surface_functional}. To do so, we recall that the normal vector can be expressed as $\hat{\textbf{n}} = \nabla b \rvert_{\partial \Gamma}$, where $b$ is the signed distance function \cite{Osher2004},
\begin{align}
    b(\textbf{x}) = \left\{ \begin{array}{c}
    -d(\textbf{x},\partial \Gamma) \hspace{0.68cm} \textbf{x} \in \Gamma \\
    0 \hspace{2.5cm} \textbf{x} \in \partial \Gamma \\
    d(\textbf{x},\partial \Gamma) \hspace{1.0cm} \textbf{x} \not \in \Gamma
    \end{array}\right. ,
\end{align}
and $d(\textbf{x},\partial \Gamma)$ is the shortest distance from $\textbf{x}$ to any point on $\partial \Gamma$. This can be seen by noting that $\hat{\textbf{n}}$ points outward, in the direction of increasing $b(\textbf{x})$, and the shortest path between a point near $\partial \Gamma$ and $\partial \Gamma$ will be along the normal direction. As $b(\textbf{x})$ measures Euclidian distance, $\nabla b$ has unit length.  

\begin{figure}
    \centering
    \includegraphics[width=0.6\textwidth]{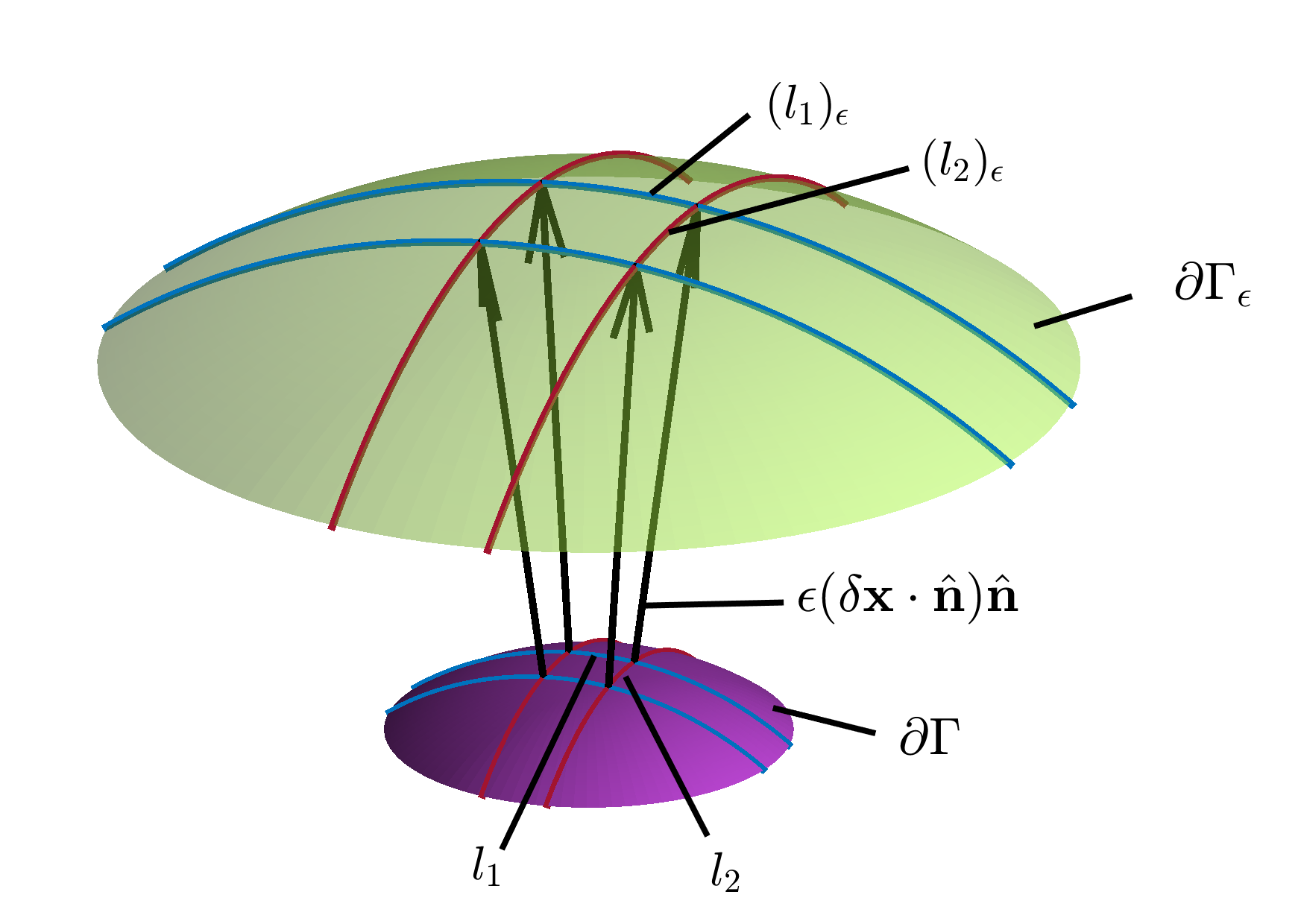}
    \caption{A local orthogonal basis is formed by the principal directions on $\partial \Gamma$, shown as the blue and red lines, with curvatures $\kappa_1$ and $\kappa_2$, respectively. The unperturbed surface area element bounded by the principal directions is given by $dA = l_1 l_2$. Upon a normal displacement of magnitude $\epsilon \delta \textbf{x} \cdot \hat{\textbf{n}}$, the new area element is given by $(dA)_{\epsilon} = l_1 l_2 (1 + \kappa_1 \epsilon \delta \textbf{x} \cdot \hat{\textbf{n}})(1 + \kappa_2\epsilon \delta \textbf{x} \cdot \hat{\textbf{n}})$, so the linear change in the area element is $\delta A = (dA)2H \delta \textbf{x} \cdot \hat{\textbf{n}}$, where $H = \frac{\kappa_1 + \kappa_2}{2}$ is the mean curvature.}
    \label{fig:surface_displacement}
\end{figure}

We can now apply the divergence theorem to write \eqref{eq:surface_functional} as 
\begin{align}
    J_2(\Gamma) = \int_{\Gamma} d^3 x \, \nabla \cdot \left(j_2(\Gamma)\nabla b(\Gamma)  \right).
\end{align}
We apply the transport theorem for volume-integrated functionals \eqref{eq:transport_theorem_volume} to obtain,
\begin{align}
    \delta J_2(\Gamma;\delta \textbf{x}) =
    \int_{\partial \Gamma} d^2 x \, \left[\delta \textbf{x} \cdot \hat{\textbf{n}} \left(\hat{\textbf{n}} \cdot \nabla j_2 + j_2 \nabla^2 b \right) + \nabla b \cdot \nabla \delta b(\Gamma;\delta \textbf{x}) + \delta j_2(\Gamma;\delta \textbf{x}) \right].
\end{align}
We can interchange shape and spatial derivatives\footnote{Under the assumption of sufficient smoothness, spatial and shape derivatives can be shown to commute by noting that $\textbf{x}$ and $\Gamma$ are independent variables (Chapter 6 in \cite{Choi2006}).} to see that $\nabla b \cdot \nabla \delta b = \frac{1}{2} \delta \left(\nabla b \cdot \nabla b\right) = 0$, as $\nabla b$ will remain a unit vector. We can also recognize that the mean curvature, $H$, is related to the normal vector by $H = \frac{1}{2} \nabla_{\partial \Gamma} \cdot \hat{\textbf{n}}$, where $\nabla_{\partial \Gamma} \cdot \textbf{f} = \nabla \cdot \textbf{f}  - \hat{\textbf{n}} \cdot \left(\nabla \textbf{f}\right)\cdot \hat{\textbf{n}}$ is the tangential divergence operator. (Sometimes $H$ is defined with the opposite sign.) For surface-integrated functionals we therefore obtain the following shape derivative,
\begin{align}
    \delta J_2(\Gamma;\delta \textbf{x}) = \int_{\partial \Gamma} d^2 x \, \left[\delta j_2(\Gamma;\delta \textbf{x})  +\left(\hat{\textbf{n}} \cdot \nabla j_2 + 2 H j_2  \right) \delta \textbf{x} \cdot \hat{\textbf{n}} \right].
    \label{eq:transport_theorem_surface}
\end{align}
The first term accounts for the Eulerian change to $j_2$, while the second and third terms account for the motion of the boundary. As one would expect, an outward perturbation of a surface with large mean curvature leads to a large change in the area. See Figure \ref{fig:surface_displacement} for a physical picture.

We can already see from \eqref{eq:transport_theorem_volume} and \eqref{eq:transport_theorem_surface} that the shape derivatives of volume and surface-integrated functionals involve integrals over the boundary. It may appear that to understand the form of these shape derivatives, we will need to specify the structure of $j_1(\Gamma)$ and $j_2(\Gamma)$. However, we can make a more general statement about shape derivatives of \textit{any} form. The Hadamard-Zolesio structure theorem \cite{Hadamard1908,Delfour2011} states that the shape derivative of a general functional of the domain $\Gamma$ with sufficient smoothness can be expressed as,
\begin{align}
    \delta J(\Gamma;\delta \textbf{x}) = \int_{\partial \Gamma} d^2 x \, \delta \textbf{x} \cdot \hat{\textbf{n}} \mathcal{G},
    \label{eq:structure_theorem}
\end{align}
where $\mathcal{G}$ is called the shape gradient. This is an example of the Riesz representation theorem, which (roughly) states that any linear functional can be expressed as an inner product with an element of the appropriate space (Chapter 4 in \cite{Rudin2006}). The shape derivative is a linear functional of the normal perturbation to the boundary, $\delta \textbf{x} \cdot \hat{\textbf{n}}$, and can be expressed as a surface integral with the shape gradient. This form is especially powerful for computation, as the deformation field only needs to be defined on the boundary, and the derivative can be written in terms of a surface integral rather than a volume integral. Intuitively, linear changes to a functional only depend on normal perturbations of the boundary. If the shape gradient can be determined, then for any possible deformation field, $\delta \textbf{x}$, the corresponding change to the functional $\delta J(\Gamma;\delta \textbf{x})$, is known. We can think of $\mathcal{G}$ as being a measure of the \textit{local} sensitivity: regions of increased $|\mathcal{G}|$ correspond to regions of increased sensitivity of $J(\Gamma)$ with respect to normal perturbations.

For stellarator optimization, we are also interested in functionals which depend on the shape of a set of filamentary lines, $C = \{C_k\}$. We expect that perturbations of the coils in the tangential direction will not result in a linear change to the functional. We can, therefore, write the shape derivative in a form analogous to the structure theorem \eqref{eq:structure_theorem} by the Riesz representation theorem,
\begin{align}
    \delta f(C;\delta \textbf{x}_{C_k}) = \sum_k \oint_{C_K} dl \, \delta \textbf{x}_{C_k} \times \hat{\textbf{t}} \cdot \bm{\mathcal{G}}_k,
    \label{eq:structure_theorem_coil}
\end{align}
where $\hat{\textbf{t}}$ is the tangent vector, integration is taken along each coil, and the sum is taken over all coils. As a curve has two independent directions perpendicular to the tangent vector, the shape gradient is now a vector, $\bm{\mathcal{G}}_k$. Its direction indicates the direction of perturbation which leads to the largest increase in the functional, and its magnitude indicates the level of sensitivity to a given perturbation.

To motivate this form of the coil shape gradient, we consider the example of the magnetic field computed from the Biot-Savart law applied to a set of filamentary coils $\{ C_k \}$,
\begin{align}
    \textbf{B}(\textbf{x},C) = \frac{\mu_0}{4\pi}\sum_k I_{C_k} \oint_{C_k} dl \,  \frac{\hat{\textbf{t}}(l)\times(\textbf{x}-\textbf{x}_k(l))}{|\textbf{x}-\textbf{x}_k(l)|^3},
\end{align}
where $\textbf{x}_k$ is the position along the $k$th coil and $\hat{\textbf{t}} = \textbf{x}_k'(l)$ is the unit tangent vector. The shape derivative of the magnetic field can now be computed with respect to a coil perturbation field $\delta \textbf{x}$ by considering the perturbation of a general closed line integral $Q_L(C) = \oint_C dl \, Q(C)$ \citep{Antonsen1982,Landreman2018},
\begin{align}
    \delta Q_L(C;\delta \textbf{x}) &= \oint_C dl \,  \left( \delta \textbf{x} \cdot \left(-\bm{\kappa} Q + \left(\textbf{I}-\hat{\textbf{t}}\hat{\textbf{t}} \right)\cdot \nabla Q\right) + \delta Q(C;\delta \textbf{x}) \right),
    \label{eq:perturbation_line_integral_ch2}
\end{align}
where $\bm{\kappa}(l) = \hat{\textbf{t}}'(l)$ is the curvature vector.

Upon application of this identity and integration by parts, we obtain,
\begin{multline}
    \delta \textbf{B}(\textbf{x},C;\delta \textbf{x}_k) = \\
    \frac{\mu_0}{4\pi} \sum_k \oint_{C_k} dl \, \delta \textbf{x}_k \times \hat{\textbf{t}}(l) \cdot 
  \Bigg( - \frac{\textbf{I}}{|\textbf{x}-\textbf{x}_k(l)|^3}
  + 3 (\textbf{x}-\textbf{x}_k(l))\frac{  (\textbf{x}-\textbf{x}_k(l)) }{|\textbf{x}-\textbf{x}_k(l)|^5}\Bigg),
  \label{eq:biot_savart_integral}
\end{multline}
where $\textbf{I}$ is the identity tensor. Thus the shape derivative of a figure of merit that depends on the vacuum magnetic field through the Biot-Savart law can be expressed in the coil shape gradient form \eqref{eq:structure_theorem_coil}. In Chapter \ref{ch:adjoint_MHD} we will show explicit examples of other figures of merit that can be expressed in this form.

\subsection{Parameter derivatives}
\label{sec:parameter_derivatives}

In practice, it may be convenient to describe a shape by a set of parameters, $\Omega$. We can relate the shape derivative and shape gradient defined in the previous Section to derivatives with respect to such parameters. 

Suppose that we have a surface described by a set of parameters, $\Omega$. For example, in the context of stellarator equilibrium calculations, the plasma boundary is often described by a set of Fourier coefficients of the cylindrical coordinates, $\{R_{m,n}^c,Z_{m,n}^s\}$,
\begin{subequations}
\begin{align}
    R &= \sum_{m,n} R_{m,n}^c \cos(m \theta - n N_P \phi) \\
    Z &= \sum_{m,n} Z_{m,n}^s \sin(m \theta - nN_P \phi).
\end{align}
\label{eq:rmnc_zmns}
\end{subequations}
Here $\theta$ is a poloidal angle, $\phi$ is a toroidal angle, and the configuration is assumed to possess stellarator symmetry, which implies that $R(-\theta,-\phi) = R(\theta,\phi)$ and $Z(-\theta,-\phi) = - Z(\theta,\phi)$ \cite{Dewar1998}. The number of periods is $N_P$, representing the discrete rotational symmetry of the equilibrium (Section 12 in \cite{Imbert2019}). This is the representation of the boundary shape used in the VMEC code \cite{Hirshman1983}.

In this case, we can compute the shape derivative corresponding to perturbations of each parameter, $\delta \textbf{x} = \left(\partial \textbf{x}(\Omega)/\partial \Omega_i\right) \delta \Omega_i $
\begin{align}
    \delta J(\Gamma(\Omega);\delta \textbf{x}) = \partder{J(\Gamma(\Omega))}{\Omega_i} \delta \Omega_i,
    \label{eq:parameter_derivatives_surface}
\end{align}
by expression our functional as a function of the parameters. We apply the structure theorem \eqref{eq:structure_theorem} to obtain the following expression,
\begin{align}
    \partder{J(\Gamma(\Omega))}{\Omega_i} = \int_{\partial \Gamma} d^2 x \, \partder{\textbf{x}(\Omega)}{\Omega_i} \cdot \hat{\textbf{n}} \mathcal{G}.
    \label{eq:linear_system_surfaces}
\end{align}
Given $\partial J(\Gamma(\Omega))/\partial \Omega_i$ and $\partial \textbf{x}(\Omega)/\partial \Omega_i$, we can consider this to be a linear system for $\mathcal{G}$. For numerical calculation, the above can be discretized using a collocation method or by expanding $\mathcal{G}$ in a set of basis functions. Often the linear system is not square, in which case an SVD or QR decomposition can be used. 

Now suppose that our coils are described by a set of parameters, $\Omega$. For example, the Cartesian components of the filamentary line can be described by a Fourier series,
\begin{subequations}
\begin{align}
x^k &= \sum_m X^{kc}_m \cos(m \theta) + X^{ks}_m\sin(m \theta) \\
y^k &= \sum_m Y^{kc}_m \cos(m \theta) + Y^{ks}_m \sin(m \theta) \\
z^k &= \sum_m Z^{kc}_m \cos(m \theta) + Z^{ks}_m \sin(m \theta),
\end{align}
\end{subequations}
where $\theta \in[0,2\pi]$ is an angle parameterizing each curve. Again we compute the shape derivative corresponding to perturbations of each parameter, $\delta \textbf{x}_{C_k} = \left(\partial \textbf{x}_{C_k}(\Omega)/\partial \Omega_i \right) \delta \Omega_i$,
\begin{align}
    \delta f(C;\delta \textbf{x}_{C_k}) = \partder{f(\{C_k(\Omega)\})}{\Omega_i} \delta \Omega_i,
    \label{eq:parameter_derivatives_coil}
\end{align}
to obtain,
\begin{align}
    \partder{f(C)}{\Omega_i} = \sum_k \oint_{C_k} dl \, \partder{\textbf{x}_{C_k}(\Omega)}{\Omega_i} \times \hat{\textbf{t}} \cdot \bm{\mathcal{G}}_k.
    \label{eq:linear_system_coil}
\end{align}
As with the case of functionals of surfaces, we can consider the above to be a linear system for $\bm{\mathcal{G}}_k$ that can be solved numerically. 

An overview of this method and examples of its application for figures of merit relevant for stellarator optimization are provided in 
\cite{Landreman2018}. 

\subsection{Discussion and applications}
\label{sec:shape_optimization_discussion}

The shape derivatives computed in this Section are quite general, applying to any functional of surfaces, volumes, or lines. For some problems we will be able to use the expressions for the shape derivatives, \eqref{eq:transport_theorem_volume} and \eqref{eq:transport_theorem_surface}, to obtain an explicit expression for the shape gradient. For example, if we consider the volume functional, \eqref{eq:volume_functional} with $j_1 = 1$, then we see from \eqref{eq:transport_theorem_volume} that the shape gradient will be $\mathcal{G} = 1$. If we consider the surface functional, \eqref{eq:surface_functional} with $j_2 = 1$, then we see from \eqref{eq:transport_theorem_surface} that the shape gradient will be $\mathcal{G} = 2H$. However, for many functionals, this type of explicit calculation is not possible. We are often interested in functionals which depend on solutions of a PDE, in which case we can compute the shape gradient by solving an additional PDE, known as an adjoint equation. We describe the adjoint method in more detail in the following Section. 

For other problems, it may be more convenient to compute the shape derivative from parameter derivatives, as in \eqref{eq:parameter_derivatives_surface} and \eqref{eq:parameter_derivatives_coil}, rather than applying the transport theorems. The shape gradient can then be inferred by solving the corresponding linear systems, \eqref{eq:linear_system_surfaces} and \eqref{eq:linear_system_coil}. Sometimes these parameter derivatives can be obtained analytically or with an adjoint method; otherwise, they are obtained with a finite-difference method. 

As the shape gradient measures the local sensitivity of a figure of merit to perturbations of a shape, we can use it to quantify the uncertainty in a figure of merit given a distribution of small perturbations to the shape. As shown in \cite{Landreman2018}, the plasma surface or coil shape gradient can be used to determine the allowable deformations of a shape given a permissible change to a figure of merit. Suppose a figure of merit $f$ has an allowable deviation $\Delta f$ (in either direction). If we define a local tolerance for the $k$th coil as,
\begin{align}
    T_k(l) = \frac{w_k(l) \Delta f}{\sum_{k'} \oint dl \, w_{k'}(l') |\bm{\mathcal{G}}_{k'}(l')|},
\end{align}
such that the perturbation amplitude $|\delta \textbf{x}_{C_k}(l) \times \hat{\textbf{t}}(l)| \le T_k(l)$ along the $k$th coil, then the the change of the figure of merit will be,
\begin{align}
    |\delta f\left(C;\delta \textbf{x}_{C_k}\right)| \le \sum_k \oint_{C_k} dl \, |\delta \textbf{x}_{C_k} \times \hat{\textbf{t}} \cdot \bm{\mathcal{G}}_k| \le \sum_k \oint_{C_k} dl \, T_k|\bm{\mathcal{G}}_k| = \Delta f,
\end{align}
upon application of the triangle inequality. Here $w_{k}(l)$ is a weight function which allows for the distribution of tolerance to be non-uniform along the coil. In identifying such a tolerance we have relied on a local approximation of the function, considering small-amplitude perturbations such that a linear approximation is valid.

Similarly, a tolerance with respect to perturbations of a surface can be defined with respect to the surface shape gradient,
\begin{align}
    T = \frac{w \Delta f}{\int_{\partial \Gamma} d^2 x \, w \mathcal{G}},
\end{align}
where $w$ is a weight function defined on the surface $\partial \Gamma$. For example, we could consider the tolerance of a figure of merit that depends on the position of the plasma boundary, $S_P$. If we constrain perturbations of the surface such that $|\delta \textbf{x} \cdot \hat{\textbf{n}}| \le T$, then we find that the corresponding change to the figure of merit is $\delta f \le \Delta f$. However, the deformation of a magnetic surface is not a quantify that can be directly experimentally controlled, requiring equilibrium reconstruction methods \cite{Hanson2013}. 

A more practically relevant quantity is computed from the sensitivity to perturbations of the magnetic field, $S_B$, defined through,
\begin{align}
    \delta f(S_P;\delta \textbf{x}) = \langle \mathcal{G} \rangle_{\psi} \delta V(\delta \textbf{x}) + \int_{S_P} d^2 x \, S_B \delta \textbf{B}(\delta \textbf{x}) \cdot \hat{\textbf{n}},
\end{align}
where $\delta V$ and $\delta \textbf{B}$ are the perturbations to the volume enclosed by $S_P$ and magnetic field resulting from a surface displacement of $\delta \textbf{x}$ and $\langle \dots \rangle_{\psi}$ is the flux-surface average \eqref{eq:flux_surface_average_appA}.

The quantity $S_B$, which quantifies the local sensitivity to perturbations of the magnetic field, is computed from the shape gradient as,
\begin{align}
    \textbf{B} \cdot \nabla S_B = \langle \mathcal{G} \rangle_{\psi} - \mathcal{G}.
\end{align}
A tolerance with respect to magnetic field perturbations can then be constructed as,
\begin{align}
    T_B = \frac{w \Delta f}{\int_{S_P} d^2 x \, w |S_B|},
\end{align}
for a chosen weight function $w$, such that if the normal magnetic perturbations satisfy $|\delta \textbf{B} \cdot \hat{\textbf{n}}| \le T_B$, then $\delta f \le \Delta f$. The tolerance with respect to magnetic perturbations can inform allowable coil deformations, location of trim coils, and position of current leads. In this way, important engineering tolerances are inferred, addressing objective 4 from Section \ref{sec:challenges_outlook}.

\section{Adjoint methods}

An adjoint method is a numerical method for the efficient calculation of derivatives of an objective function that depends on the solution to some set of equations, known as the forward system. At the heart of the adjoint method is the adjoint equation, in which the adjoint of the linearized forward operator appears in addition to an inhomogeneous term that depends on the objective function of interest. 

There are other instances in which the adjoint operator may become useful. An adjoint Fokker-Planck equation is used to compute the quasilinear generation of current by RF waves \cite{Antonsen1982} or to study runaway electron dynamics \cite{Liu2016}. An adjoint gyrokinetic equation can also be used to analyze the evolution of free energy \cite{Landreman2015}. Finally, adjoint operators are used to predict and correct discretization error \cite{Giles1999,Pierce2004} and perform efficient grid adaptation \cite{Venditti1999}. In this Chapter, we focus our attention on adjoints for efficient derivative calculations.

Adjoint methods were introduced by the optimal control theory community in the 1960s \cite{Kelley1960,Gavrilovic1963}, and were later adopted by the fluid dynamics community \cite{Pironneau1974}. They have since been popularized for aeronautical design \cite{Jameson1998}, car aerodynamics \cite{Othmer2014}, geophysics \cite{Plessix2006}, and nuclear fission reactor design \cite{Gandini1990}. Aside from the body of work associated with this Thesis, there is only one other example of the use of adjoint methods in fusion sciences: for the shape optimization of tokamak divertors based on adjoint fluid equations \cite{Dekeyser2014a,Dekeyser2014b,Dekeyser2014c,Dekeyser2014}. We refer to several introductory articles on adjoint methods \cite{Giles2000,Plessix2006,Allaire2015}.

We begin our overview of adjoint methods with its application for objective functions that depend on the solution of finite-dimensional, discrete linear systems in Section \ref{sec:linear_systems}. We will then generalize to objective functions that depend on the solution of infinite-dimensional, possibly nonlinear systems in Section \ref{sec:lagrangian}. The two approaches are compared in Section \ref{sec:discrete_continuous}. 

\subsection{Discrete approach}
\label{sec:linear_systems}

Suppose we would like to solve the optimization problem,
\begin{align}
    \min_{\Omega} f(\Omega,\overrightarrow{\textbf{x}}),
    \label{eq:optimization}
\end{align}
where $\overrightarrow{\textbf{x}}$ is the solution of a linear system,
\begin{align}
    \overleftrightarrow{\textbf{A}}(\Omega)\overrightarrow{\textbf{x}} = \overrightarrow{\textbf{b}}(\Omega).
    \label{eq:linear_system}
\end{align}
Here $\overleftrightarrow{\textbf{A}}$ is an $N \times N$ matrix and $\overrightarrow{\textbf{x}}$ and $\overrightarrow{\textbf{b}}$ are $N \times 1$ column vectors. Let $\Omega = \{ \Omega_i \}_{i=1}^{N_{\Omega}}$ be a set of design parameters defining our optimization space. 
To minimize \eqref{eq:optimization} with a gradient-based method, we compute the derivative with respect to $\Omega$ using the chain rule,
\begin{align}
    \der{f(\Omega,\overrightarrow{\textbf{x}}(\Omega))}{\Omega} = \partder{f(\Omega,\overrightarrow{\textbf{x}})}{\Omega} + \left(\partder{f(\Omega,\overrightarrow{\textbf{x}})}{\overrightarrow{\textbf{x}}}\right)^T \partder{\overrightarrow{\textbf{x}}(\Omega)}{\Omega}.
    \label{eq:combined_derivative}
\end{align}
Here $\partial f(\Omega,\overrightarrow{\textbf{x}})/\partial \overrightarrow{\textbf{x}}$ is the gradient of $f$ with respect to $\overrightarrow{\textbf{x}}$, a column vector. To evaluate $\partial \overrightarrow{\textbf{x}}(\Omega)/\partial \Omega$, we must compute linear perturbations of \eqref{eq:linear_system},
\begin{align}
    \partder{\overleftrightarrow{\textbf{A}}(\Omega)}{\Omega} \overrightarrow{\textbf{x}}(\Omega) + \overleftrightarrow{\textbf{A}}(\Omega) \partder{\overrightarrow{\textbf{x}}(\Omega)}{\Omega} = \partder{ \overrightarrow{\textbf{b}}(\Omega)}{\Omega}. 
    \label{eq:perturbed_linear}
\end{align}
We schematically evaluate the perturbation to the solution as,
\begin{align}
    \partder{\overrightarrow{\textbf{x}}(\Omega)}{\Omega} = \overleftrightarrow{\textbf{A}}(\Omega)^{-1} \left(\partder{ \overrightarrow{\textbf{b}}(\Omega)}{\Omega}-\partder{\overleftrightarrow{\textbf{A}}(\Omega)}{\Omega} \overrightarrow{\textbf{x}}(\Omega)\right). 
\end{align}
Inserting the result into \eqref{eq:combined_derivative}, we obtain
\begin{multline}
         \der{f(\Omega,\overrightarrow{\textbf{x}}(\Omega))}{\Omega} = \partder{f(\Omega,\overrightarrow{\textbf{x}})}{\Omega}\\ + \left(\partder{f(\Omega,\overrightarrow{\textbf{x}})}{\overrightarrow{\textbf{x}}}\right)^T \left(\overleftrightarrow{\textbf{A}}(\Omega)^{-1} \left(\partder{ \overrightarrow{\textbf{b}}(\Omega)}{\Omega}-\partder{\overleftrightarrow{\textbf{A}}(\Omega)}{\Omega} \overrightarrow{\textbf{x}}(\Omega)\right)\right).
         \label{eq:forward_sensitivity}
\end{multline}
This approach to computing the derivative, the forward-sensitivity method, requires computing $N_{\Omega}+1$ solutions to a linear system of size $N \times N$: we must solve \eqref{eq:linear_system} once for $\overrightarrow{{\textbf{x}}}$, and we must solve,
\begin{align}
    \overleftrightarrow{\textbf{A}}(\Omega_i) \overrightarrow{\textbf{y}} = \partder{ \overrightarrow{\textbf{b}}(\Omega)}{\Omega_i}-\partder{\overleftrightarrow{\textbf{A}}(\Omega)}{\Omega_i} \overrightarrow{\textbf{x}}(\Omega_i),
    \label{eq:forward_sensitivity_system}
\end{align}
for $\overleftrightarrow{\textbf{y}}$ once for each $\Omega_i$.

By rearranging parentheses, \eqref{eq:forward_sensitivity} is equivalent to,
\begin{multline}
         \der{f(\Omega,\overrightarrow{\textbf{x}}(\Omega))}{\Omega} = \partder{f(\Omega,\overrightarrow{\textbf{x}})}{\Omega}\\ + \left(\left(\overleftrightarrow{\textbf{A}}(\Omega)^{T}\right)^{-1}\partder{f(\Omega,\overrightarrow{\textbf{x}})}{\overrightarrow{\textbf{x}}}\right)^T  \left(\partder{ \overrightarrow{\textbf{b}}(\Omega)}{\Omega}-\partder{\overleftrightarrow{\textbf{A}}(\Omega)}{\Omega} \overrightarrow{\textbf{x}}(\Omega)\right),
\end{multline}
where we have noted that the transpose and inverse operations can be interchanged for any invertible matrix. Thus we can see that if we compute the solution to the following adjoint equation,
\begin{align}
   \overleftrightarrow{\textbf{A}}(\Omega)^{T} \overleftrightarrow{\textbf{z}} = \partder{f(\Omega,\overrightarrow{\textbf{x}})}{\overrightarrow{\textbf{x}}},
   \label{eq:adjoint_system}
\end{align}
then we can compute the derivative of the objective function in a more convenient way,
\begin{align}
         \der{f(\Omega,\overrightarrow{\textbf{x}}(\Omega))}{\Omega} = \partder{f(\Omega,\overrightarrow{\textbf{x}})}{\Omega} + \overrightarrow{\textbf{z}}^T  \left(\partder{ \overrightarrow{\textbf{b}}(\Omega)}{\Omega}-\partder{\overleftrightarrow{\textbf{A}}(\Omega)}{\Omega} \overrightarrow{\textbf{x}}(\Omega)\right).
         \label{eq:combined_derivative_adjoint}
\end{align}
This method for computing the derivative, known as the adjoint method, only requires two solutions of a linear system of size $N \times N$: \eqref{eq:linear_system} and \eqref{eq:adjoint_system}. In general, the partial derivatives of $\overrightarrow{\textbf{b}}(\Omega)$ and $\overleftrightarrow{\textbf{A}}(\Omega)$ can be computed analytically. In this way, no approximations are made in obtaining \eqref{eq:combined_derivative_adjoint}. The power of this approach becomes apparent in high-dimensional spaces: the adjoint method requires only two solutions of such linear systems, while the forward-sensitivity method requires $N_{\Omega}+1$ solutions. Approximating the derivative with a finite-difference method also requires at least $N_{\Omega} + 1$ solutions, depending on the size of the stencil.

The approach presented in this Section can be understood as a linear algebra trick. We want to solve a linear system for many right-hand sides, as in \eqref{eq:forward_sensitivity_system}. Moreover, we are only interested in a specific inner product with these solutions, \eqref{eq:forward_sensitivity}. As we are allowed to interchange the transpose and inverse operations, we arrive at the adjoint form \eqref{eq:adjoint_system}. If the partial derivatives of $\overleftrightarrow{\textbf{A}}(\Omega)$ and $\overrightarrow{\textbf{b}}(\Omega)$ can be computed analytically, and the adjoint equation is solved exactly, then no approximations are made here. In this sense, we can consider the adjoint-based derivative to be the \textit{exact} analytic derivative. In practice, there may be a small amount of error introduced due to the finite tolerance of the linear solve.

\subsubsection{Computational complexity comparison}
\label{sec:complexity_comparison}

We now compare the computational complexity of the forward-sensitivity method, the finite-difference method, and the adjoint method for computing the derivative. Here we will ignore any cost associated with constructing $\overleftrightarrow{\textbf{A}}(\Omega)$, $\overrightarrow{\textbf{b}}(\Omega)$, or their derivatives. For some matrix types (e.g. sparse) the number of required operations may be reduced from what is given here, but we simply try to estimate the relative costs. The flop counts for matrix computations can be found in standard references such as \cite{Trefethen1997}.

For both the forward and adjoint sensitivity methods, we must form the right-hand side of \eqref{eq:forward_sensitivity_system} for each $\Omega_i$, each of which requires a matrix-vector product and a vector-vector sum for a combined cost of $\approx 2N^2 + N$ flops. The forward-sensitivity method requires solving \eqref{eq:forward_sensitivity_system} $N_{\Omega}$ times. For example, an $LU$ factorization method can be used, which requires $\approx \frac{2}{3}N^3$ flops. Once the factorization is known, solving the system \eqref{eq:perturbed_linear} via backward substitution costs $\approx 2N^2$ flops for each $\Omega_i$. Once $\partial \overrightarrow{\textbf{x}}/\partial \Omega$ is obtained, $N_{\Omega}$ vector-vector products must be performed to obtain the derivatives of $f$ as in \eqref{eq:forward_sensitivity}, each which requires $2N$ flops. Thus the composite number of flops is $\approx 4N_{\Omega} N^2 + \frac{2}{3}N^3$. With a finite-difference method, the total cost of computing $\partial \overrightarrow{\textbf{x}}/\partial \Omega$ requires at least $\approx \frac{2}{3}N_{\Omega}N^3$ flops, assuming that the linear solve is the most expensive step and a one-sided stencil is used. 

Alternatively, the adjoint method for computing the derivative requires two linear solves. If an $LU$ factorization method is used, then the matrix factorization of $\overleftrightarrow{\textbf{A}} = \overleftrightarrow{\textbf{L}}\overleftrightarrow{\textbf{U}}$ can be reused to solve the adjoint system \eqref{eq:adjoint_system}, as $\overleftrightarrow{\textbf{A}}^T = \overleftrightarrow{\textbf{U}}^T\overleftrightarrow{\textbf{L}}^T$ where $\overleftrightarrow{\textbf{U}}^T$ is lower-triangular and $\overleftrightarrow{\textbf{L}}^T$ is upper-triangular. Thus the cost of computing the two solutions requires $\approx \frac{2}{3} N^3 + 4N^2$ flops. Once the adjoint solution is obtained, $N_{\Omega}$ matrix-vector products and vector-vector sums must be computed in \eqref{eq:combined_derivative_adjoint} each with cost $\approx 2 N^2 + N$ flops. Again, $N_{\Omega}$ vector-vector products are required, each of which requires $\approx 2 N$ flops. Thus the total complexity is $\approx 2N_{\Omega} N^2 + \frac{2}{3} N^3$ flops, assuming large $N$. A summary of these approximate flop counts is given in Table \ref{table:complexity}.

We see that the adjoint method provides modest savings over the forward-sensitivity method when $N_{\Omega}$ is comparable to $N$. However, for many problems the assumptions made in this Section do not apply. In particular, if $\overleftrightarrow{\textbf{A}}$ is sparse, $\overleftrightarrow{\textbf{L}}$ and $\overleftrightarrow{\textbf{U}}$ will be generally be dense, in which case the matrix-vector multiplication that appears on the right-hand-side of \eqref{eq:combined_derivative_adjoint} will be significantly cheaper than backsubstitution to solve \eqref{eq:forward_sensitivity_system}, and there will be a more significant savings with the application of the adjoint method over the forward-sensitivity method. For very large matrices it may be impractical to $LU$ factorize $\overleftrightarrow{\textbf{A}}$. Instead, a preconditioner may be factorized, and the linear system is solved with a Krylov subspace iterative method. Again for such systems, solving the factorized system will be significantly more expensive than matrix-vector multiplication.

In comparison with finite differences, the adjoint method offers a reduction of complexity by $\mathcal{O}(N_{\Omega})$. The accuracy of the finite-difference method depends on the size of the stencil and choice of step size. While a wider stencil provides a more accurate derivative, it increases the number of required function evaluations. The step size must also be chosen carefully to avoid the introduction of noise: a large step size will introduce nonlinearity, while a small step size will introduce round-off error. For these reasons, the adjoint method is preferable over a finite-difference method. 

\renewcommand{\arraystretch}{1.3}
\begin{table}
\centering
\begin{tabular}{|c|c|c|}
\hline 
     Forward Sensitivity & Finite difference & Adjoint \\
     \hline \hline 
     $4 N_{\Omega} N^2 + \frac{2}{3} N^3$ & $\frac{2}{3} N_{\Omega} N^3 $ & $2 N_{\Omega} N^2 + \frac{2}{3} N^3 $ \\
     \hline 
\end{tabular}
    \caption{Approximate flop counts for the forward-sensitivity, finite-difference, and adjoint method for calculation of the derivative.}
    \label{table:complexity}
\end{table}

\subsection{Continuous approach}
\label{sec:lagrangian}

The adjoint method presented in the previous Section applies only to functions that depended on the solution of a linear system in a finite-dimensional space. We now generalize this result to obtain an adjoint equation in an infinite-dimensional space. Often in optimization, we are interested in an objective function which depends on the solution of a PDE, 
\begin{align}
    L(\Omega,u) = 0,
    \label{eq:PDE}
\end{align}
such as the MHD equilibrium equations \eqref{eq:MHD_equilibrium}. Here $L$ is some linear or nonlinear operator, and $u$ is an unknown. We are optimizing with respect to a set of parameters, $\Omega$, which may generally be infinite-dimensional; for example, $\Omega$ may describe the shape of some domain. Our differential operator may depend on these parameters. We assume that $u$ is a member of some Hilbert space, $\mathcal{H}$, which possesses an inner product structure denoted by $\langle . \, , . \rangle$. If this PDE is linear, then the discretized form of this problem can generally be written as \eqref{eq:linear_system}, and the adjoint equation can be obtained \textit{after} discretization as described in the previous Section. The method described in this Section will allow us to get an adjoint equation \textit{before} discretization.

 We can consider $u$ to depend on $\Omega$ through the solution to \eqref{eq:PDE}. We perform linear perturbations about the base state \eqref{eq:PDE} corresponding to perturbations of $\Omega$,
\begin{align}
    \delta L(\Omega,u;\delta \Omega) + \delta L \left(\Omega,u;\delta u(\Omega;\delta \Omega) \right) = 0.
    \label{eq:linearized_pde}
\end{align}
Our objective function, $f(\Omega,u)$, is some linear or nonlinear scalar functional of $\Omega$ and $u$. Linear perturbations of $f(\Omega,u)$ can generally be written as an inner product with $\delta u$,
\begin{align}
    \delta f(\Omega,u;\delta u) = \left \langle \widetilde{f}, \delta u\right \rangle.
\end{align}
This is another example of the Riesz representation theorem: as $\delta f$ is a linear functional of $\delta u$, we can express it as an inner product with $\widetilde{f} \in \mathcal{H}$. 

We are interested in computing linear perturbations to $f$ such that $u(\Omega)$ satisfies the PDE. The constrained problem is expressed through the objective function, $f(\Omega, u(\Omega))$, whose derivative with respect to $\Omega$ is computed to be,
\begin{align}
  \delta f(\Omega,u(\Omega);\delta \Omega) = \delta f(\Omega,u;\delta \Omega) + \left\langle \widetilde{f}, \delta u(\Omega;\delta \Omega) \right\rangle,
    \label{eq:fhat_derivative}
\end{align}
and $\delta u(\Omega;\delta \Omega)$ satisfies \eqref{eq:linearized_pde}. This is an analogous expression to \eqref{eq:forward_sensitivity} in the discrete linear case. Computing the derivative in this way requires \textit{many} solutions of a PDE: one solution of the initial base state \eqref{eq:PDE} and one solution of \eqref{eq:linearized_pde} for each perturbation of the optimization parameters, $\delta \Omega$.

A more efficient method of computing these derivatives is by application of Lagrange multipliers, enforcing \eqref{eq:PDE} as a constraint. We now define the corresponding Lagrangian as, 
\begin{align}
    \mathcal{L}(\Omega,\widetilde{u},\widetilde{\lambda}) = f(\Omega,\widetilde{u}) + \left\langle \widetilde{\lambda}, L(\Omega,\widetilde{u})\right\rangle, 
    \label{eq:lagrangian}
\end{align}
where $\widetilde{\lambda} \in \mathcal{H}$ is a Lagrange multiplier. In the above expression, $\widetilde{u} \in \mathcal{H}$ but it does not necessarily satisfy \eqref{eq:PDE}, hence the distinction by the tilde. If $\mathcal{L}$ is stationary with respect to $\widetilde{\lambda}$, then $\widetilde{u}$ is a weak solution of the PDE, indicated by $u$. If $\mathcal{L}$ is stationary with respect to $\widetilde{u}$, then $\widetilde{\lambda}$ will satisfy the weak form of an adjoint PDE, at which point we denote $\widetilde{\lambda}$ by $\lambda$. If $\mathcal{L}$ is stationary with respect to both $\widetilde{u}$ and $\widetilde{\lambda}$, or $\widetilde{u} = u$ and $\widetilde{\lambda}=\lambda$, then derivatives of $\mathcal{L}$ with respect to $\Omega$ are equal to derivatives of $f$ with respect to $\Omega$,
\begin{align}
   \delta \mathcal{L}(\Omega,\widetilde{u},\widetilde{\lambda};\delta \Omega)  \rvert_{\widetilde{u} = u,\widetilde{\lambda}=\lambda} = \delta f(\Omega,u(\Omega);\delta \Omega).
\end{align}
We will show this directly in a moment. 

We now look for a stationary point of $\mathcal{L}$ with respect to $\widetilde{u}$,
\begin{align}
  \delta \mathcal{L}(\Omega,\widetilde{u}, \widetilde{\lambda};\delta \widetilde{u}) = \left \langle \widetilde{f},\delta \widetilde{u} \right \rangle + \left \langle \widetilde{\lambda}, \delta L(\Omega,\widetilde{u};\delta \widetilde{u})  \right \rangle  = 0.
\end{align}
We note that $\delta L(\Omega,\widetilde{u};\delta \widetilde{u})$ is a linear functional of $\delta \widetilde{u}$, so we can write this schematically as,
\begin{align}
    \delta L(\Omega,\widetilde{u};\delta \widetilde{u}) = \hat{L}(\Omega,\widetilde{u})\delta u,
\end{align}
where $\hat{L}(\Omega,\widetilde{u})$ is a linear operator. The adjoint of an operator $A$, which we denote by $A^{\dagger}$, is defined by $\langle A y, x \rangle = \langle y, A^{\dagger}x  \rangle$ for $x,y \in \mathcal{H}$. Thus we can rewrite the above as,
\begin{align}
      \delta \mathcal{L}(\Omega,\widetilde{u},\widetilde{\lambda};\delta \widetilde{u}) = \left \langle \widetilde{f} + \hat{L}(\Omega,\widetilde{u}) ^{\dagger} \widetilde{\lambda}, \delta \widetilde{u} \right \rangle  = 0.
\end{align}
This is a weak form of the adjoint PDE,
\begin{align}
   \widetilde{f} +  \hat{L}(\Omega,\widetilde{u})^{\dagger} \lambda = 0.
   \label{eq:adjoint_lagrangian}
\end{align}
We indicate its solution by $\lambda$, as it corresponds with a stationary point of $\mathcal{L}$ with respect to $\widetilde{u}$. We now see that if $\widetilde{u}$ satisfies \eqref{eq:PDE} and $\widetilde{\lambda}$ satisfies \eqref{eq:adjoint_lagrangian}, then derivatives of $f$ with respect to $\Omega$ are equal to derivatives of $\mathcal{L}$ with respect to $\Omega$,
\begin{align}
 \delta \mathcal{L}(\Omega,\widetilde{u},\widetilde{\lambda};\delta \Omega) \rvert_{\widetilde{u}=u,\widetilde{\lambda}=\lambda} &= \delta f(\Omega,u;\delta \Omega) + \left \langle \lambda, \delta L(\Omega,u;\delta \Omega) \right \rangle \nonumber \\
 &= \delta f(\Omega,u;\delta \Omega) - \left \langle \lambda, \delta L(\Omega,u;\delta u(\Omega;\delta \Omega) \right \rangle,
\end{align}
where we have used \eqref{eq:linearized_pde}. If we now apply the adjoint condition and enforce that $\lambda$ satisfy the adjoint PDE \eqref{eq:adjoint_lagrangian}, then we indeed obtain \eqref{eq:fhat_derivative}, as desired. 

The adjoint method for computing the derivative of $f$ with respect to the parameters $\Omega$ is,
\begin{align}
    \delta f(\Omega,u(\Omega);\delta \Omega) = \delta \mathcal{L}(\Omega,\widetilde{u},\widetilde{\lambda};\delta \Omega)  \rvert_{\widetilde{u} = u,\widetilde{\lambda}=\lambda} = \delta f(\Omega,u;\delta \Omega) + \left \langle \lambda, \delta L(\Omega,u;\delta \Omega)\right \rangle.
    \label{eq:derivative_lagrangian}
\end{align}
This is the continuous analogue of \eqref{eq:combined_derivative_adjoint}. The first term corresponds with the explicit dependence of $f$ on $\Omega$, while the second term corresponds with the dependence through $u$. 

Note that, if \eqref{eq:PDE} is satisfied, then we can choose $\lambda$ to be whatever we would like, as the second term in the Lagrangian functional \eqref{eq:lagrangian} will always vanish. For some problems, other choices for $\lambda$ may be convenient, although \eqref{eq:derivative_lagrangian} will no longer hold. In Chapter \ref{ch:adjoint_MHD}, a slightly different choice for the adjoint variable will be made. Rather than being a stationary point, boundary terms remain in the expression for $\delta \mathcal{L}(\Omega,u,\lambda;\delta u)$ (see \eqref{eq:adjoint_lambda_iota}-\eqref{eq:adjoint_lambda_current} and \eqref{eq:adjoint_coil_1}-\eqref{eq:adjoint_coil_2}).

In practice, the infinite-dimensional optimization space may be approximated by a discrete set of parameters, $\Omega = \{\Omega_i\}_{i=1}^{N_{\Omega}}$. Thus with the solution of only two PDEs, the forward \eqref{eq:PDE} and adjoint \eqref{eq:adjoint_lagrangian} problems, we obtain the derivative of our objective function with respect to an arbitrary number of parameters. An alternative is the forward-sensitivity method, using \eqref{eq:linearized_pde} and \eqref{eq:fhat_derivative}, which requires $N_{\Omega}$ linear PDE solution and one (possibly) nonlinear PDE solutions, \eqref{eq:PDE}. 

The finite-difference method requires at least $N_{\Omega}+1$ (possibly) nonlinear PDE solutions, depending on the size of the stencil. Thus the adjoint method provides a significant advantage when $N_{\Omega}$ is large, assuming that the PDE solve is expensive in comparison with other operations, such as performing the inner products. It is not straightforward to compare the complexity of these methods as in Section \ref{sec:complexity_comparison} as the flop count will depend on the numerical methods used to solve a PDE. However, we can see that the adjoint method provides a reduction in the number of required PDE solves by $\mathcal{O}(N_{\Omega})$ over both the forward-sensitivity and finite-difference methods.

Of course, both the forward and adjoint PDEs are typically solved numerically by approximation in a finite-dimensional space. The accuracy of the derivative computed with the adjoint method will, therefore, depend on the tolerance to which the base state and adjoint PDEs are solved in addition to the discrepancy between the infinite-dimensional inner product and its finite-dimensional approximation. 

\subsection{Comparison of discrete and continuous approaches}
\label{sec:discrete_continuous}

We now see that there are two general strategies to the application of the adjoint method: obtaining the adjoint before discretization, the continuous adjoint approach, or obtaining the adjoint after discretization, the discrete approach. There are relative merits to each. With the discrete adjoint method, the accuracy of the derivative only depends on the tolerance to which the forward and adjoint systems are solved. On the other hand, with the continuous method, it also depends on the discretization error of the PDE due to the difference between the infinite-dimensional inner product and its finite-dimensional approximation. The two approaches must agree in the limit of infinite resolution. In practice, the difference between the two is relatively small, though it has been suggested that the discrepancy between the continuous and discrete gradients may become important near a local minimum \cite{Dekeyser2014}, where the gradient obtained from the continuous approach may not be a descent direction of the discretized problem. 

The continuous approach offers the advantage that the adjoint equation can be derived independently of the choice of discretization; thus, if the adjoint equation has a significantly different structure from the forward equation, a distinct discretization scheme can be applied. It also may offer further insight into the structure of the adjoint equations and its boundary conditions. For this reason, the continuous approach may be preferable in the presence of shocks or singularities \cite{Giles2000}, as we demonstrate in Chapter \ref{ch:linearized_mhd}. For both approaches, the resulting adjoint equation is linear. Implementation of the discrete method is sometimes more straightforward, as the adjoint and forward operators have the same eigenvalues, so the same numerical linear algebra methods can typically be used to solve both problems. As we will see in Chapter \ref{ch:adjoint_neoclassical}, if an $LU$ factorization method is used to solve the linear system, then the factorization of the matrix or its preconditioner can be reused to solve the discrete adjoint problem. There is not a clear consensus in the literature as to which approach is preferable, and the choice usually depends on the application of interest. 

\subsection{Discussion and applications}

With an adjoint method, optimization within a high-dimensional space is no longer a significant challenge. An adjoint-based derivative provides a reduction of computational complexity over finite differences by approximately the optimization dimension, $N_{\Omega}$, as summarized in Table \ref{table:complexity}. Given that the cost of computing the gradient becomes comparable to the cost of the forward solve, we can easily take advantage of gradient-based optimization methods. For line-search gradient-based methods, each iteration reduces to a \textit{one-dimensional} line search once a descent direction is identified \cite{Nocedal2006}. Therefore with adjoint methods, high-dimensional, non-convex optimization becomes feasible, allowing us to address objectives 2 and 3 from Section \ref{sec:challenges_outlook}. 

\section{Conclusions}

In the following Chapters, we will demonstrate the application of shape calculus and adjoint methods for several problems arising in stellarator optimization. In Chapter \ref{ch:adjoint_winding_surface} we describe a discrete adjoint method for the optimization of coil shapes based on the current potential method described in Section \ref{sec:current_potential}. With the derivatives obtained from the adjoint method, we compute a shape gradient with respect to perturbations of the coil-winding surface, allowing us to identify regions where figures of merit become sensitive to coil perturbations. In Chapter \ref{ch:adjoint_neoclassical}, we compare a continuous and discrete adjoint method for computing geometric derivatives of several neoclassical quantities. These geometric derivatives allow us to compute a sensitivity function for local magnetic field strength perturbations that is analogous to the shape gradient. In Chapter \ref{ch:adjoint_MHD}, we describe a continuous adjoint method for computing the shape gradient of quantities that depend on MHD equilibrium solutions. These shape gradients can be used for equilibrium optimization of the plasma boundary or coil shapes and sensitivity analysis. For this application, the adjoint equation contains singular behavior, so a distinct discretization and solution scheme are required, discussed in Chapter \ref{ch:linearized_mhd}.
\renewcommand{\thechapter}{3}

\chapter{Adjoint winding surface optimization}
\label{ch:adjoint_winding_surface}

In this Chapter, we apply the linear adjoint approach described in Section \ref{sec:linear_systems} for the optimization of coil shapes. We assume that coils are confined to a winding surface using the current potential method introduced in Section \ref{sec:current_potential}. The application of the adjoint method will allow us to efficiently optimize in the space of the geometry of the coil-winding surface and study the sensitivity to local perturbations using the shape gradient. 

The material in this Chapter has been adapted from \cite{Paul2018} with permission.

\section{Introduction}

In the traditional stellarator optimization method, coils are designed to produce a target outer plasma boundary. The plasma boundary is separately optimized for various physics quantities, including magnetohydrodynamic (MHD) stability, neoclassical confinement, and profiles of rotational transform and pressure \cite{Nuhrenberg1988}. The coil shapes are then optimized such that one of the magnetic surfaces approximately matches the desired plasma surface. In general, the desired plasma configuration cannot be produced exactly due to engineering constraints on the coil complexity. Additional difficulty is introduced by the ill-posedness of solving Laplace's equation numerically in the vacuum region for a prescribed normal magnetic field on the plasma boundary \cite{Merkel1987,Boozer2000}.

In addition to the minimization of the magnetic field error, several factors should be considered in the design of coil shapes. The winding surface upon which the currents lie should be sufficiently separated from the plasma surface to allow for neutron shielding to protect the coils, the vacuum vessel, and a divertor system. In a reactor, the coil-plasma distance is closely tied to the tritium-breeding ratio and overall cost of electricity, as it determines the allowable blanket thickness. The coil-plasma distance was targeted in the ARIES-CS study to reduce machine size \cite{Guebaly2008}. In practice, the minimum feasible coil-plasma separation is a function of the desired plasma shape. Concave regions (such as the bean-shaped W7-X cross-section) are especially challenging to produce \cite{Landreman2016} and require the winding surface to be near the plasma surface. While decreasing the inter-coil spacing minimizes ripple fields, increasing coil-coil spacing allows adequate space for removal of blanket modules, heat transport plumbing, diagnostics, and support structures. The curvature of a coil should be below a certain threshold to allow for the finite thickness of the conducting material and to avoid prohibitively high manufacturing costs. The length of each coil should also be considered, as the expense will grow with the amount of conducting material that needs to be produced. For these reasons, identifying coils with suitable engineering properties can impact the size and cost of a stellarator device. 

Most coil design codes have assumed the coils to lie on a closed toroidal winding surface enclosing the desired plasma surface. In NESCOIL \cite{Merkel1987}, the currents on this surface are determined by minimizing the integral-squared normal magnetic field on the target plasma surface. The current density is computed using a stream function approach, where the current potential on the winding surface is decomposed in Fourier harmonics. The optimization takes the form of a least-squares problem that can be solved with the solution of a single linear system. The coil filament shapes are then obtained from the contours of the current potential. Because it is guaranteed to find a global minimum, NESCOIL is often used in the preliminary stages of the design process \cite{Spong2010, Ku2011, Drevlak2013}. NESCOIL was used for the initial coil configuration studies for NCSX \cite{Pomphrey2001}, and the W7-X coils were designed using an extension of NESCOIL, which modified the winding surface geometry for quality of magnetic surfaces and engineering properties of the coils \cite{Beidler1990}. However, the inversion of the Biot-Savart integral by NESCOIL is fundamentally ill-posed, resulting in solutions with amplified noise. The REGCOIL \cite{Landreman2017} approach addresses this problem with Tikhonov regularization. Here the surface-average-squared current density, corresponding to the squared-inverse distance between coils, is added to the objective function. With the addition of this regularization term, REGCOIL can simultaneously increase the minimum coil-coil distances and improve the reconstruction of the desired plasma surface over NESCOIL solutions. In this Chapter, we build on the REGCOIL method to optimize the current distribution in three dimensions. The current distribution on a single winding surface is computed with REGCOIL, and the winding surface geometry is optimized to reproduce the plasma surface with fidelity and improve the engineering properties of the coil shapes. 

Other nonlinear coil optimization tools exist which evolve discrete coil shapes rather than continuous surface current distributions. Drevlak's ONSET code \cite{Drevlak1998} optimizes coils within limiting inner and outer coil surfaces. The COILOPT \cite{Strickler2002,Strickler2004} code, developed for the design of the NCSX coil set \cite{Zarnstorff2001}, optimizes coil filaments on a winding surface which is allowed to vary. COILOPT++ \cite{Brown2015} improved upon COILOPT by defining coils using splines, which enables one to straighten modular coils to improve access to the plasma. The need for a winding surface was eliminated with the FOCUS \cite{Zhu2018} code, which represents coils as three-dimensional space curves. The FOCUS approach employs analytic differentiation for gradient-based optimization, as we do in this Chapter. As the design of optimal coils is central to the development of an economical stellarator, it is important to have several approaches. The current potential method could have several advantages, including the possible implementation of adjoint methods. Furthermore, the complexity of the nonlinear optimization is reduced over other approaches, as the current distribution on the winding surface is efficiently and robustly computed by solving a linear system. By optimizing the winding surface, it is possible to gain insight into what features of plasma surfaces require coils to be close to the plasma, and what features allow coils to be placed farther away \cite{Landreman2016}.

Parallels can be drawn between the design of stellarator coils and 
the design of magnetic resonance imaging (MRI) coils. MRI gradient coils which lie on a cylindrical winding surface must provide a specified spatial variation in the magnetic field within a region of interest. This inverse problem is often solved with a linear least-squares system by minimizing the squared departure from the desired field at specified points with respect to the current in differential surface elements \cite{Turner1993}. This method is comparable to the NESCOIL \cite{Merkel1987} approach for stellarator coil design. Gradient coil design was improved by the addition of a regularization term related to the integral-squared current density \cite{Forbes2005} or the integral-squared curvature \cite{Forbes2001}, comparable to the REGCOIL approach. The adjoint method has been applied to compute the sensitivity of an objective function with respect to the current potential on the MRI winding surface. Here the Biot-Savart law is written in terms of a matrix equation using the least-squares finite element method, and the adjoint of this matrix is inverted to compute the derivatives \cite{Jia2014}. As the adjoint formalism has proven fruitful in this field, we anticipate that it could have similar applications in the closely-related field of stellarator coil design. 

In the Sections that follow, we present a new method for the design of the coil-winding surface using adjoint-based optimization. An adjoint solve is performed to obtain gradients of several figures of merit, the integral-squared normal magnetic field on the plasma surface and root-mean-squared current density on the winding surface, with respect to the Fourier components describing the coil surface. A brief overview of the REGCOIL approach is given in Section \ref{section_REGCOIL}. The optimization method and objective function are described in Section \ref{sect_opt}. The adjoint method for computing gradients of the objective function is outlined in Section \ref{sect_adjoint}. Optimization results for the W7-X and HSX winding surfaces are presented in Section \ref{sect_results}. In Section \ref{sect_sensitivity} we demonstrate a method for computing local sensitivity of figures of merit to perturbations of the winding surface using the shape gradient. We discuss properties of optimized winding surface configurations in Section \ref{sect_configopt}. In Section \ref{sect_conclusions} we summarize our results and conclude. 

\section{Overview of the REGCOIL system}
\label{section_REGCOIL}

First, we review the problem of determining coil shapes once the plasma boundary and coil-winding surface have been specified. Given the winding surface geometry, our task is to obtain the surface current density, $\textbf{J}$. The divergence-free surface current density can be related to a scalar current potential $\Phi$, the stream function for $\textbf{J}$,
\begin{gather}
\textbf{J} = \hat{\textbf{n}} \times \nabla \Phi.
\end{gather}
Here $\hat{\textbf{n}}$ is the unit normal on the winding surface. The current potential $\Phi$ can be decomposed into single-valued and secular terms,
\begin{gather}
\Phi(\theta, \phi) = \Phi_{\text{sv}}(\theta,\phi) + \frac{ G \phi}{2 \pi} + \frac{I \theta}{2 \pi}.
\end{gather}
Here $\phi$ is the cylindrical azimuthal angle and $\theta$ is a poloidal angle. The quantities $G$ and $I$ are the currents linking the surface poloidally and toroidally, respectively. The single-valued term ($\Phi_{\text{sv}}$) is determined by solving the REGCOIL system. It is chosen to minimize the primary objective function,
\begin{gather}
\chi^2 = \chi^2_B + \lambda \chi^2_J.
\label{primary_objective}
\end{gather}
Here $\chi^2_B$ is the surface-integrated-squared normal magnetic field on the desired plasma surface,
\begin{gather}
\chi^2_B = \int_{S_P} d^2 x \, \left(\textbf{B} \cdot \hat{\textbf{n}} \right)^2.
\label{chi2_B}
\end{gather}
The normal component of the magnetic field on the plasma surface, $\textbf{B} \cdot \hat{\textbf{n}}$, includes contributions from currents in the plasma, current density $\textbf{J}$ on the winding surface, and currents in other external coils. The quantity $\chi^2_J$ is the surface-integrated-squared current density on the winding surface,
\begin{gather}
\chi^2_J = \int_{S_{\text{coil}}} d^2x \, |\textbf{J}|^2. 
\end{gather}
As discussed in Section \ref{sec:coil_optimization}, minimization of $\chi^2_B$ by itself ($\lambda=0$) is fundamentally ill-posed, as very different coil shapes can provide almost identical normal field on the plasma surface. (Oppositely directed currents cancel in the Biot-Savart integral.) The addition of $\chi^2_J$ to the objective function is a form of Tikhonov regularization. As we will show, minimization of $\chi^2_J$ also simplifies coil shapes. While the NESCOIL formulation relies on Fourier series truncation for regularization, the formulation in REGCOIL allows for finer control of regularization while improving engineering properties of the coil set. The regularization parameter $\lambda$ can be chosen to obtain a target maximum current density $J_{\text{max}}$, corresponding to a minimum tolerable inter-coil spacing. A 1D nonlinear root finding algorithm is typically used for this process. 

The single-valued part of the current potential $\Phi_{\text{sv}}$ is represented using a finite Fourier series,
\begin{gather}
\Phi_{\text{sv}}(\theta,\phi) = \sum_{m,n} \Phi_{m,n} \sin (m \theta - n N_P \phi),
\end{gather}
where $N_P$ is the number of periods. Only a sine series is needed if stellarator symmetry is imposed on the current density ($J(-\theta,-\phi) = J(\theta,\phi)$). As the minimization of $\chi^2$ with respect to $\Phi_{m,n}$ is a linear least-squares problem, it can be solved via the normal equations to obtain a unique solution. The Fourier amplitudes $\Phi_{m,n}$ are determined by the minimization of $\chi^2$,
\begin{gather}
\partder{\chi^2}{\Phi_{m,n}} = \partder{\chi^2_B}{\Phi_{m,n}} + \lambda \partder{\chi^2_J}{\Phi_{m,n}} = 0,
\label{regcoil_minimization}
\end{gather}
which takes the form of a linear system,
\begin{gather}
\sum_{m,n} A_{m',n';m,n} \Phi_{m,n} = b_{m',n'}.
\label{forward}
\end{gather}
We will use the notation $\overleftrightarrow{\textbf{A}}\overrightarrow{\bm{\Phi}} = \overrightarrow{\textbf{b}}$. Throughout bold-faced type with a right-facing arrow will denote the vector space of basis functions for $\Phi_{\text{sv}}$ unless otherwise noted. For additional details see \cite{Landreman2017}.

\section{Winding surface optimization}
\label{sect_opt}

We use REGCOIL to compute the distribution of current on a fixed, two-dimensional winding surface. To design coil shapes in three-dimensional space, we modify the winding surface geometry by minimizing an objective function (\ref{objective_function}). This objective function quantifies fundamental physics and engineering properties and is easy to calculate from the REGCOIL solution. Optimal coil geometries are obtained by nonlinear, constrained optimization.\footnote{The adjoint method and winding-surface optimization tools are implemented in the main branch of the REGCOIL code \href{https://github.com/landreman/regcoil}{https://github.com/landreman/regcoil}.}

\subsection{Objective function}
The cylindrical components of the winding surface are decomposed in Fourier harmonics,
\begin{subequations}
\begin{align}
R &= \sum_{m,n} R_{m,n}^c \cos(m \theta + n N_p \phi) \\
Z &= \sum_{m,n} Z_{m,n}^s \sin(m \theta + n N_p \phi),
\end{align}
\label{Fourier}
\end{subequations}
where stellarator symmetry of the winding surface is assumed ($R(-\theta,-\phi) = R(\theta,\phi)$ and $Z(-\theta,-\phi) = -Z(\theta,\phi)$). We take the Fourier components of the winding surface, $\Omega = \{R_{m,n}^c, Z_{m,n}^s\}$, as our optimization parameters and assume that the desired plasma surface is held fixed. Throughout, $\Omega$ displayed with a subscript index will refer to a single Fourier component, while in the absence of a subscript, it refers to the set of Fourier components. For a given winding surface geometry, $\Omega$, and desired plasma surface, the current potential $\Phi (\Omega)$ can be determined by solving the REGCOIL system to obtain a solution which both reproduces the desired plasma surface with fidelity and maximizes coil-coil distance, as described in Section \ref{section_REGCOIL}.

We define an objective function, $f$, which will be minimized with respect to $\Omega$,
\begin{gather}
f(\Omega, \overrightarrow{\bm{\Phi}}(\Omega))  = \chi^2_B(\Omega, \overrightarrow{\bm{\Phi}}(\Omega)) - \alpha_V V_{\text{coil}}^{1/3}(\Omega) + \alpha_{\mathcal{S}} \mathcal{S}(\Omega) + \alpha_J \norm{\textbf{J}}_2(\Omega, \overrightarrow{\bm{\Phi}}(\Omega)).
\label{objective_function}
\end{gather}
The coefficients $\alpha_V$, $\alpha_{\mathcal{S}}$, and $\alpha_J$ are positive constants that weigh the relative importance of the terms in $f$. We take $\chi^2_B$ (\ref{chi2_B}) as our proxy for the desired physics properties of the plasma surface. The normal magnetic field depends on $\overrightarrow{\bm{\Phi}}$, the single-valued current potential on the surface, and $\Omega$, the geometric properties of the coil-winding surface. The quantity $V_{\text{coil}}$ is the total volume enclosed by the coil-winding surface,
\begin{gather}
V_{\text{coil}} = \int_{S_{\text{coil}}} d^3 x.
\end{gather}
We use $V_{\text{coil}}^{1/3}$ as a proxy for the coil-plasma separation. Our objective function decreases with increasing $V_{\text{coil}}$, as we desire a winding surface which allows for increased coil-plasma separation. This minimizes coil ripple and provides increased access for neutral beams and diagnostics. We recognize that increasing $V_{\text{coil}}$ implies increased coil length and experiment size, which may not always be desired.

The quantity $\mathcal{S}$ is a measure of the spectral width of the Fourier series describing the coil-winding surface \cite{Hirshman1985},
\begin{gather}
\mathcal{S} = \sum_{m,n} m^{p} \left( (R_{m,n}^c)^2 + (Z_{m,n}^s)^2 \right).
\label{spectral_width}
\end{gather}
Smaller values of $\mathcal{S}$ correspond to Fourier spectra which decay rapidly with increasing $m$. We take advantage of the non-uniqueness of the representation in (\ref{Fourier}) to obtain surface parameterization which are more efficient. As $\chi^2_B$, $\norm{\textbf{J}}_2$, and $V_{\text{coil}}$ are coordinate-independent, these terms remain unchanged if the surface is reparameterized ($\theta$ is redefined). Minimization of $\mathcal{S}$ removes this zero-gradient direction in parameter space.
We use a typical value of $p=2$. One could also remove the redundancy in the definition of $\theta$ by using the unique and spectrally condensed representation of Hirshman and Breslau \cite{Hirshman1998} or by solving the nonlinear constraint equation of Hirshman and Meier \cite{Hirshman1985} once the optimal surface has been obtained.

The quantity $\norm{\textbf{J}}_2 = \sqrt{\chi^2_J/A_{\text{coil}}}$ is the 2-norm of the current density, 
where $A_{\text{coil}}$ is the winding surface area,
\begin{gather}
A_{\text{coil}} = \int_{\text{coil}} d^2 x \, .
\end{gather}
Although we are using a current potential approach rather than directly optimizing coil shapes, including $\norm{\textbf{J}}_2$ in the objective function allows us to obtain coils with good engineering properties. Derivatives of coil-specific metrics (such as curvature) could be computed from the current potential if desired. For example, consider $N$ contours beginning at equally-spaced toroidal angles $\phi_0^i$ and $\theta_0 = 0$. The $i^{\text{th}}$ contour is defined by functions $\theta_i(s)$ and $\phi_i(s)$ for parameter $s$, where $\partial\Phi/\partial s = 0$. The derivatives of coil metrics which depend on $\textbf{x}(\theta_i(s),\phi_i(s))$, could be computed with the adjoint method which will be described in Section \ref{sect_adjoint}. 
As the direct targeting of coil metrics introduces additional arbitrary weights in the objective function and the solution to another adjoint equation must be obtained to compute its gradient, we instead include $\norm{\textbf{J}}_2$ in our objective function.

\begin{figure}
\begin{subfigure}[b]{0.32\textwidth}
\includegraphics[trim=11cm 4cm 9cm 3cm,clip,width=1.0\textwidth]{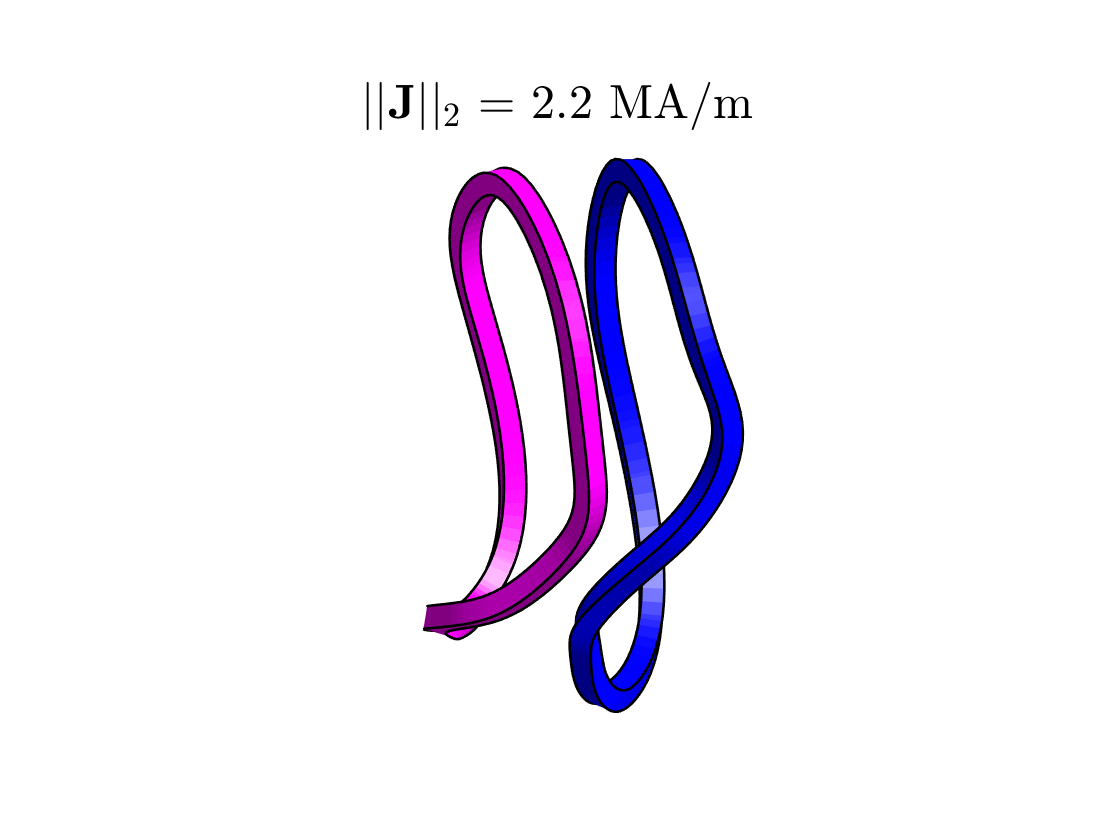}
\caption{}
\end{subfigure}
\begin{subfigure}[b]{0.32\textwidth}
\includegraphics[trim=11cm 4cm 9cm 3cm,clip,width=1.0\textwidth]{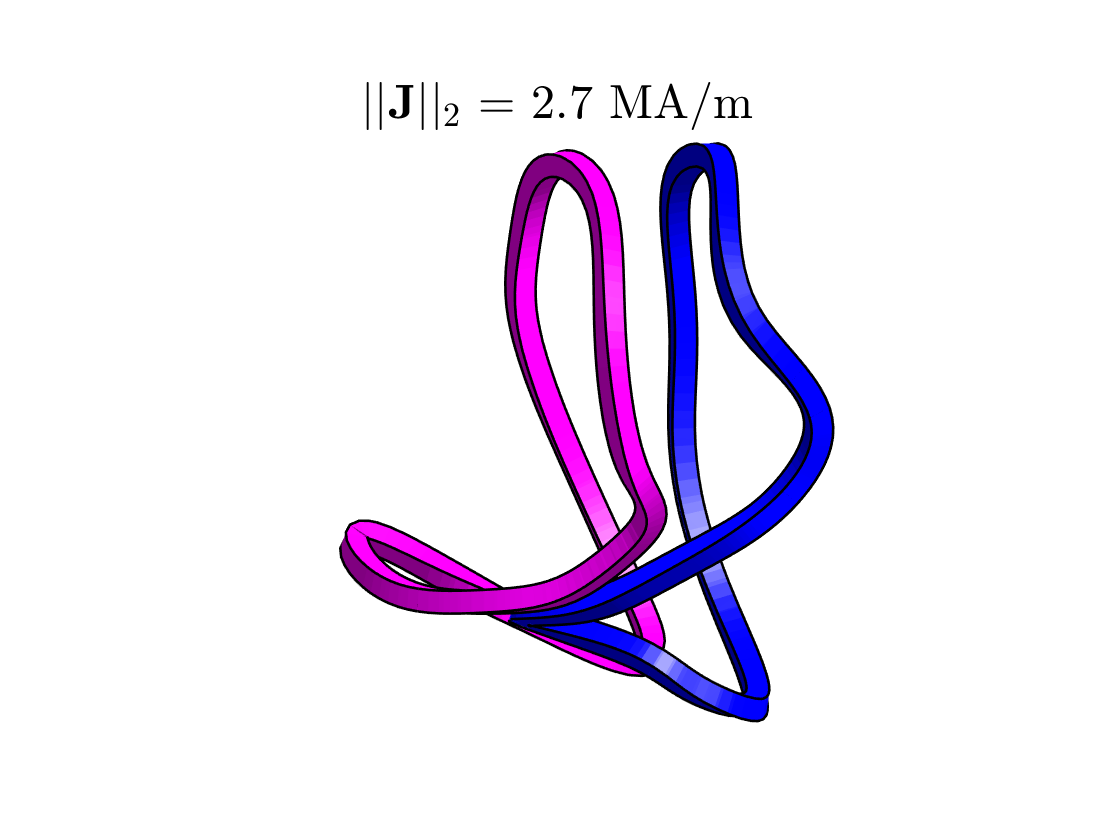}
\caption{}
\end{subfigure}
\begin{subfigure}[b]{0.32\textwidth}
\includegraphics[trim=11cm 4cm 9cm 3cm,clip,width=1.0\textwidth]{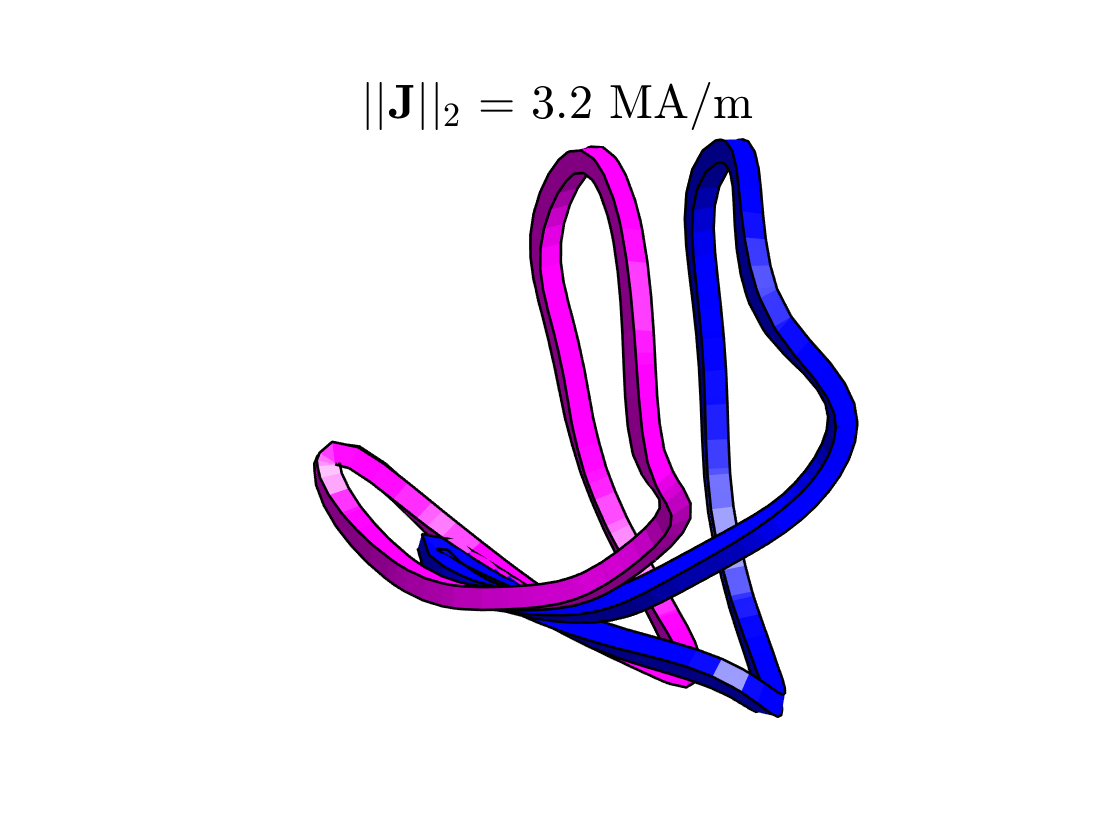}
\caption{}
\end{subfigure}
\caption{Two non-planar W7-X coils (corresponding to the two leftmost coils in Figure \ref{w7x_coils}) computed with REGCOIL using the actual W7-X winding surface. The regularization parameter $\lambda$ is chosen to achieve the shown values of $\norm{\textbf{J}}_2$. As $\norm{\textbf{J}}_2$ increases, the average length, toroidal extent, and curvature increase. Figure adapted from \cite{Paul2018} with permission.}
\label{rmsKcoilcompare}
\end{figure}

To demonstrate this correlation between $\norm{\textbf{J}}_2$
and coil shape complexity, we compute the coil set on the actual W7-X winding surface using REGCOIL. The regularization parameter $\lambda$ is varied to achieve several values of $\norm{\textbf{J}}_2$. Coil shapes are obtained from the contours of $\Phi$. In Figure \ref{rmsKcoilcompare}, two of the W7-X non-planar coils computed in this way are shown, and the corresponding coil metrics are given in Table \ref{rmsKcoilmetrics}. (These correspond to the two leftmost coils in Figure \ref{w7x_coils}.) We consider the average and maximum length $l$, toroidal extent $\Delta \phi$, curvature $\kappa$, and the minimum coil-coil distance $d_{\text{coil-coil}}^{\text{min}}$. The average, maximum, and minimum are taken over the set of 5 unique coils.
The coil shapes become more complex as $\norm{\textbf{J}}_2$ increases, quantified by increasing $\kappa$ and $\Delta \phi$ and decreasing $d_{\text{coil-coil}}^{\text{min}}$.  
Here the curvature, $\kappa$, of a three-dimensional parameterized curve, $\textbf{x}(t)$, is,
\begin{gather}
\kappa = \frac{\left\rvert \textbf{x}'(t) \times \textbf{x}''(t) \right\rvert}{\left\rvert \textbf{x}'(t) \right\rvert^3} .
\label{curvature_of_curve}
\end{gather}
We have compared coil shapes on a single winding surface, finding them to become simpler as $\norm{\textbf{J}}_2$ decreases. As $\norm{\textbf{J}}_2 = \left(\chi^2_J/A_{\text{coil}}\right)^{1/2}$, we would find similar trends with $\chi^2_J$. We have chosen to include $\norm{\textbf{J}}_2$ in the objective function as it is normalized by $A_{\text{coil}}$, so it is a more useful quantity for comparison of coil shapes on different winding surfaces. 

\begin{table} 
\centering
\renewcommand{\arraystretch}{1.4}
\begin{tabular} { | c | c | c | c |}
\hline
$\norm{\textbf{J}}_2$ [MA/m] & 2.20 & 2.70 & 3.20 \\ \hline
$J_{\text{max}}$ [MA/m] & 4.55 & 9.50 & 29.1 \\
$\chi^2_B$ [T$^2$ m$^2$] & 1.89 & $5.25 \times 10^{-3}$ & $2.10 \times 10^{-3}$ \\
Average $l$ [m] & 8.03 & 9.18 & 9.81 \\
Max $l$ [m] & 8.26 & 10.5 & 11.8 \\
Average $\Delta \phi$ [rad.] & 0.146 & 0.222 & 0.253\\
Max $\Delta \phi$ [rad.] & 0.161 & 0.282 & 0.372\\
Average $\kappa$ [m$^{-1}$] & 1.04 & 1.29  & 1.32 \\
Max $\kappa$ [m$^{-1}$] & 2.54 & 20.3 & 56.1 \\
$d_{\text{coil-coil}}^{\text{min}}$ [m] & 0.353 & 0.182 & 0.0758 \\ \hline
\end{tabular}
\caption{Comparison of metrics for coils computed with REGCOIL using the actual W7-X winding surface. Average and max are evaluated for the set of 5 unique coils. The regularization parameter $\lambda$ is varied to achieve these values of $\norm{\textbf{J}}_2$.  Table adapted from \cite{Paul2018} with permission.}
\label{rmsKcoilmetrics}
\end{table}

To minimize $f$, the relative weights in (\ref{objective_function}) ($\alpha_V$, $\alpha_{\mathcal{S}}$, and $\alpha_J$) are chosen such that each of the terms in the objective function have similar magnitudes, though much tuning of these parameters is required to obtain results which simultaneously improve the physics properties (decrease $\chi^2_B$) and engineering properties (increase $V_{\text{coil}}$ and $d_{\text{coil-coil}}^{\text{min}}$, decrease $\kappa$ and $\Delta \phi$).

\subsection{Optimization constraints}
\label{sect_constraint}

Minimization of $f$ is performed subject to the inequality constraint $d_{\text{min}} \geq d_{\text{min}}^{\text{target}}$. Here $d_{\text{min}}$ is the minimum distance between the coil-winding surface and the plasma surface,
\begin{gather}
d_{\text{min}} = \min_{\theta,\phi} \left( d_{\text{coil-plasma}} \right) = \min_{\theta,\phi} \left( \min_{\theta_p, \phi_p} \, \abs{ \textbf{x}_{C} - \textbf{x}_{P} }  \right),
\end{gather}
and $d_{\text{min}}^{\text{target}}$ is the minimum tolerable coil-plasma separation. The quantities $\theta_p$ and $\phi_p$ are poloidal and toroidal angles on the plasma surface, $\textbf{x}_{P}$ and $\textbf{x}_{C}$ are the position vectors on the plasma and winding surface, and $d_{\text{coil-plasma}}$ is the coil-plasma distance as a function of $\theta$ and $\phi$. 

The maximum current density $J_{\text{max}}$ is also constrained,
\begin{gather}
J_{\text{max}} = \max_{\theta,\phi} \, J .
\end{gather}
This roughly corresponds to a fixed minimum coil-coil spacing. This constraint is enforced by fixing $J_{\text{max}}$ to obtain the regularization parameter $\lambda$ in the REGCOIL solve, so we avoid the need for an equality constraint or the inclusion of $J_{\text{max}}$ in the objective function. Rather, $\overrightarrow{\bm{\Phi}}(\Omega)$ is determined such that $J_{\text{max}}$ is fixed. The inequality-constrained nonlinear optimization is performed using the NLOPT \cite{NLOPT} software package using a conservative convex separable quadratic approximation (CCSAQ) \cite{Svanberg2002}. While there are several gradient-based inequality-constrained algorithms available, we choose to use CCSAQ as it is relatively insensitive to the bound constraints imposed on the optimization parameters. We recognize that there are many possible combinations of constraints, objective functions, and regularization conditions that could be used. For example, $\norm{\textbf{J}}_2$ could be fixed to determine $\lambda$ while $J_{\text{max}}$ could be included in the objective function. We found that the formulation we have presented produces the best coil shapes. 

\section{Derivatives of $f$ and the adjoint method}
\label{sect_adjoint}
We must compute derivatives of $f$ with respect to the geometric parameters $\Omega$ in order to use gradient-based optimization methods. The spectral width $\mathcal{S}$ and the volume $V_{\text{coil}}$ are explicit functions of $\Omega$, so their analytic derivatives can be obtained. On the other hand, $\chi^2_B$ and $\norm{\textbf{J}}_2$ depend both explicitly on coil geometry and on $\bm{\Phi}(\Omega)$. One approach to obtain the derivatives of these quantities could be to solve the REGCOIL linear system $N_{\Omega} +1$ times, taking a finite-difference step in each Fourier coefficient. However, if $N_{\Omega}$ is large, the computational cost of this method could be prohibitively expensive. Instead, we will apply the adjoint method to compute derivatives. This technique will be demonstrated below. 

The derivative of $\chi^2_B$ can be computed using the chain rule, 
\begin{gather}
\partder{\chi^2_B(\Omega, \overrightarrow{\bm{\Phi}}(\Omega))}{\Omega_{m,n}} = \partder{\chi^2_B(\Omega,\overrightarrow{\bm{\Phi}})}{\Omega_{m,n}}  + \partder{\chi^2_B(\Omega,\overrightarrow{\bm{\Phi}})}{\overrightarrow{\bm{\Phi}}} \cdot \partder{\overrightarrow{\bm{\Phi}}(\Omega)}{\Omega_{m,n}},
\label{sensitivity_der}
\end{gather}
where $\overrightarrow{\bm{\Phi}}(\Omega)$ is understood to vary with $\Omega$ such that (\ref{forward}) is satisfied. The dot product is a contraction over the current potential basis functions, $\{\Phi_{m,n}\}$. We can compute $\partial \overrightarrow{\bm{\Phi}}(\Omega)/ \partial \Omega_{m,n}$ by differentiating the linear system (\ref{forward}) with respect to $\Omega_{m,n}$, 
\begin{gather}
\partder{\overleftrightarrow{\textbf{A}}(\Omega)}{\Omega_{m,n}} \overrightarrow{\bm{\Phi}} + \overleftrightarrow{\textbf{A}} \partder{\overrightarrow{\bm{\Phi}}(\Omega)}{\Omega_{m,n}} = \partder{\overrightarrow{\textbf{b}}(\Omega)}{\Omega_{m,n}},
\label{eq:perturbed_linear_system}
\end{gather}
and formally solving this equation to obtain,
\begin{gather}
\partder{\bm{\Phi}(\Omega)}{\Omega_{m,n}} = \overleftrightarrow{\textbf{A}}^{-1} \left( \partder{\overrightarrow{\textbf{b}}(\Omega)}{\Omega_{m,n}} - \partder{\overleftrightarrow{\textbf{A}}(\Omega)}{\Omega_{m,n}} \overrightarrow{\bm{\Phi}} \right).
\label{linear_sensitivity}
\end{gather}
Equation (\ref{linear_sensitivity}) is inserted into  (\ref{sensitivity_der}),
\begin{align}
\partder{\chi^2_B(\Omega, \overrightarrow{\bm{\Phi}}(\Omega))}{\Omega_{m,n}} = \partder{\chi^2_B(\Omega,\overrightarrow{\bm{\Phi}})}{\Omega_{m,n}}
+ \partder{\chi^2_B(\Omega,\overrightarrow{\bm{\Phi}})}{\overrightarrow{\bm{\Phi}}} \cdot \left[ \overleftrightarrow{\textbf{A}}^{-1} \left( \partder{\overrightarrow{\textbf{b}}(\Omega)}{\Omega_{m,n}} - \partder{\overleftrightarrow{\textbf{A}}(\Omega)}{\Omega_{m,n}} \overrightarrow{\bm{\Phi}} \right) \right].
\end{align}
This expression could be evaluated by solving the linear system \eqref{eq:perturbed_linear_system} for $\partial \overrightarrow{\bm{\Phi}}/\partial \Omega_{m,n}$ and performing the inner product with $\partial \chi^2_B/ \partial \overrightarrow{\bm{\Phi}}$. However, the computational cost of this method scales similarly to that of finite differencing, as described in Section \ref{sec:complexity_comparison}. Instead, we can exploit the adjoint property of the operator to obtain,
\begin{align}
\partder{\chi^2_B(\Omega, \overrightarrow{\bm{\Phi}}(\Omega))}{\Omega_{m,n}} = \partder{\chi^2_B(\Omega,\overrightarrow{\bm{\Phi}})}{\Omega_{m,n}}
+ \left[ \left(\overleftrightarrow{\textbf{A}}^{-1}\right)^{T} \partder{\chi^2_B(\Omega,\overrightarrow{\bm{\Phi}})}{\overrightarrow{\bm{\Phi}}}\right] \cdot \left( \partder{\overrightarrow{\textbf{b}}(\Omega)}{\Omega_{m,n}} - \partder{\overleftrightarrow{\textbf{A}}(\Omega)}{\Omega_{m,n}} \overrightarrow{\bm{\Phi}} \right).
\end{align}
For any invertible matrix, $\left( \overleftrightarrow{\textbf{A}}^{-1} \right)^T = \left( \overleftrightarrow{\textbf{A}}^{T} \right)^{-1}$. Hence we can instead solve a linear system involving the matrix $\overleftrightarrow{\textbf{A}}^{T}$ to compute an adjoint variable $\overrightarrow{\textbf{q}}$, defined as the solution of
\begin{gather}
\overleftrightarrow{\textbf{A}}^{T} \overrightarrow{\textbf{q}} = \partder{\chi^2_B(\Omega,\overrightarrow{\bm{\Phi}})}{\overrightarrow{\bm{\Phi}}}.
\label{adjoint}
\end{gather}
Rather than compute a finite-difference derivative for each $\Omega_{m,n}$ or solve a linear system to compute each $\partial \overrightarrow{\bm{\Phi}}/\partial \Omega_{m,n}$ as in (\ref{linear_sensitivity}), we solve two linear systems: the forward (\ref{forward}) and adjoint (\ref{adjoint}). The adjoint equation is similar to the forward equation ($\overleftrightarrow{\textbf{A}}^T$ has the same dimensions and eigenspectrum as $\overleftrightarrow{\textbf{A}}$), so the same computational tools can be used to solve the adjoint problem. We then perform an inner product with $\overrightarrow{\textbf{q}}$ to obtain the derivatives with respect to each $\Omega_{m,n}$,
\begin{gather}
\partder{\chi^2_B(\Omega, \overrightarrow{\bm{\Phi}}(\Omega))}{\Omega_{m,n}} = \partder{\chi^2_B(\Omega,\overrightarrow{\bm{\Phi}})}{\Omega_{m,n}}  + \overrightarrow{\textbf{q}} \cdot \left( \partder{\overrightarrow{\textbf{b}}(\Omega)}{\Omega_{m,n}} - \partder{\overleftrightarrow{\textbf{A}}(\Omega)}{\Omega_{m,n}} \bm{\Phi} \right).
\label{adjointsensitivity}
\end{gather}
The derivatives $\partial \overrightarrow{\textbf{b}}/\partial \Omega_{m,n}$, $\partial \overleftrightarrow{\textbf{A}}/\partial \Omega_{m,n}$, $\partial \chi^2_B/\partial \Omega_{m,n}$, and
$\partial \chi^2_B/\partial \overrightarrow{\bm{\Phi}}$ can be computed analytically. In the above discussion, the regularization parameter $\lambda$ has been assumed to be fixed. A similar method can be used if a $\lambda$ search is performed to obtain a target $J_{\text{max}}$ (see Appendix \ref{lambda_search}). The same method is used to compute derivatives of $\norm{\textbf{J}}_2$. 

We note that adjoint methods provide the most significant reduction in computational cost when the linear solve is expensive. For the REGCOIL system, this is not the case, as the cost of constructing $\overleftrightarrow{\textbf{A}}$ and $\overrightarrow{\textbf{b}}$ exceeds that of the solve. We have implemented OpenMP multithreading for the construction of $\partial \overleftrightarrow{\textbf{A}}/\partial \Omega$ and $\partial \overrightarrow{\textbf{b}}/\partial \Omega$ such that the cost of computing the gradients via the adjoint method is cheaper than computing finite-difference derivatives serially.

The constraint functions, $d_{\text{min}}$ and $J_{\text{max}}$, must also be differentiated with respect to $\Omega_{m,n}$. As $d_{\text{min}}$ is defined in terms of the minimum function, we approximate it using the smooth log-sum-exponent function \cite{Boyd2004},
\begin{gather}
d_{\text{min, lse}} = - \frac{1}{q} \log \left( \frac{\int_{S_{C}} d^2 x_C \, \int_{S_P} d^2 x_P \, \exp \left( - q \abs{\textbf{x}_{C} - \textbf{x}_{P}} \right) }{\int_{S_C} d^2 x_C \, \int_{S_P} d^2 x_P \, } \right).
\label{lse_d}
\end{gather}
This function can be analytically differentiated with respect to $\Omega_{m,n}$. As $q$ approaches infinity, $d_{\text{min, lse}}$ approaches $d_{\text{min}}$. For $q$ very large, the function obtains very sharp gradients. A typical value of $q = 10^4$ m$^{-1}$ was used. The log-sum-exponent function is also used to approximate $J_{\text{max}}$, as described in Appendix \ref{lambda_search}.

\FloatBarrier
\section{Winding surface optimization results}
\label{sect_results}
\FloatBarrier

\subsection{Trends with optimization parameters}

\begin{figure}
\centering
\includegraphics[trim=5cm 0cm 2cm 0cm,width=0.8\textwidth]{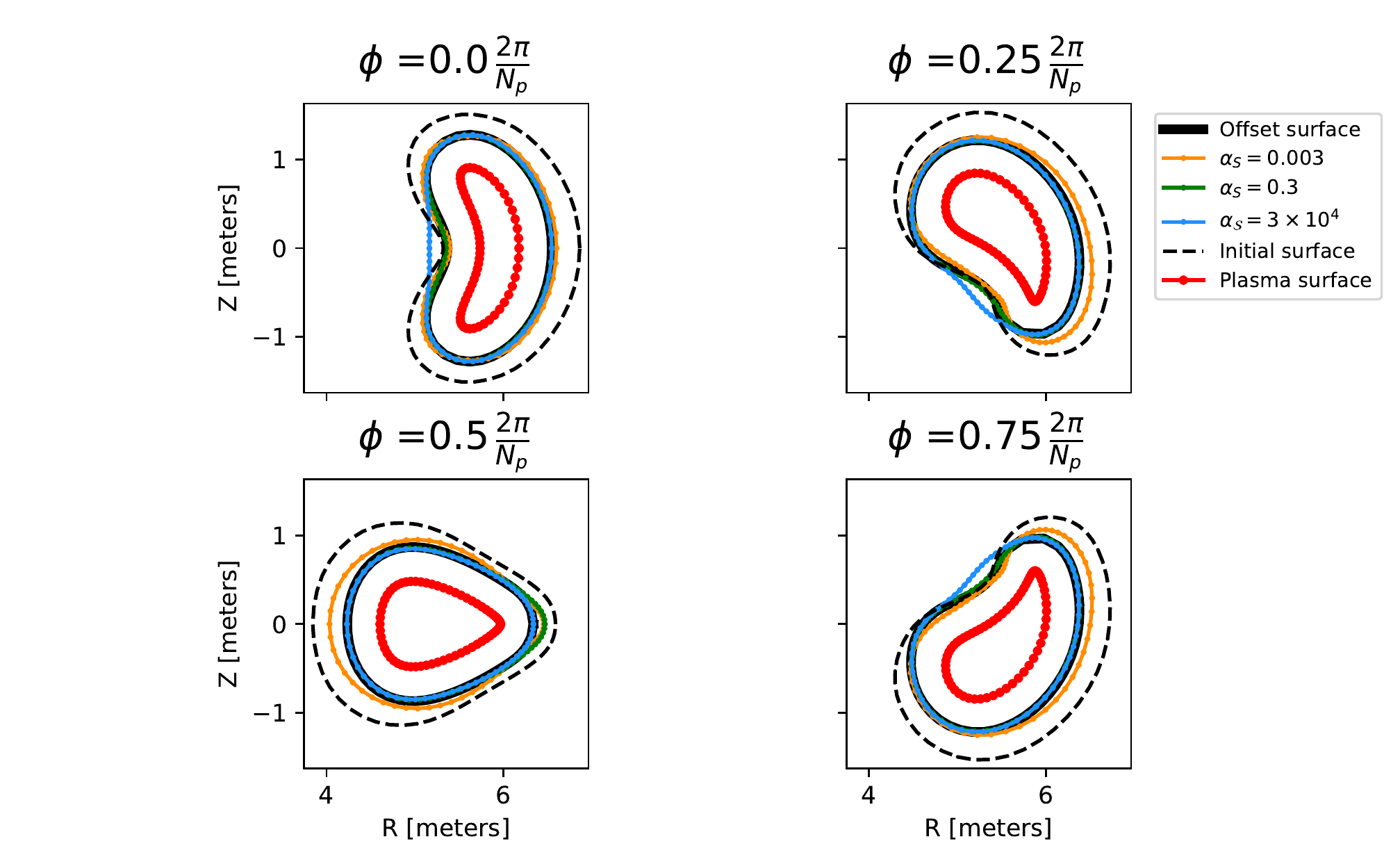}
\caption{Optimized winding surfaces obtained with $\alpha_V = \alpha_J = 0$ and the values of $\alpha_{\mathcal{S}}$ shown. The actual W7-X winding surface is used as the initial surface in the optimization (black dashed). As $\alpha_{\mathcal{S}}$ increases, the magnitude of the spectral-width term in the objective function increases, and the winding surface approaches a cylindrical torus with a minimal Fourier spectrum. For moderately small values of $\alpha_{\mathcal{S}}$, the winding surface approaches a uniform offset surface from the plasma surface (black solid). Figure adapted from \cite{Paul2018} with permission.}
\label{alpha2_scan}
\end{figure}

\begin{figure}
\centering
\includegraphics[trim=2cm 0cm 0cm 0cm,clip,width=1.0\textwidth]{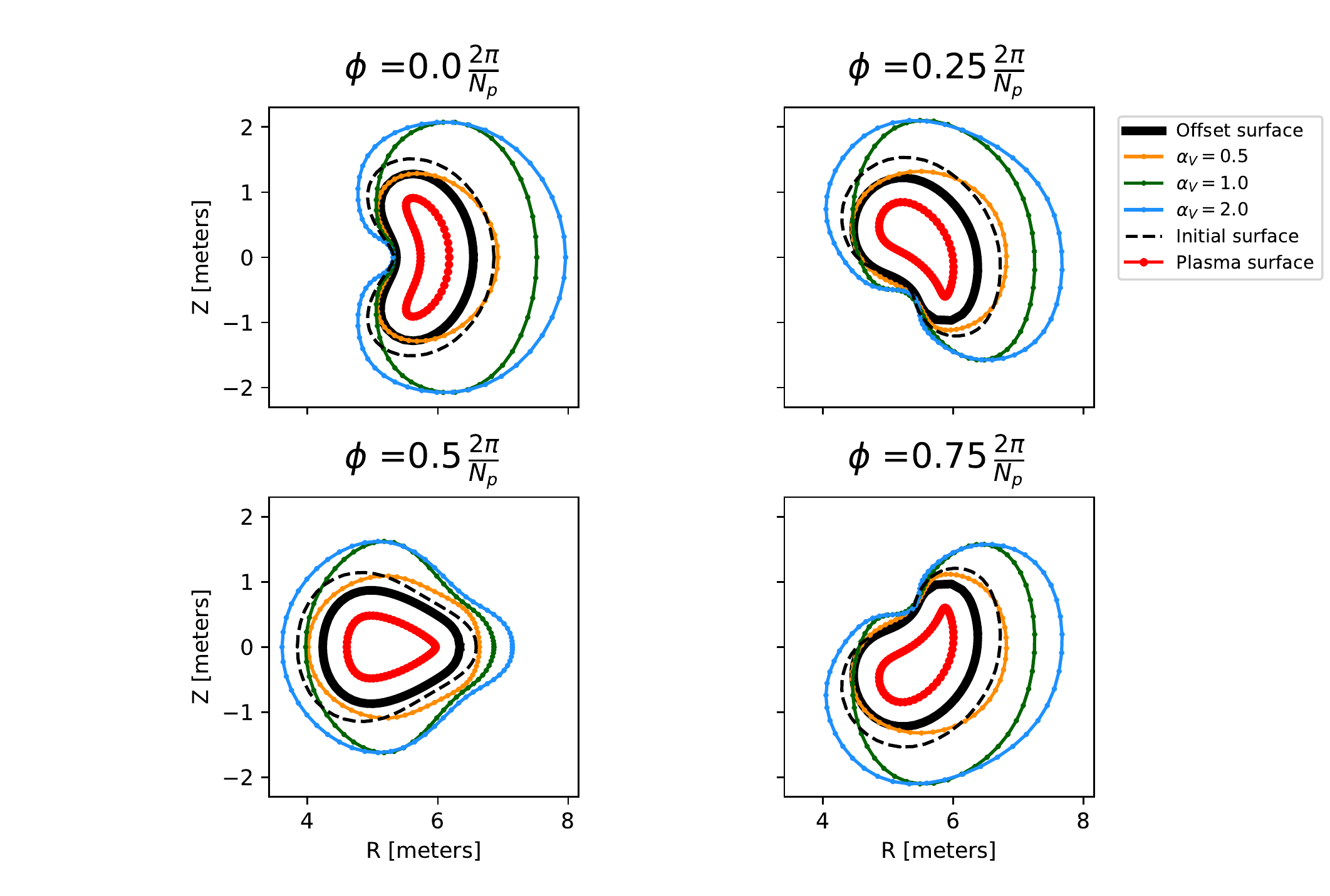}
\caption{Optimized winding surfaces obtained with $\alpha_{\mathcal{S}} = 0.3$, $\alpha_J = 0$, and the values of $\alpha_V$ shown. The actual W7-X winding surface is used as the initial surface in the optimization (black dashed). As $\alpha_V$ increases, $d_{\text{coil-plasma}}$ increases on the outboard side while it remains fixed in the concave region. Figure adapted from \cite{Paul2018} with permission.}
\label{alpha1_scan}
\end{figure}

Beginning with the actual W7-X winding surface, we perform scans over the coefficients $\alpha_V$ and $\alpha_{\mathcal{S}}$ in the objective function (\ref{objective_function}). The plasma surface was obtained from a fixed-boundary VMEC solution that predated the coil design and is free from modular coil ripple. The constraint target is set to be the minimum coil-plasma distance on the initial winding surface, $d_{\text{min}}^{\text{target}}=0.37$ m. The cross-sections of the optimized surfaces in the poloidal plane are shown in Figures \ref{alpha2_scan} and \ref{alpha1_scan} along with the last-closed flux surface (red), a constant offset surface at $d_{\text{min}}^{\text{target}}$ (black solid), and the initial winding surface (black dashed).

We perform a scan over $\alpha_{\mathcal{S}}$ with $\alpha_V = \alpha_J = 0$. For optimal values of $\alpha_{\mathcal{S}}$, the addition of the spectral width term should simply reparameterize the surface, eliminating the zero-gradient direction in parameter space. Thus we expect that when $\chi^2_B$ is the only other term in the objective function, the winding surface should collapse to a constant offset surface. When $\alpha_{\mathcal{S}}$ is too large, the surface shape changes to favor a condensed Fourier series. When $\alpha_{\mathcal{S}}$ is too small, the optimization may terminate prematurely in a local minimum due to the non-uniqueness of the representation. Indeed we find that with increasing $\alpha_{\mathcal{S}}$, the winding surface approaches a torus with a circular cross-section, which has a minimal Fourier spectrum. At moderately small values of $\alpha_{\mathcal{S}}$ ($\sim0.3$) the surface approaches a constant offset surface at $d_{\text{min}}^{\text{target}}$, as $\chi^2_B$ is dominant in objective function. For very small values of $\alpha_{\mathcal{S}}$ ($\sim 0.003$), we find that the optimization terminates at a point relatively close to the initial surface, and the resulting winding surface deviates from a constant offset surface. An intermediate value of $\alpha_{\mathcal{S}} = 0.3$ was chosen for the following optimizations of the W7-X winding surface.

A scan over $\alpha_V$ is performed at fixed $\alpha_{\mathcal{S}} = 0.3$ and $\alpha_J = 0$ such that the spectral width does not greatly increase. As $\alpha_V$ increases, $d_{\text{coil-plasma}}$ increases significantly on the outboard side while it remains fixed in the inboard concave regions. This trend is not surprising, as concave plasma shapes have been shown to be inefficient to produce with coils \cite{Landreman2016}. 
Interestingly, the winding surface obtains a somewhat pointed shape at the triangle cross-section ($\phi = 0.5$ $2\pi/N_p$), becoming elongated at the tip of the triangle and ``pinching" toward the plasma surface at the edges. 

\subsection{Optimal W7-X winding surface}
\label{w7x_results}

\begin{figure}
\centering
\includegraphics[trim=3cm 0cm 0cm 0cm,clip,width=1.0\textwidth]{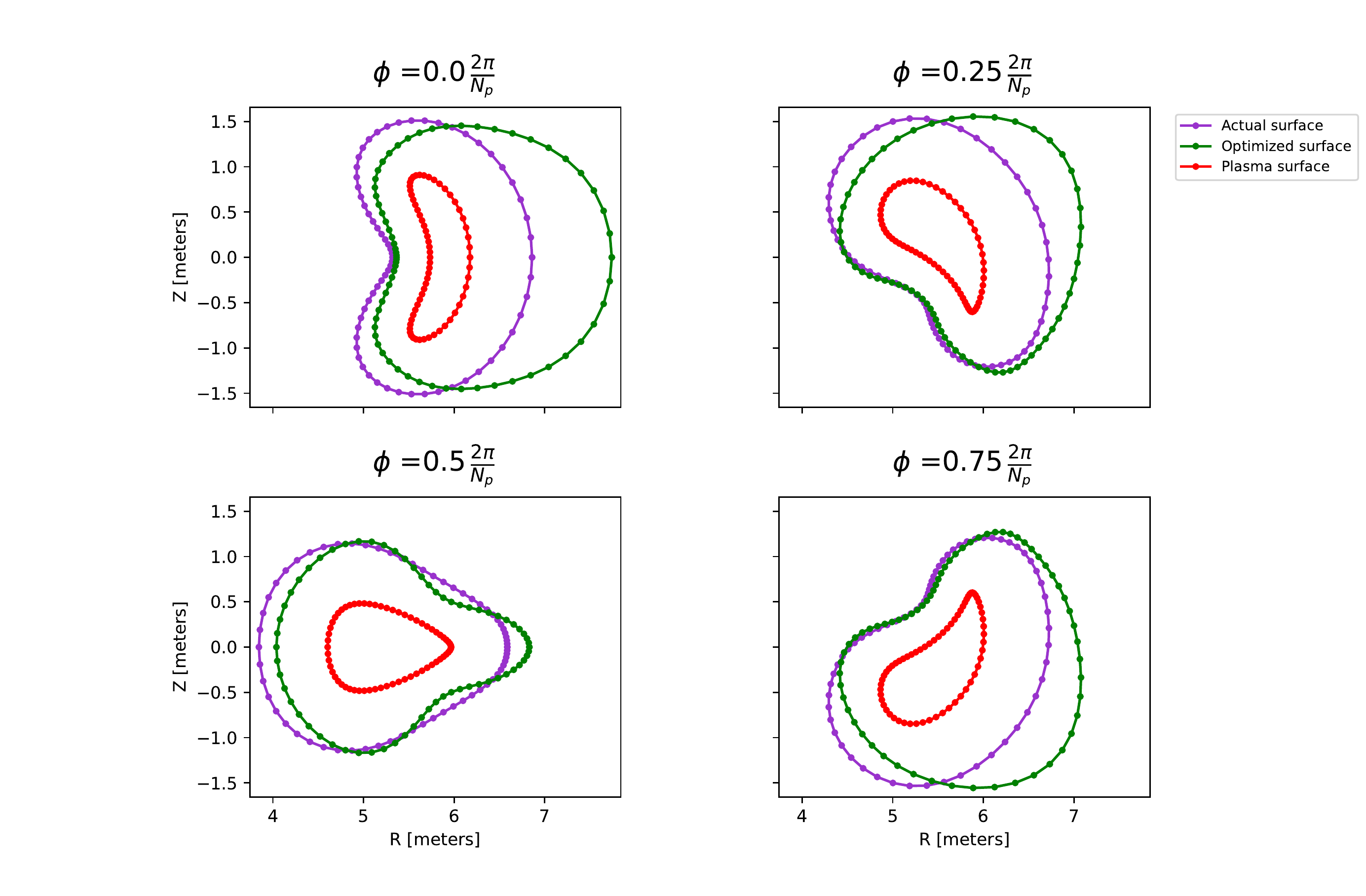}
\caption{The actual W7-X coil-winding surface and plasma surface are shown with our optimized winding surface. In comparison with the actual surface, the optimized surface reduces $\chi^2_B$ by 52\% and increases $V_{\text{coil}}$ by 22\%. Figure adapted from \cite{Paul2018} with permission.}
\label{w7x_surf}
\end{figure}

\begin{figure}
\centering
\includegraphics[width=0.8\textwidth]{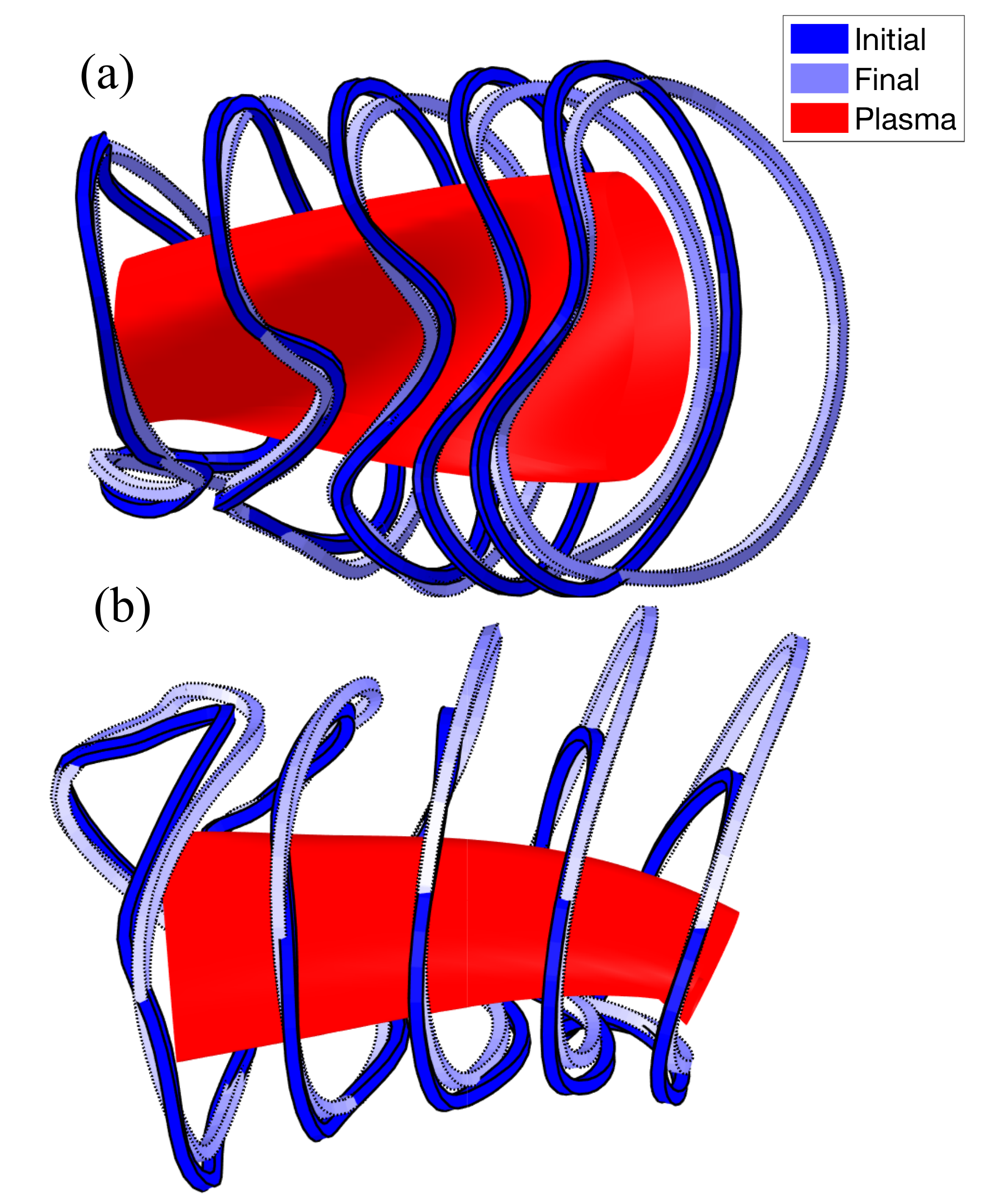}
\caption{Comparisons of coil set computed with REGCOIL using the actual W7-X winding surface (dark blue) and the optimized surface (light blue). Figure reproduced from \cite{Paul2018} with permission.}
\label{w7x_coils}
\end{figure}

We now include nonzero $\alpha_J$ and attempt a comprehensive optimization. The $J_{\text{max}}$ constraint is selected such that the metrics ($l$, $\kappa$, and $\Delta \phi$) of the coils computed on the initial surface roughly match those of the actual non-planar coil set. The coil-plasma distance constraint  $d_{\text{min}}^{\text{target}}$ is set to be the minimum $d_{\text{coil-plasma}}$ on the initial winding surface. Parameters $\alpha_V = 0.5$, $\alpha_{\mathcal{S}} = 0.24$, and $\alpha_J = 1.6\times10^{-6}$ were used in the objective function. Optimization was performed over 118 Fourier coefficients $\big(\abs{n} \leq 4$ and $m \leq 6$ in (\ref{Fourier})$\big)$ and the objective function was evaluated a total of 5165 times to reach the optimum ($1.5\times 10^4$ linear solves rather than $6.1\times10^5$ required for finite-difference derivatives). The optimal surface and coil set are shown in Figures \ref{w7x_surf} and \ref{w7x_coils}, and the corresponding metrics are shown in Table \ref{table_w7x}. We find a solution which increases $V_{\text{coil}}$ by 22\% and decreases $\chi^2_B$ by 52\% over the initial winding surface. (Note that it is numerically impossible to obtain a current distribution that exactly reproduces the plasma surface, so $\chi^2_B$ is nonzero when computed from the REGCOIL solution on the initial winding surface.) In addition, the optimized coil set features a smaller average and maximum $\Delta \phi$ and $\kappa$ and larger $d_{\text{coil-coil}}^{\text{min}}$. The length of the coils increases to accommodate for the increase in $V_{\text{coil}}$. Again we find that the increase in $V_{\text{coil}}$ is most pronounced in the outboard convex regions while $d_{\text{coil-plasma}}$ is maintained in the concave regions of the bean-shaped cross-sections. The ``pinching" feature of the winding surface is again present in the triangle cross-section ($\phi = 0.5 \, 2\pi/N_p$).  

It should be noted that the decrease in $d_{\text{coil-plasma}}$ at the bottom and top of the bean cross-section ($\phi = 0$) might interfere with the current W7-X divertor baffles. However, the increase in volume on the outboard side would allow for increased flexibility for the neutral beam injection duct \cite{Rust2011}. We have performed this optimization to show that a winding surface could be constructed that increases $V_{\text{coil}}$ (and thus the average $d_{\text{coil-plasma}}$), improves coil shapes, and decreases $\chi^2_B$. If further engineering considerations were necessary, these could be implemented. The surface we have obtained is optimal with respect to the engineering considerations and constraints we have imposed, which differ from those of the W7-X team \cite{Beidler1990}. Thus the direct comparison between our method and those of \cite{Beidler1990} cannot be made based on these results.

\begin{table} 
\centering
\renewcommand{\arraystretch}{1.4}
\begin{tabular} {| c | c | c | c |}
\hline
 & Initial & Optimized & Actual coil set \\ \hline
$\chi^2_B$ [T$^2$m$^2$] & 0.115 & 0.0711 & \\
$V_{\text{coil}}$[m$^3$] & 156 & 190 & \\
$\norm{\textbf{J}}_2$ [MA/m] & 2.21 & 2.16 & \\
$J_{\text{max}}$ [MA/m] & 7.70 & 7.70 & \\
Average $l$ [m] & 8.51 & 8.95 & 8.69  \\
Max $l$ [m] & 8.84 & 9.14 & 8.74 \\
Average $\Delta \phi$ [rad.] & 0.190 & 0.179 & 0.198 \\
Max $\Delta \phi$ [rad.] & 0.222 & 0.197 & 0.208 \\
Average $\kappa$ [m$^{-1}$] & 1.21 & 1.10 & 1.20 \\
Max $\kappa$ [m$^{-1}$] & 9.01 & 4.84 & 2.59 \\
$d_{\text{coil-coil}}^{\text{min}}$ [m] & 0.223 & 0.271 & 0.261 \\ \hline
\end{tabular}
\caption{Comparison of metrics of the actual W7-X winding surface and our optimized surface. We also show metrics of the coil set computed on the winding surfaces using REGCOIL and the metrics for the actual W7-X nonplanar coils. Regularization in REGCOIL is chosen such that the coil metrics computed on the initial surface roughly match those of the actual coil set. Coil complexity improves from the initial to the final surface (decreased average and max $\Delta \phi$ and $\kappa$, increased $d_{\text{coil-coil}}^{\text{min}}$). The average and max $l$ increases to allow for the increase in $V_{\text{coil}}$. Table adapted from \cite{Paul2018} with permission.}
\label{table_w7x}
\end{table}

\subsection{Optimal HSX winding surface}
\label{hsx_results}

\begin{figure}
\centering
\includegraphics[trim=2cm 0cm 0cm 0cm,clip,width=1\textwidth]{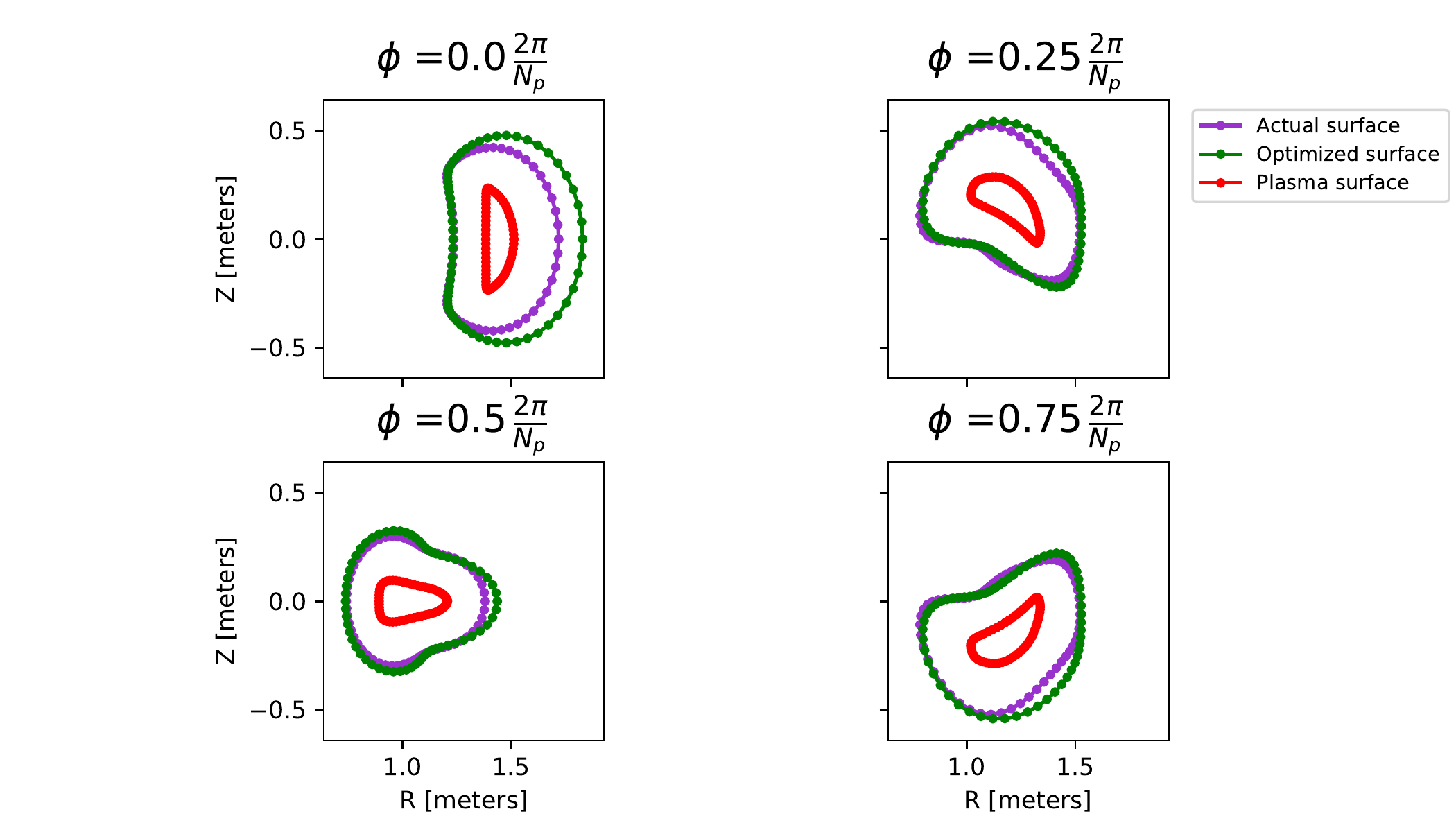}
\caption{The actual HSX coil-winding surface and plasma surface are shown with our optimized winding surface. In comparison with the actual surface, the optimized surface has decreased $\chi^2_B$ by 4\% and increased $V_{\text{coil}}$ by 18\%. Figure adapted from \cite{Paul2018} with permission.}
\label{hsx_surf}
\end{figure}

\begin{figure}
\centering
\includegraphics[width=0.8\textwidth]{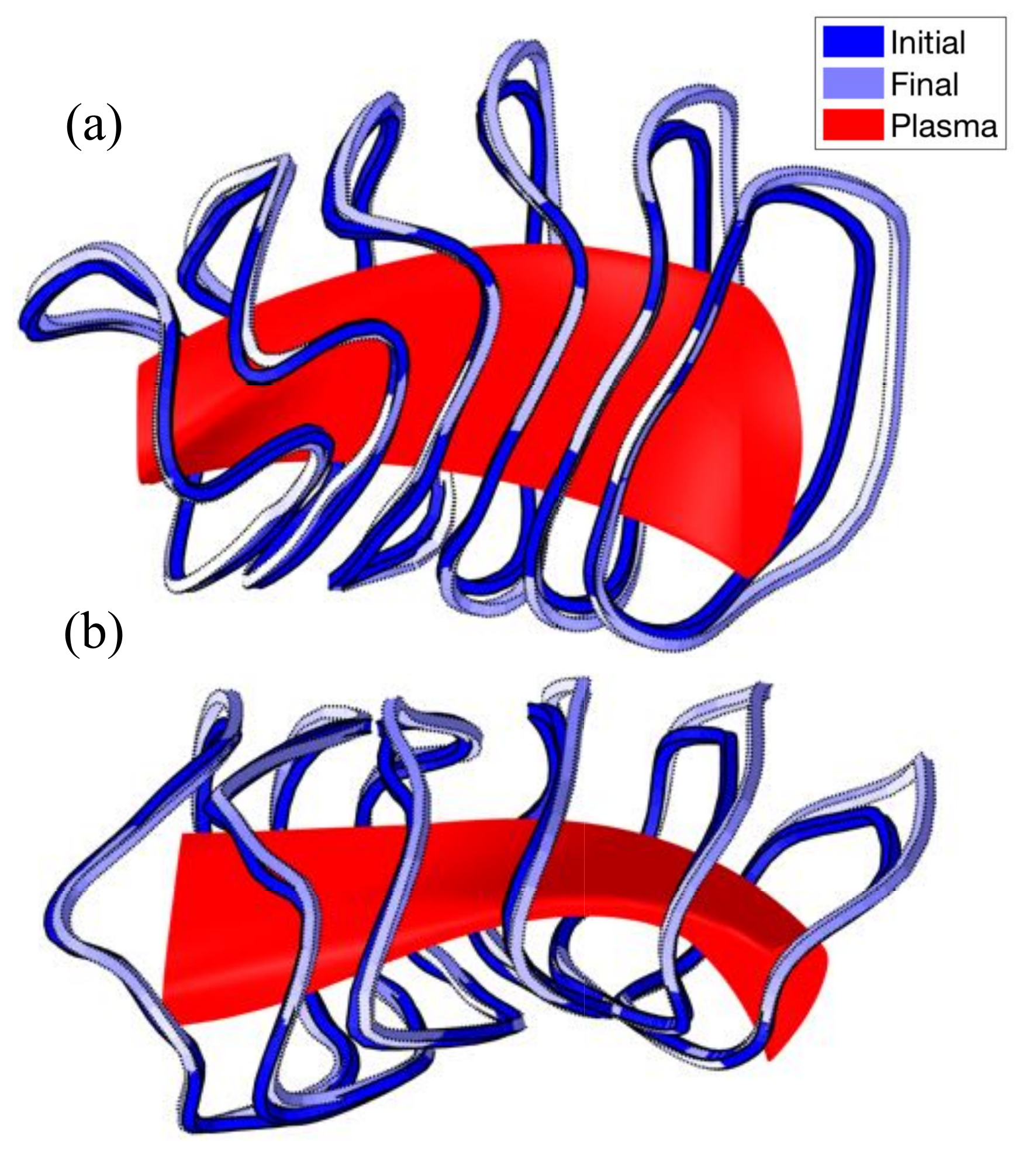}
\caption{The coils obtained from REGCOIL using the actual HSX winding surface (dark blue) and optimized surface (light blue). Figure reproduced from \cite{Paul2018} with permission.}
\label{hsx_coils}
\end{figure}

We perform the same procedure for the optimization of the HSX winding surface. Parameters $\alpha_V = 3.13\times 10^{-4}$, $\alpha_{\mathcal{S}} = 0$, and $\alpha_J = 3\times 10^{-10}$ were used in the objective function. We found that the spectral width term was not necessary to obtain a satisfying optimum in this case. The initial winding surface was taken to be a toroidal surface on which the actual modular coils lie. The plasma equilibrium used is a fixed-boundary VMEC solution without coil ripple. Optimization was performed over 100 Fourier coefficients $\big(\abs{n} \leq 5$ and $m \leq 4$ in (\ref{Fourier})$\big)$ and the objective function was evaluated a total of 560 times to reach the optimum ($1.7\times10^3$ linear solves rather than $5.7\times10^4$ required for forward-difference derivatives). The coil-plasma distance constraint was set to be $d_{\text{min}}^{\text{target}} = 0.14$ m, the minimum coil-plasma distance on the actual winding surface. The optimal surface and coil set are shown in Figures \ref{hsx_surf} and \ref{hsx_coils}, and the corresponding coil metrics are shown in Table \ref{table_hsx}. We find a solution that increases $V_{\text{coil}}$ by 18\% and decreases $\chi^2_B$ by 4\% over the initial winding surface. The coil set computed with REGCOIL using the optimized surface appears qualitatively similar to that computed with the initial surface but with increased $d_{\text{coil-plasma}}$ on the outboard side. The average and maximum $\Delta \phi$ and $\kappa$ decreased while $d_{\text{coil-coil}}^{\text{min}}$ was increased for the coil set computed on the optimal surface in comparison to that of the initial surface. As was observed in the W7-X optimization (Figure \ref{w7x_surf}), the optimized HSX winding surface obtains a somewhat pinched shape near the triangle cross-section ($\phi = 0.5 \, 2\pi/N_p$). 

\begin{table} 
\centering
\renewcommand{\arraystretch}{1.4}
\begin{tabular} {| c | c | c | c |}
\hline
& Initial & Optimized & Actual coil set \\ \hline
$\chi^2_B$ [T$^2$m$^2$] & $1.53 \times 10^{-5}$ & $1.47\times 10^{-5}$ & \\
$V_{\text{coil}}$[m$^3$] & 2.60 & 3.07 & \\
$\norm{\textbf{J}}_2$ [MA/m] & 0.956 & 0.891 & \\
$J_{\text{max}}$ [MA/m] & 1.84 & 1.84 & \\
Average $l$ [m] & 2.26 & 2.39 & 2.24  \\
Max $l$ [m] & 2.49 & 2.46 & 2.33 \\
Average $\Delta \phi$ [rad.] & 0.372 & 0.365 & 0.362 \\
Max $\Delta \phi$ [rad.] & 0.530 & 0.505 & 0.478 \\
Average $\kappa$ [m$^{-1}$] & 5.15 & 4.80 & 5.05 \\
Max $\kappa$ [m$^{-1}$] & 33.4 & 25.8 & 11.7 \\
$d_{\text{coil-coil}}^{\text{min}}$ [m] & 0.0850 & 0.0853 & 0.0930 \\ \hline
\end{tabular}
\caption{Comparison of metrics of the actual HSX winding surface and our optimized surface. We also show metrics of the coil set computed on the winding surfaces using REGCOIL and the metrics for the actual HSX modular coils. Regularization in REGCOIL is chosen such that the coil metrics computed on the initial surface roughly match those of the actual coil set. Coil complexity improves from the initial to the final surface (decreased  average and max $\Delta \phi$ and $\kappa$, increased $d_{\text{coil-coil}}^{\text{min}}$). The average and max $l$ increases to allow for the increase in $V_{\text{coil}}$. Table adapted from \cite{Paul2018} with permission.}  
\label{table_hsx}
\end{table}

\FloatBarrier
\section{Local winding surface sensitivity}
\label{sect_sensitivity}
\FloatBarrier

With the adjoint method we have computed derivatives of the objective function with respect to Fourier components of the winding surface, $\partial f/\partial \Omega$. While this representation of derivatives is convenient for gradient-based optimization, the sensitivity to local displacements of the surface is obscured. Alternatively, it is possible to represent the sensitivity of $f$ with respect to normal displacements of surface area elements of a given winding surface $S_C$,
\begin{gather}
\delta f(S_C; \delta \textbf{x}) = \int_{S_C} d^2 x \,  \, \mathcal{G} \delta \textbf{x} \cdot \hat{\textbf{n}}.
\label{shapederivative}
\end{gather}
The shape gradient and shape derivatives are described in detail in Section \ref{sec:shape_optimization}.
As both $\chi^2_B$ and $\norm{\textbf{J}}_2$ are defined in terms of surface integrals over the winding surface, it can be shown that the shape derivative of these functions 
can be written in the Hadamard form \cite{Novotny2013}. The shape gradients $\mathcal{G}_{\chi^2_B}$ and $\mathcal{G}_{\norm{\textbf{J}}_2}$ can be computed from the Fourier derivatives ($\partial \chi^2_B/\partial \Omega$ and $\partial \norm{\textbf{J}}_2/\partial \Omega$) using a singular value decomposition method \cite{Landreman2018}. Here the perturbations $\delta f$ and $\delta \textbf{x}$ are written in terms of the Fourier derivatives, and $\mathcal{G}$ is also represented in a finite Fourier series,
\begin{gather}
\partder{f(\Omega)}{\Omega_{m,n}} = \int_{S_C} d^2 x \, \left( \sum_{m,n} \mathcal{G}_{m,n} \cos(m\theta+n N_p \phi) \right) \partder{\textbf{x}(\Omega)}{\Omega_{m,n}} \cdot \hat{\textbf{n}}.
\label{SVD_sensitivity}
\end{gather}
After discretizing in $\theta$ and $\phi$, (\ref{SVD_sensitivity}) takes the form of a (generally not square) matrix equation which can be solved using the Moore-Penrose pseudoinverse to obtain $\mathcal{G}_{m,n}$. 

\begin{figure} 
\centering
\includegraphics[trim=3cm 0cm 0cm 0cm,clip,width=1.0\textwidth]{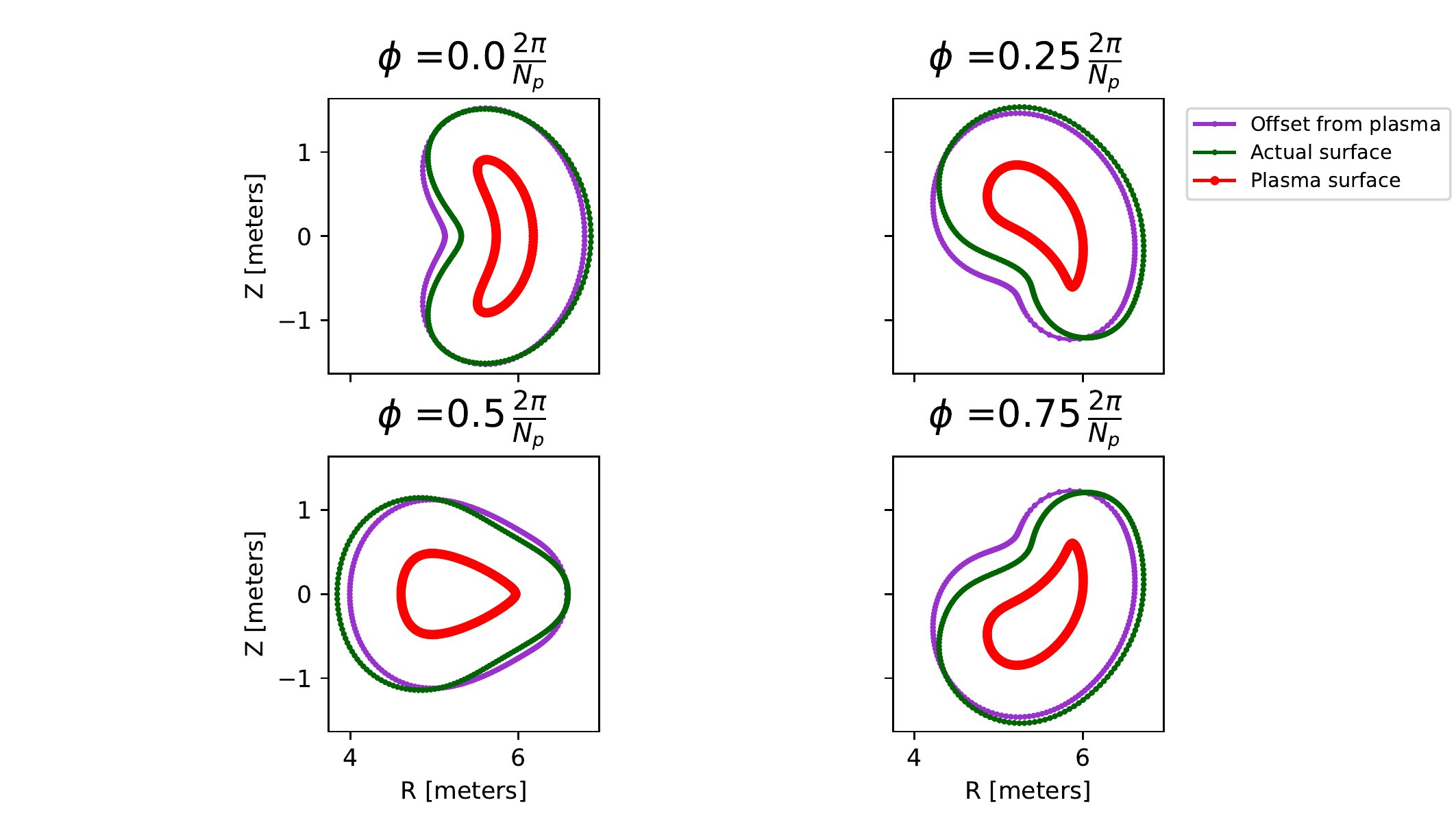}
\caption{The cross-sections of the two winding surfaces used to compute $\mathcal{G}_{\chi^2_B}$ and $\mathcal{G}_{\norm{\textbf{J}}_2}$ are shown in the poloidal plane. Figure adapted from \cite{Paul2018} with permission.}
\label{w7x3surf}
\end{figure}

\begin{figure}
\centering
\begin{subfigure}[b]{0.49\textwidth}
\includegraphics[trim=7cm 8cm 6cm 10cm, clip,width=0.8\textwidth]{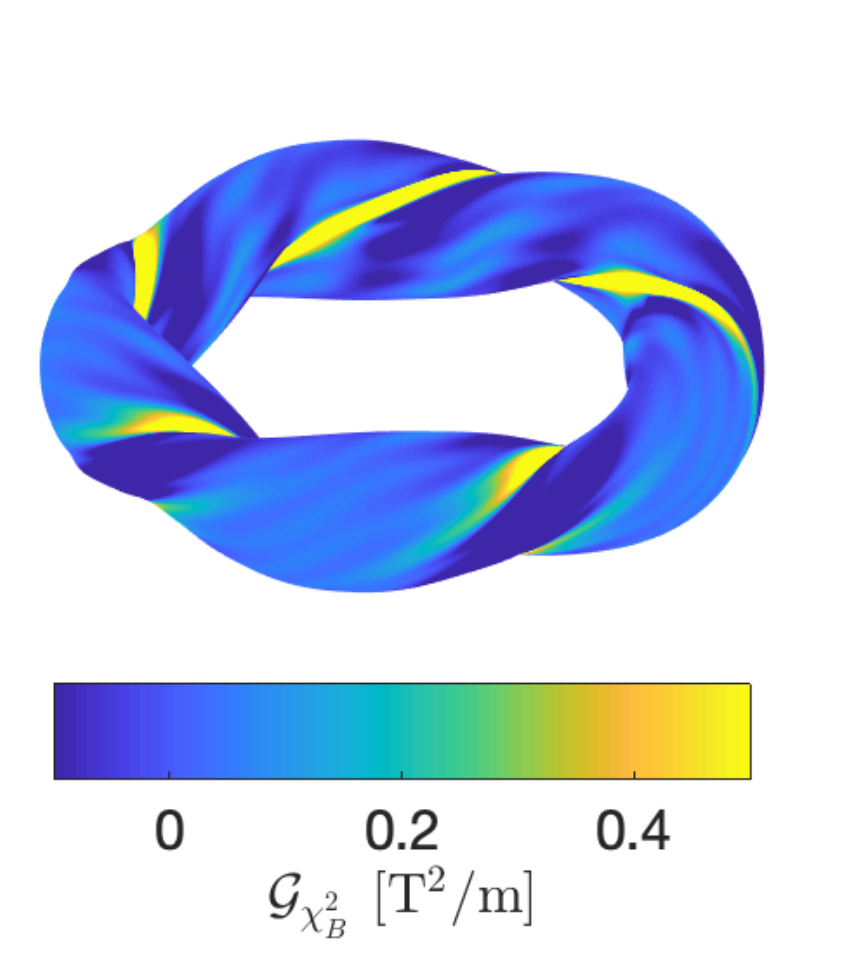}
\caption{Offset from plasma}
\end{subfigure}
\begin{subfigure}[b]{0.49\textwidth}
\includegraphics[trim=7cm 8cm 6cm 10cm, clip,width=0.8\textwidth]{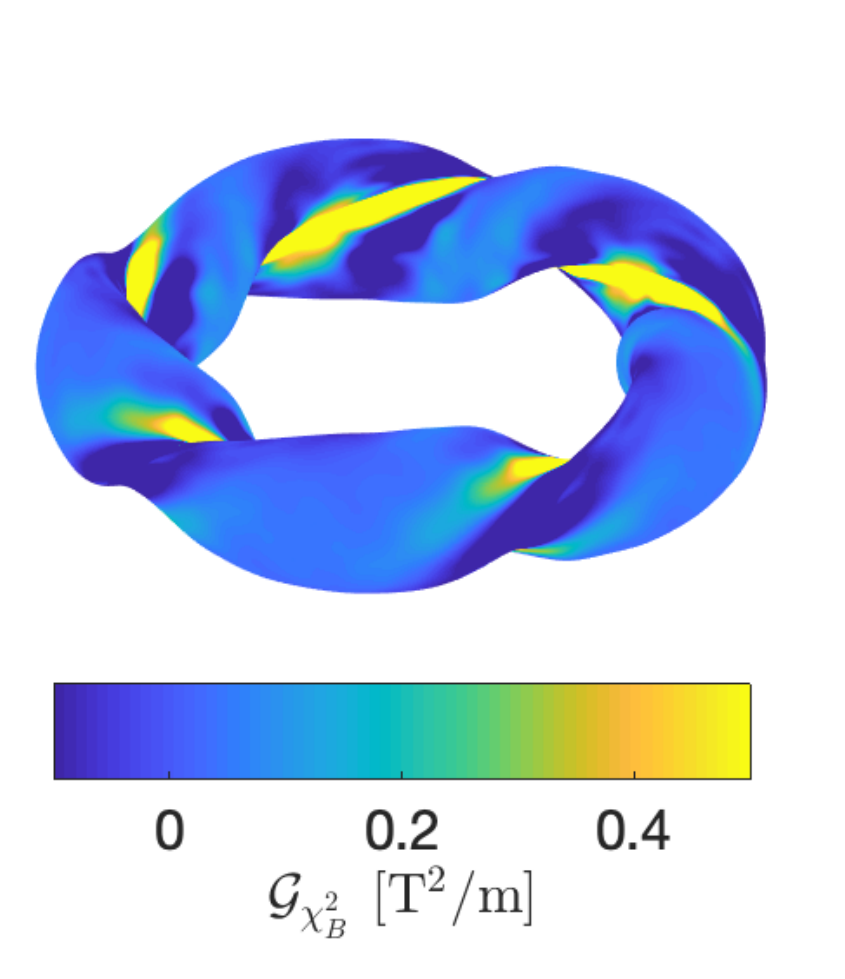}
\caption{Actual}
\end{subfigure}
\begin{subfigure}[b]{0.49\textwidth}
\includegraphics[trim=7cm 8cm 6cm 10cm, clip,width=0.8\textwidth]{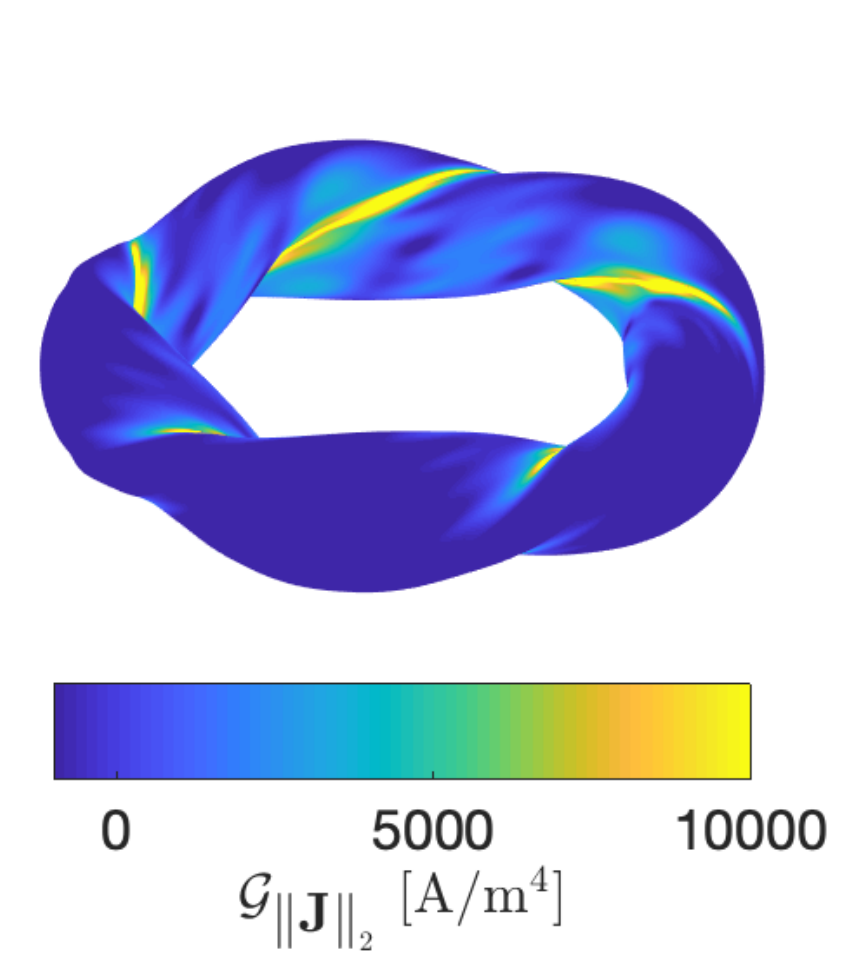}
\caption{Offset from plasma}
\end{subfigure}
\begin{subfigure}[b]{0.49\textwidth}
\includegraphics[trim=7cm 8cm 6cm 10cm, clip,width=0.8\textwidth]{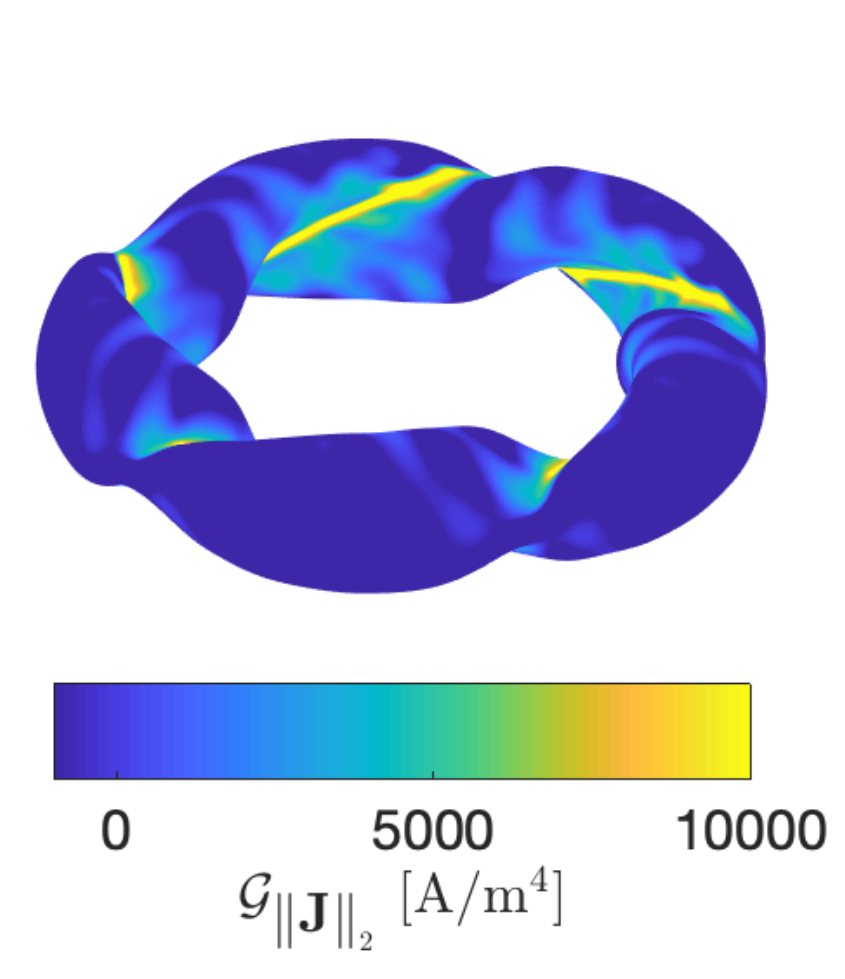}
\caption{Actual}
\end{subfigure}
\caption{Shape gradient for $\chi^2_B$ ((a) and (b)) $||\textbf{J}||_2$ ((c) and (d)). These functions are computed using the W7-X plasma surface and a uniform offset winding surface from the plasma surface with $d_{\text{coil-plasma}} = 0.61$ m ((a) and (c)) and the actual winding surface ((b) and (d)). The region of increased $\mathcal{G}_{\chi^2_B}$ corresponds with concave regions of the plasma surface (Figure \ref{fig:P_2}). Regions of large positive $\norm{\textbf{J}}_2$ correspond to regions with increased $J$ (Figure \ref{w7x_K}). Figure adapted from \cite{Paul2018} with permission.}
\label{w7x_S}
\end{figure}

We compute $\mathcal{G}_{\chi^2_B}$ and $\mathcal{G}_{\norm{\textbf{J}}_2}$ (Figure \ref{w7x_S}) at fixed $\lambda$. These quantities are computed on the actual W7-X winding surface and a surface uniformly offset from the plasma surface with $d_{\text{coil-plasma}} = 0.61$ m (the area-averaged $d_{\text{coil-plasma}}$ over the actual surface). We consider surfaces that are equidistant from the plasma surface on average as $\mathcal{G}$  scales inversely with $A_{\text{coil}}$. The poloidal cross-sections of these surfaces are shown in Figure \ref{w7x3surf}. For each surface $\lambda$ is chosen to achieve $J_{\text{max}} = 7.7$ MA/m as was used in Section \ref{w7x_results}. On both surfaces we observe a narrow region featuring a large positive $\mathcal{G}_{\chi^2_B}$, indicating that $d_{\text{coil-plasma}}$ should decrease at that location in order that $\chi^2_B$ decreases. This corresponds to locations on the plasma surface with significant concavity (Figure \ref{fig:P_2}). 
The maximum $\mathcal{G}_{\chi^2_B}$ occurs at $\phi = 0.15 \, 2\pi/N_p$ on both surfaces (Figure \ref{w7x_surf}). 
In comparison with this region, the magnitude of $\mathcal{G}_{\chi^2_B}$ is relatively small over the majority of the area of the surfaces shown, demonstrating that engineering tolerances might be more relaxed in these locations. There is also a region of negative $\mathcal{G}_{\chi^2_B}$ near $\phi = \pi/N_p$ and $\theta =0$. This is the ``tip" of the triangle-shaped cross-section, where $d_{\text{coil-plasma}}$ was increased over the course of the optimization (Figures \ref{alpha2_scan}, \ref{alpha1_scan}, and \ref{w7x_surf}). We find that $\mathcal{G}_{\chi^2_B}$ computed on the actual winding surface has similar trends to that computed on the surface uniformly offset from the plasma. This indicates that the shape gradient depends on the specific geometry of the winding surface. We have computed $\mathcal{G}_{\chi^2_B}$ for several other winding surfaces with varying $d_{\text{coil-plasma}}$. Regardless of the winding surface chosen, we observe increased sensitivity in the concave regions. 

\begin{figure}
\centering
\begin{subfigure}[b]{0.49\textwidth}
\includegraphics[trim=7cm 8cm 6cm 10cm, clip,width=0.8\textwidth]{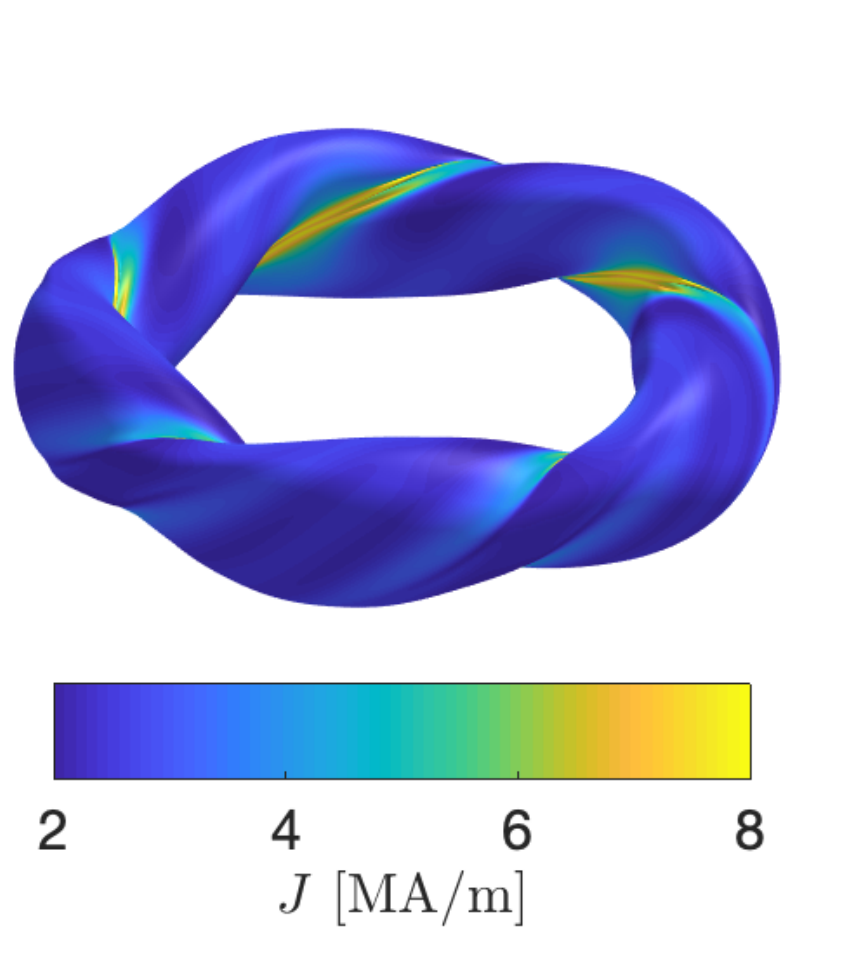}
\caption{Offset from plasma}
\end{subfigure}
\begin{subfigure}[b]{0.49\textwidth}
\includegraphics[trim=7cm 8cm 6cm 10cm, clip,width=0.8\textwidth]{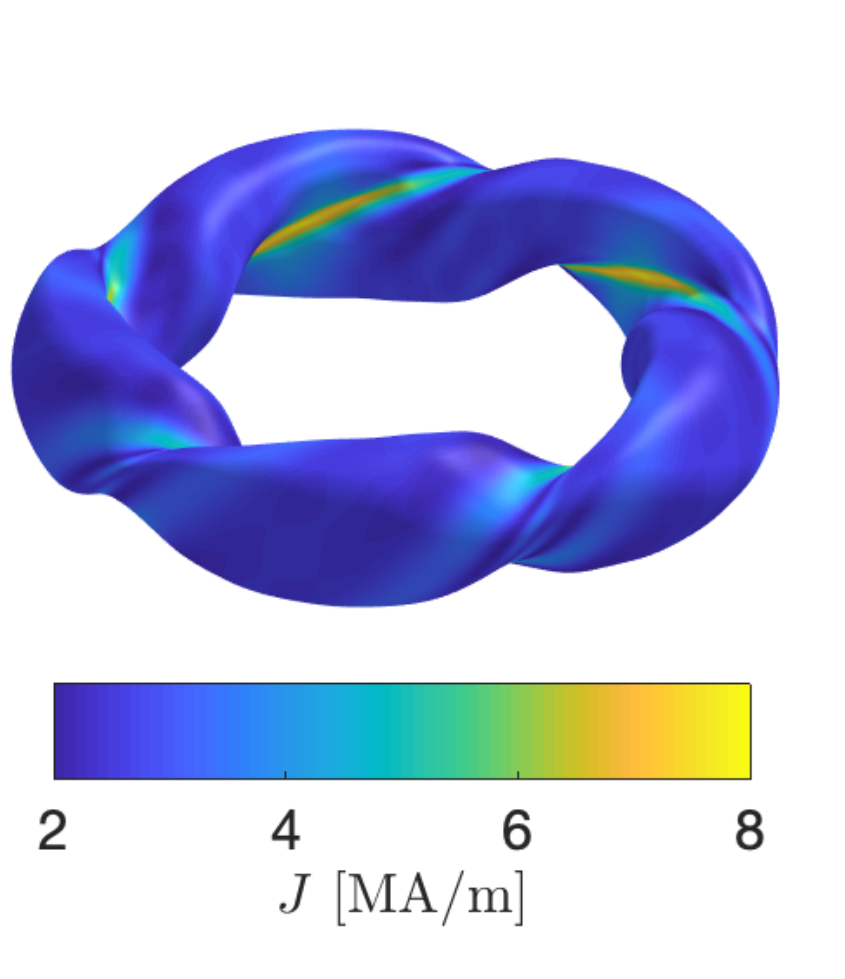}
\caption{Actual}
\end{subfigure}
\caption{Current density magnitude, $J$, computed from REGCOIL using the W7-X plasma surface and (a) a uniform offset winding surface from the plasma surface with $d_{\text{coil-plasma}} = 0.61$ m and (b) the actual winding surface. Figure adapted from \cite{Paul2018} with permission.}
\label{w7x_K}
\end{figure}

The quantity $\mathcal{G}_{\norm{\textbf{J}}_2}$ roughly quantifies how coil complexity changes with normal displacements of the coil surface. In view of Figure \ref{w7x_K}, the locations of large $\mathcal{G}_{\norm{\textbf{J}}_2}$ overlap with areas of increased $J$. On the actual winding surface, the maximum of $\mathcal{G}_{\norm{\textbf{J}}_2}$ occurs near the location of the closest approach between coils (two rightmost coils in Figure \ref{w7x_coils}(a)). The shape gradients $\mathcal{G}_{\norm{\textbf{J}}_2}$ and $\mathcal{G}_{\chi^2_B}$ have very similar trends. The concave regions of the plasma surface are difficult to produce with external coils, resulting in increased coil complexity and $J$. Therefore, $\norm{\textbf{J}}_2$ is most sensitive to displacements of the coil-winding surface in these regions. 

We recognize several ways that the shape gradient technique could be improved to provide more relevant diagnostics for experimental design. With a winding surface representation, the shape gradient does not allow for calculation of the sensitivity to lateral coil displacements. Also, our analysis does not account for field ripple due to the finite number of coils. Although Figure \ref{w7x_S} indicates that the coils should move toward the plasma to reduce the field error, the ripple fields might be significant with a filamentary model. A similar calculation could be performed using the filamentary coil sensitivity techniques presented in Section \ref{sec:shape_optimization} and discussed further in Chapter \ref{ch:adjoint_MHD}. Finally, $\chi^2_B$ does not account for the sensitivity to resonant fields that could cause the formation of islands, though there is ongoing work toward computing the shape gradient for such a metric \cite{Geraldini2019}.

Sensitivity studies on NCSX similarly found that coil errors on the inboard side in regions of small $d_{\text{coil-plasma}}$ had a significant effect on flux surface quality \cite{Williamson2005}. The necessity of small $d_{\text{coil-plasma}}$ for bean-shaped plasmas has been noted in many coil optimization efforts \cite{Strickler2002,Guebaly2008} and has been demonstrated by evaluating the singular value decomposition of the discretized Biot-Savart integral operator \cite{Landreman2016}. We can identify these regions where the fidelity of the plasma surface requires tighter tolerance on coil positions using the shape gradient. 

\section{Metrics for configuration optimization}
\label{sect_configopt}

The results presented here and in \cite{Landreman2016} indicate that the concave regions of the surface are both the regions where a small coil-plasma distance is required and the sensitivity to the winding surface position is highest. The regions of concavity can be determined by considering the principal curvatures of the plasma surface. Let $\hat{\textbf{n}}(\textbf{x}_0)$ represent the normal vector at the plasma surface at some point $\textbf{x}_0$, and let $A_n$ represent a plane that includes this normal vector. The intersection of the plane and the surface makes a curve $\textbf{x}(l)$, which has curvature $\kappa_0$ at the point $\textbf{x}_0$, as calculated from (\ref{curvature_of_curve}). The two principal curvatures $\kappa_1$ and $\kappa_2$ represent the maximum and minimum curvatures, $\kappa_0$, from all possible planes $A_n$. We choose the convention for the principal curvatures such that convex curves have positive curvature and concave curves have negative curvatures. Therefore, small values of the second principal curvature, $\kappa_2$, represent regions on the surface where the concavity is increased.

\begin{figure}
\centering
\begin{subfigure}[b]{0.49\textwidth}
\includegraphics[trim=6cm 8cm 6cm 10cm, clip,width=0.8\textwidth]{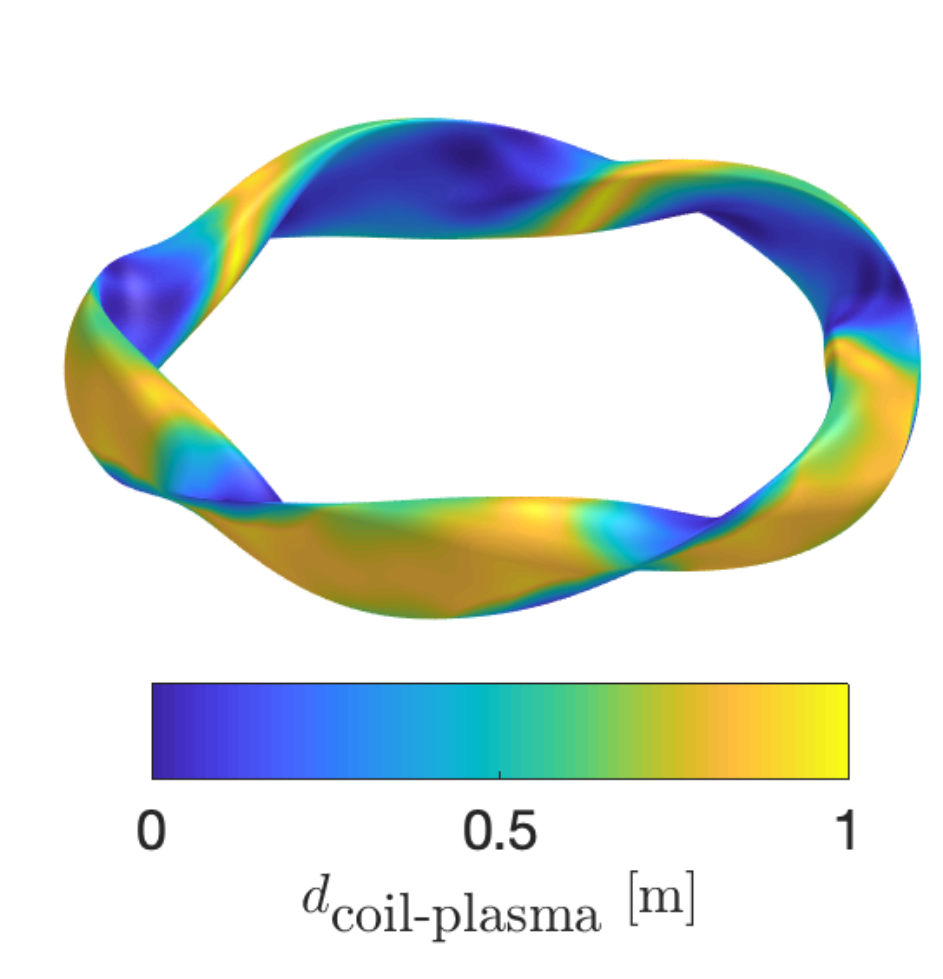}
\caption{}
\end{subfigure}
\begin{subfigure}[b]{0.49\textwidth}
\includegraphics[trim=6cm 8cm 6cm 10cm, clip,width=0.8\textwidth]{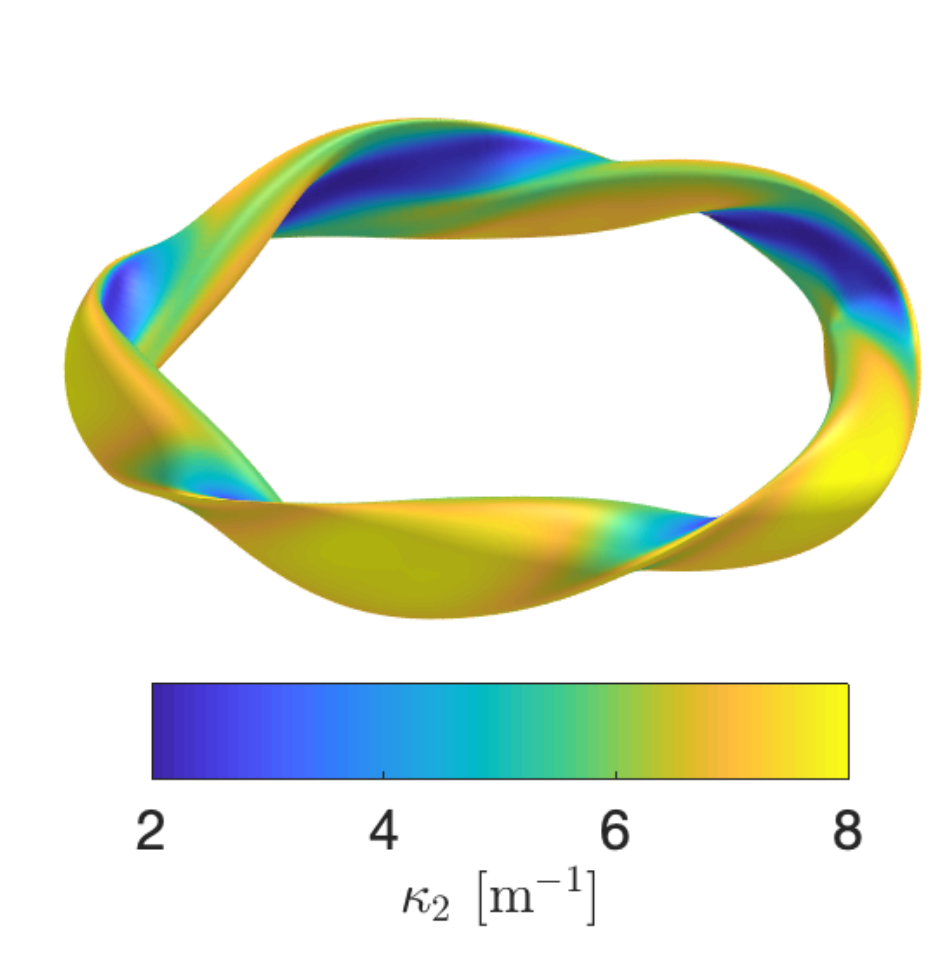}
\caption{}
\label{fig:P_2}
\end{subfigure}
\caption{(a) The minimum distance between the W7-X plasma surface and the optimized winding surface obtained in Section \ref{w7x_results} and (b) the second principle curvature $\kappa_2$ are shown as a function of location on the plasma surface. Locations of large negative $\kappa_2$ coincide with regions where the optimization resulted in small $d_{\text{coil-plasma}}$. Figure adapted from \cite{Paul2018} with permission.}
\label{dminandp2}
\end{figure}

The second principal curvature for the W7-X plasma surface is shown in Figure \ref{fig:P_2}. Although $\kappa_2$ and the shape gradients are evaluated on different surfaces, we note that regions of high concavity (negative $\kappa_2$) coincide with regions of large, positive $\mathcal{G}$ (Figure \ref{w7x_S}).
The regions of high concavity also correspond to the regions where the optimization procedure tends to place the winding surface closest to the plasma (Figure \ref{dminandp2}). We recognize that our winding surface optimization accounts for several engineering considerations in addition to reproducing the desired plasma surface. However, for a wide range of parameters the winding surfaces we obtain feature small $d_{\text{coil-plasma}}$ in the bean-shaped cross-sections (Figures \ref{alpha2_scan} and \ref{alpha1_scan}). Thus $\kappa_2$, which is exceedingly fast to compute, may serve as a target for optimization of the plasma configuration. By minimizing the regions of high concavity, it may be possible to find stellarator equilibria that are more amenable to coils that are positioned farther from the plasma. Any increase in the minimal distance between the plasma and the coils has implications for the size of a reactor, where $d_{\text{coil-plasma}}$ is set by the required blanket width. Similar metrics are considered in the ROSE code, such as the integrated absolute value of the Gaussian curvature and integrated absolute value of the maximum curvature \cite{Drevlak2018}.

\FloatBarrier
\section{Conclusions}
\label{sect_conclusions}

We have outlined a new method for the optimization of the stellarator coil-winding surface using a continuous current potential approach. Rather than evolving filamentary coil shapes, we use REGCOIL to obtain the current density on a winding surface and optimize the winding
surface using analytic gradients of the objective function. We have shown that we can indirectly improve the coil curvature and toroidal extent by targeting the root-mean-squared current density in our objective function (Figure \ref{rmsKcoilcompare}). This approach offers several potential advantages over other nonlinear coil optimization tools.
\begin{enumerate}
\item The difficulty of the optimization is reduced by the application of the REGCOIL method, which takes the form of a linear least-squares system. The optimal coil shapes on a given winding surface can thus be efficiently and robustly computed. 
\item By fixing the maximum current density to obtain the regularization in REGCOIL, we eliminate the need to implement an additional equality constraint or arbitrary weight in the objective function. 
\item By using REGCOIL to compute coil shapes on a given surface, we can apply the adjoint method for computing derivatives (Section \ref{sect_adjoint}). This allows us to reduce the number of function evaluations required during the nonlinear optimization by a factor of $\approx 50$. 
\item Given the critical role coil design plays in the stellarator optimization process, it is important to have many tools that approach the problem from different angles. Our approach differs from the other available nonlinear coil optimization applications \cite{Drevlak1998,Strickler2002,Strickler2004,Brown2015,Zhu2018} as we optimize a continuous current potential. 
\end{enumerate} 
We have demonstrated this method by optimizing coils for W7-X and HSX (Sections \ref{w7x_results} and \ref{hsx_results}). We find that we can simultaneously decrease the integral-squared error in reproducing the plasma surface, increase the volume contained within the winding surface, maintain the minimum coil-plasma distance, and improve the coil metrics over REGCOIL solutions computed on the initial winding surfaces (Tables \ref{table_w7x} and \ref{table_hsx}). Several features of these optimized winding surfaces are noteworthy. While the coil-plasma distance must be small in concave regions, it can increase greatly on the outboard, convex side of the bean cross-section. At triangle-shaped cross-sections, the winding surface obtains a somewhat ``pinched" appearance (Figures \ref{alpha1_scan}, \ref{w7x_surf}, and \ref{hsx_surf}). A similar W7-X winding surface shape has been obtained with the ONSET code (see ref. \cite{Drevlak1998}, Figure 5). Further work is required to understand this behavior.

There are several limitations to this approach that should be noted. First, we have applied a local nonlinear optimization algorithm. This is a reasonable choice if the initial condition is close to a global optimum. Second, we currently have not added coil-specific metrics to our objective function (for example, curvature or length). This could be implemented if necessary for engineering purposes. 

We should also note that this application does not allow for the full benefits of adjoint methods. While adjoint methods significantly reduce CPU time if the solve is the computational bottleneck, this is not the case for the REGCOIL system. Other applications that are dominated by the linear solve CPU time would see increased benefits from the implementation of an adjoint method, as will be seen in the following Chapters.

We demonstrate a technique for visualization of shape derivatives in real space rather than Fourier space. This shape gradient describes how an objective function changes with respect to normal displacements of the winding surface. We apply this technique to visualize the derivatives of the integral-squared normal field on the plasma surface and the root-mean-squared current density for the W7-X plasma surface and two winding surfaces (Figure \ref{w7x_S}). This diagnostic identifies the concave regions as being very sensitive to the positions of coils, as has been observed from previous coil optimization efforts. We will continue to gain insight from the shape gradient concept in Chapters \ref{ch:adjoint_neoclassical}, \ref{ch:adjoint_MHD}, and \ref{ch:linearized_mhd}.

\renewcommand{\thechapter}{4}

\chapter{Adjoint-based optimization of neoclassical properties}
\label{ch:adjoint_neoclassical}

Several critical quantities for stellarator design arise from neoclassical physics, the kinetic theory of collisional transport in the presence of magnetic field gradients and curvature. This so-called neoclassical transport results from the random-walk of charged particles as they exhibit guiding center motion. Due to the complicated guiding center orbits present in a 3D field, neoclassical transport is generally enhanced in a stellarator. One of the primary goals of stellarator optimization is to reduce this transport. Furthermore, the bootstrap current, driven by collisional processes, should be minimized in low-shear designs or if an island divertor system is to be used. These neoclassical properties are described by solutions of the drift-kinetic equation (DKE),
\begin{align}
    \left(v_{||} \hat{\textbf{b}} + \textbf{v}_{\text{d}} \right)\cdot \nabla f = C(f),
    \label{eq:drift_kinetic_intro}
\end{align}
where $f$ is the distribution function, $v_{||} = \textbf{v} \cdot \hat{\textbf{b}}$ is the parallel component of the velocity, $\textbf{v}_{\text{d}}$ is the guiding center drift velocity, and $C$ is the collision operator. The DKE is obtained from the Fokker-Planck equation under the assumption that the plasma is strongly magnetized such that \eqref{eq:drift_kinetic_intro} describes length scales much longer than the gyroradius and frequencies much smaller than the gyrofrequency. We have taken the equilibrium limit, assuming time scales longer than the gyroperiod but shorter than the transport time scale on which the profiles relax. In this Chapter we make an additional assumption of local thermodynamic equilibrium, such that $f \approx f_M$, a Maxwellian distribution (defined in Section \ref{sec:dke}), to lowest order. This assumption is valid in stellarator configurations, provided that the collisionless orbits are sufficiently confined and the collision frequency is not too low \cite{Tribaldos2005,Calvo2017}. The departure from a Maxwellian, $f_1$, is driven by gradients in $f_M$ due to variations in the density, temperature, and electrostatic potential. The drift-kinetic equation is described in many references, including Chapter 7 in \cite{Helander2005} and  \cite{Hazeltine1973,Helander2014}.
 
  In this Chapter, we will apply both the discrete and continuous adjoint methods described in Chapter \ref{ch:mathematical_fundamentals} to efficiently compute derivatives of functions that depend on such solutions of the drift kinetic equation. This analysis will allow us to efficiently optimize the local magnetic field for several neoclassical quantities in addition to analyzing their sensitivity to changes in the magnetic field. 

The material in this Chapter has been adapted from \cite{Paul2019}.

\section{Introduction}

Neoclassical transport is governed by solutions of the drift kinetic equation (DKE) \eqref{eq:DKE} from which moments (e.g., radial fluxes and bootstrap current) are computed. The DKE local to a flux surface can be solved numerically \citep{Landreman2014,Belli2015}. However, this four-dimensional problem is expensive to solve within an optimization loop, especially in low-collisionality regimes for which increased pitch-angle resolution is required to resolve the collisional boundary layer. 

Therefore, it is sometimes desirable to consider an analytic reduction of the DKE. Under the assumption of low collisionality, a bounce-averaged DKE can be considered \citep{Beidler1995,Calvo2018}. While bounce-averaging can significantly reduce the computational cost by decreasing the spatial dimensionality, this approach typically requires restrictions on the geometry, such as closeness to omnigeneity or a model magnetic field. Additional reduction of the DKE can be made in low-collisionality regimes, resulting in semi-analytic expressions. For example the effective ripple, $\epsilon_{\text{eff}}$ \citep{Nemov1999}, quantifies the geometric dependence of the $1/\nu$ radial transport ($\nu$ is the collision frequency) and has been widely used during optimization studies \citep{Zarnstorff2001,Ku2008,Henneberg2019}. (The effective ripple will be discussed further in Chapter \ref{ch:adjoint_MHD} and Appendix \ref{app:1_over_nu}.) 
The $1/\nu$ regime, though, is only relevant when $E_r$ is small enough that the typical poloidal rotation frequency is much smaller than the typical collision frequency \cite{Ho1987}, which is not always an experimentally-relevant regime. A low-collisionality semi-analytic bootstrap current model \citep{Shaing1989} is also commonly adopted for stellarator design \citep{Beidler1990,Hirshman1999}. However, this analytic expression is known to be ill-behaved near rational surfaces. Furthermore, benchmarks with numerical solutions of the DKE in the low-collisionality limit have been shown to differ significantly from the semi-analytic model \citep{Beidler2011,Kernbichler2016}. Any analytic reduction of the DKE implies additional assumptions, such as on the collisionality, size of $E_r$, or on the magnetic geometry.


Due to the limitations of bounce-averaged and semi-analytic models, there are benefits to computing neoclassical quantities using numerical solutions to the DKE without approximation. With the numerical methods currently used for stellarator optimization, this approach becomes computationally challenging within an optimization loop. Due to their fully three-dimensional nature, optimization of stellarator geometry requires navigation through high-dimensional spaces, such as the space of the shape of the outer boundary of the plasma or the shapes of electromagnetic coils. The number of parameters required to describe these spaces, $N$, is often quite large ($\mathcal{O}(10^2)$). Knowledge of the gradient of the objective function with respect to these parameters can significantly improve the convergence to a local minimum. Once a descent direction is identified, each iteration reduces to a one-dimensional line search. Gradient-based optimization with the Levenberg-Marquardt algorithm in the STELLOPT code \citep{Strickler2004} has been widely used in the stellarator community and led to the design of NCSX \citep{Reiman1999}.

 Although derivative information is valuable, numerically computing the derivative of a figure of merit $f$ (for example, with finite-difference derivatives) can be prohibitively expensive, as $f$ must be evaluated $\mathcal{O}(N)$ times. For neoclassical optimization, this implies solving the DKE $\mathcal{O}(N)$ times; thus including finite-collisionality neoclassical quantities in the objective function is often impractical. In this Chapter, we describe an adjoint method for neoclassical optimization. With this method, the computation of the derivatives of $f$ with respect to $N$ parameters has cost comparable to solving the DKE twice, thus making the inclusion of these quantities possible within an optimization loop. In this Chapter, we obtain derivatives of neoclassical figures of merit with respect to local geometric parameters on a surface rather than the outer boundary or coil shapes. However, the geometric derivatives we compute provide an important step toward adjoint-based optimization of MHD equilibria, as discussed in Section \ref{sec:equilibria_opt} and Chapter \ref{ch:adjoint_MHD}.


In Section \ref{sec:dke}, we provide an overview of the numerical solution of the DKE local to a flux surface. In Section \ref{sec:adjoint_approach} the adjoint neoclassical method is described. The continuous and discrete approaches for this problem are presented, and their implementation and benchmarks are discussed in Section \ref{sec:implementation}. The adjoint method is used to compute derivatives of moments of the neoclassical distribution function with respect to local geometric quantities. The derivative information can be used to identify regions of increased sensitivity to magnetic perturbations, as discussed in Section \ref{sec:local_sensitivity}. We demonstrate adjoint-based optimization in Section \ref{sec:vacuum_opt} by locally modifying the field strength on a flux surface. A discussion of the application of this method for optimization of MHD equilibria is presented in \ref{sec:equilibria_opt}. 
Finally, the adjoint method is applied to accelerate the calculation of the ambipolar electric field in Section \ref{sec:ambipolarity}.  

\section{Drift kinetic equation}
\label{sec:dke}

The local drift kinetic equation is,
\begin{equation}
\left(v_{||} \hat{\textbf{b}} + \textbf{v}_E \right) \cdot\nabla f_{1s} - C_s(f_{1s}) = -\textbf{v}_{\text{m}s} \cdot \nabla \psi \partder{f_{Ms}}{\psi},
\label{eq:DKE_ch2}
\end{equation}
Here $\hat{\textbf{b}} = \textbf{B}/B$ is a unit vector in the direction of the magnetic field, $v_{||} = \textbf{v}\cdot \hat{\textbf{b}}$ is the parallel component of the velocity, and $2\pi \psi$ is the toroidal flux. The Fokker-Planck collision operator is $C_s(f_{1s})$, linearized about a  Maxwellian $f_{Ms} = n_sv_{ts}^{-3} \pi^{-3/2} e^{-v^2/v_{ts}^2}$ where $v_{ts} = \sqrt{2T_s/m_s}$ is the thermal speed, $n_s$ is the density, $T_s$ is the temperature, $m_s$ is the mass, and the subscript indicates species. In \eqref{eq:DKE_ch2}, derivatives are performed holding $W_s = m_s v^2/2 +q_s \Phi$ and $\mu = v_{\perp}^2/2B$ fixed, where $v = \sqrt{\textbf{v} \cdot \textbf{v}}$ is the magnitude of velocity, $\Phi$ is the electrostatic potential, $v_{\perp} = \sqrt{v^2 - v_{||}^2}$ is the perpendicular velocity, and $q_s$ is the charge. The radial magnetic drift is,
\begin{equation}
\textbf{v}_{\text{m}s}\cdot \nabla \psi = \frac{m_s }{q_s B^2} \left( v_{||}^2 + \frac{v_{\perp}^2}{2} \right) \hat{\textbf{b}} \times \nabla B \cdot \nabla \psi,
\label{eq:radial_drift}
\end{equation}
assuming a magnetic field in MHD force balance, and $\textbf{v}_E$ is the $\textbf{E} \times \textbf{B}$ velocity,
\begin{equation}
\textbf{v}_E  = \frac{\textbf{B} \times \nabla \Phi}{B^2}.
\end{equation}
Throughout we assume $\Phi=\Phi(\psi)$ such that \eqref{eq:DKE_ch2} is linear. In \eqref{eq:DKE_ch2} we will not consider the effect of inductive electric fields, as these can be assumed to be small for stellarators without inductive current drive. We also do not consider the effects of magnetic drifts tangential to the flux surface in \eqref{eq:DKE_ch2}, as these only become important when $E_r$ is small \citep{Paul2017}. We can assume radial locality, manifested by the absence of any radial derivatives of $f_{1s}$ in \eqref{eq:DKE_ch2}, when $\nu_{*} \gg \rho_*$ \cite{Calvo2017}, where $\nu_* = \nu/(v_t/L) \ll 1$ is the normalized collision frequency for macroscopic scale length $L$ and $\rho_* =v_{t}m/(L q B)$ is the normalized gyrofrequency. Numerical solutions to \eqref{eq:DKE_ch2} are computed with the Stellarator Fokker-Planck Iterative Neoclassical Solver (SFINCS) \cite{Landreman2014} code which allows for general stellarator geometry with flux surfaces.

SFINCS solves \eqref{eq:DKE_ch2} locally on a flux surface $\psi$, a four-dimensional system. The SFINCS coordinates include two angles (poloidal angle $\theta$ and toroidal angle $\phi$), speed $X_s = v/v_{ts}$, and pitch angle $\xi_s = v_{||}/v$. Specifics about the implementation of \eqref{eq:DKE_ch2} in the SFINCS code are described in Appendix \ref{app:trajectory_models}. We will refer to two choices of implementation: the full trajectory model and the DKES trajectory model. The full trajectory model maintains $\mu$ conservation as radial coupling (terms involving $\partial f_{1s}/\partial \psi$) is dropped.  While the DKES model does not conserve $\mu$ when $E_r \neq 0$, the adjoint operator under the DKES model takes a particularly simple form, as discussed in Section \ref{sec:continuous}. This model also does not introduce any unphysical constraints on the distribution function when $E_r = 0$, as occurs for the full trajectory model \citep{Landreman2014}. These constraints motivate the introduction of particle and heat sources, which are discussed in the following Section. We will discuss details of the implementation of the DKE in the SFINCS code, as these need to be considered in arriving at the adjoint equation. However, the adjoint neoclassical approach is quite general and could be implemented in other drift-kinetic codes with slight modification.

From solutions of \eqref{eq:DKE_ch2}, several neoclassical quantities are computed, including the flux-surface averaged parallel flow,
\begin{align}
V_{||,s} = \frac{\left\langle B \int d^3 v \, f_{1s} v_{||} \right\rangle_{\psi}}{n_s \langle B^2 \rangle_{\psi}^{1/2}},
\label{eq:parallel_flow}
\end{align}
the radial particle flux,
\begin{align}
    \Gamma_s = \left \langle \int d^3 v \, \left(\textbf{v}_{\text{m}s} \cdot \nabla \rho \right) f_{1s} \right \rangle_{\psi},
    \label{eq:particle_flux}
\end{align}
and the radial heat flux (sometimes referred to as an energy flux),
\begin{align}
    Q_s = \left \langle \int d^3 v \, \frac{m_sv^2}{2} \left(\textbf{v}_{\text{m}s} \cdot \nabla \rho \right) f_{1s} \right \rangle_{\psi}.
    \label{eq:heat_flux}
\end{align}
Here the flux-surface average of a quantity $A$ is,
\begin{subequations}
\begin{align}
    \langle A \rangle_{\psi} &= \frac{\int_{0}^{2\pi} d \theta \int_0^{2\pi} d \phi \, \sqrt{g} A}{V'(\psi)} \\
    V'(\psi) &= \int_{0}^{2\pi} d \theta \int_0^{2\pi} d \phi \, \sqrt{g},
\end{align}
\label{eq:flux_surface_average_ch4}
\end{subequations}
and $\sqrt{g} = \left( \nabla \psi \times \nabla \theta \cdot \nabla \phi \right)^{-1}$ is the Jacobian. We will also consider species-summed quantities including the bootstrap current, $J_b = \sum_s q_s n_s V_{||,s}$, the radial current, $J_r = \sum_s q_s \Gamma_s$, and the total heat flux, $Q_{\text{tot}} = \sum_s Q_s$. Here the effective normalized radius is $\rho = \sqrt{\psi/\psi_0}$, where $2\pi \psi_0$ is the toroidal flux at the boundary.

\subsection{Sources and constraints}
\label{sec:sources}
To avoid unphysical constraints on $f_{1s}$ implied by the moment equations of \eqref{eq:DKE_ch2} in the presence of a non-zero $E_r$ \citep{Landreman2014}, particle and heat sources are added to the DKE \eqref{eq:dke_model}, 
\begin{gather}
    \mathbb{L}_{0s}f_{1s} - C_s (f_{1s}) - f_{Ms} \left(X_s^2 - \frac{5}{2}\right) S_{1s}^f(\psi) - f_{Ms}\left(X_s^2-\frac{3}{2}\right) S_{2s}^f(\psi) = \mathbb{S}_{0s},
\end{gather}
where $S_{1s}^f(\psi)$ and $S_{2s}^f(\psi)$ are unknowns such that $S_{1s}^f$ provides a particle source and $S_{2s}^f$ provides a heat source. The collisionless trajectory operator in SFINCS coordinates is,
\begin{gather}
    \mathbb{L}_{0s} = \dot{\textbf{x}} \cdot \nabla + \dot{X}_s \partder{}{X_s} + \dot{\xi}_s \partder{}{\xi_s},
    \label{eq:L_0s}
\end{gather}
and the inhomogeneous drive term is $\mathbb{S}_{0s} = - (\textbf{v}_{\text{m}s} \cdot \nabla \psi) \partial f_{Ms}/\partial \psi$. The source functions are determined via the requirement that $\langle \int d^3 v \, f_{1s} \rangle_{\psi} = 0$ and $\langle \int d^3 v \, X_s^2 f_{1s}\rangle_{\psi} = 0$ (i.e. $f_{1s}$ does not provide net density or pressure). So, the following system of equations is solved,
\begin{gather}
    \underbrace{\left[ \begin{array}{ccc}
       \mathbb{L}_{0s} -C_s  & - f_{Ms} (X_s^2-\frac{5}{2}) & -f_{Ms} (X_s^2-\frac{3}{2}) \\
       \mathbb{L}_{1s}  & 0 & 0 \\
       \mathbb{L}_{2s}  & 0 & 0 
    \end{array}
    \right]}_{\mathbb{L}_s} \underbrace{\left[
    \begin{array}{c}
    f_{1s} \\
    S_{1s}^{f} \\
    S_{2s}^{f} 
    \end{array}
    \right]}_{F_s} = \underbrace{\left[
    \begin{array}{c}
    \mathbb{S}_{0s}  \\
    0 \\
    0
    \end{array}
    \right]}_{\mathbb{S}_s}.
    \label{eq:dke_array}
\end{gather}
The velocity-space averaging operations are denoted $\mathbb{L}_{1s}f_{1s} = \langle \int d^3 v \, f_{1s} \rangle_{\psi}$ and $\mathbb{L}_{2s}f_{1s} = \langle \int d^3 v \, f_{1s} X_s^2 \rangle_{\psi}$. The full multi-species system can be written as,
\begin{gather}
\left[
\begin{array}{c}
\mathbb{L}_{1} \\
\vdots \\
\mathbb{L}_{N_{\text{species}}}
\end{array}
\right]
\left[ 
\begin{array}{c}
F_{1} \\
\vdots \\
F_{N_{\text{species}}}
\end{array}
\right] = \left[ 
\begin{array}{c}
\mathbb{S}_{1} \\
\vdots \\
\mathbb{S}_{N_{\text{species}}}
\end{array}
\right].
\label{eq:dke_species_array}
\end{gather}
Here the linear systems corresponding to each species as in \eqref{eq:dke_array} are coupled through the collision operator. We use the following notation to refer to the above system,
\begin{gather}
    \mathbb{L} F = \mathbb{S}.
    \label{eq:linear}
\end{gather}

\section{Adjoint approach}
\label{sec:adjoint_approach}

The goal of the adjoint neoclassical approach is to compute derivatives of a moment of the distribution function efficiently, $\mathcal{R}$ (e.g., $V_{||,s}, \Gamma_s, Q_s, J_b, J_r, Q_{\text{tot}})$, with respect to many parameters. Consider a set of parameters, $\Omega = \{ \Omega_i\}_{i=1}^{N_{\Omega}}$, on which $\mathcal{R}$ depends. Computing a forward-difference derivative with respect to $\Omega$ requires $  N_{\Omega} + 1$ solutions of \eqref{eq:linear}. With the adjoint approach, $\partial \mathcal{R}/\partial \Omega$ can be computed with one solution of \eqref{eq:linear} and one solution of a linear adjoint equation of the same size as \eqref{eq:linear}. Thus if $N_{\Omega}$ is very large and the solution to \eqref{eq:linear} is computationally expensive to obtain, the adjoint approach can reduce the cost by $N_{\Omega}$. For stellarator optimization, it is desirable to compute derivatives with respect to parameters that describe the magnetic geometry. In fully three-dimensional geometry, $N_{\Omega}$ is $\mathcal{O}(10^2)$ and solving \eqref{eq:linear} is the most expensive part of computing $\mathcal{R}$ (rather than constructing the linear system or taking a moment of the distribution function). The discretized linear system is typically very large ($N \sim 10^5-10^6$ for the calculations shown in the Chapter) and sparse. Thus matrix-matrix products are significantly less expensive than the linear solve, which is performed with a preconditioned Krylov iterative method. Consequently, the adjoint method provides a factor of $N_{\Omega} \sim 10^2$ savings over both the forward sensitivity and finite-difference methods, as described in Section \ref{sec:complexity_comparison}. The adjoint method also allows us to avoid additional round-off or truncation error arising from finite-difference derivatives. In what follows, we consider $\Omega$ to be a set of parameters describing the magnetic geometry, which will be specified in Section \ref{sec:implementation}.

We compute the derivatives of $\mathcal{R}$ using two approaches. In the first approach, we define an inner product that involves integrals over the distribution function, and an adjoint operator is obtained with respect to this inner product. This is the continuous approach introduced in Section \ref{sec:lagrangian}. In the second approach, we consider the DKE after discretization, defining an adjoint operator with respect to the Euclidean dot product. This is the discrete approach introduced in Section \ref{sec:linear_systems}. While these approaches should provide identical results within discretization error, the advantages and drawbacks of each method will be discussed at the end of Section \ref{sec:discrete}. 

\subsection{Continuous approach}
\label{sec:continuous}
Let $F = \{F_s\}_{s=1}^{N_{\text{species}}}$ be the set of unknowns computed with SFINCS before discretization, denoted by the column vector in \eqref{eq:dke_species_array} with $F_s$ given by \eqref{eq:dke_array}. That is, $F$ consists of a set of $N_{\text{species}}$ distribution functions over $(\theta,\phi,X_s,\xi_s)$ and their associated source functions. We define an inner product between two such quantities in the following way,
\begin{align}
\langle F,G \rangle = \sum_s \left \langle \int d^3 v \, \frac{f_{1s} g_{1s}}{f_{Ms}} \right \rangle_{\psi} + S_{1s}^f S_{1s}^g + S_{2s}^f S_{2s}^g. 
\label{eq:inner_product}
\end{align}
Here the superscript on $S_{1s}$ and $S_{2s}$ denotes the distribution function with which the source functions are associated and the sum is over species. The space of continuous functions, $F$, of this form such that $\langle F,F \rangle$ is bounded will be denoted by $\mathcal{H}$. It can be seen that \eqref{eq:inner_product} is indeed an inner product, as it satisfies  conjugate symmetry ($\langle G,F \rangle = \langle F,G \rangle$ 
$\forall F,G \in \mathcal{H}$), linearity ($\langle F + G,H \rangle = \langle F,H \rangle+ \langle G,H \rangle$ $\forall F,G,H \in \mathcal{H}$ and $\langle F, a G \rangle = a\langle F, G \rangle$ $\forall F,G \in \mathcal{H}$, $a \in \mathbb{R}$), and positive definiteness ($\langle F, F \rangle \geq 0$ and $\langle F,F \rangle = 0$ only if $F = 0$ $\forall F \in \mathcal{H}$) \citep{Rudin2006}. 
This implies that if $\mathcal{H}$ is finite-dimensional, then for any linear operator $L$ there exists a unique adjoint operator $L^{\dagger}$ such that $\langle LF,G \rangle = \langle F, L^{\dagger}G \rangle$ for all $F, G \in \mathcal{H}$. While here $\mathcal{H}$ is not finite-dimensional, we will show that such an adjoint operator exists for this inner product. 

Note that the norm associated with this inner product $|| F || = \sqrt{\langle F,F \rangle}$ is similar to the free energy norm, 
\begin{gather}
    W = \sum_s \left \langle \int d^3 v \, \frac{T_s f_{1s}^2}{2f_{Ms}} \right \rangle_{\psi},
\end{gather}
which obeys a conservation equation in gyrokinetic theory \citep{Krommes1994,Abel2013,Landreman2015}. The choice of inner product \eqref{eq:inner_product} is advantageous, as the linearized Fokker-Planck collision operator becomes self-adjoint for species linearized about Maxwellians with the same temperature. In what follows, we assume that all included species are of the same temperature. This assumption could be lifted, with a modification to the collision operator that appears in the adjoint equation (Appendix \ref{app:collision}). This assumption is not necessary when using the discrete approach (Section \ref{sec:discrete}). 

Consider a moment of the distribution function $\mathcal{R} \in \{ V_{||,s}, \Gamma_s, Q_s, J_b, J_r, Q_{\text{tot}}\}$, which can be written as an inner product with a vector $\widetilde{\mathcal{R}} \in \mathcal{H}$,
\begin{gather}
    \mathcal{R} = \langle F, \widetilde{\mathcal{R}} \rangle,
    \label{eq:inner_product_R}
\end{gather}
according to \eqref{eq:inner_product}.
For example, 
\begin{gather}
    \widetilde{J_r} = \left[ \begin{array}{c}
    q_s \textbf{v}_{\text{m}s} \cdot \nabla \psi f_{Ms} \\
    0 \\
    0
    \end{array}
    \right]_{s=1}^{N_{\text{species}}},
\end{gather}
where the column structure corresponds with that in \eqref{eq:dke_array} and \eqref{eq:dke_species_array}.

We are interested in computing the derivative of $\mathcal{R}$ with respect to a set of parameters, $\Omega = \{\Omega_i\}_{i=1}^{N_{\Omega}}$ such that the DKE is satisfied. Computing such a derivative with the forward sensitivity method requires that we compute $\partial F(\Omega)/\partial \Omega_i$ from the linearized DKE,
\begin{align}
    \partder{\mathbb{L}(\Omega)}{\Omega_i} F + \mathbb{L} \partder{F(\Omega)}{\Omega_i} = \partder{\mathbb{S}(\Omega)}{\Omega_i},
\end{align}
for each $\Omega_i$ and evaluate the derivative using the chain rule,
\begin{align}
    \partder{\mathcal{R}(\Omega,F(\Omega))}{\Omega_i} = \partder{\mathcal{R}(\Omega,F)}{\Omega_i} + \left \langle \widetilde{\mathcal{R}},\partder{F(\Omega)}{\Omega_i} \right \rangle . 
\end{align}
We see that the forward sensitivity method requires solutions of $N_{\Omega}$ linear systems of the same dimension as the DKE \eqref{eq:linear}. 

To avoid this additional computational cost, we instead apply the adjoint method by constructing the Lagrangian functional, enforcing \eqref{eq:linear} as a constraint,
\begin{align}
    \mathcal{L}(\Omega,F,\lambda^{\mathcal{R}}) = \mathcal{R}(\Omega,F) + \left \langle \lambda^{\mathcal{R}}, \mathbb{L}F - \mathbb{S} \right \rangle.
\end{align}
Here $\lambda^{\mathcal{R}}$ is the Lagrange multiplier. We obtain the adjoint equation by finding a stationary point of $\mathcal{L}$ with respect to $F$,
\begin{align}
    \delta \mathcal{L}(\Omega,F,\lambda^{\mathcal{R}};\delta F) = \langle \delta F, \widetilde{\mathcal{R}} \rangle + \left \langle \lambda^{\mathcal{R}}, \mathbb{L} \delta F \right \rangle = 0.
\end{align}
We can now use the adjoint property to express the above as,
\begin{align}
    \delta \mathcal{L}(\Omega,F,\lambda^{\mathcal{R}};\delta F) = \langle \delta F,  \widetilde{\mathcal{R}}  + \mathbb{L}^{\dagger} \lambda^{\mathcal{R}}\rangle.
\end{align}
A stationary point of $\mathcal{L}$ with respect to $F$ corresponds to $\lambda^{\mathcal{R}}$ which satisfies the weak form of the adjoint equation,
\begin{align}
    \mathbb{L}^{\dagger} \lambda^{\mathcal{R}} + \widetilde{\mathcal{R}} = 0. 
    \label{eq:adjoint}
\end{align}
With this adjoint variable, we can now compute derivatives of $\mathcal{R}$ with respect to \textit{any} parameter by computing the corresponding perturbations of $\mathcal{L}$, 
\begin{align}
    \partder{\mathcal{R}(\Omega,F(\Omega))}{\Omega_i} = \partder{\mathcal{L}(\Omega,F,\lambda^{\mathcal{R}})}{\Omega_i} = \partder{\mathcal{R}(\Omega,F)}{\Omega_i} + \left \langle \lambda^{\mathcal{R}}, \partder{\mathbb{L}(\Omega)}{\Omega_i}F - \partder{\mathbb{S}(\Omega)}{\Omega_i} \right \rangle.
    \label{eq:derivative_adjoint}
\end{align}
The first term on the right hand side accounts for the explicit dependence on $\Omega_i$ while the second accounts for the implicit dependence on $\Omega_i$ through $F$. 
Thus, using \eqref{eq:derivative_adjoint}, the derivative with respect to $\Omega$ can be computed with the solution to two linear systems, \eqref{eq:linear} and \eqref{eq:adjoint}. The partial derivatives on the right hand side of \eqref{eq:derivative_adjoint} can be computed analytically by considering the explicit geometric dependence of $\mathcal{R}$, $\mathbb{L}$, and $\mathbb{S}$.

When $N_{\Omega}$ is large, the cost of computing $\partial \mathcal{R}/\partial \Omega$ using \eqref{eq:derivative_adjoint} is dominated not by the linear solve but by constructing $\partial \mathbb{S}/\partial \Omega$ and $\partial \mathbb{L}/\partial \Omega$ and computing the inner product. Thus the cost still scales with $N_{\Omega}$. However, we obtain a significant savings in comparison with forward-difference derivatives, as shown in Section \ref{sec:implementation}.

The adjoint operator for each species takes the following form,
\begin{gather}
    \mathbb{L}_s^{\dagger} = \left[ \begin{array}{c c c}
        \mathbb{L}_{0s}^{\dagger} -C_s & f_{Ms} & f_{Ms} X_s^2 \\
        \mathbb{L}_{1s}^{\dagger} & 0 & 0 \\
        \mathbb{L}_{2s}^{\dagger} & 0 & 0  
    \end{array} \right],
    \label{eq:L_dagger}
\end{gather}
where $\mathbb{L}_{1s}^{\dagger} = 5/2\mathbb{L}_{1s}-\mathbb{L}_{2s}$ and $\mathbb{L}_{2s}^{\dagger} = 3/2 \mathbb{L}_{1s} - \mathbb{L}_{2s}$. The same column structure is used as for the forward operator \eqref{eq:dke_species_array},  $\mathbb{L}^{\dagger} = \{ \mathbb{L}_s^{\dagger} \}_{i=1}^{N_{\text{species}}}$. The quantity $\mathbb{L}_{0s}^{\dagger}$ satisfies $\langle \int d^3 v \,  g_{1s} \mathbb{L}_{0s} f_{1s}/f_{Ms}  \rangle_{\psi} = \langle \int d^3 v \, f_{1s} \mathbb{L}_{0s}^{\dagger} g_{1s}/f_{Ms} \rangle_{\psi}$ and depends on which trajectory model is applied. The expression \eqref{eq:L_dagger} can be verified by noting that 
\begin{align}
   \langle  \mathbb{L} F, G \rangle 
   &= \sum_s \left \langle \frac{f_{1s}\left((\mathbb{L}^{\dagger}_{0s} - C_s )g_{1s} + f_{Ms} \left( S_{1s}^g + S_{2s}^g X_s^2\right)\right)}{f_{Ms}}\right \rangle_{\psi} + S_{1s}^f\mathbb{L}_{1s}^{\dagger} g_{1s} + S_{2s}^f \mathbb{L}_{2s}^{\dagger} g_{1s} \nonumber \\
   &= \langle F,\mathbb{L}^{\dagger} G \rangle.
\end{align}
For the DKES trajectories the adjoint operator is,
\begin{gather}
    \mathbb{L}_{0s}^{\dagger} = - \mathbb{L}_{0s}. 
    \label{eq:dkes_adjoint}
\end{gather}
This anti-self-adjoint property is used in obtaining the variational principle which provides bounds on neoclassical transport coefficients in the DKES code \citep{Rij1989}.
For full trajectories it is,
\begin{gather}
    \mathbb{L}_{0s}^{\dagger} = -\mathbb{L}_{0s} + \frac{q_s}{T_s} \Phi'(\psi) \textbf{v}_{\text{m}s} \cdot \nabla \psi.
    \label{eq:full_adjoint}
\end{gather}
The anti-self-adjoint property does not hold for this trajectory model as the $\textbf{E} \times \textbf{B}$ drift \eqref{eq:full_ve} is no longer divergenceless. Appendix \ref{ap:adjoint_operators} contains details on obtaining these adjoint operators. 

\subsection{Discrete approach}
\label{sec:discrete}

Next, we consider the discrete adjoint approach. Let $\overrightarrow{\textbf{F}}$ be the set of unknowns computed with SFINCS after discretization of $F$. The linear DKE \eqref{eq:linear} upon discretization can then be written schematically as,
\begin{gather}
    \overleftrightarrow{\textbf{L}} \overrightarrow{\textbf{F}} = \overrightarrow{\textbf{S}}. 
    \label{eq:forward_discrete}
\end{gather}
In this case, we can define an inner product as the vector dot product,
\begin{gather}
    \langle \overrightarrow{\textbf{F}}, \overrightarrow{\textbf{G}} \rangle = \overrightarrow{\textbf{F}} \cdot \overrightarrow{\textbf{G}}.
\end{gather}
In real Euclidean space, the adjoint operator, $\left(\overleftrightarrow{\textbf{L}}\right)^{\dagger}$, which satisfies,
\begin{gather}
    \left \langle \overleftrightarrow{\textbf{L}} \overrightarrow{\textbf{F}},\overrightarrow{\textbf{G}} \right \rangle = \left \langle \overrightarrow{\textbf{F}},\left(\overleftrightarrow{\textbf{L}}\right)^{\dagger} \overrightarrow{\textbf{G}} \right \rangle
\end{gather}
is simply the transpose of the matrix, $\left(\overleftrightarrow{\textbf{L}}\right)^T$. Again, the moments of the distribution function, $\mathcal{R}$ can be expressed as an inner product with a vector $\overrightarrow{\textbf{R}}$,
\begin{gather}
    \mathcal{R} = \langle \overrightarrow{\textbf{F}}, \overrightarrow{\textbf{R}} \rangle. 
\end{gather}
Using the discrete approach, the following adjoint equation must be solved
\begin{gather}
    \left(\overleftrightarrow{\textbf{L}}\right)^T \overrightarrow{\bm{\lambda}}^{\mathcal{R}} = \overrightarrow{\textbf{R}}. 
    \label{eq:adjoint_discrete}
\end{gather}
The adjoint variable, $\overrightarrow{\bm{\lambda}}^{\mathcal{R}}$, can again be used to compute the derivative of $\mathcal{R}$ with respect to $\Omega$,
\begin{gather}
    \partder{\mathcal{R}\left(\Omega,\overrightarrow{\textbf{F}}(\Omega)\right)}{\Omega_i} = \partder{\mathcal{R}\left(\Omega,\overrightarrow{\textbf{F}}\right)}{\Omega_i} + \left \langle \overrightarrow{\bm{\lambda}}^{\mathcal{R}}, \left( \partder{\overrightarrow{\textbf{S}}(\Omega)}{\Omega_i} - \partder{\overleftrightarrow{\textbf{L}}(\Omega)}{\Omega_i} \overrightarrow{\textbf{F}} \right) \right \rangle. 
    \label{eq:adjoint_diagnostic_discrete}
\end{gather}
As with the continuous approach, the partial derivatives on the right hand side can be computed analytically. In this way, the derivative of $\mathcal{R}$ with respect to $\Omega$ can be computed with only two linear solves, \eqref{eq:forward_discrete} and \eqref{eq:adjoint_discrete}. 

In the SFINCS implementation, the DKE is typically solved with the preconditioned GMRES algorithm. In the continuous approach, a preconditioner matrix for both the forward and adjoint operator must be $LU$-factorized. Here the preconditioner matrix is the same as the full matrix but without cross-species or speed coupling. As the adjoint matrix is sufficiently different from the forward matrix, we do not obtain convergence when the same preconditioner is used for both problems. However, in the discrete approach, the $LU$-factorization for the preconditioner of the forward matrix can be reused for the preconditioner of the adjoint matrix. (If a matrix $A$ has been factorized as $A = LU$ then $A^{T} = U^T L^T$ where $U^T$ is lower triangular and $L^T$ is upper triangular). This provides a significant reduction in memory and computational cost for the discrete approach. 

Furthermore, the discrete adjoint approach provides the exact derivatives for the discretized problem.  With this method, the adjoint equation is obtained using the vector dot product and matrix transpose, which can be computed without any numerical approximation. The error in the derivatives obtained by the adjoint method is therefore only limited by the tolerance to which the linear solve is performed with GMRES. On the other hand, the continuous adjoint approach relies on a continuous inner product that must ultimately be approximated numerically. Thus the continuous approach provides the exact derivatives only in the limit that the discrete approximation of the inner product exactly reproduces the continuous inner product. Therefore we expect the results of the discrete and adjoint approaches to agree within discretization error, as will be demonstrated in Section \ref{sec:implementation}.

The continuous approach can be advantageous in that an adjoint equation may be prescribed independently of the discretization scheme.
Note that in the discrete approach, the adjoint operator is obtained from the matrix transpose of the discretized forward operator, which implies that the same spatial and velocity resolution parameters must be used for both the forward and adjoint solutions. In this Chapter, we will employ the same discretization parameters for both the adjoint and forward problems, but this restriction is not required for the continuous approach.

\section{Implementation and benchmarks}
\label{sec:implementation}

The adjoint method has been implemented in the SFINCS code\footnote{The adjoint method is implemented in the main branch of the SFINCS code \href{https://github.com/landreman/sfincs}{https://github.com/landreman/sfincs}.} using both the discrete and continuous approaches. The magnetic geometry is specified in Boozer coordinates (Appendix \ref{sec:boozer_coordinates}) such that the covariant form of the magnetic field is,
\begin{gather}
 \textbf{B} = I(\psi) \nabla \vartheta_B + G(\psi) \nabla \varphi_B + K(\psi,\vartheta_B,\varphi_B) \nabla \psi,
    \label{eq:boozer_covariant_ch2}
\end{gather}
where $I(\psi) = \mu_0 I_T(\psi)/2\pi$ and $G(\psi) = \mu_0 I_P(\psi)/2\pi$, $I_T(\psi)$ is the toroidal current enclosed by $\psi$, and $I_P(\psi)$ is the poloidal current outside of $\psi$. The contravariant form is,
\begin{gather}
    \textbf{B} = \nabla \psi \times \nabla \vartheta_B - \iota(\psi) \nabla \psi \times \nabla \varphi_B,
    \label{eq:boozer_contravariant_ch2}
\end{gather}
where $\iota(\psi)$ is the rotational transform. The Jacobian is obtained from dotting \eqref{eq:boozer_covariant_ch2} with \eqref{eq:boozer_contravariant_ch2}, 
\begin{gather}
    \sqrt{g} = \frac{G(\psi) + \iota(\psi) I(\psi)}{B^2}.
    \label{eq:jacobian}
\end{gather}
As $K(\psi,\vartheta_B,\varphi_B)$ does not appear in any of the trajectory coefficients (\eqref{eq:full_trajectories} and \eqref{eq:dkes_trajectories}), in the drive term in \eqref{eq:dke_model}, or in the geometric factors used to define the moments of the distribution function (\eqref{eq:parallel_flow}, \eqref{eq:particle_flux}, and \eqref{eq:heat_flux}), all the geometric dependence enters through $B(\psi,\vartheta_B,\varphi_B)$, $G(\psi)$, $I(\psi)$, and $\iota(\psi)$. We choose to use Boozer coordinates for these computations as it reduces the number of geometric parameters that must be considered, but the neoclassical adjoint method is not limited to this choice of coordinate system.

We approximate $B$ by a truncated Fourier series,
\begin{gather}
    B = \sum_{m,n} B_{m,n}^c \cos(m\vartheta_B-n N_P\varphi_B)
    \label{eq:B_Fourier},
\end{gather}
where the sum is taken over Fourier modes $m \leq m_{\max}$ and $|n| \leq n_{\max}$ and $N_P$ is the number of periods. In \eqref{eq:B_Fourier}, we have assumed stellarator symmetry such that $B(-\vartheta_B,-\varphi_B) = B(\vartheta_B,\varphi_B)$, and $N_p$ symmetry such that $B(\vartheta_B,\varphi_B+2\pi/N_P) = B(\vartheta_B,\varphi_B)$. Thus we compute derivatives with respect to the parameters \\ $\Omega = \{B_{m,n}^c, I(\psi), G(\psi), \iota(\psi)$\}. Additionally, derivatives with respect to $E_r$ are computed, which are used for efficient ambipolar solutions and computing derivatives of geometric quantities at  ambipolarity (Section \ref{sec:ambipolarity}) rather than at fixed $E_r$. 

\begin{figure}
\begin{center}
\begin{subfigure}[c]{0.422\textwidth}
\includegraphics[trim=1cm 6cm 7.2cm 7cm,clip,width=1.0\textwidth]{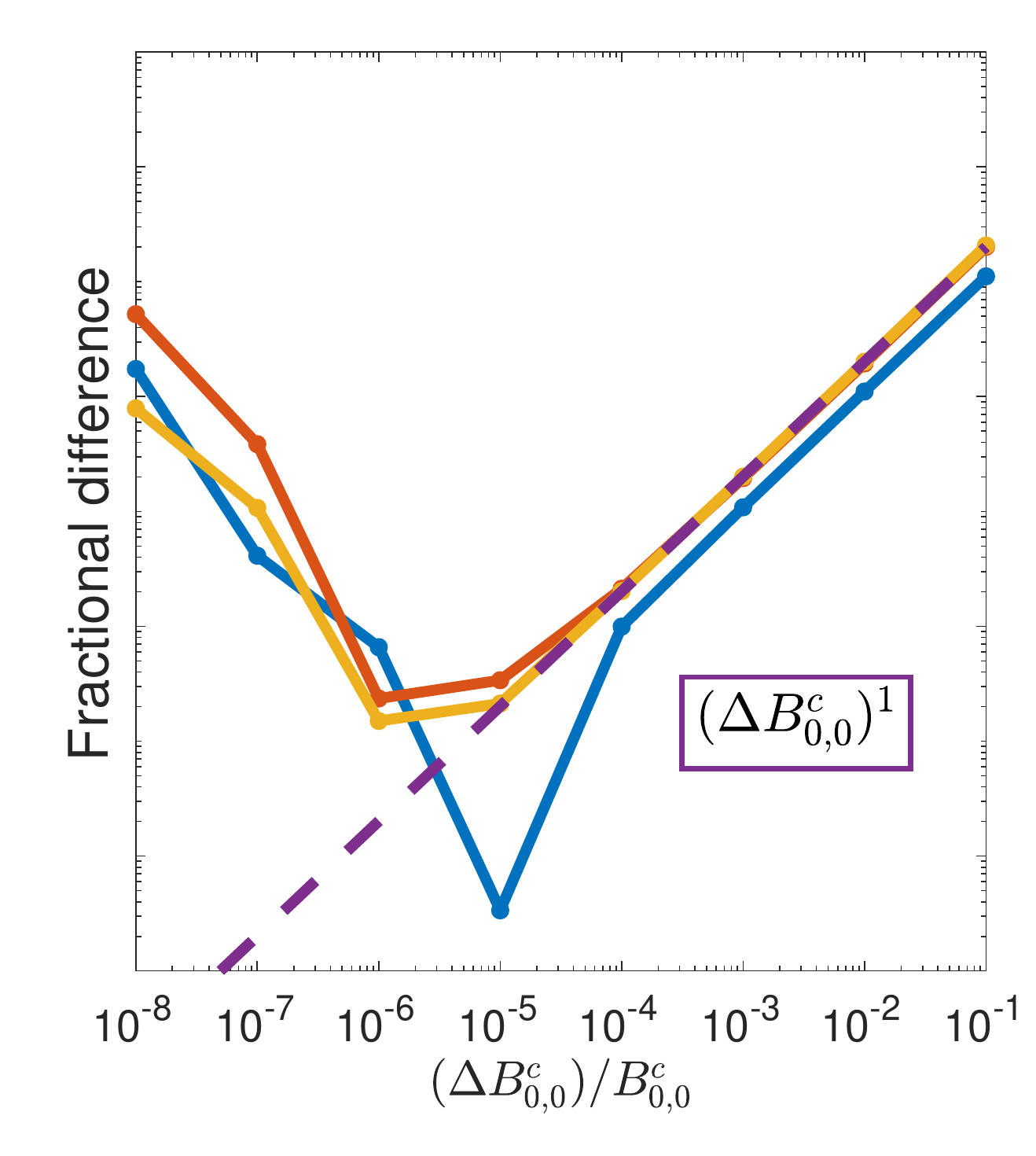}
\caption{Discrete approach}
\end{subfigure}
\begin{subfigure}[c]{0.56\textwidth}\includegraphics[trim=2cm 6cm 2cm 7cm,clip,width=1.0\textwidth]{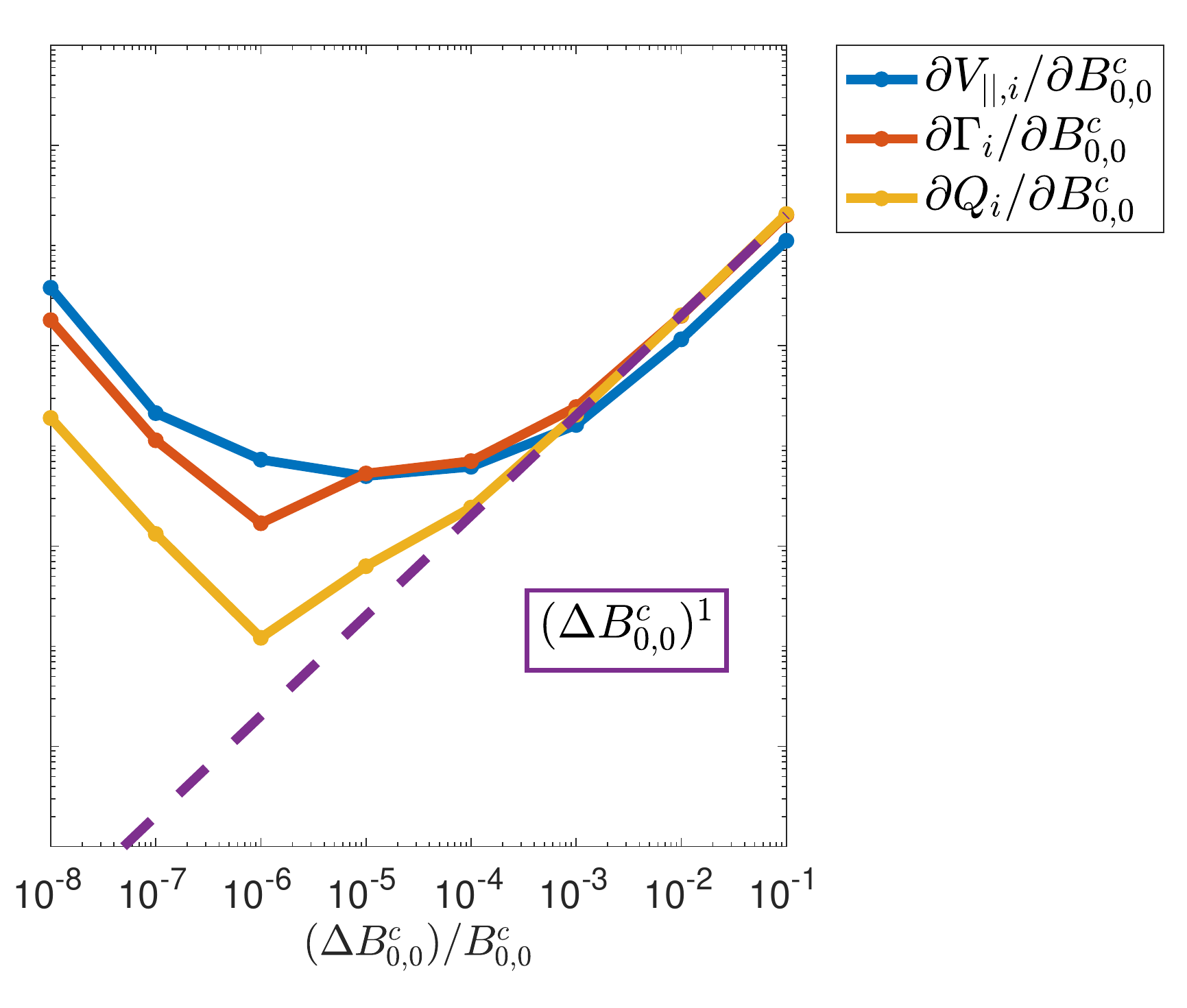}
\caption{Continuous approach}
\end{subfigure}
\caption{Fractional difference between derivatives with respect to $B_{0,0}^c$ computed with the adjoint method and with a forward-difference derivative with step size $\Delta B_{0,0}^c$. The full trajectory model was used with (a) the discrete and (b) the continuous adjoint approaches. Figure adapted from \cite{Paul2019} with permission.} 
\label{fig:benchmark_fixedEr}
\end{center}
\end{figure}

To demonstrate, we compute $\partial\mathcal{R}/\partial B_{0,0}^c$ for moments of the ion distribution function using the
discrete and continuous adjoint methods. A 3-mode model of the standard configuration W7-X geometry at $\rho = \sqrt{\psi/\psi_0} = 0.5$ is used (Table 1 in \cite{Beidler2011}),
\begin{gather}
    B = B_{0,0}^c + B_{0,1}^c \cos(N_P\varphi_B) + B_{1,1}^c \cos(\vartheta_B - N_P \varphi_B) + B_{1,0}^c \cos(\vartheta_B),
\end{gather}
where $B_{0,1}^c = 0.04645 B_{0,0}^c$, $B_{1,1}^c = -0.04351 B_{0,0}^c$, and $B_{1,0}^c = -0.01902 B_{0,0}^c$.
Electron and ion ($q_i = e$) species are included, and the derivatives are computed at the ambipolar $E_r$ with the full trajectory model. The derivatives are also computed with a forward-difference approach with varying step size $\Delta B_{0,0}^c$. In Figure \ref{fig:benchmark_fixedEr} we show the fractional-difference between $\partial \mathcal{R}/\partial B_{0,0}^c$ computed using the adjoint method and with forward-difference derivatives. We see that at large values of $\Delta B_{0,0}^c$, the adjoint and numerical derivatives begin to differ significantly due to discretization error from the forward-difference approximation. The fractional error decreases proportional to $\Delta B_{0,0}^c$ as expected until the rounding error begins to dominate \citep{Sauer2012} when $\Delta B_{0,0}^c/B_{0,0}^c$ is approximately $10^{-4}$, where $B_{0,0}^c$ is the value of the unperturbed mode. The discrete and continuous approaches show qualitatively similar trends. However, the minimum fractional difference is lower in the discrete approach due to the additional discretization error that arises with the continuous approach. With sufficient resolution parameters (41 $\theta$ grid points, 61 $\phi$ grid points, 85 $\xi$ basis functions, and 7 $X$ basis functions), the fractional error of the continuous approach is $\leq 0.1 \%$ and should not be significant for most applications. We find similar agreement for other derivatives and with the DKES trajectory model.

\begin{figure}
    \centering
    \begin{subfigure}[b]{0.49\textwidth}
    \includegraphics[trim=1cm 6.5cm 2cm 7cm,clip,width=1.0\textwidth]{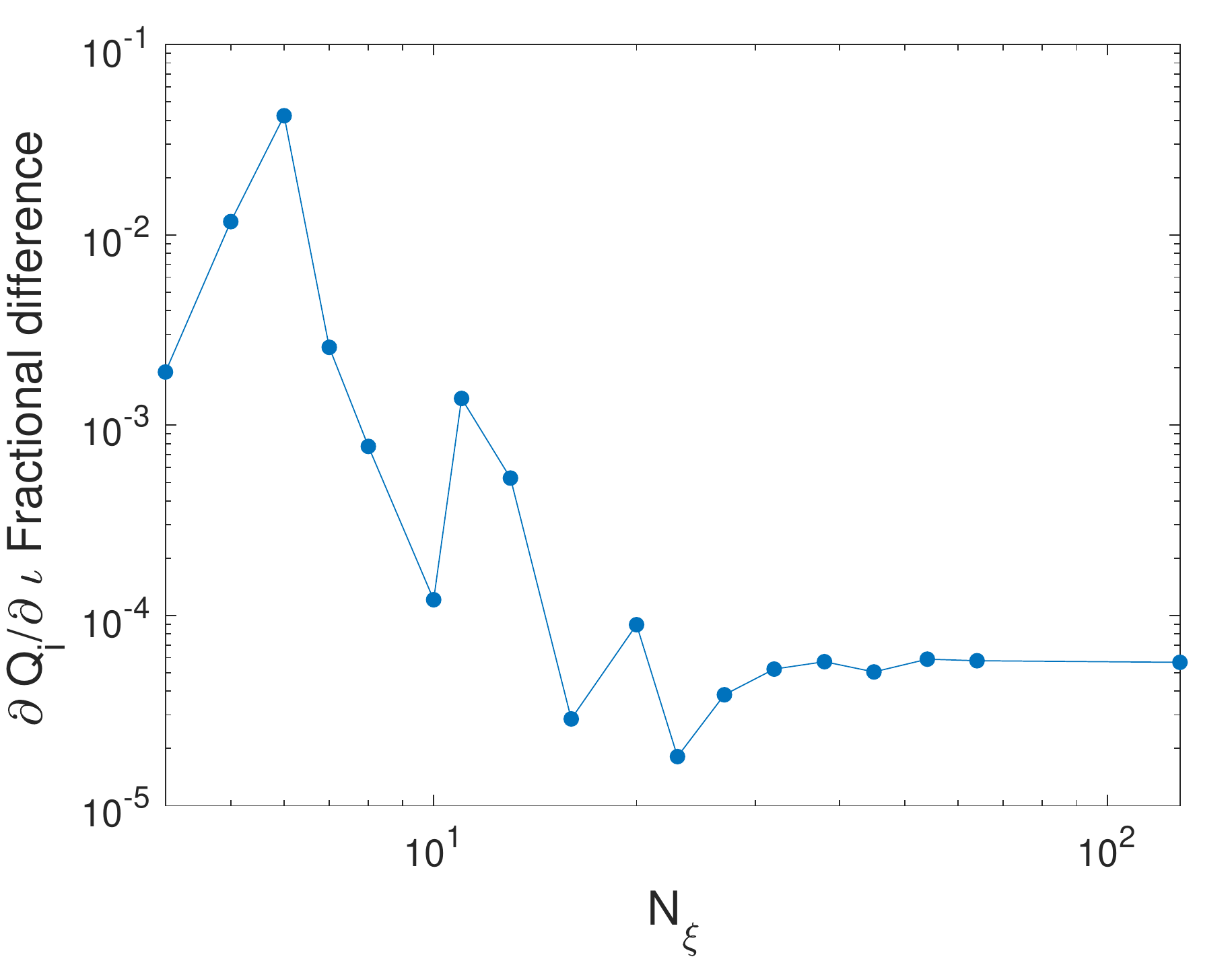}
    \caption{}
    \label{fig:continuous_discrete}
    \end{subfigure}
    \begin{subfigure}[b]{0.49\textwidth}
    \includegraphics[trim=1cm 6.5cm 2cm 7cm,clip,width=1.0\textwidth]{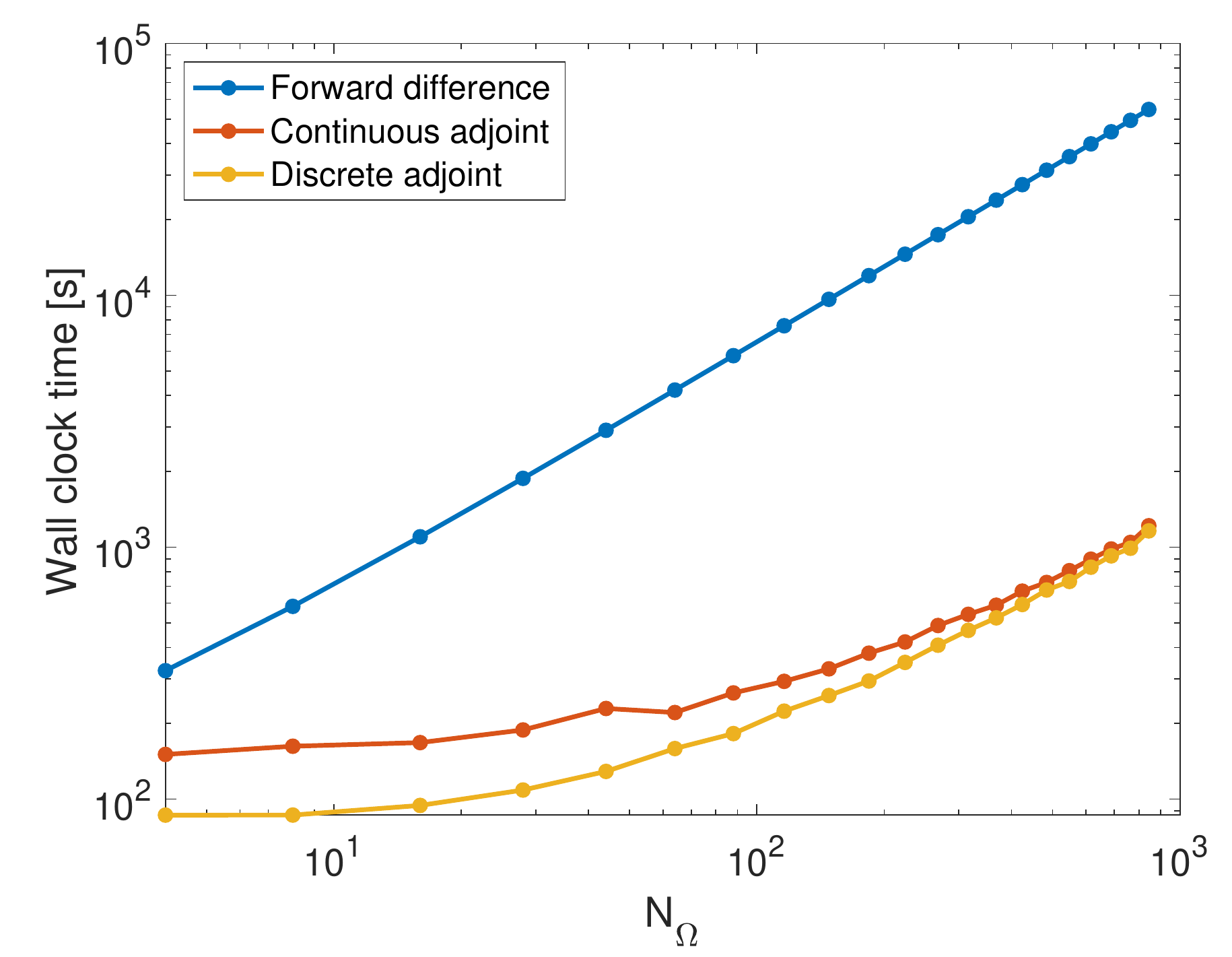}
    \caption{}
    \label{fig:computational_time}
    \end{subfigure}
    \caption{(a) The fractional difference between $\partial Q_i/\partial \iota$ computed with the continuous and discrete approaches converges with the number of pitch angle Legendre modes, $N_{\xi}$. (b) Comparison of the computational cost of computing $\partial \mathcal{R}/\partial \Omega$ with forward-difference derivatives and the adjoint approach as a function of $N_{\Omega}$, the number of parameters in the gradient. Figure reproduced from \cite{Paul2019} with permission.}
    \label{fig:discrete_continuous_scan}
\end{figure}

To demonstrate that the discrete and continuous methods indeed produce the same derivative information, we compute the fractional difference between the derivatives computed with the two methods as a function of the resolution parameters. As an example, in Figure \ref{fig:continuous_discrete} we show the fractional difference in $\partial Q_i/\partial \iota$, where $Q_i$ is the radial ion heat flux, as a function of the number of Legendre polynomials used for the pitch angle discretization, $N_{\xi}$, keeping the other resolution parameters fixed. As $N_{\xi}$ is increased, the fractional differences converge to a finite value, approximately $10^{-4}$, due to the discretization error in the other resolution parameters. Similar resolution parameters are required for the convergence of the moment itself, $Q_i$, and its derivative computed with the continuous method, $\partial Q_i/\partial \iota$. Convergence of $Q_i$ within 5\% is obtained with $N_{\xi} = 38$, similar to that required for the convergence of $\partial Q/\partial \iota$, as can be seen in Figure \ref{fig:continuous_discrete}.

In Figure \ref{fig:computational_time}, we compare the cost of calculating derivatives of one moment with respect to $N_{\Omega}$ parameters using the continuous and discrete adjoint methods and forward-difference derivatives. All computations are performed on the Edison computer at NERSC using 48 processors, and the elapsed wall time is reported. Here we include the cost of solving the linear system and computing diagnostics $N_{\Omega} + 1$ times for the forward-difference approach, and the cost of solving the forward and adjoint linear systems and computing diagnostics for the adjoint approaches. The cost of the continuous approach is slightly more than that of the discrete approach due to the cost of factorizing the adjoint preconditioner. However, at large $N_{\Omega}$ the cost of computing diagnostics for the adjoint approach (e.g., computing $\partial \mathbb{S}/\partial \Omega$ and $\partial \mathbb{L}/\partial \Omega$ and performing the inner product in \eqref{eq:derivative_adjoint}) dominates that of solving the adjoint linear system; thus the discrete and continuous approaches become comparable in cost. In this regime, the adjoint approach provides speed-up by a factor of approximately $50$. 

\section{Applications of the adjoint method}
\label{sec:applications_ch3}

\subsection{Local magnetic sensitivity analysis}
\label{sec:local_sensitivity}

With the adjoint method, it is possible to compute derivatives of a moment of the distribution function with respect to the Fourier amplitudes of the field strength, $\{\partial \mathcal{R}/\partial B_{m,n}^c\}$. Rather than consider sensitivity in Fourier space, we would like to compute the sensitivity to \textit{local} perturbations of the field strength. We now quantify the relationship between these two representations of sensitivity information.

Consider the Gateaux functional derivative \citep{Delfour2011} of $\mathcal{R}$ with respect to $B$,
\begin{gather}
    \delta \mathcal{R}(B(\textbf{x});\delta B) = \lim_{\epsilon \rightarrow 0} \frac{\mathcal{R}(B(\textbf{x}) + \epsilon \delta B(\textbf{x}))-\mathcal{R}(B(\textbf{x}))}{\epsilon}. 
    \label{eq:functional_derivative}
\end{gather}
Here the field strength is perturbed at fixed $I(\psi)$, $G(\psi)$, and $\iota(\psi)$. As $\delta \mathcal{R}(B(\textbf{x});\delta B)$ is a linear functional of $\delta B$, by the Riesz representation theorem \citep{Rudin2006}, $\delta \mathcal{R}$ can be expressed as an inner product with $\delta B$ and some element of the appropriate space. The function $\delta B$ is defined on a flux surface, $\psi$; thus it is sensible to express $\delta \mathcal{R}$ in the following way, 
\begin{gather}
    \delta \mathcal{R}(B(\textbf{x});\delta B) = \left \langle S_{\mathcal{R}} \delta B(\textbf{x}) \right \rangle_{\psi}. 
    \label{eq:magnetic_sensitivity}
\end{gather}
Here $\delta \mathcal{R}$ quantifies the change in the moment $\mathcal{R}$ associated with a local perturbation to the field strength, $\delta B(\textbf{x})$. The function $S_{\mathcal{R}}$ is analogous to the shape gradient introduced in Section \ref{sec:shape_optimization}, which will be discussed further in Section \ref{sec:equilibria_opt}.

Suppose that $B$ is stellarator symmetric and $N_P$ symmetric. If $E_r = 0$, then $S_{\mathcal{R}}$ must also possess stellarator and $N_P$ symmetry (Appendix \ref{app:symmetry}). However, when $E_r \neq 0$, $S_{\mathcal{R}}$ is no longer guaranteed to have stellarator symmetry. Nonetheless, it may be desirable to ignore the stellarator-asymmetric part of $S_{\mathcal{R}}$ if an optimized stellarator-symmetric configuration is desired. For the remainder of this Chapter, we will make this assumption, though the analysis could be extended to consider the effect of breaking of stellarator symmetry. A truncated Fourier series can approximate the quantity $S_{\mathcal{R}}$ under these assumptions,
\begin{gather}
    S_{\mathcal{R}} = \sum_{m,n} S_{m,n} \cos(m \vartheta_B - n N_P\varphi_B),
\end{gather}
where the sum is taken over $m \leq m_{\max}$ and $|n| \leq n_{\max}$. The quantity $\delta B(\textbf{x})$ can be written in terms of perturbations to the Fourier coefficients,
\begin{gather}
    \delta B(\textbf{x}) = \sum_{m,n} \delta B_{m,n}^c \cos(m \vartheta_B - n N_P \varphi_B),
\end{gather}
and now $\delta \mathcal{R}$ can be written in terms of these perturbations to the Fourier coefficients,
\begin{gather}
\delta \mathcal{R} = \sum_{m,n} \partder{\mathcal{R}}{B_{m,n}^c} \delta B_{m,n}^c. 
\end{gather}
In this way, \eqref{eq:magnetic_sensitivity} can be expressed as a linear system,
\begin{gather}
    \partder{\mathcal{R}}{B_{m,n}^c} = \sum_{m',n'} D_{m,n;m',n'} S_{m',n'},
\end{gather}
where, 
\begin{multline}
    D_{m,n;m',n'} =\\
    V'(\psi)^{-1} \int_{0}^{2\pi} d \vartheta_B \int_0^{2\pi} d \varphi_B \, \sqrt{g} \cos(m \vartheta_B - n N_P \varphi_B) \cos(m' \vartheta_B - n' N_P \varphi_B).
\end{multline}
If the same number of modes is used to discretize $\delta \mathcal{R}$ and $S_{\mathcal{R}}$, then the linear system is square.

\begin{figure}
\centering
\begin{subfigure}[b]{0.49\textwidth}
\includegraphics[trim=2cm 0cm 2cm 3cm,clip,width=1.0\textwidth]{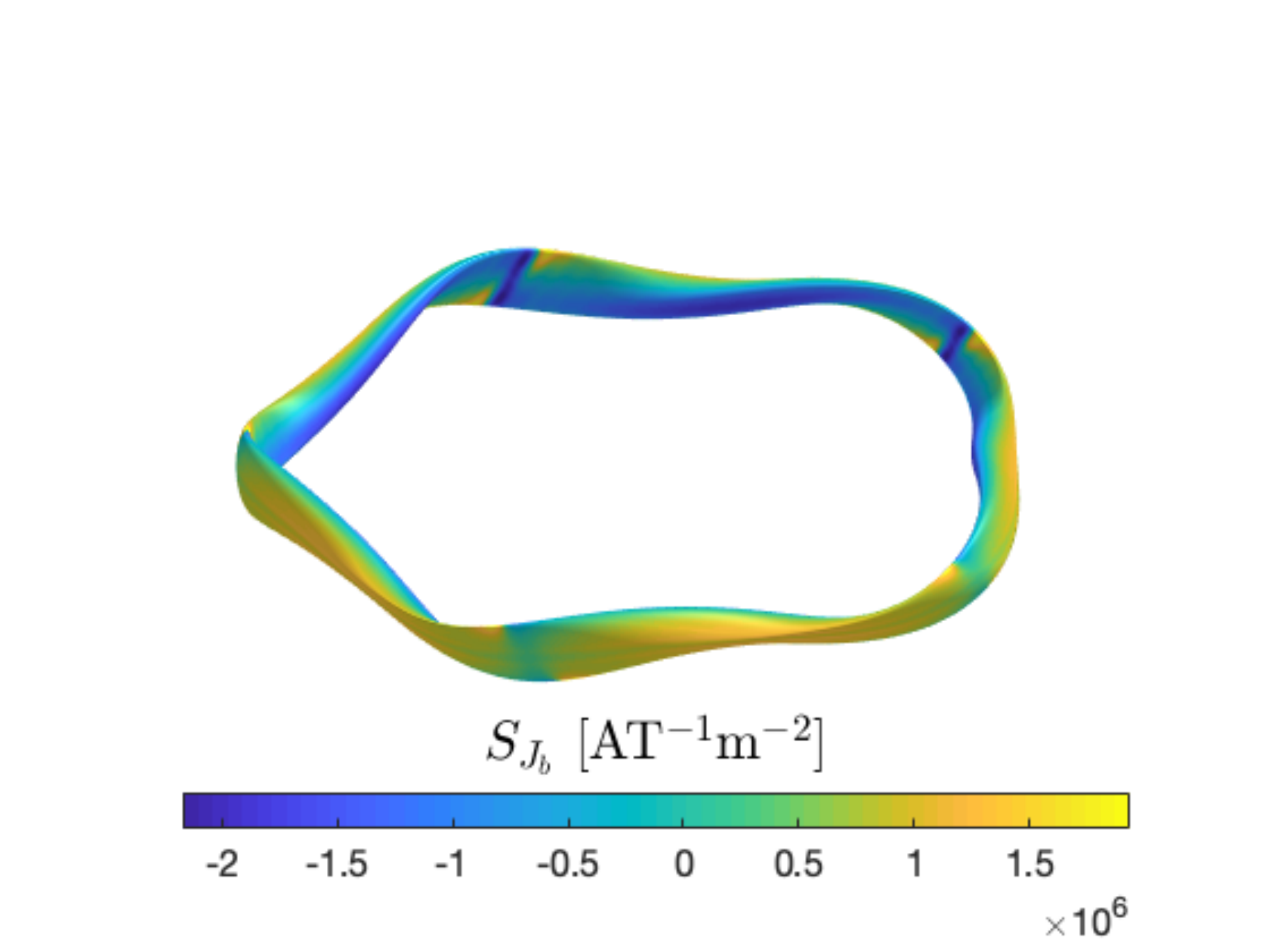}
\caption{}
\label{fig:bootstrap_local_sensitivity}
\end{subfigure}
\begin{subfigure}[b]{0.49\textwidth}
\includegraphics[trim=2cm 0cm 2cm 3cm,clip,width=1.0\textwidth]{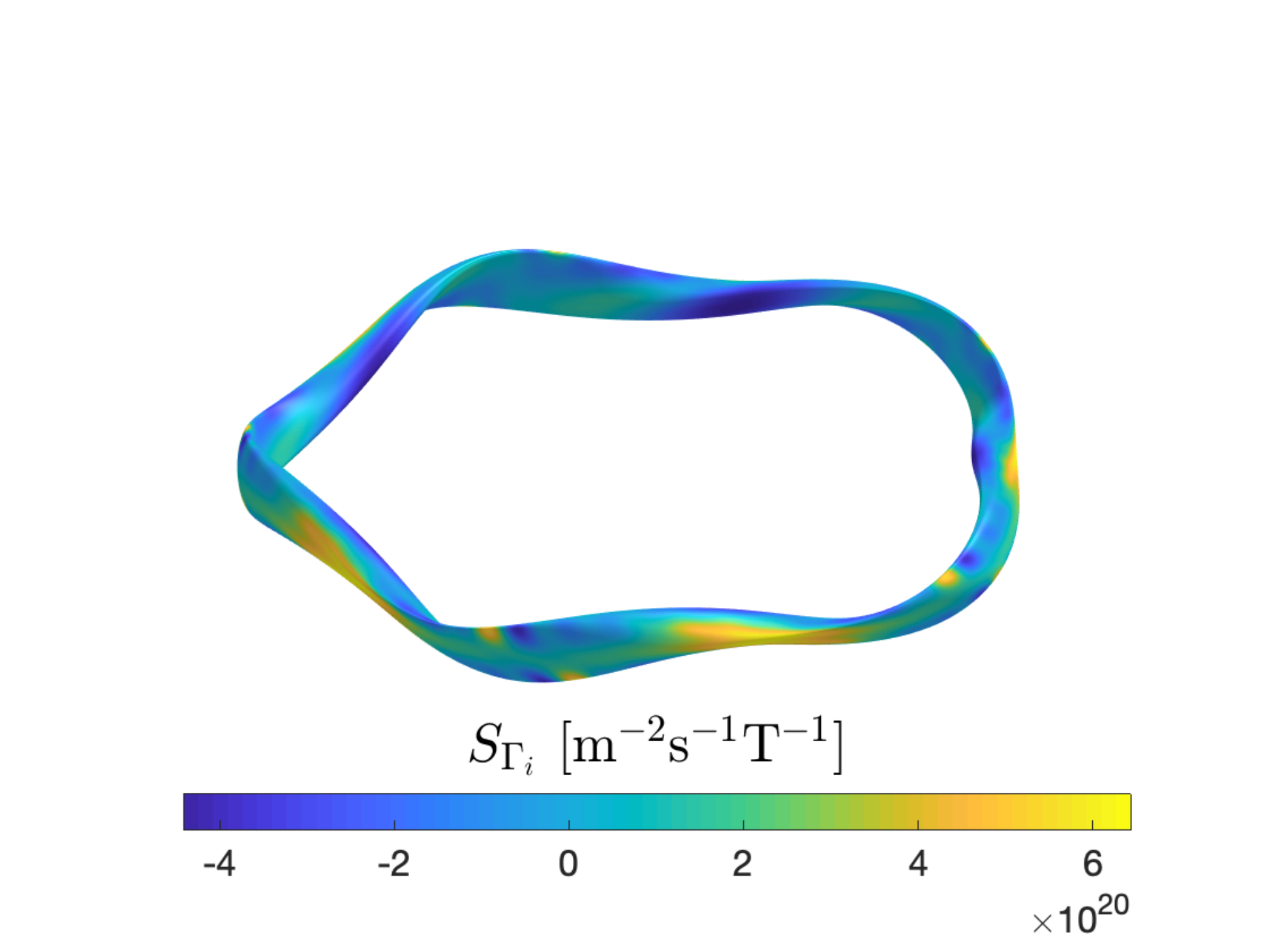}
\caption{}
\label{fig:particleFlux_sensitivity}
\end{subfigure}
\begin{subfigure}[b]{0.49\textwidth}
\includegraphics[trim=3cm 7cm 3cm 10cm,clip,width=1.0\textwidth]{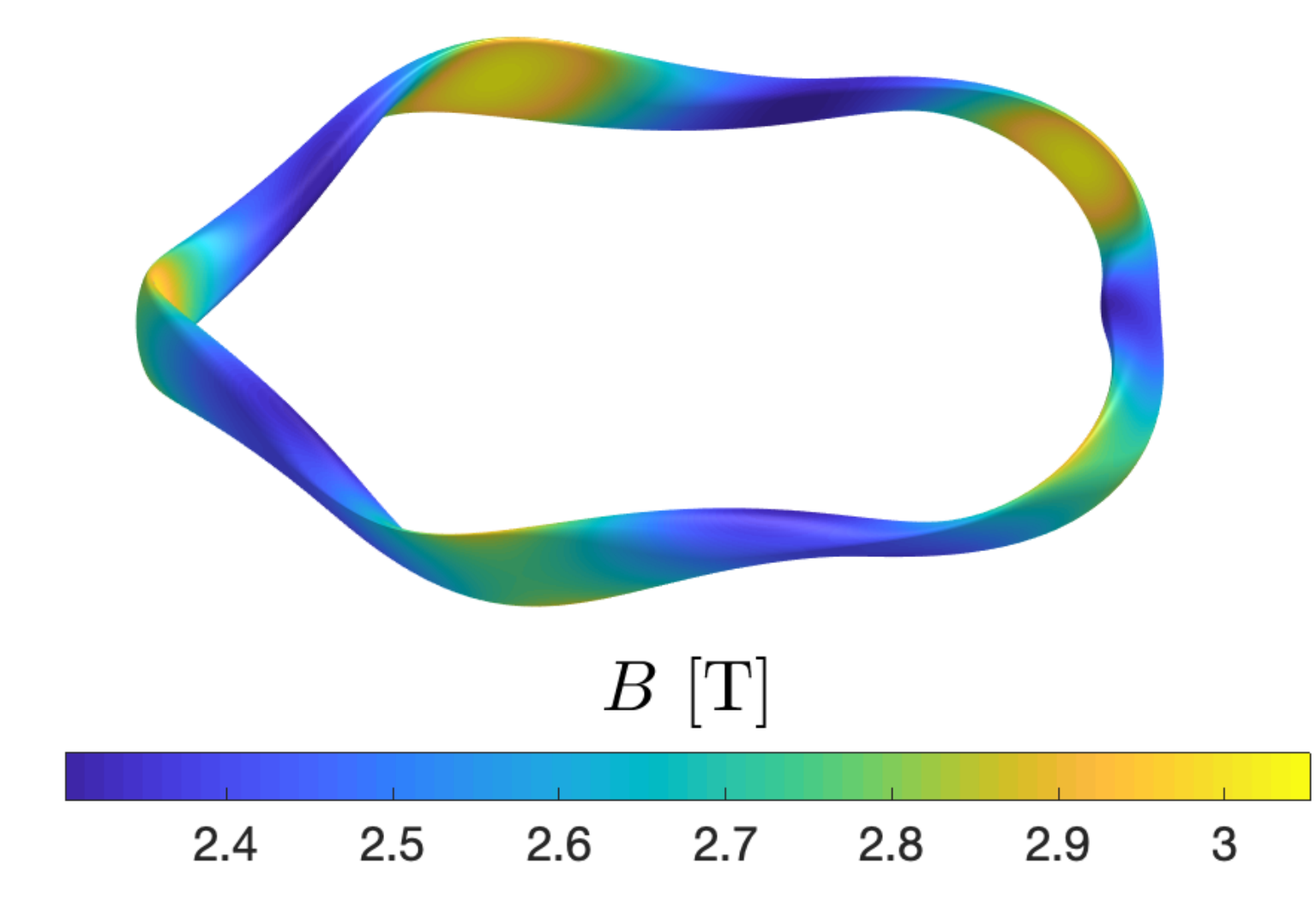}
\caption{}
\label{fig:w7x_field_strength}
\end{subfigure}
\caption{(a) The local magnetic sensitivity function for the bootstrap current, $S_{J_b}$, is shown for the W7-X standard configuration. Positive values indicate that increasing the field strength at a given location will increase $J_b$ through \eqref{eq:magnetic_sensitivity}. (b) The local sensitivity function for the ion particle flux, $S_{\Gamma_i}$. (c) The magnetic field strength on the $\rho = 0.7$ surface. Figure adapted from \cite{Paul2019} with permission.}
\end{figure}

In contrast to derivatives with respect to the Fourier modes of $B$, the sensitivity function, $S_{\mathcal{R}}$, is a spatially local quantity, quantifying the change in a figure of merit resulting from a local perturbation of the field strength. In this way, $S_{\mathcal{R}}$ can inform where perturbations to the magnetic field strength can be tolerated. The sensitivity function could be related directly to a local magnetic tolerance, as described in Section \ref{sec:shape_optimization_discussion}. In contrast with the work in \cite{Landreman2018}, here we are considering perturbations to the field strength on any flux surface rather than at the plasma boundary. However, $S_{\mathcal{R}}$ still provides insight into where trim coils should be placed or coil displacements can be tolerated without sacrificing desired neoclassical properties. The sensitivity function can also be used for gradient-based optimization in the space of the field strength on a flux surface, as demonstrated in Section \ref{sec:vacuum_opt}.

We compute $S_{J_b}$ for the W7-X standard configuration at $\rho = 0.70$, shown in Figure \ref{fig:bootstrap_local_sensitivity}. We use a fixed-boundary equilibrium that preceded the coil design and does not include coil ripple, and the full equilibrium is used rather than the truncated Fourier series considered in Section \ref{sec:implementation}. The same resolution parameters are used as in Section \ref{sec:implementation}, and derivatives with respect to $B_{m,n}^c$ are computed for $m_{\max} = n_{\max} = 20$. The largest modes for this configuration are the helical curvature $B_{1,1}^c$, the toroidal curvature $B_{1,0}^c$, and the toroidal mirror $B_{0,1}^c$. We find that $S_{J_b}$ is large and negative on the inboard side, indicating that increasing the magnitude of the toroidal curvature component of $B$ would lead to an increase in $J_b$. This result is in agreement with previous analysis \cite{Maassberg1993},
which found that at low collisionality, the bootstrap current coefficients depend strongly on the toroidal curvature. Additionally, we note a localized region of strong sensitivity on the inboard side near the bean-shaped cross-section. Experimental \cite{Dinklage2018} and numerical \cite{Geiger2010} evidence indicates that the magnitude of the bootstrap current is increased in the lower mirror-ratio configuration of W7-X, where the mirror-ratio is defined as $(B_{\text{max}} - B_{\text{min}})/(B_{\text{max}} + B_{\text{min}})$. Our result appears to be consistent with these observations: we note that the localized region of strongly positive $S_{J_b}$ is near the maximum of the magnetic field strength (Figure \ref{fig:w7x_field_strength}), indicating that increasing the mirror-ratio would lead to a decrease in the magnitude of bootstrap current, as $J_b<0$ for this configuration. 

In Figure \ref{fig:particleFlux_sensitivity} is the sensitivity function for the ion particle flux, $S_{\Gamma_i}$, computed for the same configuration using $m_{\max} = 20$ and $n_{\max} = 20$. We find that the particle flux is more sensitive to perturbations on the outboard side in localized regions, while on the inboard side the sensitivity is relatively small in magnitude.

\subsection{Gradient-based optimization}

\subsubsection{Optimization of the magnetic field strength}
\label{sec:vacuum_opt}

As a second demonstration of the adjoint neoclassical method, we consider optimizing in the space of the field strength on a surface, taking $\Omega = \{B_{m,n}^c\}$. As Boozer coordinates are used, the covariant form \eqref{eq:boozer_covariant_ch2} satisfies $(\nabla \times \textbf{B}) \cdot \nabla \psi = 0$ and the contravariant form  \eqref{eq:boozer_contravariant_ch2} satisfies $\nabla \cdot \textbf{B} = 0$. As we will artificially modify the field strength while keeping other geometry parameters fixed, the resulting field will not necessarily satisfy both of these conditions with both the covariant and contravariant forms. While there is no guarantee that the resulting field strength will be consistent with a global equilibrium solution, it provides insight into how local changes to the field strength can impact neoclassical properties. As a second step, the outer boundary could be optimized to match the desired field strength on a single surface. In Section \ref{sec:equilibria_opt}, we discuss how the derivatives computed in this Chapter could be coupled to the optimization of an MHD equilibrium. 

\begin{figure}
    \centering
    \begin{subfigure}[b]{0.49\textwidth}
    \includegraphics[trim=1cm 6cm 2cm 6cm,clip,width=1.0\textwidth]{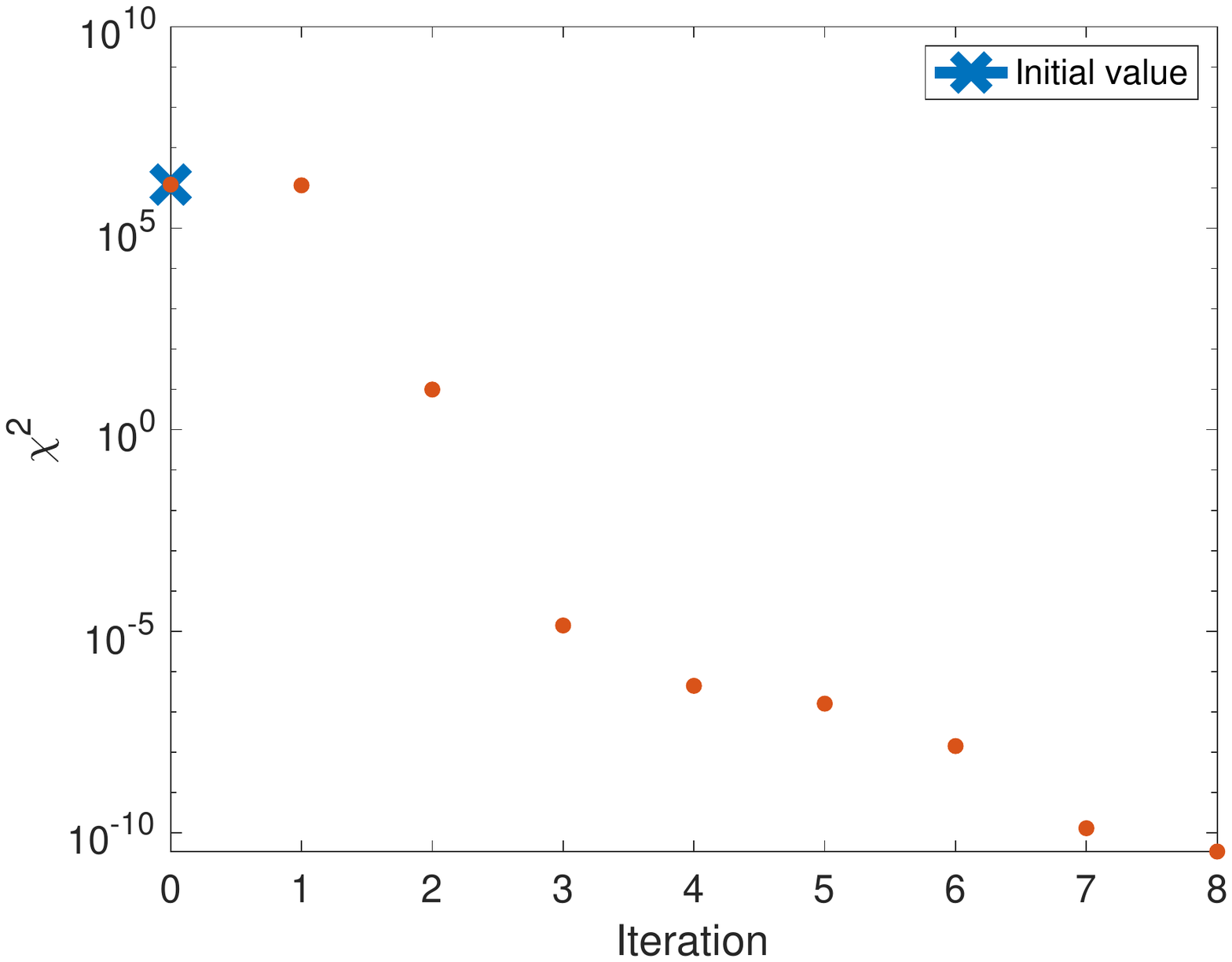}
    \caption{}
    \label{fig:bfgs_convergence}
    \end{subfigure}
    \begin{subfigure}[b]{0.49\textwidth}
    \includegraphics[trim=1cm 6cm 2cm 6cm,clip,width=1.0\textwidth]{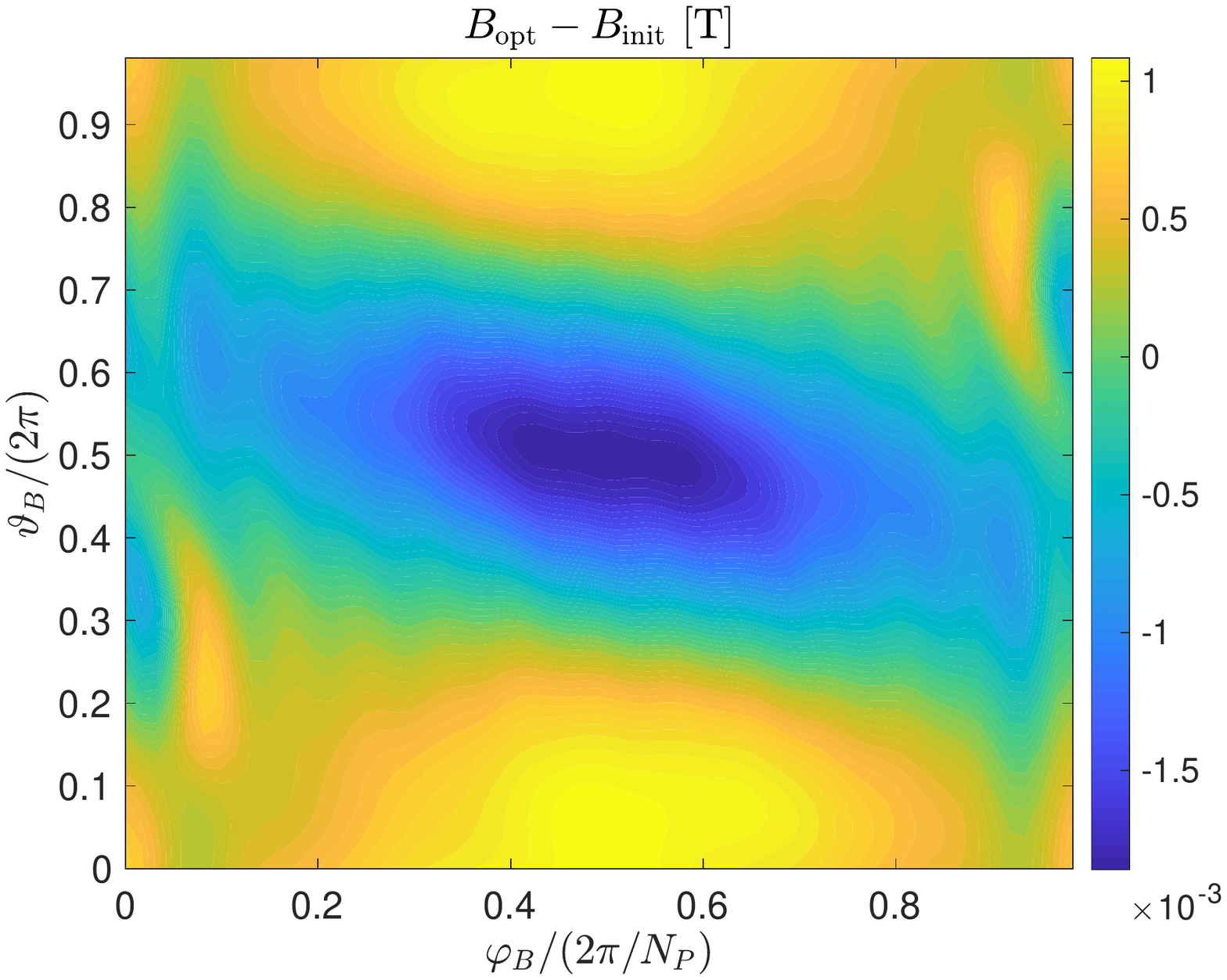}
    \caption{}
    \label{fig:B_opt}
    \end{subfigure}
    \caption{(a) Convergence of $\chi^2 = J_b^2$ for optimization over $\Omega = \{B_{m,n}^c\}$ with an adjoint-based BFGS method. (b) The change in field strength from the initial to optimized configuration. Figure adapted from \cite{Paul2019} with permission.}
    \label{fig:opt}
\end{figure}

We perform optimization with a BFGS quasi-Newton method (Chapter 6 in \cite{Nocedal2006}) using an objective function $\chi^2 = J_b^2$, implemented in the \texttt{sfincs\_adjoint} branch of the STELLOPT code. A backtracking line search is used at each iteration to find a step size that satisfies a condition of sufficient decrease of $\chi^2$. We use the same equilibrium as in Section \ref{sec:local_sensitivity}, retaining modes $m \leq 12$ and $|n| \leq 12$, and compute derivatives with respect to these modes. Convergence to $\chi^2 \leq 10^{-10}$ was obtained within 8 BFGS iterations (28 function evaluations), as shown in Figure \ref{fig:bfgs_convergence}. The difference in field strength between the initial and  optimized configuration, $B_{\text{opt}}-B_{\text{init}}$, is shown in Figure \ref{fig:B_opt}. As expected from the analysis in Section \ref{sec:local_sensitivity}, the field strength increased on the outboard side and decreased on the inboard side in comparison with $B_{\text{init}}$. (Note that $J_b <0$.)

\subsubsection{Optimization of MHD equilibria}
\label{sec:equilibria_opt}
The local sensitivity function, $S_{\mathcal{R}}$, along with $\partial \mathcal{R}/\partial I$, $\partial \mathcal{R}/\partial G$, and $\partial \mathcal{R}/\partial \iota$, can be used to determine how perturbations to the outer boundary of the plasma, $S_P$, result in perturbations to $\mathcal{R}$. This is quantified through the idea of the shape gradient, introduced in Section \ref{sec:shape_optimization}. The partial derivatives of $\mathcal{R}$ can be computed with the adjoint method outlined in Section \ref{sec:adjoint_approach}, and the shape gradient can be obtained with only one additional MHD equilibrium solution through the application of another adjoint method. 

Consider a figure of merit which is integrated over the toroidal confinement volume, $V_{P}$,
\begin{gather}
    f_{\mathcal{R}}(S_P) = \int_{V_{P}} d^3 x \, w(\psi) \mathcal{R}(\psi),
\end{gather}
where $w(\psi)$ is a weighting function. That is, SFINCS is run on a set of $\psi$ surfaces within $V_P$ and the volume integral is computed numerically. Here we consider $S_P$ to be the plasma boundary used for a fixed-boundary MHD equilibrium calculation. From the Hadamard-Zolesio structure theorem (Section \ref{sec:shape_optimization}), the perturbation to $f_{\mathcal{R}}$ resulting from normal perturbation to $S_{P}$ can be written in the following form,
\begin{gather}
    \delta f_{\mathcal{R}}(S_P;\delta \textbf{x}) = \int_{S_{P}} d^2 x \, \left( \delta \textbf{x} \cdot \hat{\textbf{n}} \right) \mathcal{G},
\end{gather}
under certain assumptions of smoothness \citep{Delfour2011}. This can be thought of as another instance of the Riesz representation theorem, as $\delta f_{\mathcal{R}}$ is a linear functional of $\delta \textbf{x}$. Here $\hat{\textbf{n}}$ is the outward unit normal on $S_P$ and $\delta \textbf{x}$ is a vector field describing the perturbation to the surface. Intuitively, only normal perturbations to $S_P$ result in a change to $f_{\mathcal{R}}$. The shape gradient is $\mathcal{G}$, which quantifies the contribution of a local normal perturbation of the boundary to the change in $f_{\mathcal{R}}$. The shape gradient can be used for fixed-boundary optimization of equilibria or analysis of sensitivity to perturbations of magnetic surfaces. It can be computed using a second adjoint method, where a perturbed MHD force balance equation is solved with the addition of a bulk force that depends on derivatives computed from the neoclassical adjoint method. This will be described in detail in Chapter \ref{ch:adjoint_MHD}. While the continuous neoclassical adjoint method described in this Chapter arises from the self-adjointness of the linearized Fokker-Planck operator, the adjoint method for MHD equilibria arises from the self-adjointness of the MHD force operator. In practice, these two adjoint methods could be coupled by first computing an MHD equilibrium solution, computing neoclassical transport and its geometric derivatives from this equilibrium with the neoclassical adjoint method, and passing these derivatives back to the equilibrium code to compute the shape gradient with the perturbed MHD adjoint method. In this way, derivatives of neoclassical quantities with respect to the shape of the outer boundary are computed with only two equilibrium solutions and two DKE solutions. 

Rather than solve an additional adjoint equation, the outer boundary could be optimized by numerically computing derivatives of $\{B_{m,n}^c(\psi),G(\psi),I(\psi)\}$ with respect to the double Fourier series describing the outer boundary shape in cylindrical coordinates, $\{R_{m,n}^c, Z_{m,n}^s\}$, using a finite-difference method. This could be done using the STELLOPT code \citep{Spong1998,Reiman1999} with BOOZ\_XFORM \citep{Sanchez2000} to perform the coordinate transformation. For example, if the rotational transform is held fixed in the VMEC equilibrium calculation \citep{Hirshman1983}, the derivative of a moment, $\mathcal{R}$, with respect to a boundary coefficient, $R_{m,n}^c$, can be computed as,
\begin{gather}
    \partder{\mathcal{R}(\psi)}{R_{m,n}^c(\psi)} = \sum_{m',n'}\partder{\mathcal{R}(\psi)}{B_{m',n'}^c(\psi)} \partder{B_{m',n'}^c(\psi)}{R_{m,n}^c(\psi)} + \partder{\mathcal{R}(\psi)}{G(\psi)}\partder{G(\psi)}{R_{m,n}^c(\psi)} + \partder{\mathcal{R}(\psi)}{I(\psi)}\partder{I(\psi)}{R_{m,n}^c(\psi)},
\end{gather}
where $\partial \mathcal{R}(\psi)/\partial B_{m,n}^c(\psi)$, $\partial \mathcal{R}(\psi)/\partial G(\psi)$, and $\partial \mathcal{R}(\psi)/\partial I(\psi)$ are computed with the neoclassical adjoint method and $\partial B_{m,n}^c(\psi)/\partial R_{m,n}^c(\psi)$, $\partial G(\psi)/\partial R_{m,n}^c(\psi)$, and \\
$\partial I(\psi)/\partial R_{m,n}^c(\psi)$ are computed with finite-difference derivatives using STELLOPT. Similarly, derivatives of $\{B_{m,n}^c(\psi),G(\psi),I(\psi)\}$ could be computed with respect to coil parameters using a free-boundary equilibrium solution, allowing for direct optimization of neoclassical quantities with respect to coil shapes. The neoclassical calculation with SFINCS is typically significantly more expensive than the equilibrium calculation (for the geometry discussed in Section \ref{sec:local_sensitivity} fixed-boundary VMEC took 54 seconds while SFINCS took 157 seconds on 4 processors of the NERSC Edison computer). As such, combining adjoint-based with finite-difference derivatives can still result in a significant computational savings. 

\subsection{Ambipolarity}
\label{sec:ambipolarity}

As stellarators are not intrinsically ambipolar, the radial electric field is not truly an independent parameter. The ambipolar $E_r$ must be obtained which satisfies the condition $J_r(E_r) = 0$. The application of adjoint-based derivatives for computing the ambipolar solution is discussed in Section \ref{sec:ambipolar_sol}. An adjoint method to compute derivatives with respect to geometric parameters at fixed ambipolarity is discussed in Section \ref{sec:deriv_ambipolarity}.

\subsubsection{Accelerating ambipolar solve}
\label{sec:ambipolar_sol}

 A nonlinear root-finding algorithm must be used to compute the ambipolar $E_r$. This root-finding can be accelerated with derivative information, such as with a Newton-Raphson method \citep{Press2007}. The derivative required, $\partial J_r/\partial E_r$, can be computed with the discrete or continuous adjoint method as described in Section \ref{sec:adjoint_approach} with the replacement $\Omega_i \rightarrow E_r$, considering $\mathcal{R} = J_r$.
 
We implement three nonlinear root finding methods: Brent's method \citep{Brent2013}, the Newton-Raphson method, and a hybrid between the bisection and Newton-Raphson methods \citep{Press2007}. Brent's method guarantees at least linear convergence by combining quadratic interpolation with bisection and does not require derivatives. The Newton-Raphson method can provide quadratic convergence under certain assumptions but in general is not guaranteed to converge. If an iterate lies near a stationary point or a poor initial guess is given, the method can fail. For this reason, we implement the hybrid method, which combines the possible quadratic convergence properties of Newton-Raphson with the guaranteed linear convergence of the bisection method. Both Brent's method and the hybrid method require the root to be bracketed and therefore may require additional function evaluations to obtain the bracket.

\begin{figure}
    \centering
    \includegraphics[trim=1cm 6cm 2cm 7.5cm,clip,width=0.49\textwidth]{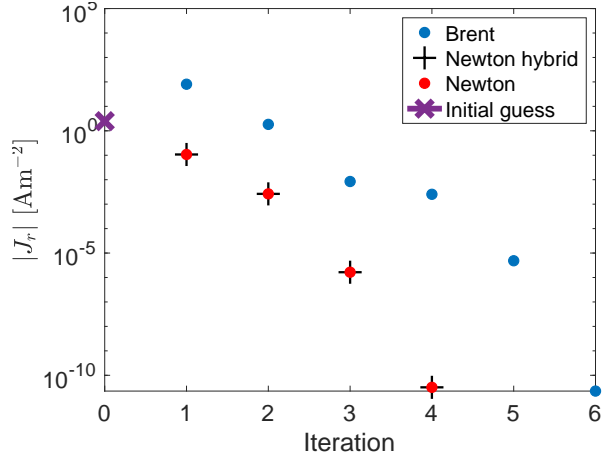}
    \caption{The ambipolar root is  obtained with Brent, Newton-Raphson, and Newton hybrid nonlinear root solvers. The derivatives obtained with the adjoint method provide better convergence properties for the Newton methods. Figure adapted from \cite{Paul2019} with permission.}
    \label{fig:root_finding}
\end{figure}

We compare these methods in Figure \ref{fig:root_finding}, using the W7-X standard configuration considered in Section \ref{sec:local_sensitivity} with the full trajectory model and the discrete adjoint approach, beginning with an initial guess of $E_r = -10$ kV/m with bounds at $E_r^{\min} = -100$ kV/m and $E_r^{\max} =$ 100 kV/m. The root is located at $E_r =-3.84$ kV/m. For this example, the hybrid and Newton methods had nearly identical convergence properties. However, the Newton method is less expensive as it does not require $J_r$ to be evaluated at the bounds of the interval. The Newton method provides a 22\% savings in wall clock time over Brent's method to obtain the root within the same tolerance. 

In the above discussion, we have assumed that there is only one stable root of interest. Of course, a given configuration may possess several roots, especially if the ions and electrons are in different collisionality regimes \citep{Hastings1985}. Multiple roots can be obtained by performing several root solves with different initial values and brackets, which could be trivially parallelized. Thus the adjoint method could still provide an acceleration in this more general case.

\subsubsection{Derivatives at ambipolarity}
\label{sec:deriv_ambipolarity}

The adjoint method described in Section \ref{sec:adjoint_approach} assumes that $E_r$ is held constant when computing derivatives with respect to $\Omega$. However, $E_r$ cannot truly be determined independently from geometric quantities, as the ambipolar solution should be recomputed as the geometry is altered. It is therefore desirable to compute derivatives at fixed ambipolarity (fixed $J_r = 0$) rather than at fixed $E_r$. This is performed by solving an additional adjoint equation, 
\begin{gather}
    \mathbb{L}^{\dagger} \lambda^{J_r} + \widetilde{J_r} = 0,
    \label{eq:J_r_adjoint}
\end{gather}
in the continuous approach or,
\begin{gather}
    \left( \overleftrightarrow{\textbf{L}} \right)^T \overrightarrow{\bm{\lambda}}^{J_r} = \overrightarrow{\textbf{J}_r},
    \label{eq:J_r_adjoint_discrete}
\end{gather}
in the discrete approach. Details are described in Appendix \ref{app:ambipolar}. 

It should be noted that by computing derivatives at ambipolarity, we assume that a given moment $\mathcal{R}$ is a differentiable function of the geometry at fixed $J_r = 0$. That is, this method cannot be applied to cases in which a stable root disappears as the geometry varies. As this will occur at a stationary point of $J_r(E_r)$, this situation could be avoided within an optimization loop by computing derivatives at constant $E_r$ rather than constant $J_r$ if $|\partial J_r/\partial E_r|$ falls below a given threshold at ambipolarity.

\begin{figure}
    \centering
    \begin{subfigure}[b]{0.49\textwidth}
    \includegraphics[trim=0.7cm 5.8cm 2cm 5.0cm,clip,width=1.0\textwidth]{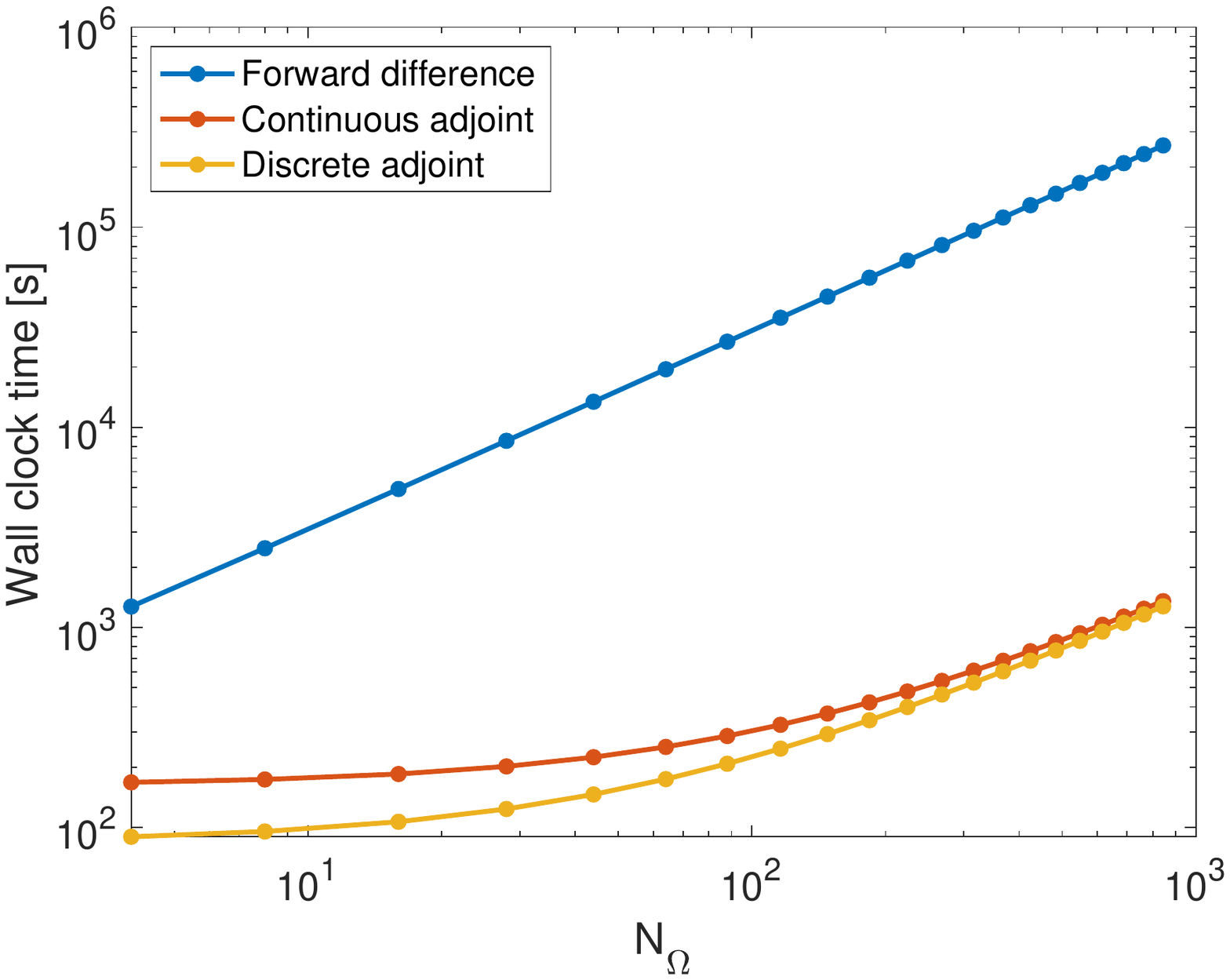}
    \caption{}
    \label{fig:cost_adjoint}
    \end{subfigure}
    \begin{subfigure}[b]{0.49\textwidth}
    \includegraphics[trim=1cm 6cm 2cm 7cm,clip,width=1.0\textwidth]{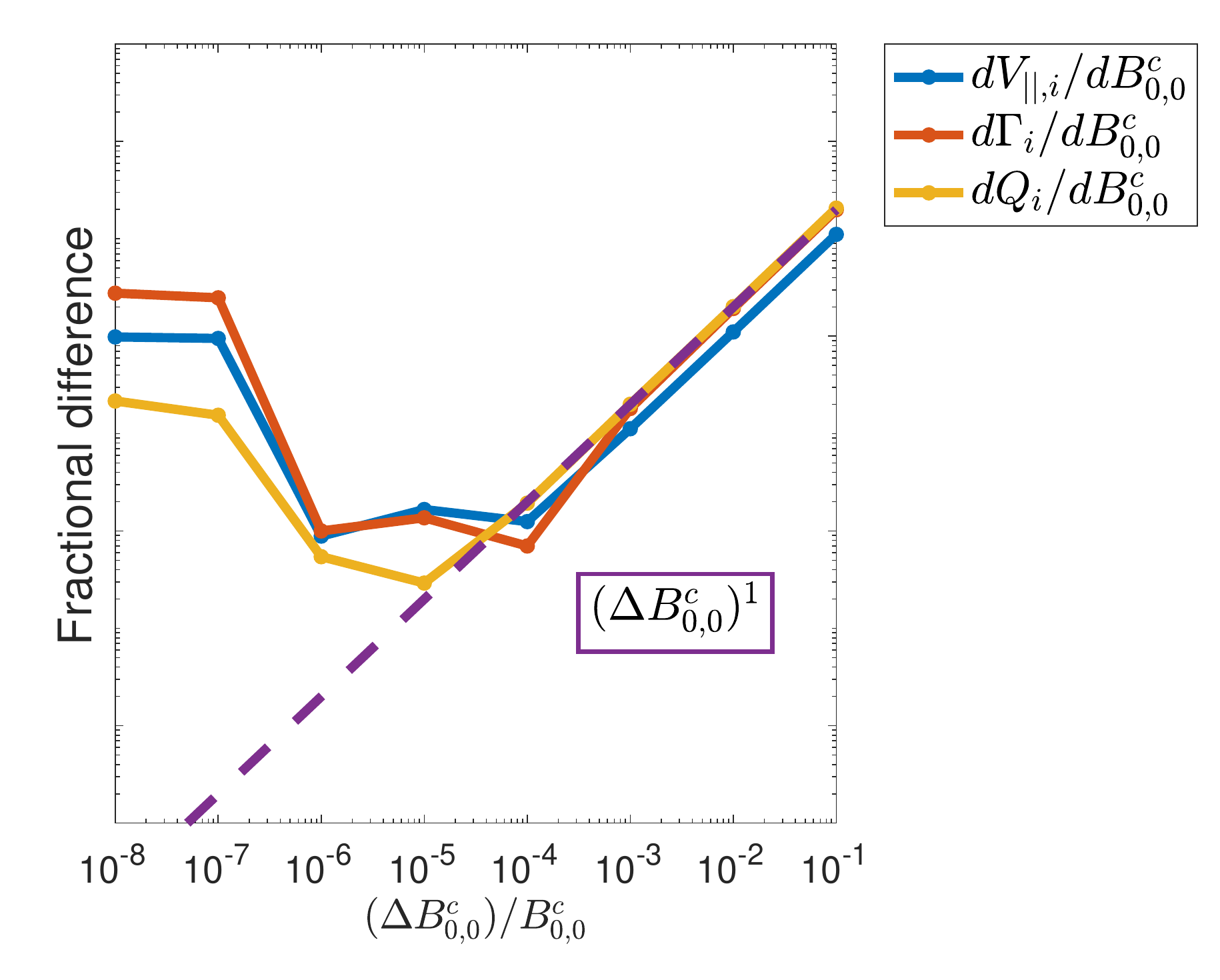}
    \caption{}
    \label{fig:ambipolar_benchmark}
    \end{subfigure}
    \caption{(a) The cost of computing the gradient $\partial \mathcal{R}/\partial \Omega$ at ambipolarity scales with $N_{\Omega}$, the number of parameters in $\Omega$. (b) The fractional difference between $\partial \mathcal{R}/\partial B_{0,0}^c$ at constant ambipolarity obtained with the adjoint method and with finite-difference derivatives. Figure adapted from \cite{Paul2019} with permission.}
\end{figure}

Although an additional adjoint solve is required, this method of computing derivatives at ambipolarity is advantageous as several linear solves are typically needed to obtain the ambipolar root. A comparison of the computational cost between the adjoint method and the forward-difference method for derivatives at ambipolarity is shown in Figure \ref{fig:cost_adjoint}. Here the full trajectory model is used, and the results for both the discrete and continuous adjoint methods are shown. For the finite-difference derivative, the ambipolar solve is performed with Brent's method at each step in $\Omega$. As in Figure \ref{fig:computational_time}, we find that for large $N_{\Omega}$, the cost of the continuous and discrete approaches are essentially the same, as the cost is no longer dominated by the linear solve. When computing the derivatives at ambipolarity, both adjoint methods decrease the cost by a factor of approximately $200$ for large $N_{\Omega}$. 

In Figure \ref{fig:ambipolar_benchmark} we show a benchmark between derivatives at ambipolarity, \\
$(\partial \mathcal{R}/\partial B_{0,0}^c)_{J_r}$, computed with the discrete adjoint method and with forward-difference derivatives. For the forward-difference method, the Newton solver is used to obtain the ambipolar $E_r$ as $B_{0,0}^c$ is varied. As the forward difference step size $\Delta B_{0,0}^c$ decreases, the fractional difference again decreases proportional to $\Delta B_{0,0}^c$ until it reaches a minimum when $\Delta B_{0,0}^c/B_{0,0}^c$ is approximately $10^{-4}$. In comparison with Figure \ref{fig:benchmark_fixedEr}, we see that the minimum fractional difference is slightly larger at fixed ambipolarity than at fixed $E_r$, as the tolerance parameters associated with the Newton solver introduce an additional source of error to the forward-difference approach.

\begin{figure}
    \centering
    \begin{subfigure}[b]{0.435\textwidth}
    \includegraphics[trim=1cm 6cm 5cm 6cm,clip,width=1.0\textwidth]{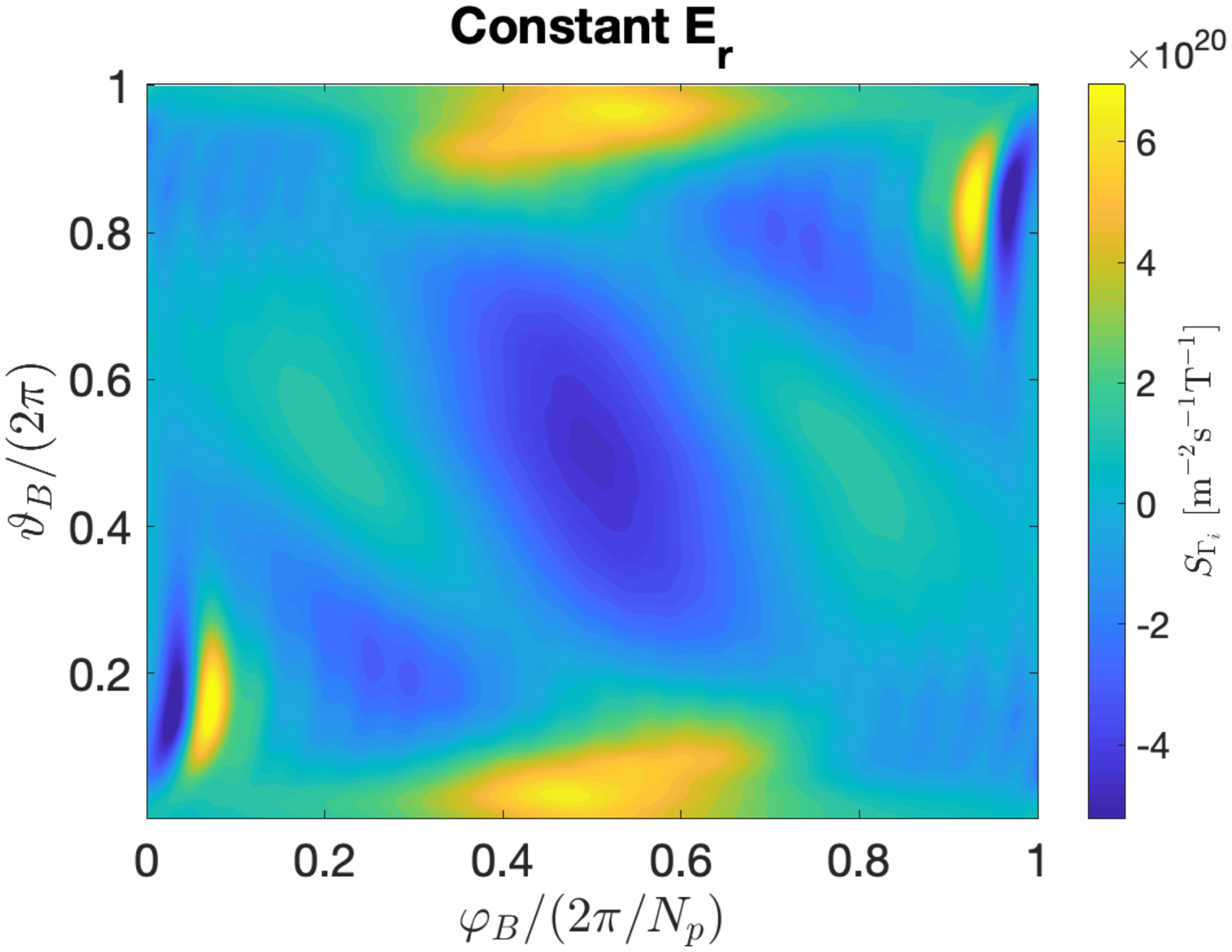}
    \caption{}
    \label{fig:S_const_Er_particle}
    \end{subfigure}
    \begin{subfigure}[b]{0.49\textwidth}
    \includegraphics[trim=3cm 6cm 1cm 6cm,clip,width=1.0\textwidth]{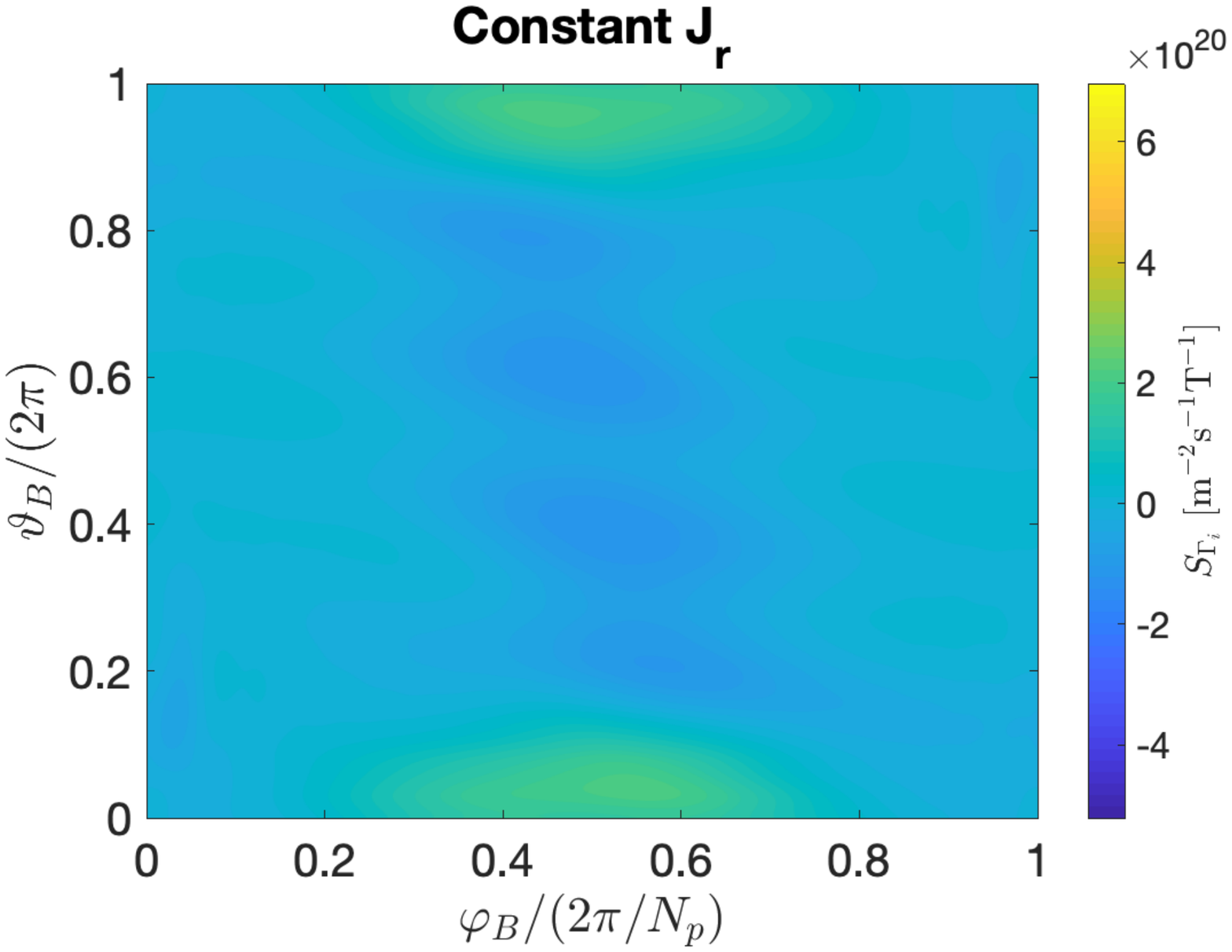}
    \caption{}
    \label{fig:S_const_Jr_particle}
    \end{subfigure}
    \begin{subfigure}[b]{0.435\textwidth}
    \includegraphics[trim=1cm 6cm 5cm 6cm,clip,width=1.0\textwidth]{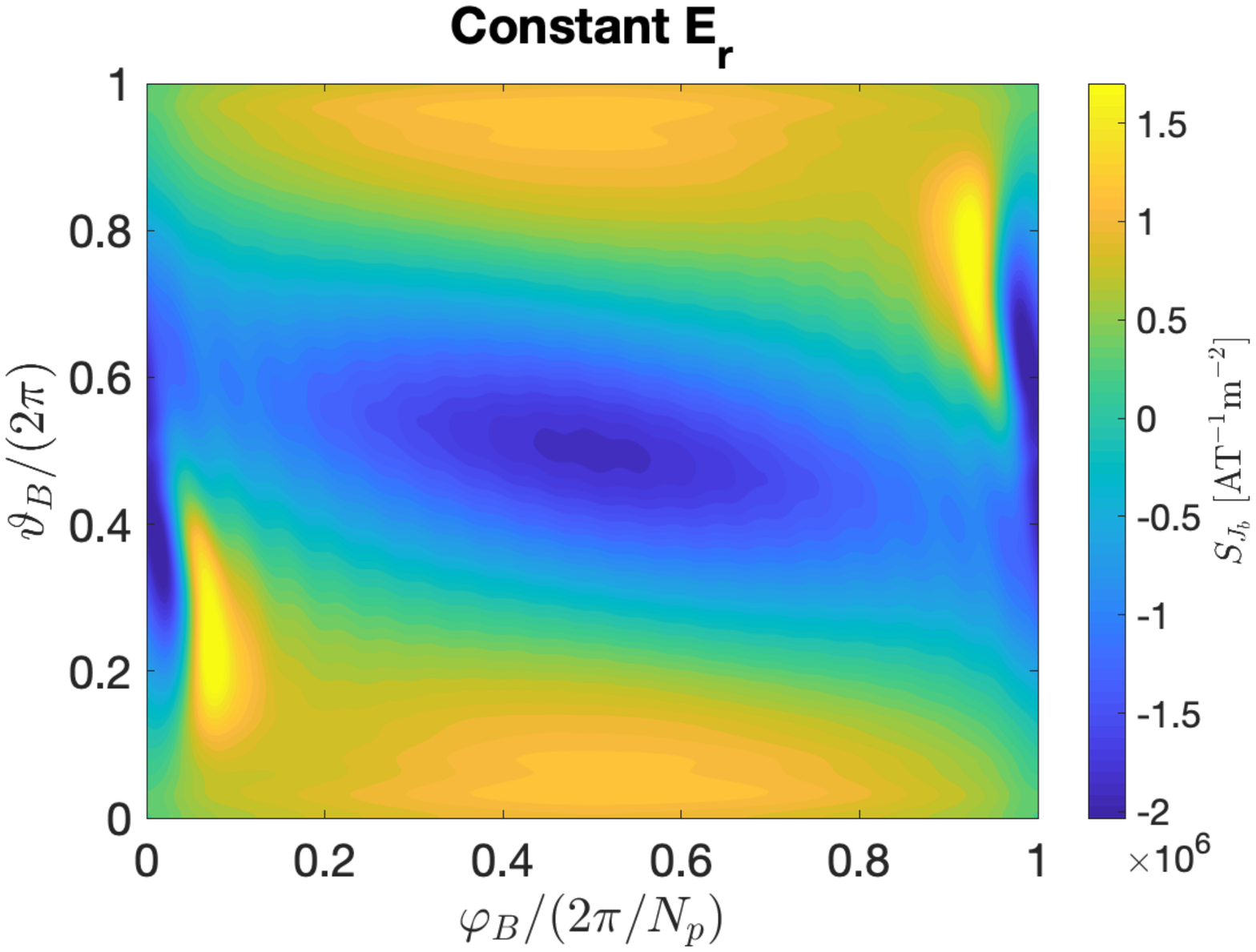}
    \caption{}
    \label{fig:S_const_Er_bootstrap}
    \end{subfigure}
    \begin{subfigure}[b]{0.49\textwidth}
    \includegraphics[trim=3cm 6cm 1cm 6cm,clip,width=1.0\textwidth]{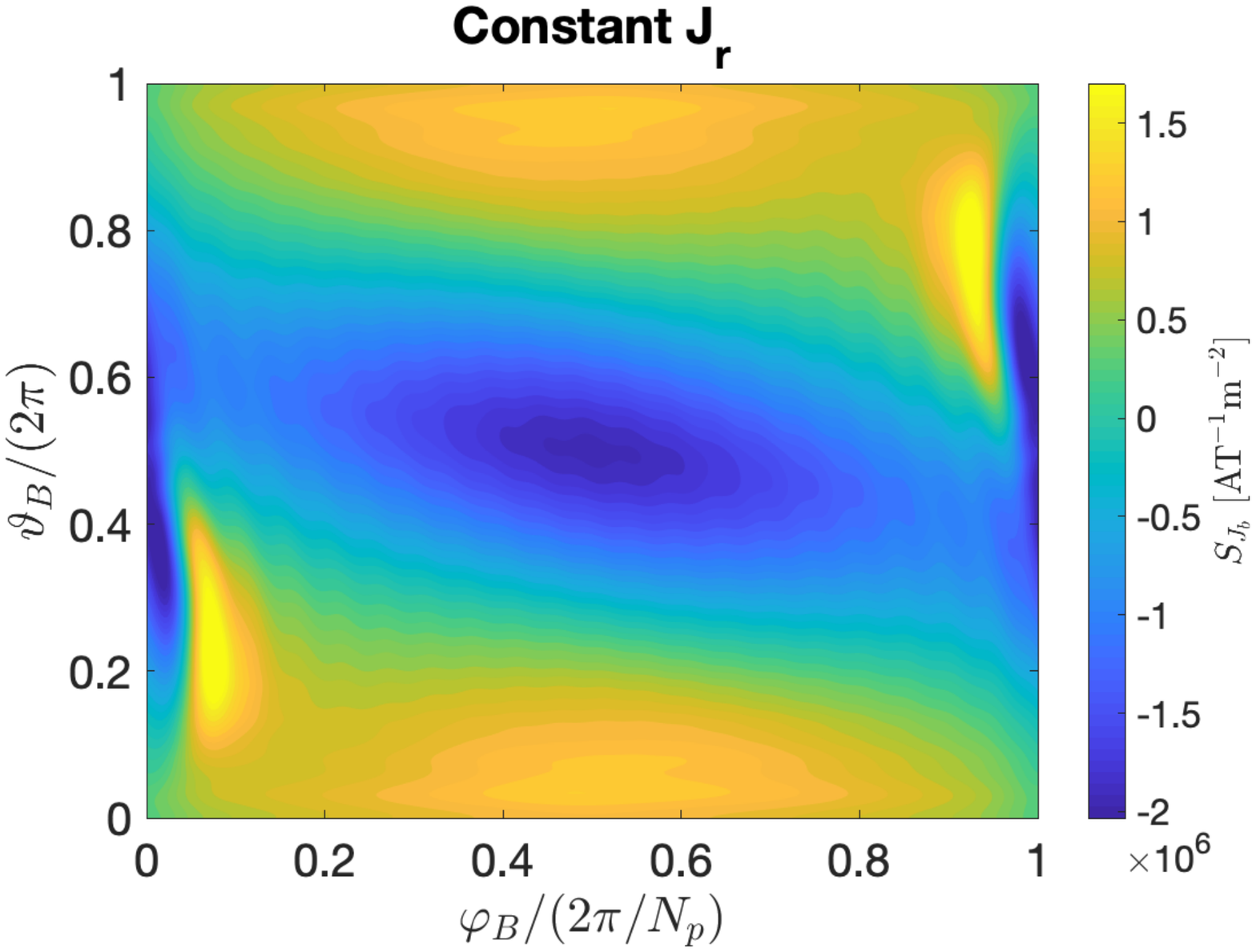}
    \caption{}
    \label{fig:S_const_Jr_bootstrap}
    \end{subfigure}
    \caption{The sensitivity function for the ion particle flux, $S_{\Gamma_i}$, is computed at (a) constant $E_r$ and (b) constant $J_r$. Similarly, $S_{J_b}$ is computed at (c) constant $E_r$ and (d) constant $J_r$. Figure adapted from \cite{Paul2019} with permission.}
\end{figure}

In Figures \ref{fig:S_const_Er_particle} and \ref{fig:S_const_Jr_particle} we compare the sensitivity function for the particle flux, $S_{\Gamma_i}$, computed using derivatives at constant $E_r$ with that computed at constant $J_r$. Here derivatives are computed using the discrete adjoint method with full trajectories, and the sensitivity function is constructed as described in Section \ref{sec:local_sensitivity}. The configuration and numerical parameters are the same as described in Section \ref{sec:local_sensitivity}. At constant $J_r$, the large region of increased sensitivity on the outboard side that appears at constant $E_r$ remains, though the overall magnitude of the sensitivity decreases. Thus it may be important to account for the effect of the ambipolar $E_r$ when optimizing for radial transport. In Figures \ref{fig:S_const_Er_bootstrap} and \ref{fig:S_const_Jr_bootstrap} we perform the same comparison for $S_{J_b}$, finding the derivatives at fixed $E_r$ and at fixed $J_r$ to be virtually identical. This is to be expected, as numerical calculations of neoclassical transport coefficients for W7-X have found that the bootstrap coefficients are much less sensitive to $E_r$ than those for the radial transport (Figures 18 and 26 in \cite{Beidler2011}). Furthermore, the bootstrap current in the $1/\nu$ regime is independent of $E_r$, and the finite-collisionality correction is small for optimized stellarators, such as W7-X \citep{Helander2017}. Therefore, the ambipolarity corrections to the derivatives are less important for $J_b$ than for the radial transport.

\section{Conclusions}
We have described a method by which moments $\mathcal{R}$ of the neoclassical distribution function can be differentiated efficiently with respect to many parameters. The adjoint approach requires defining an inner product from which the adjoint operator is obtained. We consider two choices for this inner product. One choice corresponds with computing the adjoint of the linear operator after discretization, and the other corresponds with computing it before discretization. In the case of the former, the Euclidean dot product can be used, and in the case of the latter, an inner product whose corresponding norm is similar to the free energy norm \eqref{eq:inner_product} is defined. In Section \ref{sec:implementation}, we show that these approaches provide the same derivative information within discretization error, as expected. Both methods provide a reduction in computational cost by a factor of approximately $50$ in comparison with forward-difference derivatives when differentiating with respect to many ($\mathcal{O}(10^2)$) parameters. In Section \ref{sec:deriv_ambipolarity} the adjoint method is extended to compute derivatives at ambipolarity. This method provides a reduction in cost by a factor of approximately $200$ over a forward-difference approach. We have implemented this method in the SFINCS code, 
and similar methods
could be applied to other drift kinetic solvers.

In this Chapter, we consider derivatives with respect to geometric quantities that enter the DKE through Boozer coordinates. However, the adjoint neoclassical method we have described is much more general, allowing for many possible applications. For example, derivatives of the radial fluxes with respect to the temperature and density profiles could be used to accelerate the solution of the transport equations using a Newton method \citep{Barnes2010}. The transport solution could furthermore be incorporated into the optimization loop to self-consistently evolve the macroscopic profiles in the presence of neoclassical fluxes. Rather than simply optimizing for minimal fluxes, an objective function such as the total fusion power could be considered \citep{Highcock2018}, with optimization accelerated by adjoint-based derivatives.

Another application of the continuous adjoint formulation is the correction of discretization error. The same solution obtained in Section \ref{sec:continuous} can be used to quantify and correct for the error in a moment, $\mathcal{R}$, providing similar accuracy to that computed with a higher-order stencil or finer mesh without the associated cost. This method has been applied in the field of computational fluid dynamics by solving adjoint Euler equations \citep{Venditti1999,Pierce2004} and could prove useful for efficiently obtaining solutions of the DKE in low-collisionality regimes.

In Section \ref{sec:vacuum_opt}, we have shown an example of adjoint-based neoclassical optimization, where the optimization space is taken to be the Fourier modes of the field strength on a surface, $\{B_{m,n}^c\}$. While optimization within this space is not necessarily consistent with a global equilibrium solution, it demonstrates the adjoint neoclassical method for efficient optimization. In Section \ref{sec:equilibria_opt}, two approaches to self-consistently optimize MHD equilibria are discussed. Further discussion and demonstration will be provided in Chapter \ref{ch:adjoint_MHD}.

In Appendix \ref{app:symmetry} we show that when $E_r = 0$ and the unperturbed geometry is stellarator symmetric, the sensitivity functions for moments of the distribution function are also stellarator symmetric. However, when $E_r \neq 0$ this is no longer true. This implies that obtaining minimal neoclassical transport in the $\sqrt{\nu}$ regime may require breaking of stellarator symmetry. In this Chapter, we have ignored the effects of stellarator symmetry-breaking, though we hope to extend this work to study these effects in the future.

\renewcommand{\thechapter}{5}

\chapter{Adjoint shape gradient for MHD equilibria}
\label{ch:adjoint_MHD}

Most stellarator optimization to date has assumed that the magnetic field satisfies the MHD equilibrium equations with either a fixed or free-boundary approach, as detailed in Section \ref{sec:equilibrium_optimization}. If a gradient-based optimization approach is applied, derivatives of quantities that depend on the equilibrium solutions must be computed with respect to the shapes of the filamentary coils or plasma boundary. In this Chapter, we demonstrate an adjoint approach for obtaining the coil or surface shape gradient of such functions. With the shape gradient efficiently computed, shape derivatives with respect to \textit{any} shape perturbation can be calculated.

The material in this Chapter has been adapted with permission from \cite{Antonsen2019} and \cite{Paul2020}.

\section{Introduction}

Several figures of merit quantifying confinement must be considered in the numerical optimization of stellarator MHD equilibrium. These figures of merit describing a configuration depend on the shape of the outer plasma boundary or the shape of the electro-magnetic coils. It is thus desirable to obtain derivatives with respect to these shapes for optimization of equilibria or identification of sensitivity information. These so-called shape derivatives can be computed by directly perturbing the shape, recomputing the equilibrium, and computing the resulting change to a figure of merit that depends on the equilibrium solution. However, this direct finite-difference approach requires recomputing the equilibrium for each possible perturbation of the shape. For stellarators whose geometry is described by a set of $N_{\Omega}\sim 10^2$ parameters, this requires $N_{\Omega}$ solutions to the MHD equilibrium equations. Despite this computational complexity, gradient-based optimization of stellarators has proceeded with the direct approach (e.g. \cite{Reiman1999,Ku2008,Proll2015}). 

As the target optimized configuration can never be realized exactly, an analysis of the sensitivity to perturbations, such as errors in coil fabrication or assembly, is central to the success of a stellarator. Tight tolerances have proven to be a significant driver of the cost of stellarator experiments \citep{Strykowsky2009,Klinger2013}; thus an improvement to the algorithms used to conduct sensitivity studies can have a substantial impact on the field. In studies of the coil tolerances for flux surface quality of LHD \citep{Yamazaki} and NCSX \citep{Brooks2003,Williamson2005}, perturbations of several distributions were manually applied to each coil. Sensitivity analysis can also be performed with analytic derivatives. Numerical derivatives with respect to tilt angle and coil translation of the CNT coils have been used to compute the sensitivity of the rotational transform on axis \citep{Hammond2016}. Analytic derivatives have recently been applied to study coil sensitivities of the CNT stellarator by considering the eigenvectors of the Hessian matrix \citep{Zhu2018}. Thus, in addition to gradient-based optimization, derivatives with respect to shape can be applied to sensitivity analysis. 

The shape gradient quantifies the change in a figure of merit associated with a local perturbation to a shape. Thus, if the shape gradient can be obtained, the shape derivative with respect to \textit{any} perturbation is known (more precise definitions of the shape derivative and gradient are given in Sections \ref{sec:shape_optimization} and \ref{sec:shape_calculus}). The shape gradient representation can be computed from parameter derivatives by solving a small linear system (Sections \ref{sec:parameter_derivatives}). However, computing parameter derivatives can often be computationally expensive, as numerical derivatives require evaluating the objective function at least $N_{\Omega}+1$ times if one-sided finite-difference derivatives are used, or $2N_{\Omega}$ times for centered differences. As computing the objective function often involves solving a linear or nonlinear system, such as the MHD equilibrium equations, this implies solving the system of equations $\ge N_{\Omega}+1$ times. Numerical derivatives also introduce additional noise, and the finite-difference step size must be chosen carefully. 

Rather than use parameter derivatives, in this Chapter we will use an adjoint method to compute the shape gradient. This is sometimes termed adjoint shape sensitivity or adjoint shape optimization, which has its origins in aerodynamic engineering and computational fluid dynamics \citep{Pironneau1974,Glowinski1975}. As with adjoint methods for parameter derivatives, this technique only requires the solution of two linear or nonlinear systems of equations. This technique has been applied to magnetic confinement fusion for the design of tokamak divertor shapes by solving forward and adjoint fluid equations \citep{Dekeyser2012,Dekeyser2014a,Dekeyser2014b}. As stellarators require many parameters to describe their shape, adjoint shape sensitivity could significantly decrease the cost of computing the shape gradient. If one is optimizing in the space of parameters describing the boundary of the plasma or the shape of coils, the shape gradient representation obtained from the adjoint method can be converted to parameter derivatives upon multiplication with a small matrix (Section \ref{sec:shape_optimization}). 

We begin in Section \ref{sec:shape_calculus} with a brief review of shape calculus concepts in the context of MHD equilibria. In Section \ref{sec:adjoint_relation}, the fundamental adjoint relations for perturbations to MHD equilibria are derived and discussed. These relations take a form that is similar to that of transport coefficients that are related by Onsager symmetry \citep{Onsager1931a,Onsager1931b}. Specifically, perturbations to the equilibrium are characterized as a set of generalized responses to a complementary set of generalized forces. The responses and forces can be thought of as being related by a  matrix operator, which is symmetric.  The resulting relations among forces and responses can be used to compute the shape gradient of functions of the equilibria with respect to displacements of the plasma boundary or the coil shapes. In Section \ref{sec:continuous_adjoint_method}, the continuous adjoint method that takes advantage of the generalized self-adjointness relations is discussed. Several applications to stellarator figures of merit will be demonstrated in Section \ref{sec:applications}. 

Although the adjoint relations are based on the equations of linearized MHD, we perform numerical calculations in this Chapter with nonlinear MHD solutions with the addition of a small perturbation. Demonstration is performed using nonlinear stellarator MHD equilibrium codes based on a variational principle, VMEC \cite{Hirshman1983} and ANIMEC \citep{Cooper19923d}. We obtain expressions for the shape gradients of the volume-averaged $\beta$ (Section \ref{sec:surf_Beta}), rotational transform (Section \ref{sec:iota}), vacuum magnetic well (Section \ref{sec:vacuum_well}), magnetic ripple (Section \ref{sec:ripple}), effective ripple in the $1/\nu$ neoclassical regime \citep{Nemov1999} where $\nu$ is the collision frequency (Section \ref{sec:epsilon_eff}), and departure from quasi-symmetry (Section \ref{sec:quasisymmetry}). Finally, we demonstrate that the adjoint method for neoclassical optimization outlined in Chapter \ref{ch:adjoint_neoclassical} can be coupled with a linearized adjoint MHD solution to compute derivatives of several neoclassical quantities with respect to the shape of the plasma boundary (Section \ref{sec:neoclassical}). We present calculations of the shape gradient with the adjoint approach for the volume-averaged $\beta$, rotational transform, and vacuum magnetic well figures of merit, which do not require modification to VMEC. The calculation for the magnetic ripple is computed with a minor modification of the ANIMEC code. The adjoint force balance equations needed to compute the shape gradient for the other figures of merit require the addition of a bulk force that will necessitate further modification of an equilibrium or linearized MHD code. Numerical calculations for these figures of merit will, therefore, not be presented in this Chapter.

\section{Shape calculus review}
\label{sec:shape_calculus}

We now review shape calculus fundamentals introduced in Chapter \ref{ch:mathematical_fundamentals} in the context of functions that depend on MHD equilibrium quantities. Consider a functional, $F(S_P)$, that depends implicitly on the plasma boundary, $S_P$, through the solution to the fixed-boundary MHD equilibrium equations (Section \ref{sec:equilibrium_calculation}) with boundary condition $\textbf{B} \cdot \hat{\textbf{n}}|_{S_P} = 0$ where $\hat{\textbf{n}}$ is the outward unit normal on $S_P$. We define a functional integrated over the plasma volume, $V_P$,
\begin{gather}
    f(S_P) = \int_{V_P} d^3 x \, F(S_P),
    \label{eq:f}
\end{gather}
where $S_P$ is the boundary of $V_P$. Consider a vector field describing displacements of the surface, $\delta \textbf{x}$, and a displaced surface $S_{P,\epsilon} = \{ \textbf{x}_0 + \epsilon \delta \textbf{x} : \textbf{x}_0 \in S_P\}$. The shape derivative of $F$ is defined as,
\begin{gather}
    \delta F(S_P;\delta \textbf{x}) = \lim_{\epsilon \rightarrow 0} \frac{F( S_{P,\epsilon}) - F(S_P)}{\epsilon}.
\end{gather}
The shape derivative of $f$ is defined by the same expression with $F\to f$. Under certain assumptions of smoothness of $\delta F$ with respect to $\delta \textbf{x}$, the shape derivative of the volume-integrated quantity, $f$, can be written in the following way (Section \ref{sec:shape_optimization}),
\begin{gather}
    \delta f(S_P;\delta \textbf{x}) = \int_{V_P} d^3 x \, \delta F(S_P;\delta \textbf{x}) + \int_{S_P} d^2 x \, \delta \textbf{x} \cdot \hat{\textbf{n}} F.
    \label{eq:transport_theorem}
\end{gather}
The first term accounts for the Eulerian perturbation to $F$ while the second accounts for the motion of the boundary. This is referred to as the transport theorem for domain functionals and will be used throughout this Chapter to compute the shape derivatives of figures of merit of interest.

According to the Hadamard-Zolesio structure theorem \citep{Delfour2011}, the shape derivative of a functional of $S_P$ (not restricted to the form of \eqref{eq:f}) can be written in the following form,
\begin{gather}
    \delta f(S_P;\delta \textbf{x}) = \int_{S_P} d^2 x \, \delta \textbf{x} \cdot \hat{\textbf{n}} \mathcal{G},
    \label{eq:shape_gradient_ch5}
\end{gather}
assuming $\delta f$ exists for all $\delta \textbf{x}$ and is sufficiently smooth. 
In the above expression, $\mathcal{G}$ is the shape gradient. This is an instance of the Riesz representation theorem, which states that any linear functional can be expressed as an inner product with an element of the appropriate space \citep{Rudin2006}. As the shape derivative of $f$ is linear in $\delta \textbf{x}$, it can be written in the form of \eqref{eq:shape_gradient_ch5}. Intuitively, the shape derivative does not depend on tangential perturbations to the surface. The shape gradient can be computed from derivatives with respect to the set of parameters, $\Omega$, used to discretize $S_P$,
\begin{gather}
    \partder{f(\Omega)}{\Omega_i} = \int_{S_P} d^2 x \, \partder{\textbf{x}(\Omega)}{\Omega_i} \cdot \hat{\textbf{n}} \mathcal{G}.
    \label{eq:shape_gradient_system}
\end{gather}
For example, $\Omega = \{ R_{m,n}^c, Z_{m,n}^s \}$ could be assumed, where these are the Fourier coefficients \eqref{eq:rmnc_zmns_ch5} in a cosine and sine representation of the cylindrical coordinates $(R,Z)$ of $S_P$.
Upon discretization of the right-hand side on a surface, the above takes the form of a linear system that can be solved for $\mathcal{G}$ \citep{Landreman2018}. However, this approach requires performing at least one additional equilibrium calculation for each parameter with a finite-difference approach.

The shape gradient can also be computed with respect to perturbations of currents in the vacuum region. We now consider $f$ to depend on the shape of a set of filamentary coils, $C = \{ C_k \}$, through a free-boundary solution to the MHD equilibrium equations (Section \ref{sec:equilibrium_calculation}). We consider a vector field of displacements to the coils, $\delta \textbf{x}_{C}$. The shape derivative of $f$ can also be written in shape gradient form,
\begin{gather}
    \delta f(C;\delta \textbf{x}_{C}) = \sum_k \oint_{C_k} dl \,  \delta \textbf{x}_{C_k} \cdot  \widetilde{\bm{\mathcal{G}}}_k ,
    \label{eq:coil_shape_gradient_ch5}
\end{gather}
where $\widetilde{\bm{\mathcal{G}}}_k$ is the shape gradient for coil $k$, $C_k$ is the line integral along coil $k$, and the sum is taken over coils. Again, $\widetilde{\bm{\mathcal{G}}}_k$ can be computed from derivatives with respect to a set of a parameters describing coil shapes \eqref{eq:coil_discretization}, analogous to \eqref{eq:shape_gradient_system}. Note that we have defined the shape gradient in a slightly different way here than that introduced in Chapter 2 \eqref{eq:structure_theorem_coil} (without the cross with $\hat{\textbf{t}}$), although we will find in this Chapter that $\widetilde{\bm{\mathcal{G}}}_k$ is perpendicular to $\hat{\textbf{t}}$ for the functionals under consideration. We distinguish the shape gradient as defined in \eqref{eq:coil_shape_gradient_ch5} from that defined in \eqref{eq:structure_theorem_coil} with a tilde.

To avoid the cost of direct computation of the shape gradient, we apply an adjoint approach. The shape gradient is thus obtained without perturbing the plasma surface or coil shapes directly, but instead by solving an additional adjoint equation that depends on the figure of merit of interest. We perform the calculation with the direct approach to demonstrate that the same derivative information is computed with either method. 

\section{Adjoint relations for MHD equilibria}
\label{sec:adjoint_relation}

The goal of this Section is to generalize the well-known self-adjointness \cite{Bernstein1958} of the MHD force operator,
\begin{align}
    \int_{V_P} d^3 x \, \left(\bm{\xi}_1 \cdot \textbf{F}[\bm{\xi}_2]  - \bm{\xi}_2 \cdot \textbf{F}[\bm{\xi}_1] \right) - \frac{1}{\mu_0} \int_{S_P} d^2 x \, \hat{\textbf{n}} \cdot \left( \bm{\xi}_2 \delta \textbf{B}[\bm{\xi}_1] \cdot \textbf{B} - \bm{\xi}_1 \delta \textbf{B}[\bm{\xi}_2] \cdot \textbf{B} \right) = 0,
\label{eq:self_adjointness_ch5}
\end{align}
to allow for perturbations of interest for stellarator optimization. In this expression, the perturbed magnetic field is expressed in terms of the displacement vector,
\begin{align}
    \delta \textbf{B}[\bm{\xi}_{1,2}] = \nabla \times \left(\bm{\xi}_{1,2} \times \textbf{B} \right),
    \label{eq:delta_B_mhd}
\end{align}
which follows from the assumption that the rotational transform is fixed by the perturbation (flux-freezing). The MHD force operator,
\begin{align}
     \textbf{F}[\bm{\xi}_{1,2}] = \frac{\left(\nabla \times \delta \textbf{B}[\bm{\xi}_{1,2}] \right) \times \textbf{B}}{\mu_0} + \frac{\left(\nabla \times \textbf{B} \right) \times \delta \textbf{B}[\bm{\xi}_{1,2}]}{\mu_0} - \nabla \left(\delta p[\bm{\xi}_{1,2}] \right),
     \label{eq:force_operator_ch5}
\end{align}
is a linearization of the MHD equilibrium equation,
\begin{align}
   \frac{ \left( \nabla \times \textbf{B} \right) \times \textbf{B}}{\mu_0} = \nabla p,
   \label{eq:force_balance}
\end{align}
with boundary condition,
\begin{align}
\textbf{B} \cdot \hat{\textbf{n}}\rvert_{S_P} = 0,
\label{eq:mhd_bc}
\end{align}
under the assumption that the magnetic field is perturbed according to \eqref{eq:delta_B_mhd} and the pressure is perturbed according to,
\begin{align}
    \delta p[\bm{\xi}_{1,2}] = - \bm{\xi}_{1,2} \cdot \nabla p - \gamma p \nabla \cdot \bm{\xi}_{1,2},
\end{align}
where $\gamma$ is the adiabatic index. As $\bm{\xi}$ describes the motion of field lines, modes which perturb the plasma boundary exhibit non-zero $\bm{\xi} \cdot \hat{\textbf{n}} \rvert_{S_P}$. The self-adjointness provides a relationship between two perturbations about an MHD equilibrium state described by \eqref{eq:force_balance}-\eqref{eq:mhd_bc}. This relation is incredibly valuable for ideal MHD stability analysis, forming the basis for the energy principle.

As described in Section \ref{sec:lagrangian}, when formulating a continuous adjoint approach, the adjoint of the linearized operator appearing in the forward PDE must be obtained. However, we cannot directly apply the self-adjointness relation from MHD stability theory \eqref{eq:self_adjointness_ch5} for the stellarator optimization problem. While MHD perturbations assume fixed rotational transform, stellarator optimization is often performed instead at fixed toroidal current. While the MHD self-adjointness relation allows for perturbations of the plasma boundary, we would also like to consider linearized equilibrium states corresponding to perturbations of coils in the vacuum region. We now form the appropriate generalized self-adjointness relations corresponding to fixed-boundary perturbations (applied perturbations to the plasma boundary) and free-boundary perturbations (applied perturbations to electro-magnetic coils). Even though the boundary shape changes in the former case, we refer to it as ``fixed boundary" since the equilibrium code is run in fixed-boundary mode, and since the associated adjoint problem will turn out to have no boundary perturbation. 

The resulting expressions will allow us to relate the ``direct perturbations," those corresponding to a linearized equilibrium state associated with the direct perturbation of the plasma boundary or coil shapes, and ``adjoint perturbations," with which we can compute the shape gradient efficiently. The adjoint perturbation will correspond to the change in the equilibrium when an additional bulk force acts on the plasma or the toroidal current profile is changed. For the adjoint perturbation, there is no change to the outer flux surface in the fixed-boundary case or to the coil currents in the free-boundary case. In this Section, we will show that aspects of the direct and adjoint changes are related to each other in a manner similar to Onsager symmetry. Thus, it will be shown that by calculating the adjoint perturbation, with a judiciously chosen added force or change in the toroidal current profile, the solution to the direct problem can be determined.

We consider equilibria in which the magnetic field in the plasma can be expressed in terms of scalar functions $\psi(\textbf{x}),\chi(\psi),\vartheta(\textbf{x})$, and $\varphi(\textbf{x})$,
\begin{gather}
\textbf{B} = \nabla \psi \times \nabla \vartheta - \nabla \chi \times \nabla \varphi = \nabla \psi \times \nabla \alpha,
\label{eq:magnetic_contravariant}
\end{gather}
where ($\psi$,$\vartheta$,$\varphi$) form any magnetic coordinate system (Appendix \ref{sec:magnetic_coordinates}). We will regard $\psi$ as labeling the flux surfaces and consider toroidal geometries for which,
\begin{gather}
    \alpha = \vartheta - \iota(\psi) \varphi,
    \label{eq:field_line}
\end{gather}
label field lines in a flux surface, where $\vartheta$ is a poloidal angle, $\varphi$ is a toroidal angle, and $\iota(\psi) = \chi'(\psi)$ is the rotational transform, with $\chi(\psi)$ being the poloidal flux function.  
With these definitions, the magnetic flux passing toroidally through a poloidally closed curve of constant $\psi$ is $2\pi \psi$, and the flux passing poloidally between the magnetic axis and the surface of constant $\psi$ is $2\pi \chi(\psi)$. Thus, we assume that good flux surfaces exist and leave aside the issues of islands and chaotic field lines. In addition to the representation of the magnetic field, we assume that MHD force balance \eqref{eq:force_balance} is satisfied with a scalar pressure, $p(\psi)$.

As mentioned, we will consider two cases, a fixed-boundary case in which the shape of the outer flux surface is prescribed, and a free-boundary case for which outside the plasma, whose surface is defined by a particular value of toroidal flux, the force balance equation (\ref{eq:force_balance}) does not apply, but rather, the magnetic field is determined by Ampere's law,
\begin{align}
    \nabla \times \textbf{B} = \mu_0 \textbf{J} 
    \label{eq:Ampere},
\end{align}
with a given current density $\textbf{J}_C$, representing current flowing outside the confinement region. The fixed-boundary and free-boundary equations are discussed in detail in Section \ref{sec:equilibrium_calculation}.

From (\ref{eq:force_balance}) it follows that current density stream-lines also lie in the $\psi =$ constant surfaces. The toroidal current passing through a surface, $\mathcal{S}_T(\psi)$ (Figure \ref{fig:toroidal_flux}), whose perimeter is a closed poloidal loop at constant $\psi$ is given by,
\begin{gather}
    I_T(\psi) = \int_{\mathcal{S}_T(\psi)} d^2 x \, \hat{\textbf{n}} \cdot \textbf{J} = \int_{\mathcal{S}_T(\psi)} d \psi \, d \vartheta \, \sqrt{g} \nabla \varphi \cdot \textbf{J},
    \label{eq:toroidal_curr}
\end{gather}
where $\sqrt{g}^{-1} = \nabla \psi \times \nabla \vartheta \cdot \nabla \phi$.

Equations \cref{eq:magnetic_contravariant,eq:field_line,eq:field_line,eq:force_balance,eq:Ampere,eq:toroidal_curr} describe our base equilibrium configuration. We now consider small changes in the equilibrium that are assumed to yield a second equilibrium state of the same form as 
\cref{eq:magnetic_contravariant}, but with new functions such that $\textbf{B}'=\nabla \psi' \times \nabla \vartheta' - \nabla \chi'(\psi')\times \nabla \varphi'$. Each of the primed variables is assumed to differ from the corresponding unprimed variables by a small amount (e.g. $\psi' = \psi + \delta \psi(\textbf{x})$). The perturbed magnetic field can then be expressed $\textbf{B}'=\textbf{B}+\delta \textbf{B}$, where,
\begin{gather}
    \delta \textbf{B} = \nabla \delta \psi \times \nabla \vartheta + \nabla \psi \times \nabla \delta \vartheta - \nabla \chi(\psi) \times \nabla \delta \varphi - \nabla \left( \iota(\psi) \delta \psi + \delta \chi(\psi)\right) \times \nabla \varphi.  
    \label{eq:delta_B_1}
\end{gather}
We write the perturbed poloidal flux as the sum of a term resulting from the perturbation of toroidal flux at fixed rotational transform, $\iota(\psi) \delta \psi$, and a term representing the perturbed rotational transform, $\delta \chi(\psi)$.  
Thus, we can regroup the terms in \cref{eq:delta_B_1} as follows,
\begin{gather}
    \delta \textbf{B} = \nabla \times \left( \delta \psi \nabla \vartheta - \iota(\psi) \delta \psi \nabla \varphi - \delta \vartheta \nabla \psi + \delta \varphi \nabla \chi(\psi) \right) - \nabla \delta \chi(\psi) \times \nabla \varphi.
    \label{eq:delta_B2}
\end{gather}
The group of terms in parentheses 
in \cref{eq:delta_B2} corresponds to perturbations of the magnetic field allowed by ideal MHD, which is constrained by the ``frozen-in law", and which preserves the rotational transform, ($\delta \iota(\psi) =0$). The last term in \cref{eq:delta_B2} allows for changes in the rotational transform, ($\delta \iota(\psi) = \chi'(\psi)$). 
Note also that the expression in parentheses in \cref{eq:delta_B2} can be written as a sum of terms parallel to $\nabla\psi$ and $\nabla\alpha$, and hence it is perpendicular to $\textbf{B}$. 
The group of terms in parentheses 
in \cref{eq:delta_B2} can thus be expressed in terms of a vector potential that is perpendicular to the equilibrium magnetic field, while the last term in \cref{eq:delta_B2} can be represented in terms of a vector potential in the toroidal direction, which thus has a component parallel to the equilibrium field. We can therefore write $\delta \textbf{B}[\bm{\xi},\delta \chi(\psi)] = \nabla \times \delta \textbf{A}[\bm{\xi},\delta \chi(\psi)]$, where,
\begin{gather}
    \delta \textbf{A}[\bm{\xi},\delta \chi(\psi)] = \bm{\xi} \times \textbf{B} - \delta \chi(\psi) \nabla \varphi. 
    \label{eq:delta_A1}
\end{gather}
Here, the variable $\bm{\xi}$ can be taken to be perpendicular to the applied magnetic field, as the perturbed magnetic field,
\begin{align}
    \delta \textbf{B}[\bm{\xi},\delta \chi(\psi)] = \nabla \times \left(\bm{\xi} \times \textbf{B}\right) - \delta \chi'(\psi)\nabla \psi \times \nabla \varphi ,
    \label{eq:delta_B}
\end{align}
does not depend on $\bm{\xi} \cdot \hat{\textbf{b}}$. We emphasize that this departs from the typical assumption made in ideal MHD stability theory that $\nabla \cdot \bm{\xi} = 0$.

We define a vector field of the displacement of a field line, $\delta \textbf{x}$, such that the perturbation to the field line label $\alpha = \vartheta - \iota(\psi) \varphi$ and toroidal flux satisfy,
\begin{subequations}
\begin{align}
    \delta \psi + \delta \textbf{x} \cdot \nabla \psi &= 0 \\
    \delta \alpha + \delta \textbf{x} \cdot \nabla \alpha &= 0,
\end{align}
\end{subequations}
and $\delta \textbf{x} \cdot \textbf{B} = 0$. Noting that $\delta \alpha = \delta \vartheta - \iota(\psi) \delta \varphi - \left( \iota'(\psi) \delta \psi + \delta \chi'(\psi) \right)\varphi $, we find,
\begin{align}
    \delta \textbf{x} &= \bm{\xi} + \frac{\hat{\textbf{b}} \times \nabla \delta  \chi(\psi)}{B} \varphi,
    \label{eq:delta_r}
\end{align}
which follows from \eqref{eq:delta_B2}. As one would expect, in the limit $\delta \chi(\psi) = 0$, we recover the MHD displacement vector. 

As the pressure profile is often assumed to be held fixed during a configuration optimization, we assume that the local pressure changes such that $p(\psi)$ is unchanged,
\begin{align}
    \delta p[\bm{\xi}] = - \bm{\xi} \cdot \nabla p,
    \label{eq:delta_p}
\end{align}
which follows from \eqref{eq:delta_r}. We would similarly like to consider direct perturbations that fix the toroidal current. The change in toroidal current flowing through the perturbed surface is computed using \eqref{eq:transport_theorem} by expressing \eqref{eq:toroidal_curr} as a volume integral,
\begin{gather}
    \delta I_T(\psi) = \int_{\partial \mathcal{S}_T(\psi)} d \vartheta \, \sqrt{g}  \bm{\xi} \cdot \nabla \psi  \textbf{J} \cdot \nabla \varphi + \int_{\mathcal{S}_T(\psi)} d \psi d \vartheta \, \sqrt{g}  \delta \textbf{J}[\bm{\xi},\delta \chi(\psi)] \cdot \nabla \varphi, 
    \label{eq:delta_I}
\end{gather}
where $\mathcal{S}_T(\psi)$ is a surface at constant toroidal angle (Figure \ref{fig:toroidal_flux}) bounded by the $\psi$ surface and $\partial \mathcal{S}_T(\psi)$ is the boundary of such surface, a closed poloidal loop. The perturbed current density is $\delta \textbf{J} [\bm{\xi},\delta \chi(\psi)] = \nabla \times \delta \textbf{B}[\bm{\xi},\delta \chi(\psi)]$. Here the first term accounts for the displacement of the flux surface and the second term accounts for the change in toroidal current density. 

A linearized equilibrium state satisfies,
\begin{gather}
     \textbf{F}[\bm{\xi},\delta \chi(\psi)] +  \delta \textbf{F} = 0,
    \label{eq:perturbed_force_balance}
\end{gather}
where $\delta \textbf{F}$ is an additional perturbed force to be prescribed and $\textbf{F}[\bm{\xi},\delta \chi(\psi)]$ is the generalized force operator,
\begin{align}
  \textbf{F}[\bm{\xi},\delta \chi(\psi)] =   \delta \textbf{J}[\bm{\xi},\delta \chi(\psi)] \times \textbf{B} + \textbf{J} \times \delta \textbf{B} [\bm{\xi},\delta \chi(\psi)]- \nabla\delta p[\bm{\xi}].
  \label{eq:general_force_operator}
\end{align}

We now consider two distinct perturbations of the equilibrium of the type described by \cref{eq:delta_A1,eq:delta_B,eq:delta_p,eq:delta_I,eq:perturbed_force_balance,eq:general_force_operator}, which we denote with subscripts 1 and 2. In general, variables with subscript 1 will be associated with the direct perturbation, and those with subscripts 2 will be associated with the adjoint perturbation. We then form the quantity,
\begin{gather}
    U_T = \int_{V_T} d^3 x \, \left( \delta \textbf{J}_1 \cdot \delta \textbf{A}_2 - \delta \textbf{J}_2 \cdot \delta \textbf{A}_1\right) = 0,
    \label{eq:Ut}
\end{gather}
where we use the notation $\delta \textbf{J}_{1,2} = \delta \textbf{J}[\bm{\xi}_{1,2},\delta \chi_{1,2}(\psi)]$ and $\delta \textbf{A}_{1,2} = \delta \textbf{A}[\bm{\xi}_{1,2},\delta \chi_{1,2}(\psi)]$ and the integral is, for the time being, over all space. 
The above is seen to vanish by expressing $\delta \textbf{J}_{1,2}$ in terms of $\delta \textbf{B}_{1,2}$ using Ampere's law \eqref{eq:Ampere} and applying the divergence theorem. 

We now express the volume integral in \cref{eq:Ut} as the sum of three terms,
\begin{gather}
    U_T = U_P + U_B + U_C = 0.
    \label{UT}
\end{gather}
Here $U_P$ is the contribution from the plasma volume, integrated just up to the plasma-vacuum boundary. For this term we represent the vector potentials using \cref{eq:delta_A1},
\begin{gather}
    U_P = \int_{V_P} d^3 x \, \left( \delta \textbf{J}_1 \cdot \left( \bm{\xi}_2 \times \textbf{B} - \delta \chi_2(\psi) \nabla \varphi \right) - \delta \textbf{J}_2 \cdot \left( \bm{\xi}_1 \times \textbf{B} - \delta \chi_{1}(\psi) \nabla \varphi \right) \right).
    \label{eq:Upa}
\end{gather}
To evaluate \cref{eq:Upa} we use the perturbed force balance relation \eqref{eq:perturbed_force_balance}.

The term $U_B$ comes from integrating over a thin layer at the plasma-vacuum boundary. At the boundary, the difference between the perturbed and unperturbed current density has the character of a current sheet due to the displacement of the outermost flux surface. This effective current sheet causes a jump in the tangential components of the perturbation to the magnetic fields at the surface. This jump implies that care must be taken in evaluating the perturbed magnetic fields at the surface as they have different values on either side of the plasma-vacuum surface. However, the vector potential is continuous at the plasma-vacuum boundary. Thus, we write,
\begin{gather}
    U_B = \int_{S_P} \frac{d^2 x}{| \nabla \psi |} \left( \bm{\xi}_1 \cdot \nabla \psi \textbf{J} \cdot \delta \textbf{A}_2 - \bm{\xi}_2 \cdot \nabla \psi \textbf{J} \cdot \delta \textbf{A}_1 \right), 
\end{gather}
where the vector potentials are expressed as in \cref{eq:delta_A1}. Using this expression for the vector potentials and expressing the surface integral as an integral over the toroidal and poloidal angles gives,
\begin{gather}
    U_B = \int_{S_P} d \vartheta d \varphi \, \sqrt{g} \textbf{J} \cdot \nabla \varphi \left( - \bm{\xi}_1 \cdot \nabla \psi \delta \chi_2(\psi) + \bm{\xi}_2 \cdot \nabla \psi \delta \chi_{1}(\psi) \right). 
    \label{eq:Ub}
\end{gather}
Here we note the terms in the vector potential coming from the MHD displacement cancel. 

Last, the quantity $U_C$ represents the contribution from the integral over the volume outside the plasma where only the coil currents need to be included,
\begin{gather}
    U_C = \int_{V_V} d^3 x \, \left( \delta \textbf{J}_{C_1} \cdot \delta \textbf{A}_{V_2} - \delta \textbf{J}_{C_2} \cdot \delta \textbf{A}_{V_1} \right),
    \label{eq:Uc}
\end{gather}
where $\delta \textbf{A}_{V_{1,2}}$ is the change in the vacuum vector potential, and $\delta \textbf{J}_{C_{1,2}}$ is the change in the coil current density. 

Combining $U_P$, $U_B$, and $U_C$ gives the following relation appropriate to the free-boundary case $U_T = U_P + U_B + U_C = 0$, or
\begin{multline}
    \int_{V_P} d^3 x \, \left(\bm{\xi}_1 \cdot \textbf{F}_2   - \bm{\xi}_2 \cdot \textbf{F}_1 \right) + 2\pi \int_{V_P} d \psi \left( \delta \chi_{1}(\psi) \delta I_{T,2}'(\psi)-\delta \chi_{2}(\psi) \delta I_{T,1}'(\psi) \right) \\ + \int_{V_V} d^3 x \, \left( \delta \textbf{J}_{C_1} \cdot \delta \textbf{A}_{V_2} - \delta \textbf{J}_{C_2} \cdot \delta \textbf{A}_{V_1} \right) = 0,
    \label{eq:free_boundary}
\end{multline}
where we use the notation $\textbf{F}_{1,2} = \textbf{F}[\bm{\xi}_{1,2},\delta \chi_{1,2}(\psi)]$. This is the generalized free-boundary adjoint relation. The steps leading to \cref{eq:free_boundary} are outlined in Appendix \ref{app:adjoint_relation}. When the coil currents are confined to filaments, the integral over the vacuum region can be expressed in terms of changes to the coil currents, fluxes through the coils, and integrals along the coils, 
\begin{align}
  \int_{V_V}d^3 x \,  \delta \textbf{J}_{C_{1,2}} \cdot \delta \textbf{A}_{V_{2,1}}  = \sum_{k} \left(\delta \Phi_{C_{2,1,k}} \delta I_{C_{1,2,k}} + I_{C_k} \oint_{C_k} dl \, \delta \textbf{x}_{1,2,C_k}(\textbf{x}) \cdot \hat{\textbf{t}} \times \delta \textbf{B}_{2,1} \right).
    \label{eq:coil_perturb}
\end{align}
Here $\delta \Phi_{C_{k}}$ and $\delta I_{C_{k}}$ are the change in magnetic flux through and change in current in coil $k$, respectively, and $I_{C_k}$ is the current through the unperturbed coil. The unit tangent vector along $C_k$ is $\hat{\textbf{t}}$, and $\delta \textbf{x}_{C_k}$ is a vector field of perturbations to the $k$th coil. The above expression is obtained upon application of Stokes theorem and the expression for the perturbation of a line integral \eqref{eq:perturbation_line_integral_ch2}.

A similar relation can be obtained in the fixed-boundary case. Here the integral over the plasma volume \eqref{eq:Upa} can be written as a surface integral by applying the divergence theorem,
\begin{gather}
    U_P = \frac{1}{\mu_0}\int_{S_P} d^2 x \, \hat{\textbf{n}} \cdot \left( \delta \textbf{B}_1 \times \delta \textbf{A}_2 - \delta \textbf{B}_2 \times \delta \textbf{A}_1 \right).
    \label{eq:Up_bound}
\end{gather}
Again, following steps outlined in Appendix \ref{app:adjoint_relation}, this may be rewritten in the following form, 
\begin{multline}
    \int_{V_P} d^3 x \, \left(\bm{\xi}_1 \cdot  \textbf{F}_2   - \bm{\xi}_2 \cdot  \textbf{F}_1 \right)
    - 2\pi \int_{V_P} d \psi \, \left( \delta I_{T,2}(\psi) \delta \chi_{1}'(\psi) - \delta I_{T,1}(\psi) \delta \chi_{2}'(\psi) \right) \\
    - \frac{1}{\mu_0} \int_{S_P} d^2 x \, \hat{\textbf{n}} \cdot \left( \bm{\xi}_2 \delta \textbf{B}_1 - \bm{\xi}_1 \delta \textbf{B}_2  \right) \cdot \textbf{B} = 0. 
    \label{eq:fixed_boundary}
\end{multline}
The fixed-boundary adjoint relation can also be obtained by applying the self-adjointness \eqref{eq:self_adjointness_ch5} of the MHD force operator (Appendix \ref{appendix:self-adjointness}). If the second term in \eqref{eq:fixed_boundary} is integrated by parts in $\psi$, we see that the fixed and free-boundary adjoint relations share the terms involving the products of displacements with bulk forces and perturbed fluxes with perturbed toroidal currents. The integral over the vacuum region in \eqref{eq:free_boundary} is replaced by an integral over the plasma boundary and a boundary term from the integration by parts in $\psi$ in \eqref{eq:fixed_boundary}. 

We now have two integral relations between perturbations 1 and 2, \cref{eq:free_boundary,eq:fixed_boundary}. They have a common form in that they each are the sum of three integrals: the first involving 
forces and displacements, the second involving the toroidal current and poloidal flux profiles, and the third involving the manner in which the plasma boundary is prescribed. In \cref{eq:free_boundary}, the free-boundary case, the changes in coil current densities are specified. In \cref{eq:fixed_boundary}, the fixed-boundary case, the displacement of the outer flux surface is prescribed. Equations \cref{eq:free_boundary,eq:fixed_boundary} can also be viewed as the difference in sums of generalized forces and responses.  For example, in \cref{eq:free_boundary} we can consider the quantities $\delta \textbf{F}$, $\delta \chi(\psi)$, $\delta \textbf{J}_C$ as forces and $\bm{\xi}$, $\delta I_T'(\psi)$, $\delta \textbf{A}_V$ as responses.  The fact that the sum of the products of direct forces and adjoint responses less the products of adjoint forces and direct responses vanishes is similar to the relation between forces and fluxes related by Onsager symmetry \cite{Onsager1931a,Onsager1931b}. In the case of Onsager symmetry, this relation follows from the self-adjoint property of the collision operator. In this case, the symmetry follows from the generalized self-adjointness relation.

\section{Continuous adjoint method}
\label{sec:continuous_adjoint_method}

We now demonstrate how these relations \cref{eq:free_boundary} and \cref{eq:fixed_boundary} can be used to compute the shape gradient efficiently with a continuous adjoint method. 

\subsection{Fixed-boundary}
Consider a general figure of merit which involves a volume integral over the plasma domain,
\begin{align}
    f(S_P,\textbf{B}) = \int_{V_P} d^3 x \, F(\textbf{B}),
\end{align}
where $F(\textbf{B})$ depends on the plasma surface through the fixed-boundary MHD equilibrium equations (Table \ref{table:fixed_boundary}). We are interested in computing perturbations of $f$ such that \eqref{eq:force_balance} is satisfied. This constraint is enforced using the following Lagrangian functional,
\begin{align}
    \mathcal{L}(S_P,\textbf{B},\bm{\xi}_2) = f(S_P,\textbf{B}) + \int_{V_P} d^3 x \,  \bm{\xi}_2 \cdot \left(\frac{\left(\nabla \times \textbf{B}\right) \times \textbf{B}}{\mu_0} - \nabla p \right),
    \label{eq:lagrangian_ch5}
\end{align}
where $\bm{\xi}_2$ is a Lagrange multiplier and we have defined our inner product to be a volume integral over the domain. To obtain the adjoint equation that $\bm{\xi}_2$ must satisfy, we compute the functional derivative of \eqref{eq:lagrangian_ch5} with respect to $\textbf{B}$, where we note that perturbations to the magnetic field satisfy \eqref{eq:delta_B}. As $\delta f\left(S_P,\textbf{B};\delta \textbf{B}[\bm{\xi}_1,\delta \chi_1(\psi)]\right)$ is a linear functional of $\bm{\xi}_1 \in V_P$, $\delta \chi_1'(\psi)$, and $\bm{\xi}_1 \cdot \hat{\textbf{n}} \rvert_{S_P}$,
from the Riesz representation theorem, the functional derivative of $f$ with respect to $\textbf{B}$ is expressed as,
\begin{align}
    \delta f\left(S_P,\textbf{B};\delta \textbf{B}_1\right) = \int_{V_P} d^3 x \,  \bm{\xi}_1 \cdot \textbf{L}_1 + \int_{V_P} d \psi \,  \chi_1'(\psi)L_2(\psi)
    + \int_{S_P} d^2 x \, \bm{\xi}_1 \cdot \hat{\textbf{n}} L_3,
    \label{eq:functional_derivative_B}
\end{align}
for some quantities $\textbf{L}_1$, $L_2$, and $L_3$. The functional derivative of $\mathcal{L}$ is now,
\begin{multline}
    \delta \mathcal{L} \left(S_P, \textbf{B},\bm{\xi}_2;\delta \textbf{B}_1\right) = \int_{V_P} d^3 x \,\left( \bm{\xi}_1 \cdot \textbf{L}_1 + \bm{\xi}_2 \cdot \textbf{F}_1 \right) \\
    + \int_{V_P} d \psi \, \delta \chi_1'(\psi)L_2(\psi) + \int_{S_P} d^2 x \, \bm{\xi}_1 \cdot \hat{\textbf{n}} L_3,
    \label{eq:functional_derivative_L}
\end{multline}
where $\textbf{F}_1 = \textbf{F}[\bm{\xi}_1,\delta \chi_1(\psi)]$ is the generalized force operator associated with the direct perturbation \eqref{eq:general_force_operator}. We apply the fixed-boundary self-adjointness relation \eqref{eq:fixed_boundary} to obtain,
\begin{multline}
    \delta \mathcal{L} \left(S_P, \textbf{B},\bm{\xi}_2;\delta \textbf{B}_1\right) = \int_{V_P} d^3 x \,  \bm{\xi}_2 \cdot \left(\textbf{L}_1 + \textbf{F}_1\right) \\
    + \int_{V_P} d \psi \, \left(\delta \chi_1'(\psi)L_2(\psi)
    - 2\pi  \delta I_{T,2}\delta \chi_1'(\psi) + 2\pi  \delta I_{T,1}(\psi)\delta \chi_{2}'(\psi) \right) \\
    + \int_{S_P} d^2 x \, \left[ \bm{\xi}_1 \cdot \hat{\textbf{n}} \left(L_3 + \frac{\textbf{B} \cdot \delta \textbf{B}_2}{\mu_0} \right)
    - \bm{\xi}_2\cdot\hat{\textbf{n}}  \frac{\textbf{B} \cdot \delta \textbf{B}_1}{\mu_0} \right],
    \label{eq:lagrangian_B}
\end{multline}
where $\textbf{F}_2 = \textbf{F}[\bm{\xi}_2,\delta \chi_{2}(\psi)]$ is the generalized bulk force associated with the adjoint perturbation \eqref{eq:general_force_operator}, $\delta I_{T,2}(\psi)$ is the adjoint toroidal current perturbation, and $\delta \chi_{2}(\psi)$ is the adjoint poloidal flux perturbation.

If the direct problem is computed with fixed rotational transform, then $\delta \chi_1(\psi) = 0$, and the adjoint variable (Lagrange multiplier) is chosen to satisfy the linearized equilibrium problem,
\begin{subequations}
\begin{align}
  \textbf{F}[\bm{\xi}_2,\delta \chi_{2}(\psi)]  + \textbf{L}_1 &= 0 \\
    \hat{\textbf{n}} \cdot \bm{\xi}_2 \rvert_{S_P} &= 0 \\
     \delta \chi_{2}'(\psi) &= 0,
\end{align}
\label{eq:adjoint_lambda_iota}
\end{subequations}
such that the above functional derivative \eqref{eq:lagrangian_B} vanishes, except for the final term that is already in the desired Hadamard form \eqref{eq:shape_gradient_ch5}. If instead the direct problem is computed with fixed toroidal current, then $\delta I_{T,1}(\psi) = 0$ and the adjoint variable is chosen to satisfy,
\begin{subequations}
\begin{align}
  \textbf{F}[\bm{\xi}_2,\delta \chi_{2}(\psi)]  + \textbf{L}_1 &= 0 \\
    \hat{\textbf{n}} \cdot \bm{\xi}_2 \rvert_{S_P} &= 0 \\
    \delta I_{T,2}(\psi) &= \frac{L_2}{2\pi}.
\end{align}
\label{eq:adjoint_lambda_current}
\end{subequations}
The shape derivative of $\mathcal{L}$ with respect to boundary perturbation $\bm{\xi}_1$ is now computed to be,
\begin{multline}
    \delta \mathcal{L}\left( S_P,\textbf{B},\bm{\xi}_2;\bm{\xi}_1 \right) = \int_{S_P} d^2 x \, \bm{\xi}_1 \cdot \hat{\textbf{n}} \left( F + L_3 \right) + \int_{V_P} d^3 x \,  \bm{\xi}_1 \cdot \textbf{L}_1 \\
    + \int_{V_P} d\psi \,  \delta \chi_1'(\psi)L_2(\psi)
    + \delta \left(\int_{V_P} d^3 x \,  \bm{\xi}_2 \cdot \left(\frac{\left(\nabla \times \textbf{B}\right) \times \textbf{B}}{\mu_0} - \nabla p \right) \right),
\end{multline}
where the first term is evaluated using the transport theorem \eqref{eq:transport_theorem}. The notation in the final term indicates a shape derivative with respect to boundary perturbation $\bm{\xi}_1$. The above expression can be evaluated more easily by using the generalized adjoint relation \eqref{eq:fixed_boundary}, applying the conditions placed on the adjoint state \eqref{eq:adjoint_lambda_iota} or \eqref{eq:adjoint_lambda_current},
\begin{align}
     \delta \mathcal{L}\left( S_P,\textbf{B},\bm{\xi}_2;\bm{\xi}_1\right) = \int_{S_P} d^2 x \,  \hat{\textbf{n}} \cdot \bm{\xi}_1 \left(F + L_3 + \frac{\textbf{B} \cdot \delta \textbf{B}_2}{\mu_0}\right). 
\end{align}
So we identify the shape gradient to be,
\begin{align}
    \mathcal{G} = \left(F + L_3 + \frac{\textbf{B} \cdot \delta \textbf{B}_2}{\mu_0}\right)_{S_P}.
\end{align}
Thus by solving a linearized equilibrium problem corresponding to the addition of a bulk force for $\delta \textbf{B}[\bm{\xi}_2,\delta \chi_{2}(\psi)]$, we can compute the shape derivative with respect to \textit{any} boundary perturbation using the above shape gradient. 

\subsection{Free-boundary}

We now consider free-boundary perturbations. Consider a general figure of merit which involves a volume integral over the plasma domain,
\begin{align}
    f(C,\textbf{B}) = \int_{V_P} d^3 x \, F(\textbf{B}),
\end{align}
where $F(\textbf{B})$ depends on the coil shapes $C = \{ C_k \}$ through the free-boundary MHD equilibrium equations (Table \ref{table:free_boundary}). We are interested in computing perturbations of $f$ such that \eqref{eq:force_balance} is satisfied, which we enforce with the Lagrangian functional, \begin{align}
    \mathcal{L}(C,\textbf{B},\bm{\xi}_2) = f(C,\textbf{B}) + \int_{V_P} d^3 x \,  \bm{\xi}_2 \cdot \left(\frac{\left(\nabla \times \textbf{B}\right) \times \textbf{B}}{\mu_0} - \nabla p \right).
    \label{eq:lagrangian2_ch5}
\end{align}
In this case, $\delta f(C,\textbf{B};\delta \textbf{B}[\bm{\xi}_1,\delta \chi_1(\psi)])$ is a linear functional of $\bm{\xi}_1 \in V_P$, $\delta \chi_1(\psi)$, and the boundary perturbation $\bm{\xi}_1 \cdot \hat{\textbf{n}} \rvert_{S_P}$ resulting from a coil perturbation $\delta \textbf{x}_{1,C_k} \times \hat{\textbf{t}}$. (While in the fixed-boundary case, we considered $\delta f$ to be a linear functional of $\delta \chi_1'(\psi)$, for the free-boundary case it is more convenient to consider it to be a linear functional of $\delta \chi(\psi)$.) By the Riesz representation theorem,
\begin{align}
    \delta f\left(C,\textbf{B};\delta \textbf{B}[\bm{\xi}_1,\delta \chi_1(\psi)]\right) = \int_{V_P} d^3 x \,  \bm{\xi}_1 \cdot \textbf{L}_1 + \int_{V_P} d \psi \,  \chi_1(\psi)L_2(\psi)+ \int_{S_P} d^2 x \, \bm{\xi}_1 \cdot \hat{\textbf{n}} L_3,
\end{align}
for some quantities $\textbf{L}_1$, $L_2(\psi)$, and $L_3$. The functional derivative of $\mathcal{L}$ is now,
\begin{multline}
    \delta \mathcal{L} \left(C, \textbf{B},\bm{\xi}_2;\delta \textbf{B}[\bm{\xi}_1,\delta \chi_1(\psi)]\right) = \int_{V_P} d^3 x \,\left( \bm{\xi}_1 \cdot \textbf{L}_1 + \bm{\xi}_2 \cdot \textbf{F}_1 \right) \\
    + \int_{V_P} d \psi \, \delta \chi_1(\psi)L_2(\psi) + \int_{S_P} d^2 x \, \bm{\xi}_1 \cdot \hat{\textbf{n}} L_3.
\end{multline}
We apply the free-boundary relation \eqref{eq:free_boundary} to obtain,
\begin{multline}
        \delta \mathcal{L} \left(C, \textbf{B},\bm{\xi}_2;\delta \textbf{B}[\bm{\xi}_1,\delta \chi_1(\psi)]\right) = \int_{V_P} d^3 x \,  \bm{\xi}_1\cdot \left( \textbf{L}_1 + \textbf{F}_2\right)\\
        + \int_{V_P} d \psi \, \left(\delta \chi_1(\psi)L_2(\psi)
    - 2\pi \delta I_{T,1}'(\psi)\delta \chi_{2}(\psi)   + 2\pi  \delta I_{T,2}'(\psi)\delta \chi_1(\psi) \right) \\
    + \sum_{k} I_{C_k} \oint_{C_k} dl \, \left(\delta \textbf{x}_{1,C_k}(\textbf{x}) \times \delta \textbf{B}_2 - \delta \textbf{x}_{2,C_k}(\textbf{x}) \times \delta \textbf{B}_1\right)\cdot \hat{\textbf{t}} +
    \int_{S_P} d^2 x \, \bm{\xi} \cdot \hat{\textbf{n}} L_3,
\end{multline}
where we have considered perturbations to currents in the vacuum region corresponding to displacements of the filamentary coils without change to their currents. If the direct problem is computed with fixed rotational transform, then $\delta \chi_1(\psi) = 0$, and the adjoint variable is chosen to satisfy,
\begin{subequations}
\begin{align}
  \textbf{F}[\bm{\xi}_2,\delta \chi_{2}(\psi)]  + \textbf{L}_1 &= 0 \\
    \delta \chi_{2}(\psi) &= 0 \\
    \delta \textbf{x}_{2,C_k} \times \hat{\textbf{t}} &= 0,
\end{align}
\label{eq:adjoint_coil_1}
\end{subequations}
such that the above functional derivative vanishes, except for the terms involving integrals over $S_P$ or the filamentary coils. If instead the direct problem is computed with fixed toroidal current, then $\delta I_{T,1}(\psi) = 0$ and the adjoint variable is chosen to satisfy,
\begin{subequations}
\begin{align}
  \textbf{F}[\bm{\xi}_2,\delta \chi_{2}(\psi)]  + \textbf{L}_1 &= 0 \\
    \delta I_{T,2}(\psi) &= \frac{L_2}{2\pi} \\
    \delta \textbf{x}_{2,C_k} \times \hat{\textbf{t}} &= 0.
\end{align}
\label{eq:adjoint_coil_2}
\end{subequations}
The shape derivative of $\mathcal{L}$ is now computed to be,
\begin{multline}
    \delta \mathcal{L} \left(C, \textbf{B},\bm{\xi}_2;\delta \textbf{x}_{1,C_k}\right) = \int_{V_P} d^3 x \,\left( \bm{\xi}_1 \cdot \textbf{L}_1  \right) + \delta \left(\int_{V_P} d^3 x \, \bm{\xi}_2 \cdot \left(\frac{(\nabla \times \textbf{B}) \times \textbf{B}}{\mu_0} - \nabla p \right) \right) \\
    + \int_{V_P} d \psi \, \delta \chi_1(\psi)L_2(\psi) + \int_{S_P} d^2 x \, \bm{\xi}_1 \cdot \hat{\textbf{n}} \left(L_3 + F\right),
\end{multline}
where the notation $\delta(\dots)$ indicates a shape derivative with respect to coil displacement $\delta \textbf{x}_{1,C_k}$. We can now simplify the above expression using the free-boundary relation \eqref{eq:free_boundary} and the conditions placed on the adjoint variable, \eqref{eq:adjoint_coil_1} or \eqref{eq:adjoint_coil_2}. We now obtain,
\begin{align}
    \delta \mathcal{L}(C,\textbf{B},\bm{\xi}_2;\delta \textbf{x}_{1,C_k}) = \int_{S_P} d^2 x \, \bm{\xi}_1 \cdot \hat{\textbf{n}} \left(L_3 + F \right) + \sum_k I_{C_k} \oint_{C_k} dl \, \delta \textbf{x}_{1,C_k}\times \delta \textbf{B}_2 \cdot \hat{\textbf{t}},
\end{align}
where it is understood that $\bm{\xi}_1$ is the perturbation to the boundary arising from the coil perturbation $\delta \textbf{x}_{1,C_k}$. The first term can equivalently be expressed in terms of displacements of the coil shapes using the virtual casing principle \cite{Lazerson2012}, though in this Chapter for simplicity we will consider figures of merit such that $(L_3+ F)_{S_P}$ vanishes.

Some examples of these continuous adjoint methods are discussed in the following Sections. 

\section{Applications}
\label{sec:applications}

In this Section we will consider figures of merit which depend on the shape of the outer boundary of the plasma (Sections \ref{sec:surf_Beta}, \ref{sec:surf_iota}, \ref{sec:surf_vacuum_well}, and \ref{sec:surf_ripple}) and on the shape of the electro-magnetic coils (Sections \ref{sec:coil_iota} and \ref{sec:coil_vacuum_well}). The shape gradients of these figures of merit will be computed using both a direct method and an adjoint method, to demonstrate that the adjoint method produces identical results to the direct method but at much lower computational expense. For other figures of merit (Sections \ref{sec:epsilon_eff}-\ref{sec:neoclassical}) the calculation is not possible with existing codes, but a discussion of the adjoint linearized equilibrium equations is presented. 

\subsection{Volume-averaged $\beta$}
\label{sec:beta}

Consider a figure of merit, the volume-averaged $\beta$,
\begin{align}
f_{\beta} = \frac{f_P}{f_B}
\label{eq:beta},
\end{align}
where,
\begin{align}
    f_{P} = \int_{V_p} d^3 x \, p(\psi),
\end{align}
and,
\begin{align}
    f_B = \int_{V_p} d^3 x \, \frac{B^2}{2\mu_0}.
\end{align}
(This definition of volume-averaged $\beta$ is the one employed in the VMEC code \citep{Hirshman1983}.)  
While $f_{\beta}$ is a figure of merit not often considered in stellarator shape optimization, we include this calculation to demonstrate the adjoint approach, as its shape gradient can be computed without modifications to an equilibrium code.

\subsubsection{Surface shape gradient}
\label{sec:surf_Beta}

We consider direct perturbations about an equilibrium with fixed rotational transform,
\begin{subequations}
\begin{align}
    \textbf{F}[\bm{\xi}_1,\delta \chi_1(\psi)] &= 0 \\
    \bm{\xi}_1 \cdot \hat{\textbf{n}} \rvert_{S_P} &= \delta \textbf{x} \cdot \hat{\textbf{n}} \rvert_{S_P} \\
    \delta \chi_1'(\psi) &= 0.
\end{align}
\label{eq:direct_fixed_iota}
\end{subequations}
The differential change in $f_P$ associated with displacement $\bm{\xi}_1$ is,
\begin{gather}
    \delta f_P(S_P;\bm{\xi}_1) = -\int_{V_P} d^3 x \, \bm{\xi}_1 \cdot \nabla p + \int_{S_P} d^2 x \, \bm{\xi}_1 \cdot \hat{\textbf{n}} p(\psi),
\end{gather}
which follows from the transport theorem \eqref{eq:transport_theorem}. The first term accounts for the change in $p$ at fixed position due to the motion of the flux surfaces, and the second term accounts for the motion of the boundary. The differential change in $f_B$ associated with $\bm{\xi}_1$ is,
\begin{align}
    \delta f_B(S_P;\bm{\xi}_1) = -\frac{1}{\mu_0}\int_{V_P} d^3 x \, \left( B^2 \nabla \cdot \bm{\xi}_1 + \bm{\xi}_1 \cdot \nabla \left( B^2 + \mu_0 p \right) \right) + \frac{1}{2\mu_0}\int_{S_P} d^2 x \, \bm{\xi}_1 \cdot \hat{\textbf{n}} B^2,
    \label{eq:df_B}
\end{align}
where we have noted that the perturbation to the magnetic field strength at fixed position is given by,
\begin{align}
    \delta B = -\frac{1}{B} \left(B^2 \nabla \cdot \bm{\xi}_1 + \bm{\xi}_1 \cdot \nabla \left( B^2 + \mu_0 p\right) + \delta \chi_1'(\psi) \textbf{B} \cdot \left(\nabla \psi \times \nabla \varphi \right) \right).
    \label{eq:delta_mod_B}
\end{align}
The first term in \eqref{eq:df_B} corresponds with the change in $f_B$ due to the perturbation to the field strength, while the second term accounts for the motion of the boundary. Applying the divergence theorem we obtain,
\begin{gather}
    \delta f_B(S_P;\bm{\xi}_1) = -\int_{V_P} d^3 x \, \bm{\xi}_1 \cdot \nabla p - \frac{1}{2\mu_0 }\int_{S_P} d^2 x \,   \bm{\xi}_1 \cdot \hat{\textbf{n}} B^2. 
\end{gather}

The differential change in $f_{\beta}$ associated with displacement $\bm{\xi}_1$ satisfies,
\begin{align}
    \frac{\delta f_{\beta}(S_P;\bm{\xi}_1)}{f_{\beta}} =  \int_{S_P} d^2 x \, \bm{\xi}_1 \cdot \hat{\textbf{n}} \left(\frac{p(\psi)}{f_P} +  \frac{B^2}{2 \mu_0 f_B}\right)-\left( \frac{1}{f_P} - \frac{1}{f_B} \right) \int_{V_P} d^3 x \, \bm{\xi}_1 \cdot \nabla p.
    \label{eq:f_Beta1}
\end{align}
The first term on the right of \cref{eq:f_Beta1} is already in the form of a shape gradient. To evaluate the second term, we turn to the adjoint problem, choosing,
\begin{subequations}
\begin{align}
    \textbf{F}[\bm{\xi}_2,\delta \chi_2(\psi)]-\nabla  p &= 0 \\
    \bm{\xi}_2 \cdot \hat{\textbf{n}} \rvert_{S_P} &= 0 \\
    \delta \chi'_2(\psi) &= 0.
\end{align}
\end{subequations}
That is, we add a bulk force corresponding to the equilibrium pressure gradient. This additional force produces a proportional change in magnetic field at the boundary and thus from \cref{eq:fixed_boundary}, we find,
\begin{gather}
    \frac{\delta f_{\beta}(S_P;\bm{\xi}_1)}{f_{\beta}} = \int_{S_P} d^2 x \, \bm{\xi}_1 \cdot \hat{\textbf{n}} \left( \frac{p(\psi)}{f_P} + \frac{B^2}{2\mu_0 f_B} + \left( \frac{1}{f_P} - \frac{1}{f_B}  \right) \frac{\delta \textbf{B}_2 \cdot \textbf{B} }{\mu_0} \right). 
\end{gather}
Thus, we can obtain the shape gradient without perturbing the shape of the surface,
\begin{gather}
    \mathcal{G} = f_{\beta} \left( \frac{p(\psi)}{f_P} + \frac{B^2}{2 \mu_0 f_B} + \left( \frac{1}{f_P} - \frac{1}{f_B}  \right) \frac{\delta \textbf{B}_2 \cdot \textbf{B} }{\mu_0} \right)_{S_P}. 
    \label{eq:beta_adjoint}
\end{gather}
In practice, the adjoint magnetic field is approximated from a nonlinear equilibrium solution by adding a small perturbation to the pressure of magnitude $\Delta_P$, $p' = (1 + \Delta_P)p$. A forward-difference approximation is used to obtain,
\begin{align}
    \delta \textbf{B}_2 \approx \frac{\textbf{B}(p + \Delta_P p) - \textbf{B}(p)}{\Delta_P},
    \label{eq:adjoint_forward_difference}
\end{align}
where $\textbf{B}(p)$ is the magnetic field evaluated with pressure $p(\psi)$.

A similar expression can be obtained for equilibria for which the rotational transform is allowed to vary, but the toroidal current is held fixed ($\delta I_{T,1} = 0$). In this case, 
\begin{subequations}
\begin{align}
    \textbf{F}[\bm{\xi}_2,\delta \chi_2(\psi)]-\nabla  p &= 0 \\
    \bm{\xi}_2 \cdot \hat{\textbf{n}} \rvert_{S_P} &= 0 \\
    \delta I_{T,2}(\psi) &= -I_T(\psi) \left( 1/f_P - 1/f_B\right)^{-1} \left(1/f_B \right).
\end{align}
\end{subequations}
The shape gradient can then be obtained from \eqref{eq:beta_adjoint}. 

\begin{figure}
    \centering
    \begin{subfigure}[b]{0.8\textwidth}
    \centering
    \includegraphics[trim ={1.5cm 8cm 1.5cm 10.5cm},clip,width=1.0\textwidth]{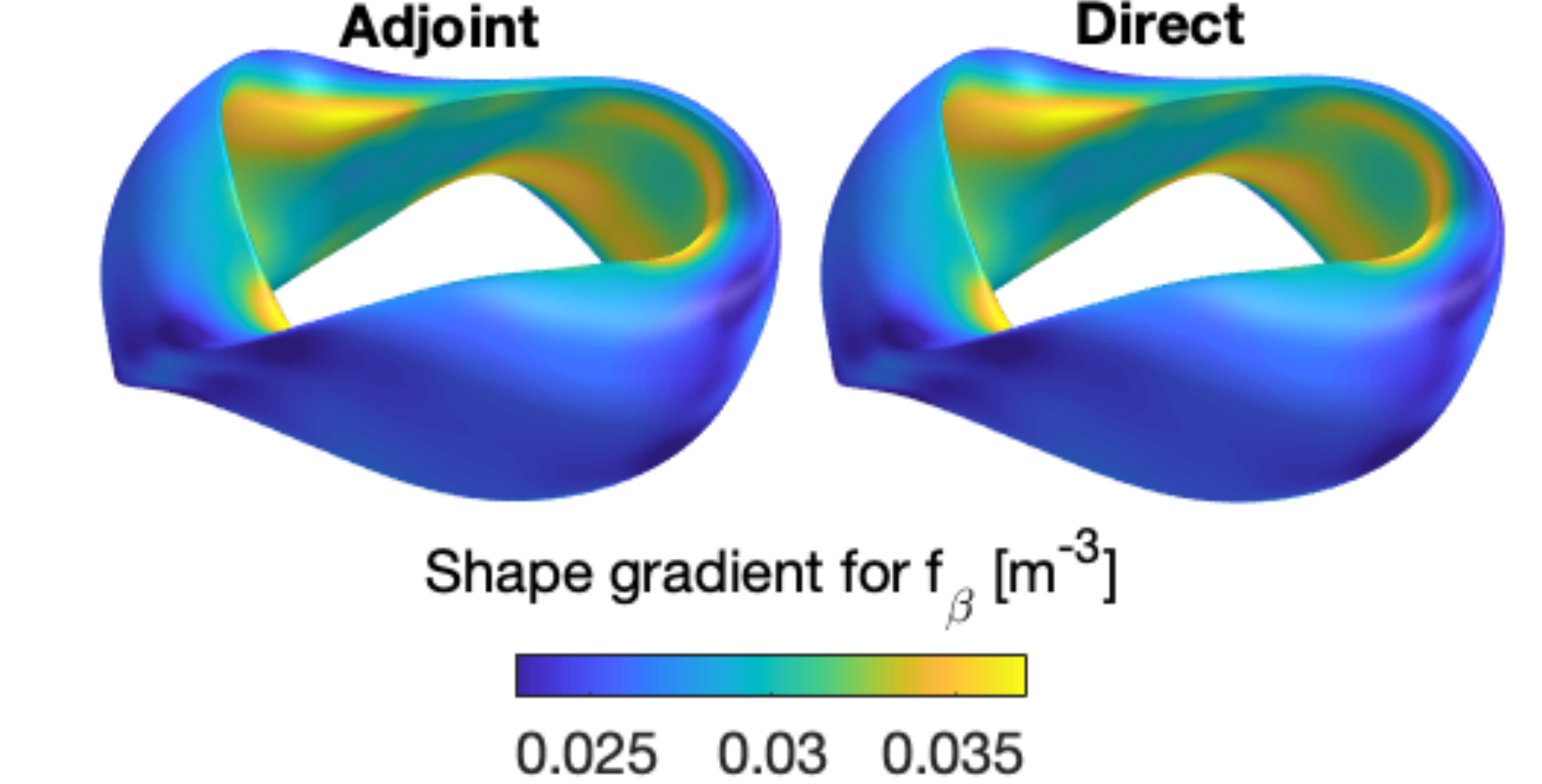}
    \caption{}
    \label{fig:beta_torus}
    \end{subfigure}
    \begin{subfigure}[b]{0.49\textwidth}
    \centering
    \includegraphics[trim={1cm 6cm 1cm 6.5cm},clip,width=1.0\textwidth]{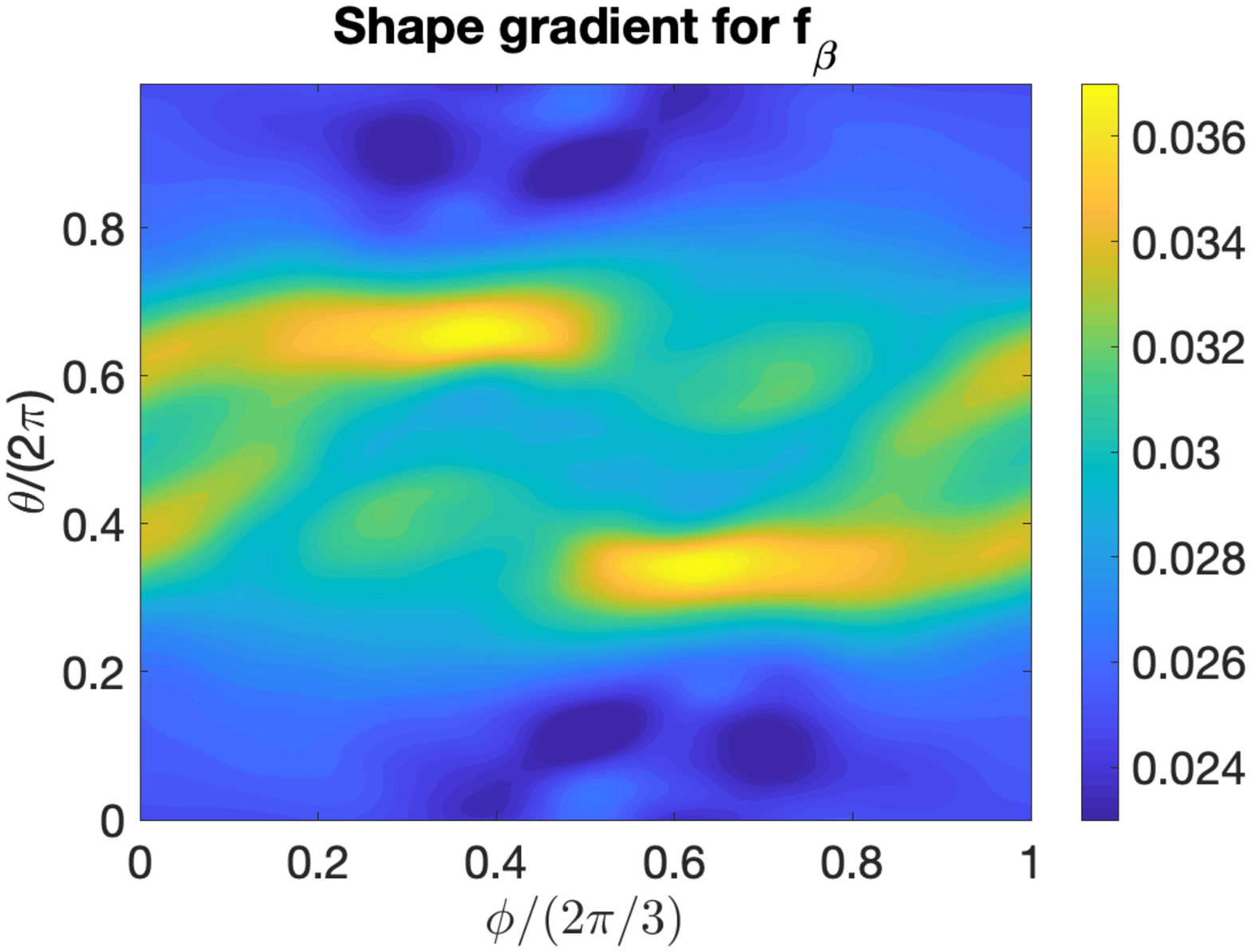}
    \caption{}
    \end{subfigure}
    \begin{subfigure}[b]{0.48\textwidth}
    \centering
    \includegraphics[trim={1cm 6cm 1cm 6.5cm},clip,width=1.0\textwidth]{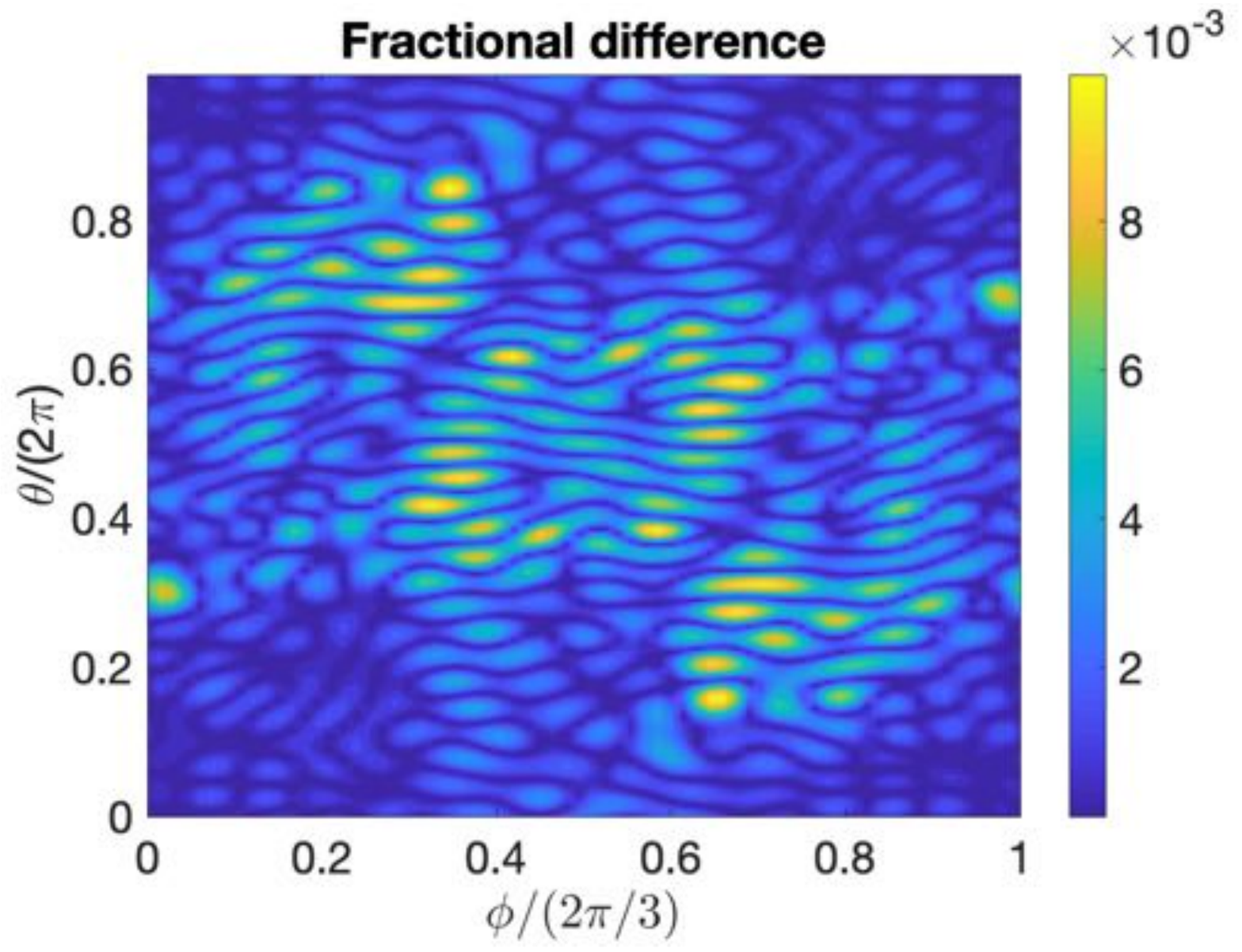}
    \caption{}
    \label{fig:beta_residual}
    \end{subfigure}
    \begin{subfigure}[b]{0.48\textwidth}
    \centering
    \includegraphics[trim={1cm 6.5cm 1cm 7cm},clip,width=1.0\textwidth]{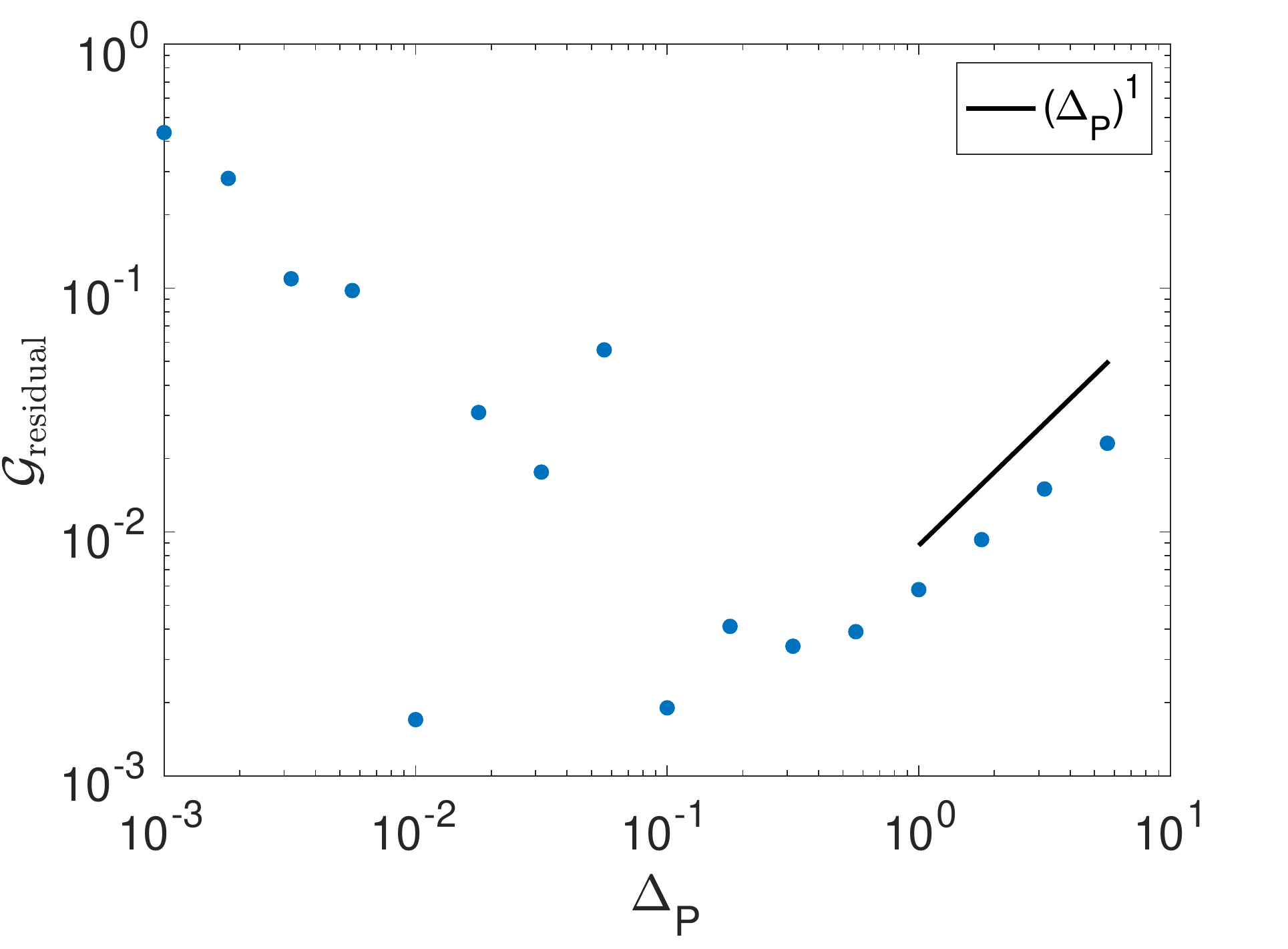}
    \caption{}
    \label{fig:delta_scan}
    \end{subfigure}
    \caption{(a) The shape gradient for $f_{\beta}$ \cref{eq:beta} computed using the adjoint solution \cref{eq:beta_adjoint} (left) and using parameter derivatives (right). (b) The shape gradient computed with the adjoint solution in the $\phi-\theta$ plane, the VMEC \cite{Hirshman1983} poloidal and toroidal angles (not magnetic coordinates). (c) The fractional difference \eqref{eq:residual} between the shape gradient obtained with the adjoint solution and with parameter derivatives. (d) The fractional difference \eqref{eq:residual} depends on the scale of the perturbation added to the adjoint force balance equation, $\Delta_P$. Figure adapted from \cite{Antonsen2019} with permission.}
    \label{fig:beta_shape_gradient}
\end{figure}

To demonstrate, we use the NCSX LI383 equilibrium \citep{Zarnstorff2001}. The pressure profile was perturbed with $\Delta_P = 0.01$ to compute the adjoint field. The unperturbed and adjoint equilibria are computed with the VMEC code \citep{Hirshman1983}. The shape gradient obtained with the adjoint solution, $\mathcal{G}_{\text{adjoint}}$, and that obtained with the direct approach, $\mathcal{G}_{\text{direct}}$, are shown in Figure \ref{fig:beta_torus}. Positive values of the shape gradient indicate that $f_{\beta}$ increases if a normal perturbation is applied at a given location as indicated by \eqref{eq:shape_gradient_ch5}. For the direct approach parameter derivatives with respect to the Fourier harmonics  describing the plasma boundary $(\partial f_{\beta}/\partial R_{m,n}^c, \partial f_{\beta}/\partial Z_{m,n}^s)$, where $R_{m,n}^c$ and $Z_{m,n}^s$ are defined through,
\begin{subequations}
\begin{align}
    R &= \sum_{m,n} R_{m,n}^c \cos(m \theta - n N_P \phi) \\
    Z &= \sum_{m,n} Z_{m,n}^s \sin(m \theta - nN_P \phi),
\end{align}
\label{eq:rmnc_zmns_ch5}
\end{subequations}
are computed with a centered 4-point stencil for $m \leq 15$ and $|n| \leq 9$ using a polynomial fitting technique. The centered-difference calculation is performed using a dedicated branch of the STELLOPT code. The shape gradient is obtained using the method outlined in Chapter \ref{ch:mathematical_fundamentals}. The fractional difference between the two methods,
\begin{gather}
\mathcal{G}_{\text{residual}} = \frac{|\mathcal{G}_{\text{adjoint}}-\mathcal{G}_{\text{direct}}|}{\sqrt{\int_{S_P} d^2 x \, \mathcal{G}_{\text{adjoint}}^2/\int_{S_P} d^2 x }},
\label{eq:residual}
\end{gather}
is shown in Figure \ref{fig:beta_residual}, where the surface-averaged value of $\mathcal{G}_{\text{residual}}$ is $1.7\times10^{-3}$. We note that the number of required equilibrium calculations for the direct shape gradient calculation depends on the Fourier resolution and finite-difference stencil chosen. In this Chapter we present the number of function evaluations required in order for the adjoint and direct shape gradient calculations to agree within a few percent. As the Fourier resolution is increased, the results of the adjoint and direct methods converge to each other.

The parameter $\Delta_P$ must be chosen carefully, as the perturbation must be large enough that the result is not dominated by round-off error, but small enough that nonlinear effects do not become important. The relationship between $\mathcal{G}_{\text{residual}}$ and $\Delta_P$ is shown in Figure \ref{fig:delta_scan}. Here $\mathcal{G}_{\text{direct}}$ is computed using the parameters reported above such that convergence is obtained. We find that $\mathcal{G}_{\text{residual}}$ decreases as $\left(\Delta_P\right)^1$ until $\Delta_P \approx 0.5$, at which point round-off error begins to dominate. This scaling is to be expected, as $\delta \textbf{B}_2$ is computed with a forward-difference derivative with step size $\Delta_P$.

For this and the following examples, the computational cost of transforming the parameter derivatives to the shape gradient was negligible compared to the cost of computing the parameter derivatives. 
The direct approach used 2357 calls to VMEC while the adjoint approach only required two. 
It is clear that the adjoint method yields nearly identical derivative information to the direct method but at a substantially reduced computational cost. 

The residual difference is nonzero due to several sources of error, including discretization error in VMEC. As a result of the assumption of nested magnetic surfaces, MHD force balance \eqref{eq:force_balance} is not satisfied exactly, but a finite force residual is introduced. Error is also introduced by computing $\delta \textbf{B}_2$ with the addition of a small perturbation to a nonlinear equilibrium calculation rather than from a linearized MHD solution.

In Figure \ref{fig:beta_shape_gradient} we find that $f_{\beta}$ is everywhere positive. This reflects the fact that the toroidal flux enclosed by $S_P$ is fixed. As perturbations which displace the plasma surface outward increase the surface area of a toroidal cross-section, the toroidal field must correspondingly decrease, thus increasing $f_{\beta}$. We find that the shape gradient is increased in regions of large field strength, as indicated by the second term in \eqref{eq:beta_adjoint}.

\subsection{Rotational transform}
\label{sec:iota}


Consider a figure of merit, the average rotational transform in a radially localized region,
\begin{gather}
    f_{\iota} = \int_{V_P} d \psi \, \iota(\psi) w(\psi).
    \label{eq:iota}
\end{gather}
Here $w(\psi)$ is a normalized weighting function,
\begin{gather}
    w(\psi) = \frac{e^{-(\psi-\psi_m)^2/\psi_w^2}}{\int_{V_P} d \psi \, e^{-(\psi-\psi_m)^2/\psi_w^2}},
    \label{eq:weighting}
\end{gather}
and $\psi_m$ and $\psi_w$ are parameters defining the center and width of the Gaussian weighting, respectively.

\subsubsection{Surface shape gradient}
\label{sec:surf_iota}

We consider direct perturbations about an equilibrium such that the toroidal current is fixed and the rotational transform is allowed to vary,
\begin{subequations}
\begin{align}
\textbf{F}[\bm{\xi}_1,\delta \chi_1(\psi)] &= 0 \\
    \bm{\xi}_1 \cdot \hat{\textbf{n}} |_{S_P} &= \delta \textbf{x} \cdot \hat{\textbf{n}} |_{S_P}  \\
    \delta I_{T,1}(\psi) &= 0.
\end{align}
\label{eq:direct_fixed_current}
\end{subequations}
The differential change of $f_{\iota}$ associated with perturbation $\bm{\xi}_1$ is,
\begin{gather}
    \delta f_{\iota}(S_P;\bm{\xi}_1) =  \int_{V_P} d \psi \, \delta \chi_1'(\psi) w(\psi).
    \label{eq:iota_deriv}
\end{gather}
For the adjoint problem, we prescribe,
\begin{subequations}
\begin{align}
    \textbf{F}[\bm{\xi}_2,\delta \chi_2(\psi)] &= 0 \\
        \bm{\xi}_2 \cdot \hat{\textbf{n}} \rvert_{S_P} &= 0 \\
    \delta I_{T,2} &= w(\psi).
\end{align}
\end{subequations}
This additional current produces a proportional change in the magnetic field at the boundary; thus using \cref{eq:fixed_boundary}, we obtain the following,
\begin{gather}
    \delta f_{\iota}(S_p;\bm{\xi}_1) = \frac{1}{2\pi \mu_0} \int_{S_P}d^2 x \, \hat{\textbf{n}} \cdot \bm{\xi}_1 \delta \textbf{B}_2 \cdot \textbf{B}.
\end{gather}
So, we can obtain the shape gradient from the adjoint solution,
\begin{gather}
    \mathcal{G} = \left(\frac{\delta \textbf{B}_2 \cdot \textbf{B}}{2\pi \mu_0}\right)_{S_P}.
    \label{eq:iota_adjoint}
\end{gather}
Note that the computation of the shape derivative of the rotational transform on a single surface, $\psi_m$, with the adjoint approach would require a delta-function current perturbation, $\delta I_{T,2} = \delta (\psi-\psi_m)$. As this type of perturbation is difficult to resolve in a numerical computation, the use of the Gaussian envelope allows the shape derivative of the rotational transform in a localized region of $\psi_m$ to be computed. 

\begin{figure}
    \centering
    \begin{subfigure}[b]{0.8\textwidth}
    \centering
    \includegraphics[trim ={1.5cm 8cm 1.5cm 10cm},clip,width=1.0\textwidth]{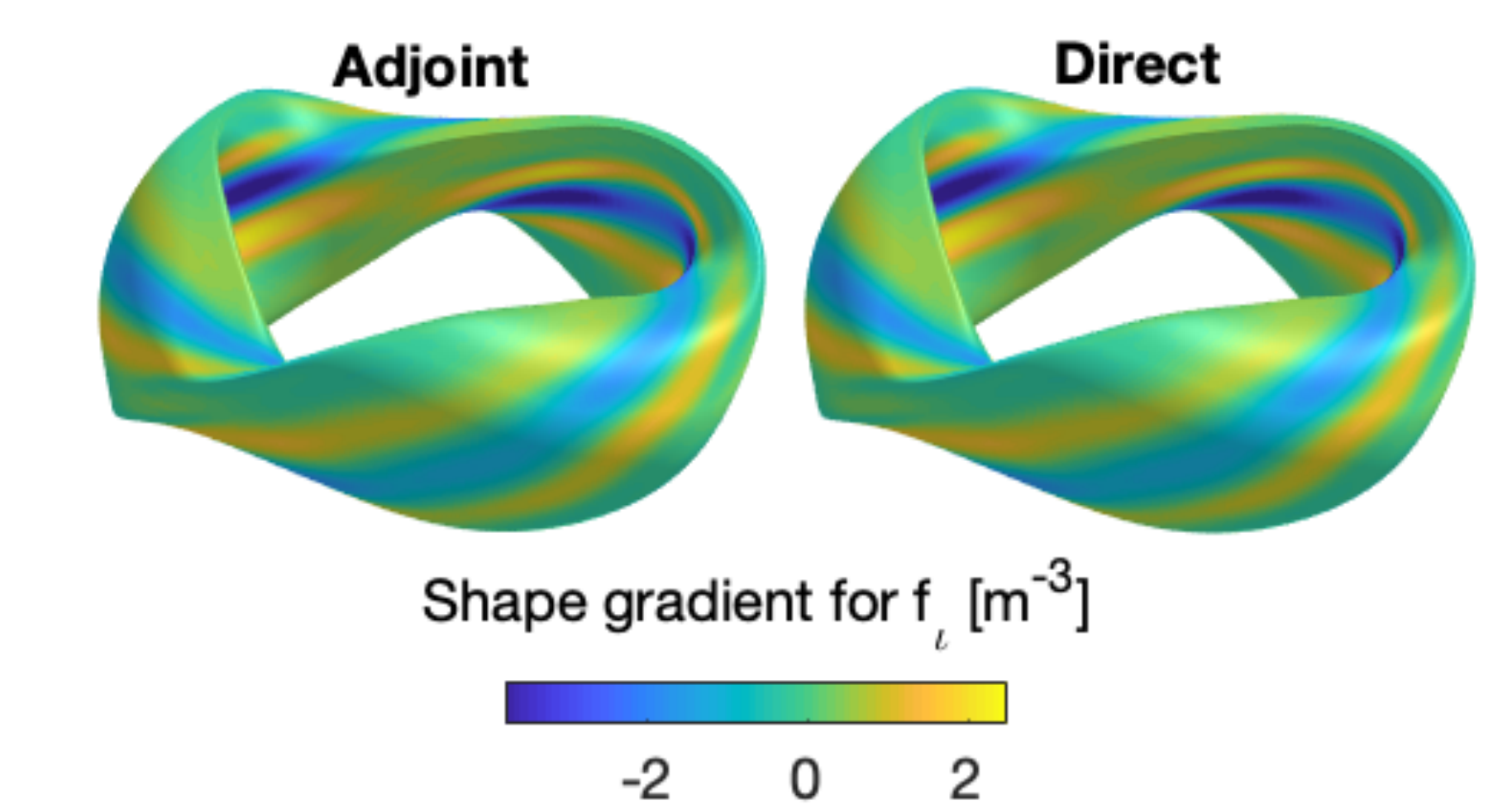}
    \caption{}
    \label{fig:iota_torus}
    \end{subfigure}
    \begin{subfigure}[b]{0.49\textwidth}
    \includegraphics[trim={1cm 6cm 1cm 6cm},clip,width=1.0\textwidth]{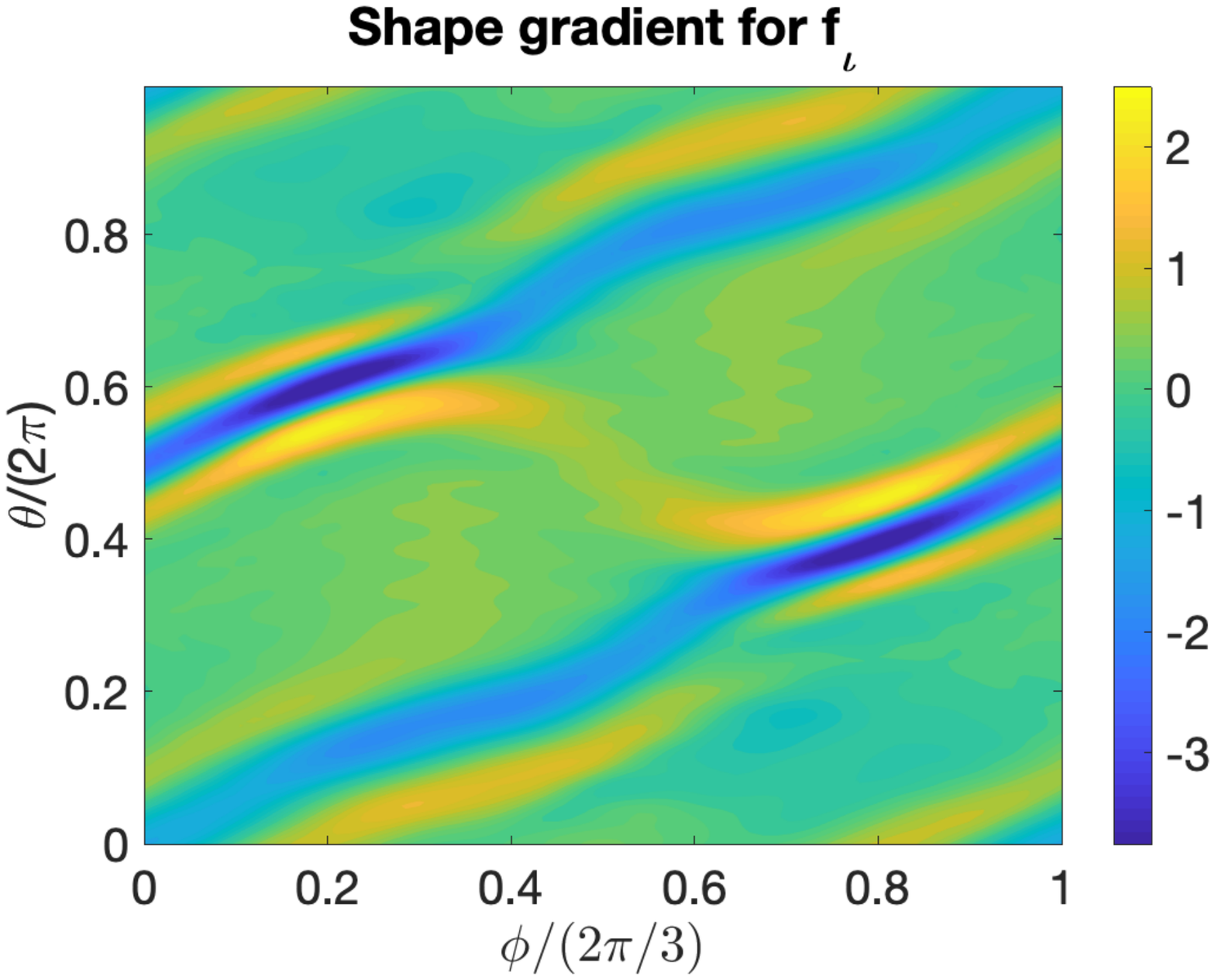}
    \caption{}
    \end{subfigure}
    \begin{subfigure}[b]{0.48\textwidth}
    \includegraphics[trim={1cm 6cm 1cm 6cm},clip,width=1.0\textwidth]{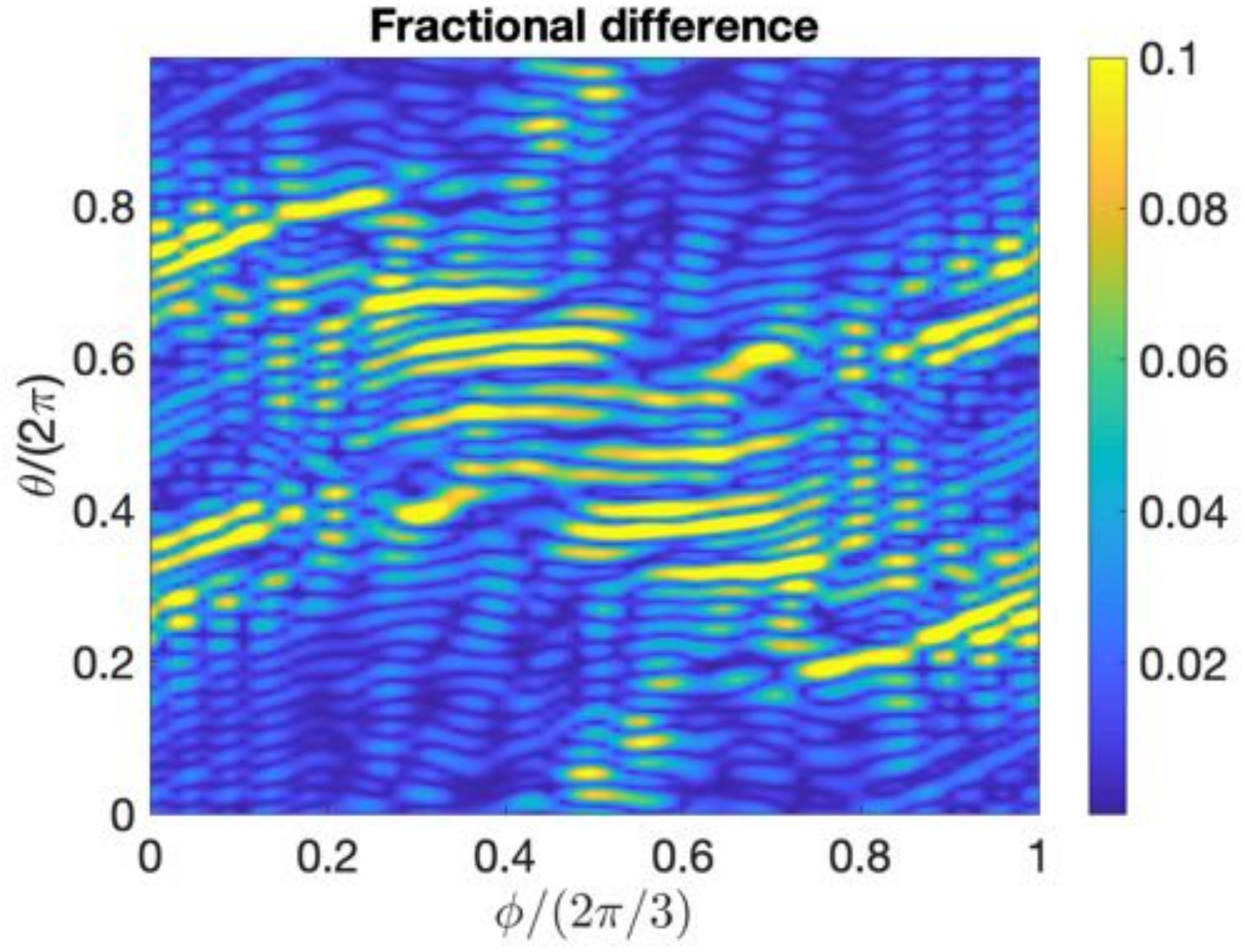}
    \caption{}
    \label{fig:iota_residual}
    \end{subfigure}
    \caption{(a) The shape gradient for $f_{\iota}$ \cref{eq:iota} computed using the adjoint solution \cref{eq:iota_adjoint} (left) and using parameter derivatives (right). (b) The shape gradient computed with the adjoint solution in the $\phi - \theta$ plane, the VMEC \cite{Hirshman1983} poloidal and toroidal angles (not magnetic coordinates).
    (c) The fractional difference \eqref{eq:residual} between the shape gradient obtained with the adjoint solution and with parameter derivatives. Again, the results are essentially indistinguishable, as expected. Figure adapted from \cite{Antonsen2019} with permission.
    } 
    \label{fig:iota_shape_gradient}
\end{figure}

To demonstrate, we use the NCSX LI383 equilibrium. We again apply a forward-difference approximation \eqref{eq:adjoint_forward_difference} of the adjoint solution, characterized by amplitude $\Delta_I = 715$ A. The parameters of the weight function are taken to be $\psi_m = 0.1 \psi_0$, and $\psi_w = 0.05 \psi_0$. The shape gradient obtained with the adjoint solution and with the direct approach are shown in Figure \ref{fig:iota_torus}. 
For the direct approach, the shape gradient is computed from parameter derivatives with respect to the Fourier harmonics of the boundary \eqref{eq:rmnc_zmns} using an 8-point stencil with $m \le 18$ and $|n| \le 12$. The fractional difference, $\mathcal{G}_{\text{residual}}$, between the two approaches is shown in Figure \ref{fig:iota_residual}, with a surface-averaged value of $2.7\times10^{-2}$. The direct approach used 7401 calls to VMEC, while the adjoint only required two. Again, it is apparent that the adjoint method allows the same derivative information to be computed at a much lower computational cost. 
   
We find that over much of the surface, the shape gradient is close to zero. A region of large negative shape gradient occurs in the concave region of the plasma surface with adjacent regions of large positive shape gradient. This indicates that ``pinching'' the surface in this region, making it more concave, would increase $\iota$ near the axis. 

\subsubsection{Coil shape gradient}
\label{sec:coil_iota}

The shape gradient of $f_{\iota}$ can also be computed with a free-boundary approach. We consider perturbations about an equilibrium with fixed toroidal current,
\begin{subequations}
\begin{align}
   \textbf{F}[\bm{\xi}_1,\delta \chi_1(\psi)] &= 0 \\
   \delta I_{T,1}(\psi) &= 0,
\end{align}
\end{subequations}
with specified perturbation to the coil shapes, $\delta \textbf{x}_{C_1} \times \hat{\textbf{t}}$. We prescribe the adjoint problem,
\begin{subequations}
\begin{align}
    \textbf{F}[\bm{\xi}_2,\delta \chi_2(\psi)] &= 0 \\
    \delta \textbf{x}_{C_2} \times \hat{\textbf{t}} &= 0 \\
    \delta I_{T,2}(\psi) &= w(\psi),
\end{align}
\end{subequations}
where $w(\psi)$ is given by \eqref{eq:weighting}. Using \cref{eq:iota_deriv} and \cref{eq:free_boundary} and noting that $\delta I_{T,2}(\psi)$ vanishes at the plasma boundary and on the axis, we find,
\begin{gather}
    \delta f_{\iota}(C;\delta \textbf{x}_C) = \frac{1}{2\pi}\int_{V_V} d^3 x \, \delta \textbf{J}_{C_1} \cdot \delta \textbf{A}_{V_2}.
\end{gather}
Using \cref{eq:coil_perturb}, this can be written in terms of changes in the positions of coils in the vacuum region,
\begin{gather}
    \delta f_{\iota}(C;\delta \textbf{x}_C) = \frac{1}{2\pi} \sum_k \left(  I_{C_k} \oint_{C_k} dl \, \delta \textbf{x}_{C_k} (\textbf{x}) \cdot \hat{\textbf{t}} \times \delta \textbf{B}_2 \right).
    \label{eq:iota_derivative_adjoint}
\end{gather}
When computing the coil shape gradient, the current in each coil is fixed. In arriving at \cref{eq:iota_derivative_adjoint}, we assume that $\delta I_{C_{1,k}} = 0$.  The coil shape gradient is thus
\begin{gather}
    \widetilde{\bm{\mathcal{G}}}_k = \frac{I_{C_k} \hat{\textbf{t}} \times \delta \textbf{B}_2}{2\pi} \bigg \rvert_{C_k}.
    \label{eq:iota_coil_adjoint}
\end{gather}
As anticipated, $\widetilde{\bm{\mathcal{G}}}_k$ has no component in the direction tangent to the coil. The adjoint magnetic field is computed with a forward-difference approximation \eqref{eq:adjoint_forward_difference} with step size $\Delta_I = 5.7\times 10^5$ A. Evaluating the shape gradient requires computing the adjoint magnetic field at the unperturbed coil locations in the vacuum region. This can be performed with the DIAGNO code \citep{Gardner1990,Lazerson2012}, which employs the virtual casing principle.

\begin{figure}
    \centering
    \includegraphics[trim={2cm 9cm 4cm 8cm},clip,width=0.8\textwidth]{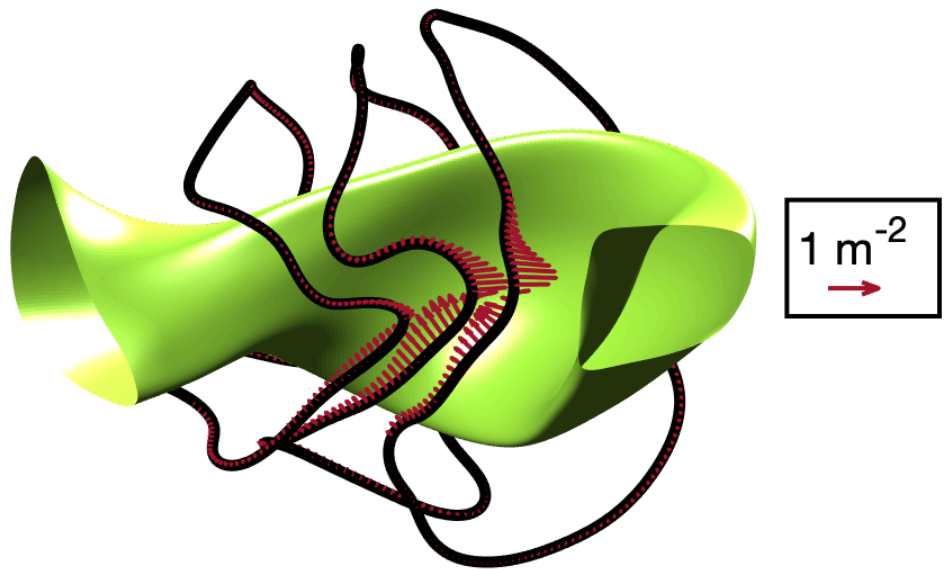}
    \caption{The coil shape gradient for $f_{\iota}$ \cref{eq:iota} computed using the adjoint solution \cref{eq:iota_coil_adjoint} for each of the 3 unique coil shapes (black). The arrows indicate the direction of $\widetilde{\bm{\mathcal{G}}}_k$, and their length indicates the local magnitude relative to the reference arrow shown. The arrows are not visible on this scale on the outboard side. Figure reproduced from \cite{Antonsen2019} with permission.
    }
    \label{fig:iota_coil_analytic}
\end{figure}

\begin{figure}
    \centering
    \begin{subfigure}[b]{0.8\textwidth}
    \centering
    \includegraphics[trim={0cm 7cm 0cm 7cm},clip,width=1.0\textwidth]{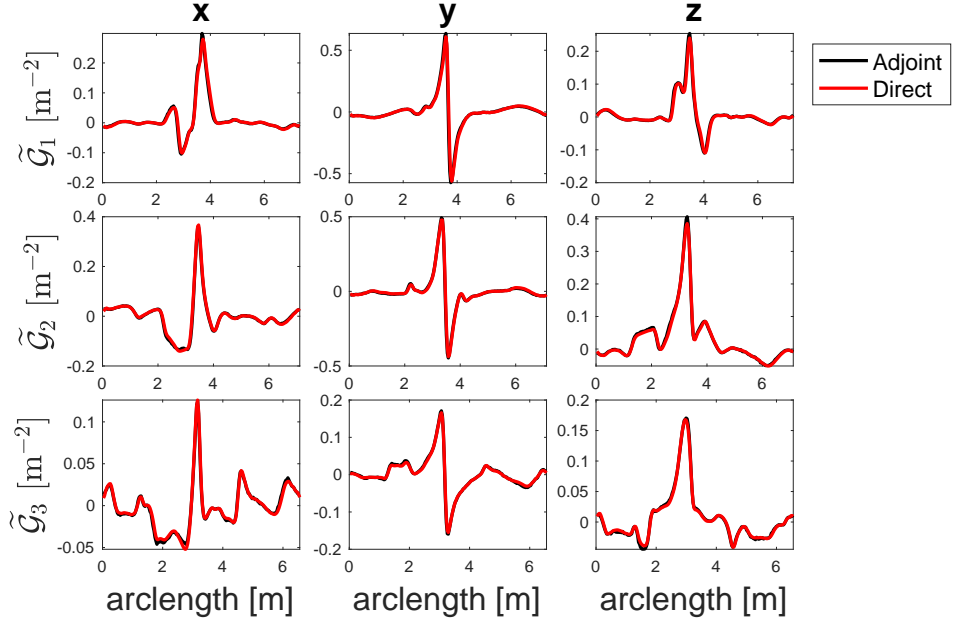}
    \caption{}
    \label{fig:iota_coil_comparison}
    \end{subfigure}
    \begin{subfigure}[b]{0.8\textwidth}
    \includegraphics[trim={1cm 7cm 1cm 7cm},clip,width=1.0\textwidth]{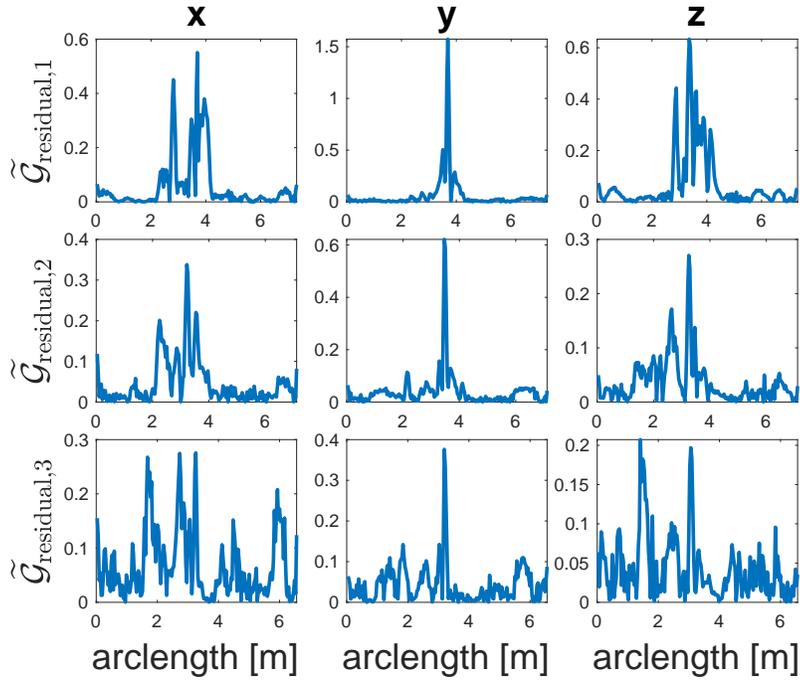}
    \caption{}
    \label{fig:iota_coil_residual}
    \end{subfigure}
    \caption{(a) The Cartesian components of the coil shape gradient for each of the 3 unique modular NCSX coils computed with the adjoint and direct approaches. 
    (b) The fractional difference \eqref{eq:residual_coil} between the shape gradient computed with the adjoint approach and the direct approach is plotted for each Cartesian component and each of the 3 unique coils. Figure adapted from \cite{Antonsen2019} with permission.}
    \label{fig:iota_residual_coil}
\end{figure}

To demonstrate, we use the NCSX stellarator LI383 equilibrium. The toroidal current profile was perturbed with $\psi_m = 0.1 \psi_0$ and $\psi_w = 0.05 \psi_0$. The shape gradient is computed for each of the three unique modular coils per half period of the C09R00 coil set\footnote{https://princetonuniversity.github.io/STELLOPT/VMEC\%20Free\%20Boundary\%20Run} \citep{Williamson2005}, keeping the planar coils fixed. The result obtained with the adjoint solution, $\widetilde{\bm{\mathcal{G}}}_{\text{adjoint},k}$, is shown in Figure \ref{fig:iota_coil_analytic}. The shape gradient is also computed with the direct approach, $\widetilde{\bm{\mathcal{G}}}_{\text{direct},k}$. For the direct approach, the Cartesian components of each coil are Fourier discretized as,
\begin{subequations}
\begin{align}
    x_k = \sum_m X_{m}^{kc} \cos(m \theta) + X_{m}^{ks} \sin(m \theta) \\
    y_k = \sum_m Y_{m}^{kc} \cos(m \theta) + Y_{m}^{ks} \sin(m \theta) \\ 
    z_k = \sum_{m} Z_{m}^{kc} \cos(m \theta) + Z_m^{ks} \sin(m \theta),
\end{align}
\label{eq:coil_discretization}
\end{subequations}
where $\theta \in [0,2\pi]$ parameterizes each filament and $k$ denotes each coil shape. The numerical derivative with respect to these parameters are computed for $m \leq 45$ using an 8-point stencil. In Figure \ref{fig:iota_coil_comparison} the Cartesian components of the shape gradient computed with the adjoint approach, $\widetilde{\mathcal{G}}_{\text{adjoint},k}^l$, and with the direct approach, $\widetilde{\mathcal{G}}_{\text{direct},k}^l$, are shown for each coil, where $l\in \{x,y,z\}$. The arrows indicate the direction and magnitude of $\widetilde{\bm{\mathcal{G}}}_k$ such that if a coil were deformed in the direction of $\widetilde{\bm{\mathcal{G}}}_k$, $f_{\iota}$ would increase according to \eqref{eq:coil_shape_gradient_ch5}. The direct approach used 6553 calls to VMEC, while the adjoint only required two. In Figure \ref{fig:iota_coil_residual} the fractional difference between the results obtained with the two methods,
\begin{gather}
    \widetilde{\mathcal{G}}_{\text{residual},k}^l = \frac{|\widetilde{\mathcal{G}}_{\text{adjoint},k}^l - \widetilde{\mathcal{G}}_{\text{direct},k}^l|}{\sqrt{\oint_{C_k} dl \, \left(\widetilde{\mathcal{G}}_{\text{adjoint},k}^l\right)^2/\oint_{C_k} dl}},
    \label{eq:residual_coil}
\end{gather}
is plotted. The line-averaged values of $\widetilde{\mathcal{G}}_{\text{residual}}^l$ are $6.1\times 10^{-2}$ for coil 1, $3.8 \times 10^{-2}$ for coil 2, and $4.8 \times 10^{-2}$ for coil 3.

From Figure \ref{fig:iota_coil_analytic}, we see that the sensitivity of $f_{\iota}$ to coil displacements is much higher in regions where the coils are close to the plasma surface. The shape gradient points toward the plasma surface in the concave region of the plasma surface, while on the outboard side the sensitivity is significantly lower, again indicating the ``pinching" effect seen in Figure \ref{fig:iota_shape_gradient}.

\subsection{Vacuum magnetic well}
\label{sec:vacuum_well}
The averaged radial (normal to a flux surface) curvature is an important metric for MHD stability \citep{Freidberg2014}, 
\begin{gather}
    \kappa_{\psi} \equiv \left\langle \bm{\kappa} \cdot \left(\partder{\textbf{x}}{\psi} \right)_{\alpha,l}\right\rangle_{\psi} = \left \langle \frac{1}{2B^2} \left(\partder{}{\psi} \left(2 \mu_0 p + B^2 \right) \right)_{\alpha,l} \right \rangle_{\psi},
\end{gather}
where the curvature is $\bm{\kappa} = \hat{\textbf{b}} \cdot \nabla \hat{\textbf{b}}$, $\hat{\textbf{b}} = \textbf{B}/B$ is a unit vector in the direction of the magnetic field and $l$ measures length along a field line. Subscripts in the above expression ($\alpha,l$) indicate quantities held fixed while computing the derivative. The flux surface average of a quantity $A$ is, \begin{gather}
    \langle A \rangle_{\psi} = \frac{\int_{-\infty}^{\infty} \frac{dl}{B}\, A}{\int_{-\infty}^{\infty} \frac{dl}{B}} =  \frac{\int_0^{2\pi} d \vartheta \int_0^{2\pi} d \varphi \, \sqrt{g} A}{V'(\psi)}.
\end{gather}
Here $V(\psi)$ is the volume enclosed by the surface labeled by $\psi$. The average radial curvature appears in the ideal MHD potential energy functional for interchange modes, and it provides a stabilizing effect when $p'(\psi) \kappa_{\psi} < 0$. As typically $p'(\psi)<0$, $\kappa_{\psi} >0$ is desirable for MHD stability. In a vacuum field, the expression for the averaged radial curvature reduces to,
\begin{gather}
    \kappa_{\psi} = - \frac{V''(\psi)}{V'(\psi)}.
\end{gather}
Thus, as volume increases with flux, $V''(\psi)<0$ is advantageous \citep{Helander2014}. The quantity $p'(\psi)V''(\psi)$ also appears in the Mercier criterion for ideal MHD interchange stability \citep{Mercier1974}. Known as the vacuum magnetic well, $V''(\psi)$ has been employed in the optimization of several stellarator configurations (e.g. \cite{Hirshman1999,Henneberg2019}).

We consider the following figure of merit,
\begin{gather}
    f_W = \int_{V_P} d\psi \, w(\psi)V'(\psi),
    \label{eq:f_w_v}
\end{gather}
where $w(\psi)$ is a radial weight function which will be chosen so that \eqref{eq:f_w_v} approximates $V''(\psi)$. This can equivalently be written as,
\begin{gather}
    f_W = \int_{V_P} d^3 x \, w(\psi).
\end{gather}

\subsubsection{Surface shape gradient}
\label{sec:surf_vacuum_well}

We consider direct perturbations about an equilibrium with fixed toroidal current \eqref{eq:direct_fixed_current}. The shape derivative of $f_W$ is computed upon application of the transport theorem \eqref{eq:transport_theorem}, noting that $\delta \psi = - \bm{\xi}_1 \cdot \nabla \psi$,
\begin{gather}
    \delta f_W(S_P;\bm{\xi}_1) = -\int_{V_P} d^3 x \,  \bm{\xi}_1 \cdot \nabla w(\psi) + \int_{S_P} d^2 x \, \, \bm{\xi}_1 \cdot \hat{\mathbf{n}} w(\psi),
    \label{eq:df_W}
\end{gather}
where we have assumed $w(\psi)$ to be differentiable. We recast the first term in \eqref{eq:df_W} as a surface integral by applying the fixed-boundary adjoint relation \eqref{eq:fixed_boundary} and prescribing the adjoint perturbation to satisfy the following,
\begin{subequations}
\begin{align}
    \textbf{F}[\bm{\xi}_2,\delta \chi_2(\psi)]  -\nabla w(\psi) &= 0 \label{eq:deltaF_vacuum} \\
    \bm{\xi}_2 \cdot \hat{\textbf{n}}|_{S_P} &= 0 \\
    \delta I_{T,2}(\psi) &= 0.
\end{align}
\end{subequations}

Upon application of \eqref{eq:fixed_boundary} we obtain the following expression for the shape gradient which depends on the adjoint solution, $\delta \textbf{B}_2$,
\begin{gather}
    \mathcal{G}_{W} = \left(w(\psi) + \frac{\delta \textbf{B}_2 \cdot \textbf{B}}{\mu_0}\right)_{S_P}.
    \label{eq:g_vacuum}
\end{gather}

\begin{figure}
    \centering
    \begin{subfigure}[b]{0.49\textwidth}
    \includegraphics[trim=3cm 3cm 3cm 6cm,clip,width=1.0\textwidth]{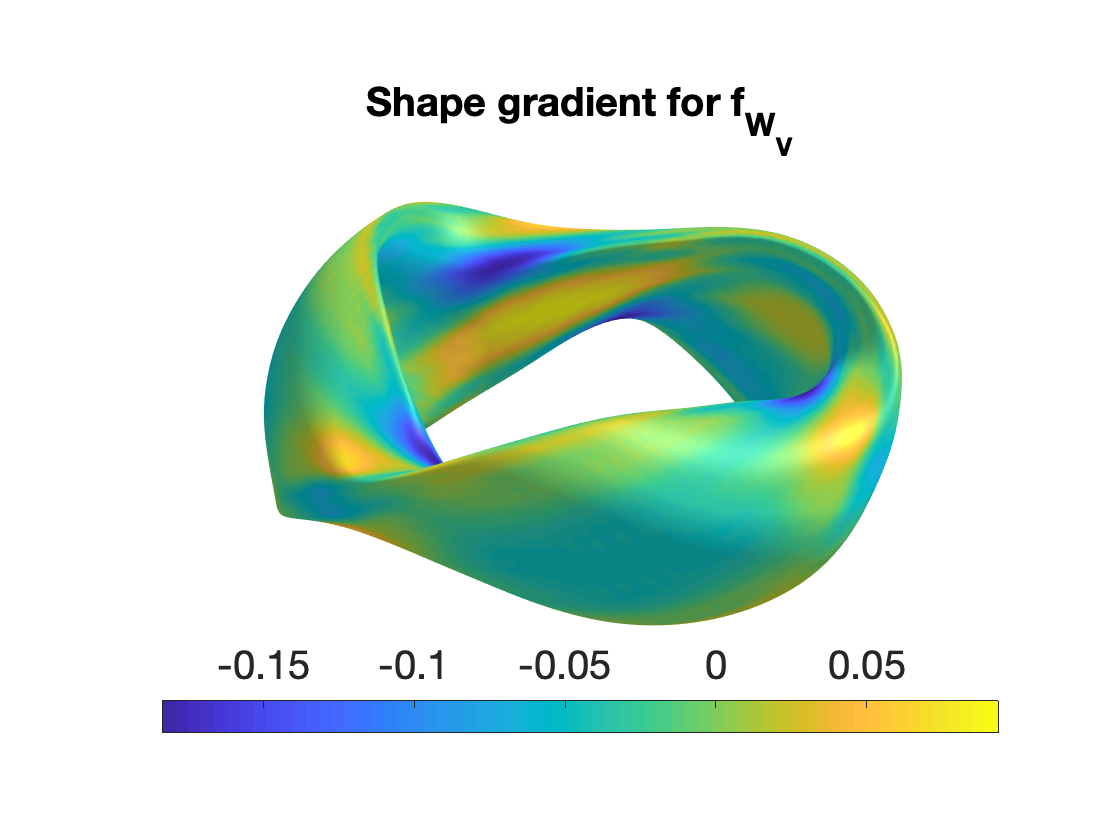}
    \caption{Adjoint}
    \label{fig:adjoint}
    \end{subfigure}
    \begin{subfigure}[b]{0.49\textwidth}
    \includegraphics[trim=3cm 3cm 3cm 6cm,clip,width=1.0\textwidth]{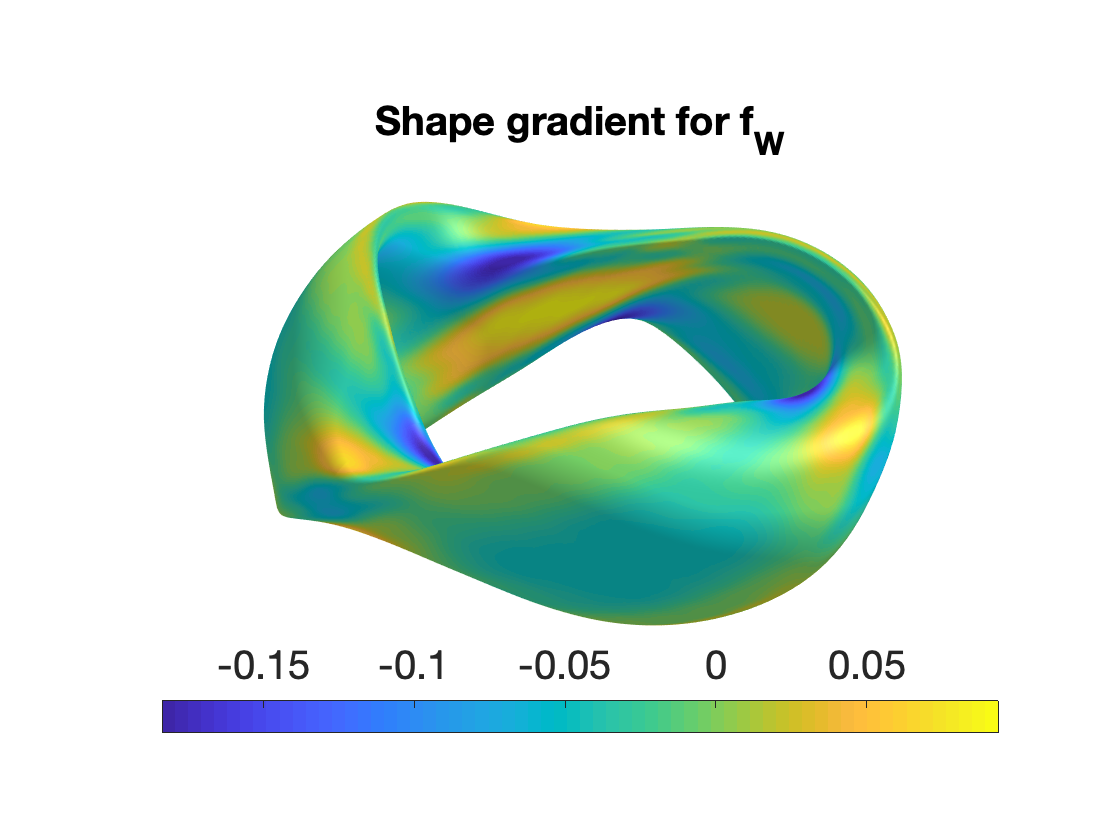}
    \caption{Direct}
    \label{fig:direct}
    \end{subfigure}
    \begin{subfigure}[b]{0.49\textwidth}
    \includegraphics[trim=1cm 6cm 1cm 6cm,clip,width=1.0\textwidth]{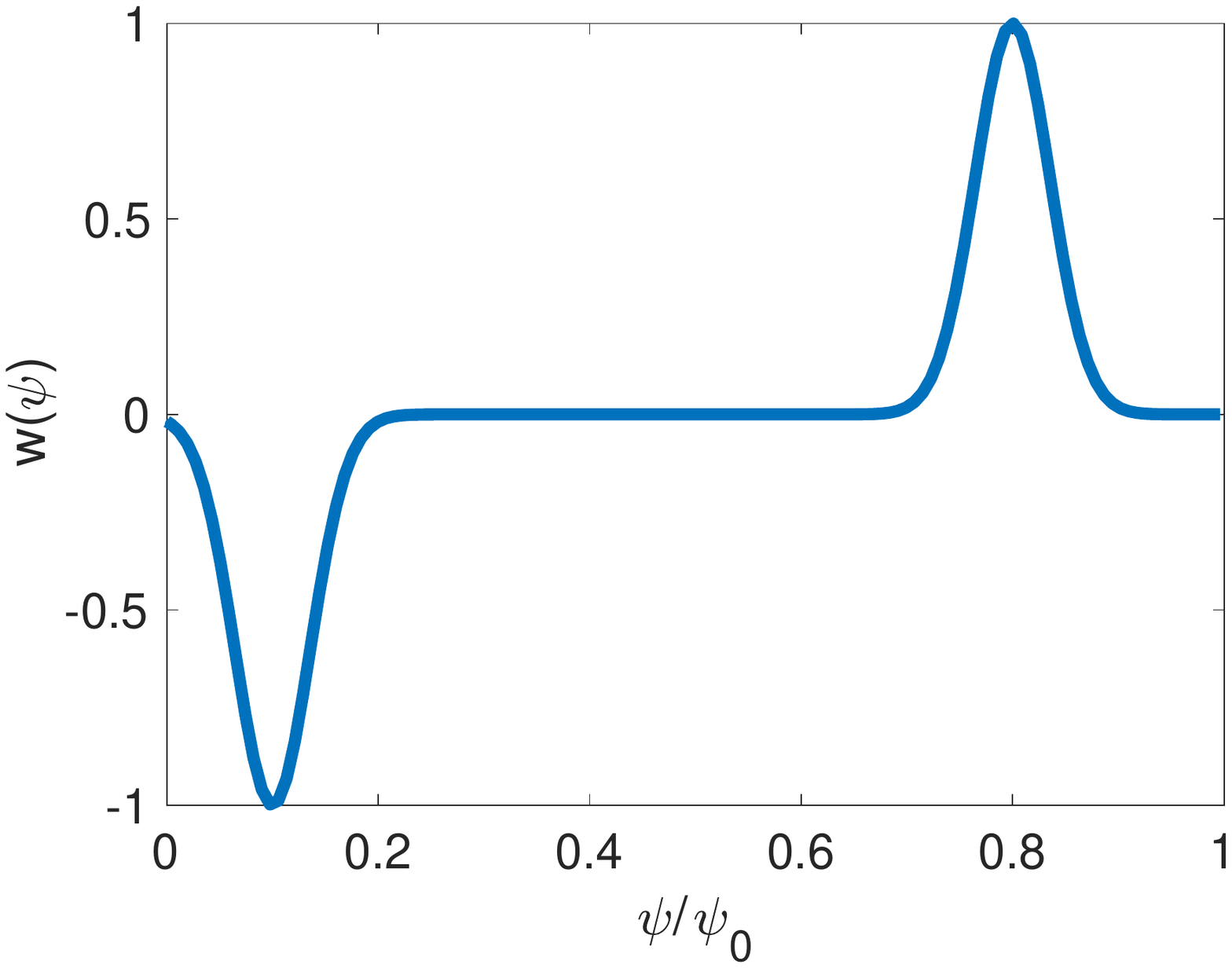}
    \caption{Weight function}
    \label{fig:weight}
    \end{subfigure}
    \caption{The shape gradient for $f_W$ \eqref{eq:f_w_v} is computed using the (a) adjoint and (b) direct approaches. (c) The weight function \eqref{eq:weight} used to compute $f_W$. Figure reproduced from \cite{Paul2020} with permission.}
    \label{fig:vacuum_well}
\end{figure}

In Figure \ref{fig:vacuum_well} we present the computation of $\mathcal{G}_{W}$ for the NCSX LI383 equilibrium \citep{Zarnstorff2001} using the the adjoint and direct approaches. We use a weight function,
\begin{gather}
    w(\psi) = \exp(-(\psi-\psi_{m,1})^2/\psi_w^2)- \exp(-(\psi-\psi_{m,2})^2/\psi_w^2),
    \label{eq:weight}
\end{gather}
such that $f_W$ remains smooth while it approximates $V'(\psi_{m,1})-V'(\psi_{m,2})$ where $\psi_{m,1} = 0.8 \psi_0$, $\psi_{m,2} = 0.1 \psi_0$, and $\psi_{w} = 0.05 \psi_0$ (Figure \ref{fig:weight}). We note that $f_W$ can be interpreted as measuring the change in volume due to the interchange of two flux tubes centered at $\psi_{m,1}$ and $\psi_{m,2}$. If $f_W>0$, this indicates that moving a flux tube radially outward will cause it to expand and lower its potential energy.

The adjoint magnetic field is computed with a forward-difference approximation \eqref{eq:adjoint_forward_difference} characterized by a step size $\Delta_P = 400$ Pa. For the direct approach, derivatives with respect to the Fourier discretization \eqref{eq:rmnc_zmns_ch5} of the boundary are computed for $m \le 20$ and $|n| \le 10$ using an 8-point centered-difference stencil with a polynomial-fitting technique. The direct approach requires 6889 calls to VMEC while the adjoint approach requires two calls. It is clear from Figure \ref{fig:vacuum_well} that the adjoint approach yields the same gradient information as the finite-difference approach, at much lower computational cost. The small difference between Figures \ref{fig:adjoint} and \ref{fig:direct} can is quantified using \eqref{eq:residual}, with a surface-averaged value of $\mathcal{G}_{\text{residual}}$ is $3.8\times 10^{-2}$. 

\subsubsection{Coil shape gradient} 
\label{sec:coil_vacuum_well}

\begin{figure}
    \centering
    \begin{subfigure}[b]{0.49\textwidth}
    \includegraphics[trim=8cm 4cm 4cm 3cm,clip,width=1.0\textwidth]{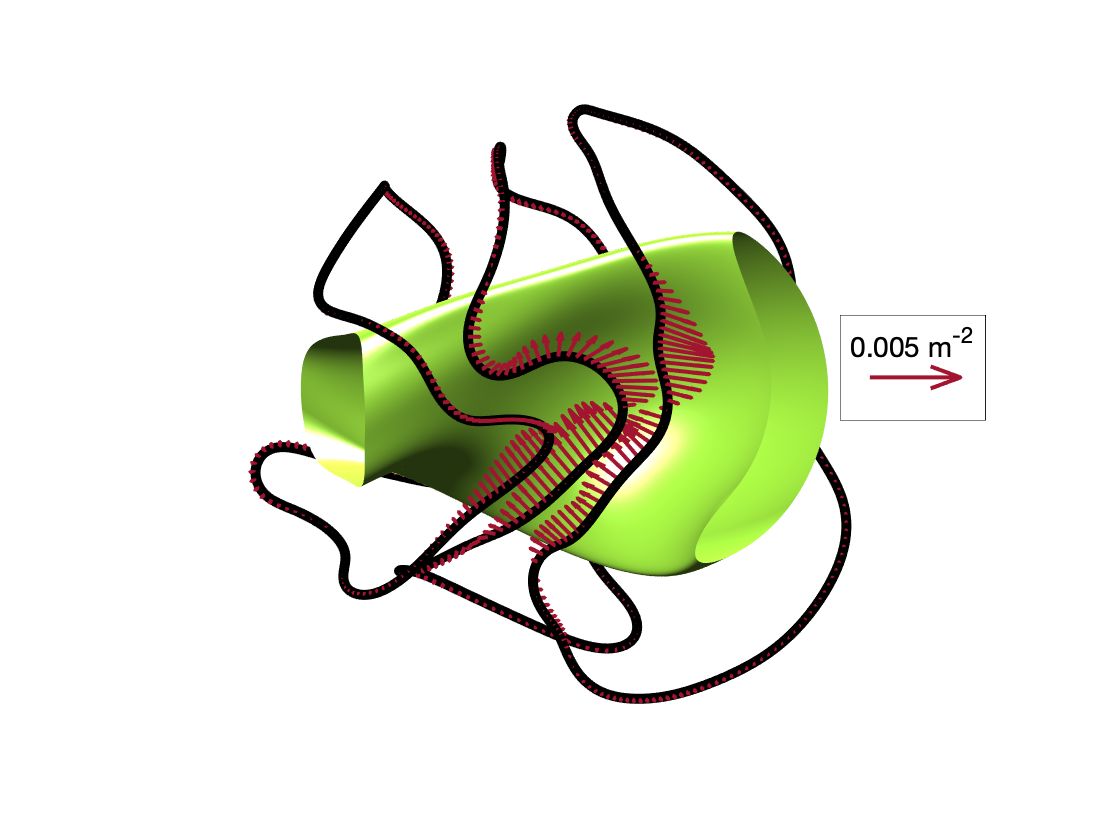}
    \caption{Adjoint}
    \end{subfigure}
    \begin{subfigure}[b]{0.49\textwidth}
    \includegraphics[trim=8cm 4cm 4cm 3cm,clip,width=1.0\textwidth]{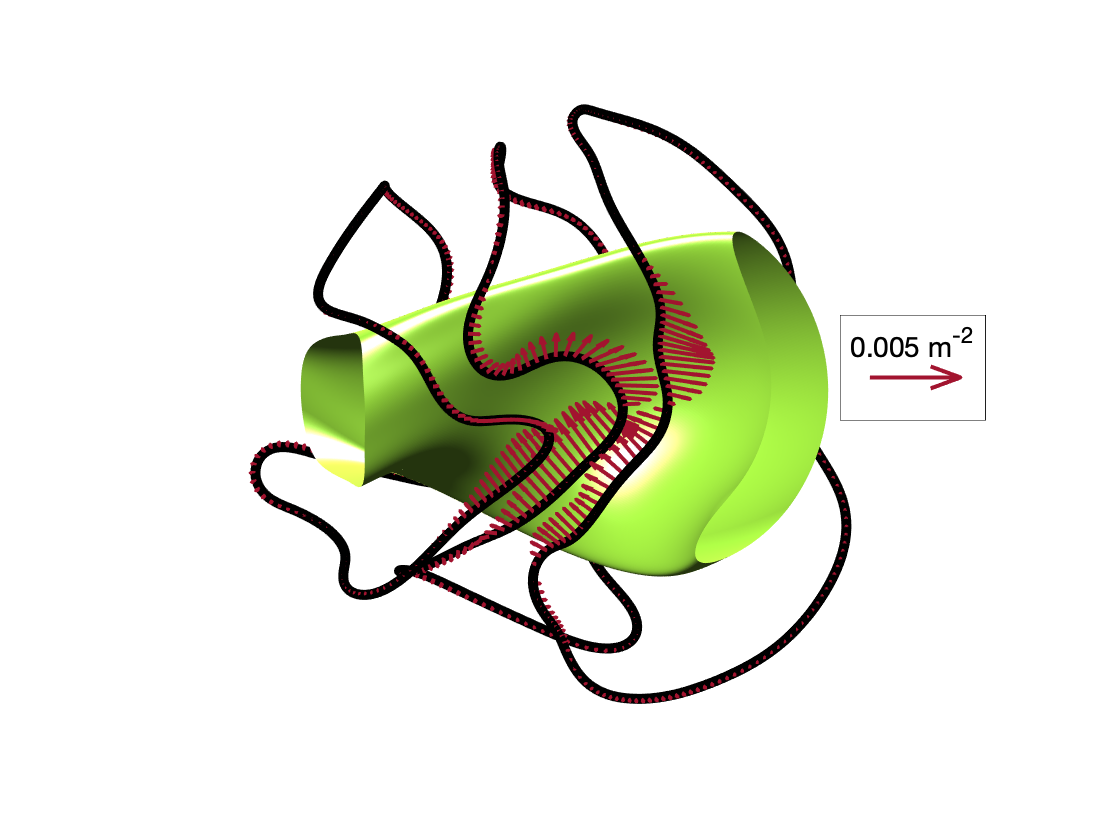}
    \caption{Direct}
    \end{subfigure}
    \caption{The coil shape gradient for $f_W$ is calculated for each of the 3 unique NCSX coil shapes. The arrows indicate the direction of $ \widetilde{\bm{\mathcal{G}}}_k$ \eqref{eq:well_coil_shape_gradient}, and their lengths indicate the magnitude scaled according to the legend. Figure reproduced from \cite{Paul2020} with permission.}
    \label{fig:coil_shape_gradient}
\end{figure}

The shape derivative of $f_W$ can also be computed with respect to a perturbation of the coil shapes. We consider perturbations about an equilibrium with fixed toroidal current, 
\begin{subequations}
\begin{align}
    \textbf{F}[\bm{\xi}_1,\delta \chi_1(\psi)] &= 0 \\
    \delta I_{T,1}(\psi) &= 0,
\end{align}
\end{subequations}
with specified perturbation to the coils shapes, $\delta \textbf{x}_{C_1} \times \hat{\textbf{t}}$. We prescribe the following adjoint perturbation,
\begin{subequations}
\begin{align}
    \textbf{F}[\bm{\xi}_2,\delta \chi_2(\psi)]  -\nabla w(\psi) &= 0 \\
    \delta\textbf{x}_{C_2} \times \hat{\textbf{t}} &= 0 \\
    \delta I_{T,2}(\psi) &= 0.
\end{align}
\end{subequations}
The same weight function \eqref{eq:weight} is applied, which decreases sufficiently fast that we can approximate $w(\psi_0) = 0$. Upon application of the free-boundary adjoint relation \eqref{eq:free_boundary}, we obtain the following coil shape gradient,
\begin{gather}
    \widetilde{\bm{\mathcal{G}}}_k = \frac{I_{C_k} \hat{\textbf{t}} \times \delta \textbf{B}_2}{\mu_0} \bigg \rvert_{C_k}.
    \label{eq:well_coil_shape_gradient}
\end{gather}
The calculation of $\widetilde{\bm{\mathcal{G}}}_k$ for each of the 3 unique coil shapes from the NCSX C09R00 coil set is shown in Figure \ref{fig:coil_shape_gradient}. A two-point centered-difference approximation of the adjoint magnetic field \eqref{eq:adjoint_forward_difference} is applied with characteristic step size $\Delta_P = 3\times 10^3$ Pa. The adjoint field is evaluated in the vacuum region using the DIAGNO code. The shape gradient is also computed with a direct approach. The Cartesian components of each coil are Fourier-discretized \eqref{eq:coil_discretization}, and derivatives are computed with respect to modes with $m \le 40$ with a 4-point centered-difference stencil. The fractional difference between the results obtained with the two approaches is quantified with \eqref{eq:residual_coil}. The line-averaged value of $\widetilde{\mathcal{G}}_{\text{residual},k}^l$ is $4.1\times 10^{-2}
$. The direct approach required 2917 VMEC calls while the adjoint only required three.

\subsection{Ripple on magnetic axis}
\label{sec:ripple}

We now consider a figure of merit which quantifies the ripple near the magnetic axis \citep{Carreras1996,Drevlak2014,Drevlak2018}.
As all physical quantities must be independent of the poloidal angle on the magnetic axis, 
this quantifies the departure from quasi-helical or quasi-axisymmetry near the magnetic axis. We define the magnetic ripple to be,
\begin{align}
    f_R &=  \int_{V_P} d^3 x \, \widetilde{f}_R,
    \label{eq:f_R}
\end{align}
with,
\begin{subequations}
\begin{align}
   \widetilde{f_R}(\psi,B) &= \frac{1}{2} w(\psi) \left( B - \overline{B} \right)^2 \\
   \overline{B} &= \frac{\int_{V_P} d^3 x \, w(\psi) B}{\int_{V_P} d^3 x \, w(\psi) } ,
\end{align}
\end{subequations}
and a weight function given by, 
\begin{align}
    w(\psi) = \exp(-\psi^2/\psi_w^2)
    \label{eq:weight_ripple},
\end{align}
with $\psi_w = 0.1 \psi_0$.

\subsubsection{Surface shape gradient}
\label{sec:surf_ripple}

We compute perturbations about an equilibrium with fixed rotational transform \eqref{eq:direct_fixed_iota}. Noting that the local perturbation to the field strength is given by \eqref{eq:delta_mod_B}, the shape derivative is computed with the transport theorem \eqref{eq:transport_theorem},
\begin{gather}
        \delta f_R(S_P;\bm{\xi}_1) = \int_{S_P} d^2 x \, \bm{\xi}_1 \cdot \hat{\textbf{n}} \widetilde{f_R} + \int_{V_P} d^3 x \, \left(\partder{\widetilde{f_R}(\psi,B)}{B} \delta B + \partder{\widetilde{f_R}(\psi,B)}{\psi} \delta \psi \right).
        \label{eq:delta_fR}
\end{gather}
We prescribe the following adjoint perturbation,
\begin{subequations}
\begin{align}
  \textbf{F}[\bm{\xi}_2,\delta \chi(\psi)]  - \nabla \cdot \mathbf{P} &= 0  \label{eq:ripple_F} \\
    \bm{\xi}_2 \cdot \hat{\textbf{n}} |_{S_P} &= 0 \\
    \delta \chi_2'(\psi) &= 0. \label{eq:ripple_iota}
\end{align}
\label{eq:ripple_adjoint}
\end{subequations}
The bulk force perturbation required for the adjoint problem is written as the divergence of an anisotropic pressure tensor, $\textbf{P} = p_{\perp} \textbf{I} + (p_{||}-p_{\perp})\hat{\textbf{b}}\hat{\textbf{b}}$ where $\textbf{I}$ is the identity tensor. The parallel and perpendicular pressures are related by the parallel force balance condition,
\begin{gather}
    \partder{p_{||}(\psi,B)}{B}  = \frac{p_{||}-p_{\perp}}{B}, 
    \label{eq:par_force_balance}
\end{gather}
which follows from the requirement that $\hat{\textbf{b}} \cdot \delta \textbf{F}_2 = 0$ \eqref{eq:perturbed_force_balance}. We take the parallel pressure to be,
\begin{gather}
    p_{||} = \widetilde{f_R}.
    \label{eq:p_||}
\end{gather}

Upon application of the fixed-boundary adjoint relation and the expression for the curvature in an equilibrium field, 
\begin{align}
    \bm{\kappa} = \frac{\nabla_{\perp} B}{B} + \frac{\nabla p}{\mu_0 B^2},
    \label{eq:curvature_equilibrium}
\end{align}
we obtain the following shape gradient,
\begin{gather}
    \mathcal{G}_R = \left( p_{\perp} +\frac{\delta \textbf{B}_2 \cdot \textbf{B}}{\mu_0} \right)_{S_P}. 
    \label{eq:well_shape_gradient}
\end{gather}
If instead the toroidal current is held fixed in the direct perturbation as in \eqref{eq:direct_fixed_current}, then the required adjoint current perturbation is given by,
\begin{align}
    \delta I_{T,2}(\psi) &= \frac{V'(\psi)}{2\pi} \left \langle \partder{\widetilde{f}_R(\psi,B)}{B} \hat{\textbf{b}} \cdot \nabla \varphi \times \nabla \psi \right \rangle_{\psi} \label{eq:ripple_I},
\end{align}
with the shape gradient unchanged. See Appendix \ref{app:axis_ripple} for details of the calculation. 

To compute the adjoint perturbation \eqref{eq:ripple_adjoint}-\eqref{eq:ripple_I}, we consider the addition of an anisotropic pressure tensor to the nonlinear force balance equation,
\begin{gather}
    \textbf{J}' \times \textbf{B}'= \nabla p' + \Delta_P \nabla \cdot \textbf{P}(\psi',B'),
    \label{eq:force_balance_animec}
\end{gather}
where $\textbf{P}(\psi',B') = p_{\perp}(\psi',B') \textbf{I} + \left(p_{||}(\psi',B')-p_{\perp}(\psi',B')\right)\hat{\textbf{b}}' \hat{\textbf{b}}'$. Here primes indicate the perturbed quantities (i.e. $B' = B + \delta B$) where unprimed quantities satisfy \eqref{eq:force_balance}. As in Section \ref{sec:vacuum_well}, the perturbation has a scale set by $\Delta_P$ which is chosen to be small enough that the response is linear. Enforcing parallel force balance from \eqref{eq:force_balance_animec} results in the following condition,
\begin{gather}
    \partder{p_{||}(\psi',B')}{B'} = \frac{p_{||}(\psi',B')-p_{\perp}(\psi',B')}{B'}.
    \label{eq:force_balance_||}
\end{gather}
If we furthermore assume that $\Delta_P \nabla \cdot \textbf{P}$ is small compared with the other terms in \eqref{eq:force_balance_animec}, we can consider it to be a perturbation to the base equilibrium \eqref{eq:force_balance}. In this way, we can apply the perturbed force balance equation \eqref{eq:perturbed_force_balance} with $\delta \textbf{F}_{2} = - \Delta_P \nabla \cdot \textbf{P}(\textbf{B})$, where $\textbf{P}$ is now evaluated with the equilibrium field which satisfies \eqref{eq:force_balance}. Thus the desired pressure tensor \eqref{eq:p_||} can be implemented by evaluating $p_{||}$ with the perturbed field such that \eqref{eq:force_balance_||} is satisfied. 

We have implemented the pressure tensor defined by  \eqref{eq:par_force_balance}-\eqref{eq:p_||} in the ANIMEC code \citep{Cooper19923d}, which modifies the VMEC variational principle to allow 3D equilibrium solutions with anisotropic pressures to be computed. The ANIMEC code has been used to model equilibria with energetic particle species using pressure tensors based on bi-Maxwellian \citep{Cooper2006} and slowing-down \citep{Cooper2005} distribution functions. 
The variational principle assumes that $p_{||}$ only varies on a surface through $B$ and can, therefore, be used to include the required adjoint bulk force. 

\begin{figure}
    \centering
    \begin{subfigure}[b]{0.49\textwidth}
    \includegraphics[trim=3cm 2cm 4cm 6cm,clip,width=1.0\textwidth]{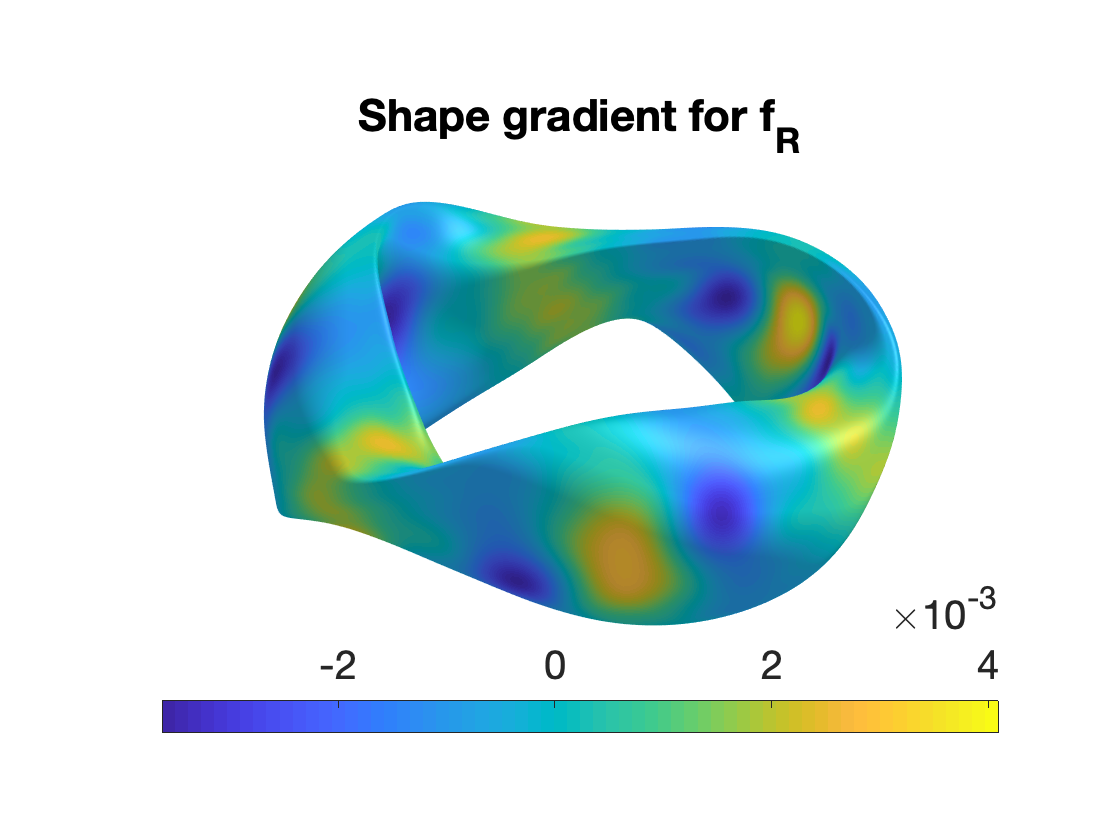}
    \caption{Adjoint}
    \end{subfigure}
    \begin{subfigure}[b]{0.49\textwidth}
    \includegraphics[trim=3cm 2cm 4cm 6cm,clip,width=1.0\textwidth]{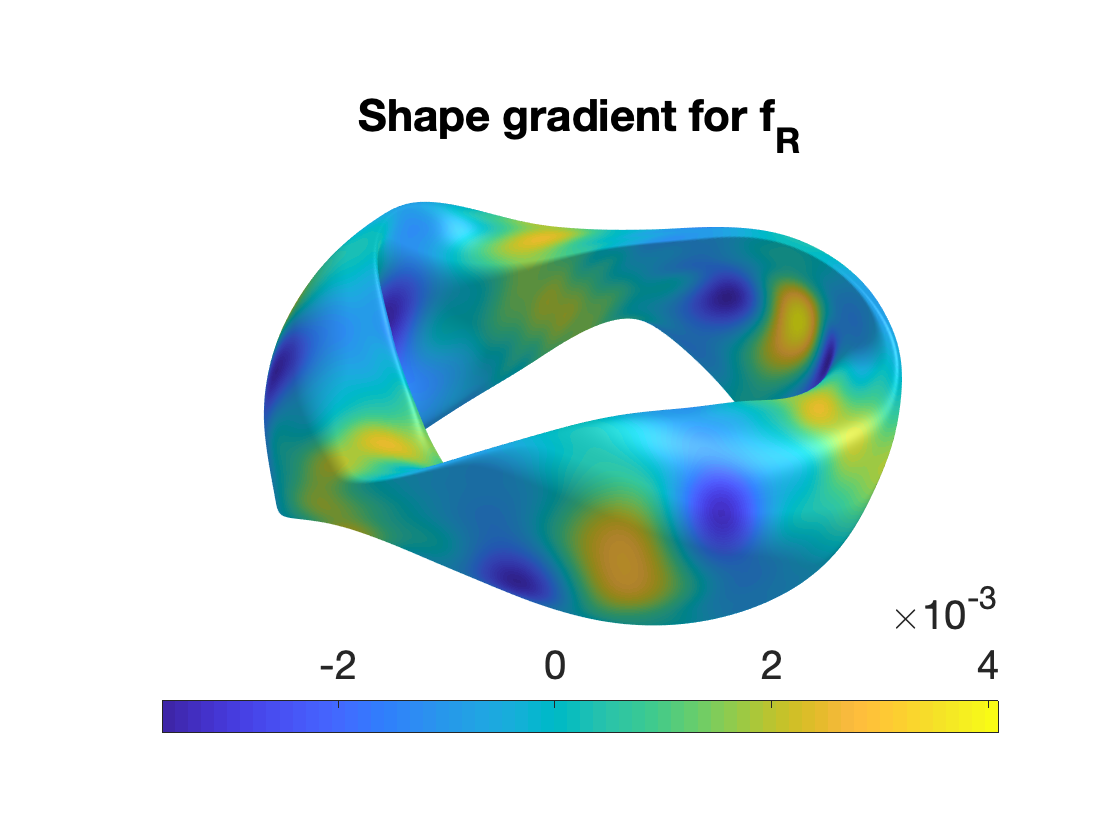}
    \caption{Direct}
    \end{subfigure}
    \begin{subfigure}[b]{0.49\textwidth}
    \includegraphics[trim=1cm 0cm 1cm 1cm,clip,width=1.0\textwidth]{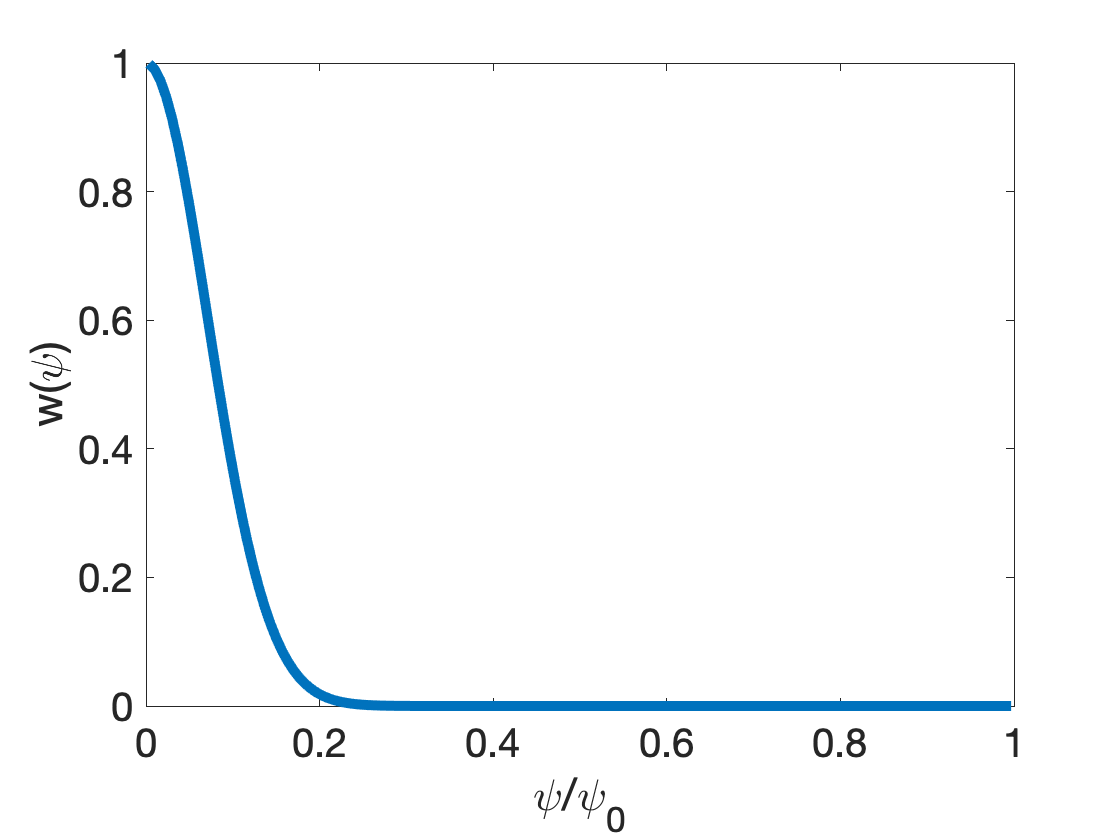}
    \caption{Weight function}
    \end{subfigure}
    \caption{The shape gradient for $f_{R}$ \eqref{eq:f_R} is computed using the (a) adjoint and (b) direct approaches with a weight function \eqref{eq:weight_ripple} shown in (c). Figure reproduced from \cite{Paul2020} with permission.}
    \label{fig:ripple}
\end{figure}

In Figure \ref{fig:ripple}, we present the computation of $\mathcal{G}_R$ for the NCSX LI383 equilibrium using the adjoint and direct approaches. For the direct approach, derivatives with respect to the Fourier discretization of the boundary \eqref{eq:rmnc_zmns_ch5} are computed for $m \le 11$ and $|n| \le 7$ using an 8-point centered-difference stencil. The adjoint field is computed from a forward-difference approximation \eqref{eq:adjoint_forward_difference} with a characteristic step size of $\Delta_P = 7.96\times 10^{3}$ Pa. The direct approach required 2761 calls to VMEC while the adjoint approach required two calls. The surface-averaged value of $\mathcal{G}_{\text{residual}}$ \eqref{eq:residual} is $3.3\times 10^{-2}$.

\subsection{Effective ripple in the $1/\nu$ regime}
\label{sec:epsilon_eff}

The effective ripple in the $1/\nu$ regime \citep{Nemov1999} is a figure of merit which has proven valuable for neoclassical optimization (e.g. \cite{Zarnstorff2001,Ku2008,Henneberg2019}). This quantity characterizes the geometric dependence of the neoclassical particle flux under the assumption of low-collisionality 
such that $\epsilon_{\text{eff}}$ is analogous to the helical ripple amplitude, $\epsilon_h$, that appears in the expression of the $1/\nu$ particle flux for a classical stellarator \citep{Galeev1979}. The following expression is obtained for the effective ripple,
\begin{gather}
    \epsilon_{\text{eff}}^{3/2}(\psi) = \frac{\pi}{4\sqrt{2}V'(\psi) \epsilon_{\text{ref}}^2} \int_{1/B_{\max}}^{1/B_{\min}} \frac{d \lambda}{\lambda} \,  \int_0^{2\pi} d \alpha \, \sum_i \frac{(\partder{}{\alpha}\hat{K}_i(\alpha,\lambda))^2}{\hat{I}_i(\alpha,\lambda)}.
\label{eq:eps_eff}
\end{gather}
Here $\lambda = v_{\perp}^2/(v^2B)$ is the pitch angle, $B_{\min}$ and $B_{\max}$ are the minimum and maximum values of the field strength on a surface labeled by $\psi$, and $\epsilon_{\text{ref}}$ is a reference aspect ratio. We have defined the bounce integrals,
\begin{subequations}
\begin{align}
    \hat{I}_i(\alpha,\lambda) &= \oint dl \, \frac{v_{||}}{Bv}\label{eq:I_hat} \\
    \hat{K}_i(\alpha,\lambda) &= \oint dl \, \frac{v_{||}^3}{Bv^3} \label{eq:K_hat},
\end{align}
\label{eq:bounce_integrals}
\end{subequations}
where the notation $\oint dl = \sum_{\sigma} \sigma \int_{\varphi_-}^{\varphi_+} d \varphi/\hat{\textbf{b}} \cdot \nabla \varphi$ indicates integration at constant $\lambda$ and $\alpha$ between successive bounce points where $v_{||}(\varphi_+) = v_{||}(\varphi_-) = 0$ and $\sigma = \text{sign}(v_{||})$. The sum in \eqref{eq:eps_eff} is taken over wells at constant $\lambda$ and $\alpha$ for $\varphi_{-,i} \in [0,2\pi)$.

We consider an integrated figure of merit,
\begin{gather}
f_{\epsilon} = \int_{V_P} d^3 x \, w(\psi) \epsilon_{\text{eff}}^{3/2}(\psi),
\label{eq:f_epsilon}
\end{gather}
where $w(\psi)$ is a radial weight function.
We perturb about an equilibrium with fixed toroidal current \eqref{eq:direct_fixed_current}.
The shape derivative of $f_{\epsilon}$ is computed to be,
\begin{gather}
    \delta f_{\epsilon}(S_P;\bm{\xi}_1) = \int_{V_P} d^3 x \, \left(\textbf{P}_{\epsilon}: \nabla \bm{\xi}_1 + \delta \chi_1'(\psi) \mathcal{I}_{\epsilon}\right)
    ,
    \label{eq:df_epsilon}
\end{gather}
where the double dot (:) indicates contraction between dyadic tensors $\textbf{A}$ and $\textbf{B}$ as $\textbf{A} : \textbf{B} = \sum_{i,j} A_{ij} B_{ji}$, with,
\begin{multline}
\mathcal{I}_{\epsilon} = \frac{\pi w(\psi)}{2\sqrt{2} \epsilon_{\text{ref}}^2} \int_{1/B_{\max}}^{1/B} \frac{d \lambda}{\lambda} \,  \\ \times \Bigg[\frac{\left(\partder{}{\alpha} \hat{K}(\alpha,\lambda,\varphi) \right)^2}{\hat{I}^2(\alpha,\lambda,\varphi)}  \left(-\varphi \textbf{B} \times \nabla \psi \cdot \nabla  \left(\frac{|v_{||}|}{vB^2} \right) + \textbf{B} \times \nabla \psi \cdot \nabla \varphi \partder{}{B} \left( \frac{|v_{||}|}{vB} \right) \right)\\ + 2\partder{}{\alpha} \left(\frac{\partder{}{\alpha} \hat{K}(\alpha,\lambda,\varphi) }{\hat{I}(\alpha,\lambda,\varphi)} \right) \left( - \varphi \textbf{B} \times \nabla \psi \cdot \nabla \left(\frac{|v_{||}|^3}{v^3 B^2} \right) +\textbf{B} \times \nabla \psi \cdot \nabla \varphi \partder{}{B} \left(\frac{|v_{||}|^3}{v^3 B} \right) \right)\Bigg],
\end{multline}
and $\textbf{P}_{\epsilon} = p_{||} \hat{\textbf{b}} \hat{\textbf{b}} + p_{\perp} (\textbf{I}-\hat{\textbf{b}}\hat{\textbf{b}})$ with,
\begin{subequations}
\begin{align}
    p_{||} = -\frac{\pi w(\psi)}{2\sqrt{2}\epsilon_{\text{ref}}^2} \int_{1/B_{\max}}^{1/B} \frac{d \lambda}{\lambda} \,  \Bigg( \frac{\left(\partder{}{\alpha} \hat{K}(\alpha,\lambda,\varphi) \right)^2}{\hat{I}^2(\alpha,\lambda,\varphi)}  \frac{|v_{||}|}{v}+2\partder{}{\alpha} \left(\frac{\partder{}{\alpha} \hat{K}(\alpha,\lambda,\varphi) }{\hat{I}(\alpha,\lambda,\varphi)} \right)\frac{|v_{||}|^3}{v^3 }\Bigg)
    \label{eq:p_perp}
\end{align}
\begin{multline}
   p_{\perp} = - \frac{\pi w(\psi)}{2\sqrt{2}\epsilon_{\text{ref}}^2} \int_{1/B_{\max}}^{1/B} \frac{d \lambda}{\lambda} \,  \Bigg( \frac{\left(\partder{}{\alpha} \hat{K}(\alpha,\lambda,\varphi) \right)^2}{\hat{I}^2(\alpha,\lambda,\varphi)}  \left(\frac{\lambda vB}{2|v_{||}|} + \frac{|v_{||}|}{v} \right) \\ +2\partder{}{\alpha} \left(\frac{\partder{}{\alpha} \hat{K}(\alpha,\lambda,\varphi) }{\hat{I}(\alpha,\lambda,\varphi)} \right)\left( \frac{3\lambda |v_{||}|B}{2v} + \frac{|v_{||}|^3}{v^3} \right) \Bigg).
    \label{eq:p_par}
\end{multline}
\label{eq:p_ripple}
\end{subequations}
Derivatives are computed assuming $\epsilon_{\text{ref}}$ is held constant. The bounce integrals are defined with respect to $\varphi$ such that $\hat{I}(\alpha,\lambda,\varphi) = \hat{I}_i$ if $\varphi \in [\varphi_{-,i},\varphi_{+,i}]$ and $\hat{I}(\alpha,\lambda,\varphi) = 0$ if $\lambda B(\alpha,\varphi) > 1$. The same convention is used for $\hat{K}(\alpha,\lambda,\varphi)$. 
We prescribe the following adjoint perturbation,
\begin{subequations}
\begin{align}
    \textbf{F}[\bm{\xi}_2,\delta \chi_2(\psi)] - \nabla \cdot \textbf{P}_{\epsilon} &= 0 \\
    \bm{\xi}_2 \cdot \hat{\textbf{n}}\rvert_{S_P} &= 0 \\
     \delta I_{T,2}(\psi) &= \frac{V'(\psi)}{2 \pi}   \langle \mathcal{I}_{\epsilon}\rangle_{\psi}.
\end{align}
\end{subequations}
The adjoint bulk force must be consistent with parallel force balance from \eqref{eq:perturbed_force_balance}, which is equivalent to the condition,
\begin{gather}
    \nabla_{||} p_{||} = \frac{\nabla_{||} B}{B} (p_{||}-p_{\perp}).
\end{gather}
This can be shown to be satisfied by \eqref{eq:p_ripple}, noting that the $\lambda$ integrand vanishes at $1/B$ such that there is no contribution from the parallel gradient acting on the bounds of the integral. There is also no contribution to the parallel gradient from the bounce-integrals, as $|v_{||}|$ vanishes at points of non-zero gradient of $\hat{I}(\alpha,\lambda,\varphi)$ and $\hat{K}(\alpha,\lambda,\varphi)$.

Upon application of the fixed-boundary adjoint relation \eqref{eq:fixed_boundary} and integration by parts, we obtain the following expression for the shape gradient,
\begin{gather}
    \mathcal{G}_{\epsilon} = \left(p_{\perp} + \frac{\delta \textbf{B} \cdot \textbf{B}}{\mu_0}\right)_{S_P}.
\end{gather}
See Appendix \ref{app:1_over_nu} for details of the calculation. The approach demonstrated in this Section could be extended to compute the shape gradients of other figures of merit involving bounce integrals, such as the $\Gamma_c$ metric for energetic particle confinement \citep{Nemov2005} or the variation of the parallel adiabatic invariant on a flux surface \citep{Drevlak2014}.

\subsection{Departure from quasi-symmetry}
\label{sec:quasisymmetry}

Quasi-symmetry is desirable as it ensures collisionless confinement of guiding centers. This property follows when the field strength depends on a linear combination of the Boozer angles, $B(\psi,\vartheta_B,\varphi_B) = B(\psi,M\vartheta_B-N\varphi_B)$ for fixed integers $M$ and $N$ \citep{Nuhrenberg1988,Boozer1995} (Appendix \ref{sec:quasisymmetry}). Several stellarator configurations have been optimized to be close to quasi-symmetry (e.g., \cite{Reiman1999,Drevlak2013,Henneberg2019,Liu2018}) by minimizing the amplitude of symmetry-breaking Fourier harmonics of the field strength. We will consider a figure of merit that does not require a Boozer coordinate transformation; instead, we use a general set of magnetic coordinates $(\psi,\vartheta,\varphi)$ to define our figure of merit.

In Boozer coordinates \citep{Boozer1981,Helander2014} ($\psi,\vartheta_B,\varphi_B$) the covariant form for the magnetic field is,
\begin{gather}
    \textbf{B} = I(\psi) \nabla \vartheta_B + G(\psi) \nabla \varphi_B + K(\psi,\vartheta_B,\varphi_B) \nabla \psi.
    \label{eq:boozer_covariant}
\end{gather}
Here $G(\psi) = \mu_0 I_P(\psi)/(2\pi)$, where $I_P(\psi)$ is the poloidal current outside the $\psi$ surface. The poloidal current can be computed using Ampere's law and expressed as an integral over a surface labeled by $\psi$, $S_P(\psi)$,
\begin{align}
    I_P(\psi) &= \frac{1}{\mu_0} \int_0^{2\pi} d \varphi \,  \textbf{B} \cdot \partder{\textbf{x}}{\varphi}
    = -\frac{1}{2\pi \mu_0} \int_{S_P(\psi)} d^2 x \, \textbf{B} \cdot \nabla \vartheta \times \hat{\textbf{n}}.
    \label{eq:poloidal_current}
\end{align}
The quantity $I(\psi) = \mu_0 I_T(\psi)/(2\pi)$, where $I_T(\psi)$ is the toroidal current inside the $\psi$ surface \eqref{eq:toroidal_curr}. We quantify the departure from quasi-symmetry in the following way,
\begin{gather}
    f_{QS} = \frac{1}{2}\int_{V_P} d^3 x \, w(\psi) \left(\textbf{B} \times \nabla \psi \cdot \nabla B -  F(\psi)\textbf{B}\cdot \nabla B\right)^2.
    \label{eq:f_QS}
\end{gather}
Here $w(\psi)$ is a radial weight function and,
\begin{gather}
    F(\psi) = \frac{(M/N)G(\psi) + I(\psi)}{(M/N)\iota(\psi)-1}.
\end{gather}
If $f_{QS} = 0$, then the field is quasi-symmetric with mode numbers $M$ and $N$ \citep{Helander2014}, which can be shown using the covariant \eqref{eq:magnetic_contravariant} and contravariant \eqref{eq:boozer_covariant} representations of the magnetic field assuming $B=B(\psi,M\vartheta_B-N\varphi_B)$ for fixed $M$ and $N$. Note that $f_{QS}$ quantifies the symmetry in Boozer coordinates but can be evaluated in any flux coordinate system.

We consider perturbation about an equilibrium with fixed toroidal current \eqref{eq:direct_fixed_current}. The perturbations to the Boozer poloidal covariant component is computed using the transport theorem \eqref{eq:transport_theorem}, 
\begin{align}
     \delta G(\psi) &= -\frac{1}{4\pi^2} \int_{S_P(\psi)} d^2 x \, \left(\nabla \cdot \left( \textbf{B} \times \nabla \vartheta \right) \bm{\xi}_1 \cdot \hat{\textbf{n}} + \delta \textbf{B} \times \nabla \vartheta \cdot \hat{\textbf{n}} \right).
     \label{eq:delta_G_1}
\end{align}
In arriving at \eqref{eq:delta_G_1} we have used the fact that spatial derivatives commute with shape derivatives. The first term accounts for the unperturbed current density through the perturbed boundary, and the second accounts for the perturbed current density through the unperturbed boundary. The contribution from the perturbation to the poloidal angle can be shown to vanish. Upon application of \eqref{eq:delta_B} we obtain, noting that $\int_{S_P(\psi)} d^2 x \, A = V'(\psi) \langle A |\nabla \psi | \rangle_{\psi}$ for any quantity $A$,
\begin{multline}
     \delta G(\psi) = 
     -\frac{V'(\psi)}{4\pi^2} \Bigg \langle \bm{\xi}_1 \cdot \nabla \psi \nabla \cdot (\textbf{B} \times \nabla \vartheta )\\
     - \frac{1}{\sqrt{g}} \partder{\textbf{x}}{\varphi} \cdot \nabla \times \left(\bm{\xi}_1 \times \textbf{B} \right) - \frac{\delta \chi_1'(\psi)}{ \sqrt{g}^{2}} \partder{\textbf{x}}{\varphi} \cdot \partder{\textbf{x}}{\vartheta} \Bigg \rangle_{\psi} \label{eq:delta_G},
\end{multline}
Applying the transport theorem \eqref{eq:transport_theorem}, the shape derivative of $f_{QS}$ takes the form, 
\begin{multline}
    \delta f_{QS}(S_P;\bm{\xi}_1) =\frac{1}{2} \int_{S_P} d^2 x \, \bm{\xi}_1 \cdot \hat{\textbf{n}} \mathcal{M}^2 w(\psi) + \frac{1}{2} \int_{V_P} d^3 x \, w'(\psi) \delta \psi \mathcal{M}^2 \\
      + \int_{V_P} d^3 x \, w(\psi) \mathcal{M} \left( \delta \textbf{B} \cdot \bm{\mathcal{A}} + \bm{\mathcal{S}} \cdot \nabla \delta B + \textbf{B} \times \nabla \delta \psi \cdot \nabla B  -  \frac{\delta G(\psi) \textbf{B} \cdot \nabla B}{\iota(\psi)-(N/M)}  \right) 
      \\ + \int_{V_P} d^3 x \, w(\psi) \mathcal{M} \left( \frac{F(\psi)}{\iota(\psi)-(N/M)} \delta \chi_1'(\psi) \textbf{B} \cdot \nabla B -  \delta \psi F'(\psi)\textbf{B} \cdot \nabla B \right),
      \label{eq:df_QS1}
\end{multline}
where $\mathcal{M} = \textbf{B} \times \nabla \psi \cdot \nabla B - F(\psi) \textbf{B} \cdot \nabla B$, $\bm{\mathcal{A}} = \nabla \psi \times \nabla B - F(\psi) \nabla B$, and $\bm{\mathcal{S}} = \textbf{B} \times \nabla \psi - F(\psi) \textbf{B}$. After several steps outlined in Appendix \ref{app:qs}, the shape derivative can be written in the following way,
\begin{gather}
    \delta f_{QS}(S_P;\bm{\xi}_1) = \int_{V_P} d^3 x \, \left(\bm{\xi}_1 \cdot \bm{\mathcal{F}}_{QS} + \delta \chi_1'(\psi) \mathcal{I}_{QS} \right) + \int_{S_P} d^2 x \, \bm{\xi}_1 \cdot \hat{\textbf{n}} \mathcal{B}_{QS},
    \label{eq:df_QS}
\end{gather}
with,
\begin{subequations}
\begin{multline}
    \bm{\mathcal{F}}_{QS} =\frac{1}{2} \nabla_{\perp} \left(w(\psi) \mathcal{M}^2 \right)
    + \left((\hat{\textbf{b}} \times \nabla \psi) \nabla_{||}B + F(\psi) \nabla_{\perp} B \right) w(\psi) \textbf{B} \cdot \nabla \mathcal{M} 
    \\ + \textbf{B} \times (\nabla \times (\nabla \psi \times \nabla B)) w(\psi) \mathcal{M} -B\nabla_{\perp} \left(w(\psi) \bm{\mathcal{S}} \cdot \nabla \mathcal{M}  \right)   + \bm{\kappa} Bw(\psi) \bm{\mathcal{S}} \cdot \nabla \mathcal{M}  \\ - \nabla \psi \nabla B \cdot \nabla \times \left(w(\psi) \mathcal{M} \textbf{B} \right) +\frac{1}{4\pi^2} \Bigg(- \nabla_{\perp} \left( \frac{w(\psi) V'(\psi) \langle \mathcal{M} \textbf{B} \cdot \nabla B \rangle_{\psi}}{(\iota(\psi)-(N/M))} \right) \left( \textbf{B} \cdot \nabla \psi \times \nabla \vartheta \right) \\
    +\frac{w(\psi) V'(\psi) \langle \mathcal{M} \textbf{B} \cdot \nabla B \rangle_{\psi}}{\iota(\psi)-(N/M)}\left(\nabla \psi\nabla \cdot \left( \textbf{B} \times \nabla \vartheta \right) - \textbf{B} \times \nabla \times \left(\nabla \psi \times \nabla \vartheta \right) \right) 
    \Bigg)
    \label{eq:F_QS}
\end{multline}
\begin{multline}
    \mathcal{B}_{QS} = -\frac{1}{2}w(\psi) \mathcal{M}^2 + B w(\psi) \bm{\mathcal{S}} \cdot \nabla \mathcal{M} - w(\psi) \mathcal{M} \nabla B \times \textbf{B} \cdot \nabla \psi \\ + \frac{w(\psi)V'(\psi) \langle \mathcal{M}\textbf{B} \cdot \nabla B \rangle_{\psi}}{4\pi^2(\iota(\psi)-(N/M))} \left(\textbf{B} \cdot \nabla \psi \times \nabla \vartheta  \right)
    \label{eq:B_QS}
\end{multline}
\begin{multline}
    \mathcal{I}_{QS} = - w(\psi) \mathcal{M} \nabla \psi \times \nabla \varphi \cdot \bm{\mathcal{A}} + w(\psi) \left(\bm{\mathcal{S}} \cdot \nabla \mathcal{M}\right) \hat{\textbf{b}} \cdot \nabla \psi \times \nabla \varphi  \\ + \frac{w(\psi) \mathcal{M} \textbf{B} \cdot \nabla B  }{\iota(\psi)-(N/M)}\left(F(\psi) -\left \langle \frac{V'(\psi)}{4\pi^2\sqrt{g}^2} \partder{\textbf{x}}{\varphi} \cdot \partder{\textbf{x}}{\vartheta} \right \rangle_{\psi} \right). 
    \label{eq:I_QS}
\end{multline}
\end{subequations}
In \eqref{eq:F_QS}, $\nabla_{||} = \hat{\textbf{b}} \cdot \nabla$ and $\nabla_{\perp} = \nabla- \hat{\textbf{b}} \nabla_{||}$ are the parallel and perpendicular gradients. 

We can now prescribe an adjoint perturbation which satisfies,
\begin{subequations}
\begin{align}
    \textbf{F}[\bm{\xi}_2,\delta \chi_2(\psi)] + \bm{\mathcal{F}}_{QS} &= 0 \\
    \bm{\xi}_2 \cdot \hat{\textbf{n}} |_{S_P} &= 0 \\
    \delta I_{T,2}(\psi) &= \frac{V'(\psi)}{2\pi}  \langle \mathcal{I}_{QS} \rangle_{\psi}.
\end{align}
\end{subequations}
We note that $\bm{\mathcal{F}}_{QS}$ satisfies the parallel force balance condition ($\hat{\textbf{b}} \cdot \bm{\mathcal{F}}_{QS}=0$) implied by \eqref{eq:perturbed_force_balance}. Upon application of the fixed-boundary adjoint relation we obtain the following shape gradient, \begin{gather}
    \mathcal{G}_{QS} = \left(\mathcal{B}_{QS} + \frac{\delta \textbf{B}_2 \cdot \textbf{B}}{\mu_0} \right)_{S_P}.
\end{gather}

\subsection{Neoclassical figures of merit}
\label{sec:neoclassical}

In Section \ref{sec:epsilon_eff}, we considered a figure of merit that quantifies the geometric dependence of the neoclassical particle flux in the $1/\nu$ regime. In applying this model, several assumptions are imposed, such as a small radial electric field, $E_r$, low collisionality, and a simplified pitch-angle scattering collision operator. In this Section, we consider a more general neoclassical figure of merit arising from a moment of the local drift kinetic equation, allowing for optimization at finite collisionality and $E_r$. It is assumed here that the collision time is comparable to the bounce time but shorter than the time needed to complete a magnetic drift orbit. In Chapter \ref{ch:adjoint_neoclassical}, an adjoint method is demonstrated for obtaining derivatives of neoclassical figures of merit with respect to local geometric quantities on a flux surface. The adjoint method described in this Section will extend these results, such that shape derivatives with respect to the plasma boundary can be computed.

Consider the following figure of merit,
\begin{gather}
    f_{NC} = \int_{V_P} d^3 x \, w(\psi) \mathcal{R}(\psi).
    \label{eq:f_NC}
\end{gather}
Here $\mathcal{R}(\psi)$ is a flux surface averaged moment of the neoclassical distribution function, $f_{1}$, which satisfies the local drift kinetic equation (DKE),
\begin{gather}
    (v_{||} \hat{\textbf{b}} + \textbf{v}_E)\cdot \nabla f_{1} - C(f_{1}) = - \textbf{v}_{\text{m}} \cdot \nabla \psi \partder{f_{M}}{\psi},
    \label{eq:DKE}
\end{gather}
where $\bm{v}_E = \textbf{E}\times \textbf{B}/B^2$ is the $\textbf{E} \times \textbf{B}$ drift velocity, $\textbf{v}_{\text{m}} \cdot \nabla \psi$ is the radial magnetic drift velocity \eqref{eq:radial_drift}, $f_{M}$ is a Maxwellian \eqref{eq:Maxwellian}, and $C$ is the linearized Fokker-Planck operator. For example, $\mathcal{R}$ can be taken to be the bootstrap current,
\begin{gather}
    J_{b} = \sum_s \frac{\langle B \int d^3 v \, f_{1s} v_{||} \rangle_{\psi}}{n_s\langle B^2 \rangle_{\psi}^{1/2}},
\end{gather}
where the sum is taken over species. We note that the geometric dependence that enters the DKE when written in Boozer coordinates only arises through the quantities $\{B, G(\psi), I(\psi), \iota(\psi) \}$. Thus for simplicity, Boozer coordinates will be assumed throughout this Section. 

The perturbation to $\mathcal{R}(\psi)$ at fixed toroidal current \eqref{eq:direct_fixed_current} can be written as,
\begin{gather}
    \delta \mathcal{R}(\psi) = \langle S_{\mathcal{R}} \delta B \rangle_{\psi} + \partder{\mathcal{R}(\psi)}{G(\psi)} \delta G(\psi)
+ \partder{\mathcal{R}(\psi)}{\iota (\psi)} \delta \chi_1'(\psi).
\end{gather}
Here $S_{\mathcal{R}}$ is a local sensitivity function which quantifies the change to $\mathcal{R}$ associated with a perturbation of the field strength $\delta B$ defined in the following way. Consider the perturbation to $\mathcal{R}$ resulting from a change in the field strength at fixed $G(\psi)$, $I(\psi)$, and $\iota(\psi)$. The functional derivative of $\mathcal{R}(\psi)$ with respect to $B(\textbf{x})$ can be expressed as,
\begin{gather}
    \delta \mathcal{R}(\delta B;B(\textbf{x})) = \left \langle S_{\mathcal{R}} \delta B(\textbf{x}) \right\rangle_{\psi}.
\end{gather}
This is another instance of the Riesz representation theorem: $\delta \mathcal{R}$ is a linear functional of $\delta B$, with the inner product taken to be the flux surface average. Thus $S_{\mathcal{R}}$ can be thought of as analogous to the shape gradient \eqref{eq:shape_gradient_ch5}.

The quantities $\{S_{\mathcal{R}},\partial \mathcal{R}(\psi)/\partial G(\psi),\partial \mathcal{R}(\psi)/\partial \iota(\psi) \}$ can be computed with the adjoint method described in Chapter \ref{ch:adjoint_neoclassical} with the SFINCS code \cite{Landreman2014}. Here we consider SFINCS to be run on a set of surfaces such that \eqref{eq:f_NC} can be computed numerically. The derivatives computed by SFINCS will appear in the additional bulk force required for the adjoint perturbed equilibrium. We consider perturbations of an equilibrium at fixed toroidal current \eqref{eq:direct_fixed_current}. The shape derivative of $f_{NC}$ can be computed on application of the transport theorem \eqref{eq:transport_theorem},
\begin{multline}
    \delta f_{NC}(S_P;\bm{\xi}_1) = \int_{S_P} d^2 x \, \bm{\xi}_1 \cdot \hat{\textbf{n}} w(\psi) \mathcal{R}(\psi) + \int_{V_P} d^3 x \,  \delta \psi \partder{}{\psi} \left(w(\psi) \mathcal{R}(\psi)\right)\\
    + \int_{V_P} d^3 x \, 
    w(\psi) \left(\partder{\mathcal{R}(\psi)}{G(\psi)}\delta G(\psi)  + \partder{\mathcal{R}(\psi)}{\iota(\psi)}\delta \chi_1'(\psi) + \left\langle S_R \delta B  \right\rangle_{\psi} \right).
    \label{eq:deltaf_NC}
\end{multline}
After several steps outlined in Appendix \ref{app:nc}, the shape derivative is written in the following form,
\begin{gather}
    \delta f_{NC} (S_P;\bm{\xi}_1) = \int_{V_P} d^3 x \, \left(\bm{\xi}_1 \cdot \bm{\mathcal{F}}_{NC} + \delta \chi_1'(\psi) \mathcal{I}_{NC} \right) + \int_{S_P} d^3 x \, \bm{\xi}_1 \cdot \hat{\textbf{n}} \mathcal{B}_{NC},
    \label{eq:df_NC}
\end{gather}
with,
\begin{subequations}
\begin{align}
    \bm{\mathcal{F}}_{NC} &=  -\nabla( \mathcal{R}(\psi) w(\psi)) -\nabla \psi (\nabla \times \textbf{B}) \cdot \nabla \vartheta \partder{\mathcal{R}(\psi)}{G(\psi)} w(\psi) \frac{B^2 \sqrt{g}}{\langle B^2 \rangle_{\psi}} \nonumber \\
    &+ \frac{ w(\psi)}{\langle B^2 \rangle_{\psi}} \partder{\mathcal{R}(\psi)}{G(\psi)} \textbf{B} \times \nabla \times \left(\partder{\textbf{x}}{\varphi}B^2 \right) + G(\psi)B^2\nabla \left(\frac{w(\psi)}{\langle B^2 \rangle_{\psi}} \partder{\mathcal{R}(\psi)}{G(\psi)} \right)  \nonumber \\
    &- \bm{\kappa} w(\psi) S_{\mathcal{R}} B + B \nabla_{\perp} (w(\psi) S_{\mathcal{R}})
    \label{eq:F_NC} \\
     \mathcal{B}_{NC}
    &= w(\psi) \mathcal{R}(\psi)  -\frac{w(\psi) B^2}{\langle B^2 \rangle_{\psi}} \partder{\mathcal{R}(\psi)}{G(\psi)} G(\psi)- w(\psi) S_{\mathcal{R}} B \label{eq:B_NC} \\
    \mathcal{I}_{NC} &=   \partder{\mathcal{R}(\psi)}{G(\psi)} \frac{w(\psi) B^2}{\langle B^2 \rangle_{\psi}\sqrt{g}} \partder{\textbf{x}}{\varphi} \cdot \partder{\textbf{x}}{\vartheta} + w(\psi) \partder{\mathcal{R}(\psi)}{\iota(\psi)} - w(\psi) S_{\mathcal{R}} \hat{\textbf{b}} \cdot \nabla \psi \times \nabla \varphi
    \label{eq:I_NC}.
\end{align}
\end{subequations}

We consider the following adjoint perturbation,
\begin{subequations}
\begin{align}
    \textbf{F}[\bm{\xi}_2,\delta \chi_2(\psi)]+ \bm{\mathcal{F}}_{NC} &= 0 \\
    \bm{\xi}_2 \cdot \hat{\textbf{n}} \rvert_{S_P} &= 0 \\
    \delta I_{T,2}(\psi) &= \frac{V'(\psi)}{2\pi} \langle \mathcal{I}_{NC} \rangle_{\psi}.
\end{align}
\end{subequations}
The adjoint bulk force $\bm{\mathcal{F}}_{NC}$ is chosen to satisfy parallel force balance required by \eqref{eq:perturbed_force_balance}. Upon application of the fixed-boundary adjoint relation we obtain the shape gradient,
\begin{gather}
    \mathcal{G}_{NC} = \left(\mathcal{B}_{NC} + \frac{\delta \textbf{B}_2 \cdot \textbf{B}}{\mu_0 }\right)_{S_P}.
\end{gather}

\section{Conclusions}

We have obtained a relationship between 3D perturbations of MHD equilibria that is a consequence of the self-adjoint property of the MHD force operator. The relation allows for the efficient computation of shape gradients for either the outer plasma surface using the fixed-boundary adjoint relation  \eqref{eq:fixed_boundary} or for coil shapes using the free boundary adjoint relation \eqref{eq:free_boundary}. The computation of the shape gradient of several stellarator figures of merit has been demonstrated with both the adjoint and direct approach. The application of the adjoint relation provides an $\mathcal{O}(N_{\Omega})$ reduction in CPU hours required in comparison with the direct method of computing the shape gradient, where $N_{\Omega}$ is the number of parameters used to describe the shape of the outer boundary or the coils. For fully 3D geometry, $N_{\Omega}$ can be $10^2-10^3$. Thus, the application of adjoint methods can significantly reduce the cost of computing the shape gradient for gradient-based optimization or local sensitivity analysis. 

We have demonstrated that the self-adjointness relations (Section \ref{sec:adjoint_relation}) can be implemented to efficiently compute the shape gradient of figures of merit relevant for stellarator configuration optimization. The shape gradient is obtained by solving an adjoint perturbed force balance equation that depends on the figure of merit of interest. For the volume-averaged $\beta$ and vacuum well parameter (Sections \ref{sec:beta} and \ref{sec:vacuum_well}), the additional bulk force required for the adjoint problem is simply the gradient of a function of flux, and so it can be implemented by adding a perturbation to the pressure profile. For the magnetic ripple on axis (Section \ref{sec:ripple}), the required bulk force takes the form of the divergence of a pressure tensor that only varies on a surface through the field strength. As the ANIMEC code currently treats this type of pressure tensor, this adjoint bulk force is implemented with a minor modification to the code. Computing the shape gradient of $\epsilon_{\text{eff}}^{3/2}$ with the adjoint approach also requires the addition of the divergence of a pressure tensor. However, this pressure tensor varies on a surface through the field line label due to the bounce integrals that appear \eqref{eq:p_ripple}. Thus the variational principle used by the ANIMEC code cannot be easily extended for this application. Similarly, the shape gradients for the quasi-symmetry (Section \ref{sec:quasisymmetry}) and neoclassical (Section \ref{sec:neoclassical}) figures of merit require an adjoint bulk force that is not in the form of the divergence of a pressure tensor. This provides an impetus for the development of a flexible perturbed MHD equilibrium code that could enable these calculations. While several 3D ideal MHD stability codes exist \citep{Anderson1990,Schwab1993,Strumberger2016}, only the CAS3D code has been modified in order to perform perturbed equilibrium calculations \citep{Nuhrenberg2003,Boozer2006}. A discussion of such linear equilibrium calculations for adjoint-based shape gradient evaluations is presented in Chapter \ref{ch:linearized_mhd}.


It should be noted that the adjoint approach we have outlined can not yield an exact analytic shape gradient, as error is introduced through the approximation of the adjoint solution. Throughout, we have assumed the existence of magnetic surfaces as the 3D equilibrium is perturbed. Therefore a code such as VMEC or ANIMEC, which minimizes an energy subject to the constraint that surfaces exist, is suitable. Generally VMEC solutions do not satisfy \eqref{eq:force_balance} exactly \citep{Nuhrenberg2009}, as they do not account for the formation of islands or current singularities associated with rational surfaces. Furthermore, the parameters $\Delta_P$ and $\Delta_I$ introduce additional numerical noise. As demonstrated in Section \ref{sec:surf_Beta}, these parameters must be small enough that nonlinear effects do not become important yet large enough that round-off error does not dominate. We have demonstrated that the typical difference between the shape gradient obtained with the adjoint method and that computed directly from numerical derivatives is $\lesssim 5\%$. These errors should not be significant for applying the shape gradient to an analysis of engineering tolerances. The discrepancy between the true shape gradient and that obtained numerically, with the adjoint approach or with finite-difference derivatives, may become problematic as one nears a local minimum during gradient-based optimization, as the resulting shape gradient may not provide an actual descent direction. This furthermore motivates the development of a perturbed equilibrium code that could eliminate this source of noise.

As demonstrated, this adjoint approach for functions of MHD equilibria is quite flexible and can be applied to many quantities of interest. Because of the demonstrated efficiency in comparison with the direct approach to computing shape gradients, we anticipate many further applications of this method.
\renewcommand{\thechapter}{6}

\chapter{Linearized equilibrium solutions}
\label{ch:linearized_mhd}

As discussed in Chapter \ref{ch:adjoint_MHD}, the application of the adjoint approach for computing the shape gradient of functions of MHD equilibria requires solutions of linearized MHD equilibrium equations. In the examples presented thus far, these linearized solutions were approximated by adding a small perturbation to a nonlinear MHD equilibrium, such as a perturbation to the prescribed toroidal current or pressure profiles. This approximation introduces error associated with the choice of the amplitude of the perturbation and limits the types of objective functions that can be treated. In this Chapter, we discuss an approach to compute the necessary linearized equilibrium solutions based on a variational method. 

\section{Introduction}

There are several existing techniques for computing linearized ideal MHD equilibria. As will be shown directly in the following Section, a linearized equilibrium state is a stationary point of an energy functional. This energy functional is related to the potential energy that appears in ideal MHD stability analysis, $W_P[\bm{\xi}] = - \frac{1}{2} \int_{V_P} d^3 x \, \bm{\xi} \cdot \textbf{F}[\bm{\xi}]$, where $\bm{\xi}$ is the displacement vector and $\textbf{F}[\bm{\xi}]$ is the MHD force operator \eqref{eq:force_operator}. For this reason, ideal MHD stability codes can be augmented for perturbed equilibrium calculations. One approach is based on the Direct Criterion of Newcomb (DCON) code \cite{Glasser2016}, which minimizes the potential energy by solving an Euler-Lagrange equation for the displacement vector. This method has been extended with the Ideal Perturbed Equilibrium Code (IPEC) \cite{Park2007,Park2007b}, which couples applied plasma boundary perturbations to perturbations of currents in the vacuum region. This code models axisymmetry-breaking perturbations on tokamak equilibria for the study of mode-locking \cite{Ferraro2019} and neoclassical toroidal viscosity (NTV) \cite{Logan2013}. Modification of DCON is currently underway to enable stability calculations for stellarators with stepped-pressure equilibria \cite{Glasser2018}. 

    The Code for the Analysis of the MHD Stability of 3D Equilibria (CAS3D) has similarly been modified for perturbed MHD equilibrium calculations. To evaluate ideal MHD stability, CAS3D solves an eigenvalue problem to obtain a minimum of $W_P[\bm{\xi}]/W_K[\bm{\xi}]$, where $W_K[\bm{\xi}] = \frac{1}{2} \int_{V_P} d^3 x \, \rho |\bm{\xi}|^2$ is the kinetic energy associated with the displacement vector $\bm{\xi}$ and $\rho$ is the density. As perturbed equilibria are stationary points of an energy functional similar to $W_P[\bm{\xi}]$, not $W_P[\bm{\xi}]/W_K[\bm{\xi}]$, such stability codes based on eigenvalue calculations need to be modified in order to compute perturbed equilibrium states. The CAS3D code allows the option to normalize $W_P[\bm{\xi}]$ by a modified energy functional such that perturbed equilibrium states can be computed \cite{Nuhrenberg2003,Boozer2006}. This technique has been used to study the effect of boundary perturbations on magnetic island width \cite{Nuhrenberg2009}.
    
     While several 3D MHD stability codes exist \cite{Schwab1993,Anderson1990,Strumberger2016}, they cannot be directly used to compute perturbed equilibrium states relevant for stellarator optimization problems. For stability studies, it is often sufficient to consider only symmetry-breaking modes (modes that break period symmetry or stellarator symmetry), while optimization is typically performed assuming preservation of symmetry. Furthermore, none of the existing codes enable the addition of a general bulk force perturbation as is required for our adjoint approach.
     
     There are additional limitations that motivate us to consider the development of an independent linearized equilibrium code. The DCON and CAS3D\footnote{This assumption is made in the original version of CAS3D \cite{Schwab1993}. There exists the option to retain the terms in the energy functional involving $\nabla \cdot \bm{\xi}$ in a more recent version \cite{Nuhrenberg2020}.} approaches minimize their respective energy functionals assuming that the displacement vector is divergenceless. This assumption implies that\footnote{This arises from noting  $\langle \nabla \cdot \bm{\xi} \rangle_{\psi} = V'(\psi)^{-1} d/d\psi \left(V'(\psi) \langle\bm{\xi} \cdot \nabla \psi\rangle_{\psi} \right)$, thus $V'(\psi)\langle\bm{\xi} \cdot \nabla \psi\rangle_{\psi}$ must be a constant. As $\bm{\xi} \cdot \nabla \psi$ must vanish at the origin due to regularity while $V'(\psi)$ is finite at the origin, the quantity $V'(\psi)\langle \bm{\xi} \cdot \nabla \psi \rangle_{\psi} = 0$.} $ \langle \bm{\xi} \cdot \nabla \psi \rangle_{\psi}$ vanishes \cite{Lortz1975,Schwab1993}, where $\langle \dots \rangle_{\psi}$ is the flux-surface average \eqref{eq:flux_surface_average_appA}. This places a significant restriction on $\xi^{\psi} \equiv \bm{\xi} \cdot \nabla \psi$ that cannot generally be satisfied in addition to the Euler-Lagrange equation. Therefore, modes that are constrained by $\langle \xi^{\psi} \rangle_{\psi}= 0$ cannot be included in the Euler-Lagrange equation. In axisymmetry, this disallows the toroidal mode number $n = 0$. In stellarator geometry with discrete $N_P$-symmetry, this disallows modes where $n$ is an integer multiple of $N_P$ (sometimes called the $N = 0$ mode family \cite{Schwab1993}). This assumption is valid for stability problems, as such modes corresponding to fixed-boundary perturbations are always stable \cite{Schwab1993}. However, for stellarator optimization and tolerance calculations, these modes cannot be ignored. Rather than assume that $\nabla \cdot \bm{\xi} = 0$, for adjoint calculations it is much more convenient to assume that $\bm{\xi} \cdot \textbf{B} = 0$, which enables the inclusion of these modes. Finally, the postprocessing of results differs significantly between stability and perturbed equilibria applications. The development of such a 3D perturbed equilibrium code could substantially reduce the computational complexity of gradient-based optimization by enabling the application of the adjoint approach to many critical objective functions. Such a tool would also allow for the analysis of the response of an equilibrium to boundary perturbations without resorting to a full nonlinear calculation. This capability would improve fixed-boundary optimization when an adjoint method is not available for sensitivity and tolerance studies.
     
     In Section \ref{sec:variational_approach}, we present the proposed method to compute linearized equilibrium states with the addition of an arbitrary bulk force. This method is based on a variational principle similar to that used in the DCON code. In Section \ref{sec:screw_pinch_analysis}, we analyze the behavior of classes of modes of the displacement vector in the simplified geometry of a screw pinch. In this way, we highlight key numerical challenges and proposed solution methods. Finally, in Section \ref{sec:tokamak_shape_gradient}, we demonstrate this method for the computation of the shape gradient of a figure of merit of interest for stellarator optimization.

\section{Variational approach for linearized equilibrium solutions}
\label{sec:variational_approach}

We consider a base equilibrium magnetic field satisfying MHD force balance,
\begin{align}
    (\nabla \times \textbf{B}) \times \textbf{B} = \mu_0 \nabla p,
    \label{eq:mhd_force_balance_ch6}
\end{align}
with prescribed pressure $p(\psi)$ and rotational transform $\iota(\psi)$. We would like to compute linearizations about this state satisfying,
\begin{align}
    \textbf{F}[\bm{\xi}] + \delta \textbf{F} = 0,
    \label{eq:perturbed_equilibrium}
\end{align}
where the MHD force operator is
\begin{align}
    \textbf{F}[\bm{\xi}] = \frac{\left(\nabla \times \delta \textbf{B}[\bm{\xi}] \right) \times \textbf{B}}{\mu_0} + \frac{\left(\nabla \times \textbf{B} \right) \times \delta \textbf{B}[\bm{\xi}]}{\mu_0} - \nabla \left(\delta p[\bm{\xi}] \right),
    \label{eq:force_operator}
\end{align}
and $\delta \textbf{F}$ is a bulk force perturbation. The perturbed magnetic field can be expressed in terms of the displacement vector $\bm{\xi}$,
\begin{align}
    \delta \textbf{B}[\bm{\xi}] = \nabla \times \left(\bm{\xi} \times \textbf{B} \right),
\end{align}
under the assumption that the rotational transform $\iota(\psi)$ is preserved by the perturbation. In this Chapter, we will not consider the effect of perturbations to the rotational transform, although such effects are necessary to compute the shape gradient of certain figures of merit. Assuming the pressure profile is fixed by the perturbation, then we can also express the perturbation to the local pressure in terms of the displacement vector,
\begin{align}
        \delta p[\bm{\xi}] = - \bm{\xi} \cdot \nabla p. 
\end{align}
The linearized force balance equation is solved subject to a boundary condition,
\begin{align}
    \bm{\xi} \cdot \hat{\textbf{n}} \big \rvert_{S_P} = \delta \textbf{x} \cdot\hat{\textbf{n}},
    \label{eq:xi_boundary_condition}
\end{align}
for a prescribed boundary perturbation $\delta \textbf{x} \cdot \hat{\textbf{n}}.$ We can express this PDE \eqref{eq:perturbed_equilibrium} with boundary condition \eqref{eq:xi_boundary_condition} in an equivalent variational form involving the energy functional,
\begin{align}
     W[\bm{\xi}] = \int_{V_P} d^3 x \, \bm{\xi} \cdot \left( \textbf{F}[\bm{\xi}] + 2 \delta \textbf{F} \right) + \frac{1}{\mu_0}\int_{S_P} d^2 x \, \hat{\textbf{n}} \cdot \left(\bm{\xi}  \delta \textbf{B}[\bm{\xi}]\right) \cdot \textbf{B}.
     \label{eq:variational_principal}
\end{align}
Stationary points of $W[\bm{\xi}]$ subject to the boundary condition \eqref{eq:xi_boundary_condition} are equivalent to solutions of \eqref{eq:perturbed_equilibrium}. While \eqref{eq:perturbed_equilibrium} is a coupled set of PDEs involving two components of the displacement vector, the application of the variational principle will allow us to arrive at an Euler-Lagrange equation that is a coupled set of ODEs for one component of the displacement vector.

We now demonstrate that stationary points of \eqref{eq:variational_principal} with respect to $\bm{\xi}$ subject to the boundary condition \eqref{eq:xi_boundary_condition} indeed correspond with solutions of \eqref{eq:perturbed_equilibrium}. We perform the first variation with respect to $\bm{\xi}$,
\begin{multline}
    \delta W[\bm{\xi};\delta \bm{\xi}] = \int_{V_P} d^3 x \, \left(\delta \bm{\xi} \cdot \left(\textbf{F}[\bm{\xi}] + 2 \delta \textbf{F} \right) + \bm{\xi} \cdot \textbf{F}[\delta \bm{\xi}]\right) \\
    + \frac{1}{\mu_0}\int_{S_P} d^2 x \, \hat{\textbf{n}} \cdot \left(\delta \bm{\xi}  \delta \textbf{B}[\bm{\xi}] + \bm{\xi}  \delta \textbf{B}[\delta \bm{\xi}]\right) \cdot \textbf{B}.
\end{multline}
We now apply the self-adjointness of the MHD force operator \eqref{eq:self_adjointness_ch5}, repeated here for convenience, 
\begin{align}
    \int_{V_P} d^3 x \, \left(\bm{\xi}_1 \cdot \textbf{F}[\bm{\xi}_2]  - \bm{\xi}_2 \cdot \textbf{F}[\bm{\xi}_1] \right)  - \frac{1}{\mu_0} \int_{S_P} d^2 x \, \hat{\textbf{n}} \cdot \left( \bm{\xi}_2 \delta \textbf{B}[\bm{\xi}_1] \cdot \textbf{B} - \bm{\xi}_1 \delta \textbf{B}[\bm{\xi}_2] \cdot \textbf{B} \right) = 0,
   \label{eq:self_adjointness_ch6}
\end{align}
to obtain,
\begin{align}
    \delta W[\bm{\xi};\delta \bm{\xi}] = 2\int_{V_P} d^3 x \, \left(\delta \bm{\xi} \cdot \left(\textbf{F}[\bm{\xi}] + \delta \textbf{F} \right)\right), 
\end{align}
where the boundary term vanishes due to \eqref{eq:xi_boundary_condition}. As $\delta W[\bm{\xi};\delta \bm{\xi}]$ must vanish for any $\delta \bm{\xi}$, we obtain \eqref{eq:perturbed_equilibrium} as our Euler-Lagrange equation. Thus stationary points of $W[\bm{\xi}]$ correspond with solutions of \eqref{eq:perturbed_equilibrium}. 

We can now obtain a simplified Euler-Lagrange equation from manipulations of our energy functional \eqref{eq:variational_principal}. A vector identity is applied in order to obtain,
\begin{multline}
    W[\bm{\xi}] = \int_{V_P} d^3 x \, \bigg[ - \frac{\delta \textbf{B}[\bm{\xi}] \cdot \delta \textbf{B}[\bm{\xi}]}{\mu_0} + \bm{\xi} \cdot \textbf{J} \times \delta \textbf{B}[\bm{\xi}]
     + \bm{\xi} \cdot \nabla \left(\bm{\xi}  \cdot  \nabla p\right) + 2\bm{\xi} \cdot \delta \textbf{F} \bigg].
     \label{eq:energy_functional_ch6}
\end{multline}
The energy functional now does not depend on second derivatives of the displacement vector. This form of the energy functional is further simplified in Appendix \ref{app:coefficient_matrices}. We apply another vector identity to obtain,
\begin{multline}
        W[\bm{\xi}] = \int_{V_P} d^3 x \, \bigg[ - \frac{\delta \textbf{B}[\bm{\xi}] \cdot \delta \textbf{B}[\bm{\xi}]}{\mu_0} + \bm{\xi} \cdot \textbf{J} \times \delta \textbf{B}[\bm{\xi}]
        - \left(\bm{\xi} \cdot \nabla p\right) \nabla \cdot \bm{\xi} + 2\bm{\xi} \cdot \delta \textbf{F} \bigg] \\ - \int_{S_P} d^2 x \, \bm{\xi}\cdot \hat{\textbf{n}} \bm{\xi} \cdot \nabla p .
\end{multline}
We can drop this boundary term, as variations that respect the boundary condition \eqref{eq:xi_boundary_condition} will automatically make it vanish. We note that this energy functional is the same (to within overall constants) as (12) in \cite{Glasser2016} if $\gamma = 0$, though we have allowed for the inclusion of an additional bulk force. 

Minimization of $W[\bm{\xi}]$ is performed upon expressing the magnetic field in a magnetic coordinate system (Appendix \ref{sec:magnetic_coordinates}),
\begin{align}
    \textbf{B} = \nabla \psi \times \nabla \vartheta - \iota(\psi) \nabla \psi \times \nabla \varphi. 
\end{align}
From the assumption that $\bm{\xi} \cdot \textbf{B} = 0$, in such a coordinate system, the energy functional only depends on the radial,
\begin{align}
    \xi^{\psi} = \bm{\xi} \cdot \nabla \psi,
\end{align}
and in-surface, 
\begin{align}
    \xi^{\alpha} = \bm{\xi} \cdot \left(\nabla \vartheta - \iota(\psi) \nabla \varphi\right),
\end{align}
components of the displacement vector. Furthermore, we note that no radial derivatives of $\xi^{\alpha}$ appear in the energy functional, as we can express the perturbed magnetic field as,
\begin{align}
    \delta \textbf{B} =\nabla \xi^{\alpha} \times \nabla \psi + \nabla \times \left(\xi^{\psi}\left(\iota(\psi) \nabla \varphi - \nabla \vartheta \right) \right).
\end{align}
Upon further manipulations of the energy functional (Appendix \ref{app:coefficient_matrices}), we also note that $\xi^{\alpha}$ only appears under derivatives with respect to $\vartheta$ and $\varphi$ in the first three terms of the energy functional \eqref{eq:energy_functional_ch6}. Given certain constraints on the bulk force perturbation that can always be satisfied (Appendix \ref{app:force_perturbation constraint}), we are free to choose $\int_0^{2\pi} d \vartheta \int_0^{2\pi} d \varphi \, \xi^{\alpha} = 0$ on all surfaces. This reflects the fact that constant shifts of $\xi^{\alpha}$ on a surface do not change the perturbed magnetic field.

We express the radial component of the displacement vector in a Fourier series,
\begin{align}
    \xi^{\psi}(\psi,\vartheta,\varphi) &= \sum_{m,n} \left( \xi^{\psi c}_{m,n}(\psi) \cos(m \vartheta - n \varphi ) + \xi_{m,n}^{\psi s}(\psi) \sin(m \vartheta - n \varphi) \right) \\
    &= \bm{\Xi}_{\psi} \cdot \bm{\mathcal{F}}^{\psi}. \nonumber
    \label{eq:displacement_basis}
\end{align}
Here $\bm{\Xi}^{\psi}$ is interpreted as a vector of Fourier amplitudes and $\bm{\mathcal{F}}^{\psi}$ is a vector of the Fourier basis functions. We similarly expand $\xi^{\alpha}$ in a Fourier series,
\begin{align}
    \xi^{\alpha}&= \sum_{m,n;\max(|m|,|n|)\ne0} \left(\xi_{m,n}^{\alpha c}(\psi) \sin(m \vartheta - n \varphi) + \xi_{m,n}^{\alpha s}(\psi) \cos(m \vartheta - n \varphi) \right) \\
    &= \bm{\Xi}_{\alpha} \cdot \bm{\mathcal{F}}^{\alpha}. \nonumber
\end{align}
As we are free to shift $\xi^{\alpha}$ by a constant on each surface, we can take the $m =0$, $n =0$ mode of $\xi^{\alpha}$ to vanish.
If the equilibrium geometric quantities have a definite parity with respect to $\vartheta$ and $\varphi$ and the prescribed boundary perturbation and bulk force perturbation maintains this parity, then $\xi^{\psi}$ will have the same parity as the equilibrium and $\xi^{\alpha}$ will have the opposite parity. For example, if the equilibrium is stellarator symmetric \cite{Dewar1998} (the cylindrical coordinates satisfy $R(\psi,-\vartheta,-\varphi) = R(\psi,\vartheta,\varphi)$ and $Z(\psi,-\vartheta,-\varphi) = -Z(\psi,\vartheta,\varphi)$) and this parity is maintained by the perturbation, only the cosine series is needed for $\xi^{\psi}$ and the sine series is needed for $\xi^{\alpha}$. We will assume stellarator symmetry for the remainder of this Chapter for simplicity of the presentation.

We similarly express the bulk force perturbation in a magnetic coordinate system,
\begin{align}
        \delta \textbf{F} = \delta F_{\psi} \nabla \psi + \delta F_{\alpha}\left( \nabla \vartheta - \iota(\psi) \nabla \varphi \right).
\end{align}
This results from the parallel force balance condition \eqref{eq:perturbed_equilibrium}, which implies that $\delta \textbf{F} \cdot \hat{\textbf{b}} = 0$.

The energy functional can be expressed schematically as,
\begin{multline}
    W[\bm{\Xi}_{\psi},\bm{\Xi}_{\alpha}] = \int_{V_P} d \psi  \, \bigg[\bm{\Xi}_{\psi}^{'}(\psi) \cdot \left(  \textbf{A}_{\psi' \psi'} \bm{\Xi}_{\psi}^{'}(\psi)  \right) +\bm{\Xi}_{\psi} \cdot \left( \textbf{A}_{\psi \psi} \bm{\Xi}_{\psi} + \textbf{A}_{\psi \psi'} \bm{\Xi}_{\psi}^{'}(\psi) + \textbf{I}_{\psi}\right) \\
    + \bm{\Xi}_{\alpha} \cdot \left( \textbf{A}_{\alpha \alpha}\bm{\Xi}_{\alpha} + \textbf{A}_{\alpha \psi'} \bm{\Xi}_{\psi}^{'}(\psi) + \textbf{A}_{\alpha \psi} \bm{\Xi}_{\psi} + \textbf{I}_{\alpha} \right) \bigg],
\end{multline}
upon integration over $\vartheta$ and $\varphi$. Explicit forms for the coefficient matrices are provided in Appendix \ref{app:coefficient_matrices}.

We now perform variations with respect to the in-surface component,
\begin{align}
        \delta W[\bm{\Xi}_{\psi},\bm{\Xi}_{\alpha};\delta \bm{\Xi}_{\alpha}] = \int_{V_P} d \psi  \,  \delta \bm{\Xi}_{\alpha} \cdot \bigg[2\textbf{A}_{\alpha \alpha} \bm{\Xi}_{\alpha} + \textbf{A}_{\alpha \psi'} \bm{\Xi}_{\psi}^{'}(\psi)+\textbf{A}_{\alpha \psi} \bm{\Xi}_{\psi}
         + \textbf{I}_{\alpha} \bigg],
\end{align}
where we have noted that $\textbf{A}_{\alpha \alpha}$ can be made symmetric due to the self-adjointness of the MHD force operator. (The explicit form given in Appendix \ref{app:coefficient_matrices} is evidently symmetric.)
Thus the in-surface component can be expressed in terms of the radial component of the displacement vector using the corresponding Euler-Lagrange equation,
\begin{align}
   2\textbf{A}_{\alpha \alpha} \bm{\Xi}_{\alpha} + \textbf{A}_{\alpha \psi'} \bm{\Xi}_{\psi}^{'}(\psi) + \textbf{A}_{\alpha \psi} \bm{\Xi}_{\psi}  + \textbf{I}_{\alpha} = 0.
\end{align}
As shown in Appendix \ref{app:coefficient_matrices},  $\textbf{A}_{\alpha \alpha}$ is invertible, so we find the reduced energy functional to be,
\begin{multline}
    W[\bm{\Xi}_{\psi}] = \int_{V_P} d \psi \bigg[ \bm{\Xi}_{\psi} \cdot \left( \textbf{C}_{\psi \psi} \bm{\Xi}_{\psi} + \textbf{C}_{\psi \psi'} \bm{\Xi}_{\psi}^{'}(\psi) + \textbf{K}_{\psi} \right) \\
    + \bm{\Xi}_{\psi}^{'}(\psi) \cdot \left(\textbf{C}_{\psi' \psi'} \bm{\Xi}_{\psi}^{'}(\psi) + \textbf{K}_{\psi'} \right) - \frac{1}{4} \textbf{I}_{\alpha} \cdot \textbf{A}_{\alpha \alpha}^{-1} \textbf{I}_{\alpha} \bigg], 
\end{multline}
with,
\begin{subequations}
\begin{align}
    \textbf{C}_{\psi \psi} &= \textbf{A}_{\psi \psi} - \frac{1}{4} \textbf{A}_{\alpha \psi} ^{T} \textbf{A}_{\alpha \alpha}^{-1} \textbf{A}_{\alpha \psi} \\
    \textbf{C}_{\psi \psi'} &= \textbf{A}_{\psi \psi'} - \frac{1}{2} \textbf{A}_{\alpha \psi}^{T} \textbf{A}_{\alpha \alpha}^{-1} \textbf{A}_{\alpha \psi'} \\
    \textbf{C}_{\psi' \psi'} &= \textbf{A}_{\psi' \psi'} - \frac{1}{4} \textbf{A}_{\alpha \psi'}^{T} \textbf{A}_{\alpha \alpha}^{-1} \textbf{A}_{\alpha \psi'} \\
    \textbf{K}_{\psi} &= \textbf{I}_{\psi} - \frac{1}{2} \textbf{A}_{\alpha \psi}^{T} \textbf{A}_{\alpha \alpha}^{-1} \textbf{I}_{\alpha} \\
    \textbf{K}_{\psi'} &= - \frac{1}{2} \textbf{A}_{\alpha \psi'}^{T} \textbf{A}_{\alpha \alpha}^{-1} \textbf{I}_{\alpha}.
\end{align} 
\end{subequations}
We now perform variations with respect to $\bm{\Xi}_{\psi}$, 
\begin{multline}
    \delta W[\bm{\Xi}_{\psi};\delta \bm{\Xi}_{\psi}] = \int_{V_P} d\psi \,  \delta \bm{\Xi}_{\psi} \cdot \bigg[ 2\textbf{C}_{\psi \psi} \bm{\Xi}_{\psi} + \textbf{C}_{\psi \psi'} \bm{\Xi}_{\psi}^{'}(\psi) + \textbf{K}_{\psi} \\
    - \der{}{\psi} \left( \textbf{C}_{\psi \psi'}^T \bm{\Xi}_{\psi} + 2 \textbf{C}_{\psi' \psi'} \bm{\Xi}_{\psi}^{'}(\psi) + \textbf{K}_{\psi'} \right) \bigg],
\end{multline}
to obtain the following Euler-Lagrange equation,
\begin{align}
   2\textbf{C}_{\psi \psi} \bm{\Xi}_{\psi} + \textbf{C}_{\psi \psi'} \bm{\Xi}_{\psi}^{'}(\psi) + \textbf{K}_{\psi}
    - \der{}{\psi} \left( \textbf{C}_{\psi \psi'}^T \bm{\Xi}_{\psi} + 2 \textbf{C}_{\psi' \psi'} \bm{\Xi}_{\psi}^{'}(\psi) + \textbf{K}_{\psi'} \right) = 0.
\end{align}
We define our vector of unknowns as,
\begin{align}
    \overrightarrow{\textbf{u}} = \left[\begin{array}{c} \bm{\Xi}_{\psi} \\
    \textbf{C}_{\psi \psi'}^T \bm{\Xi}_{\psi} + 2\textbf{C}_{\psi' \psi'} \bm{\Xi}_{\psi}'(\psi)
    \end{array} \right],
\end{align}
so that our Euler-Lagrange equation takes the form, $\overleftrightarrow{\textbf{L}}_1 \overrightarrow{\textbf{u}} +\overleftrightarrow{\textbf{L}}_2 \overrightarrow{\textbf{u}}'(\psi) + \overrightarrow{\textbf{b}} = 0$,
with,
\begin{subequations}
\begin{align}
    \overleftrightarrow{\textbf{L}}_1 &= \left[ \begin{array}{c c} 
    \textbf{C}_{\psi \psi'}^T & -\textbf{I} \\
    2 \textbf{C}_{\psi \psi} & 0
    \end{array} \right] \\
    \overleftrightarrow{\textbf{L}}_2 &= \left[ \begin{array}{c c} 
    2 \textbf{C}_{\psi'\psi'} & 0\\
    \textbf{C}_{\psi \psi'} & -\textbf{I}
    \end{array} \right] \\
    \overrightarrow{\textbf{b}} &= \left[\begin{array}{c} 0 \\
    \textbf{K}_{\psi} - \textbf{K}_{\psi'}'(\psi)\end{array} \right].
\end{align}
\end{subequations}
Currently this is an implicit system of differential equations. When $\overleftrightarrow{\textbf{L}}_2$ is invertible, this system can be transformed into an explicit system of ODEs. If $\det\left(\textbf{C}_{\psi' \psi'} \right) = 0$ at a point $\psi = \psi_s$ and $\textbf{C}_{\psi' \psi'}^{-1} \sim 1/(\psi - \psi_s)$ to leading order near $\psi_s$, then $\psi_s$ is a regular singular point. At such points, additional care must be taken in obtaining numerical solutions to the Euler-Lagrange equation. In analogy with regular singular points of an uncoupled ODE, power series solutions can be constructed near $\psi_s$ using a matrix form of Frobenius analysis (Chapter 4 in \cite{Coddington1955}). As discussed in \cite{Glasser2016}, for the Euler-Lagrange equation under consideration, such singular points occur when $\psi=0$, $\iota = 0$, or $m \iota(\psi) - n = 0$ for any $m$ and $n$ included in the spectrum for $\xi^{\psi}$ and $\xi^{\alpha}$. This singular behavior is discussed in more detail in Section \ref{sec:screw_pinch_analysis}.

This coupled set of second-order ODEs is solved with a boundary condition of $\bm{\Xi}_{\psi}(0) = 0$ and $\bm{\Xi}_{\psi}(\psi_0)$ specified according to the prescribed boundary perturbation,
\begin{align}
    \xi^{\psi c}_{m,n}(\psi_0) = \frac{\int_0^{2\pi} d \vartheta \int_0^{2\pi} d \varphi \, \delta \textbf{x} \cdot \nabla \psi \cos(m \vartheta - n \varphi)}{\int_0^{2\pi} d \vartheta \int_0^{2\pi} d \varphi \,\cos(m \vartheta - n \varphi)^2},
\end{align}
where $\psi_0$ is the flux label on the plasma boundary $S_P$. As $\nabla \psi$ vanishes at the origin, we require that $\bm{\Xi}_{\psi}(0) = 0$ such that the displacement vector remains finite.

The approach presented in this Section is very similar to that of the DCON approach, with several important distinctions. (1) Rather than assuming $\nabla \cdot \bm{\xi} = 0$, we have assumed $\hat{\textbf{b}} \cdot \bm{\xi}$. This allows us to include $n = 0$ modes in our displacement vector in axisymmetry and $n$ that are an integer multiple of the number of periods in $N_P$ symmetry. (2) We have allowed for the inclusion of a general bulk force, given it is consistent with the conventions we have adopted for our displacement vector ($\hat{\textbf{b}} \cdot \bm{\xi} = 0$ and $\xi^{\alpha c}_{0,0} = 0$). (3) DCON solves an initial value problem by integrating a set of linearly-independent solutions that are regular at the axis. We instead solve a BVP. (4) Our treatment of singular surfaces differs slightly from that of DCON, as is described in Section \ref{sec:screw_pinch_n_m}.

\section{Screw pinch analysis}
\label{sec:screw_pinch_analysis}

To further analyze the behavior of the solutions to the linearized equilibrium equations, we will consider the simplified geometry of a one-dimensional screw pinch. A screw pinch is an infinite cylindrical device with field lines that lie on surfaces of constant radius $r$. The field lines generally have both a toroidal ($\hat{\textbf{z}}$) and poloidal ($\hat{\bm{\theta}}$) component. We assume a cylindrical coordinate system with $\hat{\textbf{r}} \times \hat{\bm{\theta}} \cdot \hat{\textbf{z}} = 1$ where all equilibrium quantities only depend on $r$. The infinite length of a screw pinch is approximated by a cylindrical torus with major radius $R_0 \gg 1$,
\begin{align}
    \textbf{B} = \psi'(r) \left(\frac{\hat{\textbf{z}}}{r} + \iota(r)\frac{\hat{\bm{\theta}}}{R_0} \right).
\end{align}
Here $\psi(r)$ is the toroidal flux label,
\begin{align}
   2\pi \psi(r) = \int_0^{2\pi} d \theta \int_0^r dr' \, r' \textbf{B} \cdot \hat{\textbf{z}},
\end{align}
and $\iota(r)$ is the rotational transform,
\begin{align}
   \iota(r) = R_0\frac{\textbf{B} \cdot \nabla \theta}{\textbf{B} \cdot \nabla z},
\end{align}
the number of poloidal rotations of the field line through a $z$ displacement of $2\pi R_0$. We note that $\theta$ and $z/R_0$ are magnetic coordinates for this system. The MHD force balance equation \eqref{eq:mhd_force_balance_ch6} for this geometry becomes,
\begin{align}
    \der{}{r} \left( \mu_0 p(r) + \frac{1}{2r^2} \left(\psi'(r)\right)^2 \right) + \frac{\iota(r)\psi'(r)}{ r R_0^2}
    \der{}{r} \left( r \iota(r) \psi'(r) \right) = 0,
    \label{eq:screw_pinch_force_balance}
\end{align}
where $\iota(\psi)$, $p(\psi)$ and $\psi_0 \equiv \psi(r = 1)$ are prescribed. The solution is obtained for $r \in [0,1]$ with the boundary condition $\psi(r = 0)=0$.

Due to the toroidal and poloidal symmetry of this equilibrium, each of the Fourier modes of the displacement vector decouple from each other, and we can consider each mode independently. Although the Euler-Lagrange equation is solved for $\xi^{\psi}(\psi)$, it is more straightforward to analyze the nature of the solutions in terms of $\xi^r(r) = \bm{\xi} \cdot \nabla r$. Thus we will discuss the Euler-Lagrange equation in terms of modes of $\xi^r$,
\begin{align}
    \left(\xi^{rc}_{m,n}\right)''(r) = B_1(r)\left(\xi^{rc}_{m,n}\right)'(r) + B_2(r) \xi^{rc}_{m,n}(r) + B_3(r). 
    \label{eq:screw_pinch_Euler_Lagrange}
\end{align}
We consider a bulk force perturbation of the form,
\begin{align}
   \delta \textbf{F} = \sum_{m,n}
    \delta F^{m,n}_{rc}(r) \cos\left(m \theta - n\frac{z}{R_0}\right) \hat{\textbf{r}} +  \delta F^{m,n}_{\alpha s}(r) \sin\left(m \theta - n \frac{z}{R_0} \right) \left(\frac{1}{r}\hat{\bm{\theta}}- \frac{\iota(r)}{R_0} \hat{\textbf{z}}\right),
\end{align}
and a boundary condition given by,
\begin{align}
    \xi^r(1) = \sum_{m,n}\xi_{m,n}^{rc}(1) \cos\left(m \theta - n \frac{z}{R_0}\right).
\end{align}


\subsection{$m = 0$, $n = 0$ mode}
\label{sec:m_0_n_0}

We begin with a discussion of the $m = 0$, $n = 0$ mode. The coefficients appearing in the Euler-Lagrange equation \eqref{eq:screw_pinch_Euler_Lagrange} become,
\begin{subequations}
\begin{align}
    B_1(r) &= \frac{R_0^2 - r^2 \iota(r)(\iota(r) + 2 r \iota'(r))}{r(R_0^2 + r^2 \iota(r)^2)} - \frac{2 \psi''(r)}{\psi'(r)} \\
    B_2(r) &= \frac{(3 R_0^2 - r^2 \iota(r)^2) \psi'(r) - 2 r R_0^2 \psi''(r)}{r^2 (R_0^2 + r^2 \iota(r)^2) \psi'(r)} \\
        B_3(r) &= - \mu_0 \frac{r^2  \delta F^{0,0}_{rc}(r)}{(1 + r^2 \iota(r)^2/R_0^2) \psi'(r)^2}.
\end{align}
\label{eq:m_0_n_0_coefficients}
\end{subequations}
We note that the Euler-Lagrange equation exhibits regular singular behavior at $r = 0$. To study the regular singular behavior near the axis in more detail, we expand the toroidal flux as,
\begin{align}
    \psi(r) = \frac{\psi_2}{2} r^2 + \mathcal{O}(r^4),
\end{align}
where $\psi_2$ is some constant, which follows from noting that $\psi(r)$ must be even in $r$ from \eqref{eq:screw_pinch_force_balance}. From the indicial equation for the homogeneous problem with $B_3(r) = 0$, we find the leading order behavior to be $\xi^{rc}_{0,0}(r) \sim r^{\pm 1}$ near the origin. The negative root will be excluded given our boundary condition on the axis; thus, we expect a smooth solution for the radial displacement vector. The leading order behavior of the inhomogeneous problem will depend on the bulk force perturbation of interest. 

We first demonstrate a perturbed equilibrium with an imposed boundary perturbation and no force perturbation,
\begin{align}
    \xi_{0,0}^{rc}(1) &= 1 &
   \delta F^{0,0}_{rc}(r) &= 0. 
   \label{eq:m_0_n_0_boundary_perturbation}
\end{align}
The boundary value problem is solved with MATLAB's bvp4c routine,\footnote{https://www.mathworks.com/help/matlab/ref/bvp4c.html} which employs an implicit Runge-Kutta method with adaptive mesh refinement \cite{Kierzenka2001}. Given that the coefficients become singular on the axis, the axis is not included on the computational grid, and the inner boundary condition is imposed at a point near the axis, $\psi_{\text{min}}$. For the calculations in this Chapter, we use $\psi_{\text{min}} \sim 10^{-10}-10^{-8}$. (While some numerical methods for BVPs do not require the evaluation of the ODE at the boundary points, such as finite-difference or collocation methods, our numerical method requires evaluation at the origin.)

\begin{figure}
    \centering
    \begin{subfigure}[b]{0.49\textwidth}
    \includegraphics[trim=1cm 6cm 2cm 7cm,clip,width=1.0\textwidth]{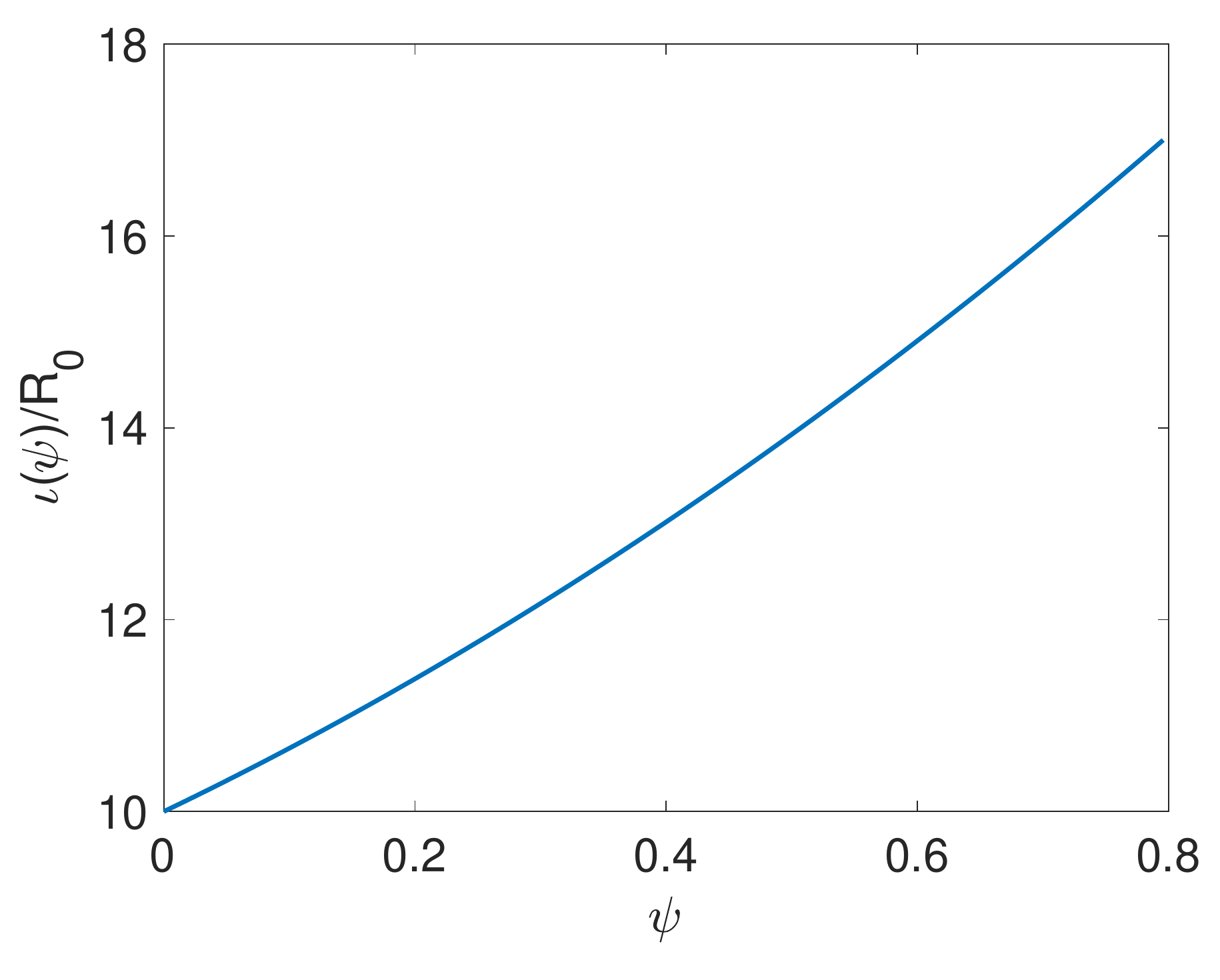}
    \caption{}
    \end{subfigure}
    \begin{subfigure}[b]{0.49\textwidth}
    \includegraphics[trim=1cm 6cm 2cm 7cm,clip,width=1.0\textwidth]{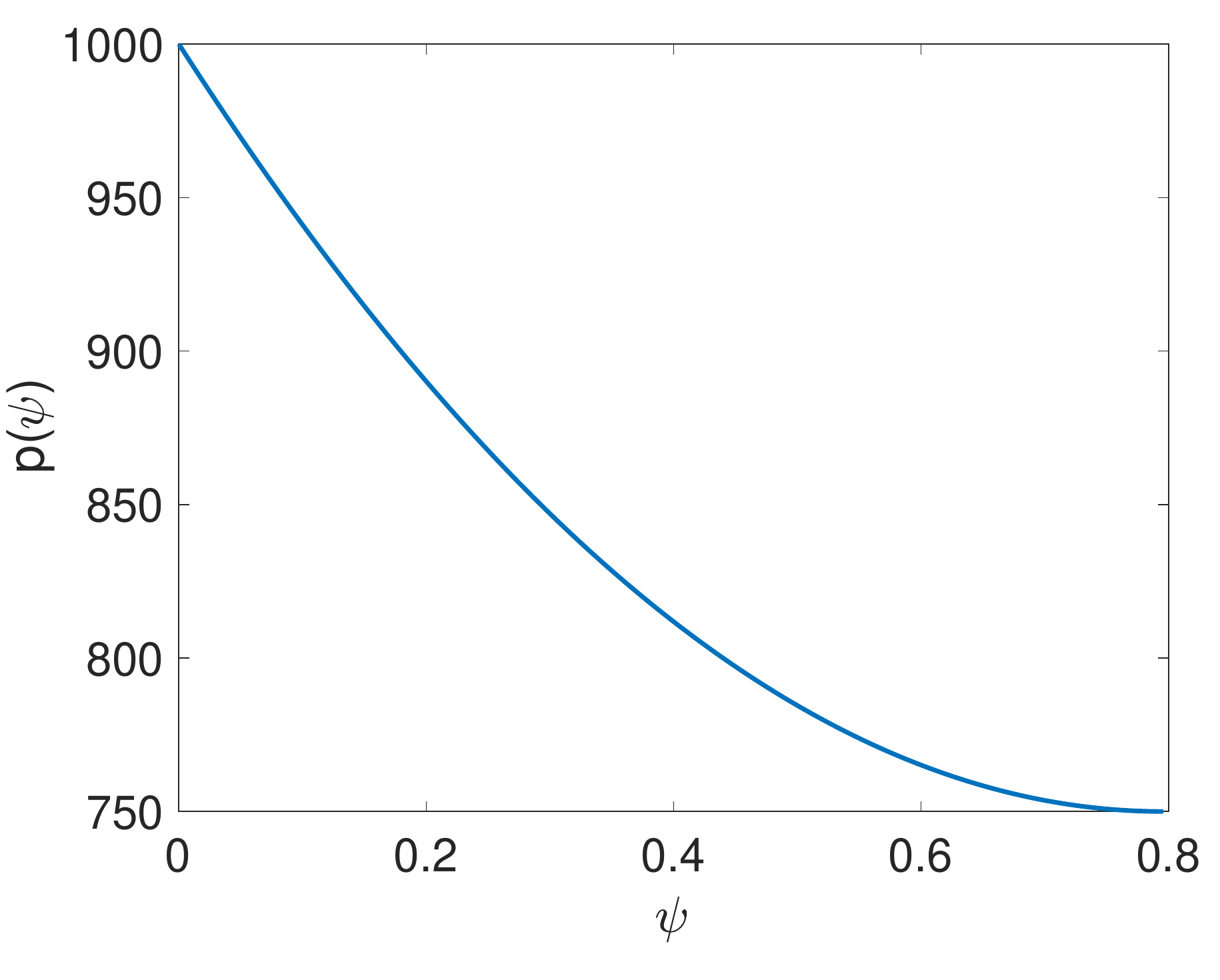}
    \caption{}
    \end{subfigure}
    \begin{subfigure}[b]{0.49\textwidth}
    \includegraphics[trim=1cm 6cm 2cm 7cm,clip,width=1.0\textwidth]{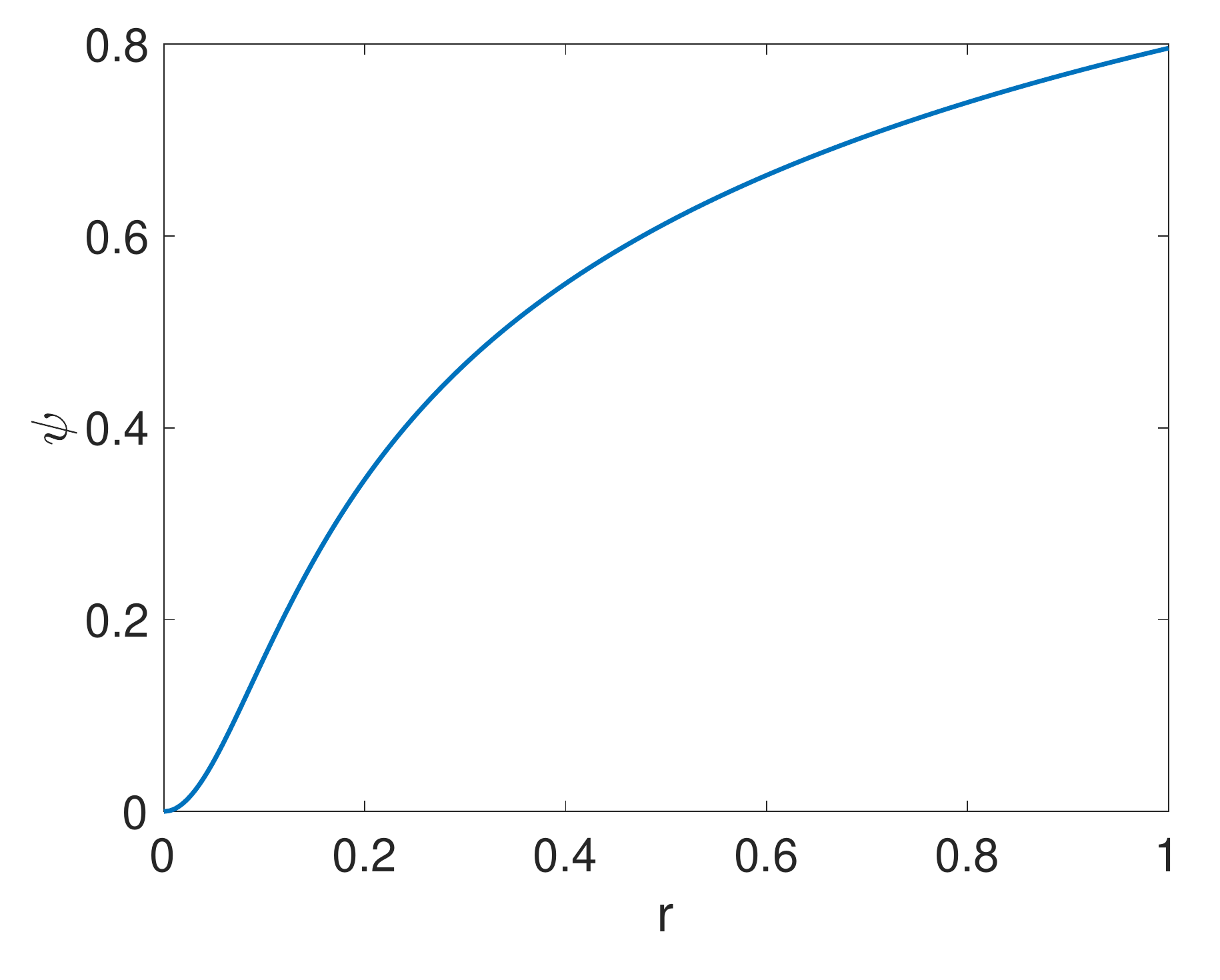}
    \caption{}
    \end{subfigure}
    \caption{Equilibrium (a) rotational transform and (b) pressure profiles used for screw pinch calculations. (c) Equilibrium flux computed with these profiles. }
    \label{fig:equilibrium_screw_pinch}
\end{figure}

The Euler-Lagrange equation is computed for a VMEC \cite{Hirshman1983} equilibrium, approximating a screw pinch by imposing a large aspect ratio boundary,
\begin{align}
    R(\psi_0,\theta_b) &= R_0 + a \cos(\theta_b) & Z(\psi_0,\theta_b) &= a \sin(\theta_b),
\end{align}
with $a = 1$ and $R_0 = 10^3$. The angle $\theta_b\in[0,2\pi]$ is used to parameterize the boundary.  The profiles are taken to be $p(\psi) = 10^3 - 5\times 10^2 \left(\psi/\psi_0\right) + 2.5 \times 10^2 (\psi/\psi_0)^2$ and  $\iota(\psi) = 10^4 + 5 \times 10^3 (\psi/\psi_0) + 2 \times 10^3 (\psi/\psi_0)^2$. The equilibrium flux and profiles are presented in Figure \ref{fig:equilibrium_screw_pinch}.

\begin{figure}
    \centering
    \includegraphics[width=0.8\textwidth]{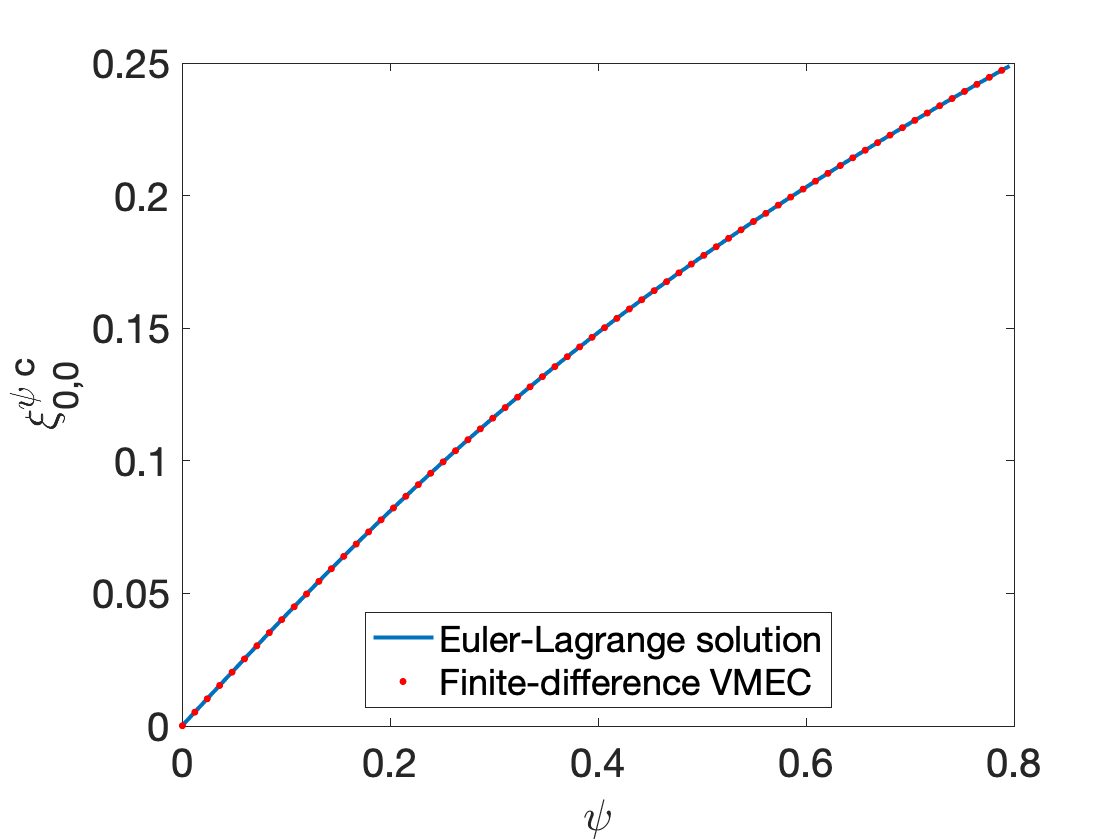}
    \caption{Benchmark of screw pinch $m = 0$, $n = 0$ mode with applied boundary perturbation \eqref{eq:m_0_n_0_boundary_perturbation}. The solution of the Euler-Lagrange equation \eqref{eq:screw_pinch_Euler_Lagrange} with coefficients \eqref{eq:m_0_n_0_coefficients}  is compared with a finite-difference VMEC calculation.}
    \label{fig:boundary_perturbation_00}
\end{figure}

We compare the numerical solution of the Euler-Lagrange equation with the displacement vector computed from finite-difference calculations with the nonlinear VMEC code.
We impose a perturbed boundary of the form,
\begin{align}
    \delta R(\psi_0,\theta_b) &= \Delta \cos(\theta_b) & \delta Z(\psi_0,\theta_b) &= \Delta  \sin(\theta_b).
\end{align}
We apply a two-point centered difference derivative with a step size of $\Delta = 10^{-2}$. The resulting displacement vector is computed from,
\begin{align}
    \xi^{\psi}(\psi,\vartheta) = \delta R(\psi,\vartheta) \partder{\psi(R,Z)}{R} + \delta Z(\psi,\theta) \partder{\psi(R,Z)}{Z},
\end{align}
where $\delta R(\psi,\vartheta)$ and $\delta Z(\psi,\vartheta)$ are the measured changes in the cylindrical coordinates at fixed flux label and straight field line poloidal angle. The result of the calculation is shown in Figure \ref{fig:boundary_perturbation_00}, where we observe good agreement between the finite-difference and Euler-Lagrange results with a volume-averaged error,
\begin{align}
 \Delta_{V} =  \frac{\int_{V_P} d^3 x \left(\xi^r_{\text{VMEC}} - \xi^r_{\text{Euler-Lagrange}}\right)^2}{\int_{V_P}d^3 x \, \left(\xi^r_{\text{VMEC}}\right)^2},
 \label{eq:euler_lagrange_error}
\end{align}
of $2.79\times 10^{-5}$.

\begin{figure}
    \centering
    \includegraphics[width=0.8\textwidth]{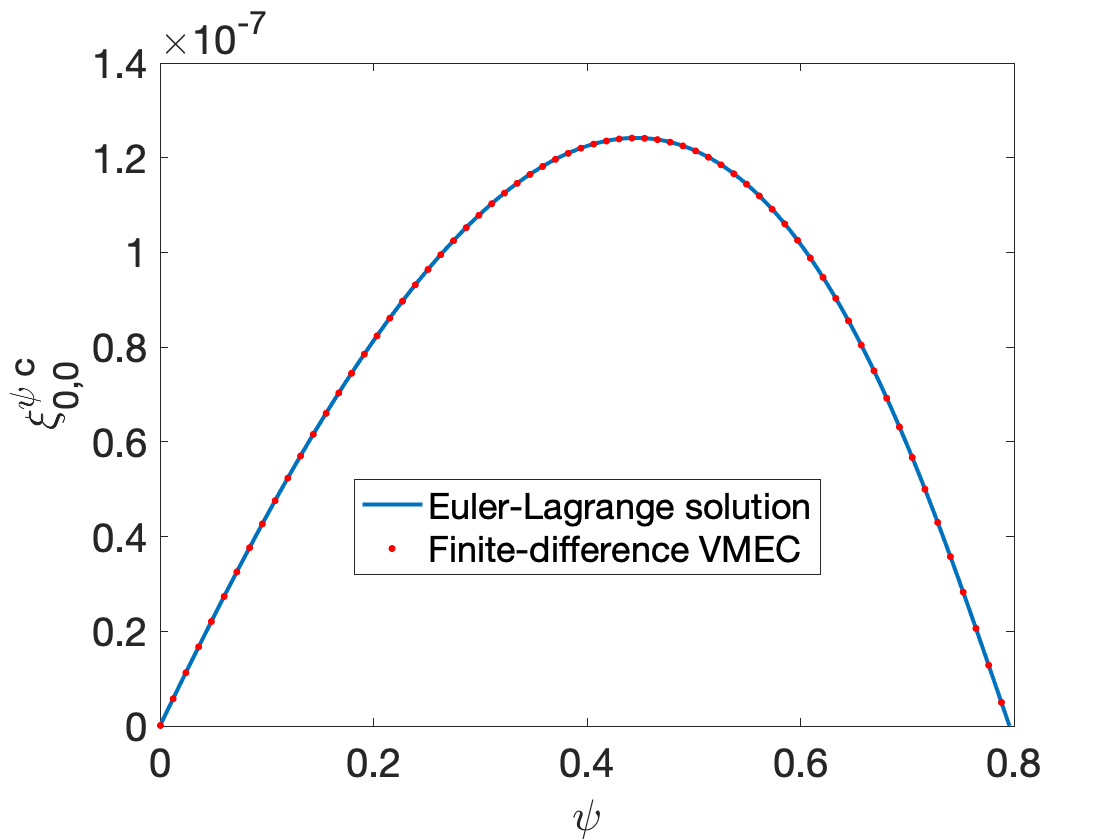}
    \caption{Benchmark of screw pinch $m = 0$, $n = 0$ mode with applied pressure perturbation \eqref{eq:m_0_n_0_pressure_perturbation}. The solution of the Euler-Lagrange equation \eqref{eq:screw_pinch_Euler_Lagrange} with coefficients \eqref{eq:m_0_n_0_coefficients} is compared with a finite-difference VMEC calculation.}
    \label{fig:pressure_perturbation_00}
\end{figure}

We next consider a perturbed equilibrium state corresponding to the addition of a bulk force in the form of the gradient of a scalar pressure perturbation,
\begin{align}
    \xi_{0,0}^{rc}(1) &= 0 &
    \delta F^{0,0}_{rc}(r) &= - \delta p'(r).
    \label{eq:m_0_n_0_pressure_perturbation}
\end{align}
This type of bulk force perturbation is necessary to compute the shape gradient for the vacuum magnetic well and beta figures of merit discussed in Chapter \ref{ch:adjoint_MHD}. We take $\delta p(r) = p(r)$, the unperturbed pressure profile. The Euler-Lagrange solution is compared with a finite-difference VMEC calculation,
\begin{align}
    \delta p(\psi) = \Delta p(\psi),
\end{align}
computed with a two-point centered-difference stencil of amplitude $\Delta = 10^{-2}$. The resulting displacement vectors are displayed in Figure \ref{fig:pressure_perturbation_00}, where we again observe good agreement between the linearized solution and its approximation with a finite-difference derivative of the nonlinear solution. The volume-averaged fractional difference \eqref{eq:euler_lagrange_error} between the solutions is found to be $1.18 \times 10^{-4}$.

\subsection{$n = 0$, $m \ne 0$ modes}
We next consider the behavior of the $n = 0$, $m \ne 0$ modes. The coefficients appearing the Euler-Lagrange equation \eqref{eq:screw_pinch_Euler_Lagrange} are, 
\begin{subequations}
\begin{align}
    B_1(r) &= - \frac{1}{r} - \frac{2 \iota'(r)}{\iota(r)} - \frac{2 \psi''(r)}{\psi'(r)} \\
    B_2(r) &= \frac{m^2 -1}{r^2} \\
    B_3(r) &= - \mu_0 R_0^2\frac{m \delta F^{m,0}_{rc} + \delta \left(F^{m,0}_{\alpha s}\right)'(r)}{m \iota(r)^2 \psi'(r)^2}.
\end{align}
\label{eq:m_1_n_0_coefficients}
\end{subequations}
In addition to the regular singular point on the axis, we note that the coefficients become singular when $\iota(r) = 0$. This class of equilibria is typically not of interest, so we will not consider this type of singularity.
Expanding the displacement vector as a power series near the origin, we find the leading order behavior of the homogeneous solution to be $\xi^{rc}_{m,0} \sim r^{-1 \pm m}$. As $\psi(r) \sim r^2$ to leading order near the axis, we note that $\xi^{\psi c}_{m,0} \sim \psi^{\pm |m|/2}$. In order to satisfy the boundary condition at $\psi = 0$, the minus solution is excluded. As $\xi^{\psi c}_{m,0}(\psi)$ becomes non-smooth at the origin, additional care must be taken in obtaining the numerical solution. We find that the accuracy is improved by solving the BVP on a grid in $\sqrt{\psi}$ rather than $\psi$, as the solution is expected to be a smooth function of $\sqrt{\psi}$ ($\xi^{\psi c}_{m,0}(\sqrt{\psi}) \sim \left(\sqrt{\psi}\right)^{m}$). To ensure the accuracy of the coefficients near the axis, we additionally employ a near-axis expansion of the equilibrium equations to $\mathcal{O}(r^6)$ (Appendix \ref{app:axis_expansion}). The incorporation of the near-axis solution becomes important when linearizing about equilibria computed with the VMEC code, which exhibits poor resolution near the magnetic axis. 

To demonstrate this method, we perform a benchmark of the homogeneous problem with an $m = 1$ boundary perturbation, 
\begin{align}
    \xi_{1,0}^{rc}(1) &= 1 & \delta F^{1,0}_{rc}(r) &= 0.
    \label{eq:m_1_n_0_boundary_perturbation}
\end{align}
The same equilibrium profiles are used as those in Section \ref{sec:m_0_n_0}. We perform a benchmark between solutions of the Euler-Lagrange equation and finite-difference approximations with VMEC equilibria. A boundary perturbation of the form,
\begin{align}
    \delta R(\psi_0,\theta_b) &= \Delta\cos(2 \theta_b) &
    \delta Z(\psi_0,\theta_b) &= \Delta \sin(2 \theta_b),
\end{align}
is imposed. The amplitude of the perturbation is taken to be $\Delta = 10^{-2}$, and the perturbed equilibrium state is computed with a two-point centered-difference stencil. 

\begin{figure}
    \centering
    \includegraphics[width = 0.8\textwidth]{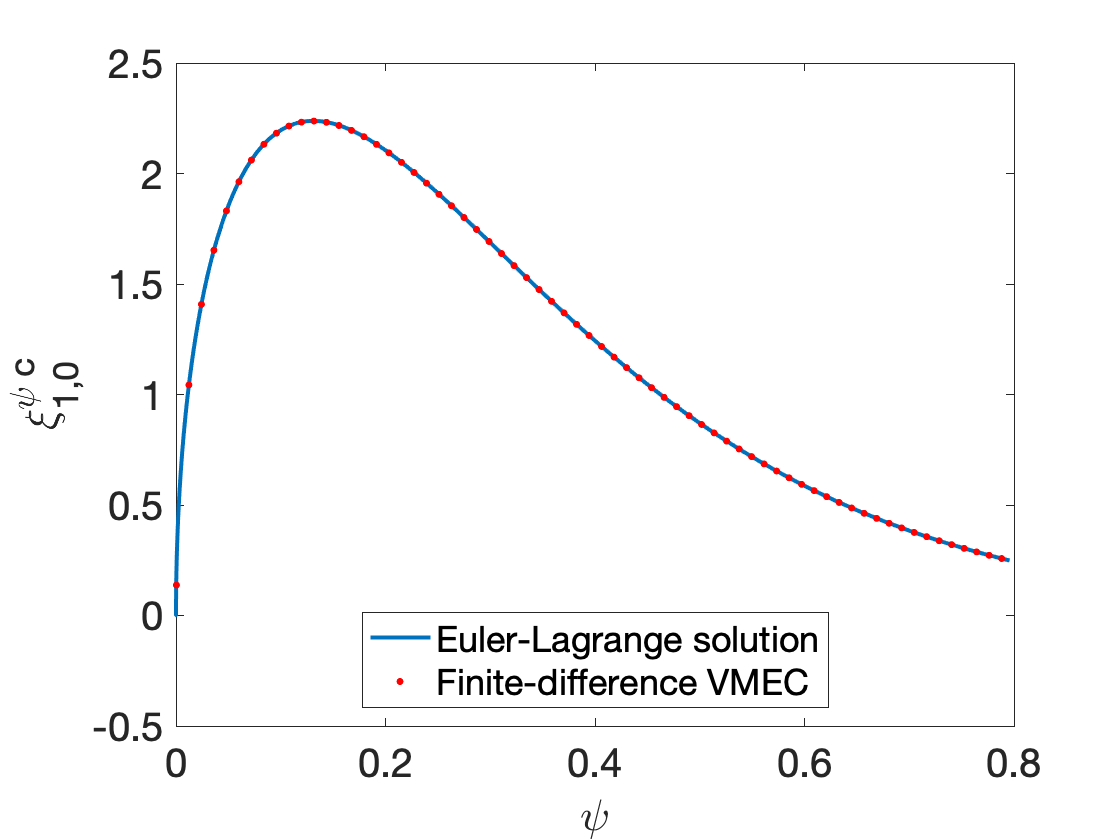}
    \caption{Benchmark of screw pinch $m = 1$, $n = 0$ mode with applied boundary perturbation \eqref{eq:m_1_n_0_boundary_perturbation}. The solution of the Euler-Lagrange equation \eqref{eq:screw_pinch_Euler_Lagrange} with coefficients \eqref{eq:m_1_n_0_coefficients} is compared with a finite-difference VMEC calculation.}
    \label{fig:euler_lagrange_m_1}
\end{figure}

The resulting displacement vector is presented in Figure \ref{fig:euler_lagrange_m_1}. We indeed find that the displacement vector has very sharp derivatives near the origin, though our numerical method can reproduce the solution obtained from VMEC. The volume-averaged fractional error between the solutions is found to be $\Delta_V = 5.67\times 10^{-4}$.

\subsection{$m = 0$, $n \ne 0$ modes}

We next consider the $m = 0$, $n \ne 0$ modes, for which the coefficients of the Euler-Lagrange equation take the form, 
\begin{subequations}
\begin{align}
    B_1(r) &= \frac{1}{r} - \frac{2 \psi''(r)}{\psi'(r)} \\
    B_2(r) &= \frac{3}{r^2} + \frac{n^2}{R_0^2} - \frac{2}{R_0^2}\iota(r) \left( \iota(r) + r \iota'(r) \right) - \frac{2(1 + \frac{r^2}{R_0^2} \iota(r)^2)}{r \psi'(r)} \psi''(r) \\
    B_3(r) &= -\mu_0\frac{r \left(n r \delta F^{0,n}_{rc}(r) + r \iota(r) \left(\delta F^{0,n}_{\alpha s}\right)'(r) + \delta F^{0,n}_{\alpha s}\left(2 \iota(r) + r \iota'(r) \right) \right)}{n \psi'(r)^2}. 
\end{align}
\label{eq:coefficients_n}
\end{subequations}
Although the ODE exhibits a regular singular point at the axis, we expect regular behavior of the homogenous solution near the origin, as the indicial equation implies that $\xi^{rc}_{0,n}(r) \sim r$. 

\subsubsection{Analytic solutions}
\begin{figure}
    \centering
    \includegraphics[width=0.8\textwidth]{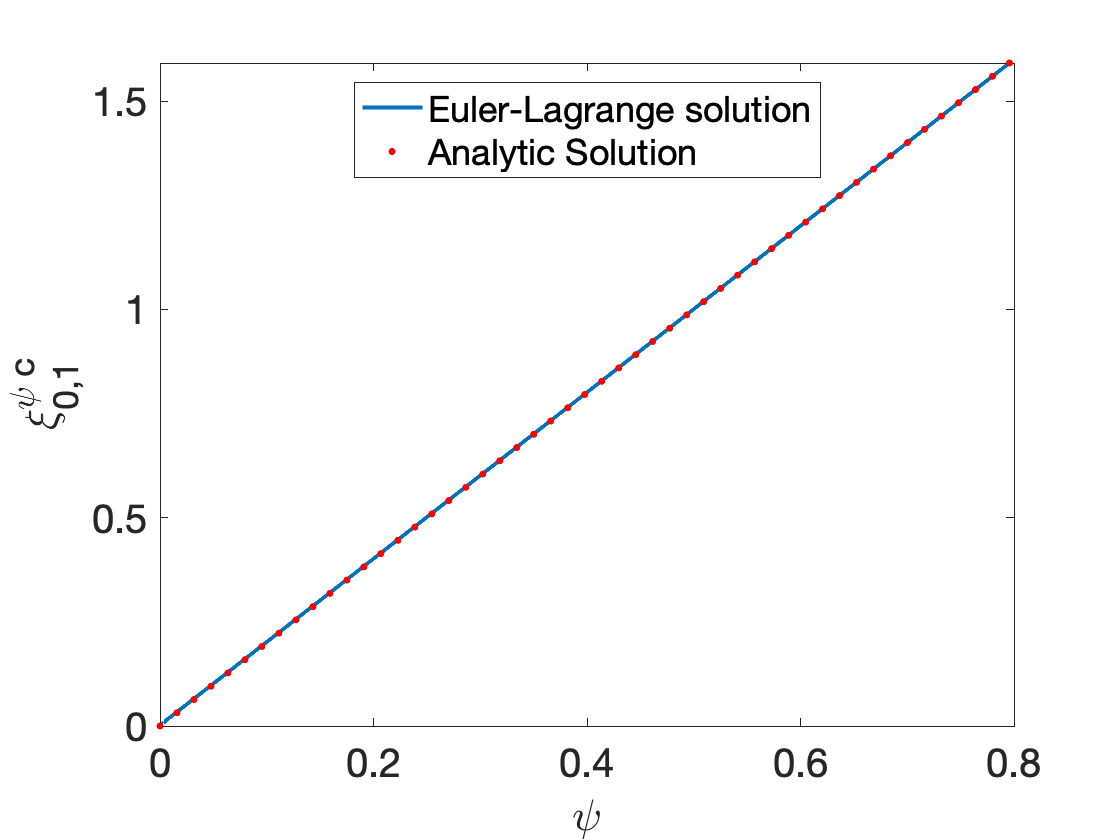}
    \caption{Benchmark of screw pinch $m = 0$, $n = 1$ mode with an applied boundary perturbation \eqref{eq:n_boundary_perturbation}. The solution of the Euler-Lagrange equation \eqref{eq:screw_pinch_Euler_Lagrange} with coefficients \eqref{eq:coefficients_n} is compared with an analytic solution \eqref{eq:analytic_boundary_n}.}
    \label{fig:analytic_n_boundary}
\end{figure}

We can compare numerical solutions of the Euler-Lagrange equation with an analytic solutions in certain limits. Assuming $\iota = 0$ and $p = 0$, we find that the equilibrium flux \eqref{eq:screw_pinch_force_balance} satisfies $\psi(r) = \psi_0 r^2$. We consider a perturbed equilibrium problem corresponding to a boundary perturbation and no force perturbation, 
\begin{align}
    \xi_{0,n}^{rc}(1) &= 1 &
    \delta F_{rc}^{0,n}(r) &= 0 .
    \label{eq:n_boundary_perturbation}
\end{align}
In this case, we recover the modified Bessel equation,
\begin{align}
   \frac{n^2r^2}{R_0^2} \left(\xi^{rc}_{0,n}\right)''\left(\frac{nr}{R_0}\right) + \frac{n r}{R_0} \left(\xi^{rc}_{0,n}\right)'\left(\frac{nr}{R_0}\right) - \left(1 + \frac{n^2 r^2}{R_0^2}\right) \xi^{rc}_{0,n}\left(\frac{nr}{R_0}\right) = 0.
\end{align}
The two solutions are $I_1(n r/R_0)$ and $K_1(nr/R_0)$, the modified Bessel functions of the first and second kind. As the solution must be finite at the origin we find,
\begin{align}
    \xi^{rc}_{0,n}(r) = \frac{I_1 \left( \frac{n r}{R_0} \right)}{I_1 \left(\frac{n}{R_0}\right)}.
    \label{eq:analytic_boundary_n}
\end{align}
A comparison between the $n = 1$ Euler-Lagrange solution and analytic solution is given in Figure \ref{fig:analytic_n_boundary}. The volume-averaged fractional error between the solutions is $\Delta_V = 1.22 \times 10^{-3}$.

\begin{figure}
    \centering
    \includegraphics[width=0.8\textwidth]{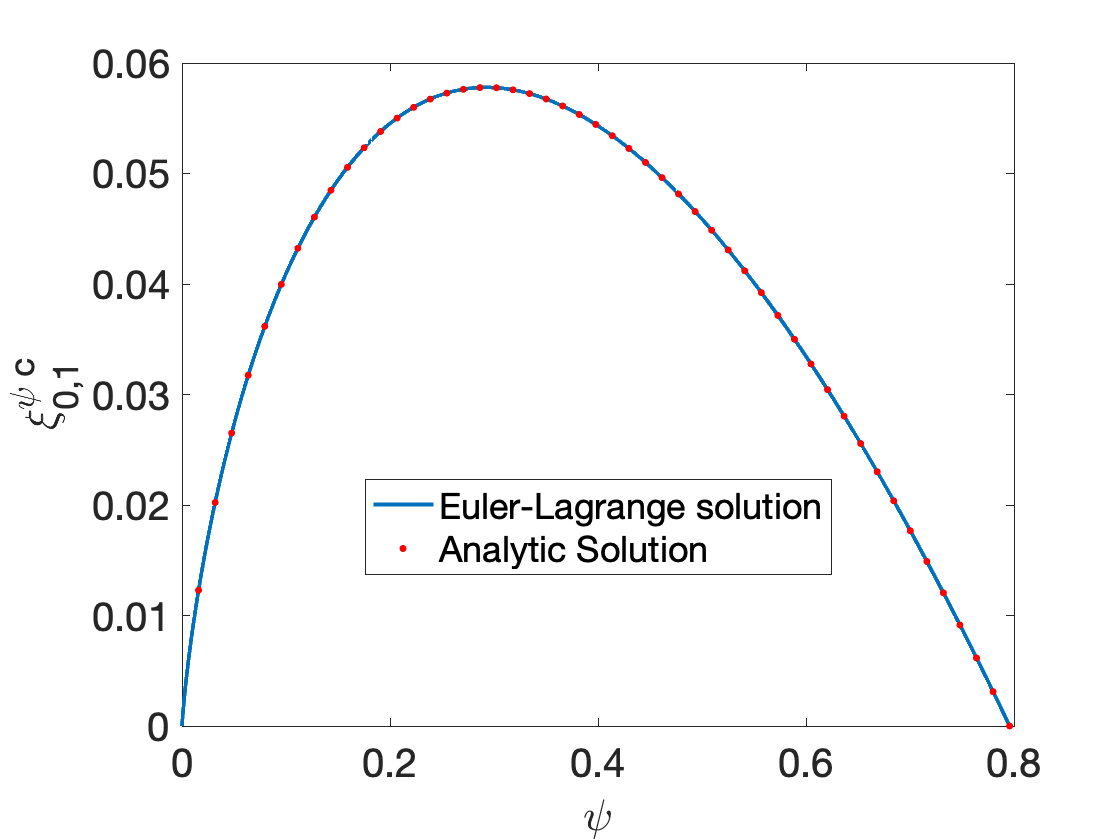}
    \caption{Benchmark of screw pinch $m = 0$, $n = 1$ mode with a bulk force perturbation $\delta F^{0,1}_{rc} = 1/r$. The solution of the Euler-Lagrange equation \eqref{eq:screw_pinch_Euler_Lagrange} with coefficients \eqref{eq:coefficients_n} is compared with an analytic solution \eqref{eq:analtyic_force_n}.}
    \label{fig:n_analytic_force}
\end{figure}

We now consider the inhomogeneous problem with a bulk force given by $\delta F^{0,n}_{rc}(r) =1/(r \mu_0)$. In this case, our Euler-Lagrange equation takes the form of an inhomogeneous modified Bessel equation,
\begin{align}
   \frac{n^2r^2}{R_0^2} \left(\xi^{rc}_{0,n}\right)''\left(\frac{nr}{R_0}\right) + \frac{n r}{R_0} \left(\xi^{rc}_{0,n}\right)'\left(\frac{nr}{R_0}\right) - \left(1 + \frac{n^2 r^2}{R_0^2}\right) \xi^{rc}_{0,n}\left(\frac{nr}{R_0}\right) + \frac{r}{\left(2 \psi_0\right)^2} = 0.
\end{align}
The solution satisfying the BVP is given by, 
\begin{multline}
    \xi^{rc}_{0,n}(r) =\frac{ R_0}{\left(2\psi_0\right)^2 r n^2 I_1\left(\frac{n}{R_0}\right)} \Bigg(r I_1\left(\frac{nr}{R_0} \right) \left(-R_0 + nK_1\left(\frac{n}{R_0} \right) \right) \\
    + I_1 \left(\frac{n}{R_0} \right) \left(R_0 - nr K_1 \left(\frac{nr}{R_0}\right) \right)\Bigg).
    \label{eq:analtyic_force_n}
\end{multline}
We note that $xK_1(x) \sim 1 + \left(A + B \log(x)\right)x^2$ for constants $A$ and $B$ near $x=0$, so our displacement vector is not smooth. We find that the numerical solution depends very sensitively on the accuracy of the coefficients, and it becomes useful to employ the axis expansion described in Appendix \ref{app:axis_expansion}. We compare the resulting numerical and analytic Euler-Lagrange solutions in Figure \ref{fig:n_analytic_force}. The volume-averaged fractional error \eqref{eq:euler_lagrange_error} between the numerical Euler-Lagrange solution and analytic solution is $\Delta_V = 6.14 \times 10^{-5}$.

\subsection{$m \ne 0$, $n \ne 0$ modes}
\label{sec:screw_pinch_n_m}

Finally, we consider modes with $m \ne 0$ and $n \ne 0$, for which the Euler-Lagrange coefficients take the form,
\begin{subequations}
\begin{align}
    B_1(r) &= - \frac{1}{r} + \frac{2n^2 r}{n^2 r^2 + m^2 R_0^2} + \frac{2m \iota'(r)}{n-m \iota(r)} - \frac{2 \psi''(r)}{\psi'(r)} \\
    B_2(r) &= \frac{2 n^2 r \mu_0 p'(r)}{(n-m\iota(r))^2 \psi'(r)^2} + \frac{n^2 (-3 + 2m^2) + \frac{n^4 r^2}{R_0^2} + \frac{m^2 (m^2-1) R_0^2}{r^2} + \frac{4 n^3}{n-m \iota(r)}}{n^2 r^2 + m^2 R_0^2} \\
    B_3(r) &= -\mu_0 \frac{n^2 r^2 + m^2 R_0^2}{(n-m \iota(r))^2 \psi'(r)^2} \delta F^{m,n}_{rc} - \mu_0\frac{ m R_0^2 + n r^2 \iota(r)}{(n-m \iota(r))^2 \psi'(r)^2} \left(\delta F^{m,n}_{\alpha s}\right)'(r)  \\
    &- \mu_0 \frac{nr \left(-2 mn R_0^2 + 2 (n^2 r^2 + 2 m^2 R_0^2) \iota(r) + (n^2 r^3 + m^2 r R_0^2) \iota'(r) \right)}{(n^2 r^2 + m^2 R_0^2)(n-m \iota(r))^2 (\psi'(r))^2} \delta F_{\alpha s}^{m,n}. \nonumber
\end{align}
\label{eq:coefficients_n_m}
\end{subequations}
By expanding the solution in a power series, we note the behavior of the solution varies as $\xi^{rc}_{m,n} \sim r^{m-1}$ near the origin. Thus, as for modes with $n = 0$ and $m \ne 0$, $\xi^{\psi}$ will vary with fractional powers of $\psi$. The numerical treatment of these modes benefits from accurate calculations of the coefficients with the near-axis expansion. In addition to the regular singular point at $r = 0$, we note that there will also be a singular point on surfaces where $\iota(r) = n/m$. 

One method to treat singular surfaces relies on a series expansion of the displacement vector within a boundary layer near the singularity. The method of Frobenius yields two independent solutions of the second-order ODE,
\begin{align}
    \xi^r_{\text{series}}(r) = A_1 \xi^{r,1}(r) + A_2 \xi^{r,2}(r),
\end{align}
near a resonant surface at $r = r_s$. A numerical solution of the ODE, $\xi^r_{\text{num}}(r)$ is integrated from the axis to the beginning of the boundary layer at $r = r_s - r_b$. The two constants, $A_1$ and $A_2$, are fixed by matching the numerical solution and its derivative at $r_s - r_b$. The series solution is then evaluated at the other edge of the boundary layer at $r_s + r_b$. The numerical solution is integrated to the plasma boundary at $r = 1$ using the initial conditions $\xi^r_{\text{num}}(r_s + r_b) = \xi^r_{\text{series}}(r_s + r_b)$ and $\left(\xi^r_{\text{num}}\right)'(r_s + r_b) = \left(\xi^r_{\text{series}}\right)'(r_s + r_b)
$. A shooting method is used to solve the BVP. This technique is similar to that used in the DCON \cite{Glasser2016} code. However, in DCON only one independent series solution is considered, as the other is not an element of the required function space for the generalized Newcomb crossing criteria. 

While the above method can reproduce the singular behavior of the Euler-Lagrange equation, as will be demonstrated shortly, it is not always desirable to include such singular behavior in the Euler-Lagrange solutions. If the perturbed current density varies as $\sim1/(r-r_s)$ near the rational surface, this will drive infinite classical transport \cite{Helander2014}, which is unphysical. An alternative is to smooth the coefficients artificially as, 
\begin{subequations}
\begin{align}
    B_1^{\text{smooth}}(r) &= B_1(r) \text{sign}(n-m \iota(r))\frac{n-m \iota(r)}{\sqrt{(n-m \iota(r))^2 + \epsilon}} \\
    B_2^{\text{smooth}}(r) &= B_2(r) \frac{(n-m \iota(r))^2}{(n-m \iota(r))^2 + \epsilon},
\end{align}
\label{eq:coefficient_smoothing}
\end{subequations}
where $\epsilon \ll 1$ is a scalar chosen to account for the smoothing by classical diffusion. When $\epsilon \rightarrow 0$, the Euler-Lagrange equation remains unchanged. For small but finite $\epsilon$, the coefficients are only modified in the vicinity of $r_s$. This is similar to a technique used in the IPEC \cite{Park2009} code.






\subsubsection{Analytic solution near singular surfaces}

To study the solutions of the Euler-Lagrange equation with $m \ne 0$ and $n \ne 0$ further, we consider a limit in which analytic solutions can be obtained. We will take $p'(\psi) = 0$ and $\iota(r) = \iota_1 r$ where $\iota_1$ is a constant.  In this case the force-balance equation \eqref{eq:screw_pinch_force_balance} gives us the following expression for the flux in terms of hypergeometric functions,
\begin{align}
    \psi(r) = \frac{r^2 \psi_0 \prescript{}{2}{F}_1\left(\frac{1}{2};\frac{3}{4};\frac{3}{2};-\frac{r^4 \iota_1^2}{R_0^2} \right)}{\prescript{}{2}{F}_1\left(\frac{1}{2};\frac{3}{4};\frac{3}{2};- \frac{\iota_1^2}{R_0^2} \right)}.
\end{align}
We define a variable $r_s = n/(m\iota_1)$ such that a singular surface occurs at $r = r_s$. The coefficients of the homogeneous problem can be expressed as,
\begin{subequations}
\begin{align}
    B_1(r) &= \frac{3}{r - r_s} - \frac{5 r_s}{r^2 - r r_s} - \frac{6}{r + r^5 \iota_1^2/R_0^2} - \frac{2 R_0^2}{r R_0^2 + r^3 r_s^2 \iota_1^2} \\
    B_2(r) &= \frac{1+m^2}{r^2} + \frac{4}{r_s r - r^2} + \frac{m^2 r_s^2 \iota_1^2}{R_0^2} + \frac{2R_0^2(r + r_s)}{r^2(r-r_s) (R_0^2 + r^2 r_s^2 \iota_1^2)}.
\end{align}
\end{subequations}
In the limit of small shear, $\epsilon_{\iota} = \iota_1 r_s^2/R_0 \ll 1$, we can approximate the coefficients as,
\begin{subequations}
\begin{align}
    B_1(r) &= \frac{3 r_s - 5 r}{r (r - r_s)} + \mathcal{O} \left(\epsilon_{\iota}^2 \right) \\
    B_2(r) &= \frac{m^2-1}{r^2} + \mathcal{O}\left(\epsilon_{\iota}^2\right).
\end{align}
\end{subequations}
In practice we choose a very small value for this expansion parameter ($\epsilon_{\iota}\sim 10^{-4}$) so that dropping the higher order terms is a very good approximation. For the $m = 2$, $n = 1$ mode subject to a boundary perturbation,
\begin{align}
    \xi_{2,1}^{rc}(1) &= 1 & \delta F^{2,1}_{rc}(r) &= 0,
    \label{eq:boundary_perturbation_n_m}
\end{align}
we have the analytic solution,
\begin{align}
    \xi^{rc}_{2,1}(r) = r\text{Re}\left[\frac{ \prescript{}{2}{F}_1\left(3-\sqrt{7};3+\sqrt{7},5,\frac{r}{r_s}\right)}{\prescript{}{2}{F}_1\left(3-\sqrt{7};3+\sqrt{7},5,\frac{1}{r_s}\right)}\right].
    \label{eq:analytic_n_m}
\end{align}

\begin{figure}
    \centering
    \includegraphics[width=0.8\textwidth]{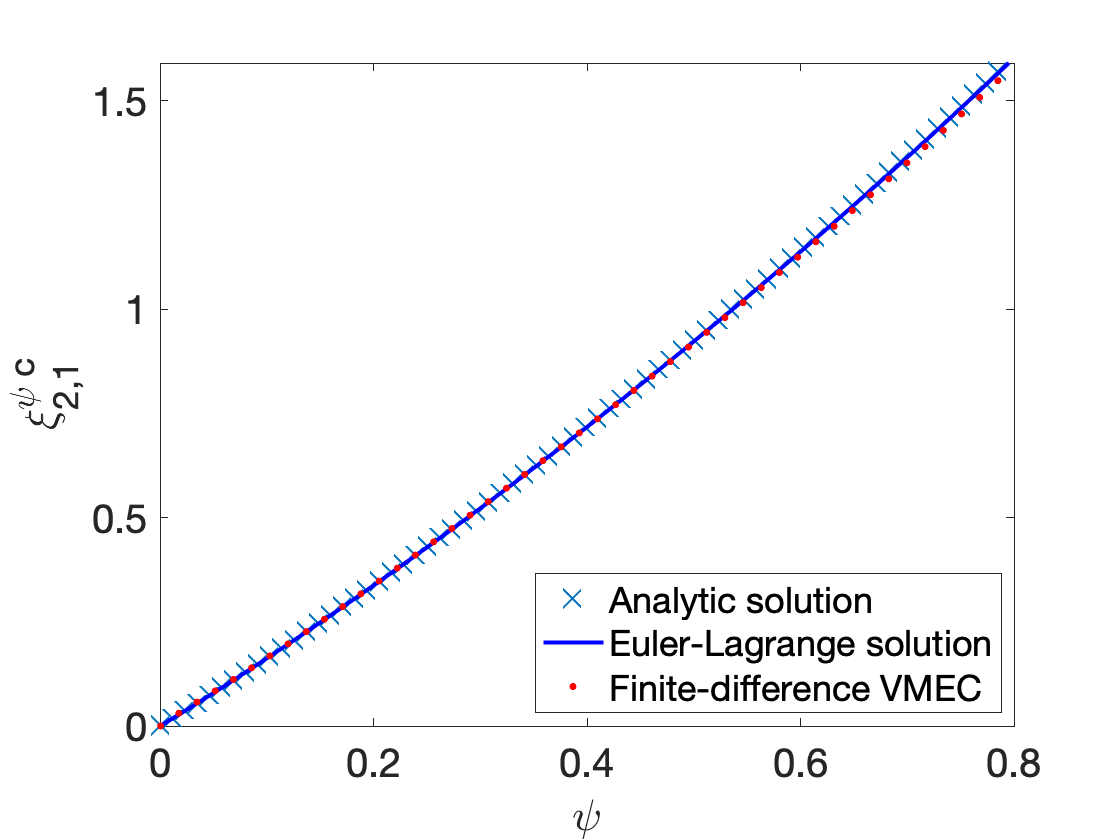}
    \caption{Benchmark of screw pinch $m = 2$, $n = 1$ mode with a boundary perturbation \eqref{eq:boundary_perturbation_n_m}. The solution of the Euler-Lagrange equation \eqref{eq:screw_pinch_Euler_Lagrange} with coefficients \eqref{eq:coefficients_n_m} is compared with an analytic solution \eqref{eq:analytic_n_m} and a finite-difference calculation from VMEC. This equilibrium does not contain a resonant surface within the volume.}
    \label{fig:no_singularity}
\end{figure}

We first consider the case in which $r_s = 2$ such that a singular surface does not appear within the volume. We compare the numerical solution of the Euler-Lagrange equation with a finite-difference calculation with VMEC. We impose a boundary perturbation of the form,
\begin{subequations}
\begin{align}
    \delta R(\psi_0,\theta_b,\phi) &= \Delta  \cos(3\theta_b - \phi) \\
    \delta Z(\psi_0,\theta_b,\phi) &= \Delta \sin(3\theta_b - \phi),
\end{align}
\label{eq:boundary_n_m}
\end{subequations}
where $\phi$ is the geometric toroidal angle. The perturbed field is computed with a two-point centered-difference stencil with amplitude $\Delta = 10^{-4}$. The results of the calculations are shown in Figure \ref{fig:no_singularity}. We note that the Euler-Lagrange solution agrees well with the analytic solution, with a volume-averaged difference of $\Delta_V = 1.86 \times 10^{-3}$, but there is a small discrepancy between the VMEC solution and the analytic solution near the edge, with a volume-averaged difference of $\Delta_V = 9.60\times 10^{-3}$. One possible source of this error is the treatment of singularities by the VMEC code. While recent results have indicated that VMEC equilibria can exhibit $1/x$-like behavior near rational surfaces \cite{Lazerson2016,Mikhailov2019}, the numerical solution is not truly singular on such surfaces, and very large numerical resolution is necessary in order to see behavior resembling a singularity. Therefore, we do not expect the displacement vector computed with finite-difference VMEC to agree with the Euler-Lagrange solution. Although for this equilibrium, $\iota$ does not resonate with the harmonics of the displacement vector, it may resonate with other modes present in the nonlinear equilibrium.

Next we consider an equilibrium with a singular surface in the volume, $r_s = 0.5$. The Euler-Lagrange equation is solved with both the power-series method, which captures the singular nature of the solution, and the coefficient smoothing method \eqref{eq:coefficient_smoothing} with several values of $\epsilon$. Again, we compare with a finite-difference VMEC solution with a boundary perturbation given by \eqref{eq:boundary_n_m}. With the power-series method, we find agreement between the Euler-Lagrange and analytic solutions. As expected, the solutions with smoothed coefficients do not reproduce the analytic expression. However, neither of these approaches approximates the VMEC solution well. Although the VMEC equilibrium is fairly well-resolved (701 flux surfaces, $10^{-12}$ force tolerance, $m \le 4$ poloidal modes, $|n| \le 4$ toroidal modes), we do not observe a response near $r = r_s$. We may need to consider a revised treatment of the singularity to match the behavior from VMEC better.


\begin{figure}
    \centering
    \includegraphics[width=0.8\textwidth]{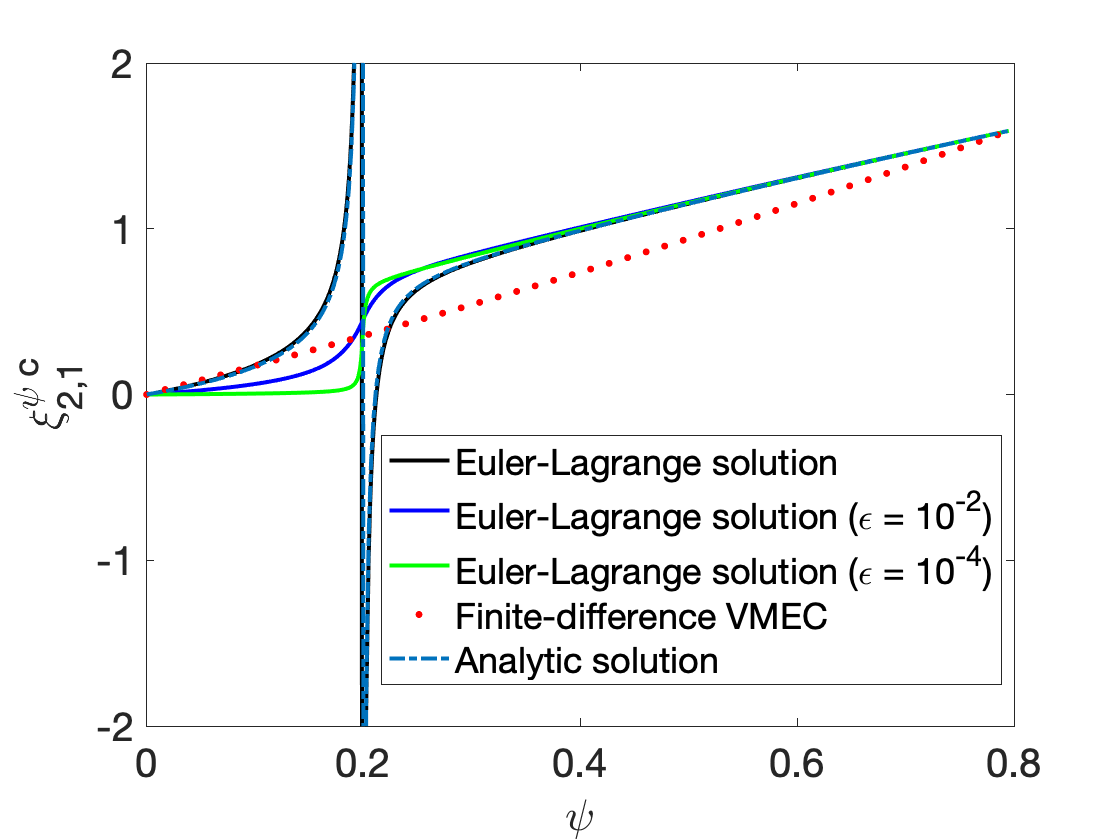}
    \caption{Benchmark of screw pinch $m = 2$, $n = 1$ mode with a boundary perturbation \eqref{eq:boundary_perturbation_n_m}. The solution of the Euler-Lagrange equation \eqref{eq:screw_pinch_Euler_Lagrange} with coefficients \eqref{eq:coefficients_n_m} is compared with an analytic solution \eqref{eq:analytic_n_m} and a finite-difference calculation from VMEC. This equilibrium contains a resonant surface at $r = 0.5$ ($\psi = 0.20$).}
    \label{fig:n_m_sin}
\end{figure}

\section{Tokamak shape gradient}
\label{sec:tokamak_shape_gradient}

We will now demonstrate the linearized equilibrium technique to compute the shape gradient of the vacuum magnetic well figure of merit discussed in Chapter \ref{ch:adjoint_MHD},
\begin{align}
    f_W(S_P) = \int_{V_P} d^3 x \, w(\psi),
    \label{eq:vacuum_magnetic_well}
\end{align}
with,
\begin{align}
    w(\psi) = \exp(-(\psi-\psi_{m,1})^2/\psi_w^2)- \exp(-(\psi-\psi_{m,2})^2/\psi_w^2),
\end{align}
where $\psi_{m,1} = 0.9\psi_0$, $\psi_{m,2} = 0.1\psi_0$, and $\psi_w = 0.05\psi_0$. The shape gradient of $f_W$ is obtained with an adjoint approach by computing a perturbed equilibrium state corresponding to the addition of a bulk force with no displacement of the boundary,
\begin{align}
  \delta \textbf{x} \cdot \nabla \psi &= 0 &  \delta \textbf{F} = - \nabla w(\psi).
  \label{eq:tok_well_adjoint}
\end{align}
The resulting perturbed field, $\delta \textbf{B}[\bm{\xi}]$, is used to compute the shape gradient,
\begin{align}
    \mathcal{G} = \frac{\delta \textbf{B}[\bm{\xi}] \cdot \textbf{B}}{\mu_0} \bigg \rvert_{S_P}.
\end{align}
We perform this calculation for an axisymmetric configuration with a plasma boundary given by,
\begin{subequations}
\begin{align}
    R(\psi_0,\theta_b) &= R_0 + a \cos(\theta_b) + b \cos(2 \theta_b) \\
    Z(\psi_0,\theta_b) &= a \sin(\theta_b) - b \sin(2 \theta_b),
\end{align}
\label{eq:tokamak_boundary}
\end{subequations}
with $R_0=3$, $a = 1$, and $b = 0.1$. Owing to its toroidal symmetry, all of the toroidal modes of the displacement vector decouple. Given the toroidal symmetry of the bulk force perturbation, we only need to consider the $n = 0$ modes. Therefore, the only singular point of the Euler-Lagrange equation is at the origin. As before, the magnetic axis is not included on the computational grid, and the coupled BVP is solved with the bvp4c routine. The radial displacement vector is computed retaining modes $m \le 30$.

\begin{figure}
    \centering
    \begin{subfigure}[b]{0.49\textwidth}
    \includegraphics[trim=4cm 3cm 4cm 3cm,clip,width=1.0\textwidth]{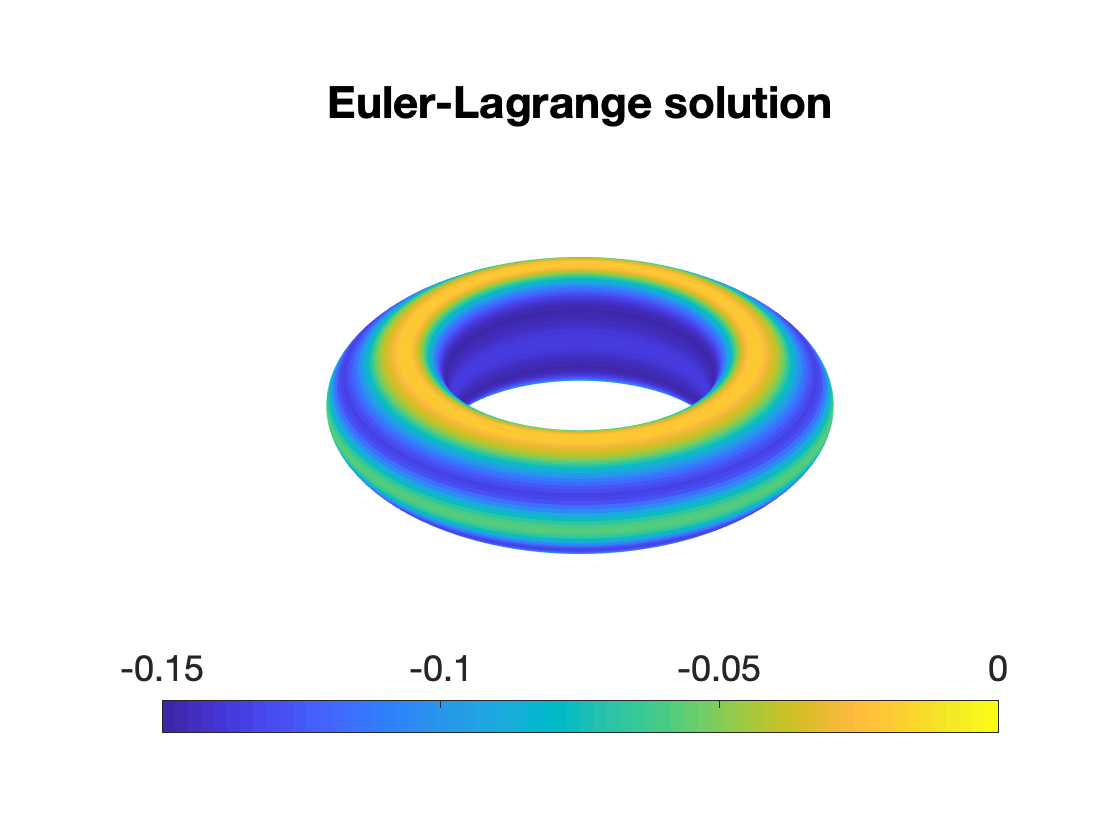}
    \caption{}
    \end{subfigure}
    \begin{subfigure}[b]{0.49\textwidth}
    \includegraphics[trim=4cm 3cm 4cm 3cm,clip,width=1.0\textwidth]{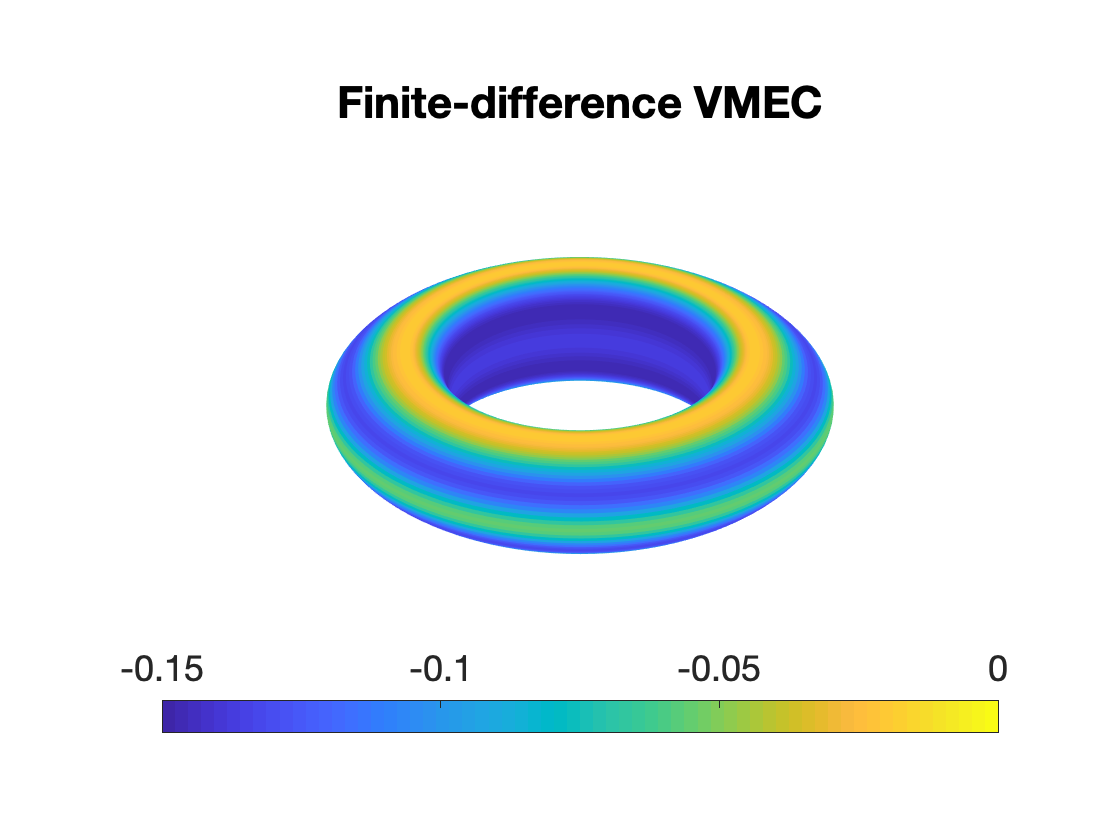}
    \caption{}
    \end{subfigure}
    \begin{subfigure}[b]{0.49\textwidth}
    \includegraphics[trim=0cm 0cm 3cm 1cm,clip,width=1.0\textwidth]{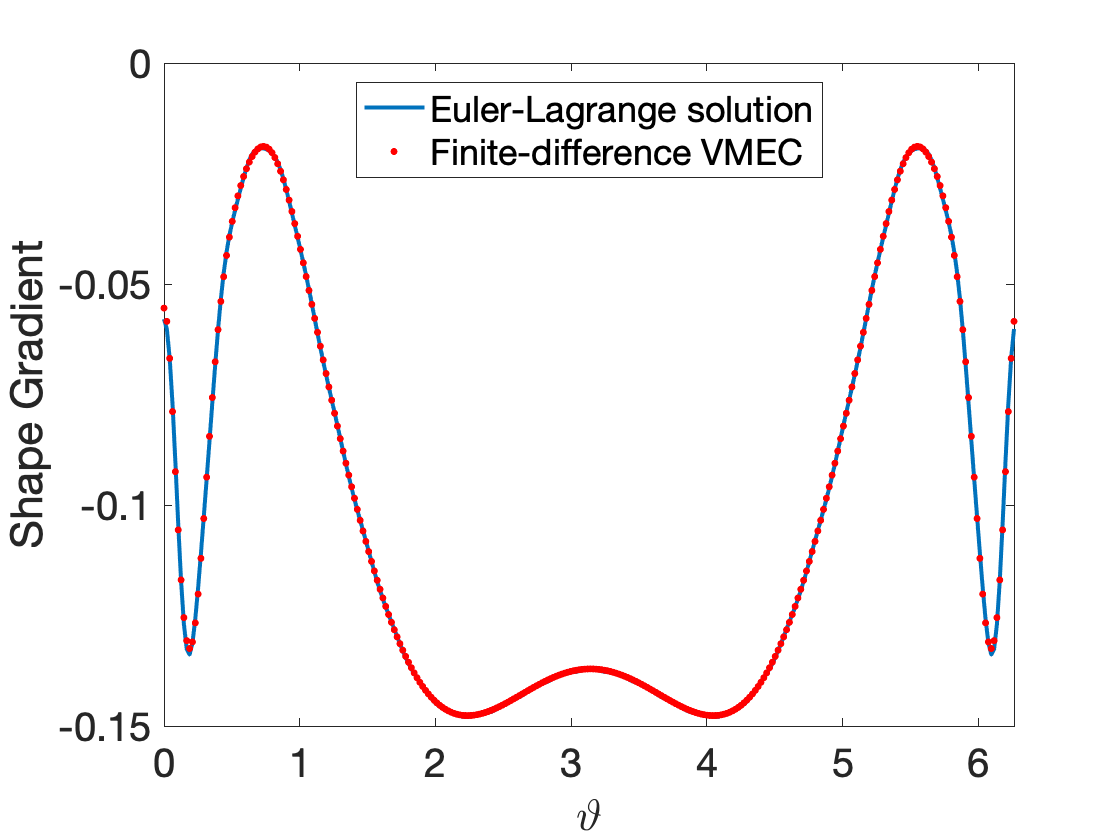}
    \caption{}
    \end{subfigure}
    \caption{The shape gradient of the vacuum magnetic well \eqref{eq:vacuum_magnetic_well} is computed for a tokamak equilibrium with triangularity \eqref{eq:tokamak_boundary} with the solution of the Euler-Lagrange equation corresponding to the adjoint problem \eqref{eq:tok_well_adjoint} and a finite-difference approximation of the adjoint problem with VMEC \eqref{eq:well_tok_vmec}.}
    \label{fig:tokamak_shape_gradient}
\end{figure}

The resulting shape gradient obtained from the Euler-Lagrange solution is shown in Figure \ref{fig:tokamak_shape_gradient} along with that computed with a finite-difference VMEC calculation,
\begin{align}
    \delta p(\psi) =  \Delta w(\psi).
    \label{eq:well_tok_vmec}
\end{align}
A two-point centered-difference derivative is computed with magnitude $\Delta = 10$. The surface-averaged fractional difference between the Euler-Lagrange and VMEC solutions is computed to be $7.3 \times 10^{-3}$.

\section{Conclusions}

We have demonstrated a variational method for computing perturbed equilibrium states corresponding to the addition of a bulk force or boundary perturbation. We considered the simplified geometry of a screw pinch to demonstrate the behavior of each of the modes of the displacement vector. Numerical solutions of the Euler-Lagrange equation are benchmarked with finite-difference calculations of the nonlinear equilibrium code, VMEC, and with analytic solutions in certain limits. Finally, we employed this approach to compute the shape gradient of a figure of merit of interest for stellarator optimization in toroidally symmetric geometry. We aim to apply this approach for computing such shape gradients in stellarator geometry, though this task may be somewhat more challenging. In fully 3D geometry, there may exist several singular surfaces throughout a volume due to toroidal mode coupling, each of which needs to be treated carefully, 

While the Euler-Lagrange equation exhibits singular behavior at rational surfaces, the equilibria computed with the VMEC code do not appear to exhibit any singular response, as demonstrated in Section \ref{sec:screw_pinch_n_m}. If the goal is to linearize about VMEC equilibria, we therefore may not want to solve the Euler-Lagrange equation exactly, but to artificially smooth the coefficients appearing in the ODE. As an alternative, artificial viscosity could be added to the Euler-Lagrange system with the addition of a small term involving a higher-order derivative. This technique, commonly used in the fluid dynamics community \cite{Mcfadden1979,Gamba1994}, turns a singular ODE into an ODE with a singular perturbation. It remains to be demonstrated that the shape gradients obtained from Euler-Lagrange solutions including such smoothing techniques can reproduce the expected shape gradients computed with the VMEC code. 

In addition to the demonstration for three-dimensional geometry, there are several interesting extensions of the work discussed in this Chapter. As discussed in Chapter \ref{ch:adjoint_MHD}, there are several figures of merit for which the adjoint problem requires the addition of a perturbation to the prescribed toroidal current profile. This would necessitate generalizing this formulation to allow for perturbations to the magnetic field that vary the rotational transform profile. While the work in this Chapter has been applied to compute the shape gradient with respect to the plasma boundary, it may be possible to couple perturbations of the boundary to coil perturbations in order to compute the coil shape gradient. This may benefit from a method similar to that used in the IPEC code, in which the virtual casing principle is applied to couple boundary perturbations to changes in the external magnetic fields.

The further development of this linear equilibrium approach would enable the shape gradient of many additional figures of merit to be computed with an adjoint method. Even if an adjoint method is not applied, the linear equilibrium approach could prove very fruitful for gradient-based, fixed-boundary optimization. Replacing a finite-difference calculation by an analytic derivative may reduce computational cost and noise associated with the finite-difference step size, enabling more efficient sensitivity and tolerance calculations for stellarator configurations.
\renewcommand{\thechapter}{7}

\chapter{Conclusions}
\label{ch:conclusions}

In this Thesis, we have aimed to address fundamental challenges (Section \ref{sec:challenges_outlook}) associated with stellarator optimization using the adjoint method and shape sensitivity analysis:
\begin{enumerate}
    \item \textit{Coil complexity}
    \item \textit{Non-convexity}
    \item \textit{High-dimensionality}
    \item \textit{Tight engineering tolerances}.
\end{enumerate}
The adjoint method allows us to efficiently compute derivatives in the context of several problems of interest for stellarator optimization. These derivatives enable navigation through high-dimensional, non-convex spaces with gradient-based methods. We demonstrate gradient-based optimization with adjoints in Chapter \ref{ch:adjoint_winding_surface}, for the design of coil shapes with minimal complexity. Computing the shape gradient of coil metrics to perturbations of the winding surface allows us to gain intuition about features of configurations that enable simpler coils. We also demonstrate gradient-based optimization of the local magnetic geometry for finite-collisionality neoclassical properties in Chapter \ref{ch:adjoint_MHD}. While including such objective functions is typically prohibitively expensive for non-convex, high-dimensional optimization, we demonstrate convergence toward a local optimum with a minimal number of function evaluations. With this adjoint method, we also gain intuition of the sensitivity of the bootstrap current and particle fluxes to perturbations in the field strength, informing engineering tolerances. Finally, in Chapter \ref{ch:adjoint_MHD} we demonstrate an adjoint method for computing the plasma surface and coil shape gradient for functions that depend on MHD equilibrium solutions. Importantly, the coil shape gradient can be used to evaluate engineering tolerances for such figures of merit (Section \ref{sec:shape_optimization_discussion}). While it has not yet been demonstrated in this Thesis, these shape gradients can also enable efficient adjoint-based optimization, either in the space of the plasma boundary or coil shapes. As discussed in Section \ref{sec:stellarator_optimization}, the direct optimization of coil shapes may result in coils that can be more feasibly engineered than those resulting from the traditional two-step optimization.

For several problems discussed in this Thesis, it is convenient to apply the discrete adjoint method (Section \ref{sec:linear_systems}). For the winding surface optimization problem in Chapter \ref{ch:adjoint_winding_surface}, the forward problem is solved as a discrete linear system, so the discrete adjoint operator can be obtained by simply taking the matrix transpose. A similar discrete adjoint method was applied for neoclassical optimization in Chapter \ref{ch:adjoint_neoclassical}, as the discretized form of the drift-kinetic equation takes the form of a linear system in the SFINCS code.  

Physical insight into the structure of the relevant equations can inform the development of continuous adjoint methods (Section \ref{sec:lagrangian}). For the neoclassical application, the adjoint equation was obtained based on an inner product similar to the free-energy norm from gyrokinetic theory. The self-adjointness of the linear Fokker-Planck operator with respect to this inner product enabled straightforward calculation of the adjoint operator. For the MHD application, the adjoint equation is obtained by noting the self-adjointness of the MHD force operator, generalized to allow for perturbations of the rotational transform and currents in the vacuum region. Finally, in Chapter \ref{ch:linearized_mhd}, a variational method for solving the adjoint equations obtained in Chapter \ref{ch:adjoint_MHD} is presented. Here we are able to borrow a variational method from MHD stability theory to efficiently compute the adjoint equilibrium problem.

\section{Outlook}

There are several natural extensions of the work presented in this Thesis. 

\subsection{Further development of adjoint methods}

\begin{itemize}
\item The advancement of the adjoint approach for functions of MHD equilibria necessitates the further development of a linearized equilibrium code, as outlined in Chapter \ref{ch:linearized_mhd}. While we have demonstrated this technique for axisymmetric equilibria, we plan to extend it to 3D equilibria. In this way, adjoint methods for computing the shape gradient of the departure from quasi-symmetry (Section \ref{sec:quasisymmetry}), effective ripple (Section \ref{sec:epsilon_eff}), and several finite-collisionality neoclassical quantities (Section \ref{sec:neoclassical}) could be demonstrated.
\item In Chapter \ref{ch:adjoint_winding_surface}, we applied the adjoint method to compute derivatives with respect to the winding surface parameters. Similarly, we can apply the adjoint method to compute derivatives with respect to \textit{plasma} surface parameters. This would allow for the identification of plasma surfaces that do not require overly-complex coils, facilitating the incorporation of coil considerations in plasma configuration optimization \cite{Carlton2019}. Similar figures of merit (without derivative information) have been used in the ROSE code \cite{Drevlak2018}.
\end{itemize}

\subsection{Further application of derivatives}

We have not yet taken full advantage of derivative information for stellarator optimization problems. 

\begin{itemize}
\item The analysis of sensitivity and tolerances presented in this Thesis is based on a local model, using a linear approximation of a function with first derivative information. A more accurate global analysis can be computed from Monte-Carlo sampling, which typically requires many function evaluations to converge. Uncertainty quantification can be accelerated through the application of a surrogate model of the design space \cite{Wu2017} with the incorporation of the uncertainty of the data. A surrogate model is an approximation to an expensive simulation based on a small number of evaluations of the function. The number of required evaluations to build the surrogate is reduced with a gradient-enhanced Gaussian process regression model \cite{Leary2004}; thus the availability of adjoint-based gradients would enable more accurate uncertainty quantification. In addition to sensitivity analysis, once a surrogate is constructed, it can replace the expensive model during optimization, allowing for more efficient local or global optimization. 
\item In particular, one type of surrogate function of interest is a neural network, which can be trained more efficiently using derivative information. Neural networks with certain choices of activation functions are differentiable, and can therefore be optimized with gradient-based optimization techniques. Gradient-based shape optimization with neural networks has proven fruitful in the field of aerodynamics \cite{Sun2019}.
\item Optimization under uncertainty methods optimize the expected value of an objective function by performing a sample average over a distribution of possible deviations. These techniques can improve the robustness of the optimum by avoiding small local minima and obtaining solutions with reduced risk. This technique has proven effective for the optimization of coil shapes with increased tolerances \cite{Lobsien2018,Lobsien2020}, using a Monte-Carlo approach. To avoid the excessive cost of a Monte-Carlo method, a linear or quadratic approximation can be made such that the expectation value and variance can be computed with derivative information  \cite{Alexanderian2017} obtained with an adjoint method.
\end{itemize}

We look forward to the adoption of adjoint methods and shape optimization tools for many stellarator design problems.
\titleformat{\chapter}
{\normalfont\large}{Appendix \thechapter:}{1em}{}
\renewcommand{\thechapter}{A}

\chapter{Toroidal coordinate systems}
\label{app:toroidal_coordinates}

In this Appendix, we briefly review coordinate systems for describing scalar and vector fields in toroidal systems. Comprehensive introductions to this topic are provided in the textbook \cite{2012Dhaeseleer}, the review article \cite{Helander2014}, and the tutorial \cite{Imbert2019}. 

\section{Toroidal coordinates}
\begin{figure}
    \centering
    \includegraphics[width=0.6\textwidth]{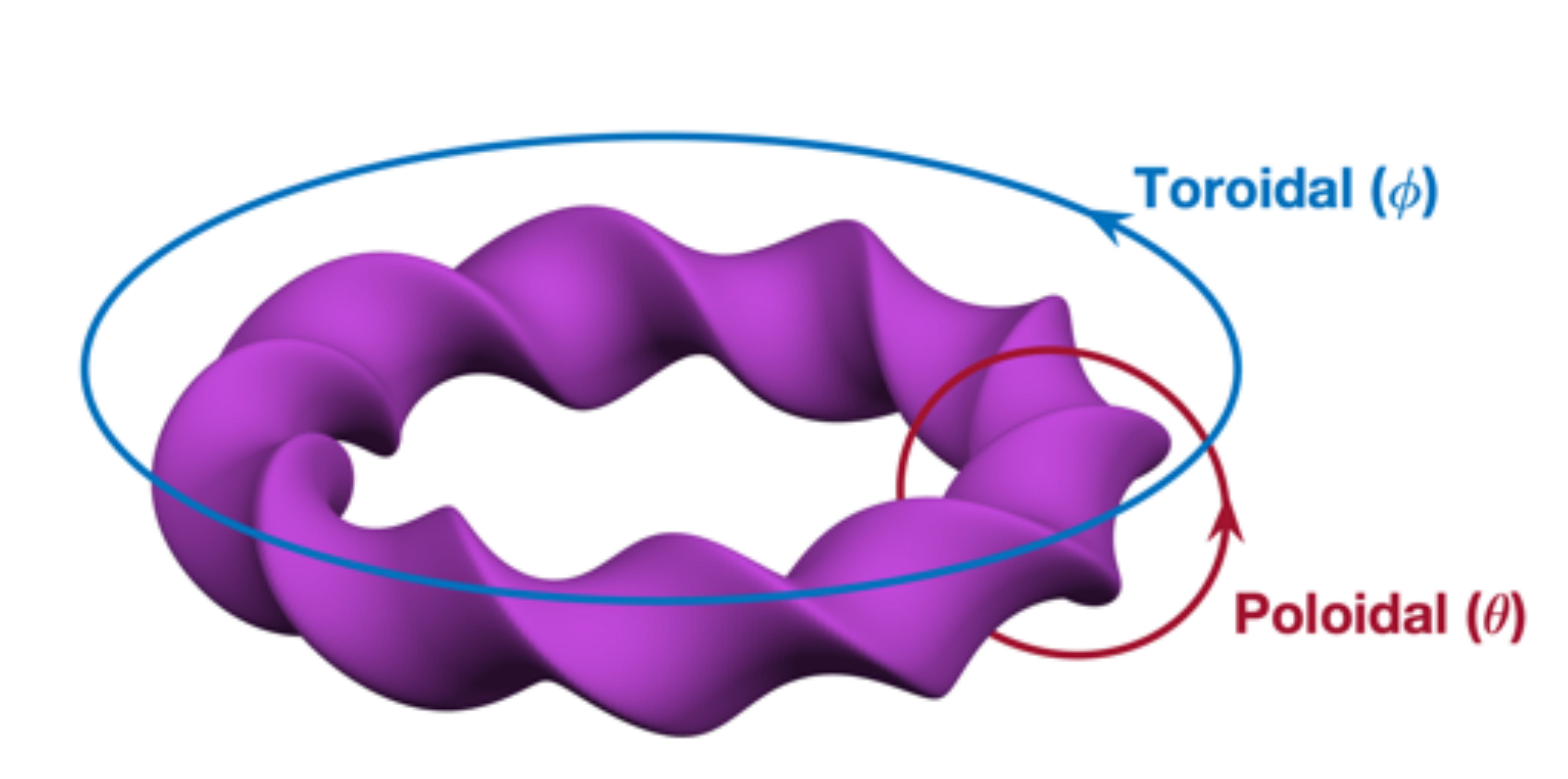}
    \caption{The position on a toroidal surface, $S$, is described by the toroidal and poloidal angles. Figure adapted from \cite{Imbert2019}.}
    \label{fig:toroidal_poloidal}
\end{figure}

In this Thesis, we often want to describe surfaces of toroidal topology or the volumes enclosed by such surfaces. We can describe the position on a toroidal surface by two angles (Figure \ref{fig:toroidal_poloidal}). A poloidal angle, denoted by $\theta$, increases by $2\pi$ upon one rotation the short way around the torus. A toroidal angle, denoted by $\phi$, increases by $2\pi$ upon one rotation the long way around the torus. 

We will consider a volume, $V$, bounded by a toroidal surface, $S$. Suppose that we use a set of continuously nested toroidal surfaces, $\Gamma(r)$, as a radial coordinate $r$, such that the position within this volume can be expressed as $\textbf{x}(r,\theta,\phi)$. A vector field, $\textbf{A}$ can be expressed in the basis of the gradients of the coordinates,
\begin{align}
    \textbf{A} = A_r \nabla r + A_{\theta} \nabla \theta + A_{\phi} \nabla \phi,
\end{align}
the covariant form, or the derivatives of the position vectors with respect to the coordinates,
\begin{align}
        \textbf{A} = A^r \partder{\textbf{x}}{r} + A^{\theta} \partder{\textbf{x}}{\theta} + A^{\phi} \partder{\textbf{x}}{\phi},
\end{align}
the contravariant form. The two basis vectors can be related through the dual relations,
\begin{align}
    \partder{\textbf{x}}{x_i} = \frac{\nabla x_j \times \nabla x_k}{\nabla x_i \cdot \nabla x_j \times \nabla x_k},
    \label{eq:dual_relation}
\end{align}
where $(x_i,x_j,x_k)=(r,\theta,\phi)$ or cyclic permutations. Such a coordinate system is generally non-orthogonal, so $\partial \textbf{x}/\partial x_i$ is not necessarily parallel to $\nabla x_i$. Several useful relations in non-orthogonal coordinate systems are summarized in Table \ref{table:non_orthogonal}. For a more detailed discussion, refer to Chapter 2 in \cite{2012Dhaeseleer}.

\begin{table}
\begin{center}
{\renewcommand{\arraystretch}{3}%
    \begin{tabular}{|c|c|}
    \hline
    Jacobian
    & $\sqrt{g} = \left(\partder{\textbf{x}}{x_i} \times \partder{\textbf{x}}{x_j}\right) \cdot \partder{\textbf{x}}{x_k} = \left( \left(\nabla x_i \times \nabla x_j\right) \cdot \nabla x_k \right)^{-1} $ \\ \hline \hline
    Differential volume  &
    $d^3 x = |\sqrt{g}| d x_i dx_j dx_k$ \\ \hline
    Differential length & $d\textbf{x} = \sum_{i=1}^3 \partder{\textbf{x}}{x_i} d x_i$ \\ \hline
    Differential surface area (constant $x_k$) & $d^2x = |\sqrt{g}| |\nabla x_k| d x_i d x_j$ \\ \hline \hline
    Divergence of vector field & $\nabla \cdot \textbf{A} = \sum_{i=1}^3 \frac{1}{\sqrt{g}} \partder{}{x_i} \left(\sqrt{g} 
    A^i
    \right) $ \\ \hline
  Curl of vector field & $\nabla \times \textbf{A} = \sum_{k=1}^3 \frac{1}{\sqrt{g}}  \left(\partder{A_j}{x_i} - \partder{A_i}{x_j} \right) \partder{\textbf{x}}{x_k} $  \\  \hline
  Gradient of scalar & $\nabla q = \sum_{i=1}^3 \partder{q}{x_i} \nabla x_i$ \\ \hline
    \end{tabular}}
\end{center}
\caption{Summary of formulas used to describe the geometry of a non-orthogonal coordinate system $(x_1,x_2,x_3)$. In the above, $\{i,j,k\}$ is a cyclic permutation of $\{1,2,3\}$. Table adapted from \cite{Imbert2019}.}
\label{table:non_orthogonal}
\end{table}

\section{Flux coordinates}

\begin{figure}
    \centering
    \includegraphics[trim=9cm 6cm 2cm 5cm,clip,width=0.5\textwidth]{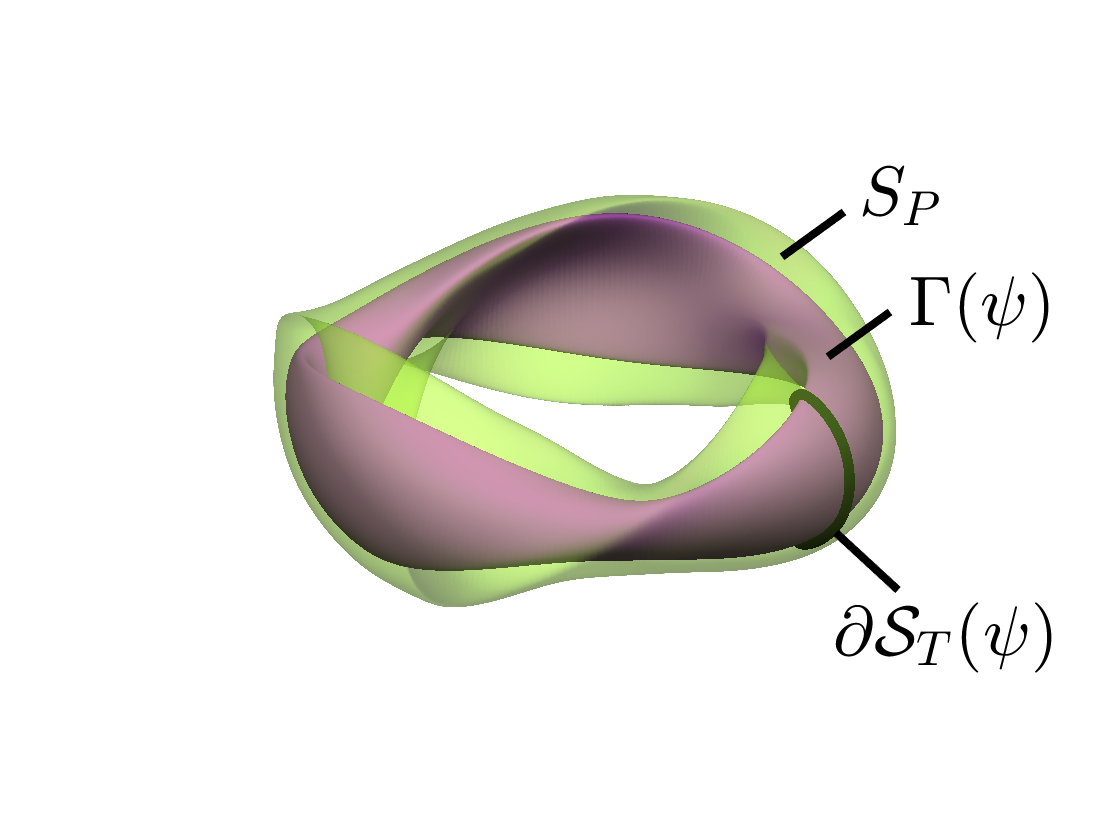}
    \caption{The plasma domain, $V_P$, is bounded by a toroidal surface, $S_P$. We make the assumption that there exists a set of toroidal magnetic surfaces, $\Gamma(\psi)$. The toroidal flux through each of these surfaces is defined by \eqref{eq:toroidal_flux} with $\mathcal{S}_T(\psi)$ an open surface bounded by a poloidally closed curve on $\Gamma(\psi)$, $\partial \mathcal{S}_T(\psi)$.}
    \label{fig:toroidal_flux}
\end{figure}

\begin{figure}
    \centering
    \includegraphics[trim=35cm 15cm 28cm 10cm,clip,width=0.6\textwidth]{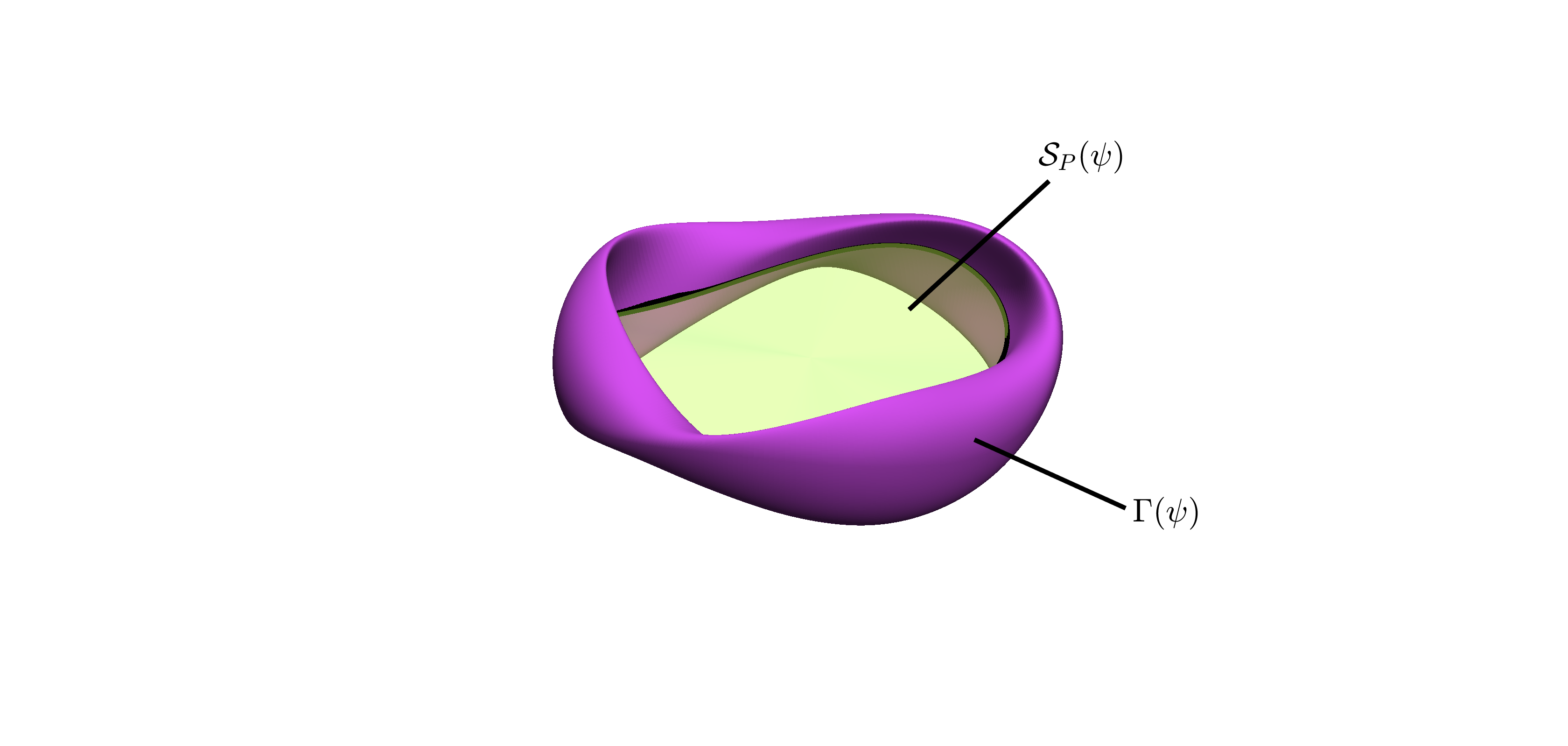}
    \caption{The poloidal flux through the magnetic surface, $\Gamma(\psi)$, is defined by \eqref{eq:poloidal_flux} with $\mathcal{S}_P(\psi)$ an open surface bounded by a toroidally closed curve on $\Gamma(\psi)$, $\partial \mathcal{S}_P(\psi)$.}
    \label{fig:poloidal_flux}
\end{figure}

If magnetic surfaces exist, indicating that the magnetic field is tangent to a set of continuously nested toroidal surfaces, we can use the toroidal flux through such surfaces as a coordinate, defined as,
\begin{align}
    2\pi \psi \equiv \int_{\mathcal{S}_T(\psi)} d^2 x \, \textbf{B} \cdot \hat{\textbf{n}}.
    \label{eq:toroidal_flux}
\end{align}
In the above expression, $\mathcal{S}_T(\psi)$ is an open surface such that $\partial \mathcal{S}_T(\psi)$  is a loop on $\Gamma(\psi)$ that closes after one poloidal rotation (Figure \ref{fig:toroidal_flux}). The unit normal is $\hat{\textbf{n}}$, often chosen to point in the direction of increasing $\phi$.
Another choice for labeling magnetic surfaces is the poloidal flux function, $\chi$, 
\begin{align}
    2\pi \chi \equiv \int_{\mathcal{S}_P(\psi)} d^2 x \, \textbf{B} \cdot \hat{\textbf{n}},
    \label{eq:poloidal_flux}
\end{align}
where $\mathcal{S}_P(\psi)$ is an open surface such that $\partial \mathcal S_P(\psi)$ is a loop on $\Gamma(\psi)$ that closes after one toroidal rotation (Figure \ref{fig:poloidal_flux}).

The rotational transform quantifies the number of poloidal turns of a field line per toroidal turn,
\begin{align}
    \iota \equiv \lim_{n \rightarrow \infty} \frac{\sum_{k = 1}^n \left(\Delta \theta \right)_k}{2 \pi n}.
\end{align}
Here $(\Delta \theta)_k$ is the change in poloidal angle in toroidal rotation $k$ and $n$ counts the toroidal turns. If flux surfaces exist, then the rotational transform can be computed from the derivative of the poloidal flux with respect to the toroidal flux,
\begin{align}
    \iota(\psi) = \chi'(\psi),
\end{align}

If a flux label, $\psi$, is used as one of the coordinates, known as a flux coordinate system, then the contravariant form for the magnetic field simplifies,
\begin{align}
    \textbf{B} = B^{\theta}\partder{\textbf{x}}{\theta} + B^{\phi} \partder{\textbf{x}}{\phi},
\end{align}
from the assumption that $\textbf{B} \cdot \nabla \psi = 0$. Given $\nabla \cdot \textbf{B} = 0$ and using \eqref{eq:dual_relation}, we can express the magnetic field as,
\begin{align}
    \textbf{B} = \nabla \psi \times \nabla \left(\theta - \iota(\psi) \phi + \lambda(\psi,\theta,\phi) \right),
    \label{eq:flux_coordinate_contravariant}
\end{align}
where $\lambda(\psi,\theta,\phi)$ is $2\pi$-periodic in $\theta$ and $\phi$ (Section 11.1 in \cite{Imbert2019}). 

In a flux-coordinate system, the flux-surface average,
\begin{align}
    \langle A \rangle_{\psi} = \frac{\int_0^{2\pi} d \theta \int_0^{2\pi} d \phi \, \sqrt{g} A}{V'(\psi)}, 
    \label{eq:flux_surface_average_appA}
\end{align}
appears in many calculations, where
\begin{align}
    V'(\psi) = \int_0^{2\pi} d \theta \int_0^{2\pi} d \phi \, \sqrt{g},
\end{align}
is the differential volume associated with a change in flux. The flux-surface average can be equivalently defined as the average over the infinitesimal volume between flux surfaces,
\begin{align}
    \left \langle A \right \rangle_{\psi} = \lim_{\Delta V \rightarrow 0} \frac{1}{\Delta V} \left(\int_{V_P(\psi) + \Delta V} d^3 x \, A - \int_{V_P(\psi)} d^3 x \, A \right),
    \label{eq:flux_surface_average_vlume}
\end{align}
where $V_P(\psi)$ is the volume enclosed by a surface labeled by $\psi$ and $V_P(\psi) + \Delta V$ is the volume of a neighboring surface. The flux-surface average is discussed in more detail in Section 4.9 of \cite{2012Dhaeseleer}.

\section{Magnetic coordinates}
\label{sec:magnetic_coordinates}

A flux coordinate system can be defined with many choices of poloidal and toroidal angles. With some choices of these angles, the contravariant expression for the magnetic field can simplify further. Given \eqref{eq:flux_coordinate_contravariant}, the definition of the poloidal and toroidal angles can be shifted to $\vartheta$ and $\varphi$ such that the magnetic field can be expressed as,
\begin{align}
    \textbf{B} = \nabla \psi \times \nabla \left(\vartheta - \iota(\psi) \varphi \right).
\end{align}
Such angles define a magnetic coordinate system. For example, one choice is $\vartheta = \theta + \lambda(\psi,\theta,\phi)$ and $\varphi = \phi$. For any choice of $\varphi$, there is a corresponding choice of $\vartheta$ that defines a magnetic coordinate system. With this choice of angles, the magnetic field lines are said to be straight in the $\vartheta-\varphi$ plane,
\begin{align}
    \der{\vartheta(l)}{\varphi(l)} = \frac{\textbf{B} \cdot \nabla \vartheta}{\textbf{B} \cdot \nabla \varphi} = \iota(\psi),
\end{align}
with a slope given by the rotational transform. Here $l$ measures length along a field line such that $df/dl = \hat{\textbf{b}} \cdot \nabla f$ for any quantity $f$, where $\hat{\textbf{b}} = \textbf{B}/B$ is the unit vector in the direction of the magnetic field.

From the covariant form for the magnetic field,
\begin{align}
    \textbf{B} = B_{\vartheta}\nabla \vartheta + B_{\varphi}\nabla \varphi + B_{\psi}\nabla \psi,
\end{align}
we can compute the net toroidal and poloidal currents enclosed by the surface labeled by $\psi$,
\begin{subequations}
\begin{align}
    I_T(\psi) &\equiv \int_{\mathcal{S}_T(\psi)} d^2 x \, \textbf{J} \cdot \hat{\textbf{n}} = \frac{1}{\mu_0} \oint_{\partial \mathcal{S}_T(\psi)} d \textbf{l} \cdot \textbf{B} = \frac{1}{\mu_0}\int_0^{2\pi} d \vartheta \, B_{\vartheta} \\
    I_P(\psi) &\equiv \int_{\mathcal{S}_P(\psi)} d^2 x \, \textbf{J} \cdot \hat{\textbf{n}} = \frac{1}{\mu_0} \oint_{\partial \mathcal{S}_P(\psi)} d \textbf{l} \cdot \textbf{B} = \frac{1}{\mu_0}\int_0^{2\pi} d \varphi \, B_{\varphi},
\end{align}
\end{subequations}
where $\mathcal{S}_T$ is defined in Figure \ref{fig:toroidal_flux} and $\mathcal{S}_P$ is defined in Figure \ref{fig:poloidal_flux}. Under the additional assumption that $\textbf{J} \cdot \nabla \psi = 0$, which follows from MHD force balance \eqref{eq:MHD_force_balance} with $p(\psi)$, we can write the covariant form as,
\begin{align}
    \textbf{B} = I(\psi) \nabla \vartheta + G(\psi) \nabla \varphi + K(\psi,\vartheta,\varphi) \nabla \psi + \nabla H(\psi,\vartheta,\varphi),
    \label{eq:covariant_magnetic}
\end{align}
where $I(\psi) = \mu_0 I_T(\psi)/(2\pi)$ and $G(\psi) = \mu_0 I_P(\psi)/(2\pi)$. See Section 2.5 in \cite{Helander2014}, Section 9.2 in \cite{Imbert2019}, and Chapter 6.5 of \cite{2012Dhaeseleer} for details. 

\section{Boozer coordinates}
\label{sec:boozer_coordinates}

As previously mentioned, there are many choices of magnetic coordinates corresponding to different choices of toroidal angle, $\varphi$. Suppose we begin with a system defined by $(\psi,\vartheta,\varphi)$ and want to transform for a system defined by $(\psi,\vartheta',\varphi')$. In order for the primed system to remain a magnetic coordinate system, we must have $\varphi' = \varphi + \gamma(\psi,\vartheta,\varphi)$ and $\vartheta' = \vartheta + \iota(\psi) \gamma(\psi,\vartheta,\varphi)$, where $\gamma(\psi,\vartheta,\varphi)$ is $2\pi$-periodic in $\vartheta$ and $\varphi$. To construct the Boozer coordinate system \cite{Boozer1980}, we will make a particular choice for $\gamma$ to simplify the covariant form for the magnetic field \eqref{eq:covariant_magnetic}. The corresponding changes to the quantities appearing in the covariant form \eqref{eq:covariant_magnetic} are
\begin{subequations}
\begin{align}
    H' &= H - \left(\iota(\psi)I(\psi) + G(\psi) \right) \gamma(\psi,\vartheta,\varphi) \\
    K' &= K + \gamma(\psi,\vartheta,\varphi) \left(\iota(\psi) I'(\psi) + G'(\psi) \right).
\end{align}
\end{subequations}
Boozer coordinates are defined such that $H' = 0$, or $\gamma(\psi,\vartheta,\varphi) = H(\psi,\vartheta,\varphi)/(\iota(\psi) I(\psi) + G(\psi))$. With this choice of transformation, we will denote $\vartheta_B = \vartheta + \iota\gamma$ and $\varphi_B = \varphi + \gamma$. The covariant form becomes,
\begin{align}
    \textbf{B} = I(\psi) \nabla \vartheta_B + G(\psi) \nabla \varphi_B + K(\psi,\vartheta_B,\varphi_B) \nabla \psi.
\end{align}
By dotting the covariant with the contravariant form, we obtain an expression for the Jacobian,
\begin{align}
    \sqrt{g} = \frac{1}{\nabla \psi \times \nabla \vartheta_B \cdot \nabla \varphi_B} = \frac{G(\psi) + \iota(\psi) I(\psi)}{B^2}.
\end{align}
We note that the Jacobian only varies on a surface through the magnetic field strength; thus each of the contravariant and covariant components of the magnetic field, except for $K(\psi,\vartheta_B,\varphi_B)$, possesses the same property. (The radial covariant component, $K(\psi,\vartheta_B,\varphi_B)$, is related to the field strength through the MHD force balance equation \eqref{eq:MHD_force_balance}.) For this reason, the Boozer coordinate system is extremely convenient for analyzing guiding center motion and neoclassical transport, as we will in Chapter \ref{ch:adjoint_neoclassical}.

\renewcommand{\thechapter}{B}

\chapter{Justification for current potential}
\label{app:current_potential}

In this Appendix, we justify the form for a continuous current density supported on a toroidal surface, $S_C$, 
\begin{align}
    \textbf{J}_C(\theta,\phi) = \hat{\textbf{n}} \times \nabla \Phi,
\end{align}
where $\hat{\textbf{n}}$ is the unit normal vector.

We consider an extension of $\textbf{J}_C$ in a neighborhood of $S_C$ of width $\Delta b$,
\begin{align}
    \widetilde{\textbf{J}}_C(b,\widetilde{\theta},\widetilde{\phi}) = \textbf{J}_C(\theta,\phi),
    \label{eq:J_extension}
\end{align}
where we define extensions of $\theta$ and $\phi$ as,
\begin{subequations}
\begin{align}
    \widetilde{\theta}(\textbf{x}) &= \theta(\textbf{x}-b(\textbf{x}) \nabla b) \\
    \widetilde{\phi}(\textbf{x}) &= \phi(\textbf{x}-b(\textbf{x}) \nabla b),
\end{align}
\end{subequations}
or a normal projection onto $S_C$. We consider $b \in [-\frac{\Delta b}{2},\frac{\Delta b}{2}]$ to be a ``thickened" region of continuous current density. We impose the constraint that $\nabla \cdot \widetilde{\textbf{J}}_C = 0$, expressed in the $(b,\widetilde{\theta},\widetilde{\phi})$ coordinate system (Table \ref{table:non_orthogonal}),
\begin{align}
    \frac{1}{\sqrt{g}}\left(\partder{\left(\sqrt{g} \widetilde{\textbf{J}}_C \cdot \nabla b\right)}{b} + \partder{\left(\sqrt{g} \widetilde{\textbf{J}}_C \cdot \nabla \widetilde{\theta}\right)}{\widetilde{\theta}} + \partder{\left(\sqrt{g} \widetilde{\textbf{J}}_C \cdot \nabla \widetilde{\phi}\right)}{\widetilde{\phi}}\right) = 0,
\end{align}
where $\sqrt{g} = \partial \textbf{x}/\partial b \cdot \left(\partial \textbf{x}/\partial \widetilde{\theta} \times \partial \textbf{x}/\partial \widetilde{\phi} \right)$
By the definition of our extension, the first term will vanish. In the limit that $\Delta b \rightarrow 0$, the divergence-free condition is expressed as,
\begin{align}
       \nabla_{\Gamma} \cdot \textbf{J}_C \equiv \frac{1}{\sqrt{g}} \left(\partder{\left(\sqrt{g} J^{\theta}\right)}{\theta} + \partder{\left(\sqrt{g} J^{\phi}\right)}{\phi}\right) = 0,
    \label{eq:surface_divergence_condition}
\end{align}
where we have expressed the current in the contravariant basis as $\textbf{J}_C = J^{\theta} \partial \textbf{x}/\partial \theta + J^{\phi} \partial \textbf{x}/\partial \phi$ and $\nabla_{\Gamma} \cdot$ is the \textit{surface divergence} (Appendix 3 in \cite{Van2007}). For a continuous current density, Ampere's law \eqref{eq:MHD_ampere} implies that $\nabla \cdot \textbf{J} = 0$. Thus the equivalent condition for a current supported on a surface is $\nabla_{\Gamma} \cdot \textbf{J}_C = 0$ \cite{Arnoldus2006}. The surface divergence of a vector field tangent to a surface $\Gamma$ ($\textbf{A} \cdot \hat{\textbf{n}} = 0$ on $\Gamma$) defined in terms of a general continuous extension, $\widetilde{\textbf{A}}$  in a neighborhood of $\Gamma$ is,
\begin{align}
 \nabla_{\Gamma} \cdot \textbf{A} \equiv \left(\nabla \cdot \widetilde{\textbf{A}}\right) \big \rvert_{\Gamma} -  \hat{\textbf{n}} \cdot \left(\nabla \widetilde{\textbf{A}} \right) \big \rvert_{\Gamma} \cdot \hat{\textbf{n}}.
    \label{eq:surface_divergence}
\end{align}
In \eqref{eq:J_extension}, we have defined our extension such that $\nabla b \cdot \left(\nabla \widetilde{\textbf{J}}_C\right) = 0 $ such that the second term in the above expression vanishes.

Given \eqref{eq:surface_divergence_condition}, we can write,
\begin{subequations}
\begin{align}
J^{\theta} &= -\frac{1}{\sqrt{g}} \partder{\Phi(\theta,\phi)}{\phi} \\
J^{\phi} &= \frac{1}{\sqrt{g}} \partder{\Phi(\theta,\phi)}{\theta},
\end{align}
where,
\begin{align}
   \Phi = \int d \theta \, \sqrt{g} J^{\phi}.
\end{align}
\end{subequations}
In other words,
\begin{align}
    \textbf{J}_C = \hat{\textbf{n}} \times \nabla \Phi.
\end{align}

\renewcommand{\thechapter}{C}

\chapter{Adjoint derivative at fixed $J_{\text{max}}$}
\label{lambda_search}

We enforce $J_{\text{max}}=$ constant in the REGCOIL solve in order to obtain the regularization parameter $\lambda$ by requiring that the following constraint be satisfied within a given tolerance,
\begin{gather}
G\left(\Omega, \overrightarrow{\bm{\Phi}}(\Omega,\lambda)\right) = J_{\text{max}}\left(\Omega, \overrightarrow{\bm{\Phi}}(\Omega, \lambda)\right) - J^{\text{target}}_{\text{max}}  = 0 .
\label{K_constraint}
\end{gather}
Here $J^{\text{target}}_{\text{max}}$ is the target maximum current density and $\overrightarrow{\bm{\Phi}}$ is chosen to satisfy the forward equation (\ref{forward}),
\begin{gather}
\overrightarrow{\textbf{F}} \left(\Omega,\overrightarrow{\bm{\Phi}},\lambda \right) = \overleftrightarrow{\textbf{A}}(\Omega,\lambda) \overrightarrow{\bm{\Phi}} - \overrightarrow{\textbf{b}}(\Omega,\lambda) = 0.
\label{forwardconstraint}
\end{gather}
A log-sum-exponent function is used to approximate the maximum function, similar to that used to approximate $d_{\text{coil-plasma}}$ (\ref{lse_d}),
\begin{gather}
J_{\text{max}} \approx J_{\text{max},\, \text{lse}} =  \frac{1}{p} \log \left( \frac{\int_{S_C} d^2 x \,  \exp\left(p J\right)}{ A_{\text{coil}} } \right).
\end{gather}
We compute the total differential of $\overrightarrow{\textbf{F}}$,
\begin{multline}
d\overrightarrow{\textbf{F}}(\Omega,\overrightarrow{\bm{\Phi}},\lambda) = \sum_{m,n} \left( \partder{\overleftrightarrow{\textbf{A}}(\Omega,\lambda)}{\Omega_{m,n}} \overrightarrow{\bm{\Phi}} - \partder{\overrightarrow{\textbf{b}}(\Omega,\lambda)}{\Omega_{m,n}} \right) d\Omega_{m,n} + \overleftrightarrow{\textbf{A}} d \overrightarrow{\bm{\Phi}} \\
+ \left(\overleftrightarrow{\textbf{A}}^K \overrightarrow{\bm{\Phi}} - \overrightarrow{\textbf{b}}^K\right) d\lambda = 0.
\end{multline}
Here $\overleftrightarrow{\textbf{A}}^K = \partial \overleftrightarrow{\textbf{A}}/\partial \lambda$ and $\overrightarrow{\textbf{b}}^K = \partial \overrightarrow{\textbf{b}}/\partial \lambda$. We left multiply by $\overleftrightarrow{\textbf{A}}^{-1}$ and solve for $d\overrightarrow{\bm{\Phi}}$ such that $d\overrightarrow{\textbf{F}}(\Omega,\overrightarrow{\bm{\Phi}},\lambda)=0$,
\begin{align}
d\overrightarrow{\bm{\Phi}} = - \sum_{m,n} \overleftrightarrow{\textbf{A}}^{-1} \left( \partder{\overleftrightarrow{\textbf{A}}(\Omega,\lambda)}{\Omega_{m,n}} \overrightarrow{\bm{\Phi}} - \partder{\overrightarrow{\textbf{b}}(\Omega,\lambda)}{\Omega_{m,n}} \right) d \Omega_{m,n} - \overleftrightarrow{\textbf{A}}^{-1} \left( \overleftrightarrow{\textbf{A}}^K \overrightarrow{\bm{\Phi}} - \overrightarrow{\textbf{b}}^K \right) d \lambda.
\label{dPhi}
\end{align}
We also compute the total differential of $G$,
\begin{gather}
dG(\Omega,\overrightarrow{\bm{\Phi}}) = \sum_{m,n} \partder{G(\Omega,\overrightarrow{\bm{\Phi}})}{\Omega_{m,n}} d\Omega_{m,n} + \partder{G(\Omega,\overrightarrow{\bm{\Phi}})}{\overrightarrow{\bm{\Phi}}} \cdot d \overrightarrow{\bm{\Phi}} = 0.
\end{gather}
Using the form for $d\overrightarrow{\bm{\Phi}}$ (\ref{dPhi}), we compute $d\lambda$ in terms of $d\Omega_{m,n}$,
\begin{multline}
d\lambda = \left( \partder{G(\Omega,\overrightarrow{\bm{\Phi}})}{\overrightarrow{\bm{\Phi}}} \cdot \left[ \overleftrightarrow{\textbf{A}}^{-1} \left( \overleftrightarrow{\textbf{A}}^K \overrightarrow{\bm{\Phi}} - \overrightarrow{\textbf{b}}^K \right) \right] \right)^{-1}\\
\times \sum_{m,n} \left( \partder{G(\Omega,\overrightarrow{\bm{\Phi}})}{\Omega_{m,n}} - \partder{G(\Omega,\overrightarrow{\bm{\Phi}})}{\overrightarrow{\bm{\Phi}}}\cdot \left[ \overleftrightarrow{\textbf{A}}^{-1} \left( \partder{\overleftrightarrow{\textbf{A}}(\Omega,\lambda)}{\Omega_{m,n}} \overrightarrow{\bm{\Phi}} - \partder{\overrightarrow{\textbf{b}}(\Omega,\lambda)}{\Omega_{m,n}} \right) \right] \right) d \Omega_{m,n}.
\label{dlambda}
\end{multline}
Using (\ref{dPhi}) and (\ref{dlambda}), the derivative of $\overrightarrow{\bm{\Phi}}$ with respect to $\Omega_{m,n}$ subject to equations (\ref{K_constraint}) and (\ref{forwardconstraint}) is given by the following expression,
\begin{multline}
\partder{\overrightarrow{\bm{\Phi}}(\Omega,\lambda(\Omega))}{\Omega_{m,n}} = - \overleftrightarrow{\textbf{A}}^{-1} \left( \partder{\overleftrightarrow{\textbf{A}}(\Omega,\lambda)}{\Omega_{m,n}} \overrightarrow{\bm{\Phi}} - \partder{\overrightarrow{\textbf{b}}(\Omega,\lambda)}{\Omega_{m,n}} \right) - \frac{\overleftrightarrow{\textbf{A}}^{-1} \left( \overleftrightarrow{\textbf{A}}^K \overrightarrow{\bm{\Phi}} - \overrightarrow{\textbf{b}}^K \right) }{ \partder{G(\Omega,\overrightarrow{\bm{\Phi}})}{\overrightarrow{\bm{\Phi}}} \cdot \left[ \overleftrightarrow{\textbf{A}}^{-1} \left( \overleftrightarrow{\textbf{A}}^K \overrightarrow{\bm{\Phi}} - \overrightarrow{\textbf{b}}^K \right) \right] } \\
\times\left( \partder{G(\Omega,\overrightarrow{\bm{\Phi}})}{\Omega_{m,n}} - \partder{G(\Omega,\overrightarrow{\bm{\Phi}})}{\overrightarrow{\bm{\Phi}}} \cdot \left[ \overleftrightarrow{\textbf{A}}^{-1} \left( \partder{\overleftrightarrow{\textbf{A}}(\Omega,\lambda)}{\Omega_{m,n}} \overrightarrow{\bm{\Phi}} - \partder{\overrightarrow{\textbf{b}}(\Omega,\lambda)}{\Omega_{m,n}} \right) \right] \right).
\end{multline}
Here $\overrightarrow{\bm{\Phi}}$ is understood to be a function of $\Omega$ and $\lambda$ through (\ref{forwardconstraint}) and $\lambda$ is understood to be a function of $\Omega$ through (\ref{K_constraint}). We use the adjoint method to avoid solving a linear system involving the operator $\overleftrightarrow{\textbf{A}}$ for each $\Omega_{m,n}$,
\begin{multline}
\partder{\overrightarrow{\bm{\Phi}}(\Omega,\lambda(\Omega))}{\Omega_{m,n}} = - \overleftrightarrow{\textbf{A}}^{-1} \left( \partder{\overleftrightarrow{\textbf{A}}(\Omega,\lambda)}{\Omega_{m,n}} \overrightarrow{\bm{\Phi}} - \partder{\overrightarrow{\textbf{b}}(\Omega,\lambda)}{\Omega_{m,n}} \right) - \frac{\overleftrightarrow{\textbf{A}}^{-1} \left( \overleftrightarrow{\textbf{A}}^K \overrightarrow{\bm{\Phi}} - \overrightarrow{\textbf{b}}^K \right) }{\partder{G(\Omega,\overrightarrow{\bm{\Phi}})}{\overrightarrow{\bm{\Phi}}} \cdot \left[ \overleftrightarrow{\textbf{A}}^{-1} \left( \overleftrightarrow{\textbf{A}}^K \overrightarrow{\bm{\Phi}} - \overrightarrow{\textbf{b}}^K \right) \right]} \\
\times \left( \partder{G(\Omega,\overrightarrow{\bm{\Phi}})}{\Omega_{m,n}} - \left[ \left( \overleftrightarrow{\textbf{A}}^T \right)^{-1} \partder{G(\Omega,\overrightarrow{\bm{\Phi}})}{\overrightarrow{\bm{\Phi}}} \right]  \cdot \left( \partder{\overleftrightarrow{\textbf{A}}(\Omega,\lambda)}{\Omega_{m,n}} \overrightarrow{\bm{\Phi}} - \partder{\overrightarrow{\textbf{b}}(\Omega,\lambda)}{\Omega_{m,n}} \right) \right). 
\label{withadjoint}
\end{multline}
We introduce a new adjoint vector $\overrightarrow{\widetilde{\textbf{q}}}$,
defined to be the solution of,
\begin{gather}
\overleftrightarrow{\textbf{A}}^T \overrightarrow{\widetilde{\textbf{q}}} = \partder{G(\Omega,\overrightarrow{\bm{\Phi}})}{\overrightarrow{\bm{\Phi}}}.
\label{adjoint_2}
\end{gather}
Equation (\ref{withadjoint}) is then used to compute the derivatives of $\chi^2_B$ with respect to $\Omega_{m,n}$,
\begin{gather}
\partder{\chi^2_B\left(\Omega,\overrightarrow{\bm{\Phi}}(\Omega,\lambda(\Omega))\right)}{\Omega_{m,n}} = \partder{\chi^2_B(\Omega,\overrightarrow{\bm{\Phi}})}{\Omega_{m,n}} + \partder{\chi^2_B(\Omega,\overrightarrow{\bm{\Phi}})}{\overrightarrow{\bm{\Phi}}} \cdot \partder{\overrightarrow{\bm{\Phi}}(\Omega,\lambda(\Omega))}{\Omega_{m,n}}.
\end{gather}
This result can be written in terms of both adjoint variables, $\overrightarrow{\textbf{q}}$ and $\overrightarrow{\widetilde{\textbf{q}}}$,
\begin{multline}
\partder{\chi^2_B\left(\Omega,\overrightarrow{\bm{\Phi}}(\Omega,\lambda(\Omega)) \right)}{\Omega_{m,n}} = \partder{\chi^2_B(\Omega,\overrightarrow{\bm{\Phi}})}{\Omega_{m,n}} - \overrightarrow{\textbf{q}} \cdot \left( \partder{\overleftrightarrow{\textbf{A}}(\Omega,\lambda)}{\Omega_{m,n}} \overrightarrow{\bm{\Phi}} - \partder{\overrightarrow{\textbf{b}}(\Omega,\lambda)}{\Omega_{m,n}} \right) \\
- \frac{ \overrightarrow{\textbf{q}} \cdot \left(\overleftrightarrow{\textbf{A}}^K \overrightarrow{\bm{\Phi}} - \overrightarrow{\textbf{b}}^K \right)}{  \overrightarrow{\widetilde{\textbf{q}}} \cdot \left( \overleftrightarrow{\textbf{A}}^K \overrightarrow{\bm{\Phi}} - \overrightarrow{\textbf{b}}^K \right) } \left( \partder{G(\Omega,\overrightarrow{\bm{\Phi}})}{\Omega_{m,n}} - \overrightarrow{\widetilde{\textbf{q}}} \cdot \left( \partder{\overleftrightarrow{\textbf{A}}(\Omega,\lambda)}{\Omega_{m,n}} \overrightarrow{\bm{\Phi}} - \partder{\overrightarrow{\textbf{b}}(\Omega,\lambda)}{\Omega_{m,n}} \right) \right).
\end{multline}
The same method is used to compute derivatives of $\norm{\textbf{J}}_2$. So, to obtain the derivatives at fixed $J_{\text{max}}$, we compute a solution to the two adjoint equations, (\ref{adjoint}) and (\ref{adjoint_2}), in addition to the forward equation, (\ref{forward}).
\renewcommand{\thechapter}{D}

\chapter{Trajectory models}
\label{app:trajectory_models}

In the SFINCS coordinate system, the DKE can be written in the following way,
\begin{equation}
\dot{\textbf{x}} \cdot \nabla f_{1s} + \dot{X}_s \partder{f_{1s}}{X_s} + \dot{\xi}_s \partder{f_{1s}}{\xi_s} - C_s(f_{1s}) = - \left( \textbf{v}_{\text{m}s} \cdot \nabla \psi \right) \partder{f_{Ms}}{\psi}.
\label{eq:dke_model}
\end{equation}
To obtain the trajectory coefficients ($\dot{\textbf{x}}$, $\dot{X}_s$, and $\dot{\xi}_s$) several approximations are made. For example, any terms that require radial coupling ($\psi$ derivatives of $f_{1s}$) cannot be retained, as this would necessitate solving a five-dimensional system.

Under the full trajectory model, the trajectory coefficients are chosen such that $\mu$ conservation is maintained as radial coupling is dropped,
\begin{subequations}
\begin{align}
 \dot{\textbf{x}} &= v_{||} \hat{\textbf{b}} + \frac{\Phi'(\psi)}{B^2}  \textbf{B} \times \nabla \psi  \\
 \dot{X}_s &= - \left( \textbf{v}_{\text{m}s} \cdot \nabla \psi \right) \frac{q_s}{2T_s X_s} \Phi'(\psi)  \\
 \dot{\xi}_s &= - \frac{1-\xi_s^2}{2B\xi_s} v_{||} \hat{\textbf{b}} \cdot \nabla B + \xi_s(1-\xi_s^2) \frac{1}{2B^3} \Phi'(\psi) \textbf{B}\times \nabla \psi \cdot \nabla B.
\end{align}
 \label{eq:full_trajectories}
\end{subequations}
Under the DKES trajectory model, the $\textbf{E} \times \textbf{B}$ velocity is taken to be divergenceless,
\begin{gather}
    \textbf{v}_E^{\text{DKES}} = \frac{\textbf{B} \times \nabla \Phi}{\langle B^2 \rangle_{\psi}},
\end{gather}
where the flux surface average of a quantity is \eqref{eq:flux_surface_average_ch4}. Under the DKES trajectory model, the trajectory coefficients are taken to be, 
\begin{subequations}
\begin{align}
\dot{\textbf{x}} &= v_{||} \hat{\textbf{b}} + \frac{1}{\langle B^2\rangle_{\psi}} \Phi'(\psi) \textbf{B} \times \nabla \psi  \\
\dot{X}_s &= 0  \\
\dot{\xi}_s &= - \frac{1-\xi_s^2}{2 B\xi_s} v_{||} \hat{\textbf{b}} \cdot \nabla B. 
\end{align}
\label{eq:dkes_trajectories}
\end{subequations}
These effective trajectories are adopted in the widely-used DKES code \citep{Hirshman1986,Rij1989}.
\renewcommand{\thechapter}{E}

\chapter{Adjoint collision operator}
\label{app:collision}

We want to find an adjoint collision operator, $C_s^{\dagger}$, that satisfies the following relation,
\begin{gather}
    \left \langle \int d^3 v \, \frac{g_{1s} C_s(f_{1s})}{f_{Ms}} \right \rangle_{\psi} = \left \langle \int d^3 v\, \frac{f_{1s} C_s^{\dagger}(g_{1s})}{f_{Ms}} \right \rangle_{\psi}.
\end{gather}
The linearized Fokker-Planck collision operator can be written as,
\begin{gather}
    C_s(f_{1s}) = \sum_{s'} C_{ss'}^L(f_{1s},f_{1s'}) = \sum_{s'} C_{ss'}(f_{1s},f_{Ms'}) + C_{ss'}(f_{Ms},f_{1s'}), 
    \label{eq:C_f_1s}
\end{gather}
where $s'$ sums over species. The first term on the right hand side of \eqref{eq:C_f_1s} is referred to as the test-particle collision operator, $C^T_{ss'}(f_{1s}) = C_{ss'}(f_{1s},f_{Ms'})$, and the second the field-particle collision operator, $C^F_{ss'}(f_{1s'}) = C_{ss'}(f_{Ms},f_{1s'})$. The test and field terms satisfy the following relations \citep{Rosenbluth1972,Sugama2009},
\begin{subequations}
\begin{align}
    \int d^3 v \, \frac{g_{1s} C_{ss'}(f_{1s}, f_{Ms'})}{f_{Ms}} &= \int d^3 v \, \frac{f_{1s} C_{ss'}(g_{1s},f_{Ms'})}{f_{Ms}} \\
    \int d^3 v \, \frac{g_{1s} C_{ss'}(f_{Ms},f_{1s'})}{f_{Ms}}  &= \frac{T_{s'}}{T_{s}} \int d^3 v \, \frac{f_{1s'}C_{s's}(f_{Ms'},g_{1s})}{f_{Ms'}}.
\end{align}
\end{subequations}
For collisions between species of the same temperature, we see that $C_{s}(f_{1s})$ is self-adjoint. The adjoint operator with respect to the inner product \eqref{eq:inner_product} is thus,
\begin{align}
  C_s^{\dagger} &= C_s^T + \sum_{s'} \frac{f_{Ms}}{f_{Ms'}} \frac{T_{s'}}{T_s} C_{s's}^F .
\end{align}
\renewcommand{\thechapter}{F}

\chapter{Adjoint collisionless trajectories}
\label{ap:adjoint_operators}

We want to find an adjoint operator, $\mathbb{L}_{0s}^{\dagger}$, that satisfies,
\begin{gather}
    \left \langle \int d^3 v \, \frac{g_{1s}\mathbb{L}_{0s} f_{1s}}{f_{Ms}} \right \rangle_{\psi} = \left \langle \int d^3 v \, \frac{f_{1s}\mathbb{L}_{0s}^{\dagger} g_{1s} }{f_{Ms}} \right \rangle_{\psi},
    \label{eq:L_0s_adjoint}
\end{gather}
for both trajectory models, where $\mathbb{L}_{0s}$ is defined in \eqref{eq:L_0s} with \eqref{eq:dkes_trajectories} for the DKES trajectories model and \eqref{eq:full_trajectories} for the full trajectory model. Throughout we use the velocity space element in SFINCS coordinates, $d^3 v = 2\pi v_{ts}^3 X_s^2 d \xi_s d X_s$.  

\subsection{DKES trajectories}

The operator under consideration is,
\begin{gather}
    \mathbb{L}_{0s} = v_{||} \hat{\textbf{b}} \cdot \nabla + \hat{\textbf{v}}_E^{\text{DKES}} \cdot \nabla - \frac{1-\xi_s^2}{2B\xi_s} v_{||} \hat{\textbf{b}} \cdot \nabla B \partder{}{\xi_s}. \label{eq:L_0s_dkes}
\end{gather}
Considering the contribution of the streaming term in \eqref{eq:L_0s_dkes} to the left hand side of \eqref{eq:L_0s_adjoint} we obtain,
\begin{gather}
    \left \langle \int d^3 v \, \frac{g_{1s} v_{||} \hat{\textbf{b}} \cdot \nabla f_{1s}}{f_{Ms}} \right \rangle_{\psi}
    = - \left \langle \int d^3 v \, \frac{f_{1s} v_{||} \textbf{B} \cdot \nabla \left(g_{1s}/B\right)}{f_{Ms}} \right \rangle_{\psi}. \label{eq:dkes_parallel_streaming}
\end{gather}
Here the identity $\langle \nabla \cdot \textbf{Q}\rangle_{\psi} = 1/V'(\psi) \partial/\partial \psi \left( V'(\psi) \langle \textbf{Q} \cdot \nabla \psi \rangle_{\psi} \right)$ for any vector $\textbf{Q}$ has been used. We next consider the contribution of the $\textbf{E} \times \textbf{B}$ drift term in \eqref{eq:L_0s_dkes},
\begin{align}
    \left \langle \int d^3 v \, \frac{g_{1s} \textbf{v}_E^{\text{DKES}} \cdot \nabla f_{1s} }{f_{Ms}} \right \rangle_{\psi} 
    &= -\left \langle \int d^3 v \, \frac{f_{1s} \textbf{v}_E^{\text{DKES}} \cdot \nabla g_{1s} }{f_{Ms}} \right \rangle_{\psi}.
    \label{eq:vE_dkes}
\end{align}
Here we have used the identity,
\begin{gather}
    \left \langle \textbf{B} \times \nabla \psi \cdot \nabla w \right \rangle_{\psi} = 0,
    \label{eq:B_times_nabla_psi}
\end{gather}
    for any $w$. We consider the contribution of the mirror-force term in \eqref{eq:L_0s_dkes}, 
\begin{align}
 \left \langle \int d^3 v \, \frac{g_{1s}  \dot{\xi}_s}{f_{Ms}} \partder{ f_{1s}}{\xi_s} \right \rangle_{\psi} 
 = -\left \langle \int d^3 v \, \frac{f_{1s} \dot{\xi}_s}{f_{Ms}} \partder{g_{1s}}{\xi_s} \right \rangle_{\psi} -\left \langle \int d^3 v \, \frac{v_{||}}{B} \hat{\textbf{b}} \cdot \nabla B \frac{g_{1s} f_{1s}}{f_{Ms}} \right \rangle_{\psi}. 
 \label{eq:dkes_mirror}
\end{align}
Combining (\ref{eq:dkes_parallel_streaming}-\ref{eq:dkes_mirror}), we obtain
\begin{gather}
    \left \langle \int d^3 v \, \frac{g_{1s} \mathbb{L}_{0s} f_{1s}}{f_{Ms}} \right \rangle_{\psi} = - \left \langle \int d^3 v \, \frac{f_{1s} \mathbb{L}_{0s} g_{1s}}{f_{Ms}} \right \rangle_{\psi}. 
\end{gather}
Therefore, in the DKES trajectory model we obtain \eqref{eq:dkes_adjoint}.

\subsection{Full trajectories}
The operator under consideration for the full model is,
\begin{multline}
    \mathbb{L}_{0s} = v_{||} \hat{\textbf{b}} \cdot \nabla + \textbf{v}_E \cdot \nabla + \frac{(1 + \xi_s^2)X_s}{2B} \textbf{v}_E \cdot \nabla B  \partder{}{X_s} \\ - \frac{1-\xi_s^2}{2B\xi_s} v_{||} \hat{\textbf{b}} \cdot \nabla B \partder{}{\xi_s}  +  \frac{\xi_s(1-\xi_s^2)}{2B}  \textbf{v}_E \cdot \nabla B \partder{}{\xi_s}.
    \label{eq:L_0s_full}
\end{multline}
The contribution to \eqref{eq:L_0s_adjoint} from the streaming term in \eqref{eq:L_0s_full} is identical to that in the case of the DKES trajectory model, \eqref{eq:dkes_parallel_streaming}. We next consider the contribution from the $\textbf{E} \times \textbf{B}$ drift term in \eqref{eq:L_0s_full}, 
\begin{align}
    \left \langle \int d^3 v \, \frac{g_{1s} \textbf{v}_E \cdot \nabla f_{1s}}{f_{Ms}} \right \rangle_{\psi}
    &= - \left \langle \int d^3 v \, \frac{f_{1s} B^2 \textbf{v}_E \cdot \nabla \left(g_{1s}/B^2\right)}{f_{Ms}} \right \rangle_{\psi}, 
    \label{eq:full_ve}
\end{align}
again using \eqref{eq:B_times_nabla_psi}. The contribution from the $\dot{X}_s$ term in \eqref{eq:L_0s_full} is, 
\begin{multline}
    \left \langle \int d^3 v \, \frac{g_{1s} \dot{X}_s }{f_{Ms}} \partder{f_{1s}}{X_s} \right \rangle_{\psi} = - \left \langle \int d^3 v \, \frac{f_{1s} \dot{X}_s }{f_{Ms}} \partder{g_{1s}}{X_s} \right \rangle_{\psi} \\ - \left \langle \int d^3 v \, (3+2X_s^2)(1+\xi_s^2) \frac{g_{1s}f_{1s}}{2f_{Ms}B} \textbf{v}_E \cdot \nabla B \right \rangle_{\psi}. 
    \label{eq:full_x}
\end{multline}
The contribution from the mirror term in \eqref{eq:L_0s_full} is the same as in the case of the DKES trajectories model \eqref{eq:dkes_mirror}. We consider the contribution from the final term in \eqref{eq:L_0s_full}, 
\begin{multline}
    \left \langle \int d^3 v \, \frac{g_{1s} \xi_s (1- \xi_s^2) \textbf{v}_E \cdot \nabla B }{2Bf_{Ms}} \partder{f_{1s}}{\xi_s} \right \rangle_{\psi} =  - \left \langle \int d^3 v \, \frac{f_{1s} \xi_s (1-\xi_s^2) \textbf{v}_E \cdot \nabla B}{2B f_{Ms}} \partder{g_{1s}}{\xi_s} \right \rangle_{\psi} \\
    - \left \langle \int d^3 v \, (1-3 \xi_s^2) \textbf{v}_E \cdot \nabla B \frac{f_{1s} g_{1s}}{2Bf_M} \right \rangle_{\psi}. 
    \label{eq:full_xi_2}
\end{multline}
Combining \eqref{eq:dkes_parallel_streaming}, \eqref{eq:full_ve}, \eqref{eq:full_x}, \eqref{eq:dkes_mirror}, and \eqref{eq:full_xi_2}, we obtain
\begin{multline}
    \left \langle \int d^3 v \, \frac{g_{1s} \mathbb{L}_{0s} f_{1s}}{f_{Ms}} \right \rangle_{\psi} = - \left \langle \int d^3 v \, \frac{f_{1s} \mathbb{L}_{0s} g_{1s}}{f_{Ms}} \right \rangle_{\psi} \\ + \Phi'(\psi) \frac{q_s}{T_s} \left \langle \int d^3 v \, \left( \textbf{v}_{\text{m}s} \cdot \nabla \psi \right) \frac{f_{1s} g_{1s}}{f_{Ms}} \right \rangle_{\psi}. 
\end{multline}
Therefore, under the full trajectory model we obtain \eqref{eq:full_adjoint}.

\renewcommand{\thechapter}{G}

\chapter{Symmetry of the sensitivity function}
\label{app:symmetry}

In this Appendix we discuss several symmetry properties of the local sensitivity function, $S_{\mathcal{R}}$, defined through \eqref{eq:magnetic_sensitivity}. The arguments that follow are similar to those in Appendix C of \cite{Landreman2018}. Throughout we will assume that $B$ is stellarator symmetric and $N_P$ symmetric. We will show that this implies $N_P$ symmetry of $S_{\mathcal{R}}$. In the limit that $E_r \rightarrow 0$, then $S_{\mathcal{R}}$ also has stellarator symmetry.

\subsection{Symmetry of $S_{\mathcal{R}}$ implied by Fourier derivatives}

First we would like to show that $S_{\mathcal{R}}$ is stellarator symmetric if and only if $\partial \mathcal{R}/\partial B_{m,n}^s = 0$ for all $m$ and $n$, where we express $B$ in a Fourier series,
\begin{gather}
    B = \sum_{m,n} B_{m, n}^c \cos(m \vartheta_B - n \varphi_B) + B_{m, n}^s \sin(m \vartheta_B - n  \varphi_B). 
\end{gather}
The perturbation, $\delta B$, is decomposed similarly. We begin with the ``if'' portion of the argument. From \eqref{eq:magnetic_sensitivity} we have, 
\begin{align}
    \partder{\mathcal{R}}{B_{m,n}^s} &= V'(\psi)^{-1} \int_0^{2\pi} d \vartheta_B \int_0^{2\pi} d \varphi_B \, \sqrt{g} S_{\mathcal{R}} \sin(m \vartheta_B - n \varphi_B).
    \label{eq:dRdBmns}
\end{align}
Suppose $\partial \mathcal{R}/\partial B_{m,n}^s = 0$ for all $m$ and $n$. The quantity $(\sqrt{g} S_{\mathcal{R}})$ can be represented as a Fourier series,
\begin{gather}
    \left( \sqrt{g} S_{\mathcal{R}} \right) = \sum_{m,n} A_{m,n}^c \cos(m \vartheta_B - n \varphi_B) + A_{m,n}^s \sin(m \vartheta_B - n \varphi_B).
    \label{eq:sqrtg_SR}
\end{gather}
From \eqref{eq:dRdBmns}, we see that $A_{m,n}^s = 0$ for all $m$ and $m$. 
Thus the quantity $(\sqrt{g}S_{\mathcal{R}})$ must be even under the transformation $(\vartheta_B,\varphi_B) \rightarrow (-\vartheta_B,-\varphi_B)$. We now note that $\sqrt{g}$ must be even from \eqref{eq:jacobian} under the assumption that $B$ is stellarator symmetric. Therefore $S_{\mathcal{R}}$ must be stellarator symmetric, assuming that $\sqrt{g}$ does not vanish anywhere, which must be the case for any well-defined coordinate transformation.

We continue with the ``only if" portion of the argument. Suppose $S_{\mathcal{R}}$ is stellarator symmetric. As $\sqrt{g}$ is also stellarator symmetric, $(\sqrt{g} S_{\mathcal{R}})$ can be expressed in a Fourier series as \eqref{eq:sqrtg_SR} with $A_{m,n}^s =0$ for all $m$ and $n$. Thus from \eqref{eq:dRdBmns} $\partial \mathcal{R}/\partial B_{m,n}^s = 0$ for all $m$ and $n$. 

We next show that if $B$ is $N_P$ symmetric, then $S_{\mathcal{R}}$ is $N_P$ symmetric if and only if $\partial \mathcal{R}/\partial B_{m,n}^c = 0$ for all $n$ that are not integer multiples of $N_P$. We begin with the ``if" portion of the argument. From \eqref{eq:magnetic_sensitivity},
\begin{gather}
    \partder{\mathcal{R}}{B_{m,n}^c} = V'(\psi)^{-1} \int_0^{2\pi} d \vartheta_B \int_0^{2\pi} d \varphi_B \, \sqrt{g} S_{\mathcal{R}} \cos(m \vartheta_B - n \varphi_B) 
    \label{eq:dRdBmnc}. 
\end{gather}
Suppose $\partial \mathcal{R}/\partial B_{m,n}^c = 0$ for all $n$ which are not integer multiples of $N_P$. Here $(\sqrt{g} S_{\mathcal{R}})$ can be expressed in a Fourier series as \eqref{eq:sqrtg_SR} with $A_{m,n}^s = 0$ for all $m$ and $n$. Inserting the Fourier series into \eqref{eq:dRdBmnc}, we find that $A_{m,n}^c = 0$ for all $n$ that are not integer multiples of $N_P$. Thus $(\sqrt{g} S_{\mathcal{R}})$ must be $N_P$ symmetric. As $\sqrt{g}$ must be $N_P$ symmetric, this implies $S_{\mathcal{R}}$ possesses the same symmetry.

Next we consider the ``only if'' portion of the argument. Suppose that $S_{\mathcal{R}}$ is $N_P$ symmetric. As $\sqrt{g}$ is also $N_P$ symmetric, then $(\sqrt{g}S_{\mathcal{R}})$ can be expressed in a Fourier series as \eqref{eq:sqrtg_SR} where the sum includes $n$ that are integer multiples of $N_P$. Inserting the Fourier series into \eqref{eq:dRdBmnc}, we find that $\partial \mathcal{R}/\partial B_{m,n}^c = 0$ for all $n$ that are not integer multiples of $N_P$. 

\subsection{Symmetry of Fourier derivatives}

To continue, we need to show that $\partial \mathcal{R}/\partial B_{m,n}^s = 0$ for all $m$ and $n$ and $\partial \mathcal{R}/\partial B_{m,n}^c = 0$ for all $n$ which are not integer multiples of $N_P$. We begin with the $N_P$ symmetry argument. We consider the symmetry of $f_{1s}$ implied by \eqref{eq:dke_model}. Under the transformation $\varphi_B\rightarrow \varphi_B + 2\pi/N_P$, we find that each of the trajectory coefficients remain unchanged, as well as the source term and collision operator. Therefore we can conclude that $f_{1s}$ is $N_P$ symmetric. We can also note that each of the $\widetilde{\mathcal{R}}$ vectors are $N_P$ symmetric, as well as $\sqrt{g}$. We consider the integrand that appears in the flux surface average in \eqref{eq:inner_product_R},
\begin{gather}
    D_s(\vartheta_B,\varphi_B) = \int d^3 v \, \frac{f_{1s} \widetilde{\mathcal{R}}_s^f \sqrt{g}}{f_{Ms}}.
\end{gather}
Here the superscript and subscript on $\widetilde{\mathcal{R}}$ denotes that we consider the unknowns corresponding to the distribution function of species $s$. We note that $D_s(\vartheta_B,\varphi_B+2\pi/N_P) = D_s(\vartheta_B,\varphi_B)$. The quantity $\mathcal{R}$ can be expressed in terms of $D_s$ as follows,
\begin{gather}
\mathcal{R} = \sum_s V'(\psi)^{-1}\int_0^{2\pi} d \varphi_B \int_0^{2\pi} d \vartheta_B \, D_s.
\label{eq:R}
\end{gather}
Next we consider the functional derivative of $\mathcal{R}$ with respect to $B$, defined as in \eqref{eq:functional_derivative}. The derivative with respect to $B_{m,n}^c$ can be thus defined as,
\begin{gather}
    \partder{\mathcal{R}}{B_{m,n}^c} = V'(\psi)^{-1} \int_0^{2\pi} d \varphi_B \int_0^{2\pi} d \vartheta_B \, \left( \sum_s \frac{\delta D_s}{\delta B} - \mathcal{R} \frac{\delta \sqrt{g}}{\delta B} \right) \cos(m \vartheta_B - n \varphi_B).
    \label{eq:C}
\end{gather}
As the functional derivative maintains the $N_P$ symmetry of $D_s$ and $\sqrt{g}$, the quantity in parenthesis in \eqref{eq:C} can be expressed in a Fourier series containing only $n$ that are integer multiples of $N_P$. Thus we see that the quantity $\partial \mathcal{R}/\partial B_{m,n}^c = 0$ for all $n$ that are not integer multiples of $N_P$. 

Next we consider a similar argument for stellarator symmetry. We begin by considering the symmetry of $f_{1s}$ implied by \eqref{eq:dke_model} in the case $E_r = 0$. Under the transformation $(\vartheta_B,\varphi_B,v_{||}) \rightarrow (-\vartheta_B,-\varphi_B,-v_{||})$, we see that both the collisionless trajectory operator and the collision operator maintain the parity of $f_{1s}$, while the source term is odd. Therefore, $f_{1s}$ must be odd under this transformation. In this case, we can write $f_{1s}$ as,
\begin{gather}
    f_{1s} = f^{-}_{a,s}(X_s,\xi_s)f^+_{b,s}(\vartheta_B,\varphi_B) + f^{+}_{a,s}(X_s,\xi_s)f^-_{b,s}(\vartheta_B,\varphi_B),
    \label{eq:f_1s}
\end{gather}
where $f^-_{a,s}(X_s, -\xi_s) = -f^-_{a,s}(X_s, \xi_s)$,  $f^+_{a,s}(X_s, -\xi_s) = f^+_{a,s} (X_s, -\xi_s)$, and analogous expressions for $f_{b,s}^+$ and $f_{b,s}^-$.

We next note that each of the $\widetilde{\mathcal{R}}^f_s$ are odd under the transformation $(\vartheta_B,\varphi_B,v_{||}) \rightarrow (-\vartheta_B,-\varphi_B,-v_{||})$. As $\sqrt{g}$ is even, then we can express $\widetilde{\mathcal{R}}_s^f\sqrt{g}$ in a similar way to \eqref{eq:f_1s},
\begin{gather}
    \widetilde{\mathcal{R}}_s^f\sqrt{g}= B_{a,s}^-(X_s,\xi_s) B_{b,s}^+(\vartheta_B,\varphi_B)+B_{a,s}^+(X_s,\xi_s)B_{b,s}^-(\vartheta_B,\varphi_B). 
\end{gather}
The integrand that appears in the flux surface average becomes,
\begin{multline}
    D_s = \int d^3 v \, f_{Ms}^{-1} \bigg( f_{a,s}^-(X_s,\xi_s) B_{a,s}^-(X_s,\xi_s) f_{b,s}^+(\vartheta_B,\varphi_B) B_{b,s}^+(\vartheta_B,\varphi_B) \\ + f_{a,s}^+(X_s,\xi_s)B_{a,s}^+(X_s,\xi_s) f_{b,s}^-(\vartheta_B,\varphi_B)B_{b,s}^-(\vartheta_B,\varphi_B) \bigg).
\end{multline}
We see that $D_s$ is even with respect to the transformation $(\vartheta_B,\varphi_B) \rightarrow (-\vartheta_B,-\varphi_B)$. 
The quantity $\mathcal{R}$ can be written as in \eqref{eq:R} and the derivative with respect to a stellarator asymmetric mode is
\begin{gather}
    \partder{\mathcal{R}}{B_{m,n}^s} = V'(\psi)^{-1} \int_0^{2\pi} d \varphi_B \int_0^{2\pi} d \vartheta_B \, \left(\sum_s \frac{\delta D_s}{\delta B} - \mathcal{R} \frac{\delta \sqrt{g}}{\delta B} \right) \sin(m \vartheta_B - n \varphi_B). 
\end{gather}
The functional derivative with respect to $B$ does not change the parity of $D_s$ or $\sqrt{g}$, thus we see that the quantity in parenthesis in the above equation is even with respect to the transformation $(\vartheta_B,\varphi_B) \rightarrow (-\vartheta_B,-\varphi_B)$. Therefore, $\partial \mathcal{R}/\partial B_{m,n}^s = 0$ for all $m$ and $n$. A similar argument cannot be made if $E_r \neq 0$, as the inhomogeneous drive term in \eqref{eq:dke_model} no longer has definite parity. However, according to the arguments in \citep{Hirshman1986b} the transport coefficients do obey this symmetry property.

\renewcommand{\thechapter}{H}

\chapter{Derivatives at ambipolarity}
\label{app:ambipolar}

In this Appendix, we derive an expression for derivatives of moments of the distribution function at fixed ambipolarity rather than fixed $E_r$ by determining the relationship between geometry parameters, $\Omega$, and $E_r$. We begin by assuming that the continuous adjoint approach outlined in Section \ref{sec:continuous} is used. The approach taken here is analogous to that used in Appendix \ref{lambda_search}, in which an additional adjoint equation is used to compute derivatives at a fixed constraint function for optimization of stellarator coil shapes. 

Consider the set of unknowns computed with SFINCS, $F$, which depends on parameters $\Omega$ and $E_r$. The total  differential of $F$ satisfies,
\begin{multline}
   \mathbb{L} dF(\Omega,E_r) = \left( \partder{\mathbb{S}(\Omega,E_r)}{E_r}-\partder{\mathbb{L}(\Omega,E_r)}{E_r} F\right) dE_r \\
   + \sum_{i=1}^{N_{\Omega}}  \left( \partder{\mathbb{S}(\Omega,E_r)}{\Omega_i} - \partder{\mathbb{L}(\Omega,E_r)}{\Omega_i}F\right)  d \Omega_i,
    \label{eq:dF}
\end{multline}
which follows from \eqref{eq:linear}. Consider $J_r(\Omega,F)$, which depends on $E_r$ through $F$. The total differential of $J_r$ can be computed,
\begin{gather} dJ_r(\Omega,F(\Omega,E_r)) = \sum_{i=1}^{N_{\Omega}} \partder{J_r\left(\Omega,F\right)}{\Omega_i} d \Omega_i + \left \langle \widetilde{J_r}, dF(\Omega,E_r) \right \rangle,
\end{gather}
which can be written using \eqref{eq:dF} and the solution to \eqref{eq:J_r_adjoint},
\begin{multline}
    dJ_r(\Omega,F(\Omega,E_r)) = \left \langle \lambda^{J_r}, \left( \partder{\mathbb{L}(\Omega,E_r)}{E_r} F -\partder{\mathbb{S}(\Omega,E_r)}{E_r}\right) \right \rangle d E_r \\
    + \sum_{i=1}^{N_{\Omega}} \left(\partder{J_r(\Omega,F)}{\Omega_i} + \left \langle \lambda^{J_r}, \left(\partder{\mathbb{L}(\Omega,E_r)}{\Omega_i} F -\partder{\mathbb{S}(\Omega,E_r)}{\Omega_i}\right) \right  \rangle  \right) d \Omega_i.
\end{multline}
By enforcing $dJ_r(\Omega,F(\Omega,E_r)) = 0$, we obtain the relationship between $E_r$ and $\Omega$ at ambipolarity,
\begin{multline}
     \partder{E_r(\Omega)}{\Omega_i} \bigg \rvert_{dJ_r = 0}  = -\left \langle \lambda^{J_r}, \left(\partder{\mathbb{L}(\Omega,E_r)}{E_r} F - \partder{\mathbb{S}(\Omega,E_r)}{E_r}\right) \right \rangle^{-1} \\
    \left(\partder{J_r(\Omega,F)}{\Omega_i} + \left \langle \lambda^{J_r}, \left(\partder{\mathbb{L}(\Omega,E_r)}{\Omega_i}F - \partder{\mathbb{S}(\Omega,E_r)}{\Omega_i}  \right)\right \rangle \right).
    \label{eq:dErdOmega}
\end{multline}
Consider a moment of the distribution function, $\mathcal{R}(\Omega,F(\Omega,E_r))$. The derivative with respect to $\Omega_i$ at fixed ambipolarity can thus be computed,
\begin{gather}
    \partder{\mathcal{R}(\Omega,F(\Omega,E_r(\Omega))}{\Omega_i} =  \partder{\mathcal{R}(\Omega,F)}{\Omega_i} + \left \langle \widetilde{\mathcal{R}}, \partder{F(\Omega,E_r(\Omega))}{\Omega_i} \right \rangle,
\end{gather}
where $E_r$ is viewed as a function of $\Omega$ through \eqref{eq:dErdOmega}. The first term corresponds to the explicit dependence on $\Omega_i$, while the second contains dependence through $F$. 
Here $\partial F(\Omega,E_r(\Omega))/\partial \Omega_i$ satisfies,
\begin{multline}
    \mathbb{L} \partder{F(\Omega,E_r(\Omega))}{\Omega_i} = \left(\partder{\mathbb{S}(\Omega,E_r)}{\Omega_i} - \partder{\mathbb{L}(\Omega,E_r)}{\Omega_i} F\right) \\
    -  \left( \partder{\mathbb{S}(\Omega,E_r)}{E_r}- \partder{\mathbb{L}(\Omega,E_r)}{E_r} F\right) \left \langle \lambda^{J_r}, \left(\partder{\mathbb{L}(\Omega,E_r)}{E_r}F -\partder{\mathbb{S}(\Omega,E_r)}{E_r} \right) \right \rangle^{-1} \\ \times \left( \partder{J_r(\Omega,F)}{\Omega_i} + \left \langle \lambda^{J_r}, \left(\partder{\mathbb{L}(\Omega,E_r)}{\Omega_i} F -\partder{\mathbb{S}(\Omega,E_r)}{\Omega_i}\right) \right \rangle \right),
    \label{eq:dFdOmega}
\end{multline}
from \eqref{eq:dF} using \eqref{eq:dErdOmega}.
Using \eqref{eq:dFdOmega} and \eqref{eq:adjoint}, we find
\begin{multline}
     \partder{\mathcal{R}(\Omega,F(\Omega,E_r(\Omega))}{\Omega_i}  = \partder{\mathcal{R}(\Omega,F)}{\Omega_i} + \left \langle \lambda^{\mathcal{R}}, \left( \partder{\mathbb{L}(\Omega,E_r)}{\Omega_i} F-\partder{\mathbb{S}(\Omega,E_r)}{\Omega_i}\right) \right \rangle
    \\ - \left \langle \lambda^{\mathcal{R}}, \left( \partder{\mathbb{L}(\Omega,E_r)}{E_r} F-\partder{\mathbb{S}(\Omega,E_r)}{E_r}\right) \right \rangle \\
    \times \frac{\left(\partder{J_r(\Omega,F)}{\Omega_i} + \left \langle \lambda^{J_r}, \left(\partder{\mathbb{L}(\Omega,E_r)}{\Omega_i} F -\partder{\mathbb{S}(\Omega,E_r)}{\Omega_i}  \right)\right \rangle \right)}{\left \langle \lambda^{J_r}, \left(\partder{\mathbb{L}(\Omega,E_r)}{E_r} F - \partder{\mathbb{S}(\Omega,E_r)}{E_r} \right) \right \rangle}. 
\end{multline}
An analogous expression can be obtained using the discrete approach,
\begin{multline}
    \partder{\mathcal{R}\left(\Omega,\overrightarrow{\textbf{F}}\left(\Omega,E_r(\Omega)\right)\right)}{\Omega_i}  = \partder{\mathcal{R}\left(\Omega,\overrightarrow{\textbf{F}}\right)}{\Omega_i}
    + \left \langle \overrightarrow{\bm{\lambda}}^{\mathcal{R}}, \left( \partder{\overrightarrow{\textbf{S}}(\Omega,E_r)}{\Omega_i}- \partder{\overleftrightarrow{\textbf{L}}(\Omega,E_r)}{\Omega_i} \overrightarrow{\textbf{F}} \right) \right \rangle \\
    - \left \langle \overrightarrow{\bm{\lambda}}^{\mathcal{R}}, \left(\partder{\overrightarrow{\textbf{S}}(\Omega,E_r)}{E_r} - \partder{\overleftrightarrow{\textbf{L}}(\Omega,E_r)}{E_r} \overrightarrow{\textbf{F}} \right) \right \rangle \\
    \times \frac{\left(\partder{J_r\left(\Omega,\overrightarrow{\textbf{F}}\right)}{\Omega_i} + \left \langle \overrightarrow{\bm{\lambda}}^{J_r}, \left( \partder{\overrightarrow{\textbf{S}}(\Omega,E_r)}{\Omega_i}- \partder{\overleftrightarrow{\textbf{L}}(\Omega,E_r)}{\Omega_i} \overrightarrow{\textbf{F}} \right)\right \rangle \right)}{\left \langle \overrightarrow{\bm{\lambda}}^{J_r}, \left(\partder{\overrightarrow{\textbf{S}}(\Omega,E_r)}{E_r} - \partder{\overleftrightarrow{\textbf{L}}(\Omega,E_r)}{E_r} \overrightarrow{\textbf{F}} \right) \right \rangle},
\end{multline}
where \eqref{eq:J_r_adjoint_discrete} has been used.

\renewcommand{\thechapter}{I}

\chapter{Derivation of generalized MHD self-adjointness relation} 
\label{app:adjoint_relation}
The quantity $U_P = U_{P_1} + U_{P_2}$ consists of two terms, accounting for changes to the vector potential due to MHD perturbations,
\begin{align}
U_{P_1} = \int_{V_P} d^3 x \, \left(\delta\textbf{J}_{1} \cdot \bm{\xi}_2 \times \textbf{B} - \delta \textbf{J}_2\cdot \bm{\xi}_1 \times \textbf{B} \right),
\end{align}
and changes to the rotational transform,
\begin{align}
    U_{P_2} =  \int_{V_P} d^3 x \, \left(\delta \chi_1(\psi) \delta \textbf{J}_2 \cdot \nabla \varphi - \delta \chi_2(\psi) \delta \textbf{J}_1 \cdot \nabla \varphi \right). 
    \label{eq:Up2}
\end{align}
The quantity $U_{P_1}$ can be expressed by using \eqref{eq:general_force_operator} and applying the divergence theorem to the pressure gradient terms, 
\begin{multline}
    U_{P_1} = \int_{V_P} d^3 x \, \bm{\xi}_2 \cdot \left( \textbf{J} \times \delta \textbf{B}_1 + \nabla p\left(\nabla \cdot \bm{\xi}_1\right) - \textbf{F}_1 \right) \\
    - \int_{V_P} d^3 x \, \bm{\xi}_1 \cdot \left( \textbf{J} \times \delta \textbf{B}_2 +\nabla p \left( \nabla \cdot \bm{\xi}_2 \right) - \textbf{F}_2 \right).
    \label{eq:Up1}
\end{multline}
We will define $\delta \widetilde{\textbf{B}}_{1,2} = \nabla \times \left(\bm{\xi}_{1,2} \times \textbf{B} \right)$ such that $\delta \textbf{B}_{1,2} = \delta \widetilde{\textbf{B}}_{1,2} - \nabla \delta \chi_{1,2}(\psi) \times \nabla \varphi$.
The terms in \eqref{eq:Up1} due to $\delta \widetilde{\textbf{B}}_{1,2}$ can be evaluated using $\textbf{J} = J_{||}\hat{\textbf{b}} + \hat{\textbf{b}}\times\nabla p/B$ and \eqref{eq:force_balance},
\begin{multline}
    \int_{V_P} d^3 x \, \left( \bm{\xi}_2 \cdot \textbf{J} \times \delta \widetilde{\textbf{B}}_1 - \bm{\xi}_1 \cdot \textbf{J} \times \delta \widetilde{\textbf{B}}_2\right) = \int_{V_P} d^3 x \, \frac{J_{||}}{B}\nabla \cdot \left( \left(\bm{\xi}_1 \times \textbf{B} \right) \times \left( \bm{\xi}_2 \times \textbf{B} \right) \right)\\ + \int_{V_P} d^3 x \, \frac{1}{B} \left( \left(\bm{\xi}_2 \cdot \nabla p \right) \hat{\textbf{b}} \cdot \delta \widetilde{\textbf{B}}_1  - \left(\bm{\xi}_1 \cdot \nabla p \right) \hat{\textbf{b}} \cdot \delta \widetilde{\textbf{B}}_2 \right).
    \label{eq:Upp}
\end{multline}
The first term in \eqref{eq:Upp} can be simplified using $\nabla \cdot \textbf{J} = 0$ and noting that the perturbation can be written as $\bm{\xi}_{1,2} = \xi^{\psi}_{1,2} \nabla \psi + \xi^{\perp}_{1,2}\hat{\textbf{b}} \times \nabla \psi$.
Applying the identity $\textbf{B} \cdot \delta \widetilde{\textbf{B}}_{1,2} = - B^2 \nabla \cdot \bm{\xi}_{1,2} - \bm{\xi}_{1,2} \cdot \nabla B^2 - \mu_0 \bm{\xi}_{1,2} \cdot \nabla p$ to the second term, the following expression can be obtained, 
\begin{multline}
    \int_{V_P} d^3 x \, \left( \bm{\xi}_2 \cdot \textbf{J} \times \delta \widetilde{\textbf{B}}_1 - \bm{\xi}_1 \cdot \textbf{J} \times \delta \widetilde{\textbf{B}}_2\right) = \\ \int_{V_P} d^3 x \, \left(\left(\nabla \cdot \bm{\xi}_2\right) \bm{\xi}_1 \cdot \nabla p - \left( \nabla \cdot \bm{\xi}_1 \right) \bm{\xi}_2 \cdot \nabla p \right).
\end{multline}
Hence we obtain the following expression for $U_{P_1}$,
\begin{multline}
    U_{P_1} = \int_{V_P} d^3 x \, \left(- \bm{\xi}_2 \cdot \textbf{F}_1 + \bm{\xi}_1 \cdot \textbf{F}_2\right) \\
    - \int_{V_P} d^3 x \, \left( \delta \chi_1'(\psi) \bm{\xi}_2 \cdot \nabla \psi  - \delta \chi_2'(\psi) \bm{\xi}_1 \cdot \nabla \psi  \right) \textbf{J} \cdot \nabla \varphi.
\end{multline}
We now consider $U_{P_2}$ defined in \eqref{eq:Up2}.
Applying \eqref{eq:delta_I} for the change in toroidal current, integrating by parts in $\psi$, and combining the expressions for $U_{P_1}$ \eqref{eq:Up1} and $U_{P_2}$ \eqref{eq:Up2}, we obtain,
\begin{multline}
    U_P = \int_{V_P} d^3 x \, \left(-\bm{\xi}_2 \cdot \textbf{F}_1 + \bm{\xi}_1 \cdot \textbf{F}_2\right) + 2\pi \int_{V_P} d \psi \, \left( \delta \chi_1(\psi) \delta I_{T,2}'(\psi) - \delta \chi_2(\psi) \delta I_{T,1}'(\psi) \right) \\
    - \int_{S_P} d^2 x \, \left(\delta \chi_1(\psi) \bm{\xi}_2 - \delta \chi_2(\psi) \bm{\xi}_1 \right) \cdot \hat{\textbf{n}} \textbf{J} \cdot \nabla \varphi.
    \label{eq:Upb}
\end{multline}
Next we combine $U_P$ \eqref{eq:Upb} with $U_B$ \eqref{eq:Ub} and $U_C$ \eqref{eq:Uc} to obtain the free-boundary adjoint relation \eqref{eq:free_boundary}.

To obtain the fixed-boundary adjoint relation, the integral over the plasma volume \eqref{eq:Upa} can be related to a surface integral by applying the divergence theorem to arrive at \cref{eq:Up_bound}. Using \eqref{eq:delta_A1} and applying several vector identities,
\begin{multline}
    U_P = -\frac{1}{\mu_0} \int_{S_P} d^2 x \, \hat{\textbf{n}} \cdot \left( \bm{\xi}_1 \delta \textbf{B}_2  - \bm{\xi}_2 \delta \textbf{B}_1 \right) \cdot \textbf{B} \\
    - \frac{1}{\mu_0} \int_{S_P} d^2 x \, \left( \delta \chi_2(\psi) \delta \textbf{B}_1 - \delta \chi_1(\psi) \delta \textbf{B}_2 \right) \cdot \nabla \varphi \times \hat{\textbf{n}}.
    \label{eq:Uc+Ub}
\end{multline}
Using \eqref{eq:Upb} and expressing the second term in \eqref{eq:Uc+Ub} as a perturbed current using \eqref{eq:delta_I}, the fixed boundary adjoint relation \eqref{eq:fixed_boundary} is obtained.
\renewcommand{\thechapter}{J}

\chapter{Alternate derivation of fixed-boundary adjoint relation}
\label{appendix:self-adjointness}

The MHD force operator,
\begin{gather}
    \textbf{F}[\bm{\xi}_{1,2}] = \textbf{J} \times \left(\nabla \times \left(\bm{\xi}_{1,2} \times \textbf{B} \right)\right) + \frac{\nabla \times \left( \nabla \times \left(\bm{\xi}_{1,2} \times \textbf{B} \right) \right) \times \textbf{B}}{\mu_0} + \nabla \left(\bm{\xi}_{1,2} \cdot \nabla p \right),
    \label{eq:force_operator_appP}
\end{gather}
possesses the following self-adjointness property \citep{Bernstein1958, Goedbloed2004},
\begin{gather}
    \int_{V_P} d^3 x \, \left( \bm{\xi}_2 \cdot \textbf{F}[\bm{\xi}_1] - \bm{\xi}_1 \cdot \textbf{F}[\bm{\xi}_2] \right) = \frac{1}{\mu_0}\int_{S_P} d^2 x \, \hat{\textbf{n}} \cdot \left( \bm{\xi}_1 \textbf{B} \cdot \delta \widetilde{\textbf{B}}_2 - \bm{\xi}_2 \textbf{B} \cdot \delta \widetilde{\textbf{B}}_1 \right) ,
    \label{eq:self_adjointness}
\end{gather}
where $\delta \widetilde{\textbf{B}}_{1,2} = \nabla \times \left(\bm{\xi}_{1,2} \times \textbf{B}\right)$ is the perturbed field corresponding to the MHD perturbations. As we consider linearized equilibrium states that preserve $p(\psi)$, the perturbed pressure satisfies $\delta p(\psi) = - \bm{\xi} \cdot \nabla p$. The force operator we adopt \eqref{eq:force_operator_appP} is the $\gamma \rightarrow 0$ limit of the more general form of the force operator \eqref{eq:force_operator_ch5}, which sometimes includes the term $\nabla \left(\gamma p \nabla \cdot \bm{\xi} \right)$.

For perturbations described by \cref{eq:delta_A1,eq:delta_B,eq:delta_p,eq:delta_I,eq:perturbed_force_balance,eq:general_force_operator}, the force operator satisfies,
\begin{gather}
    \textbf{F}[\bm{\xi}_{1,2}] = \textbf{J} \times \left( \nabla \delta \chi_{1,2}(\psi) \times \nabla \varphi \right) + \frac{\nabla \times \left( \nabla \delta \chi_{1,2}(\psi) \times \nabla \varphi \right) \times \textbf{B}}{\mu_0}  -  \delta \textbf{F}_{1,2}.
    \label{eq:force_operator_appP2}
\end{gather}
Using \cref{eq:force_operator_appP2} and several vector identities, the left hand side of \cref{eq:self_adjointness} can be written as
\begin{multline}
    \int_{V_P}d^3 x \, \left( \bm{\xi}_2 \cdot \textbf{F}[\bm{\xi}_1] - \bm{\xi}_1 \cdot \textbf{F}[\bm{\xi}_2] \right) = \int_{V_P} d^3 x \, \left( \delta \chi_1'(\psi) \bm{\xi}_2 - \delta \chi_{2}'(\psi) \bm{\xi}_1 \right)\cdot \nabla \psi \textbf{J} \cdot \nabla \varphi \\ - \frac{1}{\mu_0} \int_{V_P} d^3 x \, \nabla \psi \times \nabla \varphi \cdot \left( \delta \chi_1'(\psi) \delta \widetilde{\textbf{B}}_2 - \delta \chi_2'(\psi) \delta \widetilde{\textbf{B}}_1 \right)\\
    - \frac{1}{\mu_0} \int_{S_P} d^2 x \, \left(\bm{\xi}_2 \delta \chi_{1}'(\psi) - \bm{\xi}_1 \delta \chi_2'(\psi) \right) \cdot \hat{\textbf{n}} \left( \nabla \psi \times \nabla \varphi \cdot \textbf{B}\right) \\ - \int_{V_P} d^3 x \, \left( \bm{\xi}_2 \cdot \delta \textbf{F}_1 - \bm{\xi}_1 \cdot \delta \textbf{F}_2 \right).
    \label{eq:self_adjoint_1}
\end{multline}
In arriving at \eqref{eq:self_adjoint_1}, we use $\textbf{J} \cdot \nabla \psi = 0$, which follow from MHD force balance \eqref{eq:force_balance}.
Using \cref{eq:delta_I} to re-express the first two terms on the right-hand side,
\begin{multline}
  \int_{V_P}d^3 x \, \left( \bm{\xi}_2 \cdot \textbf{F}[\bm{\xi}_1] - \bm{\xi}_1 \cdot \textbf{F}[\bm{\xi}_2] \right) =  2 \pi \int_{V_P} d \psi \, \left( \delta I_{T,2}(\psi) \delta \chi_{1}'(\psi) - \delta I_{T,1}(\psi) \delta \chi_{2}'(\psi) \right) \\ - \frac{1}{\mu_0} \int_{S_P} d^2 x \, \left( \bm{\xi}_2 \delta \chi_{1}'(\psi)  - \bm{\xi}_1 \delta \chi_{2}'(\psi) \right) \cdot \hat{\textbf{n}} \left( \nabla \psi \times \nabla \varphi \cdot \textbf{B} \right) \\
- \int_{V_P} d^3 x \, \left( \bm{\xi}_2 \cdot \delta \textbf{F}_1 - \bm{\xi}_1 \cdot \delta \textbf{F}_2 \right).
\end{multline}
Using \cref{eq:delta_A1,eq:self_adjointness} we obtain \cref{eq:fixed_boundary}.

\renewcommand{\thechapter}{K}

\chapter{Interpretation of the displacement vector}
\label{app:displacement}
For MHD perturbations such that $\delta \textbf{B} = \nabla \times \left( \bm{\xi} \times \textbf{B} \right)$ the displacement can be interpreted as a vector describing the motion of a field lines. Thus a normal perturbation to the surface of the plasma as in \eqref{eq:shape_gradient_ch5} can be expressed in terms of the displacement vector,
\begin{gather}
    \delta f(S_P;\bm{\xi}) = \int_{S_P} d^2 x \, \mathcal{G} \bm{\xi} \cdot \hat{\textbf{n}}. 
    \label{eq:shape_gradient_surface_xi}
\end{gather}
For perturbations that allow for changes in the rotational transform it remains to be shown that a similar relation can be found. 

As we require that $\psi$ remain a flux surface label in the perturbed equilibrium, the Lagrangian perturbation to $\psi$ at fixed position is 
\begin{gather}
    \delta \psi = - \delta \textbf{x} \cdot \nabla \psi. 
\end{gather}
The perturbed magnetic field, $\textbf{B}' = \textbf{B} + \delta \textbf{B}$ must remain tangent to $\psi' = \psi + \delta \psi$ surfaces; thus to first order in the perturbation,
\begin{gather}
    0 = \textbf{B}' \cdot \nabla \psi' = \textbf{B} \cdot \nabla \delta \psi + \delta \textbf{B} \cdot \nabla \psi. 
\end{gather}
Applying the form for the perturbed field allowing for changes in the rotational transform, $\delta \textbf{B} = \nabla \times \left( \bm{\xi} \times \textbf{B} - \delta \chi (\psi) \nabla \varphi \right)$, and using several vector identities, the following condition is obtained
\begin{gather}
    \textbf{B} \cdot \nabla \left( \delta \textbf{x} \cdot \nabla \psi \right) = \textbf{B} \cdot \nabla \left( \bm{\xi} \cdot \nabla \psi \right).
\end{gather}
This implies that $\delta \textbf{x} \cdot \nabla \psi = \bm{\xi} \cdot \nabla \psi + F(\psi)$, where $F(\psi)$ is some flux function which can be determined by requiring that the perturbation to the toroidal flux as a function of $\psi$ vanishes, $\delta \Psi_T(\psi) = 0$. 

The perturbed toroidal flux through a surface labeled by $\psi$ contains two terms, corresponding to the flux of the unperturbed field through the perturbed surface and the perturbed field through the unperturbed surface, 
\begin{gather}
    \delta \Psi_T(\psi) = \int_{\partial S_T(\psi)} d \vartheta \, \sqrt{g}  \delta \textbf{x} \cdot \nabla \psi \textbf{B} \cdot \nabla \varphi + \int_{S_T(\psi)} d \psi d \vartheta \, \sqrt{g} \delta \textbf{B} \cdot \nabla \varphi. 
\end{gather}
Using the form for $\delta \textbf{B}$,  applying the divergence theorem, and noting that $\textbf{B} \cdot \nabla \varphi = \sqrt{g}^{-1}$, the following condition is obtained,
\begin{gather}
    \delta \Psi_T(\psi) = \int_0^{2\pi} d \vartheta \, \left(\delta \textbf{x} \cdot \nabla \psi - \bm{\xi} \cdot \nabla \psi \right). 
\end{gather}
By requiring that $\delta \Psi_T(\psi) = 0$, we find that $F(\psi) = 0$. Thus we can express shape gradients in the form of \eqref{eq:shape_gradient_surface_xi} even when the rotational transform is allowed to vary. 

\renewcommand{\thechapter}{L}

\chapter{Details of axis ripple calculation}
\label{app:axis_ripple}

In this Appendix, we compute the shape derivative of the finite-pressure magnetic well figure of merit from \eqref{eq:delta_fR} and show that if we impose an adjoint perturbation of the form \eqref{eq:ripple_adjoint}, the shape gradient is given by \eqref{eq:well_shape_gradient}.

We use the expression for the perturbation to the field strength \eqref{eq:delta_mod_B} and $\delta \psi = - \bm{\xi}_1 \cdot \nabla \psi$ with \eqref{eq:delta_fR} to obtain,
\begin{multline}
    \delta f_R(S_P;\bm{\xi}_1) = \int_{S_P} d^2 x \, \bm{\xi}_1 \cdot \hat{\textbf{n}} \widetilde{f_R} - \int_{V_P} d^3 x \, \partder{\widetilde{f_R}}{\psi} \bm{\xi}_1 \cdot \nabla \psi 
    \\ - \int_{V_P} d^3 x \, \partder{\widetilde{f_R}}{B} \frac{1}{B} \left( B^2 \nabla \cdot \bm{\xi}_1 + \bm{\xi}_1 \cdot \nabla \left(B^2 + \mu_0 p \right) + \delta \chi_1'(\psi) \textbf{B} \cdot \left(\nabla \psi \times \nabla \varphi \right)\right). 
\end{multline}
The third term can be integrated by parts to obtain, 
\begin{multline}
        \delta f_R(S_P;\bm{\xi}_1) = \int_{S_P} d^2 x \, \bm{\xi}_1 \cdot \hat{\textbf{n}} \left(\widetilde{f_R}-\partder{\widetilde{f_R}}{B} B \right) + \int_{V_P} d^3 x \,\left( \partder{^2\widetilde{f_R}}{B\partial \psi}B - \partder{\widetilde{f_R}}{\psi}\right) \bm{\xi}_1 \cdot \nabla \psi 
    \\ + \int_{V_P} d^3 x \,\left(- \partder{\widetilde{f_R}}{B} B\bm{\xi}_1 \cdot  \bm{\kappa}  + B \partder{^2 \widetilde{f_R}}{B^2} \bm{\xi}_1 \cdot \nabla B + \delta \chi_1'(\psi) \partder{\widetilde{f_R}}{B} \hat{\textbf{b}} \cdot \left( \nabla \varphi \times \nabla \psi \right) \right),
\end{multline}
where the expression for the curvature in an equilibrium field \eqref{eq:curvature_equilibrium} has been applied. 

We compute one term that appears in the fixed-boundary adjoint relation \eqref{eq:fixed_boundary} using the prescribed adjoint bulk force perturbation \eqref{eq:ripple_F},
\begin{multline}
   \int_{V_P} d^3 x \, \bm{\xi}_1 \cdot \textbf{F}_2 = \int_{V_P} d^3 x \, \left(-\partder{^2 p_{||}}{B \partial \psi} B + \partder{p_{||}}{\psi} \right) \bm{\xi}_1 \cdot \nabla \psi \\
    +\int_{V_P} d^3 x \,\left( \partder{p_{||}}{B} B \bm{\xi}_1 \cdot \bm{\kappa} - B  \partder{^2p_{||}}{B^2} \bm{\xi}_1 \cdot \nabla B \right), 
\end{multline}
where we have applied the parallel force balance condition \eqref{eq:par_force_balance}. 
Therefore, if we impose $p_{||} = \widetilde{f_R}$, we obtain the following expression for the shape derivative of $f_R$,
\begin{multline}
    \delta f_R(S_P;\bm{\xi}_1) = \int_{S_P} d^2 x \, \bm{\xi}_1 \cdot \hat{\textbf{n}} \left( \widetilde{f_R} - \partder{\widetilde{f_R}}{B} B \right) - \int_{V_P} d^3 x \, \bm{\xi}_1 \cdot \textbf{F}_2 \\
    + \int_{V_P} d^3 x  \, \delta \chi_1'(\psi) \partder{\widetilde{f_R}}{B} \hat{\textbf{b}} \cdot \left( \nabla \varphi \times \nabla \psi \right).
\end{multline}
Upon application of the fixed-boundary adjoint relation we obtain \eqref{eq:well_shape_gradient} with \eqref{eq:ripple_adjoint}.

\renewcommand{\thechapter}{M}

\chapter{Details of effective ripple in the $1/\nu$ regime calculation}
\label{app:1_over_nu}

Neoclassical transport in the $1/\nu$ collisionality regime is discussed in many references including \cite{Frieman1970}, \cite{Connor1974}, and \cite{Ho1987}. In this Appendix we sketch the computation of  $\epsilon_{\text{eff}}^{3/2}$ originally introduced in \cite{Nemov1999} and compute linear perturbations of $f_{\epsilon}$ \eqref{eq:f_epsilon}, showing them to take the form of \eqref{eq:df_epsilon}.

In the $1/\nu$ regime, the distribution function is ordered in the parameter $\nu_* = \nu/(v_t/L) \ll 1$, where $\nu$ is the collision frequency, the thermal speed is $v_t = \sqrt{2T/m}$ for mass $m$ and temperature $T$, and $L$ is a macroscopic scale length,
\begin{gather}
    f_{1} = f_{1}^{-1} + f_{1}^0 + \mathcal{O}(\nu_*).
\end{gather}
In velocity space we use a pitch angle coordinate
$\lambda = v_{\perp}^2/(v^2B)$, energy coordinate $\epsilon = v^2/2$, and $\sigma = \text{sign}(v_{||})$, where $v_{\perp} = \sqrt{v^2-v_{||}^2}$ is the perpendicular velocity and $v_{||} = \textbf{v} \cdot \hat{\textbf{b}}$ is the parallel velocity. We use the field line label, $\alpha$, and length along a field line, $l$, to describe location on a constant $\psi$ surface. In the $1/\nu$ regime the $\textbf{E}\times \textbf{B}$ precession frequency is assumed to be small relative to the collision frequency, so the drift kinetic equation \eqref{eq:DKE_ch2} becomes,
\begin{gather}
    v_{||} \partder{f_{1}}{l} = C(f_{1}) - \textbf{v}_{\text{m}} \cdot \nabla \psi \partder{f_{0}}{\psi},
\end{gather}
where the Maxwellian with density $n$ is, 
\begin{gather}
f_{0} = n \pi^{-3/2} v_{t}^{-3} e^{-v^2/v_{t}^2},
\label{eq:Maxwellian}
\end{gather}
and the radial magnetic drift is,
\begin{gather}
    \textbf{v}_{\text{m}} \cdot \nabla \psi = (v^2 + v_{||}^2) \frac{m}{2q B^3} \nabla \psi \times \textbf{B} \cdot \nabla B,
    \label{eq:radial_drift_appI}
\end{gather}
for charge $q$. The drift kinetic equation to $\mathcal{O}(\nu_*^{-1})$ is, 
\begin{gather}
    v_{||} \partder{f_{1}^{-1}}{l} = 0.
\end{gather}
In the trapped portion of phase space, this implies that $f_{1}^{-1}=f_{1}^{-1}(\psi,\alpha,\epsilon,\lambda)$, and in the passing portion of phase space, this implies that $f_{1}^{-1}=f_{1}^{-1}(\psi,\epsilon,\lambda,\sigma)$. The drift kinetic equation to $\mathcal{O}(\nu_*^0)$ is,
\begin{gather}
    v_{||} \partder{f_{1}^0}{l} = C(f_{1}^{-1}) - \textbf{v}_{\text{m}} \cdot \nabla \psi \partder{f_{0}}{\psi}.
    \label{eq:kinetic_equation}
\end{gather}
In the passing region, this implies that $f_{1}^{-1}$ is a Maxwellian, so it can be taken to vanish. We employ a pitch-angle scattering operator,
\begin{gather}
    C = \frac{2\nu(\epsilon) v_{||}}{B\epsilon} \partder{}{\lambda} \left(\lambda v_{||}\partder{}{\lambda} \right).
\end{gather}
The parallel streaming term in \eqref{eq:kinetic_equation} is annihilated by the bounce averaging operation,  
\begin{gather}
    0 = \langle C(f_1^{-1})\rangle_{b} - \langle \textbf{v}_{\text{m}} \cdot \nabla \psi\rangle_{b} \partder{f_0}{\psi},
    \label{eq:bounce_averaged}
\end{gather}
where the bounce average of a quantity $A$ is $\langle A \rangle_b = \tau^{-1} \oint dl \, A/v_{||}$ and the bounce time is $\tau = \oint dl \, v_{||}^{-1}.$ The bounce-averaged equation \eqref{eq:bounce_averaged} can be expressed in terms of the parallel adiabatic invariant $J = \oint dl \, v_{||}$ using the relation,
\begin{gather}
    \langle \textbf{v}_{\text{m}} \cdot \nabla \psi \rangle_{b} = \frac{m}{q\tau} \partder{J}{\alpha}.
\end{gather}
Integrating \eqref{eq:bounce_averaged} with respect to $\lambda$ we obtain,
\begin{gather}
    \partder{f_{1}^{-1}}{\lambda} =\frac{m\epsilon}{2q\lambda\nu(\epsilon)} \partder{f_{0}}{\psi} \left(\oint dl \, \frac{v_{||}}{B} \right)^{-1} \int_{1/B_{\max}}^{\lambda} d \lambda' \, \partder{J}{\alpha}.
\end{gather}
Here $B_{\max}$ is the maximum value of the field strength on the surface labeled by $\psi$.
We have used the boundary condition $\left( \oint dl \, v_{||}/B \right) \partial f_1^{-1}/\partial \lambda \rvert_{\lambda = 1/B_{\max}} =0$, as there is no flux in pitch-angle from the passing region. The integration with respect to $\lambda$ is performed to obtain, 
\begin{gather}
    \partder{f_{1}^{-1}}{\lambda} = -\frac{m}{6q\lambda \nu(\epsilon)} \partder{f_0}{\psi} \left(\oint dl \, \frac{v_{||}}{B} \right)^{-1} \partder{}{\alpha}\left(\oint dl \, \frac{v_{||}^3}{B} \right) .
    \label{eq:dfdlambda}
\end{gather}
The particle flux from $f_1^{-1}$ is obtained by multiplying \eqref{eq:kinetic_equation} by $f_{1}^{-1} (\partial f_{0}/\partial \psi)^{-1}$, integrating over velocity space, and flux surface averaging,
\begin{align}
    \left \langle \bm{\Gamma} \cdot \nabla \psi \right \rangle_{\psi} &\equiv \left \langle \int d^3 v \, f_{1}^{-1} \textbf{v}_{\text{m}} \cdot \nabla \psi \right \rangle_{\psi} 
    = \left \langle \int d^3 v \, f_{1}^{-1} C(f_{1}^{-1}) \left(\partder{f_{0}}{\psi} \right)^{-1} \right \rangle_{\psi}.
\end{align}
The velocity space integration is performed using the velocity-space Jacobian $d^3 v = 2\pi \sum_{\sigma} B \epsilon/|v_{||}| d \lambda d \epsilon$. Upon integration by parts in $\lambda$ and applying \eqref{eq:dfdlambda}, the following expression is obtained,
\begin{multline}
    \left \langle \bm{\Gamma} \cdot \nabla \psi \right \rangle_{\psi} = \\ - \frac{4\sqrt{2}\pi}{V'(\psi)} \left(\frac{m}{3q} \right)^2 \int_0^{\infty} d \epsilon \, \left(\partder{f_{0}}{\psi} \right) \frac{\epsilon^{5/2}}{\nu(\epsilon)} \int_{1/B_{\max}}^{1/B_{\min}} \frac{d \lambda}{\lambda} \,  \int_0^{2\pi} d \alpha \, \sum_i \frac{(\partder{}{\alpha}\hat{K}_i(\alpha,\lambda))^2}{\hat{I}_i(\alpha,\lambda)},
    \label{eq:particle_flux_app_I}
\end{multline}
where the bounce integrals are defined by \eqref{eq:bounce_integrals}. The sum in \eqref{eq:particle_flux_app_I} is taken over trapping regions for particles with pitch angle $\lambda$ on a field line labeled by $\alpha$ for left bounce points $\varphi_{-,i} \in[0,2\pi)$.

The parameter $\epsilon_{\text{eff}}^{3/2}$ quantifies the geometric dependence of the $1/\nu$ particle flux.
It is defined in terms of the radial particle flux in the following way \citep{Nemov1999},
\begin{align}
 \langle \bm{\Gamma} \cdot \nabla \psi \rangle_{\psi} = -32\langle |\nabla \psi | \rangle_{\psi}^2 \left(\frac{m}{3q}\right)^2 \frac{1}{B_0^2R^2}\epsilon_{\text{eff}}^{3/2} \int_0^{\infty} d \epsilon \, \left(\partder{f_{0}}{\psi} \right) \frac{\epsilon^{5/2}}{\nu(\epsilon)} . 
 \label{eq:classical_flux}
\end{align}
We take our normalizing length and field values to be such that $B_0 R = \epsilon_{\text{ref}}^{-1} \langle |\nabla \psi | \rangle_{\psi}$, where $\epsilon_{\text{ref}}$ is a reference aspect ratio. Comparing \eqref{eq:particle_flux_app_I} with \eqref{eq:classical_flux} we obtain the expression for $\epsilon_{\text{eff}}^{3/2}$ \eqref{eq:eps_eff}. The corresponding expression (29) in \cite{Nemov1999} is obtained by noting that $\hat{H}^{\text{Nemov}} = -(\partial \hat{K}/\partial \alpha) \lambda^{1/2} B_0^{3/2}$ and $\hat{I} = 2\hat{I}^{\text{Nemov}}$, where $\hat{H}^{\text{Nemov}}$ and $\hat{I}^{\text{Nemov}}$ are given in (30)-(31) of \cite{Nemov1999}.

The shape derivative of $f_{\epsilon}$ \eqref{eq:f_epsilon} is computed to be,
\begin{gather}
    \delta f_{\epsilon} (S_P;\bm{\xi}_1) = \int_{V_P} d \psi \,  w(\psi) \delta (V'(\psi)\epsilon_{\text{eff}}^{3/2}(\psi)).
\label{eq:df_epsilon2}
\end{gather}
The perturbation to the bounce integrals is computed using the following identity for the perturbation of a line integral $Q_L = \int_{l_0}^{l_L} dl \, Q$ due to displacement of the integration curve by vector field $\delta \textbf{x}$ \citep{Antonsen1982,Landreman2018},
\begin{gather}
    \delta Q_L = \int_{l_0}^{l_L} dl \,  \left( \delta \textbf{x} \cdot \left(-\bm{\kappa}Q + \left(\textbf{I}-\hat{\textbf{t}}\hat{\textbf{t}} \right)\cdot \nabla Q\right)  + \delta Q \right) + Q(l_L) \delta l_L - Q(l_0) \delta l_0,
    \label{eq:perturbation_line_integral}
\end{gather}
where $\delta Q$ is the perturbation to the integrand at fixed position, $\hat{\textbf{t}} = \textbf{x}'(l)$ is the unit tangent vector, $\bm{\kappa} = \textbf{x}''(l)$ is the curvature, and $\delta l_L$ and $\delta l_0$ are perturbations to the bounds of the integral.

We compute the perturbation to the bounce integrals to be,
\begin{subequations}
\begin{align}
    \delta \hat{I}_i &=\oint dl \, \left(-\frac{v_{||}}{vB} \bm{\kappa} \cdot \delta \textbf{x} - \left(\frac{\lambda v}{2B v_{||}}+\frac{v_{||}}{B^2v} \right) \left(\delta \textbf{x} \cdot \nabla B + \delta B \right) \right) \\
    \delta \hat{K}_i &= \oint dl \, \left(-\frac{v_{||}^3}{v^3B} \bm{\kappa} \cdot \delta \textbf{x} - \left(\frac{3\lambda v_{||}}{2Bv} + \frac{v_{||}^3}{B^2v^3} \right)\left(\delta \textbf{x} \cdot \nabla B + \delta B \right) \right), 
\end{align}
\end{subequations}
where $\delta B$ is the perturbation to the field strength \eqref{eq:delta_mod_B} and $\delta \textbf{x}$ is given by \eqref{eq:delta_r}. We note that $\delta \textbf{x} \cdot \hat{\textbf{b}} = 0$ such that the perpendicular projection, $(\textbf{I}-\hat{\textbf{t}}\hat{\textbf{t}})$, is not needed. There is no contribution due to the perturbation of the bounce points, as the integrand vanishes at these points. The expressions \eqref{eq:df_epsilon}-\eqref{eq:p_ripple} can now be obtained by writing \eqref{eq:df_epsilon2} in terms of the perturbations of the bounce integrals, using $\bm{\xi}_1 \cdot \nabla B + \delta B = - B\left(\textbf{I} - \hat{\textbf{b}} \hat{\textbf{b}} \right):\nabla \bm{\xi}_1 - \delta \chi_1'(\psi) \hat{\textbf{b}} \cdot (\nabla \psi \times \nabla \varphi)$ and 
$\bm{\kappa} \cdot \bm{\xi}_1 = - \hat{\textbf{b}} \hat{\textbf{b}}: \nabla \bm{\xi}_1$.
\renewcommand{\thechapter}{N}

\chapter{Details of departure from quasi-symmetry calculation}
\label{app:qs}

In this Appendix we compute the shape derivative of $f_{QS}$ \eqref{eq:f_QS} to obtain \eqref{eq:df_QS}-\eqref{eq:I_QS} by expressing each term in \eqref{eq:df_QS1} in the desired form. The second term in \eqref{eq:df_QS1} is expressed using $\delta \psi = - \bm{\xi}_1 \cdot \nabla \psi$,
\begin{gather}
    \frac{1}{2}\int_{V_P} d^3 x \, w'(\psi) \delta \psi \mathcal{M}^2 = -\frac{1}{2} \int_{V_P} d^3 x \,  \mathcal{M}^2 \bm{\xi}_1 \cdot \nabla w(\psi) .
\end{gather}
The third term in \eqref{eq:df_QS1} is computed upon application of \eqref{eq:delta_B}, the divergence theorem, and noting that $\mathcal{M} = \textbf{B} \cdot \bm{\mathcal{A}}$, 
\begin{multline}
    \int_{V_P} d^3 x \, w(\psi) \mathcal{M} \delta \textbf{B} \cdot \bm{\mathcal{A}} =  - \int_{S_P} d^2 x \,  \bm{\xi}_1 \cdot \textbf{\textbf{n}}
    w(\psi) \mathcal{M}^2 - \int_{V_P} d^3 x \, w(\psi) \delta \chi'_1(\psi) \mathcal{M} \nabla \psi \times \nabla \varphi \cdot \bm{\mathcal{A}} \\
    +\int_{V_P} d^3 x \, \bm{\xi}_1 \cdot \left(w(\psi) \mathcal{M} \left( \textbf{B} \times \left( \nabla \times \bm{\mathcal{A}} \right) \right) - \bm{\mathcal{A}} w(\psi) \textbf{B} \cdot \nabla \mathcal{M}  + \mathcal{M}\nabla \left( w(\psi)\mathcal{M} \right) \right).
    \label{eq:qs_2}
\end{multline}
The quantity $\bm{\mathcal{A}}$ can be projected into the perpendicular direction as $\bm{\xi}_1 \cdot \hat{\textbf{b}} = 0$, noting that,
\begin{gather}
    \hat{\textbf{b}} \times \left( \bm{\mathcal{A}} \times \hat{\textbf{b}} \right) = -(\hat{\textbf{b}} \times \nabla \psi) \nabla_{||} B - F(\psi) \nabla_{\perp} B.
\end{gather}
Similarly, any terms in \eqref{eq:qs_2} involving $\bm{\xi}_1 \cdot \nabla$ can be expressed as $\bm{\xi}_1 \cdot \nabla_{\perp}$. The corresponding terms in \eqref{eq:F_QS} are obtained using the expression for the curvature in an equilibrium field. The fourth term in \eqref{eq:df_QS1} is expressed in the following way upon application of \eqref{eq:delta_mod_B}, the divergence theorem, and noting that $\bm{\mathcal{S}} \cdot \nabla \psi = \nabla \cdot \bm{\mathcal{S}} = 0 $,
\begin{multline}
    \int_{V_P} d^3 x \, w(\psi) \mathcal{M} \bm{\mathcal{S}} \cdot \nabla \delta B = \int_{S_P} d^2 x \, \bm{\xi}_1 \cdot \mathbf{n}  B  w(\psi) \bm{\mathcal{S}} \cdot \nabla \mathcal{M}  - \int_{V_P} d^3 x \, \bm{\xi}_1 \cdot \left[B \nabla \left( w(\psi) \bm{\mathcal{S}} \cdot \nabla \mathcal{M} \right) \right]  \\
    + \int_{V_P} d^3 x \, w(\psi) (\bm{\mathcal{S}} \cdot \nabla \mathcal{M}) \left(\delta \chi'_1(\psi) \hat{\textbf{b}} \cdot(\nabla \psi \times \nabla \varphi)
    + B\bm{\xi}_1 \cdot \bm{\kappa} \right).
\end{multline}
We express terms involving $\bm{\xi}_1 \cdot \nabla$ as $\bm{\xi}_1 \cdot \nabla_{\perp}$ to obtain the corresponding terms in \eqref{eq:F_QS}. The fifth term in \eqref{eq:df_QS1} is expressed in the following way upon application of $\delta \psi = - \bm{\xi}_1 \cdot \nabla \psi$, the divergence theorem, and several vector identities,
\begin{multline}
    \int_{V_P} d^3 x \, w(\psi) \mathcal{M} \textbf{B} \times \nabla \delta \psi \cdot \nabla B = - \int_{S_P} d^2 x \, \bm{\xi}_1 \cdot \hat{\textbf{n}}  w(\psi) \mathcal{M} \nabla B \times \textbf{B} \cdot \nabla \psi  \\
     - \int_{V_P} d^3 x \, \bm{\xi}_1 \cdot \nabla \psi \nabla B \cdot \nabla \times \left( w(\psi) \mathcal{M} \textbf{B} \right).
\end{multline}
The sixth term in \eqref{eq:df_QS1} upon application of \eqref{eq:delta_G} is,
\begin{multline}
-\int_{V_P} d^3 x \, \frac{\delta G(\psi)w(\psi) \mathcal{M} \textbf{B} \cdot \nabla B }{\iota(\psi)-(N/M)} = \\
     \frac{1}{4\pi^2} \int_{S_P} d^2 x \, \frac{w(\psi)  V'(\psi) \langle \mathcal{M} \textbf{B} \cdot \nabla B \rangle_{\psi}}{(\iota(\psi)-(N/M))} \left( \textbf{B} \cdot \nabla \psi \times \nabla \vartheta \right) \bm{\xi}_1 \cdot \hat{\textbf{n}} \\
    - \frac{1}{4\pi^2} \int_{V_P} d^3 x \,  \bm{\xi}_1 \cdot  \nabla \left(\frac{w(\psi) V'(\psi) \langle \mathcal{M}\textbf{B} \cdot \nabla B \rangle_{\psi}}{(\iota(\psi)-(N/M))} \right) \textbf{B} \cdot \nabla \psi \times \nabla \vartheta \\
    + \frac{1}{4\pi^2} \int_{V_P} d^3 x \, \frac{w(\psi)V'(\psi) \langle \mathcal{M} \textbf{B} \cdot \nabla B \rangle_{\psi}}{\iota(\psi)-(N/M)} \left( \bm{\xi}_1 \cdot \left(\nabla \psi \nabla \cdot \left( \textbf{B} \times \nabla \vartheta \right) -\textbf{B} \times \nabla \times \left(\nabla \psi \times \nabla \vartheta \right)\right) \right)\\
    - \frac{1}{4\pi^2}\int_{V_P} d^3 x \, \delta \chi'_1(\psi) \frac{w(\psi)V'(\psi) \langle \mathcal{M} \textbf{B} \cdot \nabla B \rangle_{\psi}}{\sqrt{g}^2(\iota(\psi)-(N/M))}  \partder{\textbf{x}}{\varphi} \cdot \partder{\textbf{x}}{\vartheta}. 
\end{multline}
In obtaining the corresponding terms in \eqref{eq:F_QS}, terms involving $\bm{\xi}_1 \cdot \nabla$ are expressed as $\bm{\xi}_1 \cdot \nabla_{\perp}$.
The seventh term in \eqref{eq:df_QS1} is expressed using $\delta \psi =- \bm{\xi}_1 \cdot \nabla \psi$. Combining all terms, we obtain \eqref{eq:df_QS}-\eqref{eq:I_QS}.
\renewcommand{\thechapter}{O}

\chapter{Details of neoclassical figures of merit calculation}
\label{app:nc}

In this Section we compute the shape derivative of $f_{NC}$ \eqref{eq:f_NC} to obtain \eqref{eq:df_NC}-\eqref{eq:I_NC} by expressing each term in \eqref{eq:deltaf_NC} in the desired form. Throughout Boozer coordinates will be assumed.  

The second term in \eqref{eq:deltaf_NC} is expressed using $\delta \psi = - \bm{\xi}_1 \cdot \nabla \psi$. The third term in \eqref{eq:deltaf_NC} can be computed using \eqref{eq:delta_G}, noting that $V'(\psi)/(4\pi^2\sqrt{g}) = B^2/\langle B^2 \rangle_{\psi}$ in Boozer coordinates and applying the divergence theorem,
\begin{multline}
    \int_{V_P} d^3 x \, w(\psi) \partder{\mathcal{R}(\psi)}{G(\psi)} \delta G(\psi) = -\int_{V_P} d^3 x \, w(\psi) \frac{B^2 \sqrt{g}}{\langle B^2 \rangle_{\psi}} \partder{\mathcal{R}(\psi)}{G(\psi)} \bm{\xi}_1 \cdot \nabla \psi (\nabla \times \textbf{B}) \cdot \nabla \vartheta \\
    + \int_{V_P} d^3 x \,  \left( \bm{\xi}_1 \cdot \nabla \left( \partder{\mathcal{R}(\psi)}{G(\psi)} \frac{w(\psi)}{\langle B^2 \rangle_{\psi}} \right)B^2 G(\psi) + \frac{w(\psi)}{\langle B^2 \rangle_{\psi}}  \partder{\mathcal{R}(\psi)}{G(\psi)} \bm{\xi}_1 \cdot \textbf{B} \times \nabla \times \left( \partder{\textbf{x}}{\varphi}B^2\right) \right)  \\
    + \int_{V_P} d^3 x \,  \frac{w(\psi)\delta \chi_1'(\psi)B^2}{\sqrt{g}\langle B^2 \rangle_{\psi}} \partder{\mathcal{R}(\psi)}{G(\psi)} \partder{\textbf{x}}{\varphi} \cdot \partder{\textbf{x}}{\vartheta} 
    - \int_{S_P} d^2 x \, w(\psi) \frac{B^2}{\langle B^2 \rangle_{\psi}} \partder{\mathcal{R}(\psi)}{G(\psi)} G(\psi)\bm{\xi}_1 \cdot \hat{\textbf{n}}  .
\end{multline}
The fifth term in \eqref{eq:deltaf_NC} can be computed using \eqref{eq:delta_mod_B}, the divergence theorem, and the expression for the curvature in an equilibrium field \eqref{eq:curvature_equilibrium},
\begin{multline}
    \int_{V_P} d^3 x \, w(\psi) \langle S_{\mathcal{R}} \delta B \rangle_{\psi} = \int_{V_P} d^3 x \, \left(\bm{\xi}_1 \cdot \nabla \left( w(\psi) S_{\mathcal{R}}\right) B - B S_{\mathcal{R}} w(\psi) \bm{\xi}_1 \cdot \bm{\kappa} \right) \\
  -\int_{V_P} d^3 x \, \delta \chi_1'(\psi) S_{\mathcal{R}}w(\psi) \hat{\textbf{b}} \cdot \nabla\psi \times \nabla \varphi - \int_{S_P} d^2 x \, w(\psi) S_{\mathcal{R}} B \bm{\xi}_1 \cdot \hat{\textbf{n}}. 
\end{multline}
 The resulting terms can be combined to write the shape derivative in the form of \eqref{eq:df_NC}, noting that any terms involving $\bm{\xi}_1 \cdot \nabla$ can be expressed as $\bm{\xi}_1 \cdot \nabla_{\perp}$.

\renewcommand{\thechapter}{P}

\chapter{Linearized equilibrium energy functional and coefficient matrices}
\label{app:coefficient_matrices}

\section{Further simplification of energy functional}

We will now further simplify the energy functional \eqref{eq:energy_functional_ch6} using a magnetic coordinate system. Each of the contravariant components of the perturbed magnetic field are evaluated to be,
\begin{subequations}
\begin{align}
    Q^{\psi} &\equiv \delta \textbf{B}[\bm{\xi}] \cdot \nabla \psi = \frac{1}{\sqrt{g}}\left(\partder{\xi^{\psi}}{\varphi} + \iota \partder{\xi^{\psi}}{\vartheta}\right) \\
    Q^{\vartheta} &\equiv \delta \textbf{B}[\bm{\xi}] \cdot \nabla \vartheta = \frac{1}{\sqrt{g}} \left(\partder{\xi^{\alpha}}{\varphi} - \partder{\xi^{\psi} \iota}{\psi} \right) \\
    Q^{\varphi} &\equiv \delta \textbf{B}[\bm{\xi}] \cdot \nabla \varphi = -\frac{1}{\sqrt{g}} \left(\partder{\xi^{\alpha}}{\vartheta}+ \partder{\xi^{\psi}}{\psi} \right).
\end{align}
\end{subequations}
We also express the current density in the contravariant basis as,
\begin{align}
    \textbf{J} = J^{\vartheta} \partder{\textbf{x}}{\vartheta} + J^{\varphi} \partder{\textbf{x}}{\varphi}. 
\end{align}
The first term in the energy functional is expressed as,
\begin{align}
    W_1 &\equiv -\frac{1}{\mu_0} \int_{V_P} d^3 x \, \delta \textbf{B}[\bm{\xi}] \cdot \delta \textbf{B}[\bm{\xi}] \\
    &= -\frac{1}{\mu_0} \int_{V_P} d^3 x \,  \Bigg[\left(Q^{\psi}\right)^2 g_{\psi \psi} + \left(Q^{\vartheta} \right)^2 g_{\vartheta \vartheta} + \left(Q^{\varphi} \right)^2 g_{\varphi \varphi}+ 2 Q^{\psi} Q^{\vartheta} g_{\psi \vartheta}
    \Bigg] , \nonumber
\end{align}
where $g_{x_i x_j} = \partial \textbf{x}/\partial x_i \cdot \partial \textbf{x}/\partial x_j$ are the metric coefficients. Here we have assumed that $\varphi = \phi$, the geometric toroidal angle, such that $g_{\vartheta \varphi} = g_{\psi \varphi} = 0$.

The second term in the energy functional is expressed as,
\begin{align}
    W_2 &\equiv \int_{V_P} d^3 x \, \bm{\xi} \cdot \textbf{J} \times \delta \textbf{B}[\bm{\xi}] \\
    &= \int_{V_P} d^3 x \, \sqrt{g} \left(\xi^{\psi} \left(J^{\vartheta} Q^{\varphi} - J^{\varphi} Q^{\vartheta} \right) +Q^{\psi} \left( \xi^{\vartheta} J^{\varphi}  - \xi^{\varphi} J^{\vartheta}\right) \right). \nonumber
\end{align}
Here we can note that the radial component of MHD force balance yields $p'(\psi) = J^{\vartheta} - \iota(\psi) J^{\varphi}$ to write,
\begin{align}
    W_2 &= \int_{V_P} d^3 x \, \sqrt{g} \left(\xi^{\psi} \left(J^{\vartheta} Q^{\varphi} - J^{\varphi} Q^{\vartheta} \right) +Q^{\psi} \left(\xi^{\alpha} J^{\varphi} - p'(\psi) \xi^{\varphi} \right) \right).
\end{align}
The third term in the energy functional can be expressed as,
\begin{align}
    W_3 &\equiv \int_{V_P} d^3 x \, \bm{\xi} \cdot \nabla \left(\bm{\xi} \cdot \nabla p \right) \\
    &= \int_{V_P} d^3 x \, \left[ \xi^{\psi} \partder{(\xi^{\psi} p'(\psi))}{\psi} + p'(\psi) \left(\xi^{\alpha} \partder{\xi^{\psi}}{\vartheta} + \sqrt{g}Q^{\psi}\xi^{\varphi}   \right) \right]. \nonumber
\end{align}
Combining $W_2$ and $W_3$, we see that the energy functional indeed only depends on $\xi^{\alpha}$ and $\xi^{\psi}$,
\begin{multline}
    W_2 + W_3 = \int_{V_P} d^3 x \, \Bigg(\sqrt{g}\xi^{\psi} \left(J^{\vartheta} Q^{\varphi} - J^{\varphi} Q^{\vartheta} \right) + \xi^{\alpha} \textbf{J} \cdot \nabla \xi^{\psi}
    + \xi^{\psi} \partder{(\xi^{\psi} p'(\psi))}{\psi} \Bigg).
\end{multline}
We now can apply the divegernce theorem, noting that $\nabla \cdot \textbf{J} = \textbf{J} \cdot \nabla \psi = 0$, to obtain,
\begin{align}
        W_2 + W_3 = \int_{V_P} d^3 x \, \Bigg(\xi^{\psi} \left( J^{\varphi} \iota'(\psi) \xi^{\psi} - 2 \textbf{J} \cdot \nabla \xi^{\alpha}
   + \xi^{\psi} p''(\psi) \right)\Bigg).
\end{align}
We now see that the first three terms of the energy functional only depend on $\xi^{\alpha}$ through its $\vartheta$ and $\varphi$ derivatives. Furthermore, given the restriction of $\delta F_{\alpha}$ discussed in Appendix \ref{app:force_perturbation constraint}, the $m = 0$, $n = 0$ mode of $\xi^{\alpha}$ will not enter the variational principle. 

\section{Explicit forms of coefficient matrices}

We can now express the linear operators that couple the Fourier components of $\xi^{\alpha}$, $\xi^{\psi}$, and $\partial \xi^{\psi}/\partial \psi$ given the simplifications of the energy functional in the previous Section:
\begin{subequations}
\begin{align}
    \textbf{A}_{\psi'\psi'} &= -\frac{V'(\psi)}{\mu_0} \left \langle  \frac{1}{\left(\sqrt{g}\right)^2}\left(g_{\varphi \varphi} + \iota(\psi)^2 g_{\vartheta \vartheta} \right) \bm{\mathcal{F}}^{\psi} \bm{\mathcal{F}}^{\psi} \right \rangle_{\psi}  \\
    \textbf{A}_{\psi \psi} &= \frac{V'(\psi)}{\mu_0} \Bigg \langle \frac{1}{\left(\sqrt{g}\right)^2} \Bigg[- g_{\psi\psi} \Bigg(\partder{\bm{\mathcal{F}}^{\psi}}{\varphi} \partder{\bm{\mathcal{F}}^{\psi}}{\varphi} + \iota(\psi)^2 \partder{\bm{\mathcal{F}}^{\psi}}{\vartheta} \partder{\bm{\mathcal{F}}^{\psi}}{\vartheta} \Bigg)  \\
    &\hspace{0.5cm}-g_{\psi\psi} \iota(\psi) \left( \partder{\bm{\mathcal{F}}^{\psi}}{\vartheta} \partder{\bm{\mathcal{F}}^{\psi}}{\varphi} + \partder{\bm{\mathcal{F}}^{\psi}}{\varphi}\partder{\bm{\mathcal{F}}^{\psi}}{\vartheta}\right) \nonumber \\
    &\hspace{0.5cm} +  \left(\mu_0\left(\sqrt{g}\right)^2 \left(J^{\varphi}\iota'(\psi) + p''(\psi) \right) - g_{\vartheta \vartheta} \left(\iota'(\psi)\right)^2 \right) \bm{\mathcal{F}}^{\psi} \bm{\mathcal{F}}^{\psi} \nonumber \\
    &\hspace{0.5cm}+ g_{\psi \vartheta} \iota'(\psi) \left(\left(\partder{\bm{\mathcal{F}}^{\psi}}{\varphi}+ \iota(\psi) \partder{\bm{\mathcal{F}}^{\psi}}{\vartheta} \right)\bm{\mathcal{F}}^{\psi} + \bm{\mathcal{F}}^{\psi} \left(\partder{\bm{\mathcal{F}}^{\psi}}{\varphi} + \iota(\psi) \partder{\bm{\mathcal{F}}^{\psi}}{\vartheta} \right)\right)\Bigg] \Bigg \rangle_{\psi} \nonumber \\
    \textbf{A}_{\psi \psi'} &= \frac{V'(\psi)}{\mu_0} \left \langle  \frac{2\iota(\psi)}{\left(\sqrt{g}\right)^2} \Bigg[-\bm{\mathcal{F}}^{\psi} g_{\vartheta \vartheta} \iota'(\psi) + g_{\psi \vartheta} \left(\partder{\bm{\mathcal{F}}^{\psi}}{\varphi} + \iota(\psi) \partder{\bm{\mathcal{F}}^{\psi}}{\vartheta}\right)  \Bigg] \bm{\mathcal{F}}^{\psi} \right \rangle_{\psi} \\
    \textbf{A}_{\alpha \alpha} &= - \frac{V'(\psi)}{\mu_0} \left \langle  \frac{1}{\left(\sqrt{g}\right)^2} \Bigg[g_{\vartheta \vartheta} \partder{\bm{\mathcal{F}}^{\alpha}}{\varphi}\partder{\bm{\mathcal{F}}^{\alpha}}{\varphi} + g_{\varphi \varphi} \partder{\bm{\mathcal{F}}^{\alpha}}{\vartheta}\partder{\bm{\mathcal{F}}^{\alpha}}{\vartheta}  \Bigg] \right \rangle_{\psi} \\
    \textbf{A}_{\alpha \psi'} &= \frac{2 V'(\psi)}{\mu_0} \left \langle \, \frac{1}{\left(\sqrt{g}\right)^2} \Bigg[ g_{\vartheta \vartheta} \iota \partder{\bm{\mathcal{F}}^{\alpha}}{\varphi}  - g_{\varphi \varphi} \partder{\bm{\mathcal{F}}^{\alpha}}{\vartheta} \Bigg]\bm{\mathcal{F}}^{\psi} \right \rangle_{\psi}  \\
    \textbf{A}_{\alpha \psi} &= - \frac{2 V'(\psi)}{\mu_0} \Bigg \langle  \frac{1}{\left(\sqrt{g}\right)^2} \Bigg[ \left(- g_{\vartheta \vartheta} \iota'(\psi) \partder{\bm{\mathcal{F}}^{\alpha}}{\varphi} + \mu_0 \left(\sqrt{g}\right)^2 \textbf{J} \cdot \nabla \bm{\mathcal{F}}^{\alpha} \right) \bm{\mathcal{F}}^{\psi} \\
    &\hspace{0.5cm} + g_{\psi \vartheta} \partder{\bm{\mathcal{F}}^{\alpha}}{\varphi} \left(\partder{\bm{\mathcal{F}}^{\psi}}{\varphi} + \iota(\psi) \partder{\bm{\mathcal{F}}^{\psi}}{\vartheta} \right) \Bigg] \Bigg \rangle_{\psi}\nonumber \\
    \textbf{I}_{\psi} &= 2V'(\psi) \left \langle \bm{\mathcal{F}}^{\psi} \delta F_{\psi} \right \rangle_{\psi} \\
    \textbf{I}_{\alpha} &= 2V'(\psi) \left \langle \bm{\mathcal{F}}^{\alpha} \delta F_{\alpha} \right \rangle_{\psi},
\end{align}
\end{subequations}
where $\left \langle ... \right \rangle_{\psi}$ is the flux-surface average \eqref{eq:flux_surface_average_appA}.

\section{Invertibility of $\textbf{A}_{\alpha \alpha}$}

Obtaining the Euler-Lagrange solution for $\xi^{\alpha}$ requires inverting $\textbf{A}_{\alpha \alpha}$. We now show that this matrix is, in fact, negative definite and thus invertible. For any non-zero vector $\bm{\Xi}_{\alpha}$, we can write the inner product with $\textbf{A}_{\alpha \alpha}$ as,
\begin{align}
    \bm{\Xi}_{\alpha} \cdot \left( \textbf{A}_{\alpha \alpha} \bm{\Xi}_{\alpha} \right) = - \frac{1}{\mu_0}\int_0^{2\pi} d \vartheta \int_0^{2\pi} d \varphi \, \left[ \frac{g_{\vartheta \vartheta}}{\sqrt{g}} \left(\partder{\xi^{\alpha}}{\varphi} \right)^2 + \frac{g_{\varphi \varphi}}{\sqrt{g}} \left(\partder{\xi^{\alpha}}{\vartheta} \right)^2 \right].
\end{align}
We note that for a well-defined coordinate system, $g_{\vartheta \vartheta} > 0$, $g_{\varphi \varphi}>0$, and $\sqrt{g} > 0$. While either $\partial \xi^{\alpha}/\partial \varphi$ or $\partial \xi^{\alpha}/\partial \vartheta$ may vanish, they will not vanish simultaneously throughout the integrand as we have excluded the $n=0$, $m = 0$ mode. Therefore, the integrand will only vanish at isolated points. Thus the above integral is negative definite, and $\textbf{A}_{\alpha \alpha}$ is invertible throughout the volume.

\renewcommand{\thechapter}{Q}

\chapter{Constraint on bulk force perturbation}
\label{app:force_perturbation constraint}

As shown in Appendix \ref{app:coefficient_matrices}, the first three terms in the energy functional \eqref{eq:energy_functional_ch6} only depend on $\xi^{\alpha}$ through its derivatives with respect to $\vartheta$ and $\varphi$. In this Appendix, we show that it is always possible to choose the in-surface component of the bulk force perturbation, $\delta F_{\alpha}$, such that the final term in the energy functional,
\begin{align}
    W_4 \equiv \int_{V_P} d^3 x \, \xi^{\alpha} \delta F_{\alpha},
\end{align}
does not depend on $\xi^{\alpha c}_{0,0} = \frac{1}{(2\pi)^2} \int_0^{2\pi} d \vartheta \int_0^{2\pi} d \varphi \, \xi^{\alpha}$. As $\xi^{\alpha c}_{0,0}$ does not enter our variational principle, we can take it to vanish. 
The condition that $\xi^{\alpha c}_{0,0}$ does not enter $W_4$ is equivalent to requiring that, 
\begin{align}
    \langle \delta F_{\alpha} \rangle_{\psi}= 0,
    \label{eq:condition_in_surface}
\end{align}
on every surface, where $\langle \dots \rangle_{\psi}$ is the flux-surface average \eqref{eq:flux_surface_average_appA}. This follows from the surface-averaged in-surface component of the linearized force-balance equation \eqref{eq:perturbed_equilibrium},
\begin{align}
    \left \langle  \partder{\textbf{x}}{\vartheta} \cdot \textbf{F}[\bm{\xi}]\right \rangle_{\psi} = 0.
    \label{eq:fsa_force_balance}
\end{align}
This property of the MHD force operator holds for any equilibrium field that satisfies MHD force balance \eqref{eq:mhd_force_balance_ch6}. To see this we note that the flux-surface average can be defined in terms of an average over the infinitesimal volume between flux surfaces $\Delta V$  \eqref{eq:flux_surface_average_vlume}. We can now apply the self-adjointness relation \eqref{eq:self_adjointness_ch6} to simplify \eqref{eq:fsa_force_balance},
\begin{multline}
   \left \langle  \partder{\textbf{x}}{\vartheta} \cdot \textbf{F}[\bm{\xi}] \right \rangle_{\psi} =  \left \langle  \bm{\xi} \cdot \textbf{F}\left[\partder{\textbf{x}}{\vartheta}\right] \right \rangle_{\psi} \\
   + \lim_{\Delta V \rightarrow 0}\frac{1}{\mu_0 \Delta V} \left(\int_{\partial(V_P + \Delta V)} d^2 x \, \hat{\textbf{n}} \cdot \bm{\xi} \textbf{B} \cdot \delta \textbf{B}\left[\partder{\textbf{x}}{\vartheta}\right] -\int_{\partial(V_P)} d^2 x \, \hat{\textbf{n}} \cdot \bm{\xi} \textbf{B} \cdot \delta \textbf{B}\left[\partder{\textbf{x}}{\vartheta}\right] \right),
\end{multline}
where we have noted that $\hat{\textbf{n}} \cdot \partder{\textbf{x}}{\vartheta} = 0$, as $\hat{\textbf{n}} \propto \nabla \psi$. The quantity $\delta \textbf{B}\left[\partial \textbf{x}/ \partial \vartheta\right] = \nabla \times \left(\partial \textbf{x}/ \partial \vartheta \times \textbf{B}\right)$ is shown to vanish by expressing $\textbf{B}$ in contravariant form and using the dual relations \eqref{eq:dual_relation} between the contravariant and covariant basis vectors. The remaining flux-surface averaged term can also be shown to vanish,
\begin{multline}
  \left \langle \bm{\xi} \cdot \textbf{F}\left[\partder{\textbf{x}}{\vartheta}\right] \right \rangle_{\psi} = \\ \left \langle  \bm{\xi} \cdot \left( \textbf{J} \times \delta \textbf{B}\left[\partder{\textbf{x}}{\vartheta}\right] + \frac{\left(\nabla \times \delta \textbf{B}\left[\partder{\textbf{x}}{\vartheta}\right]\right) \times \textbf{B}}{\mu_0} +\nabla \left(\partder{\textbf{x}}{\vartheta} \cdot \nabla p \right) \right) \right \rangle_{\psi},
 \end{multline}
 as $\partder{\textbf{x}}{\varphi} \cdot \nabla \psi = 0$ and $\delta \textbf{B} \left[\partial \textbf{x}/\partial \vartheta \right] = 0$. 
 
 Therefore, we see that in order to satisfy linear force balance, $\delta F_{\alpha}$ must be chosen to satisfy the condition \eqref{eq:condition_in_surface}. However, this property can \textit{always} be imparted on a bulk force arising from the adjoint formulation. Consider the fixed-boundary adjoint relation \eqref{eq:fixed_boundary} without perturbations to the rotational transform,
 \begin{align}
  \int_{V_P} d^3 x \, \left( \bm{\xi}_1 \cdot  \textbf{F}_2   - \bm{\xi}_2 \cdot  \textbf{F}_1 \right)
    - \frac{1}{\mu_0} \int_{S_P} d^2 x \, \hat{\textbf{n}} \cdot \left( \bm{\xi}_2 \delta \textbf{B}[\bm{\xi}_1] \cdot \textbf{B} - \bm{\xi}_1 \delta \textbf{B}[\bm{\xi}_2] \cdot \textbf{B} \right) = 0.
 \end{align}
As $\delta \textbf{B}[\bm{\xi}]$ does not depend on $\xi^{\alpha c}_{0,0}$, we can choose to define the displacement vector such that $\xi^{\alpha c}_{0,0}=0$. This is analogous to our convention that $\bm{\xi} \cdot \textbf{B} = 0$, as $\delta \textbf{B}[\bm{\xi}]$ does not depend on the parallel component of $\bm{\xi}$. Given this convention for the displacement vector, we can note that $\langle \delta F_{\alpha,2} \rangle_{\psi}$ and $\langle \delta F_{\alpha,1} \rangle_{\psi}$ do not enter the above adjoint relation. Therefore, we are free to choose our bulk force such that the desired constraint \eqref{eq:condition_in_surface} is satisfied.

\renewcommand{\thechapter}{R}

\chapter{Near-axis expansion of screw pinch equilibria}
\label{app:axis_expansion}

The MHD force-balance equation for a screw pinch is,
\begin{align}
    \der{}{r} \left( \mu_0 p(r) + \frac{1}{2r^2} \left(\psi'(r)\right)^2 \right) + \frac{\iota(r)\psi'(r)}{R_0^2 r}
    \der{}{r} \left( r \iota(r) \psi'(r) \right) = 0.
    \label{eq:screw_pinch_force_balance_app}
\end{align}
We note that \eqref{eq:screw_pinch_force_balance_app} remains unchanged under the transformation $r \rightarrow -r$, so $\psi(r)$ must be even in $r$. Thus near the origin we can express the flux function as,
\begin{align}
    \psi(r) = \frac{\psi_2}{2} r^2 + \frac{\psi_4}{4!} r^4 + \frac{\psi_6}{6!} r^6 + \mathcal{O}(r^8),
\end{align}
under the assumption that $\psi(0) = 0$. We similarly express the rotational transform and pressure profiles in a power series near the axis,
\begin{subequations}
\begin{align}
    \iota(\psi(r)) &= \iota_0 + \iota_1 \psi(r) + \frac{\iota_2}{2} \psi(r)^2 + \frac{\iota_3}{3!} \psi(r)^3 + \mathcal{O}(\psi(r)^4) \\
    p(\psi(r)) &= p_0 + p_1 \psi(r) + \frac{p_2}{2} \psi(r)^2 + \frac{p_3}{3!} \psi(r)^3 + \mathcal{O}(\psi(r)^4) .
\end{align}
\end{subequations}
The force-balance equation to $\mathcal{O}(r)$ becomes,
\begin{align}
 \mu_0 p_1 \psi_2 + \frac{2 \iota_0^2 \psi_2^2}{R_0^2} + \frac{\psi_2 \psi_4}{3} = 0,
 \label{eq:r}
\end{align}
and to $\mathcal{O}(r^3)$ it is,
\begin{align}
    \frac{\mu_0 p_2 \psi_2^2}{2} + \frac{3 \iota_0 \iota_1 \psi_2^3}{R_0^2} + \frac{\mu_0 p_1 \psi_4}{6} + \frac{\iota_0^2 \psi_2 \psi_4}{R_0^2} + \frac{\psi_4^2}{18} + \frac{\psi_2 \psi_6}{30} = 0.
    \label{eq:r3}
\end{align}
In order to determine the power series expansion of $\psi$, we match the solution near the axis with a numerical solution for $\psi(r)$ at some chosen boundary location near the axis, $r_b$. To perform an expansion to $\mathcal{O}(r^2)$, $\psi_2$ is chosen such that 
\begin{align}
    \psi_2 = \frac{2 \psi(r_b)}{r_b^2}.
\end{align}
To perform an expansion to $\mathcal{O}(r^4)$, \eqref{eq:r} is used to express $\psi_4$ in terms of $\psi_2$, and $\psi_2$ is chosen such that $\psi_2 r_b^2/2 + \psi_4 r_b^4/4! = \psi(r_b)$,
\begin{subequations}
\begin{align}
\psi_2 &= \frac{-\mu_0 p_1 r_b^4 - 8 \psi(r_b)}{2 r_b^2 \left(\frac{r_b^2 \iota_0^2}{R_0^2} - 2 \right)} \\
    \psi_4 &= - 3 \left(\mu_0 p_1 + \frac{2 \iota_0^2 \psi_2}{R_0^2} \right).
\end{align}
\end{subequations}
To perform an expansion to $\mathcal{O}(r^6)$, \eqref{eq:r} and \eqref{eq:r3} are used to express $\psi_4$ and $\psi_6$ in terms of $\psi_2$, and $\psi_2$ is chosen such that $\psi_2 r_b^2/2 + \psi_4 r_b^4/4! + \psi_6 r_b^6/6! = \psi(r_b)$. The resulting equation for $\psi_2$ is quadratic, but only one solution is allowed in practice to ensure that $(\psi_6 r_b^6/6!)/(\psi_2 r_b^2/2 + \psi_4 r_b^4/4!) \sim r_b^2$ in the limit that $r_b \ll 1$,
\begin{subequations}
\begin{align}
\psi_2 &= -\frac{R_0^2}{12 \iota_0 \iota_1 r_b^6} \Bigg(-24r_b^2 + \frac{12 r_b^4 \iota_0^2}{R_0^2} + r_b^6 \left(\mu_0 p_2 - \frac{8 \iota_0^4}{R_0^4} \right) \\ &\hspace{0.5cm}+ r_b^2\Bigg[\left(-24 + \mu_0 p_2 r_b^4 + \frac{12 r_b^2 \iota_0^2}{R_0^2} - \frac{8 r_b^4 \iota_0^4}{R_0^4} \right)^2 \nonumber \\ &\hspace{0.5cm}+ \frac{48 r_b^2 \iota_0 \iota_1}{R_0^2} \left(p_1 r_b^4 \left(-3 + \frac{2 r_b^2 \iota_0^2}{R_0^2} \right) - 24 \psi(r_b) \right)  \Bigg]^{1/2}\Bigg) \nonumber \\
    \psi_4 &= - 3 \left(\mu_0 p_1 + 2 \frac{\iota_0^2}{R_0^2} \psi_2 \right) \\
    \psi_6 &= 15 \left(\frac{4 \mu_0 p_1\iota_0^2}{R_0^2} - \mu_0 p_2 \psi_2 + \frac{8 \iota_0^4 \psi_2}{R_0^4} - \frac{6 \iota_0 \iota_2 \psi_2^2}{R_0^2} \right).
\end{align}
\end{subequations}

\begin{figure}
    \centering
    \begin{subfigure}[b]{0.51\textwidth}
    \includegraphics[width=1.0\textwidth,trim=0cm 0cm 3cm 1cm,clip]{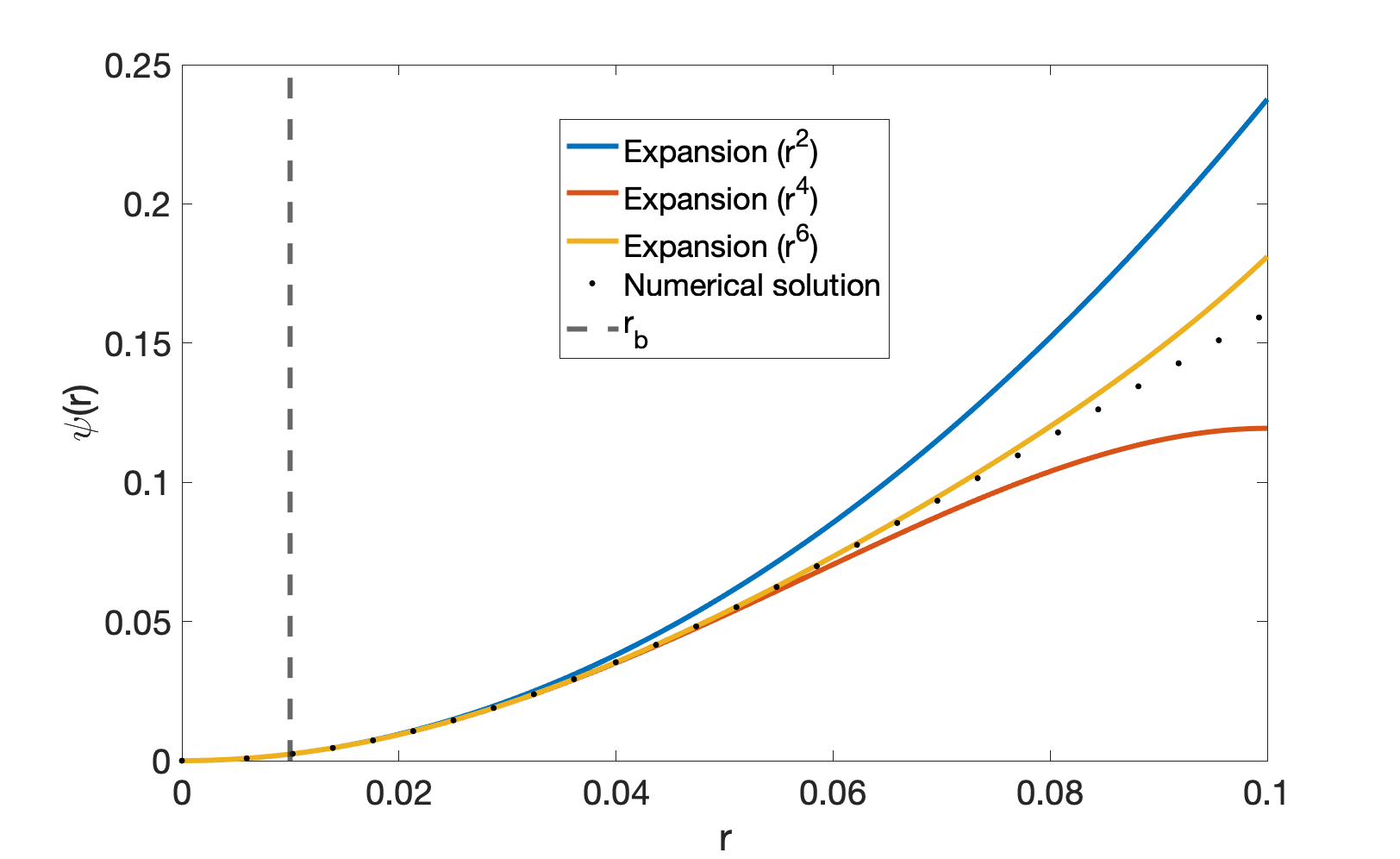}
    \caption{}
    \end{subfigure}
    \begin{subfigure}[b]{0.48\textwidth}
    \includegraphics[width=1.0\textwidth,trim=0cm 0cm 1cm 1cm,clip]{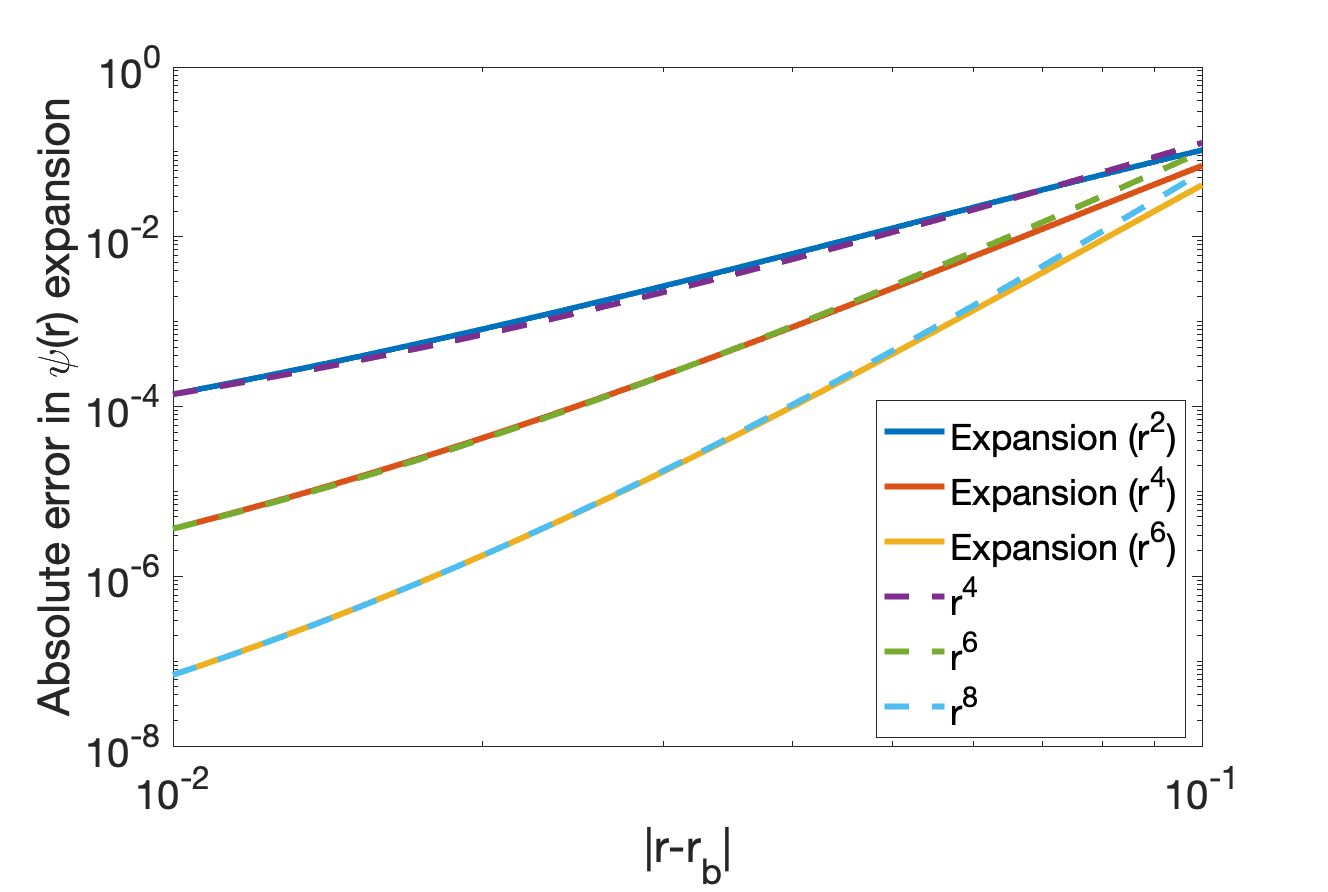}
    \caption{}
    \end{subfigure}
    \caption{(a) The axis expansion solutions to $\mathcal{O}(r^2)$, $\mathcal{O}(r^4)$, and $\mathcal{O}(r^6)$ are compared with the numerical solution of $\psi(r)$ near the axis. (b) The absolute error in the expansion is shown, $\rvert\sum_n \psi_n r^n/n! - \psi(r)\rvert$ where $\psi(r)$ is the numerical solution. As expected, the error in the axis expansion to $\mathcal{O}(r^p)$ scales as $|r-r_b|^{p+2}$ near $r = r_b$.}
    \label{fig:axis_expansion}
\end{figure}

We compare the resulting solution for $\psi$ to a numerical solution of \eqref{eq:screw_pinch_force_balance_app} using MATLAB's bvp4c routine. The solution is computed for $r \in [0,1]$ with a boundary condition of $\psi(0) = 0$ and $\psi(1) = \psi_0$. The same profiles are used as described in Section \ref{sec:m_0_n_0}. The axis expansion solution is matched with the numerical solution at $r_b = 10^{-2}$. In Figure \ref{fig:axis_expansion} we present a comparison between the numerical solution and axis expansion of $\psi(r)$. As expected, the error in the axis expansion to $\mathcal{O}(r^p)$ scales as $\sim |r-r_b|^{p+2}$ as one moves away from $r = r_b$.

\singlespacing
\bibliography{bibliography} 

\begin{thebibliography}{246}
\providecommand{\natexlab}[1]{#1}
\providecommand{\url}[1]{\texttt{#1}}
\expandafter\ifx\csname urlstyle\endcsname\relax
  \providecommand{\doi}[1]{doi: #1}\else
  \providecommand{\doi}{doi: \begingroup \urlstyle{rm}\Url}\fi

\bibitem[PPP()]{PPPLTimeline}
Princeton plasma physics laboratory - timeline.
\newblock URL \url{https://www.pppl.gov/about/history/timeline}.
\newblock date accessed: 01/03/2019.

\bibitem[Abel et~al.(2013)Abel, Plunk, Wang, Barnes, Cowley, Dorland, and
  Schekochihin]{Abel2013}
I.~Abel, G.~Plunk, E.~Wang, M.~Barnes, S.~Cowley, W.~Dorland, and
  A.~Schekochihin.
\newblock Multiscale gyrokinetics for rotating tokamak plasmas: fluctuations,
  transport and energy flows.
\newblock \emph{Reports on Progress in Physics}, 76\penalty0 (11):\penalty0
  116201, 2013.

\bibitem[Alexanderian et~al.(2017)Alexanderian, Petra, Stadler, and
  Ghattas]{Alexanderian2017}
A.~Alexanderian, N.~Petra, G.~Stadler, and O.~Ghattas.
\newblock Mean-variance risk-averse optimal control of systems governed by
  {PDEs} with random parameter fields using quadratic approximations.
\newblock \emph{SIAM/ASA Journal on Uncertainty Quantification}, 5\penalty0
  (1):\penalty0 1166--1192, 2017.

\bibitem[Allaire(2015)]{Allaire2015}
G.~Allaire.
\newblock A review of adjoint methods for sensitivity analysis, uncertainty
  quantification and optimization in numerical codes.
\newblock \emph{{Ingénieurs de l’Automobile}}, 836:\penalty0 33, 2015.

\bibitem[Almagri et~al.(1998)Almagri, Anderson, and Anderson]{Almagri1998}
A.~F. Almagri, D.~T. Anderson, and S.~F.~B. Anderson.
\newblock Design and construction of {HSX}: A helically symmetric stellarator.
\newblock In \emph{Helical System Research}. 1998.

\bibitem[Anderson(2019)]{Anderson2019}
D.~Anderson.
\newblock Personal communication, 9 2019.

\bibitem[Anderson et~al.(1990)Anderson, Cooper, Gruber, Merazzi, and
  Schwenn]{Anderson1990}
D.~V. Anderson, W.~Cooper, R.~Gruber, S.~Merazzi, and U.~Schwenn.
\newblock Methods for the efficient calculation of the {(MHD)}
  magnetohydrodynamic stability properties of magnetically confined fusion
  plasmas.
\newblock \emph{The International Journal of Supercomputing Applications},
  4\penalty0 (3):\penalty0 34, 1990.

\bibitem[Anderson et~al.(1995)Anderson, Almagri, Anderson, Matthews, Talmadge,
  and Shohet]{Anderson1995}
F.~S.~B. Anderson, A.~F. Almagri, D.~T. Anderson, P.~G. Matthews, J.~N.
  Talmadge, and J.~L. Shohet.
\newblock The {Helically Symmetric eXperiment},{(HSX)} goals, design and
  status.
\newblock \emph{Fusion Technology}, 27\penalty0 (3T):\penalty0 273--277, 1995.

\bibitem[Antonsen and Lee(1982)]{Antonsen1982}
T.~Antonsen and Y.~Lee.
\newblock Electrostatic modification of variational principles for anisotropic
  plasmas.
\newblock \emph{Physics of Fluids}, 25\penalty0 (1):\penalty0 132, 1982.

\bibitem[Antonsen et~al.(2019)Antonsen, Paul, and Landreman]{Antonsen2019}
T.~Antonsen, E.~J. Paul, and M.~Landreman.
\newblock Adjoint approach to calculating shape gradients for three-dimensional
  magnetic confinement equilibria.
\newblock \emph{Journal of Plasma Physics}, 85\penalty0 (2), 2019.

\bibitem[Arnoldus(2006)]{Arnoldus2006}
H.~F. Arnoldus.
\newblock Conservation of charge at an interface.
\newblock \emph{Optics Communications}, 265\penalty0 (1):\penalty0 52--59,
  2006.

\bibitem[Bader et~al.(2019)Bader, Drevlak, Anderson, Faber, Hegna, Likin,
  Schmitt, and Talmadge]{Bader2019}
A.~Bader, M.~Drevlak, D.~Anderson, B.~Faber, C.~Hegna, K.~Likin, J.~Schmitt,
  and J.~Talmadge.
\newblock Stellarator equilibria with reactor relevant energetic particle
  losses.
\newblock \emph{Journal of Plasma Physics}, 85\penalty0 (5), 2019.

\bibitem[Barnes et~al.(2010)Barnes, Abel, Dorland, G{\"o}rler, Hammett, and
  Jenko]{Barnes2010}
M.~Barnes, I.~Abel, W.~Dorland, T.~G{\"o}rler, G.~Hammett, and F.~Jenko.
\newblock Direct multiscale coupling of a transport code to gyrokinetic
  turbulence codes.
\newblock \emph{Physics of Plasmas}, 17\penalty0 (5):\penalty0 056109, 2010.

\bibitem[Bauer et~al.(2012)Bauer, Betancourt, and Garabedian]{Bauer2012}
F.~Bauer, O.~Betancourt, and P.~Garabedian.
\newblock \emph{A Computational Method in Plasma Physics}.
\newblock Springer Science \& Business Media, 2012.

\bibitem[Beidler et~al.(1990)Beidler, Grieger, Herrnegger, Harmeyer, Lotz,
  Maassberg, Merkel, N{\"{u}}hrenberg, Rau, Sapper, Sardei, Scardovelli,
  Schl{\"{u}}ter, and Wobig]{Beidler1990}
C.~Beidler, G.~Grieger, F.~Herrnegger, E.~Harmeyer, W.~Lotz, H.~Maassberg,
  P.~Merkel, J.~N{\"{u}}hrenberg, F.~Rau, J.~Sapper, F.~Sardei, R.~Scardovelli,
  A.~Schl{\"{u}}ter, and H.~Wobig.
\newblock Physics and engineering design for {Wendelstein VII-X}.
\newblock \emph{Fusion Technology}, 17\penalty0 (1):\penalty0 148, 1990.

\bibitem[Beidler et~al.(2011)Beidler, Allmaier, Isaev, Kasilov, Kernbichler,
  Leitold, Maassberg, Mikkelsen, Murakami, Schmidt, et~al.]{Beidler2011}
C.~Beidler, K.~Allmaier, M.~Y. Isaev, S.~Kasilov, W.~Kernbichler, G.~Leitold,
  H.~Maassberg, D.~Mikkelsen, S.~Murakami, M.~Schmidt, et~al.
\newblock Benchmarking of the mono-energetic transport coefficients—results
  from the {International Collaboration on Neoclassical Transport in
  Stellarators} ({ICNTS}).
\newblock \emph{Nuclear Fusion}, 51\penalty0 (7):\penalty0 076001, 2011.

\bibitem[Beidler and D'haeseleer(1995)]{Beidler1995}
C.~D. Beidler and W.~D. D'haeseleer.
\newblock A general solution of the ripple-averaged kinetic equation
  ({GSRAKE}).
\newblock \emph{Plasma Physics and Controlled Fusion}, 37\penalty0
  (4):\penalty0 463, 1995.

\bibitem[Belli and Candy(2015)]{Belli2015}
E.~A. Belli and J.~Candy.
\newblock Neoclassical transport in toroidal plasmas with nonaxisymmetric flux
  surfaces.
\newblock \emph{Plasma Physics and Controlled Fusion}, 57\penalty0
  (5):\penalty0 054012, 2015.

\bibitem[Berkl et~al.(1968)]{Berkl1968}
E.~Berkl et~al.
\newblock Plasma physics and controlled nuclear fusion research 1968.
\newblock In \emph{Proceedings of the 3rd International Conference
  Novosibirsk}, volume~1, 1968.

\bibitem[Bernstein et~al.(1958)Bernstein, Frieman, Kruskal, and
  Kulsrud]{Bernstein1958}
I.~Bernstein, E.~Frieman, M.~Kruskal, and R.~Kulsrud.
\newblock An energy principle for hydromagnetic stability problems.
\newblock \emph{Proceedings of the Royal Society A}, 244\penalty0
  (1236):\penalty0 17, 1958.

\bibitem[Boozer(1981)]{Boozer1981}
A.~Boozer.
\newblock Plasma equilibrium with rational magnetic surfaces.
\newblock \emph{The Physics of Fluids}, 24\penalty0 (11):\penalty0 1999, 1981.

\bibitem[Boozer(1995)]{Boozer1995}
A.~Boozer.
\newblock Quasi-helical symmetry in stellarators.
\newblock \emph{Plasma Physics and Controlled Fusion}, 37\penalty0
  (11A):\penalty0 A103, 1995.

\bibitem[Boozer(1980)]{Boozer1980}
A.~H. Boozer.
\newblock Guiding center drift equations.
\newblock \emph{The Physics of Fluids}, 23\penalty0 (5):\penalty0 904, 1980.

\bibitem[Boozer(1983)]{Boozer1983}
A.~H. Boozer.
\newblock Transport and isomorphic equilibria.
\newblock \emph{The Physics of Fluids}, 26\penalty0 (2):\penalty0 496, 1983.

\bibitem[Boozer(2000)]{Boozer2000}
A.~H. Boozer.
\newblock {Stellarator coil optimization by targeting the plasma
  configuration}.
\newblock \emph{Physics of Plasmas}, 7\penalty0 (8):\penalty0 3378, 2000.

\bibitem[Boozer(2015)]{Boozer2015}
A.~H. Boozer.
\newblock Non-axisymmetric magnetic fields and toroidal plasma confinement.
\newblock \emph{Nuclear Fusion}, 55\penalty0 (2):\penalty0 025001, 2015.

\bibitem[Boozer(2019)]{Boozer2019}
A.~H. Boozer.
\newblock Stellarators as a fast path to fusion energy.
\newblock \emph{arXiv preprint arXiv:1912.06289}, 2019.

\bibitem[Boozer and N{\"u}hrenberg(2006)]{Boozer2006}
A.~H. Boozer and C.~N{\"u}hrenberg.
\newblock Perturbed plasma equilibria.
\newblock \emph{Physics of Plasmas}, 13\penalty0 (10):\penalty0 102501, 2006.

\bibitem[Boyd and Vandenberghe(2004)]{Boyd2004}
S.~Boyd and L.~Vandenberghe.
\newblock \emph{Convex Optimization}.
\newblock Cambridge University Press, 2004.

\bibitem[Brent(2013)]{Brent2013}
R.~P. Brent.
\newblock \emph{Algorithms for Minimization Without Derivatives}.
\newblock Courier Corporation, 2013.

\bibitem[Brooks and Reiersen(2003)]{Brooks2003}
A.~Brooks and W.~Reiersen.
\newblock Coil tolerance impact on plasma surface quality for {NCSX}.
\newblock In \emph{20th IEEE/NPSS Symposium on Fusion Engineering, 2003}, page
  553. IEEE, 2003.

\bibitem[Brown et~al.(2015)Brown, Breslau, Gates, Pomphrey, and
  Zolfaghari]{Brown2015}
T.~Brown, J.~Breslau, D.~Gates, N.~Pomphrey, and A.~Zolfaghari.
\newblock Engineering optimization of stellarator coils lead to improvements in
  device maintenance.
\newblock In \emph{IEEE 26th Symposium on Fusion Engineering (SOFE)}, Austin,
  Texas, 2015.

\bibitem[Calvo et~al.(2017)Calvo, Parra, Velasco, and Alonso]{Calvo2017}
I.~Calvo, F.~I. Parra, J.~L. Velasco, and J.~A. Alonso.
\newblock The effect of tangential drifts on neoclassical transport in
  stellarators close to omnigeneity.
\newblock \emph{Plasma Physics and Controlled Fusion}, 59\penalty0
  (5):\penalty0 055014, 2017.

\bibitem[Calvo et~al.(2018)Calvo, Velasco, Parra, Alonso, and
  Garc{\'\i}a-Rega{\~n}a]{Calvo2018}
I.~Calvo, J.~L. Velasco, F.~I. Parra, J.~A. Alonso, and J.~M.
  Garc{\'\i}a-Rega{\~n}a.
\newblock Electrostatic potential variations on stellarator magnetic surfaces
  in low collisionality regimes.
\newblock \emph{Journal of Plasma Physics}, 84\penalty0 (4), 2018.

\bibitem[Canik et~al.(2007)Canik, Anderson, Anderson, Clark, Likin, Talmadge,
  and Zhai]{Canik2007}
J.~Canik, D.~Anderson, F.~Anderson, C.~Clark, K.~Likin, J.~Talmadge, and
  K.~Zhai.
\newblock Reduced particle and heat transport with quasisymmetry in the
  {Helically Symmetric Experiment}.
\newblock \emph{Physics of Plasmas}, 14\penalty0 (5):\penalty0 056107, 2007.

\bibitem[Carlton-Jones et~al.(2019)Carlton-Jones, Paul, and
  Dorland]{Carlton2019}
A.~Carlton-Jones, E.~Paul, and W.~Dorland.
\newblock Computing the shape gradient of coil complexity with respect to the
  plasma boundary with an adjoint method.
\newblock \emph{Bulletin of the American Physical Society}, 64, 2019.

\bibitem[Carreras et~al.(1996)Carreras, Lynch, and Ware]{Carreras1996}
B.~Carreras, V.~Lynch, and A.~Ware.
\newblock Configuration studies for a small-aspect-ratio tokamak stellarator
  hybrid.
\newblock Technical report, Oak Ridge National Lab., 1996.

\bibitem[Cary and Hanson(1991)]{Cary1991}
J.~R. Cary and J.~D. Hanson.
\newblock Simple method for calculating island widths.
\newblock \emph{Physics of Fluids B: Plasma Physics}, 3\penalty0 (4):\penalty0
  1006, 1991.

\bibitem[Cary and Shasharina(1997)]{Cary1997}
J.~R. Cary and S.~G. Shasharina.
\newblock Omnigenity and quasihelicity in helical plasma confinement systems.
\newblock \emph{Physics of Plasmas}, 4\penalty0 (9):\penalty0 3323, 1997.

\bibitem[Choi and Kim(2006)]{Choi2006}
K.~K. Choi and N.-H. Kim.
\newblock \emph{Structural Sensitivity Analysis and Optimization 1: Linear
  Systems}.
\newblock Springer Science \& Business Media, 2006.

\bibitem[Coddington and Levinson(1955)]{Coddington1955}
E.~A. Coddington and N.~Levinson.
\newblock \emph{Theory of Ordinary Differential Equations}.
\newblock Tata McGraw-Hill Education, 1955.

\bibitem[Connor and Hastie(1974)]{Connor1974}
J.~Connor and R.~Hastie.
\newblock Neoclassical diffusion in an $l=3$ stellarator.
\newblock \emph{Physics of Fluids}, 17\penalty0 (114):\penalty0 114, 1974.

\bibitem[Cooper et~al.(1992)Cooper, Hirshman, Merazzi, and
  Gruber]{Cooper19923d}
W.~Cooper, S.~Hirshman, S.~Merazzi, and R.~Gruber.
\newblock {3D} magnetohydrodynamic equilibria with anisotropic pressure.
\newblock \emph{Computer Physics Communications}, 72\penalty0 (1):\penalty0 1,
  1992.

\bibitem[Cooper et~al.(2005)Cooper, Hirshman, Yamaguchi, Narushima, Okamura,
  Sakakibara, Suzuki, Watanabe, Yamada, and Yamazaki]{Cooper2005}
W.~Cooper, S.~Hirshman, T.~Yamaguchi, Y.~Narushima, S.~Okamura, S.~Sakakibara,
  C.~Suzuki, K.~Watanabe, H.~Yamada, and K.~Yamazaki.
\newblock Three-dimensional anisotropic pressure equilibria that model balanced
  tangential neutral beam injection effects.
\newblock \emph{Plasma Physics and Controlled Fusion}, 47\penalty0
  (3):\penalty0 561, 2005.

\bibitem[Cooper et~al.(2006)Cooper, Graves, Hirshman, Yamaguchi, Narushima,
  Okamura, Sakakibara, Suzuki, Watanabe, Yamada, et~al.]{Cooper2006}
W.~Cooper, J.~Graves, S.~Hirshman, T.~Yamaguchi, Y.~Narushima, S.~Okamura,
  S.~Sakakibara, C.~Suzuki, K.~Watanabe, H.~Yamada, et~al.
\newblock Anisotropic pressure bi-{Maxwellian} distribution function model for
  three-dimensional equilibria.
\newblock \emph{Nuclear Fusion}, 46\penalty0 (7):\penalty0 683, 2006.

\bibitem[Coor et~al.(1958)Coor, Cunningham, Ellis, Heald, and Kranz]{Coor1958}
T.~Coor, S.~Cunningham, R.~Ellis, M.~Heald, and A.~Kranz.
\newblock Experiments on the ohmic heating and confinement of plasma in a
  stellarator.
\newblock \emph{The Physics of Fluids}, 1\penalty0 (5):\penalty0 411, 1958.

\bibitem[Dekeyser(2014)]{Dekeyser2014}
W.~Dekeyser.
\newblock \emph{Optimal Plasma Edge Configurations for Next-Step Fusion
  Reactors}.
\newblock PhD thesis, Katholieke Universiteit Leuven, 2014.

\bibitem[Dekeyser et~al.(2012)Dekeyser, Reiter, and Baelmans]{Dekeyser2012}
W.~Dekeyser, D.~Reiter, and M.~Baelmans.
\newblock Divertor design through shape optimization.
\newblock \emph{Contributions to Plasma Physics}, 52\penalty0 (5):\penalty0
  544, 2012.

\bibitem[Dekeyser et~al.(2014{\natexlab{a}})Dekeyser, Reiter, and
  Baelmans]{Dekeyser2014a}
W.~Dekeyser, D.~Reiter, and M.~Baelmans.
\newblock Automated divertor target design by adjoint shape sensitivity
  analysis and a one-shot method.
\newblock \emph{Journal of Computational Physics}, 278:\penalty0 117,
  2014{\natexlab{a}}.

\bibitem[Dekeyser et~al.(2014{\natexlab{b}})Dekeyser, Reiter, and
  Baelmans]{Dekeyser2014b}
W.~Dekeyser, D.~Reiter, and M.~Baelmans.
\newblock Optimal shape design for divertors.
\newblock \emph{International Journal of Computational Science and Engineering
  2}, 9\penalty0 (5-6):\penalty0 397, 2014{\natexlab{b}}.

\bibitem[Dekeyser et~al.(2014{\natexlab{c}})Dekeyser, Reiter, and
  Baelmans]{Dekeyser2014c}
W.~Dekeyser, D.~Reiter, and M.~Baelmans.
\newblock A one shot method for divertor target shape optimization.
\newblock \emph{Proceedings in Applied Mathematics and Mechanics}, 14\penalty0
  (1):\penalty0 1017, 2014{\natexlab{c}}.

\bibitem[Delfour and Zol\'{e}sio(2011)]{Delfour2011}
M.~C. Delfour and J.-P. Zol\'{e}sio.
\newblock \emph{Shapes and Geometries}.
\newblock Society for Industrial and Applied Mathematics, 2011.

\bibitem[Dewar and Hudson(1998)]{Dewar1998}
R.~Dewar and S.~Hudson.
\newblock Stellarator symmetry.
\newblock \emph{Physica D: Nonlinear Phenomena}, 112\penalty0 (1):\penalty0
  275--280, 1998.

\bibitem[D'haeseleer et~al.(1991)D'haeseleer, Hitchon, Callen, and
  Shohet]{2012Dhaeseleer}
W.~D. D'haeseleer, W.~N. Hitchon, J.~D. Callen, and J.~L. Shohet.
\newblock \emph{Flux Coordinates and Magnetic Field Structure: A Guide to a
  Fundamental Tool of Plasma Theory}.
\newblock Springer, 1991.

\bibitem[Dinklage et~al.(2018)Dinklage, Beidler, Helander, Fuchert,
  Maa{\ss}berg, Rahbarnia, Pedersen, Turkin, Wolf, Alonso,
  et~al.]{Dinklage2018}
A.~Dinklage, C.~Beidler, P.~Helander, G.~Fuchert, H.~Maa{\ss}berg,
  K.~Rahbarnia, T.~S. Pedersen, Y.~Turkin, R.~Wolf, A.~Alonso, et~al.
\newblock Magnetic configuration effects on the {Wendelstein 7-X} stellarator.
\newblock \emph{Nature Physics}, 14\penalty0 (8):\penalty0 855--860, 2018.

\bibitem[Drevlak(1999)]{Drevlak1999}
M.~Drevlak.
\newblock Optimization of heterogenous magnet systems.
\newblock In \emph{Proceedings of the 12th International Stellarator Workshop},
  number P1-17, 1999.

\bibitem[Drevlak et~al.(2013)Drevlak, Brochard, Helander, Kisslinger,
  Mikhailov, N\"{u}hrenberg, N\"{u}hrenberg, and Turkin]{Drevlak2013}
M.~Drevlak, F.~Brochard, P.~Helander, J.~Kisslinger, M.~Mikhailov,
  C.~N\"{u}hrenberg, J.~N\"{u}hrenberg, and Y.~Turkin.
\newblock {ESTELL: A Quasi-Toroidally Symmetric Stellarator}.
\newblock \emph{Contributions to Plasma Physics}, 53\penalty0 (6):\penalty0
  459, 2013.

\bibitem[Drevlak et~al.(2014)Drevlak, Geiger, Helander, and
  Turkin]{Drevlak2014}
M.~Drevlak, J.~Geiger, P.~Helander, and Y.~Turkin.
\newblock Fast particle confinement with optimized coil currents in the {W7-X}
  stellarator.
\newblock \emph{Nuclear Fusion}, 54\penalty0 (7):\penalty0 073002, 2014.

\bibitem[Drevlak et~al.(2018)Drevlak, Beidler, Geiger, Helander, and
  Turkin]{Drevlak2018}
M.~Drevlak, C.~Beidler, J.~Geiger, P.~Helander, and Y.~Turkin.
\newblock Optimisation of stellarator equilibria with {ROSE}.
\newblock \emph{Nuclear Fusion}, 59\penalty0 (1):\penalty0 016010, 2018.

\bibitem[El-Guebaly et~al.(2008)El-Guebaly, Wilson, Henderson, Sawan,
  Sviatoslavsky, Slaybaugh, Kiedrowski, Ibrahim, Martin, Raffray, Malang, Lyon,
  Ku, Wang, Bromberg, Merrill, Waganer, Najmabadi, and the
  Aries-CS~Team]{Guebaly2008}
L.~El-Guebaly, P.~Wilson, D.~Henderson, M.~Sawan, G.~Sviatoslavsky,
  R.~Slaybaugh, B.~Kiedrowski, A.~Ibrahim, C.~Martin, R.~Raffray, S.~Malang,
  J.~Lyon, L.~P. Ku, X.~Wang, L.~Bromberg, B.~Merrill, L.~Waganer,
  F.~Najmabadi, and the Aries-CS~Team.
\newblock {Designing ARIES-CS Compact Radial Build and Nuclear System:
  Neutronics, Shielding, and Activation}.
\newblock \emph{Fusion Science and Technology}, 54:\penalty0 747, 2008.

\bibitem[Ferraro et~al.(2019)Ferraro, Park, Myers, Brooks, Gerhardt, Menard,
  Munaretto, and Reinke]{Ferraro2019}
N.~M. Ferraro, J.-K. Park, C.~Myers, A.~Brooks, S.~Gerhardt, J.~Menard,
  S.~Munaretto, and M.~Reinke.
\newblock Error field impact on mode locking and divertor heat flux in
  {NSTX-U}.
\newblock \emph{Nuclear Fusion}, 59\penalty0 (8):\penalty0 086021, 2019.

\bibitem[Forbes and Crozier(2001)]{Forbes2001}
L.~K. Forbes and S.~Crozier.
\newblock {A novel target-field method for finite-length magnetic resonance
  shim coils: I. Zonal shims}.
\newblock \emph{Journal of Physics D: Applied Physics}, 34:\penalty0 3447,
  2001.

\bibitem[Forbes et~al.(2005)Forbes, Brideson, and Crozier]{Forbes2005}
L.~K. Forbes, M.~A. Brideson, and S.~Crozier.
\newblock {A Target-Field Method to Design Circular Biplanar Coils for
  Asymmetric Shim and Gradient Fields}.
\newblock \emph{IEEE Transactions on Magnetics}, 41\penalty0 (6):\penalty0
  2134, 2005.

\bibitem[Freidberg(2014)]{Freidberg2014}
J.~Freidberg.
\newblock \emph{Ideal MHD}.
\newblock Cambridge University Press, 2014.

\bibitem[Frieman(1970)]{Frieman1970}
E.~Frieman.
\newblock Collisional diffusion in nonaxisymmetric toroidal systems.
\newblock \emph{Physics of Fluids}, 13\penalty0 (490):\penalty0 490, 1970.

\bibitem[Galeev and Sagdeev(1979)]{Galeev1979}
A.~Galeev and R.~Sagdeev.
\newblock \emph{Theory of Neoclassical Diffusion}, volume~7 of \emph{Reviews of
  Plasma Physics}, page 257.
\newblock 1979.

\bibitem[Gamba(1994)]{Gamba1994}
I.~M. Gamba.
\newblock Viscosity approximating solutions to {ODE} systems that admit shocks,
  and their limits.
\newblock \emph{Advances in Applied Mathematics}, 15\penalty0 (2):\penalty0
  129--182, 1994.

\bibitem[Gandini(1990)]{Gandini1990}
A.~Gandini.
\newblock Importance and sensitivity analysis in assessing system reliability.
\newblock \emph{IEEE Transactions on Reliability}, 39\penalty0 (1):\penalty0
  61, 1990.

\bibitem[Garabedian(2002)]{Garabedian2002}
P.~Garabedian.
\newblock Three-dimensional stellarator codes.
\newblock \emph{Proceedings of the National Academy of Sciences}, 99\penalty0
  (16):\penalty0 10257, 2002.

\bibitem[Garabedian and McFadden(2009)]{Garabedian2009}
P.~R. Garabedian and G.~B. McFadden.
\newblock Design of the {DEMO} fusion reactor following {ITER}.
\newblock \emph{Journal of Research of the National Institute of Standards and
  Technology}, 114\penalty0 (4):\penalty0 229, 2009.

\bibitem[Gardner(1990)]{Gardner1990}
H.~Gardner.
\newblock Modelling the behaviour of the magnetic field diagnostic coils on the
  {W VII-AS} stellarator using a three-dimensional equilibrium code.
\newblock \emph{Nuclear Fusion}, 30\penalty0 (8):\penalty0 1417, 1990.

\bibitem[Gates and Delgado-Aparicio(2012)]{Gates2012}
D.~Gates and L.~Delgado-Aparicio.
\newblock Origin of tokamak density limit scalings.
\newblock \emph{Physical Review Letters}, 108\penalty0 (16):\penalty0 165004,
  2012.

\bibitem[Gates et~al.(2018)Gates, Anderson, Anderson, Zarnstorff, Spong,
  Weitzner, Neilson, Ruzic, Andruczyk, Harris, et~al.]{Gates2018}
D.~A. Gates, D.~Anderson, S.~Anderson, M.~Zarnstorff, D.~A. Spong, H.~Weitzner,
  G.~Neilson, D.~Ruzic, D.~Andruczyk, J.~Harris, et~al.
\newblock Stellarator research opportunities: a report of the {National
  Stellarator Coordinating Committee}.
\newblock \emph{Journal of Fusion Energy}, 37\penalty0 (1):\penalty0 51, 2018.

\bibitem[Gavrilovi{\'c} et~al.(1963)Gavrilovi{\'c}, Petrovi{\'c}, and
  {\v{S}}iljak]{Gavrilovic1963}
M.~Gavrilovi{\'c}, R.~Petrovi{\'c}, and D.~{\v{S}}iljak.
\newblock Adjoint method in the sensitivity analysis of optimal systems.
\newblock \emph{Journal of the Franklin Institute}, 276\penalty0 (1):\penalty0
  26, 1963.

\bibitem[Geiger et~al.(2010)Geiger, Beidler, Drevlak, Maassberg,
  N{\"u}hrenberg, Suzuki, and Turkin]{Geiger2010}
J.~Geiger, C.~Beidler, M.~Drevlak, H.~Maassberg, C.~N{\"u}hrenberg, Y.~Suzuki,
  and Y.~Turkin.
\newblock Effects of net currents on the magnetic configuration of {W7-X}.
\newblock \emph{Contributions to Plasma Physics}, 50\penalty0 (8):\penalty0
  770, 2010.

\bibitem[Geraldini and Landreman(2019)]{Geraldini2019}
A.~Geraldini and M.~Landreman.
\newblock Optimizing stellarator surfaces using magnetic island width
  sensitivity.
\newblock \emph{Bulletin of the American Physical Society}, 64, 2019.

\bibitem[Gerhardt et~al.(2005)Gerhardt, Talmadge, Canik, and
  Anderson]{Gerhardt2005}
S.~P. Gerhardt, J.~N. Talmadge, J.~M. Canik, and D.~T. Anderson.
\newblock Measurements and modeling of plasma flow damping in the {Helically
  Symmetric eXperiment}.
\newblock \emph{Physics of Plasmas}, 12\penalty0 (5):\penalty0 056116, 2005.

\bibitem[Giles and Pierce(1999)]{Giles1999}
M.~Giles and N.~Pierce.
\newblock Improved lift and drag estimates using adjoint {Euler} equations.
\newblock In \emph{14th Computational Fluid Dynamics Conference}, page 3293,
  1999.

\bibitem[Giles and Pierce(2000)]{Giles2000}
M.~B. Giles and N.~A. Pierce.
\newblock An introduction to the adjoint approach to design.
\newblock \emph{Flow, Turbulence and Combustion}, 65\penalty0 (3-4):\penalty0
  393, 2000.

\bibitem[Glasser(2016)]{Glasser2016}
A.~Glasser.
\newblock The direct criterion of {Newcomb} for the ideal {MHD} stability of an
  axisymmetric toroidal plasma.
\newblock \emph{Physics of Plasmas}, 23\penalty0 (7):\penalty0 072505, 2016.

\bibitem[Glasser(2018)]{Glasser2018}
A.~Glasser.
\newblock {DCON} for stellarators.
\newblock \emph{Bulletin of the American Physical Society}, 63, 2018.

\bibitem[Glowinski and Pironneau(1975)]{Glowinski1975}
R.~Glowinski and O.~Pironneau.
\newblock On the numerical computation of the minimum-drag profile in laminar
  flow.
\newblock \emph{Journal of Fluid Mechanics}, 72\penalty0 (2):\penalty0 385,
  1975.

\bibitem[Goedbloed and Poedts(2004)]{Goedbloed2004}
J.~H. Goedbloed and S.~Poedts.
\newblock \emph{Principles of Magnetohydrodynamics: With Applications to
  Laboratory and Astrophysical Plasmas}.
\newblock Cambridge University Press, 2004.

\bibitem[Grad(1967)]{Grad1967}
H.~Grad.
\newblock Toroidal containment of a plasma.
\newblock \emph{The Physics of Fluids}, 10\penalty0 (1):\penalty0 137, 1967.

\bibitem[Greene(1997)]{Greene1997}
J.~Greene.
\newblock A brief review of magnetic wells.
\newblock \emph{Comments on Plasma Physics and Controlled Fusion}, 17:\penalty0
  389, 1997.

\bibitem[Grieger et~al.(1992)Grieger, Lotz, Merkel, N{\"u}hrenberg, Sapper,
  Strumberger, Wobig, Burhenn, Erckmann, Gasparino, et~al.]{Grieger1992}
G.~Grieger, W.~Lotz, P.~Merkel, J.~N{\"u}hrenberg, J.~Sapper, E.~Strumberger,
  H.~Wobig, R.~Burhenn, V.~Erckmann, U.~Gasparino, et~al.
\newblock Physics optimization of stellarators.
\newblock \emph{Physics of Fluids B: Plasma Physics}, 4\penalty0 (7):\penalty0
  2081, 1992.

\bibitem[Hadamard(1908)]{Hadamard1908}
J.~Hadamard.
\newblock \emph{M{\'e}moire sur le probl{\`e}me d'analyse relatif {\`a}
  l'{\'e}quilibre des plaques {\'e}lastiques encastr{\'e}es}, volume~33.
\newblock Imprimerie Nationale, 1908.

\bibitem[Hammond et~al.(2016)Hammond, Anichowski, Brenner, Pedersen,
  Raftopoulos, Traverso, and Volpe]{Hammond2016}
K.~Hammond, A.~Anichowski, P.~Brenner, T.~S. Pedersen, S.~Raftopoulos,
  P.~Traverso, and F.~Volpe.
\newblock Experimental and numerical study of error fields in the {CNT}
  stellarator.
\newblock \emph{Plasma Physics and Controlled Fusion}, 58\penalty0
  (7):\penalty0 074002, 2016.

\bibitem[Hanson et~al.(2013)Hanson, Anderson, Cianciosa, Franz, Harris,
  Hartwell, Hirshman, Knowlton, Lao, Lazarus, et~al.]{Hanson2013}
J.~D. Hanson, D.~Anderson, M.~Cianciosa, P.~Franz, J.~Harris, G.~Hartwell,
  S.~P. Hirshman, S.~F. Knowlton, L.~L. Lao, E.~A. Lazarus, et~al.
\newblock Non-axisymmetric equilibrium reconstruction for stellarators,
  reversed field pinches and tokamaks.
\newblock \emph{Nuclear Fusion}, 53\penalty0 (8):\penalty0 083016, 2013.

\bibitem[Harafuji et~al.(1989)Harafuji, Hayashi, and Sato]{Harafuji1989}
K.~Harafuji, T.~Hayashi, and T.~Sato.
\newblock Computational study of three-dimensional magnetohydrodynamic
  equilibria in toroidal helical systems.
\newblock \emph{Journal of Computational Physics}, 81\penalty0 (1):\penalty0
  169, 1989.

\bibitem[Haslinger and M{\"a}kinen(2003)]{Haslinger2003}
J.~Haslinger and R.~A. M{\"a}kinen.
\newblock \emph{Introduction to Shape Optimization: Theory, Approximation, and
  Computation}.
\newblock Society for Industrial and Applied Mathematics, 2003.

\bibitem[Hastings et~al.(1985)Hastings, Houlberg, and Shaing]{Hastings1985}
D.~Hastings, W.~Houlberg, and K.-C. Shaing.
\newblock The ambipolar electric field in stellarators.
\newblock \emph{Nuclear Fusion}, 25\penalty0 (4):\penalty0 445, 1985.

\bibitem[Hawryluk and Zohm(2019)]{Hawryluk2019}
R.~Hawryluk and H.~Zohm.
\newblock The challenge and promise of studying burning plasmas.
\newblock \emph{Physics Today}, 72\penalty0 (12):\penalty0 34, 2019.

\bibitem[Hazeltine(1973)]{Hazeltine1973}
R.~D. Hazeltine.
\newblock Recursive derivation of drift-kinetic equation.
\newblock \emph{Plasma Physics}, 15\penalty0 (1):\penalty0 77, 1973.

\bibitem[Hegna and Nakajima(1998)]{Hegna1998}
C.~C. Hegna and N.~Nakajima.
\newblock On the stability of {Mercier} and ballooning modes in stellarator
  configurations.
\newblock \emph{Physics of Plasmas}, 5\penalty0 (5):\penalty0 1336, 1998.

\bibitem[Hegna et~al.(2018)Hegna, Terry, and Faber]{Hegna2018}
C.~C. Hegna, P.~W. Terry, and B.~J. Faber.
\newblock Theory of {ITG} turbulent saturation in stellarators: identifying
  mechanisms to reduce turbulent transport.
\newblock \emph{Physics of Plasmas}, 25\penalty0 (2):\penalty0 022511, 2018.

\bibitem[Helander(2014)]{Helander2014}
P.~Helander.
\newblock Theory of plasma confinement in non-axisymmetric magnetic fields.
\newblock \emph{Reports on Progress in Physics}, 77\penalty0 (8):\penalty0
  087001, 2014.

\bibitem[Helander and N{\"u}hrenberg(2009)]{Helander2009}
P.~Helander and J.~N{\"u}hrenberg.
\newblock Bootstrap current and neoclassical transport in quasi-isodynamic
  stellarators.
\newblock \emph{Plasma Physics and Controlled Fusion}, 51\penalty0
  (5):\penalty0 055004, 2009.

\bibitem[Helander and Sigmar(2005)]{Helander2005}
P.~Helander and D.~J. Sigmar.
\newblock \emph{Collisional Transport in Magnetized Plasmas}.
\newblock Cambridge University Press, 2005.

\bibitem[Helander and Simakov(2008)]{Helander2008}
P.~Helander and A.~Simakov.
\newblock Intrinsic ambipolarity and rotation in stellarators.
\newblock \emph{Physical Review Letters}, 101\penalty0 (14):\penalty0 145003,
  2008.

\bibitem[Helander et~al.(2012)Helander, Beidler, Bird, Drevlak, Feng, Hatzky,
  Jenko, Kleiber, Proll, Turkin, et~al.]{Helander2012}
P.~Helander, C.~Beidler, T.~Bird, M.~Drevlak, Y.~Feng, R.~Hatzky, F.~Jenko,
  R.~Kleiber, J.~Proll, Y.~Turkin, et~al.
\newblock Stellarator and tokamak plasmas: a comparison.
\newblock \emph{Plasma Physics and Controlled Fusion}, 54\penalty0
  (12):\penalty0 124009, 2012.

\bibitem[Helander et~al.(2017)Helander, Parra, and Newton]{Helander2017}
P.~Helander, F.~Parra, and S.~Newton.
\newblock Stellarator bootstrap current and plasma flow velocity at low
  collisionality.
\newblock \emph{Journal of Plasma Physics}, 83\penalty0 (2), 2017.

\bibitem[Helander et~al.(2020)Helander, Drevlak, Zarnstorff, and
  Cowley]{Helander2019}
P.~Helander, M.~Drevlak, M.~Zarnstorff, and S.~Cowley.
\newblock Stellarators with permanent magnets.
\newblock \emph{Physical Review Letters}, 124\penalty0 (9):\penalty0 095001,
  2020.

\bibitem[Hender et~al.(2007)Hender, Wesley, Bialek, Bondeson, Boozer, Buttery,
  Garofalo, Goodman, Granetz, Gribov, et~al.]{Hender2007}
T.~Hender, J.~Wesley, J.~Bialek, A.~Bondeson, A.~Boozer, R.~Buttery,
  A.~Garofalo, T.~Goodman, R.~Granetz, Y.~Gribov, et~al.
\newblock {MHD} stability, operational limits and disruptions.
\newblock \emph{Nuclear Fusion}, 47\penalty0 (6):\penalty0 S128, 2007.

\bibitem[Henneberg et~al.(2019{\natexlab{a}})Henneberg, Drevlak, and
  Helander]{Henneberg2019b}
S.~Henneberg, M.~Drevlak, and P.~Helander.
\newblock Improving fast-particle confinement in quasi-axisymmetric stellarator
  optimization.
\newblock \emph{Plasma Physics and Controlled Fusion}, 62\penalty0
  (1):\penalty0 014023, 2019{\natexlab{a}}.

\bibitem[Henneberg et~al.(2019{\natexlab{b}})Henneberg, Drevlak,
  N{\"u}hrenberg, Beidler, Turkin, Loizu, and Helander]{Henneberg2019}
S.~Henneberg, M.~Drevlak, C.~N{\"u}hrenberg, C.~Beidler, Y.~Turkin, J.~Loizu,
  and P.~Helander.
\newblock Properties of a new quasi-axisymmetric configuration.
\newblock \emph{Nuclear Fusion}, 59\penalty0 (2):\penalty0 026014,
  2019{\natexlab{b}}.

\bibitem[Highcock et~al.(2018)Highcock, Mandell, Barnes, and
  Dorland]{Highcock2018}
E.~Highcock, N.~Mandell, M.~Barnes, and W.~Dorland.
\newblock Optimisation of confinement in a fusion reactor using a nonlinear
  turbulence model.
\newblock \emph{Journal of Plasma Physics}, 84\penalty0 (2), 2018.

\bibitem[Hirsch et~al.(2008)Hirsch, Baldzuhn, Beidler, Brakel, Burhenn,
  Dinklage, Ehmler, Endler, Erckmann, Feng, et~al.]{Hirsch2008}
M.~Hirsch, J.~Baldzuhn, C.~Beidler, R.~Brakel, R.~Burhenn, A.~Dinklage,
  H.~Ehmler, M.~Endler, V.~Erckmann, Y.~Feng, et~al.
\newblock Major results from the stellarator {Wendelstein} {7-AS}.
\newblock \emph{Plasma Physics and Controlled Fusion}, 50\penalty0
  (5):\penalty0 053001, 2008.

\bibitem[Hirshman and Breslau(1998)]{Hirshman1998}
S.~P. Hirshman and J.~Breslau.
\newblock {Explicit spectrally optimized Fourier series for nested magnetic
  surfaces}.
\newblock \emph{Physics of Plasmas}, 5:\penalty0 2664, 1998.

\bibitem[Hirshman and Meier(1985)]{Hirshman1985}
S.~P. Hirshman and H.~K. Meier.
\newblock {Optimized Fourier representations for three ‐ dimensional magnetic
  surfaces}.
\newblock \emph{Physics of Fluids}, 28:\penalty0 1387, 1985.

\bibitem[Hirshman and Whitson(1983)]{Hirshman1983}
S.~P. Hirshman and J.~C. Whitson.
\newblock Steepest-descent moment method for three-dimensional
  magnetohydrodynamic equilibria.
\newblock \emph{Physics of Fluids}, 26\penalty0 (12):\penalty0 3553, 1983.

\bibitem[Hirshman et~al.(1986{\natexlab{a}})Hirshman, Shaing, and van
  Rij]{Hirshman1986b}
S.~P. Hirshman, K.~C. Shaing, and W.~I. van Rij.
\newblock Consequences of time-reversal symmetry for the electric field scaling
  of transport in stellarators.
\newblock \emph{{Physical Review Letters}}, 56\penalty0 (16):\penalty0 1697,
  1986{\natexlab{a}}.

\bibitem[Hirshman et~al.(1986{\natexlab{b}})Hirshman, Shaing, van Rij, Beasley,
  and Crume]{Hirshman1986}
S.~P. Hirshman, K.~C. Shaing, W.~I. van Rij, C.~O. Beasley, and E.~C. Crume.
\newblock {Plasma transport coefficients for nonsymmetric toroidal confinement
  systems}.
\newblock \emph{Physics of Fluids}, 29\penalty0 (9):\penalty0 2951,
  1986{\natexlab{b}}.

\bibitem[Hirshman et~al.(1999)Hirshman, Spong, Whitson, Nelson, Batchelor,
  Lyon, Sanchez, Brooks, Y.-Fu, Goldston, et~al.]{Hirshman1999}
S.~P. Hirshman, D.~A. Spong, J.~C. Whitson, B.~Nelson, D.~B. Batchelor, J.~F.
  Lyon, R.~Sanchez, A.~Brooks, G.~Y.-Fu, R.~J. Goldston, et~al.
\newblock Physics of compact stellarators.
\newblock \emph{Physics of Plasmas}, 6\penalty0 (5):\penalty0 1858, 1999.

\bibitem[Hirshman et~al.(2011)Hirshman, Sanchez, and Cook]{Hirshman2011}
S.~P. Hirshman, R.~Sanchez, and C.~Cook.
\newblock {SIESTA}: A scalable iterative equilibrium solver for toroidal
  applications.
\newblock \emph{Physics of Plasmas}, 18\penalty0 (6):\penalty0 062504, 2011.

\bibitem[Ho and Kulsrud(1987)]{Ho1987}
D.-M. Ho and R.~Kulsrud.
\newblock Neoclassical transport in stellarators.
\newblock \emph{Physics of Fluids}, 30\penalty0 (2):\penalty0 442, 1987.

\bibitem[Hofmann et~al.(1996)Hofmann, Baldzuhn, Brakel, Feng, Fiedler, Geiger,
  Grigull, Herre, Jaenicke, Kick, et~al.]{Hofmann1996}
J.~Hofmann, J.~Baldzuhn, R.~Brakel, Y.~Feng, S.~Fiedler, J.~Geiger, P.~Grigull,
  G.~Herre, R.~Jaenicke, M.~Kick, et~al.
\newblock Stellarator optimization studies in {W7-AS}.
\newblock \emph{Plasma Physics and Controlled Fusion}, 38\penalty0
  (12A):\penalty0 A193, 1996.

\bibitem[Hudson et~al.(2018)Hudson, Zhu, Pfefferl{\'e}, and
  Gunderson]{Hudson2018}
S.~Hudson, C.~Zhu, D.~Pfefferl{\'e}, and L.~Gunderson.
\newblock Differentiating the shape of stellarator coils with respect to the
  plasma boundary.
\newblock \emph{Physics Letters A}, 382\penalty0 (38):\penalty0 2732, 2018.

\bibitem[Hudson et~al.(2002)Hudson, Monticello, Reiman, Boozer, Strickler,
  Hirshman, and Zarnstorff]{Hudson2002}
S.~R. Hudson, D.~Monticello, A.~Reiman, A.~Boozer, D.~Strickler, S.~Hirshman,
  and M.~Zarnstorff.
\newblock Eliminating islands in high-pressure free-boundary stellarator
  magnetohydrodynamic equilibrium solutions.
\newblock \emph{Physical Review Letters}, 89\penalty0 (27):\penalty0 275003,
  2002.

\bibitem[Hudson et~al.(2011)Hudson, Dewar, Hole, and McGann]{Hudson2011}
S.~R. Hudson, R.~Dewar, M.~Hole, and M.~McGann.
\newblock Non-axisymmetric, multi-region relaxed magnetohydrodynamic
  equilibrium solutions.
\newblock \emph{Plasma Physics and Controlled Fusion}, 54\penalty0
  (1):\penalty0 014005, 2011.

\bibitem[Imbert-Gerard et~al.(2019)Imbert-Gerard, Paul, and Wright]{Imbert2019}
L.-M. Imbert-Gerard, E.~Paul, and A.~Wright.
\newblock An introduction to symmetries in stellarators.
\newblock \emph{arXiv preprint arXiv:1908.05360}, 2019.

\bibitem[Isaev et~al.(2003)Isaev, N{\"u}hrenberg, Mikhailov, Cooper, Watanabe,
  Yokoyama, Yamazaki, Subbotin, and Shafranov]{Isaev2003}
M.~Y. Isaev, J.~N{\"u}hrenberg, M.~Mikhailov, W.~Cooper, K.~Watanabe,
  M.~Yokoyama, K.~Yamazaki, A.~Subbotin, and V.~Shafranov.
\newblock A new class of quasi-omnigenous configurations.
\newblock \emph{Nuclear Fusion}, 43\penalty0 (10):\penalty0 1066, 2003.

\bibitem[Jameson et~al.(1998)Jameson, Martinelli, and Pierce]{Jameson1998}
A.~Jameson, L.~Martinelli, and N.~Pierce.
\newblock Optimum aerodynamic design using the {Navier-Stokes} equations.
\newblock \emph{Theoretical and Computational Fluid Dynamics}, 10\penalty0
  (1-4):\penalty0 213, 1998.

\bibitem[Jia et~al.(2014)Jia, Liu, Zaitsev, Hennig, and Korvink]{Jia2014}
F.~Jia, Z.~Liu, M.~Zaitsev, J.~Hennig, and J.~G. Korvink.
\newblock {Design multiple-layer gradient coils using least-squares finite
  element method}.
\newblock \emph{Structural and Multidisciplinary Optimization}, 49\penalty0
  (3):\penalty0 523, 2014.

\bibitem[Johnson(2014)]{NLOPT}
S.~G. Johnson.
\newblock The {NLopt} nonlinear-optimization package, May 2014.
\newblock URL \url{http://ab-initio.mit.edu/nlopt}.

\bibitem[Kelley(1960)]{Kelley1960}
H.~J. Kelley.
\newblock Gradient theory of optimal flight paths.
\newblock \emph{American Rocket Society Journal}, 30\penalty0 (10):\penalty0
  947, 1960.

\bibitem[Kernbichler et~al.(2016)Kernbichler, Kasilov, Kapper, Martitsch,
  Nemov, Albert, and Heyn]{Kernbichler2016}
W.~Kernbichler, S.~Kasilov, G.~Kapper, A.~F. Martitsch, V.~Nemov, C.~Albert,
  and M.~Heyn.
\newblock Solution of drift kinetic equation in stellarators and tokamaks with
  broken symmetry using the code {NEO-2}.
\newblock \emph{Plasma Physics and Controlled Fusion}, 58\penalty0
  (10):\penalty0 104001, 2016.

\bibitem[Kierzenka and Shampine(2001)]{Kierzenka2001}
J.~Kierzenka and L.~F. Shampine.
\newblock A {BVP} solver based on residual control and the {Maltab PSE}.
\newblock \emph{ACM Transactions on Mathematical Software (TOMS)}, 27\penalty0
  (3):\penalty0 299--316, 2001.

\bibitem[Kisslinger et~al.(1999)Kisslinger, Beidler, Harmeyer, Herrnegger,
  Wobig, and Maurer]{Kisslinger1999}
J.~Kisslinger, C.~Beidler, E.~Harmeyer, F.~Herrnegger, H.~Wobig, and W.~Maurer.
\newblock Coil system of a {Helias} reactor.
\newblock Technical report, 1999.

\bibitem[Klinger et~al.(2013)Klinger, Baylard, Beidler, Boscary, Bosch,
  Dinklage, Hartmann, Helander, Ma{\ss}berg, Peacock, et~al.]{Klinger2013}
T.~Klinger, C.~Baylard, C.~Beidler, J.~Boscary, H.~Bosch, A.~Dinklage,
  D.~Hartmann, P.~Helander, H.~Ma{\ss}berg, A.~Peacock, et~al.
\newblock Towards assembly completion and preparation of experimental campaigns
  of {Wendelstein 7-X} in the perspective of a path to a stellarator fusion
  power plant.
\newblock \emph{Fusion Engineering and Design}, 88\penalty0 (6-8):\penalty0
  461, 2013.

\bibitem[Kress et~al.(1989)Kress, Maz'ya, and Kozlov]{Kress1989}
R.~Kress, V.~Maz'ya, and V.~Kozlov.
\newblock \emph{Linear Integral Equations}, volume~82.
\newblock Springer, 1989.

\bibitem[Krommes and Hu(1994)]{Krommes1994}
J.~A. Krommes and G.~Hu.
\newblock The role of dissipation in the theory and simulations of homogeneous
  plasma turbulence, and resolution of the entropy paradox.
\newblock \emph{Physics of Plasmas}, 1\penalty0 (10):\penalty0 3211, 1994.

\bibitem[Kruskal and Kulsrud(1958)]{Kruskal1958}
M.~D. Kruskal and R.~Kulsrud.
\newblock Equilibrium of a magnetically confined plasma in a toroid.
\newblock \emph{The Physics of Fluids}, 1\penalty0 (4):\penalty0 265, 1958.

\bibitem[Ku et~al.(2008)Ku, Garabedian, Lyon, Turnbull, Grossman, Mau,
  Zarnstorff, and Team]{Ku2008}
L.~Ku, P.~Garabedian, J.~Lyon, A.~Turnbull, A.~Grossman, T.~Mau, M.~Zarnstorff,
  and A.~Team.
\newblock Physics design for {ARIES-CS}.
\newblock \emph{Fusion Science and Technology}, 54\penalty0 (3):\penalty0 673,
  2008.

\bibitem[Ku and Boozer(2011)]{Ku2011}
L.~P. Ku and A.~H. Boozer.
\newblock {New classes of quasi-helically symmetric stellarators}.
\newblock \emph{Nuclear Fusion}, 51:\penalty0 013004, 2011.

\bibitem[Landreman(2017)]{Landreman2017}
M.~Landreman.
\newblock {An improved current potential method for fast computation of
  stellarator coil shapes}.
\newblock \emph{Nuclear Fusion}, 57\penalty0 (4):\penalty0 046003, 2017.

\bibitem[Landreman and Boozer(2016)]{Landreman2016}
M.~Landreman and A.~H. Boozer.
\newblock {Efficient magnetic fields for supporting toroidal plasmas}.
\newblock \emph{Physics of Plasmas}, 23\penalty0 (3):\penalty0 032506, 2016.

\bibitem[Landreman and Paul(2018)]{Landreman2018}
M.~Landreman and E.~J. Paul.
\newblock Computing local sensitivity and tolerances for stellarator physics
  properties using shape gradients.
\newblock \emph{Nuclear Fusion}, 58\penalty0 (7):\penalty0 076023, 2018.

\bibitem[Landreman and Sengupta(2018)]{Landreman2018b}
M.~Landreman and W.~Sengupta.
\newblock Direct construction of optimized stellarator shapes. {Part 1. Theory}
  in cylindrical coordinates.
\newblock \emph{Journal of Plasma Physics}, 84\penalty0 (6), 2018.

\bibitem[Landreman et~al.(2014)Landreman, Smith, Moll{\'{e}}n, and
  Helander]{Landreman2014}
M.~Landreman, H.~M. Smith, A.~Moll{\'{e}}n, and P.~Helander.
\newblock {Comparison of particle trajectories and collision operators for
  collisional transport in nonaxisymmetric plasmas}.
\newblock \emph{Physics of Plasmas}, 21\penalty0 (4), 2014.

\bibitem[Landreman et~al.(2015)Landreman, Plunk, and Dorland]{Landreman2015}
M.~Landreman, G.~G. Plunk, and W.~Dorland.
\newblock Generalized universal instability: transient linear amplification and
  subcritical turbulence.
\newblock \emph{Journal of Plasma Physics}, 81\penalty0 (5), 2015.

\bibitem[Landreman et~al.(2019)Landreman, Sengupta, and Plunk]{Landreman2019}
M.~Landreman, W.~Sengupta, and G.~G. Plunk.
\newblock Direct construction of optimized stellarator shapes. {Part 2.
  Numerical} quasisymmetric solutions.
\newblock \emph{Journal of Plasma Physics}, 85\penalty0 (1), 2019.

\bibitem[Lazerson(2012)]{Lazerson2012}
S.~Lazerson.
\newblock The virtual-casing principle for {3D} toroidal systems.
\newblock \emph{Plasma Physics and Controlled Fusion}, 54\penalty0
  (12):\penalty0 122002, 2012.

\bibitem[Lazerson et~al.(2016)Lazerson, Loizu, Hirshman, and
  Hudson]{Lazerson2016}
S.~A. Lazerson, J.~Loizu, S.~Hirshman, and S.~R. Hudson.
\newblock Verification of the ideal magnetohydrodynamic response at rational
  surfaces in the {VMEC} code.
\newblock \emph{Physics of Plasmas}, 23\penalty0 (1):\penalty0 012507, 2016.

\bibitem[Leal(2007)]{Leal2007}
L.~G. Leal.
\newblock \emph{Advanced Transport Phenomena: Fluid Mechanics and Convective
  Transport Processes}.
\newblock Cambridge University Press, 2007.

\bibitem[Leary et~al.(2004)Leary, Bhaskar, and Keane]{Leary2004}
S.~J. Leary, A.~Bhaskar, and A.~J. Keane.
\newblock A derivative based surrogate model for approximating and optimizing
  the output of an expensive computer simulation.
\newblock \emph{Journal of Global Optimization}, 30\penalty0 (1):\penalty0
  39--58, 2004.

\bibitem[Lee et~al.(1990)Lee, Harris, and Lee]{Lee1990}
D.~Lee, J.~Harris, and G.~Lee.
\newblock Magnetic island widths due to field perturbations in toroidal
  stellarators.
\newblock \emph{Nuclear Fusion}, 30\penalty0 (10):\penalty0 2177, 1990.

\bibitem[Liu et~al.(2016)Liu, Brennan, Bhattacharjee, and Boozer]{Liu2016}
C.~Liu, D.~P. Brennan, A.~Bhattacharjee, and A.~H. Boozer.
\newblock Adjoint {Fokker-Planck} equation and runaway electron dynamics.
\newblock \emph{Physics of Plasmas}, 23\penalty0 (1):\penalty0 010702, 2016.

\bibitem[Liu et~al.(2018)Liu, Shimizu, Isobe, Okamura, Nishimura, Suzuki, Xu,
  Zhang, Liu, Huang, et~al.]{Liu2018}
H.~Liu, A.~Shimizu, M.~Isobe, S.~Okamura, S.~Nishimura, C.~Suzuki, Y.~Xu,
  X.~Zhang, B.~Liu, J.~Huang, et~al.
\newblock Magnetic configuration and modular coil design for the {Chinese First
  Quasi-Axisymmetric Stellarator}.
\newblock \emph{Plasma and Fusion Research}, 13:\penalty0 3405067, 2018.

\bibitem[Lobsien et~al.(2018)Lobsien, Drevlak, Pedersen, et~al.]{Lobsien2018}
J.-F. Lobsien, M.~Drevlak, T.~S. Pedersen, et~al.
\newblock Stellarator coil optimization towards higher engineering tolerances.
\newblock \emph{Nuclear Fusion}, 58\penalty0 (10):\penalty0 106013, 2018.

\bibitem[Lobsien et~al.(2020)Lobsien, Drevlak, Kruger, Lazerson, Zhu, and
  Pedersen]{Lobsien2020}
J.-F. Lobsien, M.~Drevlak, T.~Kruger, S.~Lazerson, C.~Zhu, and T.~S. Pedersen.
\newblock Improved performance of stellarator coil design optimization.
\newblock \emph{Journal of Plasma Physics}, 86\penalty0 (2):\penalty0
  815860202, 2020.

\bibitem[Logan et~al.(2013)Logan, Park, Kim, Wang, and Berkery]{Logan2013}
N.~C. Logan, J.-K. Park, K.~Kim, Z.~Wang, and J.~W. Berkery.
\newblock Neoclassical toroidal viscosity in perturbed equilibria with general
  tokamak geometry.
\newblock \emph{Physics of Plasmas}, 20\penalty0 (12):\penalty0 122507, 2013.

\bibitem[Lortz(1975)]{Lortz1975}
D.~Lortz.
\newblock The general “peeling” instability.
\newblock \emph{Nuclear Fusion}, 15\penalty0 (1):\penalty0 49, 1975.

\bibitem[{M. Drevlak}(1998)]{Drevlak1998}
{M. Drevlak}.
\newblock {Automated optimization of stellarator coils}.
\newblock \emph{Fusion Technology}, 33:\penalty0 106, 1998.

\bibitem[Maassberg et~al.(1993)Maassberg, Lotz, and
  N{\"u}hrenberg]{Maassberg1993}
H.~Maassberg, W.~Lotz, and J.~N{\"u}hrenberg.
\newblock Neoclassical bootstrap current and transport in optimized stellarator
  configurations.
\newblock \emph{Physics of Fluids B: Plasma Physics}, 5\penalty0 (10):\penalty0
  3728, 1993.

\bibitem[McFadden(1979)]{Mcfadden1979}
G.~B. McFadden.
\newblock An artificial viscosity method for the design of supercritical
  airfoils.
\newblock 1979.

\bibitem[Mercier and Luc(1974)]{Mercier1974}
C.~Mercier and H.~Luc.
\newblock The {MHD} approach to the problem of plasma confinement in closed
  magnetic configurations.
\newblock \emph{Lectures in Plasma Physics, Commission of the European
  Communities, Luxembourg}, 1974.

\bibitem[Merkel(1987)]{Merkel1987}
P.~Merkel.
\newblock {Solution of stellarator boundary value problems with external
  currents}.
\newblock \emph{Nuclear Fusion}, 27\penalty0 (5):\penalty0 867, 1987.

\bibitem[Mikhailov et~al.(2012)Mikhailov, Drevlak, N{\"u}hrenberg, and
  Shafranov]{Mikhailov2012}
M.~Mikhailov, M.~Drevlak, J.~N{\"u}hrenberg, and V.~Shafranov.
\newblock Medium-$\beta$ free-boundary equilibria of a quasi-isodynamic
  stellarator.
\newblock \emph{Plasma Physics Reports}, 38\penalty0 (6):\penalty0 439, 2012.

\bibitem[Mikhailov et~al.(2019)Mikhailov, N{\"u}hrenberg, and
  Zille]{Mikhailov2019}
M.~Mikhailov, J.~N{\"u}hrenberg, and R.~Zille.
\newblock Elimination of current sheets at resonances in three-dimensional
  toroidal ideal-magnetohydrodynamic equilibria.
\newblock \emph{Nuclear Fusion}, 59\penalty0 (6):\penalty0 066002, 2019.

\bibitem[Miner~Jr et~al.(2001)Miner~Jr, Valanju, Hirshman, Brooks, and
  Pomphrey]{Miner2001}
W.~H. Miner~Jr, P.~M. Valanju, S.~P. Hirshman, A.~Brooks, and N.~Pomphrey.
\newblock Use of a genetic algorithm for compact stellarator coil design.
\newblock \emph{Nuclear Fusion}, 41\penalty0 (9):\penalty0 1185, 2001.

\bibitem[Mohammadi and Pironneau(2004)]{Mohammadi2004}
B.~Mohammadi and O.~Pironneau.
\newblock Shape optimization in fluid mechanics.
\newblock \emph{Annual Review of Fluid Mechanics}, 36:\penalty0 255, 2004.

\bibitem[Murakami et~al.(2002)Murakami, Wakasa, Maassberg, Beidler, Yamada,
  Watanabe, Group, et~al.]{Murakami2002}
S.~Murakami, A.~Wakasa, H.~Maassberg, C.~Beidler, H.~Yamada, K.~Watanabe, L.~E.
  Group, et~al.
\newblock Neoclassical transport optimization of {LHD}.
\newblock \emph{Nuclear Fusion}, 42\penalty0 (11):\penalty0 L19, 2002.

\bibitem[Murakami et~al.(2004)Murakami, Yamada, Sasao, Isobe, Ozaki, Saida,
  Goncharov, Lyon, Osakabe, Seki, et~al.]{Murakami2004}
S.~Murakami, H.~Yamada, M.~Sasao, M.~Isobe, T.~Ozaki, T.~Saida, P.~Goncharov,
  J.~Lyon, M.~Osakabe, T.~Seki, et~al.
\newblock Effect of neoclassical transport optimization on energetic ion
  confinement in {LHD}.
\newblock \emph{Fusion Science and Technology}, 46\penalty0 (2):\penalty0
  241--247, 2004.

\bibitem[Mynick(2006)]{Mynick2006}
H.~Mynick.
\newblock Transport optimization in stellarators.
\newblock \emph{Physics of Plasmas}, 13\penalty0 (5):\penalty0 058102, 2006.

\bibitem[Najmabadi et~al.(2008)Najmabadi, Raffray, Abdel-Khalik, Bromberg,
  Crosatti, El-Guebaly, Garabedian, Grossman, Henderson, Ibrahim,
  et~al.]{Najmabadi2008}
F.~Najmabadi, A.~Raffray, S.~Abdel-Khalik, L.~Bromberg, L.~Crosatti,
  L.~El-Guebaly, P.~Garabedian, A.~Grossman, D.~Henderson, A.~Ibrahim, et~al.
\newblock The {ARIES-CS} compact stellarator fusion power plant.
\newblock \emph{Fusion Science and Technology}, 54\penalty0 (3):\penalty0 655,
  2008.

\bibitem[Nelson et~al.(2003)Nelson, Berry, Brooks, Cole, Chrzanowski, Fan,
  Fogarty, Goranson, Heitzenroeder, Hirshman, et~al.]{Nelson2003}
B.~Nelson, L.~Berry, A.~Brooks, M.~Cole, J.~Chrzanowski, H.-M. Fan, P.~Fogarty,
  P.~Goranson, P.~Heitzenroeder, S.~Hirshman, et~al.
\newblock Design of the {National Compact Stellarator Experiment (NCSX)}.
\newblock \emph{Fusion Engineering and Design}, 66:\penalty0 169, 2003.

\bibitem[Nemov et~al.(1999)Nemov, Kasilov, Kernbichler, and Heyn]{Nemov1999}
V.~Nemov, S.~Kasilov, W.~Kernbichler, and M.~Heyn.
\newblock Evaluation of 1/$\nu$ neoclassical transport in stellarators.
\newblock \emph{Physics of Plasmas}, 6\penalty0 (12):\penalty0 4622, 1999.

\bibitem[Nemov et~al.(2005)Nemov, Kasilov, Kernbichler, and Leitold]{Nemov2005}
V.~Nemov, S.~Kasilov, W.~Kernbichler, and G.~Leitold.
\newblock The {$\nabla B$} drift velocity of trapped particles in stellarators.
\newblock \emph{Physics of Plasmas}, 12\penalty0 (11):\penalty0 112507, 2005.

\bibitem[Nocedal and Wright(2006)]{Nocedal2006}
J.~Nocedal and S.~J. Wright.
\newblock \emph{Numerical Optimization}.
\newblock Springer, 2006.

\bibitem[Novotny and Sokolowski(2013)]{Novotny2013}
A.~A. Novotny and J.~Sokolowski.
\newblock \emph{Topological Derivatives in Shape Optimization}.
\newblock Springer, 2013.

\bibitem[N{\"u}hrenberg(2020)]{Nuhrenberg2020}
C.~N{\"u}hrenberg.
\newblock Personal communication, 4 2020.

\bibitem[N{\"u}hrenberg and Boozer(2003)]{Nuhrenberg2003}
C.~N{\"u}hrenberg and A.~H. Boozer.
\newblock Magnetic islands and perturbed plasma equilibria.
\newblock \emph{Physics of Plasmas}, 10\penalty0 (7):\penalty0 2840, 2003.

\bibitem[N{\"u}hrenberg et~al.(2009)N{\"u}hrenberg, Boozer, and
  Hudson]{Nuhrenberg2009}
C.~N{\"u}hrenberg, A.~H. Boozer, and S.~R. Hudson.
\newblock Magnetic-surface quality in nonaxisymmetric plasma equilibria.
\newblock \emph{Physical Review Letters}, 102\penalty0 (23):\penalty0 235001,
  2009.

\bibitem[N{\"u}hrenberg and Zille(1988)]{Nuhrenberg1988}
J.~N{\"u}hrenberg and R.~Zille.
\newblock Quasi-helically symmetric toroidal stellarators.
\newblock \emph{Physics Letters A}, 129:\penalty0 113, 1988.

\bibitem[N{\"u}hrenberg et~al.(1994)N{\"u}hrenberg, Lotz, and
  Gori]{Nuhrenberg1994}
J.~N{\"u}hrenberg, W.~Lotz, and S.~Gori.
\newblock Theory of fusion plasmas.
\newblock In \emph{Proceedings of the Joint Varenna-Lausanne International
  Workshop}, page~3, 1994.

\bibitem[Onsager(1931{\natexlab{a}})]{Onsager1931a}
L.~Onsager.
\newblock Reciprocal relations in irreversible processes. {I}.
\newblock \emph{Physical review}, 37\penalty0 (4):\penalty0 405,
  1931{\natexlab{a}}.

\bibitem[Onsager(1931{\natexlab{b}})]{Onsager1931b}
L.~Onsager.
\newblock Reciprocal relations in irreversible processes. {II}.
\newblock \emph{Physical review}, 38\penalty0 (12):\penalty0 2265,
  1931{\natexlab{b}}.

\bibitem[Osher et~al.(2004)Osher, Fedkiw, and Piechor]{Osher2004}
S.~Osher, R.~Fedkiw, and K.~Piechor.
\newblock Level set methods and dynamic implicit surfaces.
\newblock \emph{Applied Mechanics Review}, 57\penalty0 (3):\penalty0 B15, 2004.

\bibitem[Othmer(2014)]{Othmer2014}
C.~Othmer.
\newblock Adjoint methods for car aerodynamics.
\newblock \emph{Journal of Mathematics in Industry}, 4\penalty0 (1):\penalty0
  6, 2014.

\bibitem[Park(2009)]{Park2009}
J.-K. Park.
\newblock \emph{Ideal Perturbed Equilibria in Tokamaks}.
\newblock PhD thesis, Princeton University, 2009.

\bibitem[Park et~al.(2007{\natexlab{a}})Park, Boozer, and Glasser]{Park2007}
J.-K. Park, A.~H. Boozer, and A.~H. Glasser.
\newblock Computation of three-dimensional tokamak and spherical torus
  equilibria.
\newblock \emph{Physics of Plasmas}, 14\penalty0 (5):\penalty0 052110,
  2007{\natexlab{a}}.

\bibitem[Park et~al.(2007{\natexlab{b}})Park, Schaffer, Menard, and
  Boozer]{Park2007b}
J.-K. Park, M.~J. Schaffer, J.~E. Menard, and A.~H. Boozer.
\newblock Control of asymmetric magnetic perturbations in tokamaks.
\newblock \emph{Physical Review Letters}, 99\penalty0 (19):\penalty0 195003,
  2007{\natexlab{b}}.

\bibitem[Paul et~al.(2017)Paul, Landreman, Poli, Spong, Smith, and
  Dorland]{Paul2017}
E.~J. Paul, M.~Landreman, F.~M. Poli, D.~A. Spong, H.~M. Smith, and W.~Dorland.
\newblock Rotation and neoclassical ripple transport in {ITER}.
\newblock \emph{Nuclear Fusion}, 57\penalty0 (11):\penalty0 116044, 2017.

\bibitem[Paul et~al.(2018)Paul, Landreman, Bader, and Dorland]{Paul2018}
E.~J. Paul, M.~Landreman, A.~Bader, and W.~Dorland.
\newblock An adjoint method for gradient-based optimization of stellarator coil
  shapes.
\newblock \emph{Nuclear Fusion}, 58\penalty0 (7):\penalty0 076015, 2018.

\bibitem[Paul et~al.(2019)Paul, Abel, Landreman, and Dorland]{Paul2019}
E.~J. Paul, I.~G. Abel, M.~Landreman, and W.~Dorland.
\newblock An adjoint method for neoclassical stellarator optimization.
\newblock \emph{Journal of Plasma Physics}, 85\penalty0 (5), 2019.

\bibitem[Paul et~al.(2020)Paul, Antonsen, Landreman, and Cooper]{Paul2020}
E.~J. Paul, T.~Antonsen, M.~Landreman, and W.~A. Cooper.
\newblock Adjoint approach to calculating shape gradients for three-dimensional
  magnetic confinement equilibria. {Part 2. Applications}.
\newblock \emph{Journal of Plasma Physics}, 86\penalty0 (1):\penalty0
  905860103, 2020.

\bibitem[Pedersen et~al.(2016)Pedersen, Otte, Lazerson, Helander, Bozhenkov,
  Biedermann, Klinger, Wolf, Bosch, Wendelstein, et~al.]{Pedersen2016}
T.~S. Pedersen, M.~Otte, S.~Lazerson, P.~Helander, S.~Bozhenkov, C.~Biedermann,
  T.~Klinger, R.~C. Wolf, H.-S. Bosch, T.~Wendelstein, et~al.
\newblock Confirmation of the topology of the {Wendelstein 7-X} magnetic field
  to better than 1: 100,000.
\newblock \emph{Nature Communications}, 7:\penalty0 13493, 2016.

\bibitem[Pierce and Giles(2004)]{Pierce2004}
N.~A. Pierce and M.~B. Giles.
\newblock {Adjoint and defect error bounding and correction for functional
  estimates}.
\newblock \emph{Journal of Computational Physics}, 200:\penalty0 769, 2004.

\bibitem[Pironneau(1974)]{Pironneau1974}
O.~Pironneau.
\newblock On optimum design in fluid mechanics.
\newblock \emph{Journal of Fluid Mechanics}, 64\penalty0 (1):\penalty0 97,
  1974.

\bibitem[Pironneau(1982)]{Pironneau1982}
O.~Pironneau.
\newblock \emph{Optimal Shape Design for Elliptic Systems}.
\newblock Springer, 1982.

\bibitem[Plessix(2006)]{Plessix2006}
R.~E. Plessix.
\newblock {A review of the adjoint-state method for computing the gradient of a
  functional with geophysical applications}.
\newblock \emph{Geophysical Journal International}, 167\penalty0 (2):\penalty0
  495, 2006.

\bibitem[Plunk et~al.(2019)Plunk, Landreman, and Helander]{Plunk2019}
G.~G. Plunk, M.~Landreman, and P.~Helander.
\newblock Direct construction of optimized stellarator shapes. {Part 3.
  Omnigenity} near the magnetic axis.
\newblock \emph{Journal of Plasma Physics}, 85\penalty0 (6), 2019.

\bibitem[Pomphrey et~al.(2001)Pomphrey, Berry, Boozer, Brooks, Hatcher,
  Hirshman, Ku, Miner, Mynick, Reiersen, Strickler, and Valanju]{Pomphrey2001}
N.~Pomphrey, L.~Berry, A.~Boozer, A.~Brooks, R.~Hatcher, S.~Hirshman, L.-P. Ku,
  W.~Miner, H.~Mynick, W.~Reiersen, D.~Strickler, and P.~Valanju.
\newblock {Innovations in compact stellarator coil design}.
\newblock \emph{Nuclear Fusion}, 41:\penalty0 339, 2001.

\bibitem[Press et~al.(2007)Press, Teukolsky, Vetterling, and
  Flannery]{Press2007}
W.~H. Press, S.~A. Teukolsky, W.~T. Vetterling, and B.~P. Flannery.
\newblock \emph{Numerical Recipes: The Art of Scientific Computing}.
\newblock Cambridge University Press, 2007.

\bibitem[Proll et~al.(2015)Proll, Mynick, Xanthopoulos, Lazerson, and
  Faber]{Proll2015}
J.~Proll, H.~Mynick, P.~Xanthopoulos, S.~Lazerson, and B.~Faber.
\newblock {TEM} turbulence optimisation in stellarators.
\newblock \emph{Plasma Physics and Controlled Fusion}, 58\penalty0
  (1):\penalty0 014006, 2015.

\bibitem[Reiman et~al.(1999)Reiman, Fu, Hirshman, Ku, Monticello, Mynick, Redi,
  Spong, Zarnstorff, Blackwell, et~al.]{Reiman1999}
A.~Reiman, G.~Fu, S.~Hirshman, L.~Ku, D.~Monticello, H.~Mynick, M.~Redi,
  D.~Spong, M.~Zarnstorff, B.~Blackwell, et~al.
\newblock Physics design of a high-quasi-axisymmetric stellarator.
\newblock \emph{Plasma Physics and Controlled Fusion}, 41\penalty0
  (12B):\penalty0 B273, 1999.

\bibitem[Rosenbluth et~al.(1972)Rosenbluth, Hazeltine, and
  Hinton]{Rosenbluth1972}
M.~Rosenbluth, R.~Hazeltine, and F.~L. Hinton.
\newblock Plasma transport in toroidal confinement systems.
\newblock \emph{The Physics of Fluids}, 15\penalty0 (1):\penalty0 116, 1972.

\bibitem[Rudin(2006)]{Rudin2006}
W.~Rudin.
\newblock \emph{Real and Complex Analysis}.
\newblock Tata McGraw-Hill Education, 2006.

\bibitem[Rust et~al.(2011)Rust, Heinemann, Mendelevitch, Peacock, and
  Smirnow]{Rust2011}
N.~Rust, B.~Heinemann, B.~Mendelevitch, A.~Peacock, and M.~Smirnow.
\newblock {W7-X neutral-beam-injection : Selection of the NBI source positions
  for experiment start-up}.
\newblock \emph{Fusion Engineering and Design}, 86\penalty0 (6-8):\penalty0
  728, 2011.

\bibitem[Sakakibara et~al.(2008)Sakakibara, Watanabe, Suzuki, Narushima,
  Ohdachi, Nakajima, Watanabe, Garcia, Weller, Toi, et~al.]{Sakakibara2008}
S.~Sakakibara, K.~Watanabe, Y.~Suzuki, Y.~Narushima, S.~Ohdachi, N.~Nakajima,
  F.~Watanabe, L.~Garcia, A.~Weller, K.~Toi, et~al.
\newblock {MHD} study of the reactor-relevant high-beta regime in the {Large
  Helical Device}.
\newblock \emph{Plasma Physics and Controlled Fusion}, 50\penalty0
  (12):\penalty0 124014, 2008.

\bibitem[Sanchez et~al.(2000)Sanchez, Hirshman, Ware, Berry, and
  Spong]{Sanchez2000}
R.~Sanchez, S.~Hirshman, A.~Ware, L.~Berry, and D.~Spong.
\newblock Ballooning stability optimization of low-aspect-ratio stellarators.
\newblock \emph{Plasma Physics and Controlled Fusion}, 42\penalty0
  (6):\penalty0 641, 2000.

\bibitem[Sauer(2012)]{Sauer2012}
T.~Sauer.
\newblock \emph{Numerical Analysis}.
\newblock Pearson, 2012.

\bibitem[Schwab(1993)]{Schwab1993}
C.~Schwab.
\newblock Ideal magnetohydrodynamics: Global mode analysis of three-dimensional
  plasma configurations.
\newblock \emph{Physics of Fluids B: Plasma Physics}, 5\penalty0 (9):\penalty0
  3195, 1993.

\bibitem[Shaing et~al.(1989)Shaing, Crume~Jr, Tolliver, Hirshman, and
  Van~Rij]{Shaing1989}
K.-C. Shaing, E.~Crume~Jr, J.~Tolliver, S.~Hirshman, and W.~Van~Rij.
\newblock Bootstrap current and parallel viscosity in the low collisionality
  regime in toroidal plasmas.
\newblock \emph{Physics of Fluids B: Plasma Physics}, 1\penalty0 (1):\penalty0
  148, 1989.

\bibitem[Shimizu et~al.(2018)Shimizu, Liu, Isobe, Okamura, Nishimura, Suzuki,
  Xu, Zhang, Liu, et~al.]{Shimizu2018}
A.~Shimizu, H.~Liu, M.~Isobe, S.~Okamura, S.~Nishimura, C.~Suzuki, Y.~Xu,
  X.~Zhang, J.~Liu, B.and~Huang, et~al.
\newblock Configuration property of the {Chinese First Quasi-Axisymmetric
  Stellarator}.
\newblock \emph{Plasma and Fusion Research}, 13:\penalty0 3403123, 2018.

\bibitem[Sinclair et~al.(1970)Sinclair, Hosea, and Sheffield]{Sinclair1970}
R.~Sinclair, J.~Hosea, and G.~Sheffield.
\newblock Magnetic surface mappings by storage of phase-stabilized low-energy
  electron beams.
\newblock \emph{Applied Physics Letters}, 17\penalty0 (2):\penalty0 92, 1970.

\bibitem[Smith and Cowley(2010)]{Smith2010}
C.~L. Smith and S.~Cowley.
\newblock The path to fusion power.
\newblock \emph{Philosophical Transactions of the Royal Society A:
  Mathematical, Physical and Engineering Sciences}, 368\penalty0
  (1914):\penalty0 1091, 2010.

\bibitem[Spitzer~Jr(1951)]{Spitzer1951}
L.~Spitzer~Jr.
\newblock A proposed stellarator.
\newblock Technical report, Princeton University, NJ Forrestal Research Center,
  1951.

\bibitem[Spitzer~Jr(1952)]{Spitzer1952}
L.~Spitzer~Jr.
\newblock Magnetic fields and particle orbits in a high-density stellarator.
\newblock Technical report, Princeton University, NJ Project Matterhorn, 1952.

\bibitem[Spitzer~Jr(1958)]{Spitzer1958}
L.~Spitzer~Jr.
\newblock The stellarator concept.
\newblock \emph{The Physics of Fluids}, 1\penalty0 (4):\penalty0 253, 1958.

\bibitem[Spong and Harris(2010)]{Spong2010}
D.~A. Spong and J.~H. Harris.
\newblock {New QP / QI Symmetric Stellarator Configurations}.
\newblock \emph{Plasma and Fusion Research}, 5:\penalty0 S2039, 2010.

\bibitem[Spong et~al.(1998)Spong, Hirshman, Whitson, Batchelor, Carreras,
  Lynch, and Rome]{Spong1998}
D.~A. Spong, S.~P. Hirshman, J.~C. Whitson, D.~B. Batchelor, B.~A. Carreras,
  V.~E. Lynch, and J.~A. Rome.
\newblock ${J}$* optimization of small aspect ratio stellarator/tokamak hybrid
  devices.
\newblock \emph{Physics of Plasmas}, 5\penalty0 (5):\penalty0 1752, 1998.

\bibitem[Spong et~al.(2001)Spong, Hirshman, Berry, Lyon, Fowler, Strickler,
  Cole, Nelson, Williamson, Ware, et~al.]{Spong2001}
D.~A. Spong, S.~P. Hirshman, L.~A. Berry, J.~F. Lyon, R.~H. Fowler, D.~J.
  Strickler, M.~J. Cole, B.~N. Nelson, D.~E. Williamson, A.~S. Ware, et~al.
\newblock Physics issues of compact drift optimized stellarators.
\newblock \emph{Nuclear Fusion}, 41\penalty0 (6):\penalty0 711, 2001.

\bibitem[Stix(1998)]{Stix1998}
T.~H. Stix.
\newblock Highlights in early stellarator research at princeton.
\newblock \emph{Journal of Plasma Fusion Research Series}, 1:\penalty0 3, 1998.

\bibitem[Strickler et~al.(2002)Strickler, Berry, and Hirshman]{Strickler2002}
D.~J. Strickler, L.~A. Berry, and S.~P. Hirshman.
\newblock {Designing Coils for Compact Stellarators}.
\newblock \emph{Fusion Science and Technology}, 41\penalty0 (2):\penalty0 107,
  2002.

\bibitem[Strickler et~al.(2003)Strickler, Berry, and Hirshman]{Strickler2003}
D.~J. Strickler, L.~A. Berry, and S.~P. Hirshman.
\newblock Integrated plasma and coil optimization for compact stellarators.
\newblock Technical report, 2003.

\bibitem[Strickler et~al.(2004)Strickler, Hirshman, Spong, Cole, Lyon, Nelson,
  Williamson, and Ware]{Strickler2004}
D.~J. Strickler, S.~P. Hirshman, D.~A. Spong, M.~J. Cole, J.~F. Lyon, B.~E.
  Nelson, D.~E. Williamson, and A.~S. Ware.
\newblock Development of a robust quasi-poloidal compact stellarator.
\newblock \emph{Fusion Science and Technology}, 45\penalty0 (1):\penalty0 15,
  2004.

\bibitem[Strumberger and G{\"u}nter(2016)]{Strumberger2016}
E.~Strumberger and S.~G{\"u}nter.
\newblock {CASTOR3D}: {Linear} stability studies for {2D} and {3D} tokamak
  equilibria.
\newblock \emph{Nuclear Fusion}, 57\penalty0 (1):\penalty0 016032, 2016.

\bibitem[Strykowsky et~al.(2009)Strykowsky, Brown, Chrzanowski, Cole,
  Heitzenroeder, Neilson, Rej, and Viol]{Strykowsky2009}
R.~Strykowsky, T.~Brown, J.~Chrzanowski, M.~Cole, P.~Heitzenroeder, G.~Neilson,
  D.~Rej, and M.~Viol.
\newblock Engineering cost \& schedule lessons learned on ncsx.
\newblock In \emph{2009 23rd IEEE/NPSS Symposium on Fusion Engineering}, pages
  1--4. IEEE, 2009.

\bibitem[Sugama et~al.(2009)Sugama, Watanabe, and Nunami]{Sugama2009}
H.~Sugama, T.-H. Watanabe, and M.~Nunami.
\newblock Linearized model collision operators for multiple ion species plasmas
  and gyrokinetic entropy balance equations.
\newblock \emph{Physics of Plasmas}, 16\penalty0 (11):\penalty0 112503, 2009.

\bibitem[Sun and Wang(2019)]{Sun2019}
G.~Sun and S.~Wang.
\newblock A review of the artificial neural network surrogate modeling in
  aerodynamic design.
\newblock \emph{Proceedings of the Institution of Mechanical Engineers, Part G:
  Journal of Aerospace Engineering}, 233\penalty0 (16):\penalty0 5863--5872,
  2019.

\bibitem[Sunn~Pedersen et~al.(2017)Sunn~Pedersen, Dinklage, Turkin, Wolf,
  Bozhenkov, Geiger, Fuchert, Bosch, Rahbarnia, Thomsen, et~al.]{Pedersen2017}
T.~Sunn~Pedersen, A.~Dinklage, Y.~Turkin, R.~Wolf, S.~Bozhenkov, J.~Geiger,
  G.~Fuchert, H.-S. Bosch, K.~Rahbarnia, H.~Thomsen, et~al.
\newblock Key results from the first plasma operation phase and outlook for
  future performance in {Wendelstein 7-X}.
\newblock \emph{Physics of Plasmas}, 24\penalty0 (5):\penalty0 055503, 2017.

\bibitem[Svanberg(2002)]{Svanberg2002}
K.~Svanberg.
\newblock A class of globally convergent optimization methods based on
  conservative convex separable approximations.
\newblock \emph{SIAM Journal on Optimization}, 12\penalty0 (2):\penalty0 555,
  2002.

\bibitem[Tikhonov(1963)]{Tikhonov1963}
A.~N. Tikhonov.
\newblock On the solution of ill-posed problems and the method of
  regularization.
\newblock In \emph{Doklady Akademii Nauk}, volume 151, pages 501--504. Russian
  Academy of Sciences, 1963.

\bibitem[Trefethen and Bau~III(1997)]{Trefethen1997}
L.~N. Trefethen and D.~Bau~III.
\newblock \emph{Numerical Linear Algebra}.
\newblock Society for Industrial and Applied Mathematics, 1997.

\bibitem[Tribaldos and Guasp(2005)]{Tribaldos2005}
V.~Tribaldos and J.~Guasp.
\newblock Neoclassical global flux simulations in stellarators.
\newblock \emph{Plasma physics and controlled fusion}, 47\penalty0
  (3):\penalty0 545, 2005.

\bibitem[Turner(1993)]{Turner1993}
R.~Turner.
\newblock {Gradient coil design : A review of methods}.
\newblock \emph{Magnetic Resonance Imaging}, 11:\penalty0 903, 1993.

\bibitem[Van~Bladel(2007)]{Van2007}
J.~G. Van~Bladel.
\newblock \emph{Electromagnetic Fields}, volume~19.
\newblock John Wiley \& Sons, 2007.

\bibitem[van Rij and Hirshman(1989)]{Rij1989}
W.~I. van Rij and S.~P. Hirshman.
\newblock {Variational bounds for transport coefficients in three-dimensional
  toroidal plasmas}.
\newblock \emph{Physics of Fluids B: Plasma Physics}, 1\penalty0 (3):\penalty0
  563, 1989.

\bibitem[Venditti and Darmofal(1999)]{Venditti1999}
D.~Venditti and D.~Darmofal.
\newblock A multilevel error estimation and grid adaptive strategy for
  improving the accuracy of integral outputs.
\newblock In \emph{14th Computational Fluid Dynamics Conference}, page 3292,
  1999.

\bibitem[Wagner(1998)]{Wagner1998}
F.~Wagner.
\newblock Stellarators and optimised stellarators.
\newblock \emph{Fusion Technology}, 33\penalty0 (2T):\penalty0 67, 1998.

\bibitem[Wagner et~al.(2005)Wagner, B{\"a}umel, Baldzuhn, Basse, Brakel,
  Burhenn, Dinklage, Dorst, Ehmler, Endler, et~al.]{Wagner2005}
F.~Wagner, S.~B{\"a}umel, J.~Baldzuhn, N.~Basse, R.~Brakel, R.~Burhenn,
  A.~Dinklage, D.~Dorst, H.~Ehmler, M.~Endler, et~al.
\newblock {W7-AS}: One step of the {Wendelstein} stellarator line.
\newblock \emph{Physics of Plasmas}, 12\penalty0 (7):\penalty0 072509, 2005.

\bibitem[Weller et~al.(2006)Weller, Sakakibara, Watanabe, Toi, Geiger,
  Zarnstorff, Hudson, Reiman, Werner, N{\"u}hrenberg, et~al.]{Weller2006}
A.~Weller, S.~Sakakibara, K.~Watanabe, K.~Toi, J.~Geiger, M.~Zarnstorff,
  S.~Hudson, A.~Reiman, A.~Werner, C.~N{\"u}hrenberg, et~al.
\newblock Significance of {MHD} effects in stellarator confinement.
\newblock \emph{Fusion Science and Technology}, 50\penalty0 (2):\penalty0 158,
  2006.

\bibitem[Wesson and Campbell(2011)]{Wesson2011}
J.~Wesson and D.~J. Campbell.
\newblock \emph{Tokamaks}, volume 149.
\newblock Oxford University Press, 2011.

\bibitem[Williamson et~al.(2005)Williamson, Brooks, Brown, Chrzanowski, Cole,
  Fan, Freudenberg, Fogarty, Hargrove, Heitzenroeder, Lovett, Miller, Myatt,
  Nelson, Reiersen, and Strickler]{Williamson2005}
D.~Williamson, A.~Brooks, T.~Brown, J.~Chrzanowski, M.~Cole, H.-M. Fan,
  K.~Freudenberg, P.~Fogarty, T.~Hargrove, P.~Heitzenroeder, G.~Lovett,
  P.~Miller, R.~Myatt, B.~Nelson, W.~Reiersen, and D.~Strickler.
\newblock {Modular coil design developments for the National Compact
  Stellarator Experiment (NCSX)}.
\newblock \emph{Fusion Engineering and Design}, 75-79:\penalty0 71, 2005.

\bibitem[Wolf et~al.(2019)Wolf, Alonso, {\"A}k{\"a}slompolo, Baldzuhn,
  Beurskens, Beidler, Biedermann, Bosch, Bozhenkov, Brakel, et~al.]{Wolf2019}
R.~Wolf, A.~Alonso, S.~{\"A}k{\"a}slompolo, J.~Baldzuhn, M.~Beurskens,
  C.~Beidler, C.~Biedermann, H.-S. Bosch, S.~Bozhenkov, R.~Brakel, et~al.
\newblock Performance of {Wendelstein 7-X} stellarator plasmas during the first
  divertor operation phase.
\newblock \emph{Physics of Plasmas}, 26\penalty0 (8):\penalty0 082504, 2019.

\bibitem[Wu et~al.(2017)Wu, Wang, and Kozlowski]{Wu2017}
X.~Wu, C.~Wang, and T.~Kozlowski.
\newblock Kriging-based surrogate models for uncertainty quantification and
  sensitivity analysis.
\newblock In \emph{Proceedings of the MC-2017, International Conference on
  Mathematics Computational Methods Applied to Nuclear Science Engineering},
  2017.

\bibitem[Xanthopoulos et~al.(2014)Xanthopoulos, Mynick, Helander, Turkin,
  Plunk, Jenko, G{\"o}rler, Told, Bird, and Proll]{Xanthopoulos2014}
P.~Xanthopoulos, H.~Mynick, P.~Helander, Y.~Turkin, G.~Plunk, F.~Jenko,
  T.~G{\"o}rler, D.~Told, T.~Bird, and J.~Proll.
\newblock Controlling turbulence in present and future stellarators.
\newblock \emph{Physical Review Letters}, 113\penalty0 (15):\penalty0 155001,
  2014.

\bibitem[Yamazaki et~al.(1993)Yamazaki, Yanagi, Ji, Kaneko, Ohyabu, Satow,
  Morimoto, Yamamoto, Motojima, and {the LHD Design Group}]{Yamazaki}
K.~Yamazaki, N.~Yanagi, H.~Ji, H.~Kaneko, N.~Ohyabu, T.~Satow, S.~Morimoto,
  J.~Yamamoto, O.~Motojima, and {the LHD Design Group}.
\newblock Requirements for accuracy of superconducting coils in the {Large
  Helical Device}.
\newblock \emph{Fusion Engineering and Design}, 20:\penalty0 79--86, 1993.

\bibitem[Yoshikawa and Stix(1985)]{Yoshikawa1985}
S.~Yoshikawa and T.~Stix.
\newblock Experiments on the {Model C} stellarator.
\newblock \emph{Nuclear Fusion}, 25\penalty0 (9):\penalty0 1275, 1985.

\bibitem[Zarnstorff et~al.(2001)Zarnstorff, Berry, Brooks, Fredrickson, Fu,
  Hirshman, Hudson, Ku, Lazarus, Mikkelsen, et~al.]{Zarnstorff2001}
M.~Zarnstorff, L.~Berry, A.~Brooks, E.~Fredrickson, G.~Fu, S.~Hirshman,
  S.~Hudson, L.~Ku, E.~Lazarus, D.~Mikkelsen, et~al.
\newblock Physics of the compact advanced stellarator {NCSX}.
\newblock \emph{Plasma Physics and Controlled Fusion}, 43\penalty0
  (12A):\penalty0 A237, 2001.

\bibitem[Zhu et~al.(2018{\natexlab{a}})Zhu, Hudson, Song, and Wan]{Zhu2018}
C.~Zhu, S.~R. Hudson, Y.~Song, and Y.~Wan.
\newblock {New method to design stellarator coils without the winding surface}.
\newblock \emph{Nuclear Fusion}, 58:\penalty0 016008, 2018{\natexlab{a}}.

\bibitem[Zhu et~al.(2018{\natexlab{b}})Zhu, Hudson, Song, and Wan]{Zhu2018b}
C.~Zhu, S.~R. Hudson, Y.~Song, and Y.~Wan.
\newblock Designing stellarator coils by a modified {Newton method using
  FOCUS}.
\newblock \emph{Plasma Physics and Controlled Fusion}, 60\penalty0
  (6):\penalty0 065008, 2018{\natexlab{b}}.

\bibitem[Zhu et~al.(2019{\natexlab{a}})Zhu, Gates, Hudson, Liu, Xu, Shimizu,
  and Okamura]{Zhu2019}
C.~Zhu, D.~A. Gates, S.~R. Hudson, H.~Liu, Y.~Xu, A.~Shimizu, and S.~Okamura.
\newblock Identification of important error fields in stellarators using the
  {Hessian} matrix method.
\newblock \emph{Nuclear Fusion}, 59\penalty0 (12):\penalty0 126007,
  2019{\natexlab{a}}.

\bibitem[Zhu et~al.(2019{\natexlab{b}})Zhu, Zarnstorff, Gates, and
  Brooks]{Zhu2019b}
C.~Zhu, M.~Zarnstorff, D.~Gates, and A.~Brooks.
\newblock Designing stellarators using perpendicular permanent magnets.
\newblock \emph{arXiv preprint arXiv:1912.05144}, 2019{\natexlab{b}}.

\end{thebibliography}

\end{document}